\documentclass[11pt,nocut]{report}

\usepackage{latex_style/packages}
\usepackage{latex_style/notations}
\usepackage{psl-cover}
\usepackage{indentfirst}
\usepackage[export]{adjustbox}
\usepackage{bbm}
\usepackage{multirow}

\newcommand{\beq}{\begin{equation}}
\newcommand{\eeq}{\end{equation}}
\newcommand{\sgn}{{\rm sgn}}
\newcommand{\fint}{{\dashint}}

\newcommand{\nn}{\nonumber}

\newlength{\bibitemsep}
\newlength{\bibparskip}
\let\oldthebibliography\thebibliography
\renewcommand\thebibliography[1]{%
  \oldthebibliography{#1}%
  \setlength{\parskip}{\bibitemsep}%
  \setlength{\itemsep}{\bibparskip}%
}

\title{Exact results for active particle models:\newline
from long-range interactions to first-passage properties
}
\author{Léo Touzo}
\date{}

\glsxtrnewsymbol[description={Active Brownian particle}]{ABP}{ABP :}
\glsxtrnewsymbol[description={Active Ornstein-Uhlenbeck particle}]{AOUP}{AOUP :}
\glsxtrnewsymbol[description={Calogero-Moser}]{CaloMo}{CM :}
\glsxtrnewsymbol[description={Dyson Brownian Motion}]{DBM}{DBM :}
\glsxtrnewsymbol[description={Dean-Kawasaki equation}]{DKeq}{DK equation :}
\glsxtrnewsymbol[description={Direction reversing active Brownian particle}]{DRABP}{DRABP :}
\glsxtrnewsymbol[description={Gaussian Orthogonal Ensemble}]{GGOE}{GOE :}
\glsxtrnewsymbol[description={Gaussian Symplectic Ensemble}]{GGSE}{GSE :}
\glsxtrnewsymbol[description={Gaussian Unitary Ensemble}]{GGUE}{GUE :}
\glsxtrnewsymbol[description={Left hand side}]{LHS}{l.h.s. :}
\glsxtrnewsymbol[description={Mean first-passage time}]{MFPT}{MFPT :}
\glsxtrnewsymbol[description={Macroscopic fluctuation theory}]{MFT}{MFT :}
\glsxtrnewsymbol[description={Motility-induced phase separation}]{MIPS}{MIPS :}
\glsxtrnewsymbol[description={Mean-squared displacement}]{MSD}{MSD :}
\glsxtrnewsymbol[description={Partial differential equation}]{PDE}{PDE :}
\glsxtrnewsymbol[description={Probability Density Function}]{ProbDF}{PDF :}
\glsxtrnewsymbol[description={Right hand side}]{RHS}{r.h.s. :}
\glsxtrnewsymbol[description={Random Matrix Theory}]{rumtzq}{RMT :}
\glsxtrnewsymbol[description={Run-and-tumble particle}]{RTP}{RTP :}
\glsxtrnewsymbol[description={Stochastic differential equation}]{SDE}{SDE :}
\glsxtrnewsymbol[description={Supplementary material}]{SupMat}{SM :}
\glsxtrnewsymbol[description={Average over realizations of the noise}]{AVG}{\ensuremath{\langle \dots\rangle} :}
\glsxtrnewsymbol[description={Time derivative of \ensuremath{x(t)}}]{timeDer}{\ensuremath{\dot x(t)} :}
\glsxtrnewsymbol[description={Kronecker delta: \ensuremath{\delta_{a,b}=1} if \ensuremath{a=b} and \ensuremath{0} otherwise}]{kron}{\ensuremath{\delta_{a,b}} :}
\glsxtrnewsymbol[description={Indicator function: \ensuremath{\mathbbm{1}_{A}=1} if the condition \ensuremath{A} is satisfied and \ensuremath{0} otherwise}]{indic}{\ensuremath{\mathbbm{1}_{A}} :}
\glsxtrnewsymbol[description={Dirac Delta function}]{dirac}{\ensuremath{\delta(x)} :}
\glsxtrnewsymbol[description={Heaviside Theta function}]{ruzzzt}{\ensuremath{\Theta(x)} :}
\glsxtrnewsymbol[description={Sign of \ensuremath{x}}]{sgn}{\ensuremath{{\rm sgn}(x)} :}
\glsxtrnewsymbol[description={Error function: \ensuremath{{\rm erf}(x)=\frac{2}{\sqrt{\pi}}\int_0^x \rmd t \, e^{-t^2}}}]{laerrmbw}{\ensuremath{{\rm erf}(x)} :}
\glsxtrnewsymbol[description={Gamma function}]{ruzzt}{\ensuremath{\Gamma(x)} :}
\glsxtrnewsymbol[description={Lambert \ensuremath{W} function}]{lambw}{\ensuremath{W(x)} :}
\glsxtrnewsymbol[description={Cosine integral: \ensuremath{{\rm Ci}(x)=-\int_x^\infty dt \frac{\cos t}{t}}}]{cosint}{\ensuremath{{\rm Ci}(x)} :}
\glsxtrnewsymbol[description={Sine integral: \ensuremath{{\rm Si}(x)=\int_0^x dt \frac{\sin t}{t}}}]{sinint}{\ensuremath{{\rm Si}(x)} :}
\glsxtrnewsymbol[description={Airy function}]{laiai}{\ensuremath{\Ai(x)} :}
\glsxtrnewsymbol[description={\ensuremath{i^{th}} zero of the Airy function}]{laiaizeros}{\ensuremath{a_i} :}
\glsxtrnewsymbol[description={Polylogarithm of order \ensuremath{s}}]{lilisi}{\ensuremath{\Li_s(x)} :}
\glsxtrnewsymbol[description={Riemann zeta function}]{lildsisi}{\ensuremath{\zeta(x)} :}
\glsxtrnewsymbol[description={\ensuremath{n^{th}} Hermite polynomial: \ensuremath{H_n(x)=(-1)^n e^{x^2} \frac{d^n}{dx^n} e^{-x^2}}}]{Hermitepol}{\ensuremath{H_n(x)} :}
\glsxtrnewsymbol[description={Hypergeometric function}]{Hypergeo}{\ensuremath{_2F_2(a,b;c;z)} :}
\glsxtrnewsymbol[description={Cauchy principal value}]{lamezaax}{\ensuremath{\dashint} :}

\institute{l'\'Ecole Normale Supérieure}
\doctoralschool{Physique en \^Ile-de-France}{564}
\specialty{Physique Théorique}
\date{12 juin 2025}

\jurymember{1}{Mathieu LEWIN}{Université Paris Dauphine}{Président du jury}
\jurymember{2}{Malte HENKEL}{Université de Lorraine}{Rapporteur}
\jurymember{3}{Clément SIRE}{Université Paul Sabatier}{Rapporteur}
\jurymember{4}{Joachim KRUG}{Université de Cologne}{Examinateur}
\jurymember{5}{Vivien LECOMTE}{Université Grenoble-Alpes}{Examinateur}
\jurymember{6}{Valentina ROS}{Université Paris-Saclay}{Examinatrice}
\jurymember{7}{Pierre LE DOUSSAL}{\'Ecole Normale Supérieure}{Directeur de thèse}
\jurymember{8}{Grégory SCHEHR}{Sorbonne Université}{Co-directeur de thèse}
\jurymember{9}{Leo RADZIHOVSKY}{Université du Colorado}{Invité}

\frabstract{
Les particules actives sont des objets ou des organismes vivants qui possèdent une certaine forme d'autopropulsion. En raison de cet apport d'énergie, ces systèmes sont intrinsèquement hors d'équilibre, ce qui leur confère de nombreuses propriétés intéressantes. C'est encore plus vrai lorsqu'un grand nombre de ces entités interagissent ensemble, ce qui donne lieu à des transitions de phase d'un type entièrement nouveau. Cependant, le fait que le bruit agissant sur ces systèmes soit corrélé en temps rend difficile les études analytiques exactes. Le but de cette thèse est d'obtenir de nouveaux résultats exacts pour des modèles de particules actives en une dimension, en se concentrant sur deux aspects différents : leur comportement en présence d'interactions à longue portée et leurs propriétés de premier passage.
\\

Cette thèse est divisée en quatre parties. Dans la première partie, nous donnons un aperçu des résultats exacts existants à la fois pour les modèles de particules actives et pour les particules browniennes avec des interactions à longue portée (gaz de Riesz). Les deux parties suivantes se concentrent sur la façon dont les méthodes de ces deux domaines peuvent être combinées et étendues pour obtenir de nouveaux résultats pour les modèles de particules actives avec des interactions à longue portée. Dans la deuxième partie, nous étudions la densité de particules dans l'état stationnaire, dans la limite d'un grand nombre de particules, en utilisant une extension de l'équation de Dean-Kawasaki aux particules run-and-tumble (RTPs). Dans le cas de l'interaction de Coulomb 1D (attractive ou répulsive), nous obtenons des expressions exactes pour la densité stationnaire pour différents types de potentiels de confinement, ce qui met en lumière de nouvelles transitions de phase hors-équilibre. Certains résultats sont également obtenus pour une interaction de Coulomb 2D répulsive (log-gaz), bien que la contrainte de non-croisement des trajectoires rende l'étude plus difficile dans ce cas. Dans la troisième partie, nous nous concentrons sur les fluctuations de position d'un traceur. Dans la limite d'un bruit faible, nous calculons exactement et analysons dans différents régimes un certain nombre de fonctions de corrélation des positions des particules et des distances interparticulaires, à la fois pour le gaz de Riesz brownien et pour son équivalent actif, et nous montrons que l'activité joue un rôle important à la fois à temps court et aux petites distances. La dernière partie de cette thèse se concentre sur la dualité de Siegmund, qui relie les propriétés de premier passage d'un processus stochastique avec des frontières absorbantes à sa distribution spatiale avec des murs durs. Nous étendons cette dualité à une nouvelle classe de processus stochastiques, qui inclut les particules actives et les modèles de diffusivité diffusive.

}

\enabstract{
Active particles are objects or living organisms which possess some form of self-propulsion. Due to this input of energy, such systems are intrinsically out-of-equilibrium, which gives them many interesting properties. This is even more true when a large number of these entities interact together, leading to entirely new types of phase transitions. However, the fact that the noise driving these systems is time-correlated makes exact analytical studies difficult. The goal of this thesis is to obtain new exact results for models of active particles in one dimension, focusing on two different aspects: their behavior in the presence of long-range interactions and their first-passage properties.
\\

This thesis is divided into four parts. In the first part we give an overview of existing exact results both for active particle models and for Brownian particles with long-range interactions (Riesz gases). The next two parts focus on how methods from these two fields can be combined and extended to derive new results for models of active particles with long-range interactions. In part two, we study the density of particles in the stationary state, in the limit where the number of particles is very large, using an extension of the Dean-Kawasaki equation to run-and-tumble particles (RTPs). In the case of the 1D Coulomb interaction (attractive or repulsive), we obtain exact expressions for the stationary density for different types of confining potentials, which sheds lights on new non-equilibrium phase-transitions. Some results are also obtained for a repulsive 2D Coulomb interaction (log-gas), although the single-file constraint makes the study more difficult in this case. In part three, we focus on the fluctuations at the tagged particle level. In the limit of weak noise, we compute exactly and analyze in different regimes a variety of correlation functions of the particle positions and interparticle distances, both for the Brownian Riesz gas and for its active counterpart, and show that the activity plays an important role both at short times and at small distances. The last part of this thesis focuses on Siegmund duality, which connects the first-passage properties of a stochastic process with absorbing boundaries to its spatial distribution with hard walls. We extend this duality to a new class of stochastic processes, which includes active particles and diffusing diffusivity models.
}

\frkeywords{Physique statistique hors-équilibre, particules actives, mouvement persistant, systèmes de particules en interaction, interactions longue-portée, gaz de Riesz, propriétés de premier passage.}
\enkeywords{Non-equilibrium statistical physics, active particles, persistent motion, interacting particle systems, long-range interactions, Riesz gas, first-passage properties.}

\begin{document}

\maketitle{}
\pagenumbering{roman}

{%
	\hypersetup{linkcolor=black}
	\setcounter{tocdepth}{1}
	\tableofcontents
}

\setcounter{tocdepth}{3}
\newpage
\chapter*{Remerciements}
\addcontentsline{toc}{chapter}{Remerciements}%

Je voudrais commencer par remercier mes directeurs de thèse, Pierre Le Doussal et Grégory Schehr, pour tout le temps passé ensemble pendant ces trois années. J'ai énormément appris à leurs côtés, que ce soit sur le plan de la physique, des méthodes de calcul ou de la rédaction d'articles scientifiques (même si, avec du recul, il faut bien reconnaître que certains papiers auraient mérité d'être plus succincts). Leur disponibilité, leur bienveillance et leur passion pour la physique sont en grande partie ce qui ont fait de ces trois années de thèse une expérience aussi enrichissante et agréable.

Je tiens ensuite à remercier les rapporteurs, Malte Henkel et Clément Sire, pour le soin qu'ils ont apporté à la relecture de ce manuscrit et pour leurs nombreux commentaires utiles. Je remercie également Joachim Krug, Vivien Lecomte, Mathieu Lewin, Valentina Ros et Leo Radzihovsky pour avoir accepté de faire partie du jury.

Je voudrais ensuite remercier tout le personnel scientifique et administratif du LPENS, en particulier Benjamin Basso, qui a fait partie de mon comité de suivi, Guillaume Barraquand, Jesper Jacobsen, Manel Kriouane, Christine Chambon, Laura Baron-Ledez, Jean-François Allemand et Frédéric Chevy.
Durant ces trois années j'ai aussi passé beaucoup de temps au LPTHE, et je tiens à remercier tous ses membres pour leur accueil chaleureux. 
J'en profite pour remercier également les voisins du LPTMC avec qui j'ai pu intéragir à de nombreuses occasions et qui ont toujours été très accueillants. Je remercie également les membres du LPTMS pour leur accueil lors de mes nombreuses visites, et en particulier Claudine Le Vaou et Alberto Rosso. Je remercie aussi Satya Majumdar pour les nombreux échanges que nous avons eu au cours de ces trois années. I also thank my collaborators from ICTS Bengalore, in particular Saikat Santra and Abhishek Dhar.

J'ai également beaucoup apprécié enseigner pendant ma thèse, et je tiens à remercier Guillaume Legendre, Kris Van Houcke et Lydéric Bocquet, ainsi que Nehal Mittal, Elric Frigerio, Auriane Huyghues-Despointes et Noé Suchel, avec qui j'ai partagé cette expérience. J'ai aussi eu l'occasion de participer durant cette thèse à plusieurs écoles d'été passionantes, à Beg Rohu, à Cargèse et aux Houches, dont je tiens à remercier tous les organisateurs. Je souhaite également remercier Giulio Biroli et Frédéric Van Wijland qui m'ont donné de précieux conseils lors de ma recherche de thèse.

Je remercie tous les doctorants et post-doctorants du LPENS et du LPTHE, ainsi que ceux du LPTMS et du LPTMC que j'ai eu le plaisir de côtoyer, pour la bonne ambiance et pour les très nombreuses discussions, en lien ou non avec la physique, qui ont rempli ces années de thèse. Je dois bien sûr un remerciement particulier à Mathis Guéneau, 
pour ces interactions quotidiennes qui ont rendu la thèse plus agréable, ainsi que pour toutes les discussions intéressantes que nous avons eu au cours de ces trois années, qui m'ont beaucoup appris. Explorer ensemble l'univers merveilleux de la dualité de Siegmund a été une expérience particulièrement enrichissante, et ce projet n'aurait certainement pas été aussi loin sans son enthousiasme et son travail méticuleux. Je voudrais aussi remercier particulièrement mes autres camarades de bureau, Andreani, Diyar, Gabriele, Giuseppe, Ludwig, Marco, Misha, Paul et Yann, pour l'atmosphère de travail conviviale, ainsi que mes prédécesseurs, Alexandre, Ana, Benjamin, Bertrand, Clément, Francesco et Gabriel, pour leurs précieux conseils à différents moments de la thèse.

Je tiens aussi à remercier l'ensemble de mes amis hors du labo, en particulier Akim, Antoine, Martin, Mathieu, Valentina et Victor pour avoir rendu ces années plus agréables. Je remercie tout particulièrement l'Assiette Creuse, maison de retraite pour coccinelles, pour tous les bons moments passés ensemble et pour leur soutien pendant ces trois années.

Je voudrais enfin remercier l'ensemble de ma famille, et en particulier mes parents, mes grands-parents et ma soeur, pour leur soutien continu tout au long de mes études et dans la vie en général.

\newcounter{firstbib}

\renewcommand\bibname{Publications related to this Thesis}

\printunsrtglossary[type=symbols,title={Index of notations and abbreviations}]
\newpage
\chapter*{\vspace{-0.4cm}Résumé en français}
\addcontentsline{toc}{chapter}{Résumé en français}%


\vspace*{0.85cm}

Depuis sa création il y a plus d'un siècle, la physique statistique s'est révélée être un outil extrêmement puissant pour décrire les systèmes constitués d'un grand nombre d'entités en interaction à l'équilibre. L'une de ses plus grandes réalisations a été de permettre une compréhension précise des effets collectifs qui émergent dans ces systèmes, connus sous le nom de transitions de phase. Cependant, plus récemment, un fort intérêt s'est développé pour l'étude de systèmes {\it hors équilibre}, en particulier venant de la biologie. Dans ces systèmes, l'apport constant d'énergie rompt les hypothèses de la physique statistique à l'équilibre, ce qui rend les études analytiques extrêmement difficiles. En particulier, ces systèmes peuvent atteindre des états stationnaires qui ne sont pas décrits par la mesure de Gibbs, ce qui nécessite le développement d'outils entièrement nouveaux pour étudier leur comportement à long terme. En même temps, la capacité de ces systèmes à échapper aux lois habituelles de la physique statistique ouvre la voie à toute une série de phénomènes fascinants, tels que de nouveaux types de transitions de phase qui seraient totalement impossibles à l'équilibre.

Un exemple paradigmatique qui a attiré beaucoup d'attention au cours des trois dernières décennies est la matière active, et plus précisément l'étude des {\it particules actives} \cite{Ramaswamy2010,Bechinger,Marchetti2018}. Il s'agit de particules qui possèdent une certaine forme d'autopropulsion, conduisant à un {\it mouvement persistant}, ce qui signifie que ces particules ont une tendance plus forte à suivre des lignes droites par rapport à une particule brownienne standard. Bien que les modèles originaux aient été introduits pour étudier le comportement collectif des animaux, tels que les volées d'oiseaux ou les bancs de poissons \cite{Vicsek,CavagnaGiardina,Calovi2018}, ou pour modéliser les déplacements d'organismes vivants plus petits tels que les bactéries \cite{Berg2004}, des particules actives non organiques ont également été réalisées expérimentalement par une variété d'approches différentes, en utilisant par exemple des réactions chimiques, des instabilités électrodynamiques, des disques vibrés ou même de petits robots \cite{ErbeExperiments,JanusReview,Quincke,Chate2010,Schranz}. D'un point de vue plus mathématique, le mouvement des particules actives est modélisé par l'introduction d'un bruit non brownien qui est {\it corrélé dans le temps} (et en général non gaussien). De nombreux modèles différents ont été introduits au fil des ans, l'un d'entre eux étant la particule {\it run-and-tumble} (RTP). Ce modèle s'inspire du mouvement de bactéries telles que {\it E. Coli}, qui consiste en une série de déplacements en ligne droite, séparés par des {\it tumbles} au cours desquelles la particule se réoriente dans une direction aléatoire \cite{Berg2004}. Pour tous ces modèles, déjà au niveau d'une seule particule, la complexité du bruit permet l'apparition de nouveaux phénomènes intéressants, comme la tendance des particules actives à s'accumuler près des bords \cite{Yang2014,Uspal2015}, mais elle rend aussi les calculs analytiques particulièrement difficiles. Ces dernières années, une littérature importante s'est développée autour de ces modèles, concernant en particulier leurs états d'équilibre non-Boltzmann en présence de confinement \cite{TailleurCates2009,DKM19,LMS2020,SzamelAOUP,Bonilla2019,Wijland21, ABM2,AngelaniHardWalls}, ainsi que leurs propriétés de premier passage \cite{Targetsearch,MalakarRTP,Singh2020,Singh2022,MFPT1DABP,RTPsurvivalMori,TVB12,RBV16,SurvivalRTPDriftDeBruyne,DKM19,AngelaniMFPT,MathisMFPT,MathisMFPT2,Grange2025}. Malgré cela, de nombreuses questions restent ouvertes, même au niveau d'une seule particule.

Tout cela est encore plus vrai en présence d'interactions. En effet, il a été démontré que les systèmes de particules actives en interaction présentent une variété de nouveaux effets collectifs, tels que la transition vers un mouvement collectif en présence d'interactions d'alignement ({\it flocking}) \cite{Vicsek,TonerTu95,TonerTu98,TonerTuReview}, ou la bien connue ``séparation de phase induite par la motilité" (MIPS) en présence d'interactions répulsives à courte portée \cite{CT2015,OByrne2021}. L'élaboration d'une description analytique précise de ces phénomènes représente toutefois un défi considérable. La plupart des études théoriques concernant les systèmes de particules actives en interaction reposent soit sur des théories des champs générales dérivées d'arguments de symétrie, soit sur des équations hydrodynamiques {\it coarse-grained} nécessitant généralement un certain degré d'approximation. Jusqu'à présent, très peu de résultats exacts ont été obtenus au-delà de deux particules \cite{slowman,slowman2,Mallmin2019,KunduGap2020,Hahn2023,Metson2022,MetsonLong,LMS2021,Hahn2025}, à l'exception notable de certains modèles sur réseau avec des interactions de contact, pour lesquels des équations hydrodynamiques exactes ont été dérivées \cite{KH2018,Erignoux_derivation,Agranov2021,Agranov2022,lattice2lanes2025}, ainsi que des chaînes harmoniques de particules actives, pour lesquelles certaines fonctions de corrélation spatio-temporelle des positions des particules ont été calculées exactement \cite{PutBerxVanderzande2019,SinghChain2020,HarmonicChainRevABP,HarmonicChainRTPDhar}. 

Trouver d'autres modèles pour lesquels des résultats analytiques exacts peuvent être obtenus ferait grandement progresser notre compréhension des systèmes de particules actives en interaction. Un type de système pour lequel cela pourrait être possible, et qui jusqu'à présent a reçu très peu d'attention, est le cas des particules actives avec des interactions à longue portée (c'est-à-dire avec un potentiel d'interaction de paire qui se comporte comme une loi de puissance). Bien que ce type d'interaction ne soit généralement pas répandu dans les systèmes de particules actives réels, il peut être réalisé expérimentalement, par exemple en plaçant des colloïdes paramagnétiques soumis à un champ magnétique (générant ainsi une interaction dipôle-dipôle avec un potentiel $\sim 1/|x|^3$, $|x|$ étant la distance entre les particules) en contact avec un bain de particules actives \cite{ActiveMeltingNature2024}. On peut également penser aux interactions hydrodynamiques (c'est-à-dire médiées par le fluide environnant), bien qu'elles soient généralement plus complexes et dépendent de la vitesse et de l'orientation des particules \cite{ishikawa2007,zottl2023,Filella2018}. La principale motivation pour étudier ces systèmes est cependant théorique. En effet, les modèles de particules browniennes interagissant via des potentiels (répulsifs) en loi de puissance $\sim |x|^{-s}$, connus sous le nom de {\it gaz de Riesz} (ou gaz de Coulomb si $s=d-2$ en $d$ dimensions), ont été largement étudiés à la fois dans la littérature physique et mathématique \cite{Riesz,Lewin,SerfatyBook}, en particulier en une dimension, en partie en raison de leur connexion avec la théorie des matrices aléatoires (RMT) \cite{Mehta_book,Forrester_book,bouchaud_book}. Le modèle le plus connu de cette famille est le mouvement brownien de Dyson (DBM), ou {\it log-gas}, correspondant à une interaction coulombienne 2D $s=0$ (c'est-à-dire à un potentiel d'interaction logarithmique), qui a des liens étroits avec les ensembles de matrices gaussiennes \cite{Dyson}. D'autres cas particuliers incluent la {\it rank diffusion}, ou {\it jellium model}, $s=-1$ (voir par exemple \cite{PLDRankedDiffusion}), qui correspond à une interaction de Coulomb 1D, et le modèle de Calogero-Moser (CM) $s=2$, principalement étudié dans le contexte de la dynamique hamiltonienne, mais pour lequel la dynamique de Langevin sur-amortie a récemment été étudiée numériquement \cite{Agarwal2019}. Il est donc tentant de chercher dans le large éventail d'outils analytiques développés pour étudier ces modèles pour voir si certains d'entre eux peuvent nous aider dans notre quête pour mieux comprendre les particules actives en interaction.
\\

L'objectif de cette thèse est donc double : 
\vspace{4pt}

\noindent (i) Exploiter les méthodes développées dans le contexte des gaz de Riesz et de la théorie des matrices aléatoires pour faire progresser nos connaissances sur les systèmes de particules actives en interaction, en particulier avec des potentiels d'interaction en loi de puissance, par la dérivation de résultats exacts.

\vspace{4pt}

\noindent (ii) Dans le processus, nous apporterons également notre contribution à la littérature sur les gaz de Riesz, en particulier en étudiant comment le bruit non-brownien affecte les propriétés de ces modèles.

\vspace{4pt}

Tout au long de cette thèse, nous travaillerons en une dimension, et nous nous concentrerons principalement sur deux aspects de ces modèles, à savoir :

\vspace{4pt}

\noindent (i) la densité macroscopique des particules, en particulier dans l'état stationnaire,

\vspace{4pt}

\noindent (ii) les fluctuations spatio-temporelles des positions des particules.
\\

Cette thèse est divisée en quatre parties :
\\

Dans la partie ~\ref{part:context}, nous examinons plus en détail la littérature existante pertinente dans le cadre de cette thèse, dans les deux domaines des particules actives et des gaz de Riesz. Nous commençons au chapitre~\ref{chap:active} par présenter les modèles les plus connus de particules actives. Parmi ces modèles, la particule {\it run-and-tumble} (RTP) en 1D, qui passe d'une vitesse $\pm v_0$ à l'autre avec un taux $\gamma$, constituera l'élément de base central de la plupart des modèles étudiés dans cette thèse. Nous rappelons ensuite quelques résultats exacts existants pour les particules actives sans interactions, en particulier concernant leurs états stables non-Boltzmann en présence de confinement, ainsi que leurs propriétés de premier passage. Nous discutons ensuite dans le chapitre~\ref{chap:interactions} les systèmes de particules actives en interaction, en passant en revue les types d'interactions les plus courants ainsi que les effets collectifs les plus emblématiques qui ont été observés dans de tels systèmes. Nous donnons également plus de détails sur quelques cas pour lesquels des résultats exacts ont été obtenus (y compris les modèles sur réseau et les chaînes harmoniques mentionnés ci-dessus). En ce qui concerne les gaz de Riesz, que nous abordons dans le chapitre~\ref{chap:Riesz_review}, nous nous concentrons sur les résultats qui sont particulièrement pertinents pour la présente thèse, notamment en ce qui concerne la densité macroscopique et les fluctuations des positions des particules, en mettant l'accent sur les cas particuliers du mouvement brownien de Dyson, de la {\it rank diffusion} et du modèle de Calogero-Moser.
\\

La partie~\ref{part:density} se concentre sur l'étude de la densité des particules dans les modèles de particules actives en interaction. Dans le chapitre~\ref{chap:DeanRTP}, nous dérivons des équations hydrodynamiques exactes pour des RTP unidimensionnelles interagissant via un potentiel de paire $W(x)$, qui généralisent l'équation de Dean-Kawasaki (DK) pour les particules browniennes \cite{Dean,Kawa}. Ces équations contiennent des termes de bruit qui fournissent une description exacte des fluctuations de la densité. Dans le reste de cette partie, nous nous concentrons sur la limite où le nombre de particules $N$ est infini, de sorte que les fluctuations disparaissent. Nous utilisons ensuite ces équations pour étudier deux exemples différents : la {\it rank diffusion active} au chapitre~\ref{chap:activeRD}, correspondant à des RTPs interagissant via un potentiel de Coulomb 1D, et le {\it mouvement brownien de Dyson actif} au chapitre~\ref{chap:ADBM_Dean}, où le potentiel d'interaction est logarithmique. Dans les deux cas, des progrès analytiques significatifs peuvent être réalisés dans la détermination de l'état stationnaire hors équilibre.

Pour la {\it rank diffusion} active (chapitre~\ref{chap:activeRD}), où la force d'interaction $-W'(x)=\frac{\kappa}{N}\, \sgn(x)$ est indépendante de la distance, nous considérons à la fois le cas répulsif $\kappa>0$ et le cas attractif $\kappa=-\bar \kappa<0$. En l'absence de confinement, mais pour une interaction attractive, les particules forment un état lié stationnaire. En convertissant les équations non-locales de DK dans la limite $N\to+\infty$ en une forme locale (en s'inspirant de \cite{PLDRankedDiffusion}), nous sommes capables d'obtenir une solution analytique exacte pour la densité stationnaire, qui présente une transition de phase hors équilibre entre une phase où la densité est lisse avec un support non borné, pour $v_0>\bar \kappa$, et une phase où le support est borné, avec des ``clusters" de particules, c'est à dire des pics delta dans la densité stationnaire, se formant sur les bords, pour $v_0<\bar \kappa$. Dans le cas répulsif, les particules forment un gaz en expansion dont le comportement à temps long n'est pas très différent du cas brownien. Ces résultats peuvent être étendus en présence d'un potentiel confinant linéaire, ainsi qu'à un potentiel harmonique. Dans les deux cas, nous trouvons un diagramme de phase très riche, pour une interaction attractive et pour une interaction répulsive. Enfin, nous étudions également une extension de ce modèle où l'interaction est non-réciproque entre les particules de vitesses $+v_0$ et $-v_0$, conduisant à une densité asymétrique avec différents régimes. Dans tous ces cas, nos résultats analytiques sont confirmés par des simulations numériques.

Pour le DBM actif (chapitre~\ref{chap:ADBM_Dean}), le potentiel d'interaction $W(x)=\frac{2g}{N}\, \log |x|$ est répulsif, et nous ajoutons un potentiel de confinement harmonique $V(x)=\frac{\lambda}{2}x^2$. Dans ce cas, l'étude est rendue beaucoup plus difficile par l'échec des équations de DK. En effet, nous avons trouvé qu'en présence d'une force de répulsion qui diverge au contact, empêchant les particules de se croiser, le mouvement persistant des RTPs conduit à de fortes corrélations locales qui rompent la description hydrodynamique. Nous introduisons cependant une variante du modèle, où deux particules n'interagissent que si elles sont dans le même état (c'est-à-dire $+v_0$ ou $-v_0$), ce qui permet aux particules de se croiser. Dans ce cas, l'équation de DK fonctionne et nous permet d'étudier en détail le comportement de la densité stationnaire à grand $N$ dans les différentes limites du modèle, ainsi que de calculer récursivement ses moments. Pour la version originale du DBM actif, qui empêche les croisements de particules, nous fournissons des arguments solides, basés sur des résultats numériques, ainsi que sur l'étude analytique des fluctuations réalisée dans la partie~\ref{part:fluctuations}, permettant d'affirmer que dans la limite de grand $N$ la densité stationnaire converge vers le demi-cercle de Wigner, comme pour le DBM standard. Nous montrons que la rupture du demi-cercle nécessite un scaling $v_0/\sqrt{g\lambda}\sim\sqrt{N}$, et nous effectuons une étude numérique de la limite $v_0/\sqrt{g\lambda}\gg\sqrt{N}$, où l'effet à longue portée de l'interaction devient négligeable et où les particules ont tendance à s'agréger en grands groupes en raison de leur mouvement persistant.

Les résultats présentés dans cette partie ont conduit à la publication de \cite{ADBM1} pour le DBM actif et de \cite{activeRD1,activeRD2} pour la {\it rank diffusion} active. 
\\

Dans la partie~\ref{part:fluctuations}, nous étudions les fluctuations au niveau des positions des particules $x_i(t)$ dans des gaz de Riesz unidimensionnels de particules browniennes et actives. Dans le chapitre~\ref{chap:passiveRieszFluct}, nous considérons des particules browniennes sur un cercle de taille $L$ interagissant via un potentiel répulsif en loi de puissance $W(x)=g \, s^{-1}|x|^{-s}$, avec $s>-1$. Nous nous concentrons sur la limite des faibles températures où les particules ne subissent que de faibles déplacements autour de la configuration de l'état fondamental. Dans cette limite, en linéarisant les équations du mouvement et en inversant la matrice hessienne, nous pouvons calculer exactement les corrélations à deux points et à deux temps des positions des particules. Cela nous permet d'obtenir des expressions exactes pour une variété de fonctions de corrélation statiques et dynamiques, que nous analysons dans la limite $N,L\to+\infty$ avec une densité fixe $\rho=N/L$. Ceci nous permet de retrouver certains résultats obtenus récemment dans la littérature physique et mathématique par des méthodes complètement différentes. En particulier, nous trouvons qu'à des temps élevés $t\gg\tau=g\rho^{s+2}$, le déplacement quadratique moyen (MSD) d'une particule pendant le temps $t$ prend une échelle subdiffusive, comme $\sim\sqrt{t}$ pour $s>1$ (cas à courte portée, similaire à la diffusion sur une seule file), et comme $\sim t^{\frac{s}{1+s}}$ pour $0<s<1$ (cas à longue portée). Remarquablement, ces résultats coïncident exactement, jusqu'aux préfacteurs, avec ceux obtenus dans \cite{DFRiesz23} en utilisant la théorie des fluctuations macroscopiques (MFT). Nous obtenons également que la variance de la distance entre les particules $i$ et $i+k$ (les particules ordonnées) varie de façon sous-linéaire comme $\sim k^s$, comme cela a été montré récemment dans \cite{BoursierCLT,BoursierCorrelations} (alors qu'elle est linéaire pour $s>1$). De plus, cette méthode nous permet de calculer de nouvelles quantités telles que la covariance à temps égal des déplacements des particules ou les corrélations dynamiques de la distance entre particules.

La simplicité de cette approche nous permet de la généraliser facilement aux RTP (ou à d'autres particules actives), dans la limite d'un bruit faible (c'est-à-dire d'un petit $v_0$), ce que nous faisons dans le chapitre~\ref{chap:activeRieszFluct}. Cela ajoute une échelle de temps supplémentaire $1/\gamma$ correspondant au temps de persistance des particules actives. Nous constatons que nous retrouvons les résultats browniens à des temps élevés $t\gg1/\gamma$ et pour de grandes séparations $k\gg \hat g^{1/z_s}$, où $\hat g=(2\gamma\tau)^{-1}$ est un paramètre sans dimension qui mesure l'activité et $z_s=\min(1+s,2)$ un exposant dynamique qui caractérise à la fois le système brownien et le système actif. Cependant, l'activité joue un rôle important à court terme et pour de petites distances. En particulier, pour $t\ll 1/\gamma$, nous trouvons différents types de comportements superdiffusifs du MSD en fonction de la condition initiale que nous considérons ({\it annealed} or {\it quenched}). Pour de petites séparations $k\gg \hat g^{1/z_s}$ (lorsque $\hat g \gg 1$, c'est-à-dire pour de grands temps de persistance), nous constatons que la variance de la distance entre particules augmente plus rapidement que linéairement avec $k$ pour tout $s>0$, ce qui est un indicateur de {\it giant number fluctuations} \cite{TonerTuReview,ChateGiant,GinelliGiant,DasGiant2012,Chate2010,NarayanGiant,ZhangGiant}.

Dans le chapitre~\ref{chap:ADBMfluct}, nous étendons ces résultats à deux cas particuliers du gaz de Riesz sur l'axe réel avec un potentiel harmonique confinant : le mouvement brownien de Dyson actif ($s=0$), étudié dans le chapitre~\ref{chap:ADBM_Dean} au niveau de la densité des particules, et le modèle de Calogero-Moser actif, correspondant à $s=2$. Dans ces deux cas, nous obtenons des formes d'échelle explicites pour la covariance à deux points et à deux temps des positions des particules, qui sont exactes dans la double limite d'un bruit faible et d'un grand $N$. Dans le {\it bulk}, cette covariance scale comme $N^{-1}$ pour les deux modèles. Cependant, pour le DBM actif, nous trouvons un régime de bord distinct, qui présente un scaling différent, comme $N^{-2/3}$. L'existence d'un régime de bord avec des fluctuations plus fortes rappelle ce qui est observé pour le DBM standard \cite{Mehta_book,Forrester_book,bouchaud_book}, ainsi que pour le modèle CM avec des particules browniennes \cite{Agarwal2019}. Nous constatons cependant qu'un tel régime n'existe pas dans le modèle CM actif. Le scaling des fluctuations dans le bulk confirme les affirmations faites dans le chapitre~\ref{chap:ADBM_Dean} concernant la densité stationnaire dans le DBM actif, et suggère également un comportement similaire pour le modèle CM actif. Nous étudions également le cas du modèle CM avec des particules browniennes, fournissant une confirmation analytique des résultats numériques de \cite{Agarwal2019}.

La plupart des résultats donnés dans cette partie sont présentés dans \cite{RieszFluct}. Les résultats pour le modèle de Calogero-Moser actif ont été obtenus en collaboration avec un groupe de l'ICTS Bengalore et sont publiés dans \cite{ADBM2}.
\\

La dernière partie, partie~\ref{part:siegmund}, concerne toujours l'étude des particules actives par le biais de calculs exacts, mais est légèrement déconnectée des deux parties précédentes, puisqu'elle n'implique pas d'interactions. Elle se concentre plutôt sur les propriétés de premier passage des particules actives, qui ont attiré beaucoup d'attention ces dernières années, en partie à cause de leur pertinence dans des contextes biologiques.
Nous avons déjà mentionné plus haut le comportement particulier que les particules actives tendent à présenter en milieu confiné, à savoir leur tendance à s'accumuler près des frontières en raison de leur mouvement persistant. De manière surprenante, il s'avère que ces deux types de problèmes, plus précisément l'étude des particules actives en présence de {\it conditions aux bords absorbantes} et en présence de {\it parois dures} (entendues comme des barrières infinies de potentiel), sont en fait liés.

Dans le chapitre~\ref{chap:Exitproba}, nous considérons une RTP unidimensionnel sur un intervalle $[a,b]$, avec des {\it murs absorbants} en $a$ et en $b$, soumis à un potentiel extérieur arbitraire $V(x)$. Dans ce cadre, nous calculons explicitement la probabilité de sortie (aussi appelée {\it hitting} ou {\it splitting probability}), c'est-à-dire la probabilité que, à partir d'une position $x\in [a,b]$, la particule soit finalement absorbée en $b$ (et non en $a$). Nous constatons que cette quantité est exactement égale à la distribution cumulative des positions dans l'état stationnaire d'une RTP avec des {\it parois dures} en $a$ et $b$, avec un potentiel extérieur $-V(x)$ (que nous calculons également). Cela rappelle le cas brownien, pour lequel la probabilité de sortie
est identique à la cumulative de la distribution de Boltzmann à l'équilibre avec des parois dures en $a$ et $b$, à un remplacement $V(x)\to -V(x)$ près.

Cette relation est en fait liée à un concept plus général, connu dans la littérature mathématique sous le nom de {\it dualité de Siegmund} \cite{Siegmund}. Pour des modèles tels que le mouvement brownien ou les marches aléatoires avec des pas i.i.d. en une dimension, elle relie la distribution des positions en présence de conditions aux limites absorbantes en $a$ et $b$ à celle d'un processus dual avec des parois dures en $a$ et $b$, y compris à temps fini. Bien que l'existence d'un processus dual de Siegmund ait été prouvée pour une large classe de processus, une formulation explicite du processus dual n'est pas toujours facile à trouver. Dans le chapitre~\ref{chap:Siegmund}, après avoir passé en revue la littérature existante sur la dualité de Siegmund, nous présentons une formulation explicite du dual de Siegmund pour une grande famille de processus stochastiques continus unidimensionnels, entraînés par un bruit corrélé dans le temps. Ce résultat s'applique non seulement à tous les modèles les plus courants de particules actives, mais aussi à d'autres processus stochastiques pertinents en physique, tels que les modèles de {\it diffusing diffusivity} \cite{DiffDiffChubynsky, DiffDiffChechkin, DiffDiffJain, DiffDiffFPTSposini} et le {\it stochastic resetting} \cite{resettingPRL,resettingReview, resettingBriefReview}. Nous montrons également un résultat similaire dans le cas de marches aléatoires à temps discret et continu. Nous illustrons ces résultats par des simulations numériques, et nous discutons de la pertinence de cette dualité pour les modèles physiques, à la fois pour les calculs analytiques et numériques. 

Les résultats des chapitres~\ref{chap:Exitproba} et \ref{chap:Siegmund} ont été obtenus en collaboration avec Mathis Gu\'eneau, et sont publiés respectivement dans \cite{SiegmundShort} et dans \cite{SiegmundLong}.
\newpage

\chapter*{\vspace*{-0.4cm}Introduction, goal and overview}
\addcontentsline{toc}{chapter}{Introduction, goal and overview}%

\vspace*{0.85cm}

Since its foundation more than a century ago, statistical mechanics has proved to be an extremely powerful tool to describe systems formed by a large number of interacting entities at equilibrium. One of its greatest achievements was to allow for a precise understanding of the collective effects that emerge in such systems, known as phase transitions. More recently however, there has been a strong interest in the study of systems which are {\it out-of-equilibrium}, in particular coming from biology. In such cases, the constant input of energy breaks the hypotheses of equilibrium statistical mechanics, which makes analytical studies extremely challenging. In particular, these systems may reach steady-states which are not described by the Gibbs measure, thus requiring the development of entirely new tools to study their large time behavior. At the same time, the ability of these systems to escape the usual laws of statistical mechanics opens the way for a whole range of fascinating phenomena, such as new types of phase transitions which would be completely impossible at equilibrium.

A paradigmatic example which has attracted a lot of attention other the last three decades is active matter, and more precisely the study of {\it active particles} \cite{Ramaswamy2010,Bechinger,Marchetti2018}. These are particles which possess some form of self-propulsion, leading to a {\it persistent motion}, meaning that these particles have a stronger tendency to follow straight lines compared to a standard Brownian particle. Although the original models were introduced to study the collective behavior of animals, such as flocks of birds or schools of fish \cite{Vicsek,CavagnaGiardina,Calovi2018}, or to model the displacements of smaller living organisms such as bacteria \cite{Berg2004}, non-organic active particles have also been realized experimentally through a variety of different approaches, using for instance chemical reactions, electrodynamic instabilities, vibrated disks of even small robots \cite{ErbeExperiments,JanusReview,Quincke,Chate2010,Schranz}. From a more mathematical perspective, the motion of active particles is modeled through the introduction of non-Brownian noise which is {\it correlated in time} (and in general non-Gaussian). Many different models have been introduced over the years, one of them being the {\it run-and-tumble particle} (RTP). This model is inspired from the motion of bacteria such as {\it E. Coli}, which consists in a series of straight runs, separated by tumbling events during which the particle reorients itself in a random direction \cite{Berg2004}. For all these models, already at the level of a single particle, the complexity of the noise allows for the appearance of interesting new phenomena, such as the tendency of active particles to accumulate near boundaries \cite{Yang2014,Uspal2015}, but it also makes analytical computations particularly difficult. Other the past years, an important literature has developed around these models, regarding in particular their non-Boltzmann steady-states in the presence of confinement \cite{TailleurCates2009,DKM19,LMS2020,SzamelAOUP,Bonilla2019,Wijland21,ABM2,AngelaniHardWalls}, as well as their first-passage properties \cite{Targetsearch,MalakarRTP,Singh2020,Singh2022,MFPT1DABP,RTPsurvivalMori,TVB12,RBV16,SurvivalRTPDriftDeBruyne,DKM19,AngelaniMFPT,MathisMFPT,MathisMFPT2,Grange2025}. Despite this, many questions remain open, even at the single-particle level.

All this is even more true when we add interactions. Indeed, systems of interacting active particles have been shown to display a variety of new collective effects, such as the transition to collective motion in the presence of alignment interactions (flocking) \cite{Vicsek,TonerTu95,TonerTu98,TonerTuReview}, or the well-known motility-induced phase separation (MIPS) in the presence of short-range repulsive interactions \cite{CT2015,OByrne2021}. Developing a precise analytical description of these phenomena represents however a considerable challenge. Most theoretical studies concerning systems of interacting active particles rely either on general field theories derived from symmetry arguments, or on coarse-grained hydrodynamic equations which generally require some degree of approximation. Until now, very few exact results have been obtained beyond the two-particle case \cite{slowman,slowman2,Mallmin2019,KunduGap2020,Hahn2023,Metson2022,MetsonLong,LMS2021,Hahn2025}, with the notable exception of some lattice models with contact interactions, for which exact hydrodynamic equations have been derived \cite{KH2018,Erignoux_derivation,Agranov2021,Agranov2022,lattice2lanes2025}, as well as harmonic chains of active particles, for which some space-time correlation functions of the particle positions have been computed exactly \cite{PutBerxVanderzande2019,SinghChain2020,HarmonicChainRevABP,HarmonicChainRTPDhar}. 

Finding other models for which exact analytical results can be obtained would greatly advance our understanding of interacting active particle systems. One type of system for which this might be possible, which until now has received very little attention, is the case of active particles with long-range interactions (i.e., with a pairwise interaction potential which behaves as a power law). Although this type of interaction is generally not prevalent in real-life active particle systems, it can be realized experimentally, e.g., by placing paramagnetic colloids subjected to a magnetic field (thus generating a dipole-dipole interaction with a $\sim 1/|x|^3$ potential, $|x|$ being the distance between particles) in contact with a bath of active particles \cite{ActiveMeltingNature2024}. One may also think of hydrodynamic interactions (i.e., mediated by the surrounding fluid), although these are generally more complex and depend on the velocity and the orientation of the particles \cite{ishikawa2007,zottl2023,Filella2018}. The main motivation to study such systems is however theoretical. Indeed, models of Brownian particles interacting via pairwise (repulsive) power law potentials $\sim |x|^{-s}$, known as {\it Riesz gases} (or Coulomb gases if $s=d-2$ in $d$ dimensions), have been extensively studied both in the physics and in the mathematics literature \cite{Riesz,Lewin,SerfatyBook}, particularly in one dimension, in part due to their connection with random matrix theory (RMT) \cite{Mehta_book,Forrester_book,bouchaud_book}. The most well-known model in this family is the Dyson Brownian motion (DBM), or log-gas, corresponding to a 2D Coulomb interaction $s=0$ (i.e., to a logarithmic interaction potential), which has strong connections with the Gaussian matrix ensembles \cite{Dyson}. Other special cases include the rank diffusion, or jellium model, $s=-1$ (see, e.g., \cite{PLDRankedDiffusion}), which corresponds to a 1D Coulomb interaction, and the Calogero-Moser (CM) model $s=2$, mostly studied in the context of Hamiltonian dynamics, but for which the overdamped Langevin dynamics were recently investigated numerically \cite{Agarwal2019}. It is thus tempting to search into the broad array of analytical tools developed to study those models to see if some of them may help us in our quest to better understand interacting active particles.
\\

The goal of this thesis is thus twofold: 
\vspace{4pt}

\noindent (i) Leverage the methods developed in the context of Riesz gases and random matrix theory to advance our knowledge on interacting active particle systems, in particular with power law interaction potentials, through the derivation of exact results.

\vspace{4pt}

\noindent (ii) In the process, we will also bring our contribution to the literature on Riesz gases, in particular by studying how non-Brownian noise affects the properties of these models.

\vspace{4pt}

Throughout this thesis, we will work in one dimension, and we will mostly focus on two aspects of these models, namely:

\vspace{4pt}

\noindent (i) the macroscopic density of particles, in particular in the stationary state,

\vspace{4pt}

\noindent (ii) the space-time fluctuations of the particle positions.
\\

This thesis is divided into four parts:
\\

In Part ~\ref{part:context}, we review in more detail the existing literature, from the fields of both active particles and Riesz gases, which is relevant for this thesis. We begin in Chapter~\ref{chap:active} by introducing the most well-known models of active particles. Among these models, the run-and-tumble particle (RTP) in 1D, which switches between velocities $\pm v_0$ with a rate $\gamma$, will constitute the central building block of most models studied in this thesis. We then recall some existing exact results for non-interacting active particles, in particular concerning their non-Boltzmann steady states in the presence of confinement, as well as their first-passage properties. We then discuss in Chapter~\ref{chap:interactions} systems of interacting active particles, reviewing the most common types of interactions as well as the most emblematic collective effects that have been observed in such systems. We also provide some more details on a few cases for which exact results have been derived (including the lattice models and the harmonic chains mentioned above). Concerning Riesz gases, which we discuss in Chapter~\ref{chap:Riesz_review}, we focus on the results which are particularly relevant for the present thesis, in particular concerning the macroscopic density and the tagged particle fluctuations, with a particular emphasis on the special cases of the Dyson Brownian motion, the rank diffusion and the Calogero-Moser model.
\\

Part~\ref{part:density} focuses on the study of the particle density in models of interacting active particles. In Chapter~\ref{chap:DeanRTP}, we derive exact hydrodynamic equations for one-dimensional RTPs interacting via a pairwise potential $W(x)$, which generalize the Dean-Kawasaki (DK) equation for Brownian particles \cite{Dean,Kawa}. These equations contain noise terms which provide an exact description of the fluctuations in the density. In the rest of this part we however mostly focus on the limit where the number of particles $N$ is infinite, so that the fluctuations vanish. We then use these equations to study two different examples: the {\it active rank diffusion} in Chapter~\ref{chap:activeRD}, corresponding to RTPs interacting via a 1D Coulomb potential, and the {\it active Dyson Brownian motion} in Chapter~\ref{chap:ADBM_Dean}, where the interaction potential is logarithmic. In both cases, significant analytical progress can be made in the determination of the out-of-equilibrium stationary state.

For the active rank diffusion (Chapter~\ref{chap:activeRD}), where the interaction force $-W'(x)=\frac{\kappa}{N}\, \sgn(x)$ is independent of the distance, we consider both the repulsive case $\kappa>0$ and the attractive case $\kappa=-\bar \kappa<0$. In the absence of confinement but for an attractive interaction, the particles form a stationary bound state. By mapping the non-local DK equations in the limit $N\to+\infty$ into a local form (taking inspiration from \cite{PLDRankedDiffusion}), we are able to obtain an exact analytical solution for the stationary density, which exhibits a non-equilibrium phase transition between a phase where the density is smooth with unbounded support, for $v_0>\bar \kappa$, and a phase where the support is bounded, with clusters of particles, i.e., delta peaks in the density, forming at the edges, for $v_0<\bar \kappa$. In the repulsive case, the particles form an expanding gas whose large time behavior is not very different from the Brownian case. These results can be extended in the presence of a linear confining potential, as well as to a harmonic potential. 
In both cases, we find a very rich phase diagram for both an attractive and a repulsive interaction. Finally, we also study an extension of this model where the interaction is non-reciprocal between the particles of velocities $+v_0$ and $-v_0$, leading to an asymmetric density with different regimes. In all these cases, our analytical results are supported by numerical simulations.

For the active DBM (Chapter~\ref{chap:ADBM_Dean}), the interaction potential $W(x)=\frac{2g}{N}\, \log |x|$ is repulsive and we add a harmonic confining potential $V(x)=\frac{\lambda}{2}x^2$. In this case, the study is made significantly more difficult by the failure of the DK equations. Indeed, we found that in the presence of a repulsion force which diverges at contact, preventing the particles from passing each other, the persistent motion of the RTPs leads to strong local correlations which break the hydrodynamic description. We however introduce a variant of the model, where two particles only interact if they are in the same state (i.e., $+v_0$ or $-v_0$), thus allowing the particles to cross. In this case, the DK equation works and it allows us to study in detail the behavior of the stationary density at large $N$ in the different limits of the model, as well as to recursively compute its moments. For the original version of the active DBM, which prevents particle crossings, we provide strong evidence, based on numerical results as well as on the analytical study of the fluctuations performed in Part~\ref{part:fluctuations}, that in the large $N$ limit the stationary density converges to the Wigner semi-circle, as for the standard DBM. We show that breaking the semi-circle requires scaling $v_0/\sqrt{g\lambda}\sim\sqrt{N}$, and we perform a numerical study of the limit $v_0/\sqrt{g\lambda}\gg\sqrt{N}$, where the long-range effect of the interaction becomes negligible and where the particles tend to aggregate into large clusters due to their persistent motion.

The results presented in this part led to the publication \cite{ADBM1} for the active DBM and to \cite{activeRD1,activeRD2} for the active ranked diffusion. 
\\

In Part~\ref{part:fluctuations}, we study the fluctuations at the level of the particle positions $x_i(t)$ in one-dimensional Riesz gases of Brownian and active particles. In Chapter~\ref{chap:passiveRieszFluct}, we consider Brownian particles on a circle of size $L$ interacting via a repulsive power law potential $W(x)=g \, s^{-1}|x|^{-s}$ with $s>-1$. We focus on the small temperature limit where the particles only undergo small displacements around the equally spaced ground state configuration. In this limit, by linearizing the equations of motion and inverting the Hessian matrix, we can compute exactly the two-point two-time correlations of the particle positions. This allows us to obtain exact expressions for a variety of static and dynamical correlation functions, which we analyze in the limit $N,L\to+\infty$ with fixed density $\rho=N/L$. This allows us to recover some results obtained recently in the physics and mathematics literature via completely different methods. In particular, we find that at large times $t\gg\tau=g\rho^{s+2}$, the mean squared displacement (MSD) of a particle during time $t$ takes a subdiffusive scaling, as $\sim\sqrt{t}$ for $s>1$ (short-range case, similar to single-file diffusion), and as $\sim t^{\frac{s}{1+s}}$ for $0<s<1$ (long-range case). Remarkably, these result coincide exactly, up to prefactors, with those obtained in \cite{DFRiesz23} using macroscopic fluctuation theory. We also obtain that the variance of the distance between particles $i$ and $i+k$ (the particles being ordered) scales sublinearly as $\sim k^s$, as shown recently in \cite{BoursierCLT,BoursierCorrelations} (while it is linear for $s>1$). In addition, this method allows us to compute new quantities such as the equal time covariance of the particle displacements or the dynamical correlations of the interparticle distance.

The simplicity of this approach allows us to easily generalize to RTPs (or other active particles), in the limit of weak noise (i.e., small $v_0$), which we do in Chapter~\ref{chap:activeRieszFluct}. This adds an additional timescale $1/\gamma$ corresponding to the persistence time of the active particles. We find that we recover the Brownian results at large times $t\gg1/\gamma$ and large separations $k\gg \hat g^{1/z_s}$, where $\hat g=(2\gamma\tau)^{-1}$ is a dimensionless parameter which measures the activity and $z_s=\min(1+s,2)$ a dynamical exponent which characterizes both the Brownian and the active system. However, the activity plays an important role at short times and for small distances. In particular, for $t\ll 1/\gamma$ we find different types of superdiffusive behavior of the MSD depending on the initial condition that we consider (annealed or quenched). For small separations $k\gg \hat g^{1/z_s}$ (when $\hat g \gg 1$, i.e., for large persistence times), we find that the variance of the interparticle distance increases faster than linearly in $k$ for any $s>0$, which is an indicator of giant number fluctuations \cite{TonerTuReview,ChateGiant,GinelliGiant,DasGiant2012,Chate2010,NarayanGiant,ZhangGiant}.

In Chapter~\ref{chap:ADBMfluct}, we extend these results to two special cases of the Riesz gas on the real axis with a confining harmonic potential: the active Dyson Brownian motion ($s=0$), studied in Chapter~\ref{chap:ADBM_Dean} at the level of the particle density, and the active Calogero-Moser model, corresponding to $s=2$. In these two cases, we obtain explicit scaling forms for the two-point two-time covariance of particle positions which are exact in the double limit of weak noise and large $N$. In the bulk, this covariance scales as $N^{-1}$ for both models. However, for the active DBM we find a distinct edge regime, which exhibits a different scaling as $N^{-2/3}$. The existence of an edge regime with stronger fluctuations is reminiscent of what is observed for the standard DBM \cite{Mehta_book,Forrester_book,bouchaud_book}, as well as for the CM model with Brownian particles \cite{Agarwal2019}. We find however that there is no such regime in the active CM model. The scaling of the fluctuations in the bulk supports the statements made in Chapter~\ref{chap:ADBM_Dean} concerning the stationary density in the active DBM, and also suggests a similar behavior for the active CM model. We also study the case of the CM model with Brownian particles, providing an analytical confirmation of the numerical results of \cite{Agarwal2019}.

Most of the results given in this part are presented in \cite{RieszFluct}. The results for the active Calogero-Moser model were obtained in collaboration with a group from ICTS Bengalore and are published in \cite{ADBM2}.
\\

The last part, Part~\ref{part:siegmund}, still concerns the study of active particles via exact computations but is slightly disconnected from the two previous parts, since it does not involve interactions. Instead, it focuses on the first-passage properties of active particles, which have attracted a lot of attention in recent years, due in part to their relevance in biological contexts. 
We have already mentioned above the peculiar behavior that active particles tend to exhibit in confinement, namely their tendency to accumulate near boundaries due to their persistent motion. Surprisingly, it turns out that these two types of problems, more precisely the study of active particles in the presence of {\it absorbing boundary conditions} and in the presence of {\it hard walls} (understood as infinite barriers of potential), are actually related.

In Chapter~\ref{chap:Exitproba}, we consider a one-dimensional RTP on an interval $[a,b]$, with {\it absorbing walls} at $a$ and $b$, subjected to an arbitrary external potential $V(x)$. In this setting we compute explicitly the {\it exit probability} (also called hitting or splitting probability), i.e., the probability that, starting from some position $x\in[a,b]$, the particle eventually gets absorbed at $b$ (and not $a$). We find that this quantity is exactly equal to the cumulative distribution of positions in the stationary state of a RTP with {\it hard walls} at $a$ and $b$, with an external potential $-V(x)$ (which we also compute). This is reminiscent of the Brownian case, for which the exit probability 
is identical to the cumulative of the equilibrium Boltzmann distribution with hard walls at $a$ and $b$ up to a change $V(x)\to-V(x)$.

This relation is actually connected to a more general concept, known in the mathematics literature as {\it Siegmund duality} \cite{Siegmund}. For models such as Brownian motion or random walks with i.i.d. steps in one dimension, it relates the distribution of positions in the presence of absorbing boundary conditions at $a$ and $b$ with the one of a dual process with hard walls at $a$ and $b$, even at finite time. Although the existence of a Siegmund dual has been proved for a large class of processes, an explicit formulation of the dual is not always easy to find. In Chapter~\ref{chap:Siegmund}, after reviewing the existing literature on Siegmund duality, we present an explicit formulation of the Siegmund dual for a large family of one-dimensional continuous stochastic processes, driven by time-correlated noise. This result applies not only to all the most common models of active particles, but also to other stochastic processes which are relevant in physics, such as diffusing diffusivity models \cite{DiffDiffChubynsky, DiffDiffChechkin, DiffDiffJain, DiffDiffFPTSposini} and stochastic resetting \cite{resettingPRL,resettingReview, resettingBriefReview}. We also show a similar result in the case of discrete and continuous time random walks. We illustrate these results with numerical simulations, and we discuss the relevance of this duality for physical models, both for analytical and numerical computations. 

The results of Chapters~\ref{chap:Exitproba} and \ref{chap:Siegmund} were obtained in collaboration with Mathis Gu\'eneau, and are published in \cite{SiegmundShort} and in \cite{SiegmundLong} respectively.

\clearpage
\pagenumbering{arabic}

\part{Interacting active particles and Riesz gases: Context and motivations: }\label{part:context}

\vspace*{\fill}

\begin{center}
{\bf Abstract}
\end{center}

In this first part, we give an overview of the existing literature in two different fields which are relevant for the rest of this thesis: active particles and Riesz gases. In Chapter~\ref{chap:active}, we introduce the general topic of active matter and the most common models of active particles, before reviewing some exact results obtained for non-interacting active particles, with a focus on non-Boltzmann stationary states and first-passage properties. In Chapter~\ref{chap:interactions}, we consider systems of interacting active particles. We briefly describe the main types of interactions that have been considered and the surprising collective effects that emerge in those systems (such as MIPS and flocking). We then review in more details a few instances of such systems for which exact results have been obtained. Finally, in Chapter~\ref{chap:Riesz_review}, we leave aside active particles and consider instead Brownian particles with power-law interactions, also called Riesz gases. We provide a non-exhaustive introduction to these models, focusing on existing results which are especially relevant for this thesis, in particular concerning the particle density and the microscopic correlations, and with a particular emphasis on a few special cases: the Dyson Brownian motion, the Calogero-Moser model and the 1D Coulomb interaction. 

\vspace*{\fill}

\chapter{Active particles} \label{chap:active}

\section{General context and relevance}

Active particles can be broadly defined as systems which possess a form of self-propulsion, meaning that they can convert energy from an external source into directed motion \cite{Ramaswamy2010,Bechinger,Marchetti2018}. Many examples of such systems can be found in nature, in particular at the microscopic scale, including micro-organisms such as bacteria \cite{Berg2004}, as well as living cells \cite{cellsMarchetti}. Some macroscopic systems such as swarms of fish or flocks of birds can also be studied in this context \cite{CavagnaGiardina,Calovi2018}. However, this concept also applies to artificial systems, from micro-robots to chemically propelled particles \cite{ErbeExperiments,JanusReview,Quincke,Chate2010,Schranz}. From the point of view of statistical mechanics, these systems pose considerable conceptual and technical challenges. Indeed, due to the external input of energy they are intrinsically out-of-equilibrium, which leads to interesting properties such as breaking of time-reversal symmetry and non-Boltzmann steady-states. One of the manifestations of these effects is the peculiar way in which active particles tend to interact with obstacles and boundaries \cite{Yang2014,Uspal2015}. Even more fascinating phenomena emerge when considering large assemblies of interacting active particle: entirely new collective effects appear, such as swarming, or new types of phase separations which are completely absent in equilibrium systems. The study of such phenomena cannot be done using the standard tools of equilibrium statistical mechanics and requires the development of completely new ideas.

On a more mathematical level, active particles are driven by non-Brownian, time-correlated noise (or ``colored" noise), leading to a ``persistent" motion (meaning that the trajectories have a stronger tendency to follow straight lines). Various models of active particles have been introduced over time, which are relevant in different contexts, and the precise form of this noise varies between models. However, the presence of time-correlated noise makes the theoretical study of these models particularly challenging, since the stochastic process describing the position of the particle is non-Markovian \cite{HJ95}. Understanding how a single active particle behaves in different environments is an entire topic of research in itself, which has triggered a lot of interest in physics but also in mathematics. It is then easy to see why adding interactions to such systems gives rise to tremendously difficult problems.

Beyond its fundamental interest, both in the framework of non-equilibrium statistical mechanics and for the understanding of biological systems, the study of active particles opens the way for a variety of practical applications. For instance, micro-swimmers could be used for health care purposes, e.g., for the targeted delivery of drugs, or for the depollution of water and soils (see \cite{Bechinger} and references therein). The use of robot swarms, in particular for data collection and transport purposes, in a variety of fields such as research, industry or agriculture is another promising application \cite{Schranz}. Active particles are realized in experiments through various routes. ``Janus particles" are colloids with only half of their surface coated with a chemical component which catalyses a chemical reaction between reactants present in the solution (e.g., platinum in a solution of $H_2O_2$). The liberation of energy due to the chemical reaction then leads the particle to move in a preferred direction  \cite{ErbeExperiments,JanusReview}. We can also  mention experimental realizations based on electrostatic instabilities (such as the so-called ``Quincke rollers") \cite{Quincke}, or the use of asymmetric disks on a vibrating plate \cite{Chate2010}.

In this chapter we will review some important results concerning the study of a single active particle. As in the rest of this thesis, the emphasis will be on exact results. We begin by introducing the most well-known models of active particles and reviewing their main properties in Sec.~\ref{sec:active_models}. We then recall some existing results concerning non-Boltzmann steady-states for a single active particle with different types of confinement in Sec.~\ref{sec:1particle_potential}, as well as on the first-passage properties of these models in Sec.~\ref{sec:firstpassage}. The effect of interactions will be reviewed in details in the next chapter.

Before going on, let us mention that the study of active particles is an extremely broad field, and that many interesting topics which are less relevant for the present thesis will be omitted in this review. We briefly mention some of them here for the sake of completeness. First, since the active particle models that we will consider are particularly relevant at very small scales where inertia is generally negligible, we will focus on overdamped dynamics throughout this thesis, as it is the case in most of the literature, although the effect of inertia has recently started to be studied \cite{Lowen2020}. Some other effects which may be particularly relevant for the study of bacteria will also be omitted, in particular chemotaxis, which allows a particle to follow the concentration gradient of some chemical component by tuning its tumbling rate (see Sec.~\ref{sec:RTPdef} below) (see \cite{Cates2012} and references therein), as well as birth-death processes, which may lead to interesting effects such as pattern formation \cite{CMPT2010}. Finally, a fundamental aspect of any out-of-equilibrium process is the way in which it affects the laws of thermodynamics. In this regard, the study of active particles in the context of stochastic thermodynamics, and in particular the computation of entropy production (which is a way to quantify the distance to equilibrium) is a particularly interesting topic \cite{Chaudhuri2014,Mandal2017,Pietzonka2018,Razin2020}. The effect of activity on other thermodynamic quantities, such as the pressure, has also been studied \cite{Solon15}.

\section{Active particle models} \label{sec:active_models}

In this section we introduce the main existing models of active particles. We recall their defining equations and the context in which each model was introduced, as well as a few important results. We begin with the run-and-tumble particle, which will be the main focus of this thesis.
\\

\noindent {\it {\bf General comment on the units.} As mentioned above, all the models below are in the overdamped limit and we do not consider the effect of inertia. In addition, throughout this thesis, we will fix the unit of mass such that the mobility (i.e., the inverse of the friction coefficient) is always equal to 1. We will also fix the unit of temperature such that $k_B=1$. Thus, due to the Einstein relation, the notion of temperature identifies with the notion of diffusion coefficient, and the only remaining units relevant for our systems are the unit of distance and the unit of time.}

\subsection{The run-and-tumble particle (RTP)} \label{sec:RTPdef}

Run-and-tumble dynamics is a particular type of active dynamics which consists in a series of {\it runs}, during which the particle follows a straight line, and {\it tumbling} events, during which it changes orientation at random (see Fig.~\ref{fig:trajectories}). This stochastic process has been studied for a long time in mathematics, where it is mostly known as the {\it persistent random walk} \cite{Kac1974,Orsingher90,Masoliver1993,PRWWeiss,Masoliver2017}. 
In recent years it has been the object of a renewed interest in the context of active matter \cite{Schnitzer,Tailleur_RTP,Cates2012}. Indeed, this type of motion was found to be a good description of the behavior of bacteria such as {\it E. Coli} \cite{Berg2004}. It also has the advantage of being one of the simplest models of a stochastic process with time-correlated noise (in particular in one dimension), which allows for significant analytical progress in many situations.

The motion of a run-and-tumble particle (RTP) in arbitrary dimension can be more precisely defined as follows: the particle starts with a given orientation and follows a straight line, with a fixed velocity $v_0$. After some time $\tau$, drawn from an exponential distribution with parameter $\gamma$, it tumbles, i.e., it takes a new orientation, drawn from a uniform distribution, and follows another straight line with the same velocity $v_0$. The parameter $\gamma$ is called the {\it tumbling rate}. Note that if, in addition, some external forces are acting on the particle, its resulting velocity may differ from $v_0$. In the rest of this thesis we will refer to $v_0$ as the {\it driving velocity} to avoid any ambiguity.

\begin{figure}
    \centering
    \includegraphics[width=0.55\linewidth, trim={0 6cm 0 1cm},clip]{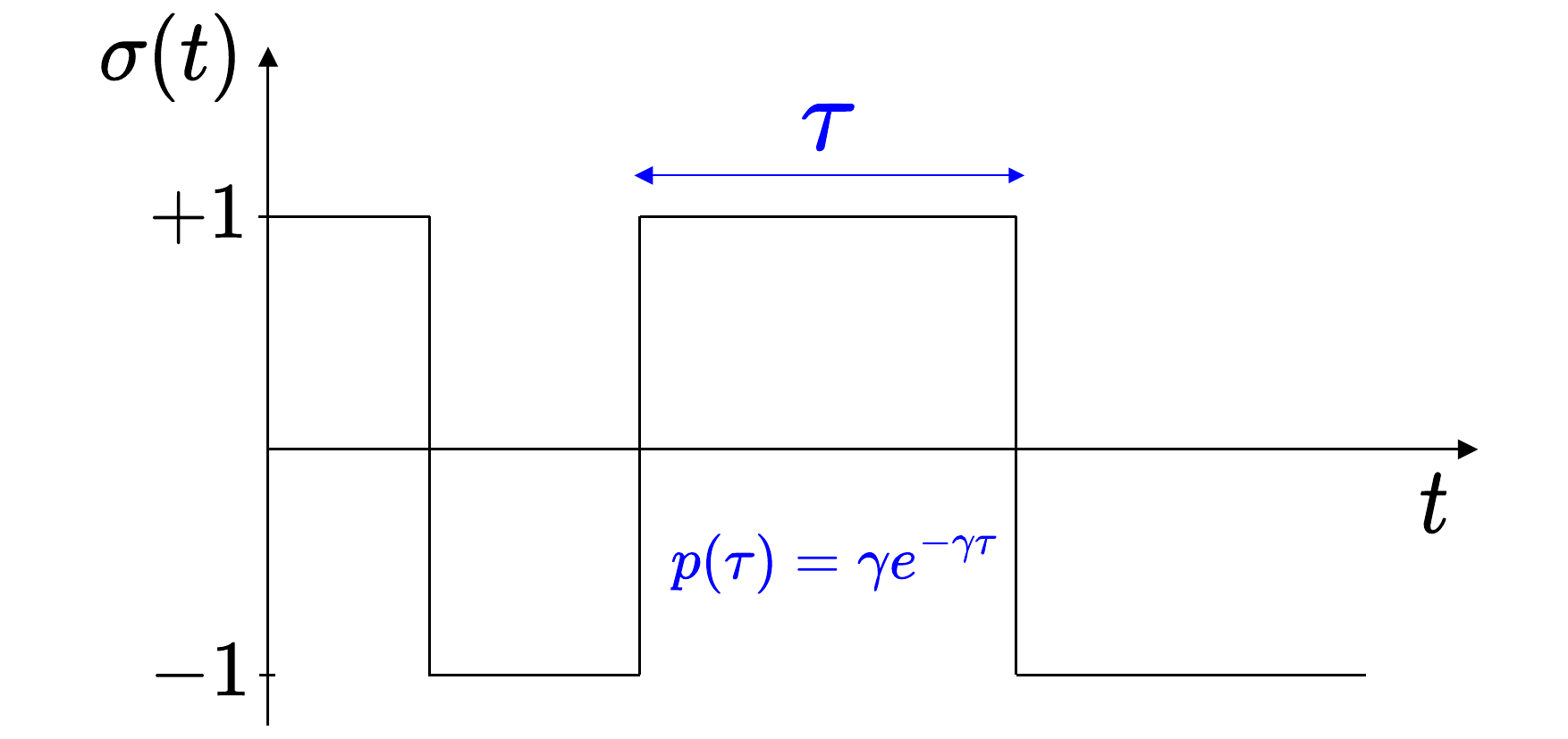}
    \caption{Example of a realization of the telegraphic noise $\sigma(t)$ plotted as a function of time. It switches between the values $+1$ and $-1$ with a rate $\gamma$.}
    \label{fig:sigma1D_example}
\end{figure}

\begin{figure}
    \centering
    \includegraphics[width=0.45\linewidth]{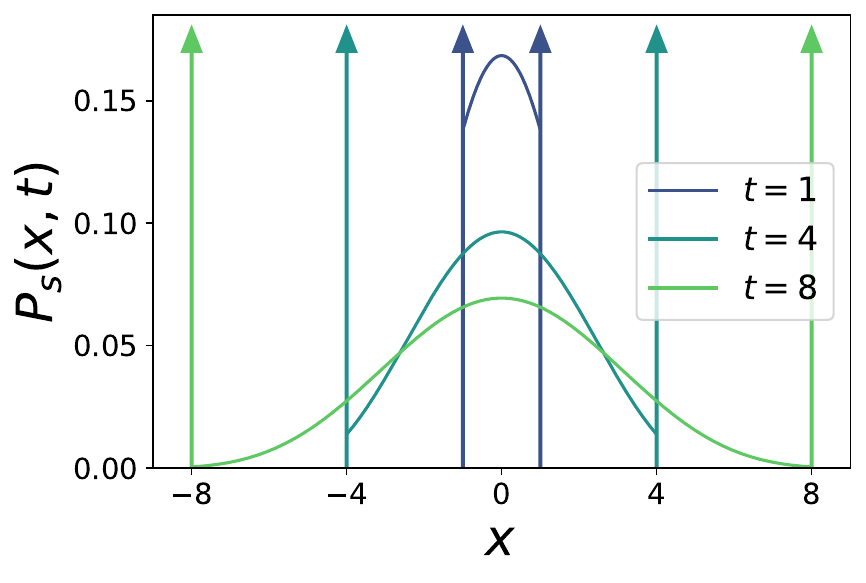}
    \caption{Propagator of a free RTP (given in \eqref{propagatorFreeRTP}) plotted at different times. The arrows represent delta functions, with weight decaying exponentially in time.}
    \label{fig:freeRTP}
\end{figure}

\begin{figure}
    \centering
    \includegraphics[width=0.35\linewidth]{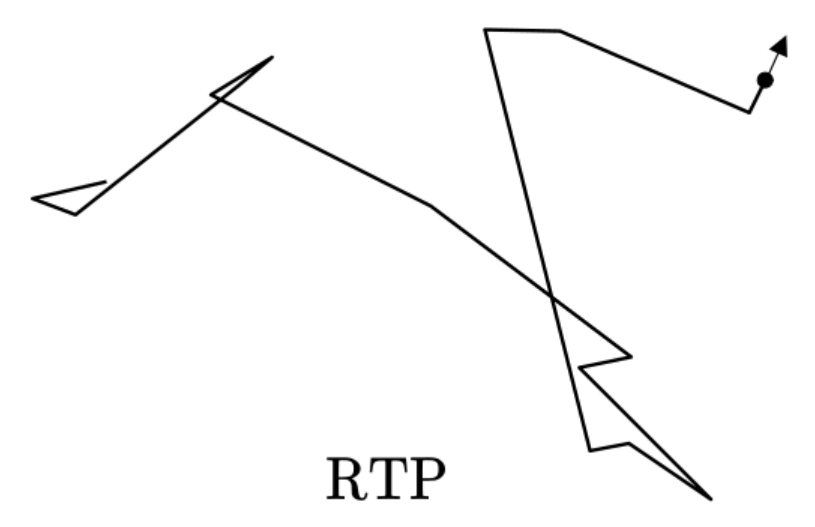}
    \hspace{0.5cm}
    \includegraphics[width=0.35\linewidth]{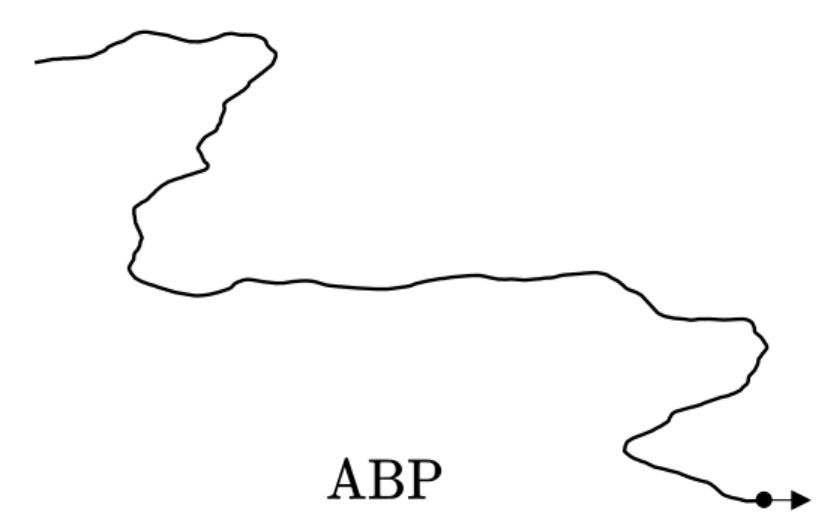}
    \caption{Examples of 2D trajectories for an RTP (left) an ABP (right). For an AOUP the trajectory looks qualitatively similar to an ABP, but the velocity is not constant along the trajectory.}
    \label{fig:trajectories}
\end{figure}

The RTP is the central building block of most of the models that will be studied in this thesis. Since we will essentially be focusing on one-dimensional systems, it is worth giving a bit more details on the 1D case. In this case, the stochastic equation of motion, in the presence of an arbitrary external potential $V(x)$, can be written
\begin{equation}\label{defRTP}
\frac{dx}{dt} = -V'(x) + v_0\, \sigma(t) \quad , \quad
\sigma(t+dt) = \begin{cases}
\sigma(t) &\text{, with probability } (1 - \gamma \, dt)\\
-\sigma(t) &\text{, with probability } \gamma\, dt 
\end{cases}\, ,
\end{equation}
where $\sigma(0)=\pm 1$. The driving noise $\sigma(t)$ is often referred to as ``telegraphic noise" (see Fig.~\ref{fig:sigma1D_example} for an illustration). It is exponentially correlated in time,
\begin{equation} \label{sigmacorr}
    \langle \sigma(t)\sigma(t') \rangle = e^{-2\gamma |t-t'|}
\end{equation}
(we will see that this is a common feature of many active particle models). Here, as in the rest of this thesis, the notation $\langle \dots \rangle$ denotes an average over the noise history. Although these correlations make the process $x(t)$ non-Markovian, it is still possible to write a Fokker-Planck equation for the density distribution of positions by taking into account the driving velocity. Indeed, the particle can be in two states, with $\sigma=+1$ and $\sigma=-1$ respectively, which we will call $+$ and $-$ state in the following. One can thus write a pair of partial differential equations describing separately the evolution of the density of $+$ particles $P_+(x,t)$ and $-$ particles $P_-(x,t)$, and add terms to take into account the tumblings between the two states which couple the two equations. This leads to
\bea \label{FP_RTP}
\partial_t P_+ &=& \partial_x[(-v_0+V'(x)) P_+] -\gamma P_+ +\gamma P_- \;, \\
\partial_t P_- &=& \partial_x[(v_0+V'(x)) P_-] -\gamma P_- +\gamma P_+ \;, \nn
\eea
which is the starting point for the derivation of most of the results presented in the next section. The first term on the right-hand side is a drift term, the particle being subjected to a total force $\pm v_0-V'(x)$ depending on its state, while the terms proportional to $\gamma$ correspond to the tumbling events. These equations can be rewritten in terms of the total particle density $P_s(x,t) = P_+(x,t)+P_-(x,t)$ and of the difference $P_d(x,t) = P_+(x,t)-P_-(x,t)$ as
\bea \label{FP_RTP2}
\partial_t P_s &=& -v_0 \partial_x P_d + \partial_x (V'(x)P_s) \;, \\
\partial_t P_d &=& -v_0 \partial_x P_s + \partial_x (V'(x)P_d) -2\gamma P_d \;. \nn
\eea
In the case of a free RTP, $V'(x)=0$, the two equations can be combined to obtain an equation for the density $P_s(x,t)$, known as the {\it telegrapher's equation},
\be \label{eqtelegraph}
(2\gamma \partial_t +\partial_{t}^2 - v_0^2 \partial_x^2)P_s = 0 \;. 
\ee
Assuming that at $t=0$ the particle is located at $x=0$ and has the same probability to be in the $+$ and $-$ states, i.e., $P_+(x,0)=P_-(x,0)=1/2$, the solution in free space reads \cite{Masoliver1993} (see Fig.~\ref{fig:freeRTP})
\bea \label{propagatorFreeRTP}
P_s(x,t) &=& \frac{e^{-\gamma t}}{2} \Bigg( \delta(x-v_0 t) + \delta (x+v_0 t) \\
&& + \frac{\gamma}{2v_0} I_0\Big(\sqrt{\gamma^2t^2 - \big(\frac{\gamma x}{v_0}\big)^2}\Big) + \frac{\gamma t}{2\sqrt{v_0^2t^2 -x^2}} I_1\Big(\sqrt{\gamma^2t^2 - \big(\frac{\gamma x}{v_0}\big)^2}\Big) \Bigg) \;, \nn
\eea
for $|x|<v_0 t$, where $I_0(x)$ and $I_1(x)$ are modified Bessel functions of the first kind. The delta peaks with a weight decaying exponentially in time correspond to the case where no tumbling event has occurred before time $t$. For a generalization to the 2D case, as well as other results for a 2D free RTP see \cite{2dRTPSantra}. At large times $\gamma t \gg 1$ and for typical displacements $x\ll v_0 t$, the solution \eqref{propagatorFreeRTP} becomes approximately equal to a Gaussian,
\be \label{diff_limitRTP}
P_s(x,t) \sim \frac{e^{-\frac{x^2}{4T_{\rm eff}t}}}{\sqrt{4\pi T_{\rm eff}t}} \quad , \quad T_{\rm eff} = \frac{v_0^2}{2\gamma} \;.
\ee
A free RTP thus behaves diffusively at large times, with an effective diffusion coefficient $T_{\rm eff}$. To observe an effect of the activity on timescales much larger than the inverse tumbling rate, one either needs to confine the particle, or to add some form of interactions. More generally, we expect a RTP to behave diffusively in the limit $\gamma\to+\infty$, where the correlations of the noise become negligible. The result above suggests that one should simultaneously take $v_0\to+\infty$, with the effective diffusion coefficient $T_{\rm eff}$ being fixed. This is the so-called {\it diffusive limit} of the RTP.

To conclude this section, let us mention a few extensions of this model which have been studied. First, as for other active particle models, Brownian noise can be added to the dynamics on top of the RTP noise. Although this often makes analytical computations more difficult, a few results have been obtained in this case, such as the free propagator and first-passage properties \cite{MalakarRTP}. Second, both the tumbling rate and the driving velocity may have different values between the $+$ and $-$ states, leading to an asymmetry in the motion of the particle. These values may also be space-dependent, in particular in the case of chemotaxis \cite{Tailleur_RTP,Cates2012,Singh2020}. The case of a random velocity, drawn from a given distribution after each tumbling (independent of the previous one) is also sometimes considered \cite{SurvivalRTPDriftDeBruyne,RTPsurvivalMori}. Another question, which may be relevant in the case of bacteria, is the duration of the tumbling events, which we have assumed here to be instantaneous, but which may also have a finite, random duration \cite{3statesBasu,Sun2024,slowman2} (this may also be relevant for mappings to two-particle models, see the next chapter). Finally, some observations suggest that the time between tumbling events for bacteria may actually not be exponential, but could be closer to a gamma distribution ($p(\tau)\propto \tau^k e^{-\gamma \tau}$) \cite{Korobkova2006,Xie2011,Theves2013}, or even a power law or log-normal distribution \cite{natureruntime,runtime2}. This of course makes analytical computations considerably more difficult, since the process describing the evolution of the noise is not anymore Markovian, and the Fokker-Planck equation may no longer be used \cite{levywalks,Detcheverry,Naftali2024}.

\subsection{The run-and-tumble particle on a lattice} \label{sec:PRWdef}

Although in this thesis we will mostly be considering models defined in continuous space, equivalents of the RTP model (or persistent random walk) can also be defined on a lattice, both in continuous and discrete time. This is particularly useful to study the effect of contact interactions (see, e.g., \cite{slowman,KH2018}). We will give some examples of results obtained for such models in the next chapter.

In continuous time, the general idea is again that a given particle can be either in the $+$ or $-$ state, and switches between the two states with a rate $\gamma$. A $+$ particle moves to the site on the right with some rate $\lambda$, while a $-$ particle jumps to the left with the same rate (with possibly some additional exclusion rule if one wants to study contact interactions between particles). Some coarse-graining operation may then be performed after an appropriate rescaling of the parameters.

One may also define a discrete time version, by saying that at each time step, the particle jumps in the same direction as the previous time step with some probability $q$ (with $q>1/2$ for positive correlations) and in the opposite direction with probability $1-q$ \cite{Larralde2020,PRWSurvivalLacroixMori}. Denoting $\Delta x$ the lattice spacing and $\Delta t$ the duration of a time-step, the continuous RTP is recovered in the limit $\Delta t\to 0$, $\Delta x\to 0$, $q\to 1$ with $v_0=\frac{\Delta x}{\Delta t}$ and $\gamma = \frac{1-q}{\Delta t}$ fixed. This model will be used as an example in Chapter~\ref{chap:Siegmund}.

\subsection{The active Brownian particle (ABP)}

Another popular model, which is more relevant for active colloids such as Janus particles, is the active Brownian particle (ABP) \cite{ABP2012,ABM,ABM2,Kurzthaler2018}. In this model, the particle self-propels with a constant velocity $v_0$, while undergoing angular diffusion with coefficient $D_R$. In two dimensions, the equations of motion involving the position ${\bf r}=(x,y)$ and the rotation angle $\phi$ read
\begin{equation}\label{defABP}
\frac{dx}{dt} = -\partial_x V(x,y) + v_0 \cos \phi(t) \quad , \quad
\frac{dy}{dt} = -\partial_y V(x,y) + v_0 \sin \phi(t) \quad , \quad
\frac{d\phi}{dt} = \sqrt{2D_R} \ \eta(t),
\end{equation}
where $V(x,y)$ is an external potential and $\eta(t)$ is a Gaussian white noise with zero mean and unit variance, $\langle \eta(t) \eta(t') \rangle = \delta(t-t')$. This model can of course be extended to higher dimensions by adding additional angles. It is less clear how it could be defined in 1D, although the 1D RTP may sometimes be referred to as a 1D ABP. One may also simply consider the projection of a 2D ABP along the $x$-axis.

As in the RTP case, the noise driving the ABP is exponentially correlated in time. Indeed, let us denote $\xi_x(t)=v_0 \cos \phi(t)$ the driving noise along the $x$ direction and assume that the velocity is initially oriented along the $x$-axis. Since $\phi(t)$ is a Gaussian variable with $\langle \phi(t)\phi(t')\rangle = 2D_R \, {\rm min}(t,t')$, one has (using that $\langle e^{iX} \rangle = e^{-\frac{1}{2}\langle X^2 \rangle}$ for any Gaussian variable $X$)
\be \label{ABPcorr}
\langle \xi_x(t) \xi_x(t') \rangle = \frac{v_0^2}{2} \big( e^{-D_R |t-t'|} + e^{-D_R (t+t'+2{\rm min}(t,t'))}\big) \underset{t,t'\to+\infty}{\simeq}  \frac{v_0^2}{2} e^{-D_R |t-t'|} \;.
\ee

Also similar to the RTP case, the process ${\bf r}(t)$ is not Markovian in itself, but the process $({\bf r}(t),\phi(t))$ is Markovian, so that one can still write a Fokker-Planck density for the joint distribution of position and rotation angle $p(x,y,\phi,t)$,
\be \label{FP_ABP}
\partial_t p = \partial_x[(-v_0 \cos\phi + V'(x,y)) p] + \partial_y[(-v_0 \sin\phi + V'(x,y)) p] + D_R \, \partial_\phi^2 p \;.
\ee

Finally, and again similar to the RTP, the large time behavior of an ABP in the absence of external potential, on timescales $t\gg D_R^{-1}$, is effectively diffusive. The effective diffusion coefficient can be easily deduced from \eqref{ABPcorr} at large times by integration,
\be
\langle x(t)^2 \rangle \simeq 2 T_{\rm eff} t \quad , \quad T_{\rm eff} = \frac{v_0^2}{2D_R} \;.
\ee
The diffusive limit of the ABP is obtained for $D_R \to +\infty$, $v_0\to+\infty$ with $T_{\rm eff}$ being fixed.

Before moving on to our last model of interest, we would like to mention a variation of this model, which has recently attracted some attention: the direction reversing active Brownian particle (DRABP) \cite{DRABP1,DRABP2}. It is a combination of the ABP and RTP models, in the sense that it undergoes both rotational diffusion and sudden shifts of direction. Different behaviors may be observed depending on how the timescales of these two effects compare with each other.

\subsection{The active Ornstein-Uhlenbeck particle (AOUP)}

The last model that we want to introduce is the active Ornstein-Uhlenbeck particle (AOUP) \cite{SzamelAOUP}. As its name suggests, the driving noise of this model follows an Ornstein-Uhlenbeck process. In one dimension, the equations of motion read
\begin{equation}\label{defAOUP}
\frac{dx}{dt} = -V'(x) + v(t) \quad , \quad
\tau \frac{dv}{dt} = -v(t) + \sqrt{2D} \, \eta(t)\, ,
\end{equation}
where $V(x)$ is an external potential, $\tau$ is the persistence time, $D$ is a  diffusion coefficient, and $\eta(t)$ a Gaussian white noise with zero mean and unit variance. The main interest of this model is that it introduces time correlations in the noise while retaining its Gaussianity. 
Indeed, as for the previously introduced models, the noise is exponentially correlated in time,
\be \label{AOUPcorr}
\langle v(t) v(t') \rangle = \frac{D}{\tau} e^{-|t-t'|/\tau} + (v(0)^2-\frac{D}{\tau}) e^{-(t+t')/\tau}  \underset{t,t'\to+\infty}{\simeq} \frac{D}{\tau} e^{-|t-t'|/\tau} \;,
\ee
while the stationary distribution of $v(t)$ is Gaussian,
\be \label{AOUP_statv}
p_{st}(v)= \sqrt{\frac{\tau}{2\pi D}} \, e^{-\frac{\tau v^2}{2D}}\, .
\ee

A Fokker-Planck equation can be written for the joint density of position and driving velocity $p(x,v,t)$,
\be \label{FP_AOUP}
\partial_t p = \partial_x[(-v + V'(x)) p] + \partial_v \big(\frac{v}{\tau} p\big)+ D \, \partial_v^2 p \;.
\ee
A particularity of this model is that in the absence of external potential, the equations of motion are exactly the same as for a Brownian particle with inertia. The difference is that any external force acting on the particle acts on $x(t)$ instead of $v(t)$. This means that for a free AOUP, the large time behavior will once again be diffusive, with an effective diffusion coefficient
\be \label{Teff_AOUP}
T_{\rm eff} = D \;.
\ee
The diffusive limit of the AOUP is obtained for $\tau \to 0$, with $D=T_{\rm eff}$ fixed.

\section{Non-Boltzmann steady-states} \label{sec:1particle_potential}

As we have seen in the previous section, on large timescales the behavior of a free active particle is indistinguishable from that of a Brownian particle (at least when looking at the typical fluctuations). The situation is however very different when considering a confined active particle. Indeed, in this case the particle generally reaches a non-equilibrium steady-state which may be very different from the Boltzmann distribution describing an equilibrium particle in the same setting. In particular, when the activity is strong, active particles tend to accumulate near the boundaries. Experimentally, the confinement of active particles can be realized through various means, from the acoustic or optical trapping of active colloids \cite{Takatori2016,Buttinoni2022}, to simply placing a robot inside a parabolic dish \cite{Dauchot2019}. In this section we review some important analytical results that have been obtained for the non-Boltzmann steady-states of the active particle models introduced in the previous section, in the presence of a confining potential, and then in the presence of hard walls.

\subsection{Active particles in a confining potential} \label{sec:confiningPot}

\noindent {\bf Run-and-tumble particle.} For a RTP in one dimension, the stationary distribution in the presence of an arbitrary confining potential has been known for a long-time \cite{Klyatskin1977,Lefever1980}. In the context of active particles, it was studied in detail in \cite{TailleurCates2009} for the harmonic case and in \cite{DKM19} for more general confining potentials (see also \cite{Sevilla} for a mapping to an equilibrium system with space-dependent temperature). The derivation is based on the Fokker-Planck equations \eqref{FP_RTP2} with $\partial_t P_s=\partial_t P_d=0$. Particular care should be taken however in the treatment of the boundary conditions. For convenience, in the following we will work with the external force $F(x)=-V'(x)$ instead of the potential.

An important difference between the telegraphic noise of RTPs and Brownian noise is that it is bounded. In the presence of an external force, this implies that some regions may be inaccessible to the particle in the stationary state. For simplicity, let us focus on the case of a convex potential, i.e., a decreasing $F(x)$. Then, assuming that $F(x)$ is also continuous and unbounded (which is the case, e.g., for a potential of the form $V(x)=\alpha |x|^p$ with $p>1$), the equation of motion \eqref{defRTP} has two fixed points $x_-$ and $x_+$ (with $x_-<x_+$), defined by
\be \label{fixedpoints_singleRTP}
F(x_-)=v_0 \quad , \quad F(x_+)=-v_0 \;,
\ee
corresponding respectively to a particle in the $-$ and in the $+$ state \cite{DKM19}. For particles at the left of $x_-$, the total force acting on the particle is always positive, and for particles at the right of $x_+$ it is always negative. Thus, in the stationary state the density can be non-zero only on the interval $[x_-,x_+]$. In addition, a $+$ particle located at $x_-$ will always move towards the right, while a $-$ particle at $x_+$ will always move towards the left. This implies the boundary conditions
\be \label{BC_singleRTP}
P_+(x_-)=P_-(x_+)=0 \;.
\ee
If the potential is not convex, then the support may be more difficult to determine and it may depend on the initial condition. The stationary density will however always be zero in regions where $|F(x)|>v_0$. See Sec.~\ref{sec:interactions_exact} for a more detailed discussion in the 2-particle case.

Using the boundary conditions \eqref{BC_singleRTP}, the equations \eqref{FP_RTP2} (with the time derivatives set to zero) can easily be integrated to obtain (see \cite{DKM19})
\be \label{eqRTPpotential}
P_s(x) = \frac{A}{v_0^2-F(x)^2} \exp \left( 2\gamma \int_0^x dy \frac{F(y)}{v_0^2-F(y)^2} \right) \quad , \quad P_d(x) = -\frac{1}{v_0} F(x) \, P_s(x) \;,
\ee
for any $x\in[x_-,x_+]$, where $A$ is a constant determined by the normalization $\int_{x_-}^{x_+} P_s(x)dx=1$.
\\

\begin{figure}
    \centering
    \includegraphics[width=0.45\linewidth]{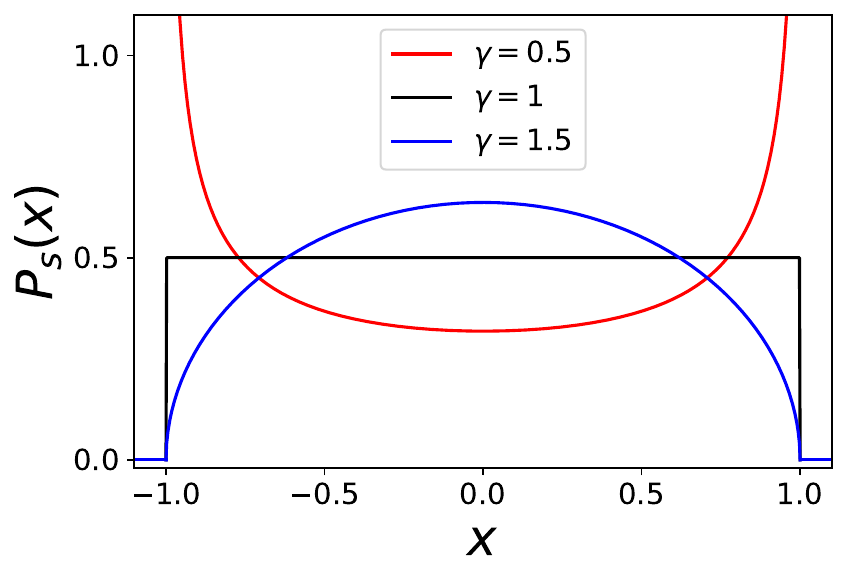}
    \caption{Probability density $P_s(x)$ for a 1D RTP in a harmonic potential, as given by \eqref{eqRTPharmonic}, for $\mu=1$, $v_0=1$ and different values of $\gamma$. For $\gamma<\mu$, the density diverges at the edges of the support.}
    \label{fig:RTPharmonic}
\end{figure}

{\it Harmonic potential.} In the case of a harmonic potential $V(x)=\frac{\mu}{2}x^2$, one has $x_\pm = \pm v_0/\mu$, and the result \eqref{eqRTPpotential} reads
\be \label{eqRTPharmonic}
P_s(x) = \frac{2}{4^{\gamma/\mu}B(\gamma/\mu,\gamma/\mu)} \frac{\mu}{v_0} \left[1-\left(\frac{\mu x}{v_0}\right)^2\right]^{\frac{\gamma}{\mu}-1} \;,
\ee
where $B(\alpha,\beta)$ is the beta function. At small persistence times, i.e., for large tumbling rates $\gamma>\mu$, this density is unimodal and vanishes at the edges $x_\pm$. However, for larger persistence time, i.e., for $\gamma<\mu$, the density becomes bimodal and diverges at the the edges (see Fig.~\ref{fig:RTPharmonic}). This second behavior is very different from what is observed in the Brownian case (where the density is Gaussian), and it reflects the tendency of active particles to accumulate near boundaries when the persistence is strong (i.e., here for small $\gamma$). Using the second equation in \eqref{eqRTPpotential}, one can also obtain the stationary densities of $+$ and $-$ particles respectively, which read
\be \label{eqRTPharmonic_sign}
P_\pm(x) = \frac{A}{2} \left(1\pm\frac{\mu x}{v_0}\right)^{\frac{\gamma}{\mu}} \left(1\mp\frac{\mu x}{v_0}\right)^{\frac{\gamma}{\mu}-1}
\ee
where $A$ is the normalization constant appearing in \eqref{eqRTPpotential}. As expected, the divergence at $x_+$ is only due to the $+$ particles and the one at $x_-$ to the $-$ particles, i.e., the particles with a driving velocity oriented towards the edge of the support.

In the harmonic case, the first moments of the particle position can be obtained by integrating the equation of motion \eqref{defRTP} and averaging over the noise. This yields $\langle x(t) \rangle=x(0)e^{-\mu t}$ and
\be
\langle x(t)^2 \rangle - \langle x(t) \rangle^2 = \begin{dcases} v_0^2 \left[ \frac{1}{\mu(\mu+2\gamma)} + \frac{2e^{-(\mu+2\gamma)t}}{4\gamma^2-\mu^2} + \frac{e^{-2\mu t}}{\mu(\mu-2\gamma)} \right] \quad \text{for } \mu \neq 2\gamma \;, \\
\frac{v_0^2}{8\gamma^2} [1-e^{-4\gamma t} -4\gamma t e^{-4\gamma t}] \hspace{3.2cm} \text{for } \mu=2\gamma \;. \end{dcases}
\ee
This suggests that the relaxation timescale is given by $\lambda_0=\min(\mu+2\gamma,2\mu)$, which is confirmed in \cite{DKM19} by a study of the time-dependent solution in Laplace space.

Concerning the stationary density, a qualitatively similar behavior is obtained for anharmonic potentials of the form $V(x)=\alpha |x|^p$ with $p>1$, with a transition between a vanishing and a diverging density at the edges of the support for some critical value of $\gamma$. Let us also mention that the case of a RTP in a harmonic potential with an additional state with zero driving velocity was also studied in \cite{3statesBasu,Sun2024}, leading to a similar behavior but with a possible additional divergence at $x=0$. In 2D, the case of a RTP with 4 possible orientations separated by $90°$ in a harmonic trap was studied in \cite{Smith2DRTP}.
\\

{\it Linear potential.} The case of a linear potential $V(x)=a|x|$ is quite different from the harmonic case. Indeed, in this case the external force $F(x)=-a \, {\rm sgn}(x)$ is constant on each half of the real line. Thus, if $a \geq v_0$, the particle will move towards $x=0$ and remain there. In this case, the steady state distribution is a single delta function at $x=0$. Conversely, if $a<v_0$, the support of the density is unbounded, and \eqref{eqRTPpotential} yields for any $x$,
\be \label{eqRTPlinear}
P_s(x) = \frac{\gamma a}{v_0^2-a^2} \exp \left( -\frac{2\gamma a}{v_0^2-a^2} |x| \right) \;.
\ee
Note that for $V(x)=\alpha |x|^p$ with $p<1$, the stationary density is always a delta at $x=0$ since the force diverges close to $x=0$ (and is oriented towards $x=0$).
\\

{\it Periodic force and sedimentation.} Before moving on, let us mention that the stationary state of a 1D RTP with periodic boundary conditions, subjected to an arbitrary external force (including a random landscape), was studied in \cite{LMS2020}. In this case, depending on the choice of the external force, the particle may have a non-zero average velocity, which was computed explicitly along with the effective diffusion coefficient at large times. Let us also mention the somewhat simpler, but nevertheless interesting setting of sedimentation, i.e., of an RTP on the positive half-line subject to a constant negative force $-f$ and with a zero-flux boundary condition at $x=0$ \cite{TailleurCates2009}. As we will see in the next section, in the case of active particles this type of boundary condition has to be treated with care, and the behavior of the density near $x=0$ depends on the way in which this boundary condition is implemented. However, away from $x=0$ the stationary density was found to decay exponentially, as in the Brownian case, but with a different characteristic length-scale $\delta_s=\frac{v_0^2-f^2}{2\gamma f}$. For $v_0\gg f$, $\delta_s$ converges to the Brownian result with diffusion coefficient $T_{\rm eff}$, $\delta_s=T_{\rm eff}/f$, while for $v_0 \to f^+$ it converges to zero, and the density collapses to a delta function at $x=0$.
\\

\noindent {\bf Active Ornstein-Uhlenbeck particle.} Computing analytically the stationary distribution of an AOUP with an external potential is generally more challenging than for an RTP. A notable exception is the case of a harmonic potential $V(x)=\frac{\mu}{2}x^2$, which due to the Gaussian nature of the AOUP can be computed quite easily \cite{SzamelAOUP}. One finds that the stationary density is Gaussian, but with an effective temperature which is different from the one of a free AOUP given in \eqref{Teff_AOUP},
\be
P_s(x) = \sqrt{\frac{\mu}{2\pi T_{\rm eff}^\mu}} \; e^{-\frac{\mu x^2}{2T_{\rm eff}^\mu}}  \quad, \quad T_{\rm eff}^{\mu} = \frac{D}{1+\mu\tau} \;.
\ee

The effect of more general external potentials was studied using an expansion at small persistence time $\tau$ (with $\tau \to 0$ corresponding to the Brownian limit) \cite{Bonilla2019,Wijland21}. 
Finally, for an AOUP the case of sedimentation is found to give exactly the same result as in the Brownian case, i.e., the density decays exponentially away from $x=0$ on a length-scale $\delta_s=T_{\rm eff}/f$, with $T_{\rm eff}$ given in \eqref{Teff_AOUP} \cite{SzamelAOUP}.
\\

\noindent {\bf Active Brownian particle.} Analytical studies of the steady state inside a confining potential are even more challenging when it comes to the ABP model, since it requires to work in at least two dimensions. In the harmonic case however, the time dependent moments were computed, allowing for a detailed analysis of the different limiting regimes \cite{ABM2}. 
As in the RTP case, a crossover from a unimodal to a bimodal distribution is observed as the activity parameter $\mu/D_R$ is increased, although in this case no divergence is found. Similar results were obtained in the presence of additional thermal noise \cite{Malakar2020,Chaudhuri2021,Franosch2022}.


\subsection{Active particles with hard walls} \label{sec:HardWalls}

Another way to confine a particle is simply to add walls, i.e., to impose zero-flux boundary conditions. However, in the case of active particles there are several ways to impose such boundary conditions, which are not equivalent. One could for instance impose reflective boundary conditions, meaning that when the particle reaches the wall its direction of motion is instantly reversed \cite{AngelaniReflecting}. However, this type of boundary condition does not account for an important characteristic of active particles, namely their tendency to accumulate near walls due to their persistent motion \cite{Yang2014,Uspal2015}. In this section, as well as in Part~\ref{part:siegmund} of this thesis, we consider instead what we will call a {\it hard wall} boundary condition, which can be understood as an infinite step of potential. This means that the particle remains stuck at the wall as long as its total velocity is oriented towards it, and it only moves away from the wall when the orientation of its velocity reverts naturally (e.g., at the next tumbling event for a RTP). We now review some analytical results that have been obtained for different active particle models in the presence of such boundary conditions.
\\

\noindent {\bf Run-and-tumble particle.} For a one-dimensional RTP in the presence of hard walls (but without any external force) an exact time-dependent solution of the Fokker-Planck equations \eqref{FP_RTP} in Laplace space was obtained in \cite{AngelaniHardWalls}. Here we focus on the determination of the steady-state. Consider a 1D RTP on an interval $[a,b]$ with hard walls at $a$ and $b$. In the absence of external force, the stationary version of \eqref{FP_RTP2} implies that $P_d(x)=cst$ and $\partial_x P_s(x)=-\frac{2\gamma}{v_0} P_d(x) = cst$. The zero-flux boundary conditions ($\partial_x P_s(x)=0$ at $x=a$ and $b$) then imply that the density is constant on $[a,b]$, with $P_+=P_-$. However, due to the persistent motion of the RTP, the density also includes delta functions at $a$ and $b$. We will denote their weights $\kappa_a$ and $\kappa_b$ respectively. The delta peak at $b$ only contains $+$ particles, while the one at $a$ only contains $-$ particles. In the stationary state, the value of $\kappa_b$ can thus be obtained by balancing the flux of $+$ particles arriving at the wall with the fraction of $+$ particles at the wall which undergo a tumbling event per unit time, i.e., $\gamma \kappa_b = v_0 P_+$, and similarly for $\kappa_a$. The only remaining step is then to determine the normalization constant using $(b-a)P_s+\kappa_a+\kappa_b=1$, which leads to
\be \label{HardwallsNoforce}
\kappa_a=\kappa_b = \frac{1}{2} \frac{1}{1+\frac{\gamma L}{v_0}} \quad , \quad P_+=P_-= \frac{1}{2} \frac{1}{L+\frac{v_0}{\gamma}} \quad , \quad L=b-a \;.
\ee

In \cite{AngelaniHardWalls}, the expression of the pressure exerted on the wall by the particle is also computed, and a generalization to boundaries which affect the tumbling-rate is also studied. Some results have also been derived for a 2D RTP. In \cite{Lee2013}, the steady-state distribution is computed for a 2D "discrete" RTP (i.e., the angle of the velocity can only take a finite number of values) inside a channel. In \cite{PireySphere2023} the effect of a spherical obstacle on the steady-state distribution of a 2D RTP is studied. In this case the stationary density is found to exhibit both a delta peak at the point of contact with the obstacle, as well as an algebraic divergence near the obstacle.
\\

\noindent {\bf AOUP and ABP.} For other active particle models, the effect of a hard wall boundary condition is more difficult to study analytically. Concerning the AOUP, some approximate analytical results have been obtained (accompanied by numerical simulations) for a ``smooth" version of the hard wall, i.e., a very steep harmonic potential \cite{hardWallsJoanny,hardWallsCaprini}. For the ABP, approximate expressions have been obtained for the stationary distribution in the presence of hard walls in different geometries \cite{Duzgun2018}.

\section{First-passage properties of active particles} \label{sec:firstpassage}

The last question that we want to address in this chapter is the effect of absorbing boundary conditions, i.e., the first-passage properties of active particles. First-passage problems are a vast topic with applications in a large variety of fields, from biology and chemistry (e.g., for the determination of reaction times), to mathematical finance, as well as computer science where random search algorithms play an important role in many applications \cite{redner,Bray2013,Metzler_book}. These questions are also strongly connected to the field of extreme value statistics, which is itself essential in many contexts \cite{Comtet2005,reviewEVSPal,livreSG}. In this section, we give a very brief overview of the type of first-passage problems that are generally considered and the methods to address them, with a focus on one-dimensional stochastic processes, before rapidly reviewing the existing results in the case of active particles.

\subsection{General definitions and Brownian case} 

Consider a generic 1D stochastic process $x(t)$, representing the position of some particle. We denote $x(0)=x$ the initial position of the particle at time $t=0$, and we place a target, or an {\it absorbing} boundary condition at some position $b$, meaning that if the particle reaches this point at some time $t$, it will remain there at any later times. The first quantity that one may be interested in is the {\it survival probability}, i.e., the probability that the particle has not yet been absorbed at time $t$, which we denote $Q_b(x,t)$. For a symmetric random walk and in the absence of external force, this quantity generally decays algebraically to zero at large times, $Q_b(x,t)\sim t^{-\theta}$, and $\theta$ is called the {\it persistence exponent} \cite{Bray2013,SirePersistence1,SirePersistence2,DerridaPersistence}. A related quantity is the {\it first-passage time distribution}, which can be computed as 
\be
\mathcal{F}_b(x,t)=-\partial_t Q_b(x,t) \;.
\ee
If they are well-defined, one may also compute the moments of this distribution, in particular the {\it mean first-passage time} (MFPT),
\be \label{def_MFPT}
T_b(x) = \int_0^\infty t \mathcal{F}_b(x,t) dt = \int_0^\infty Q_b(x,t) \;.
\ee

\begin{figure}
    \centering
    \includegraphics[width=0.45\linewidth,trim={9cm 0.5cm 10.5cm 0.5cm},clip]{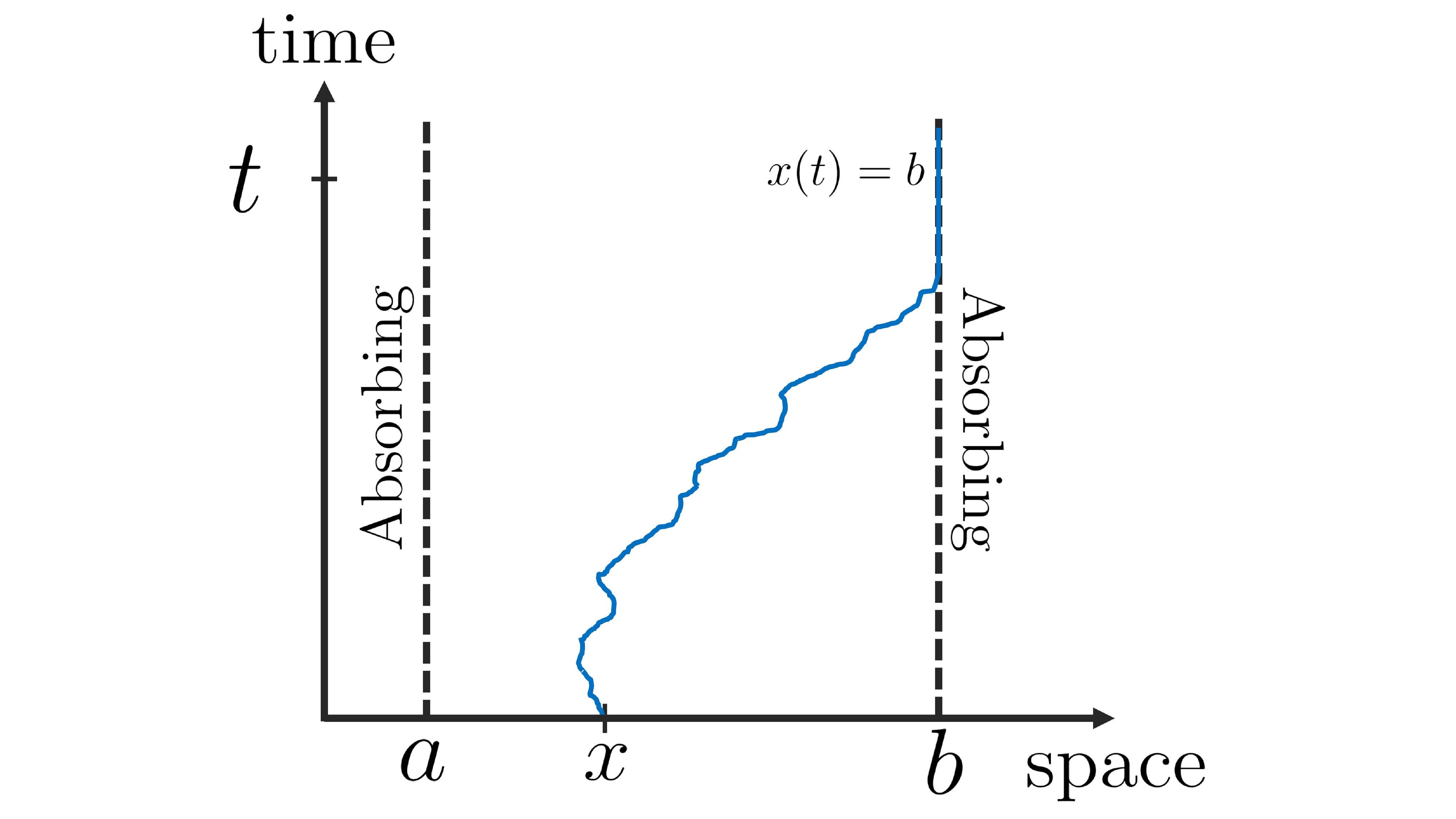}
    \caption{Example of a trajectory of a stochastic process $x(t)$ with absorbing boundary conditions at $a$ and $b$, which gets absorbed at $b$.}
    \label{fig:absorbingBCBrownian}
\end{figure}

One may also consider the same process on an interval $[a,b]$ with absorbing boundary conditions at both $a$ and $b$ (as in Fig.~\ref{fig:absorbingBCBrownian}). In this case one may still want to study the survival probability, which we denote $Q_{[a,b]}(x,t)$, and related quantities, but another question that arises is: what is the probability that the particle reaches a given boundary, e.g., $b$, before reaching $a$ ? This is called the {\it exit probability} (also known as {\it hitting} or {\it splitting} probability), and we will denote it $E_b(x)$. 

All the quantities that we have just introduced are encoded in a single quantity, namely the probability that the particle is absorbed at $b$ before time $t$~\footnote{Note that this quantity depends implicitly on the position $a$ of the other wall if it exists, although this is not made explicit in the notation.},
\be \label{defExit}
E_b(x,t) = \mathbb{P}(x(t) = b | x(0)=x) \;.
\ee
The exit probability defined previously is simply the infinite time limit of this quantity,
\be \label{rel_exit_limit}
E_b(x)=\underset{t\to+\infty}{\lim} E_b(x,t) \;,
\ee
while the survival probability can be obtained as
\be \label{rel_survival_exit}
Q_{[a,b]}(x,t) =1-E_a(x,t)-E_b(x,t) \; .
\ee
These definitions still make sense when only one boundary is present, i.e., for $a\to-\infty$. The finite time exit probability is not often studied in itself as it is generally difficult to compute, but it will be at the center of Part~\ref{part:siegmund} of this thesis. 

In the case of continuous stochastic processes, a useful method to study first-passage quantities is the backward Fokker-Planck equation \cite{redner,Bray2013,Metzler_book}. For a Brownian motion at temperature $T(x)$ (which may depend on the position), in the presence of an external potential $V(x)$, it reads, for $x\in [a,b]$ (with the It\=o convention),
\be \label{BFP}
\partial_t E_b(x,t) = -V'(x) \partial_x E_b(x,t) + T(x)\partial_x^2 E_b(x,t)  
\ee
(note that in the presence of a single absorbing wall, the survival probability $Q_b(x,t) =1-E_b(x,t)$ satisfies the same equation). The hitting probability $E_b(x)$ can be easily obtained from this equation by setting the time derivative to zero, and using the boundary conditions $E_b(a)=0$ and $E_b(b)=1$. For a constant diffusion coefficient $T(x)=T$, we obtain
\be \label{exit_brownian}
E_b(x)=\frac{\int_a^x dz\, e^{\frac{V(z)}{T}}}{\int_a^b dz\, e^{\frac{V(z)}{T}}} \;.
\ee
In the absence of external potential, it is simply linear. Interestingly, the expression \eqref{exit_brownian} is exactly the same as the cumulative of the stationary distribution of positions for a Brownian particle with hard walls at $a$ and $b$, up to a change $V(x)\to -V(x)$. This connection between absorbing boundary condition and hard walls is actually much more general. This is related to the concept of {\it Siegmund duality} \cite{Siegmund}, which will be the topic of Part~\ref{part:siegmund} of this thesis.

Let us also recall the expression for the survival probability for a Brownian particle on the interval $(-\infty,b)$ (i.e., with only an absorbing wall at $b$), in the absence of external potential,
\be \label{survival_BM_intro}
Q_b(x,t) = \text{erf}\left(\frac{b - x}{\sqrt{4T\, t}}\right) \underset{t\to+\infty}{\sim} \frac{b-x}{\sqrt{\pi T t}} \; ,
\ee
where ${\rm erf}(z)=\frac{2}{\sqrt{\pi}}\int_0^z e^{-u^2} du$ is the error function, leading to a persistence exponent $\theta=1/2$. Note that this implies that the MFPT \eqref{def_MFPT} is infinite in this case. The expression \eqref{survival_BM_intro} is derived in Appendix~\ref{survivalBM} using the method of images. As for \eqref{exit_brownian}, a connection with hard wall boundary conditions is also discussed there.

\subsection{Known results for active particles} 

As we have mentioned above, random search processes appear in many contexts in biology, from sperm cells searching for an oocyt during reproduction to animal foraging for food \cite{benichou1,benichou2}. The question of how first-passage properties are affected by active noise is thus particularly relevant, and it has attracted a lot of attention in recent years. Here we give a very brief overview of the results that have been obtained in this context (for a more detailed review see, e.g., \cite{Targetsearch}).
\\

\begin{figure}
    \centering
    \includegraphics[width=0.75\linewidth, trim={0.5cm 0.5cm 0.5cm 8.5cm},clip]{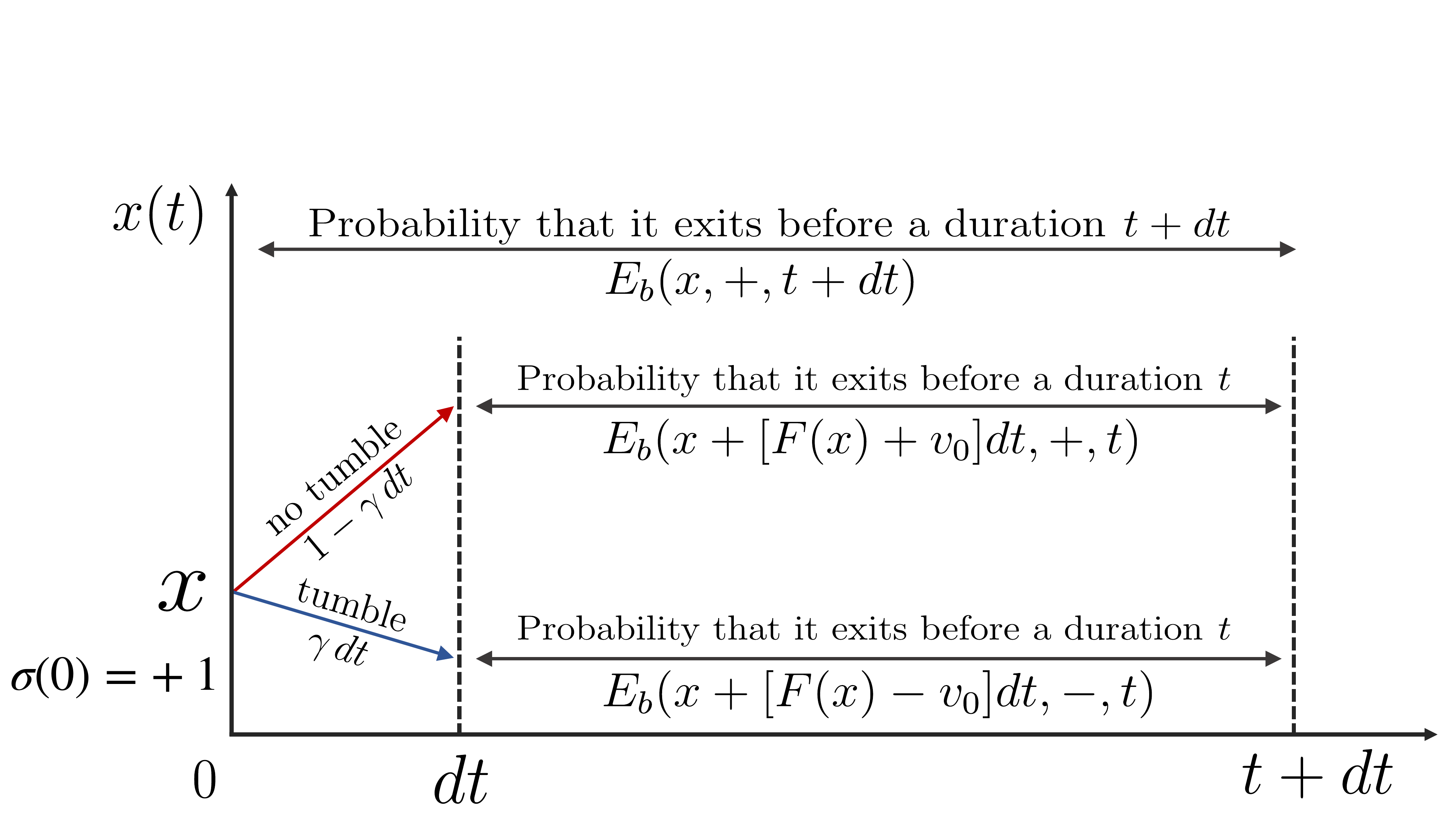}
    \caption{Schematic representation of the derivation of the backward Fokker-Planck equations for a RTP \eqref{timeBFP}.}
    \label{fig:backwardFP}
\end{figure}

\noindent {\bf Run-and-tumble particle.} Once again, it is for the RTP model that the largest number of results has been obtained, thanks to its relative simplicity. We start by re-deriving the backward Fokker-Planck equation for the exit probability at finite time for a 1D RTP, in the presence of an external force $F(x)=-V'(x)$, which will be useful in Part~\ref{part:siegmund}. In the case of an active particle, it is of course important to take into account the initial orientation of the driving velocity. We thus need to define two probabilities, depending on the initial state $\sigma(0)=\pm1$ of the particle
\begin{equation}
    E_b(x,\pm,t) = \mathbb{P}\left(x(t)=b\, |\, x(0)=x, \sigma(0)=\pm 1 \right)\, .
\end{equation}
If we assume that $\sigma(0) = +1$ or $\sigma(0)=-1$ with probability $1/2$, the exit probability at time $t$ regardless of the initial velocity is
\begin{equation}
    E_b(x,t) = \frac{1}{2}\left(E_b(x,+,t)+E_b(x,-,t)\right)\,.
\end{equation} 
We want to obtain a pair of coupled first-order differential equations for $E_b(x,\pm,t)$. Let us evolve the particle during an infinitly small time interval $[0,dt]$, and average over the possible trajectories. Suppose the RTP starts its motion at $x$, then at order $O(dt)$, for the positive state of the RTP, the only possible events are that the particle switches sign, with probability $\gamma\, dt$, and moves from $x$ to $x + \left[F(x)-v_0\right]\, dt$, or that it moves from $x$ to $x +\left[F(x)+v_0\right] \, dt$ while staying in the positive state, with probability $1-\gamma dt$ (see Fig.~\ref{fig:backwardFP}). This translates to
\begin{equation}
    E_b(x,+,t+dt)=(1-\gamma dt)\, E_b(x+\left[F(x)+v_0\right]\, dt,+, t) + \gamma dt E_b(x+\left[F(x)-v_0\right]\, dt,-, t)\, .
\end{equation}
After Taylor-expanding at order $dt$ (and repeating the operation for $E_b(x,-,t)$), we obtain the desired equations
\begin{eqnarray}\label{timeBFP}
    \partial_t E_b(x,+,t)=\left[F(x)+v_0\right]\partial_x E_b(x,+,t) +\gamma E_b(x,-,t)- \gamma E_b(x,+,t)\, , \\
    \partial_t E_b(x,-,t)=\left[F(x)-v_0\right]\partial_x E_b(x,-,t) +\gamma E_b(x,+,t)- \gamma E_b(x,-,t)\, . \nonumber
\end{eqnarray}
These equations can also be used to derive results for the exit probability $E_b(x)$ and the survival probability using \eqref{rel_exit_limit} and \eqref{rel_survival_exit}. It is important to note however that the boundary conditions are slightly less trivial than in the Brownian case. Indeed, a particle starting infinitely close to an absorbing boundary still has a non-zero probability to escape this boundary as long as its driving velocity is initially oriented in the opposite direction. One thus only has the boundary conditions $E_b(a^+,-,t)=0$ and $E_b(b^-,+,t)=0$.

For a free RTP in one dimension, the exact expression of the survival probability is well-known \cite{Masoliver92,Orsingher95,Targetsearch,MalakarRTP,MukherjeeOccupationTime}. It is more complex than in the Brownian case, with some discontinuities (in particular the survival probability remains strictly equal to one as long as $x>v_0 t$), but the large time behavior is similar, with the same persistence exponent $\theta=1/2$, although with an asymmetry between particles initialized in the $+$ and $-$ states,
\be \label{survivalFreeRTP_largetime}
Q_b(x,+,t) \underset{t\to+\infty}{\sim} \frac{b-x}{\sqrt{\pi T_{\rm eff} t}} \quad , \quad Q_b(x,-,t) \underset{t\to+\infty}{\sim} \frac{b-x+v_0/\gamma}{\sqrt{\pi T_{\rm eff} t}} \;,
\ee
where we recall that $T_{\rm eff}=\frac{v_0^2}{2\gamma}$. Note that, while $Q_b(x,+,t)$ naturally vanishes ar $x=b$, this is not the case for $Q_b(x,-,t)$ (i.e., when the initial velocity is oriented away from the wall), in agreement with the discussion above. The survival probability behaves as if the initial position of the particle was shifted away from the wall by a distance $v_0/\gamma$, which corresponds to the {\it persistence length} of the RTP (i.e., the average distance that it travels before the first tumbling event). This phenomenon also arises in the context of discrete-time random walks, where this characteristic length is called the {\it Milne extrapolation lenth} (see \cite{LMS2019} and references therein). The exit probability can be easily obtained from \eqref{timeBFP} with $F(x)=0$ by setting the time derivatives to zero \cite{MalakarRTP,Singh2020,Singh2022}. It it linear, as in the Brownian case, but with discontinuities at the boundaries since $E_b(a^+,+)>0$ and $E_b(b^-,-)<1$,
\begin{equation} \label{exitproba}
    E_b(x,+) = \frac{1+\frac{\gamma}{v_0}(x-a)}{1+\frac{\gamma}{v_0}(b-a)} \quad , \quad  E_b(x,-) = \frac{x-a}{\frac{v_0}{\gamma}+(b-a)}\, .
\end{equation}

Both the survival probability and the exit probability have also been computed in the presence of additional Brownian noise \cite{MalakarRTP}. We also mention the study of  the extremal statistics of a 1D RTP in the presence of an absorbing wall in \cite{Singh2022}, as well as a perturbative computation of the MFPT with two absorbing walls in \cite{MFPT1DABP}. Some results were also obtained in higher dimensions, in particular the survival probability of a $d$-dimensional RTP in a half-space\cite{RTPsurvivalMori}, as well as the MFPT to a target inside a bounded domain \cite{TVB12,RBV16}.

The survival probability of a 1D RTP in the presence of a constant drift as been computed exactly in \cite{SurvivalRTPDriftDeBruyne}. If the drift is oriented towards the absorbing wall, the MFPT becomes finite (as in the Brownian case), and it has been computed explicitly. For more general external potentials, the computations become more difficult and mostly the MFPT has been studied (the survival probability with a harmonic potential was studied in \cite{DKM19}, but a fully explicit solution is only available in Laplace space). A general expression for an arbitrary external potential was obtained in \cite{AngelaniMFPT} in the presence of a reflective boundary condition. In \cite{MathisMFPT}, the MFPT is studied in detail for external potentials of the form $V(x)=\alpha|x|^p$. For $p>1$, it was shown that there exists an optimal tumbling rate which minimizes the MFPT. See also \cite{MathisMFPT2,Grange2025} for recent studies of the MFPT with more general forces, as well as \cite{AngelaniMFPT,Singh2020} for the case of a space-dependent tumbling rate rate. Finally, a lot of recent studies consider the case of partially absorbing conditions, meaning that a particle in contact with the wall is only absorbed with a certain rate \cite{BressloffStickyBoundary,BressloffStickyBoundaries,AngelaniGenericBC,AngelaniOptimalEscapes,RTPpartiallyAbsorbingTarget}. We also mention that the probability for 2 RTPs on the real axis to not cross up to time $t$ was computed in \cite{LMS2019}. Contrary to the Brownian case, it cannot be mapped directly to the survival probability of a single RTP, but it still decays at large time as $t^{-1/2}$.
\\

\noindent {\bf AOUP and ABP.} For other models of active particles, obtaining exact results for the first-passage properties is generally much more difficult. Currently, only a few situations have been investigated (often requiring some approximations or some numerical steps, or the use of a perturbative approach), such as the survival probability of an ABP in a half plane \cite{ABM,Baouche2025} or in a 2D channel with a partially absorbing wall \cite{{BressloffABP}} (without external force), as well as the related problem of the escape rate of an AOUP or ABP through a potential barrier \cite{AOUPEscapeLecomte,Caraglio2024}.

\section{Conclusion}

In this chapter, we gave an overview of the existing results for active particles at the single particle level, with an emphasis on analytical methods and exact results for the non-equilibrium steady-states and first-passage properties. Already at this level, one may be surprised by the diversity of the phenomena that can be observed and by the complexity of the problems that arise. In the next chapter, we will see that an even richer variety of behaviors appears when interactions are added.

We have also hinted at the surprising connection which exists between first-passage properties and spatial distribution with hard walls in the case of Brownian motion, known as Siegmund duality, and it is natural to wonder if this may be extended to active particles. This will be the topic of Part~\ref{part:siegmund} of this thesis.

\chapter{Interacting active particles} \label{chap:interactions}

\section{Motivations and phenomenology} \label{sec:interactions_pheno}

In the previous chapter we focused on situations where a single active particle is present, or where the particles can be considered independent. There are however many ways in which active particles can interact together, from hydrodynamic interactions between active colloids or bacteria evolving in a fluid, to alignment between neighbors in animal herds. Due to the non-equilibrium nature of active particles, these interactions often lead to completely new collective phenomena which would be impossible in an equilibrium system \cite{Ramaswamy2010,Bechinger,Marchetti2018,Marchetti2013,Ramaswamy2017}. Even simple steric interactions (i.e., hard-core repulsion) may lead to a new type of phase separation.

In this section we will briefly review the main classes of interactions that have been studied, as well as the collective effects that emerge when they are present, including motility-induced phase separation (MIPS) and the transition to flocking. Since such systems are generally extremely challenging to study analytically, many theoretical studies rely either on numerical simulations, or on field theories obtained using symmetry arguments. Whenever hydrodynamic equations can be derived directly from a microscopic model via some coarse-graining procedure, this is generally at the cost of some approximations. In Sec. \ref{sec:interactions_exact} we will review in more details some models for which exact results have been obtained.

As in the single particle case, the study of interacting active particles is a very vast topic and some questions will not be covered in this chapter. In particular, we will not mention mixtures of active and passive particles, although such systems have also been shown to exhibit interesting collective behaviors \cite{Poon2012,Stenhammar2015}. Non-reciprocal interactions, which are also very relevant in the context of active particles, will be introduced in Chapter~\ref{chap:activeRD} of this thesis.

\subsection{Steric interactions and quorum sensing: the emergence of motility-induced phase separation}

Motility-induced phase separation (MIPS) is one of the most emblematic phase transitions in active matter. It corresponds to the spontaneous separation of a system of active particles into high density and low density regions \cite{CT2015,OByrne2021}. While in equilibrium systems, such transitions generally require the presence of attractive interactions between the particles, the peculiarity of MIPS is that, for active particles, short-range repulsive interactions (e.g., hard-core repulsion) are enough to observe phase separation. MIPS is therefore a purely out-of-equilibrium phase transition. The physical interpretation behind this effect is that, due to their persistent motion, active particles tend to get stuck against each other when they collide, for a time of the order of their typical reorientation time (similar to how they tend to get stuck at a wall, see Sec.~\ref{sec:HardWalls}). This leads to a slowing down of the dynamics of the particles in regions of high density, which in turn leads to the accumulation of particles in these regions. When the right conditions are met, i.e., for sufficiently strong activity and at high enough density, this feedback loop may lead to phase separation.

\begin{figure}
    \centering
    \includegraphics[width=0.4\linewidth]{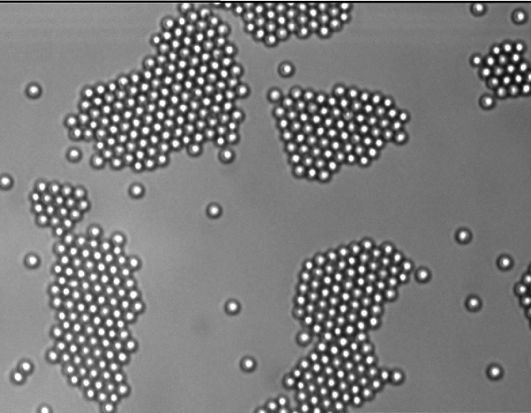}
    \hspace{0.8cm}
    \includegraphics[width=0.4\linewidth]{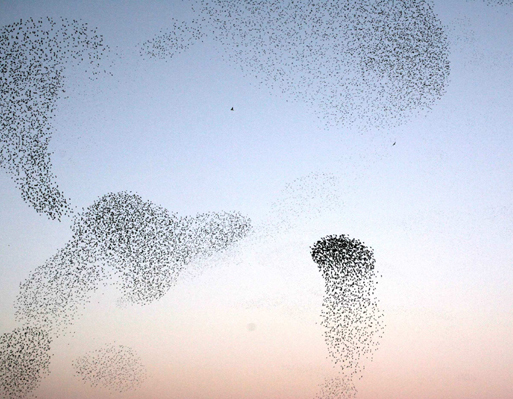}
    \caption{{\bf Left:} Motility induced phase separation observed in a solution of self-phoretic colloids (with a self-propulsion mechanism controlled by light). Figure adapted from \cite{Palacci2013}. Reprinted with permission from AAAS. {\bf Right:} Photograph of a typical flock of starlings. Figure adapted from \cite{Ballerini2008}. Copyright (2008) National Academy of Sciences, U.S.A.}
    \label{fig:MIPSflockingExp}
\end{figure}

MIPS has been observed experimentally in solutions of active colloids \cite{Buttinoni2013,Palacci2013} (see left panel of Fig.~\ref{fig:MIPSflockingExp}) and bacteria \cite{Liu2019}. It was also found to occur numerically in a variety of models \cite{FM2012,Redner2013,Levis2014,FHM2014,SG2014}. Theoretically, it was shown to arise in all the most famous models of active particles, including RTPs \cite{Tailleur_RTP}, ABPs \cite{Bialke2013}, AOUPs \cite{Wijland21}, as well as in lattice models \cite{Thom2011}, which suggests that it is a very general feature of active particles with repulsive interactions. Most theoretical studies are either based on field theories derived from symmetry arguments \cite{Wittkowski2014,Solon2018_1,Solon2018_2}, or rely on the fact that the effect of the interactions can be approximated at the coarse-grained level by introducing an effective driving velocity $v_{\rm eff}(\rho)$ which decreases with the density $\rho$ \cite{Bialke2013,OByrne2021}. In this case, one can easily show that the density in the stationary state is inversely proportional to the velocity, $\rho= c/v_{\rm eff}(\rho)$ where $c$ is a normalization constant. Starting from a uniform density $\rho_0$ perturbation of the density $\delta \rho$ then leads to a perturbation of the effective velocity $v_{\rm eff}(\rho_0+\delta \rho) \simeq v_{\rm eff}(\rho_0)+v_{\rm eff}'(\rho_0)\delta \rho$, which in turn leads to a perturbation of the density $\delta \rho'$,
\be
\rho_0 + \delta \rho' \simeq \frac{c}{v_{\rm eff}(\rho_0)+v_{\rm eff}'(\rho_0)\delta \rho} \simeq \frac{c}{v_{\rm eff}(\rho_0)} (1-\frac{v_{\rm eff}'(\rho_0)}{v_{\rm eff}(\rho_0)}\delta \rho) = \rho_0 - \rho_0\frac{v_{\rm eff}'(\rho_0)}{v_{\rm eff}(\rho_0)}\delta \rho \;.
\ee
Using this qualitative argument, we easily see that, if the effective velocity decreases fast enough as the density increases, i.e., if 
\be
-\frac{v_{\rm eff}'(\rho_0)}{v_{\rm eff}(\rho_0)} > \frac{1}{\rho_0} \;,
\ee
then the perturbation is amplified, i.e., $\delta \rho'>\delta \rho$, which leads to phase separation. A more detailed study of the approximate hydrodynamic equations allows for a precise description of the phase diagram.

Note that a density-dependent velocity may also arise in different contexts. In particular, some bacteria liberate chemicals in their environment, and adapt their movement in reaction to the chemicals emitted by other bacteria around them \cite{QuorumSensing}. This effect is called {\it quorum sensing}, and it may also lead to motility-induced phase separation \cite{CT2015,OByrne2021}.

\subsection{Alignment interactions and the flocking transition}

Historically, one of the first motivations for the study of active matter was to understand the collective motion of animals. The Vicsek model was introduced in \cite{Vicsek} as a minimal model for the formation of animal herds. It consists in an assembly of spins (in 2 or more dimensions), which move in the direction of their orientation with a fixed velocity. In addition, at each time step, each spins updates its orientation to align with the average orientation of the surrounding spins, plus some additional noise which plays the role of temperature. If the spins were fixed, this model would simply amount to the XY model, but the motion of the spins makes it out-of-equilibrium. In two dimensions, a phase transition is found to occur between a disordered phase at strong noise and low density, and an ordered phase at weak noise and high density, the average velocity playing the role of an order parameter. What makes this transition particularly interesting is that, for an equilibrium system, such a transition would be completely impossible since the Mermin-Wagner theorem prevents the existence of true long-range order in dimension $d\leq 2$ (for continuous degrees of freedom and short-range interactions). This transition was later found to be of the first order \cite{ChateGiant,Gregoire2004}.

A first theoretical description of this model was proposed by Toner and Tu in \cite{TonerTu95}. By adding a convective term to the continuum equations of the XY model, they were able to confirm analytically the existence of the transition. 
More general hydrodynamic equations allowing for the description of a wider class of models were later derived using symmetry arguments \cite{TonerTu98,TonerTuReview}. Such equations have also been derived through the coarse-graining of various microscopic models, although with some approximations \cite{Bertin2006,Bertin2009,Peshkov2014}. These microscopic models also include active particles with nematic alignment instead of polar alignment, i.e., elongated particles which tend to become parallel with each other when they collide but do not necessarily want to move in the same direction \cite{Aranson2005,Baskaran2008,Peshkov2014}.
These theoretical studies allowed to shed light on other interesting properties of the flocking transition. In particular, it was shown that the polarization of the velocity field is generally accompanied by large fluctuations of the density, and in some cases even by the formation of stripe patterns \cite{Bertin2009,Mishra2010}.

Controlled experiments have been develop to confirm these predictions, for instance using elongated rods or asymmetric disks on a vibrated plate \cite{Chate2010,Kudrolli2008,Kumar2014}. Precise observations of real animal flocks have also been performed, in particular in the case of birds, which suggest that they interact with a fixed number of neighbors rather than with individuals within a fixed distance, making the order less sensitive to perturbations \cite{Ballerini2008,Cavagna2010,Ginelli2010}.

\subsection{Interactions mediated by the environment}

Many real-life realizations of active particles evolve inside a liquid environment. The motion of the particles thus generate flows in the surrounding fluid, which in turn affects the motion of the particles, leading to effective long-range interactions between them. Many observations have shown that these interactions can have dramatic effects, such as making a phase unstable or allowing for new phases to emerge. Thus, a realistic description of these systems should take these effects into account. This is true for microscopic systems such as bacteria \cite{ishikawa2007} and active colloids \cite{zottl2023}, but also at larger scales, where it was shown for instance that hydrodynamic interactions are essential to understand the behavior of fish swarms \cite{Filella2018}. 

Of course, modeling these effects is extremely complex, as it requires coupling the equations of motion of the active particles with the hydrodynamic equations describing the fluid \cite{Hatwalne2004,Liverpool2006,Saintillan2008,Pahlavan2011,Brotto2013,Yoshinaga2017}. Sometimes these equations can be integrated out to obtain approximate expressions for the resulting effective forces \cite{Baskaran2009,Leoni2010}. These forces are generally long-range and anisotropic, and often play an important role in the alignment between the particles. Minimal models involving only two active particles have also been investigated numerically to better understand the coupling between particles induced by the medium \cite{Debnath2018,Maes_bound_state}.

Besides the velocity field of the fluid, the dynamics of the active particles may be coupled to other elements in their environment, in particular to chemical fields. This may be the case for bacteria, which as mentioned above may liberate chemicals which affect the motion of others around them, but also of Janus particles. Indeed, the motion of these particles is generated by chemical reactions involving elements present in the solution. The consumption and liberation of chemicals due to these reactions may affect the motion of other particles around them, once again leading to complex long-range interactions, sometimes called ``phoretic interactions" \cite{zottl2023,Saha2014,Pohl2014}.

\subsection{Active crystals}

To conclude this section, let us go back to the case of active particles with short-range repulsive interactions. We have mentioned that, at high activity and high density, such systems undergo a phase separation (MIPS). It has been observed that, when the density is increased even more, the low density regions progressively disappear and the system enters a crystal phase, well below the packing fraction required for such a state to appear in an equilibrium system with only hard-core repulsion \cite{Palacci2013,Reichhardt2014,Bialke2014,Menzel2014}. The melting of these ``active crystals'' has been shown  displays a very rich phenomenology, with the formation of topological defects of various types \cite{James2021,Leticia2022}.

Other models of active crystals, and more generally active solids, have also been studied. For instance, active particles connected by harmonic springs have been found to undergo very large deformations without melting \cite{Chate2023}. Other studies have focused on active versions of the vertex model (in which particles undergo deformations with an interaction energy depending on their shape), more suited for the description of cell tissues \cite{Marchetti2018,Bi2014,Bi2015,SussmanVertex,ClaussenVertex}. Recently, active crystals have also been realized experimentally using paramagnetic colloids subjected to a magnetic field (leading to a long-range repulsive interaction potential between the particles decaying as $1/r^3$), placed inside a bath of light-activated bacteria \cite{ActiveMeltingNature2024}. The observations showed that the melting of such a system is a complex process, involving several effective temperatures.

\section{Exact results} \label{sec:interactions_exact}

In the previous section we have seen that interactions in active particle systems can take many different forms and that they often lead to interesting collective effects, including new types of phase transitions which would be completely impossible at equilibrium. However, going beyond numerical simulations and studying such systems analytically is particularly difficult. Many works rely on phenomenology and symmetry arguments to obtain hydrodynamic equations describing their large-scale behavior. Even when such equations can be derived directly from the microscopic dynamics using exact coarse-graining procedures (see, e.g., \cite{Marchetti2018} in the case of ABPs), studying them generally requires some approximations (typically mean-field approximations and gradient expansions), or the use of perturbative approaches such as the renormalization group, 
and explicit solutions are rarely possible. In addition, very few studies consider the correlations at the microscopic scales.

In this section, we focus on simpler models (mostly in one dimension) for which exact results have been obtained. We start with two-particle models, which can already exhibit very rich dynamics, and provide interesting insights into the behavior of larger systems. We then review two types of many-particle models for which exact results have been obtained, namely active particles on a lattice with contact interactions, and harmonic chains of active particles. In each case, we give the main results and refer to the corresponding papers for the derivations. The models, ideas and methods presented in this section are closely related with what will be discussed in Parts~\ref{part:density} and \ref{part:fluctuations} of this thesis. In particular, in Part~\ref{part:density} we will focus on the derivation and the analysis of exact hydrodynamic equations for active particles with long-range interactions, extending the discussion below which focuses on contact interactions. In Part~\ref{part:fluctuations} we will study the tagged particle fluctuations for active particles with repulsive power law interactions, extending the results presented here for harmonic chains of active particles.

\begin{figure}
    \centering
    \includegraphics[width=0.45\linewidth,valign=t,trim={4cm 6cm 4cm 14.5cm},clip]{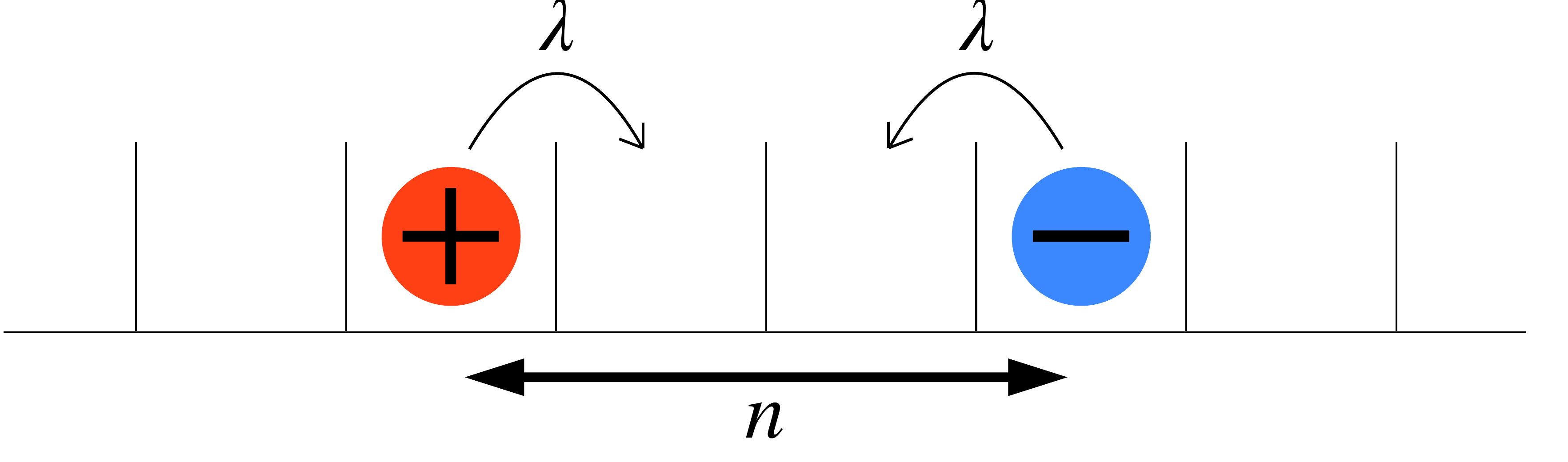}
    \hspace{0.5cm}
    \includegraphics[width=0.45\linewidth,valign=t,trim={4cm 6cm 4cm 14.5cm},clip]{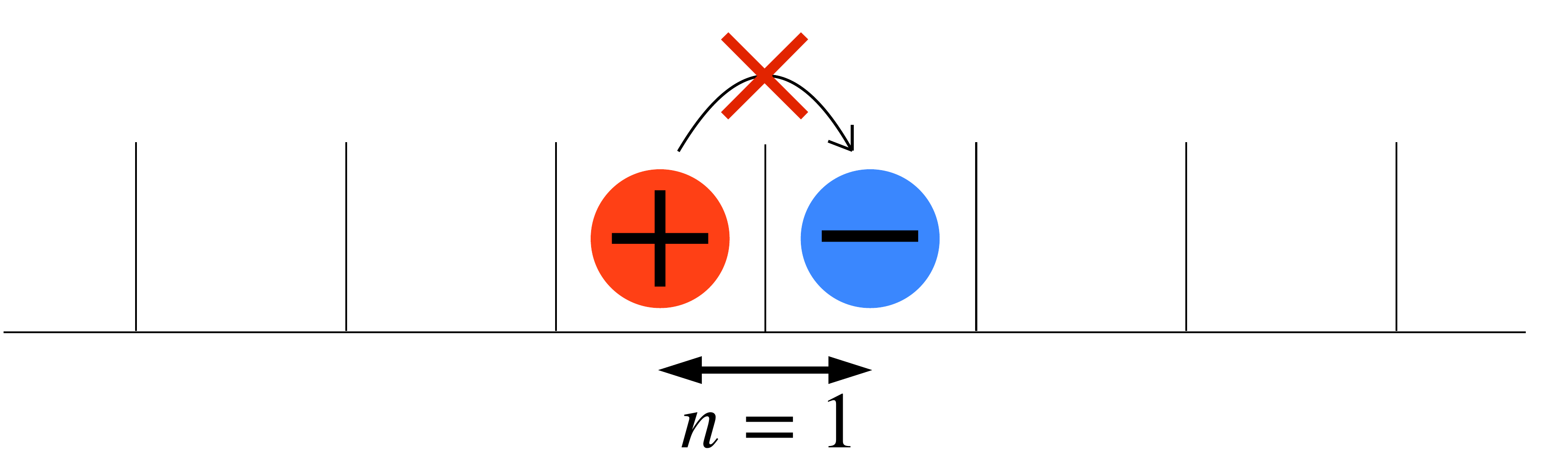}
    \caption{Schematic representation of the two-particle lattice model with exclusion interaction studied in \cite{slowman}.}
    \label{fig:lattice2part}
\end{figure}

\subsection{2-particle models}

\noindent {\bf Short-range interactions.} We start with the case of 2 RTPs on a 1D periodic lattice of size $L$ with an exclusion interaction, which was studied in \cite{slowman} (see Fig.~\ref{fig:lattice2part}). Each RTP jumps to a neighboring site with a rate $\lambda=1$, to the right for $+$ particles and to the left for $-$ particles, but only if it is empty, and remain on the same site otherwise. In addition, each particle tumbles (i.e., changes sign) with rate $\gamma$. The goal is to compute the distribution of the distance $n=i_2-i_1$ between the positions $i_1$ and $i_2$ of the two particles, in units of lattice spacing (with $1\leq n\leq L$). One can write master equations for the joint distribution of $n$ and of the states of the two particles $\sigma_1$ and $\sigma_2$, $P_{\sigma_1, \sigma_2}(n,t)$,
\bea
\partial_t P_{+,+}(n,t) &=& P_{+,+}(n-1,t) \mathbbm{1}_{n>1} + P_{+,+}(n+1,t) \mathbbm{1}_{L-n>1} \\
&& \hspace{0.05cm} + \gamma [P_{+,-}(n,t)+P_{-,+}(n,t)] -P_{+,+}(n,t)[2\gamma+\mathbbm{1}_{n>1}+\mathbbm{1}_{L-n>1}] \;, \nonumber \\
\partial_t P_{+,-}(n,t) &=& 2P_{+,-}(n+1,t) \mathbbm{1}_{L-n>1} + \gamma [P_{+,+}(n,t)+P_{-,-}(n,t)] - P_{+,-}(n,t)[2\gamma+2\mathbbm{1}_{n>1}] \;, \nonumber
\eea
where the indicator function $\mathbbm{1}_{k>1}=1$ if $k>1$ and $0$ otherwise. The equivalents for $P_{-,+}(n,t)$ and $P_{-,-}(n,t)$ are obtained from the symmetries $P_{-,-}(n,t)=P_{+,+}(n,t)$ and $P_{-,+}(n,t)=P_{+,-}(L-n,t)$. The distribution of $n$ is then given by $P(n,t)=\sum_{\sigma_1,\sigma_2} P_{\sigma_1, \sigma_2}(n,t)$. Using generating functions, it is then possible to obtain an exact solution for the stationary state \cite{slowman}. The solution presents a uniform part at large separations $n,L-n \gg 1$, similar to the diffusive case. At intermediate separations however, the solution decays exponentially, with a characteristic length $1/|\log(z)|$, $z=1+\gamma-\sqrt{\gamma(2+\gamma)}$ ($z$ decreases monotonously from $z=1$ at $\gamma=0$ to $z\to0$ as $\gamma\to+\infty$). Finally, the densities $P_{+,-}(n)$ and $P_{-,+}(n)$ also contain an additional term coming from the ``jammed" configuration where the particles are facing each other on neighboring sites. The system thus behaves as if there was some attractive interaction between the particles. 

In the small persistence limit $\gamma \gg 1$, one recovers at leading order the uniform stationary distribution of the ``passive" (i.e., diffusive) case, with the jammed contribution appearing as a first order correction, 
\be
P(n) \simeq \frac{1}{L-1} \left[ 1+\frac{1}{2\gamma} \left( \delta_{n,1} +\delta_{n,L-1}-\frac{2}{L-1} \right) + O\left(\frac{1}{\gamma^2} \right) \right] \;.
\ee
In the opposite limit $\gamma \ll 1$, the stationary distribution coincides with the stationary state that would be reached in the absence of tumblings, with a uniform distribution when $\sigma_1=\sigma_2$, while for $\sigma_1\neq \sigma_2$ the particles are always in a jammed configuration,
\be
P_{+,+}(n)=P_{-,-}(n) \simeq \frac{1}{4(L-1)} \quad , \quad P_{+,-}(n)=P_{-,+}(L-n) \simeq \frac{1}{4} \delta_{n,1} \;.
\ee

The authors also discuss the continuous limit, corresponding to the scaling limit $\gamma \to 0$, $L\to \infty$ with $\gamma L$ fixed. Introducing the physical system size $\ell$, such that the separation between the particles is $x=n\ell/L$ and the driving velocity reads $v_0=\lambda\ell/L$, one finds
\be \label{2RTPslattice_continuous}
P_{+,+}(x) = P_{-,-}(x) = \frac{\gamma+v_0(\delta(x)+\delta(\ell-x))}{4(\gamma \ell+2v_0)} \quad , \quad P_{+,-}(x)=P_{-,+}(\ell-x)=\frac{\gamma+2v_0\delta(x)}{4(\gamma \ell+2v_0)} \;.
\ee
The jammed contributions appear as delta functions in the density. The exponential parts, which have a typical size of the order of $\sim \sqrt{L}$, are not visible at the macroscopic scale and appear as contributions to the delta functions. In particular, the densities $P_{+,+}(n)$ and $P_{-,-}(n)$ now also present delta peaks. This is because when the two particles collide, the two of them need to undergo a tumbling event so that they can move away from each other.

These results were later extended to the case of a finite tumbling time in \cite{slowman2}. The relaxation dynamics was studied in detail in \cite{Mallmin2019}, allowing for the exact determination of the full eigenvalue spectrum, which revealed the existence of dynamical phase transitions. Finally, the effect of additional thermal noise was investigated in \cite{KunduGap2020}, where it was shown that the delta functions in \eqref{2RTPslattice_continuous} are replaced by exponentials. More recently, some works have focused on studying these systems in a more formal general mathematical framework \cite{Hahn2023}. Together, these results allow for a better understanding of how the persistent motion of active particles combined with short-range repulsive interactions can result in effective attractive interaction, providing a useful insight into collective phenomena such as MIPS.

A different type of contact interaction has also been considered in \cite{Metson2022,MetsonLong}, again for a model of 2 RTPs on a lattice. In this model, when a particle jumps on a site already occupied by another particle, the particle which was already on the site is displaced in the direction of the jump over a distance $m$ drawn from some distribution $\Phi(m)$, and its velocity is reversed with some probability $r$. This ``recoil interaction" mimics the dynamics observed in some micro-organisms. For an arbitrary recoil distribution $\Phi(m)$, an exact expression was derived for the stationary distribution of the inter-particle distance in the continuum limit. This result allowed to show that, depending on the choice of $\Phi(m)$ and on the persistence length $\xi=\lambda/(\gamma L)$, the recoil interaction may effectively behave either as an attractive or as a repulsive interaction.
\\

\noindent {\bf Long-range interactions.} Concerning active particles with long-range interactions, the current literature is more limited. The case of 2 RTPs on the real line interacting via an attractive interaction potential was investigated in \cite{LMS2021}. The stochastic equations of motion for the positions $x_1(t)$ and $x_2(t)$ read,
\bea
\frac{dx_1}{dt} &=& f(x_1-x_2) + v_0 \sigma_1(t) + \sqrt{2T} \, \xi_1(t) \;, \\
\frac{dx_2}{dt} &=& f(x_2-x_1) + v_0 \sigma_2(t) + \sqrt{2T} \, \xi_2(t) \;, \nonumber
\eea
where the interaction force $f(x)=-V'(x)$ is an odd function of $x$, the $\sigma_i(t)$ denote independent telegraphic noises and the $\xi_i(t)$ denote independent centered Gaussian white noises with unit variance. These equations can be rewritten in terms of the center of mass $w=(x_1+x_2)/2$ and of the inter-particle distance $y=x_1-x_2$,
\bea
\frac{dw}{dt} &=& \frac{v_0}{2} [\sigma_1(t) + \sigma_2(t)] + \sqrt{T} \, \tilde\eta(t) \;, \\
\frac{dy}{dt} &=& 2f(y) + v_0 [\sigma_1(t) - \sigma_2(t)] + \sqrt{4T} \, \eta(t) \;, \label{eqy_2RTPs}
\eea
where $\eta(t)=[\xi_1(t) - \xi_2(t)]/\sqrt{2}$ and $\tilde \eta(t)=[\xi_1(t) + \xi_2(t)]/\sqrt{2}$ are again two independent centered Gaussian white noises with unit variance. The center of mass $w(t)$ can be described as a free RTP with 3 internal states $(-v_0,0,v_0)$, which behaves diffusively at large times. By contrast, the inter-particle distance $y(t)$ can be seen as a 3-state RTP $(-2v_0,0,2v_0)$ subjected to an external force $2f(y)$ (similar to, e.g., \cite{3statesBasu}), which for a sufficiently attractive force can reach a stationary bound state at large times.

The joint distribution of $y$, $\sigma_1$ and $\sigma_2$, $P_{\sigma_1,\sigma_2}(y,t)$, obeys a set of 4 coupled Fokker-Planck equations,
\be
\partial_t P_{\sigma_1, \sigma_2} = -\partial_y \left\{[2f(y)+v_0(\sigma_1-\sigma_2)] P_{\sigma_1,\sigma_2} \right\} -2\gamma P_{\sigma_1,\sigma_2} +\gamma (P_{-\sigma_1,\sigma_2}+P_{\sigma_1,-\sigma_2}) +2T\partial_y^2 P_{\sigma_1,\sigma_2} \;,
\ee
for which we want to find the stationary solutions. As in the case above, the distribution of $y$ is obtained as $P(y,t) = \sum_{\sigma_1,\sigma_2} P_{\sigma_1,\sigma_2}(y,t)$.

For a linear interaction potential, $f(y)=-\bar c \, {\rm sgn}(y)$, the stationary distribution can be computed exactly both for $T=0$ and for $T>0$ \cite{LMS2021}. For $T=0$, if $\bar c>v_0$, the driving noise cannot compete with the attraction, and the stationary solution is a delta function, $P(y)=\delta(y)$. If instead $\bar c<v_0$, then when $\sigma_1=\sigma_2$ the separation $y(t)$ still relaxes to $y=0$ in a finite time, leading to a delta function in the density, but when $\sigma_1 \neq \sigma_2$, the particles may move away from each other at a constant velocity, which leads to an exponential part in the distribution. In the end, one finds \cite{LMS2021}
\be \label{PDF2part_linear}
P(y) = \frac{\bar c \gamma v_0^2}{v_0^4-\bar c^4} e^{-\frac{2\gamma \bar c }{v_0^2-\bar c^2}|y|} + \frac{\bar c^2}{v_0^2+\bar c^2} \delta(y) \;.
\ee
In the presence of thermal noise, $T>0$, the delta peaks disappear and the density becomes a sum of exponentials.

\begin{figure}
    \centering
    \includegraphics[width=0.49\linewidth,trim={0.5cm 0.5cm 0.5cm 0.5cm},clip]{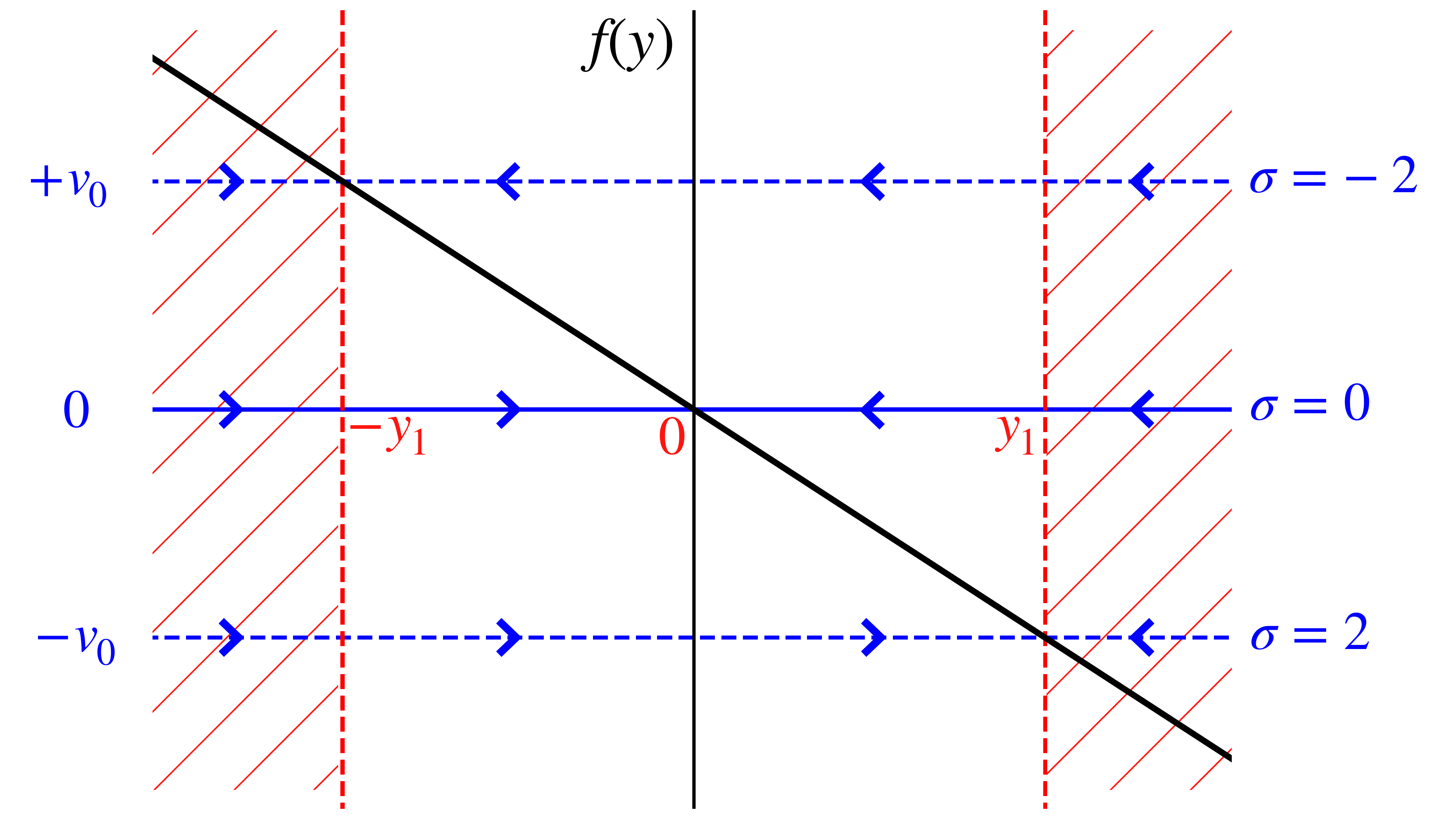}
    \includegraphics[width=0.49\linewidth,trim={0.5cm 0.5cm 0.5cm 0.5cm},clip]{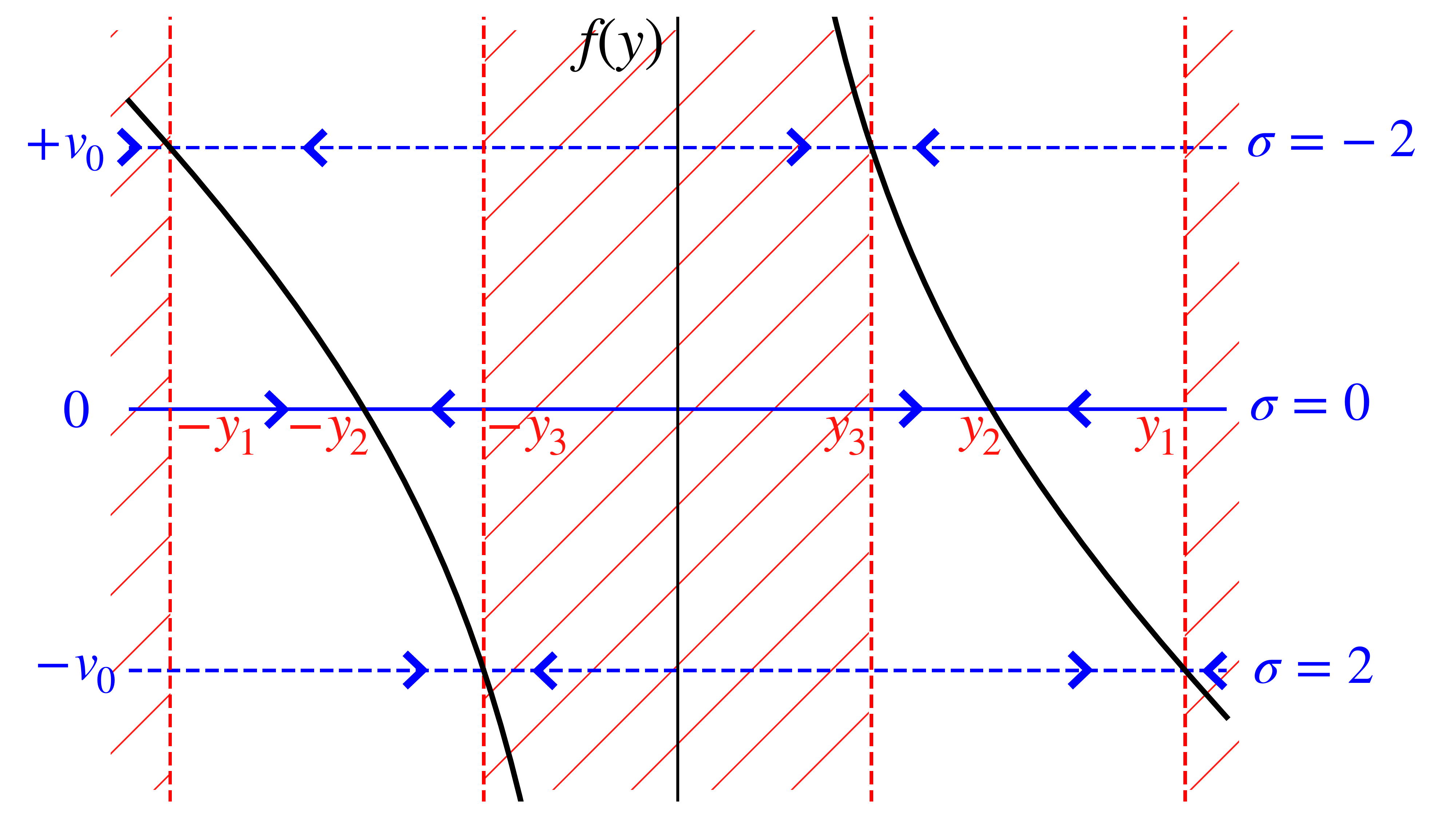}
    \caption{Dynamical diagrams for a harmonic interaction $f(y)=-\mu y$ (left), and for a force with a repulsive part, $f(y)=\frac{1}{y}-y$ (right). The total force on the effective particle with position described by $y(t)$ is $2f(y)+v_0(\sigma_1 - \sigma_2)$. Thus, the particle moves towards the right when $f(y)>-v_0\sigma/2$, when $\sigma_1-\sigma_2=\sigma\in\{-2,0,2\}$, and towards the left otherwise (see the arrows on the corresponding blue lines). The support of the stationary distribution of $y$ can be read from these arrows. In particular, the regions in which the particle moves in the same direction independently of its state, i.e., where $|f(y)|>v_0$, are inaccessible in the stationary state (regions hashed in red on the figures). In the example on the right, the support is disjoint, which implies that the two particles cannot cross.}
    \label{fig:2RTPlongrange}
\end{figure}

For a linear interaction potential, the stationary density $P(y)$ is supported by the whole real line. However, for more general interactions, and in the absence of thermal noise, the support may be non-trivial. One way to determine it is to use a dynamical diagram, as in Fig.~\ref{fig:2RTPlongrange} (this also applies to the 1 particle case, see the discussion in Sec.~\ref{sec:1particle_potential}). If the force $f(y)$ is continuously decreasing with $|f(y)|>v_0$ for large $y$, then the support is bounded. For instance, in the case of a harmonic force $f(y)=-\mu y$ (left panel of Fig.~\ref{fig:2RTPlongrange}), the support is $[-v_0/\mu,v_0/\mu]$. More complex situations leading to a disjoint support may also arise. This may be the case in particular if the interaction force has a repulsive component. For instance, for $f(y)=1/y-y$ (right panel of Fig.~\ref{fig:2RTPlongrange}), the support of $y$ is {\it a priori} included in $[-y_1,-y_3]\cup[y_3,y_1]$ where $y_1$ and $y_3$ are the positive solutions of $y^2-v_0 y -1=0$ and $y^2+v_0 y -1=0$ respectively. In this case, the stationary state depends on the initial condition, since the trajectories of the two particles cannot cross (i.e., the sign of $y(t)$ remains constant).

For a generic interaction force $f(y)$ and for $T=0$, a general second order differential equation for the stationary density $P(y)$ was derived. The treatment of the boundary conditions is non-trivial, but we refer again to \cite{LMS2021} for the details. An explicit expression for the solution, involving hypergeometric functions, was also derived in the harmonic case.
\\

Recently, the combined effect of an attractive long-range interaction and an additional jamming interaction (i.e., hard-core repulsion) was investigated in \cite{Hahn2025}. The stationary distribution of the distance $y(t)$ between 2 RTPs on the real line was computed exactly in 3 different cases, with hard-core repulsion and in the absence of thermal noise: for a linear attractive interaction, both for RTPs with instantaneous tumblings (2-state RTP) and with finite tumbling times (3-state RTP), and for a harmonic attractive interaction with instantaneous tumblings. One way to see this problem is to say that the dynamics of $y(t)$ is still described by \eqref{eqy_2RTPs} (in the instantaneous tumbling case and with $T=0$), but with a  hard wall boundary condition at $y=0$. In the linear, instantaneous case, the solution is still the sum of a delta function and an exponential term (for $\bar c<v_0$), but the delta term receives an additional contribution coming from the jammed configuration, i.e., $(\sigma_1,\sigma_2)=(-,+)$ and $y=x_1-x_2=0^+$ (assuming $x_1(0)>x_2(0)$). In the non-instantaneous case, the solution is slightly more complex, with an additional exponential term when $v_0>2\bar c$. In the harmonic case, the solution can again be expressed in terms of hypergeometric functions, but with an additional delta term coming from the jamming interaction. The relaxation dynamics were also investigated through the derivation of rigorous bounds \cite{Hahn2025}.

\subsection{Lattice models} \label{sec:lattice_hydro}

\noindent {\bf Exact hydrodynamic equations.} Deriving exact results beyond two interacting particles is particularly challenging. One possible approach is to consider the opposite limit, where the number of particles is extremely large, and to derive coarse-grained equations describing the particle density at the macroscopic scale. As we have mentioned above, this generally requires some approximations. There is however one case for which such equations have been derived exactly and used to obtain precise analytical results: lattice models with short-range interactions. In \cite{KH2018}, two such models were introduced, each one illustrating one of the most emblematic collective phenomena in active matter: MIPS and the transition to collective motion. For both models, exact hydrodynamic equations were derived using the method introduced in \cite{Erignoux_derivation}. These equations were then use to obtain the exact phase diagram using the method presented in \cite{Solon2018_1,Solon2018_2}.

\begin{figure}
    \centering
    \includegraphics[width=\linewidth,trim={0.5cm 9cm 0.5cm 18cm},clip]{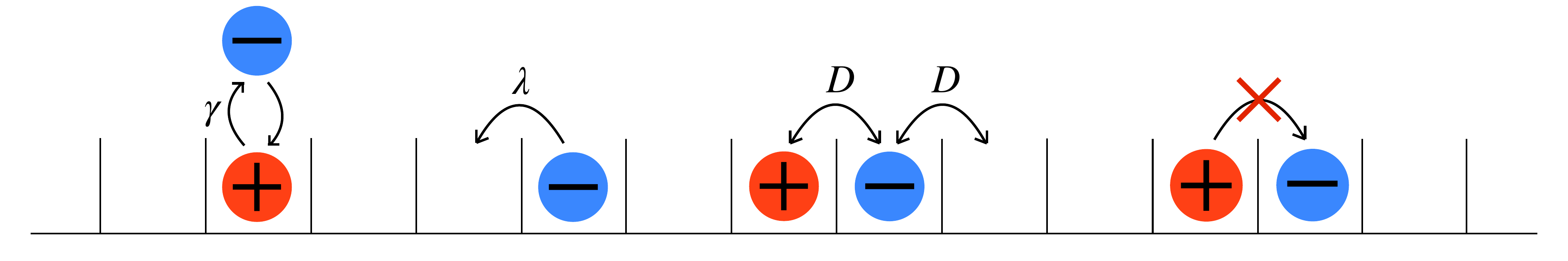}
    \caption{Schematic representation of the lattice model of Sec.~\ref{sec:lattice_hydro}.}
    \label{fig:lattice_hydro}
\end{figure}

The first model is a model of RTPs with an exclusion interaction, similar to the one described above for two particles. It consists in a one-dimensional lattice with $L$ sites and periodic boundary conditions, supporting $N=\rho_0 L$ particles with $\rho_0\in [0,1]$. Each site is characterized by its occupation number $\sigma_i\in\{-1,0,1\}$, with $\sum_i|\sigma_i|=N$. A $+$ particle ($\sigma_1=1$) jumps to the site on the right with rate $\lambda/L$ if it is empty ($\sigma_{i+1}=0$), while a $-$ particle ($\sigma_i=-1$) does the same towards the left. In addition, a particle switches sign with rate $\gamma/L^2$. Finally, two neighboring sites exchange their occupation number with rate $D$, representing a symmetric diffusive motion. Note that this allows particles to bypass each other, thus escaping a jammed configuration (but we will see that the exclusion interaction which prevents the directed jumps still slows down the dynamics sufficiently to generate a MIPS). Note also that the scaling with $L$ of the different rates was carefully chosen in order to have a well-defined hydrodynamic limit.

Denoting $\sigma_i^{\pm}(\tau)=1$ if $\sigma_i(\tau)=\pm 1$ and $0$ otherwise, the microscopic dynamics obeys the equation
\be \label{eq_lattice_discrete}
\partial_\tau \langle \sigma_i^+ \rangle = D[\langle \sigma_{i+1}^+ \rangle + \langle \sigma_{i-1}^+ \rangle -2 \langle \sigma_i^+ \rangle ] - \frac{\gamma}{L^2} [ \langle \sigma_i^+ \rangle - \langle \sigma_i^- \rangle ] + \frac{\lambda}{L} [ \langle \sigma_{i-1}^+ (1-|\sigma_i|) \rangle - \langle \sigma_i^+ (1-|\sigma_{i+1}|) \rangle ] \;,
\ee
and similarly for $\langle \sigma_i^- \rangle$. These equations are, however, not closed, since the evolution of $\langle \sigma_i^+(\tau) \rangle$ involves the correlator $\langle \sigma_i^+(\tau) (1-|\sigma_{i+1}(\tau)|) \rangle$. Closed equations can however be obtained at the macroscopic scale, after a rescaling $x=i/L$ and $t=\tau/L^2$, in terms of the coarse-grained densities,
\be \label{def_rho_lattice}
\rho^{\pm}(x,t) = \frac{1}{2L^\delta} \sum_{|i-Lx|\leq L^\delta} \sigma_i^\pm (t L^2) \;,
\ee
where $\delta\in(0,1)$ is a coarse-graining parameter. In the limit $L\to +\infty$, these equations read (see \cite{Erignoux_derivation} for a derivation), for $\sigma=\pm 1$,
\be \label{eqrhosigma_lattice}
\partial_t \rho^\sigma = D\partial_x^2 \rho^\sigma - \lambda \partial_x [ \rho^\sigma (1-\rho)] -\gamma (\rho^\sigma - \rho^{-\sigma}) \;,
\ee
where $\rho=\rho^++\rho^-$ is the total density of particles. These equations are essentially \eqref{eq_lattice_discrete} with the discrete derivatives replaced by continuous ones. However, an important point for the derivation is that, due to the scaling used, the local correlations are dominated by the diffusion. This "local equilibrium" hypothesis, valid at large $L$, is what allows to close the two-point correlations.

The equation \eqref{eqrhosigma_lattice} can be rewritten in terms of the total density $\rho$ and of the difference $m=\rho^+-\rho^-$, 
\bea \label{eqrho_lattice}
\partial_t \rho &=& D\partial_x^2 \rho - \lambda \partial_x [m(1-\rho)] \;, \\
\partial_t m &=& D\partial_x^2 m - \lambda \partial_x [\rho(1-\rho)] -2\gamma m \;.
\label{eqm_lattice}
\eea
From these equations, one can obtain the exact phase diagram of the model as a function of the two parameters, the P\'eclet number $\text{Pe}=\lambda/\sqrt{D\gamma}$, which controls the activity, and the average density $\rho_0=N/L$. Indeed, one can easily show that the homogeneous solution $\rho=\rho_0$, $m=0$ is linearly unstable when
\be
\text{Pe}^2(1-\rho_0)(2\rho_0-1)>2 \;.
\ee
This means that for $\text{Pe}>4$ (i.e., for strong activity), there exists an interval of values of $\rho_0$, $[\rho_l^s,\rho_h^s]$, where the homogeneous solution is linearly unstable (this defines the {\it spinodal} curve). Thus, a phase separation occurs between a phase of low density $\rho_g$ (gas phase), and a phase of high density $\rho_\ell$ (liquid phase). These two densities where determined exactly using the method introduced in \cite{Solon2018_1,Solon2018_2}, providing an exact equation for the {\it binodal} curve of the phase diagram. This is the first microscopic model for which the hydrodynamic description and the phase diagram of the MIPS could be determined exactly.

The second model introduced in \cite{KH2018} is similar, but replaces the exclusion interaction by an alignment interaction, i.e., there can be more than one particle per site, but the tumbling rate $\gamma/L^2$ now depends on the number of $+$ and $-$ particles on the same site, encouraging particles on the same site to align with each other. Once again, exact hydrodynamic equations similar to \eqref{eqrhosigma_lattice} can be derived, and the phase diagram of the flocking transition can be obtained exactly. We refer to \cite{KH2018} for the details. As mentioned in \cite{KH2018}, let us also add that both models can be generalized to higher dimensions, where similar exact hydrodynamic equations can be derived.
\\

\noindent {\bf Finite size fluctuations.} The above discussion focused on the limit of infinite system size, where the particle density is fully described by its mean. The effect of finite size fluctuations (in the model with exclusion interactions defined above) has later been studied, using the framework of macroscopic fluctuation theory (MFT -- see \cite{MFTreview} for a review). This was first done in \cite{Agranov2021} at the level of the typical (Gaussian) fluctuations, and later extended to the level of large deviations (taking into account the non-Gaussian nature of the noise) in \cite{Agranov2022}.

At the typical level, it was found that the fluctuations of the density can be accounted for by adding Gaussian white noise terms to the equations \eqref{eqrho_lattice}-\eqref{eqm_lattice},
\bea \label{eqrho_lattice_noise}
\partial_t \rho &=& D\partial_x^2 \rho - \lambda \partial_x [m(1-\rho)] + \sqrt{\frac{D}{L}} \, \partial_x \eta_\rho \;, \\
\partial_t m &=& D\partial_x^2 m - \lambda \partial_x [\rho(1-\rho)] -2\gamma m + \frac{1}{\sqrt{L}} \left( \sqrt{D} \, \partial_x \eta_m +2\sqrt{\gamma} \, \eta_K \right) \;,
\label{eqm_lattice_noise}
\eea
where the noise is delta correlated, $\langle \eta_p(x,t) \eta_q(x',t') \rangle = S_{p,q} \, \delta(x-x') \delta(t-t')$, with correlation matrix
\bea
&&S_{\rho,\rho} = 2\rho(1-\rho) \;, \quad S_{m,m} = 2(\rho-m^2) \;, \quad S_{K,K} = \rho \;, \nn \\
&& S_{\rho,m} = 2m(1-\rho) \;, \quad S_{\rho,K}=S_{m,K} = 0 \;.
\eea
These fluctuating hydrodynamic equations were then used to compute the two-point correlation functions of the density $\langle \delta \rho(x)\delta\rho(x')\rangle$, $\langle \delta m(x)\delta m(x')\rangle$ and $\langle \delta \rho(x) \delta m(x') \rangle$, where $\delta \rho=\rho - \rho_0$ and $\delta m = m$, as well as the dynamical functions in Fourier space. The spatial correlations decay exponentially with a characteristic length $\ell \xi$, where $\ell=\sqrt{D/\gamma}$ is the diffusion length, and
\be
\xi = \frac{1}{\sqrt{2-\text{Pe}^2(1-\rho_0)(2\rho_0-1)}} \;,
\ee
which diverges at the critical point $(\rho_0=3/4,\text{Pe}=4)$ with an exponent 1/2. See \cite{Agranov2021} for more details.

In \cite{Agranov2022}, the probability of observing a given history of the densities and of the local fluxes was expressed in a large deviation form at large $L$, taking into account the Poissonian statistics of the tumbling events. This was used to compute the large deviations of the total integrated current flowing through the system, revealing the existence of a dynamical phase transition into a traveling wave phase separated state. This method was also used to study the entropy production of the model in \cite{EntropyAgranov}.
\\

\noindent {\bf Other approaches and related models.} Recently, a variation of this model was studied, both at the level of the noiseless hydrodynamic equations and of the fluctuating hydrodynamics \cite{lattice2lanes2025}. In this version, the diffusion does not allow for the exchange of two neighboring particles. Instead, a quasi-one-dimensional geometry is considered, composed of two lanes parallel to each other. The particles can jump to the neighboring site of the other lane, provided that it is empty, with a rate which is assumed very large compared to the tumbling rate, so that the two lanes are equilibrated at all times. Thus, at the macroscopic scale the system can again be described by densities $\rho(x,t)$ and $m(x,t)$ which only depend on the position $x$ in the longitudinal direction. The fluctuating hydrodynamic equations are derived using a different method based on the evaluation of the Martin–Siggia–Rose–Janssen–de Dominicis action, and are slightly different from \eqref{eqrho_lattice}-\eqref{eqm_lattice}. 
The phase diagram of the MIPS is very similar to the previous model, and was derived exactly using the same method. The two-point correlation functions were also computed.

The generalization of this model to higher dimensions does not pose any particular issue. However, in the strictly 1D case (i.e., with a single line), the approach used predicts the same hydrodynamic equations as in the 2-lane model, but this does not agree with the numerical simulations. It is a general observation that hydrodynamic descriptions typically fail for systems of active particles with a single-file condition, i.e., when the ordering of particles is preserved. This is an issue which we also encountered during this thesis, and which will be discussed in Part~\ref{part:density}.

To conclude, let us mention that the present model in the absence of diffusion (in 1D with a true single-file constraint) was also studied using a completely different approach without coarse-graining in \cite{Dandekar2020}, where the focus was instead on the distribution of cluster and gap sizes. In the limit of large tumbling rate, a mapping to a mass transport model and a mean-field approximation allowed to compute the rate of the exponential tails of these distributions, while a coalescence-fragmentation model was shown to be a good description of the model in the opposite limit. These results were then used to derive an effective hydrodynamic description, within these approximations.

\subsection{Harmonic chains of active particles} \label{sec:harmonicChain}

There is another important category of models for which some exact results have been obtained in recent years, namely 1D chains of active particles with a harmonic interaction between nearest neighbors. For such models, the focus has been on the computation of correlation functions at the microscopic level, i.e., of the tagged particle fluctuations. This type of model was first introduced to study the effect of active noise on polymers \cite{PutBerxVanderzande2019,Samanta2016,Chaki2019}. A second motivation then came from seeing the harmonic repulsion as an approximation for a short-range repulsive interaction \cite{SinghChain2020,HarmonicChainRevABP,HarmonicChainRTPDhar}. Indeed, for Brownian particles in 1D, a nearest-neighbor harmonic interaction was shown to be a good approximation for more generic short-range pair-wise potentials with a hard-core repulsion part \cite{Lizana2010}. Previous studies of the tagged fluctuations in active particle systems with a single-file constraint revealed some interesting behaviors \cite{Galanti2013,Teomy1,Teomy2,Dolai2020,Banerjee2022}. In particular, while the large time behavior of such systems is similar to the single-file dynamics of diffusive particles \cite{Harris65,Arratia83,SingleFileMajumdar1991,SingleFileKrapivsky2014,SingleFileKrapivsky2015,TaggedSFD2015,SingleFileReview}, with a mean-squared displacement (MSD) increasing as $~\sqrt{t}$, the short-time behavior was found to be unique to active particles, with scaling forms interpolating between these two regimes. While these studies were mostly based on numerical simulations, mean-field approximations and approximate hydrodynamic descriptions, the case of harmonic chains allowed for the exact computation of the two-point two-time correlations \cite{SinghChain2020,HarmonicChainRevABP,HarmonicChainRTPDhar}.

A 1D harmonic chain of active particles consists in $N$ particles with positions $x_i(t)$ ($i=1,...,N$) obeying the following stochastic differential equation (SDE):
\be \label{eq_harmonicChain}
\frac{dx_i}{dt} = -K(2x_i - x_{i+1} - x_{i-1}) + \zeta_i(t) \;,
\ee
where the $\zeta_i(t)$ denote independent 1D RTP or AOUP noises, or 2D ABP noises projected in 1D, as defined in Sec.~\ref{sec:active_models} (for the ABP the transverse fluctuations were also studied in \cite{SinghChain2020}). Since we only consider two-point correlations, we only need the two-time correlations of the active noise, given in \eqref{sigmacorr} for RTPs, in \eqref{AOUPcorr} for AOUPs and in \eqref{ABPcorr} for ABPs (higher order moments were only studied numerically, to measure the non-Gaussianity of the fluctuations). 
Here we consider periodic boundary conditions as in \cite{SinghChain2020,HarmonicChainRevABP,HarmonicChainRTPDhar}. The cases of free \cite{PutBerxVanderzande2019} or fixed extremities \cite{HarmonicChainRTPDhar} were also considered, but in the limit $N\to+\infty$ on which we focus here this distinction becomes irrelevant. We take this opportunity to mention that the case of a harmonic chain of passive particles, where only the edge particles are subjected to active noise, was also considered (see, e.g., \cite{HarmonicBasu}), but we will not discuss it here.

\begin{figure}
    \centering
    \includegraphics[width=0.95\linewidth,trim={0.5cm 14cm 0.5cm 14cm},clip]{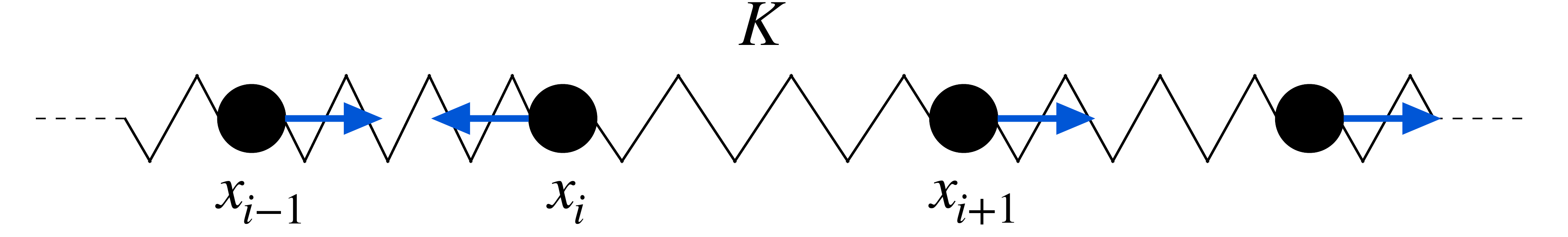}
    \caption{Schematic representation of a harmonic chain of RTPs.}
    \label{fig:harmonicChain}
\end{figure}

Before reviewing some of the main results concerning the model \eqref{eq_harmonicChain}, let us introduce the different timescales of the problem. The first one is the interaction timescale $\tau_K=1/K$. This is the timescale at which the interaction becomes relevant, meaning that for $t\ll \tau_K$, the particles behave effectively as independent free particles. The second important timescale is the persistent time of the active noise,
\be
\tau_A = \begin{cases} 1/\gamma \quad \ \; \; \text{for RTPs} \;, \\
\tau \hspace{1.12cm} \text{for AOUPs} \;, \\
1/D_R \quad \text{for ABPs} \;. \end{cases}
\ee
In the present model, the activity dominates only for $t\ll \tau_A$, while the Brownian results are recovered for $t\gg \tau_A$. We note that \cite{HarmonicChainRevABP} considers DRABPs, which adds a second activity timescale, but we will not detail this case here. Finally, if the size of the system is finite, there is an additional timescale $\tau_N=N^2/K$ characterizing the equilibration at the level of the full system. These times-scales are generally assumed to be well separated, which allows to obtain simplified expressions for the correlation functions in the different time regimes.

The general method for the computation of the two-point two-time correlations is as follows: taking a Fourier transform of \eqref{eq_harmonicChain} in space  \cite{PutBerxVanderzande2019,SinghChain2020,HarmonicChainRevABP} or in time \cite{HarmonicChainRTPDhar}, one obtains a formal expression of the positions $x_i(t)$ as a function of the $\zeta_i(t)$'s. Writing the two-point two-time covariance of the positions $\langle x_i(t)x_j(t')\rangle_c = \langle x_i(t)x_j(t')\rangle - \langle x_i(t) \rangle \langle x_j(t')\rangle $ and inserting the expressions of the noise correlations $\langle \zeta_i(t) \zeta_j(t') \rangle = \delta_{i,j}\langle \zeta_i(t) \zeta_i(t') \rangle$ given in \eqref{sigmacorr}, \eqref{AOUPcorr} and \eqref{ABPcorr}, one obtains a general expression, which can then be specialized to obtain various correlation functions, such as the mean-squared displacement (MSD) $C_0(t)$, the equal-time covariance $C_k(t)$ or the position autocorrelation $C_0(t,t')$,
\bea \label{def_C0_harmonciChain}
&&C_0(t) = \langle (x_i(t)-x_i(0))^2\rangle_c \;, \\
&&C_k(t) = \langle (x_i(t)-x_i(0))(x_{i+k}(t)-x_{i+k}(0))\rangle_c \;, \label{def_Ck_harmonciChain}\\
&&C_0(t,t') = \langle(x_i(t)-x_i(0))(x_i(t')-x_i(0))\rangle_c \label{def_C0tt_harmonciChain}
\eea
(independent of $i$ by translation invariance), and analyzed to obtain simplified expressions in the different time regimes.

It is important to note that the expressions of these correlation functions depend on the choice of initial condition. In particular, for $t\ll \tau_A$, the results depend on the initial condition for the positions $x_i(0)$, but also for the driving noises $\zeta_i(0)$. For the positions, two types of initial conditions can be considered: a {\it quenched} initial condition, where the $x_i(0)$'s are fixed to precise values (generally an equally spaced configuration), as in \cite{PutBerxVanderzande2019,SinghChain2020,HarmonicChainRevABP}, or an {\it annealed} initial condition, where the system is assumed to have already reached its stationary state, so that the $x_i(0)$'s are drawn from their stationary distribution, as in \cite{HarmonicChainRTPDhar}. Concerning the active noise, while \cite{SinghChain2020} considered different types of initial conditions for the three models of RTPs (annealed, i.e., $\sigma_i(0)=\pm 1$ with equal probability, for the RTPs, $v_i(0)=0$ for the AOUPs, and a driving velocity aligned with the $x$-axis but with a random sign for the ABPs), leading to very different short-time behaviors between the different models, it was noted in \cite{HarmonicChainRTPDhar} that, in the stationary state, the two-time correlations take the same exponential form for all three models (see \eqref{sigmacorr}, \eqref{AOUPcorr} and \eqref{ABPcorr}), implying that when taking an annealed initial condition for the active noise, a simple mapping exists between the correlation functions of these three models. Thus, in the following we focus on the RTP case $\zeta_i(t)=v_0\sigma_i(t)$, where $\sigma_i(0)=\pm 1$ with equal probability, and the terms ``quenched'' and ``annealed'' refer to the initial condition for the positions $x_i(0)$.

\begin{figure}
    \centering
    \includegraphics[width=0.7\linewidth, trim={0 9cm 0 0},clip]{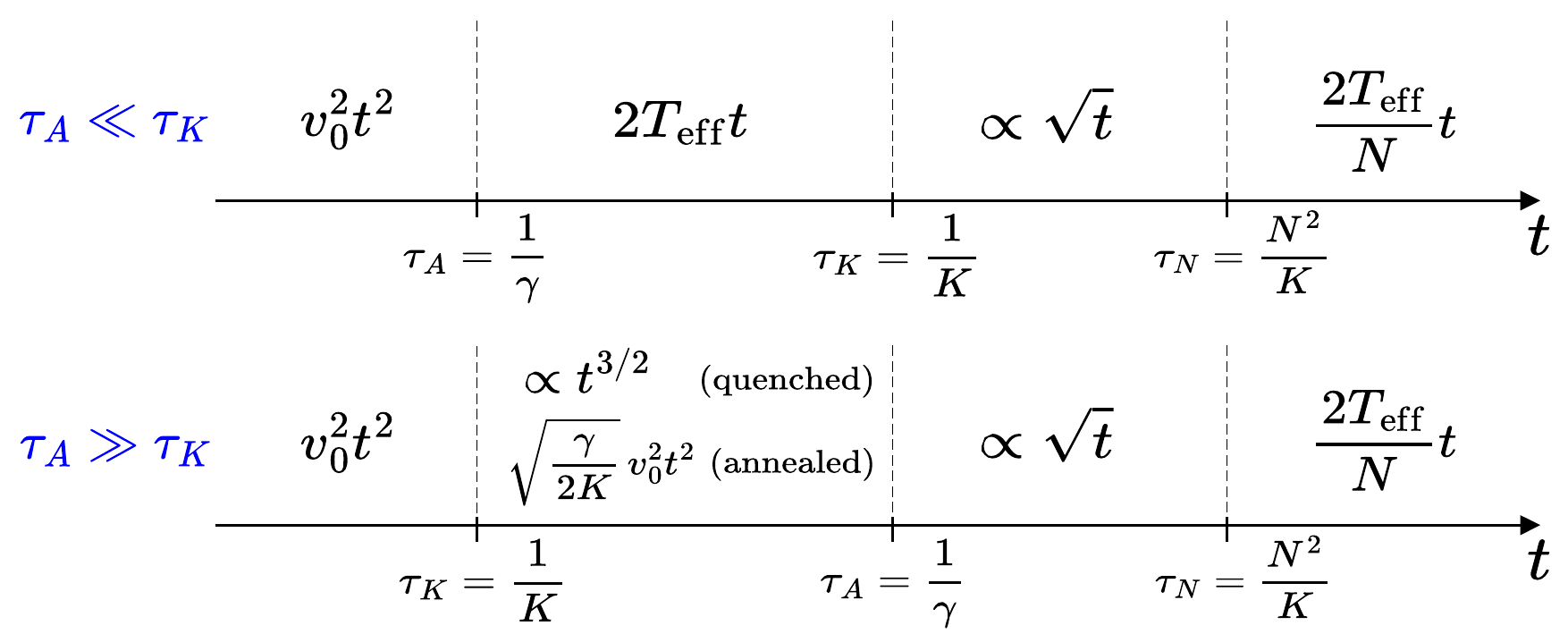}
    \caption{Representation of the different time regimes for the harmonic chain of RTPs. The full expression or the time-dependence of the mean-squared displacement $C_0(t)$ (defined in \eqref{def_C0_harmonciChain} is given for each time regime. Here $T_{\rm eff}=v_0^2/(2\gamma)$ denotes the effective temperature for a free RTP.}
    \label{fig:diagramC0_harmonic}
\end{figure}

On timescales $t\ll \tau_K$, the particles are effectively independent. For $t\ll\tau_K,\tau_A=1/\gamma$, this leads to a ballistic behavior $C_0(t) \simeq v_0^2 t^2$. If $\tau_K \gg \tau_A$, the regime $\tau_A\ll t\ll\tau_K$ corresponds to the free diffusion of independent particles, with an effective diffusion coefficient $T_{\rm eff}=v_0^2/(2\gamma)$ in the RTP case. The opposite case $\tau_K \ll \tau_A$ is more interesting, as it leads to a crossover between the regime $\tau_K \ll t \ll \tau_A$, where both the interaction and the activity are relevant, to the Brownian single-file behavior for $t \gg \tau_A,\tau_K$. For the MSD, in the quenched case, this crossover is described by the scaling form \cite{SinghChain2020}
\be \label{scaling_MSD_harmonicChain}
C_0(t) \simeq \frac{v_0^2t^{3/2}}{\pi} \sqrt{\frac{2}{K}} \, \mathcal{T}(\gamma t) \;,
\ee
where
\bea \label{scaling_MSD_harmonicChain_asympt}
\mathcal{T}(y) &\simeq& \frac{2\sqrt{\pi}}{3} (2-\sqrt{2}) + O(y) \;, \quad \text{as } y\to0\;, \\ 
&\simeq& \frac{\sqrt{\pi}}{2y} + O(y^{-2}) \;, \quad \text{as } y\to+\infty\;.
\eea
For $t\ll \tau_A=1/\gamma$, this leads to $C_0(t)\propto t^{3/2}$, while for $t\gg\tau_A$ we recover the single-file diffusion scaling $C_0(t)\propto \sqrt{t}$. In the annealed case, one finds instead that the ballistic behavior survives until $t\sim\tau_A$, but with a different prefactor, $C_0(t) \simeq \sqrt{\gamma/(2K)} \, v_0^2 t^2$, and there is a direct crossover to single-file diffusion \cite{HarmonicChainRTPDhar}. Finally, in the case where $N$ is finite (and for periodic boundary conditions), the single-file behavior saturates after a time $t\sim \tau_N=N^2/K$, and the MSD becomes dominated by the collective diffusion, i.e., the diffusion of the center of mass, $C_0(t)\sim 2T_{\rm eff}\,t/N$. The behavior of the MSD for the different time regimes is summarized in Fig.~\ref{fig:diagramC0_harmonic}.

For the equal-time covariance, in the quenched case, two different scaling forms were obtained \cite{SinghChain2020}, for $\tau_K \ll t \ll \tau_A$,
\bea \label{scaling_Ck_harmonicChain1}
C_k(t) \simeq \frac{v_0^2t^{3/2}}{\pi} \sqrt{\frac{2}{K}} \, \Omega\left(\frac{k}{\sqrt{2Kt}}\right) \;, \quad \Omega(y) \hspace{-0.25cm} &\simeq& \hspace{-0.25cm} \frac{2\sqrt{\pi}}{3} (2-\sqrt{2}) -\sqrt{\pi}(\sqrt{2}-1)y^2 + O(y^3) \;, \text{ as } y\to0\;, \nn \\ 
&\simeq& \hspace{-0.25cm} \left(\frac{2\sqrt{\pi}}{y^4} + O(y^{-6})\right) e^{-y^2/4} \;, \text{ as } y\to+\infty\;,
\eea
and for $t \gg \tau_A,\tau_K$,
\bea \label{scaling_Ck_harmonicChain2}
C_k(t) \simeq T_{\rm eff} \sqrt{\frac{2t}{\pi K}} \, \mathcal{C}\left(\frac{k}{\sqrt{2Kt}}\right) \;, \quad \mathcal{C}(y) \hspace{-0.25cm} &\simeq& \hspace{-0.25cm} 1 -\frac{\sqrt{\pi}}{2} y + O(y^2) \;, \quad \text{as } y\to0\;, \\ 
&\simeq& \hspace{-0.25cm} \left(\frac{2}{y^2} + O(y^{-4})\right) e^{-y^2/4} \;, \quad \text{as } y\to+\infty \nn
\eea
(the second one was also found in diffusive single-file models, see references in \cite{SinghChain2020}). In both cases, one recovers the MSD for $k\ll \sqrt{2Kt}$, while for large separations $k\gg \sqrt{2Kt}$, one finds a faster than exponential decay with a decay length $2\sqrt{2Kt}$. Similar scaling forms were obtained for the two-time autocorrelation $C_0(t,t')$, as well as for the full two-point two-time correlations. In particular, it was found that $C_0(t,t')$ is not a function of only $t-t'$, which is reminiscent of aging in some simple systems \cite{CugliandoloAging}. We refer to \cite{SinghChain2020} for these results, as well as for the full expressions of the scaling functions in \eqref{scaling_MSD_harmonicChain}, \eqref{scaling_Ck_harmonicChain1} and \eqref{scaling_Ck_harmonicChain2}. Some results for the statistics of the separation between two consecutive particles, $x_{i+1} -x_{i}$ were also obtained in \cite{HarmonicChainRevABP} and \cite{HarmonicChainRTPDhar}.

\section{Conclusion}

Interacting active particles have been shown to exhibit a variety of fascinating phenomena, from the motility induced phase separation in the presence of short-range repulsion, to the transition to collective motion in the presence of alignment interactions. Finding models for which exact analytical results can be obtained would greatly contribute to our understanding of these systems, but it is also a tremendous technical challenge. Until now, only a few attempts have been successful, mostly in the case of two-particle models, lattice models with contact interactions and harmonic chains.

Another setting which might allow for such results is the case of long-range interactions, e.g., via a pairwise Coulomb potential. Until now, there has been very few studies concerning such models, in particular beyond the two-particle case. One reason for this might be the apparent lack of physical motivation. Indeed, for real-life examples of active particles, long-range electromagnetic interactions rarely play a central role. Interactions which are mediated by the environment, such as hydrodynamic interactions, do have long-range effects but are also much more complex (i.e., they do not depend solely on the positions of the particles). However, recently, experiments were performed using passive colloids with long-range magnetic interactions in contact with a bath of active particles \cite{ActiveMeltingNature2024}, which proves that such systems may indeed be realized experimentally, and that they may also exhibit interesting properties. Of course, an additional argument for the study of such models is that it would also advance our understanding of active particle systems in general, in particular through the development of new analytical methods.

As already announced in the introduction, these models will constitute the main topic of this thesis. In the next chapter, we will review some of the main results that have been obtained for Brownian particles with long-range interactions, before studying how active noise affects these results in Parts~\ref{part:density} and \ref{part:fluctuations}.

\chapter{Brownian particles with long-range interactions: Dyson Brownian motion, rank diffusion and Riesz gases} \label{chap:Riesz_review}

\section{Riesz and Coulomb gases} 

Long-range interactions, such as electromagnetic interactions or gravitational interactions, play a central role in physics. The deterministic dynamics of particles interacting via inverse power law potentials has been studied for a long time \cite{Pad90,Chavanis,Plasma,Miller90}. In this chapter, we review some results concerning the stochastic dynamics of such systems in the presence of Brownian noise. Such models are known as {\it Riesz gases} and they have attracted a lot of interest other the years, both in physics and in mathematics \cite{Riesz,Lewin,SerfatyBook}. In one dimension, for $N$ particles labeled $i=1,\dots,N$, the general equations of motion for the positions $x_i(t)$ read
\be \label{def_Riesz_Brownian}
\frac{dx_i}{dt} = -\sum_{j(\neq i)} W'(x_i-x_j) - V'(x_i) + \sqrt{2T} \, \xi_i(t) \; , \quad W(x)= \begin{cases} g \, s^{-1} |x|^{-s} \quad \text{for } s\neq 0 \;, \\
-g\log |x| \quad \ \text{for } s=0  \;. \end{cases}
\ee
Here $g$ is a positive constant, $g>0$, so that the interaction is always repulsive. In this chapter we will focus on $s\geq -1$, although Riesz gases are well-defined for any $s>-2$. 

The term ``long-range'' is sometimes used to describe any interaction which does not have a characteristic decay length, i.e., mostly for any power law interaction potential. However, when studying Riesz gases, it is important to distinguish between two different situations. When $s>d$, where $d$ is the dimension of the system (here $d=1$), the interaction potential is integrable. This means that the energy is dominated by the interactions between particles which are close to each other. In this case, the interaction is called {\it short-range}. Riesz gases with short-range interactions bear many resemblances with single-file diffusive systems such as the symmetric exclusion process \cite{Harris65,Arratia83,SingleFileMajumdar1991,SingleFileKrapivsky2014,SingleFileKrapivsky2015,TaggedSFD2015,SingleFileReview}. For instance the MSD of a tagged particle scales as $\sqrt{t}$ in both cases \cite{Harris65,Arratia83,DFRiesz23}. By contrast, for $s \leq d$, the interaction potential is non-integrable and the behavior of the system can only be understood by taking into account the pairwise interactions at all distances. In this case, the interaction is called {\it long-range}. Riesz gases with long-range interactions exhibit behaviors which are very different from their short-range counterparts. More generally, the statistical mechanics of systems with long-range interactions is very rich and may exhibit peculiar phenomena such as inequivalence of ensembles (see, e.g., \cite{DauxoisPhysRep2009,Dauxois_book} for general reviews).

Beyond the physical relevance of long-range interactions in fields as diverse as astrophysics \cite{Pad90,Chavanis} or plasma physics \cite{Plasma}, Riesz gases have strong connections with many fields of mathematics, such as random matrix theory \cite{Mehta_book,Forrester_book,bouchaud_book}, Ginzburg-Landau vortices \cite{GB2012} or sphere packing problems \cite{Cohn2017,Cohn2022,Petrache2020}. Because of this, both their equilibrium \cite{Riesz,leble2017,leble2018,riesz3,Agarwal2019,jit2021,leble_loggas,BoursierCLT,BoursierCorrelations,jit2022,santra2022,dereudre2023number,Lelotte2023,Beenakker_riesz,Riesz_FCS,UsRieszCumulants} and more recently their dynamical behavior \cite{Huse_riesz,SerfatyDynamics2022,DFRiesz23,SerfatyDynamics2023,Riesz_expansion} have attracted a lot of attention. Here we will only review some aspects of these models which are relevant for this thesis. For recent reviews (from a more mathematical perspective) see, e.g., \cite{Lewin,SerfatyBook}.

Among the large class of Riesz gases, some models are of particular relevance. In particular, the {\it log-gas}, or {\it Dyson Brownian motion} (DBM), corresponding to $s=0$, is strongly connected to the Gaussian ensembles of random matrix theory (and to the Ginibre ensemble in 2D). The {\it Calogero-Moser model}, $s=2$ has also been particularly studied and has some similarities with the DBM \cite{Agarwal2019}. The case $s=d-2$ corresponds to the {\it Coulomb interaction} in $d$ dimensions. For $d=1$ it is also called the {\it rank interaction}, since the interaction force is independent of the distance and therefore only depends on the ordering of the particles. Below we will give more details about these specific models (in 1D), before returning to the general Riesz gas for a short discussion on the microscopic dynamics.
\\

\noindent {\bf Main observables of interest.} In this chapter and in the rest of this thesis, we will be mainly interested in two types of observables. To understand the behavior of the different models at the macroscopic scale, we will study the density of particles $\rho(x,t)$, which we define as
\be \label{def_density_Brownian}
\rho(x,t) = \frac{1}{N} \sum_{i=1}^N \delta(x-x_i(t)) \;.
\ee
We will mostly focus on the limit $N\to +\infty$, where the density generally becomes a smooth function. One way to study this quantity, including the finite $N$ fluctuations, is through the Dean-Kawasaki equation \cite{Dean,Kawa}. For a generic interaction potential $W(x)= N^{-1}\tilde W(x)$, and external potential $V(x)$, the DK equation reads
\be \label{Dean_eq_Brownian}
\partial_t \rho(x,t) = \partial_x \{ \rho(x,t) [ V'(x) +
\int dy \rho(y,t) \tilde W'(x-y) ]\} + T \partial^2_x \rho(x,t)  
+ \sqrt{\frac{2 T}{N}} \, \partial_x [ \sqrt{\rho(x,t) } \, \eta(x,t) ] \;,
\ee
where $\eta(x,t)$ is a Gaussian white noise with zero mean and unit variance. Note that, depending on the behavior of $\tilde W'(x)$ near $x=0$, the integral has to be treated with care. We will discuss this equation more in detail in Part~\ref{part:density}, along with its extension to RTPs.

We will also be interested in the statics and the dynamics at the microscopic level, which we will study using the variance and covariance of the particle positions $x_i$ in the stationary state, as well as the correlation functions already defined in \eqref{def_C0_harmonciChain}, \eqref{def_Ck_harmonciChain} and \eqref{def_C0tt_harmonciChain} in the context of harmonic chains of active particles, in particular the mean squared displacement $C_0(t)$. A related question which is also of particular interest is the statistics of the interparticle distances (or gaps), $x_i-x_{i+k}$.

\section{Dyson Brownian motion} \label{sec:DBM_brownian}

\subsection{Definition and connection with random matrix theory}

Due to its strong connection with random matrix theory (RMT), the Dyson Brownian motion (DBM), or log-gas, is probably the most studied Riesz gas model \cite{Mehta_book,Forrester_book,bouchaud_book,Dyson,Spohn2,TristanThese}. It was introduced by Freeman Dyson in 1962 to study the distribution of eigenvalues of the Gaussian matrix ensembles \cite{Dyson}. It is defined by setting $s=0$ in \eqref{def_DBM_Brownian}, leading to a pairwise repulsion force $1/(x_i-x_j)$. We will mostly be interested in its behavior in the presence of a harmonic external potential $V(x)=\lambda x^2/2$. In order for the density to have a finite support as $N\to+\infty$ (see discussion below), we rescale the interaction strength $g$ and the temperature $T$ with a factor $1/N$, so that the equations of motion now read (we also add a factor 2 to the interaction strength for convenience)
\be \label{def_DBM_Brownian}
\frac{dx_i}{dt} = - \lambda x_i +  \frac{2g }{N} \sum_{j (\neq i)} 
\frac{1}{x_i-x_j} + \sqrt{\frac{2 T}{N}} \, \xi_i(t) \;.
\ee
Note that our choice of scaling differs from some of the references cited below. We adapt all the expressions to match our scaling convention. At large times, the systems reaches an equilibrium state at large times, where the joint distribution of positions is given by the Gibbs measure
\be \label{DBM_Gibbs}
P(x_1,...,x_N) = \frac{1}{Z} e^{- \frac{N}{T} ( \frac{\lambda}{2} \sum_i x_i^2 + \frac{2 g}{N} \sum_{i<j} \log|x_i-x_j| ) } = \frac{e^{- \frac{N\lambda}{2T} \sum_i x_i^2}}{Z} \prod_{i<j} |x_i-x_j|^{\beta} \; , \quad \beta = \frac{2g}{T} \;.
\ee
One may recognize the joint law of eigenvalues of the Gaussian $\beta$-ensemble. Indeed, there is a deep connection between the DBM and the Gaussian matrix ensembles, which we now discuss.
\\

\noindent {\bf Gaussian ensembles.} The Gaussian Orthogonal Ensemble (GOE), Gaussian Unitary Ensemble (GUE) and Gaussian Symplectic Ensemble (GSE), defined respectively on the set of real symmetric matrices, complex Hermitian matrices and quaternionic self-adjoint matrices, are the only matrix ensembles which satisfy at the same time two important properties: their entries are {\it independent} (apart from the obvious symmetries), and they are {\it invariant by rotation}. This is why, since their introduction by Wigner in the context of nuclear physics \cite{Wigner1}, they are an essential model for structure-less data. This gives them a central role in many fields of physics, mathematics or computer science \cite{Mehta_book,Forrester_book,OxfordRMT}. In addition, many of the results concerning the eigenvalues of Gaussian matrices are universal, in the sense that they extend to a wider class of random matrices known as {\it Wigner matrices}, i.e., real symmetric, complex Hermitian or quaternionic self-adjoint matrices with independently distributed entries \cite{Erdos,ErdosYau,Pastur,Soshnikov,Sosh99,Tao_book}. 

One way to define these ensembles is by giving the probability law of the entries. For the GOE, it can be described as follows: each entry on the upper-triangular part of the matrix is an independent real centered Gaussian variable, with variance given by
\be
\langle M_{ij}^2 \rangle = \begin{cases} \frac{\sigma}{N} \quad \hspace{0.05cm} \text{if } i < j \;, \\
\frac{2\sigma}{N} \quad \text{if } i = j \;, \end{cases}
\ee
where $\sigma$ is a real number, while the remaining entries are fixed by the constraint $M_{ij}=M_{ji}$.

To see the connection with the DBM, let us consider a real symmetric matrix $M$ with entries obeying independent Ornstein-Uhlenbeck processes
\be
\frac{dM_{ij}}{dt} = -\lambda M_{ij} + \sqrt{\frac{2\lambda \sigma_{ij}}{N}} \eta_{ij}(t) \quad , \quad \sigma_{ij}=\begin{cases} \sigma \quad \; \text{if } i\neq j \;, \\ 2\sigma \quad \text{if } i=j \;, \end{cases}
\ee
for $i\leq j$, where the $\eta_{ij}(t)$ are $N(N+1)/2$ independent centered Gaussian white noises with unit variance. Assuming $M_{ij}(t=0)=0$, then at any time $t$, $M_{ij}(t)$ is a centered Gaussian variable with variance $\langle M_{ij}(t)^2 \rangle=(1-e^{-2\lambda t})\sigma_{ij}/N$. Thus, in the stationary state the matrix ${\bf M}$ belongs to the GOE with variance $\sigma/N$ for the off-diagonal entries (and with variance $(1-e^{-2\lambda t})\sigma/N$ at finite time). Using perturbation theory for the eigenvalues, one can show that the eigenvalues $x_i(t)$ ($i=1,...,N$) obey the following SDE (see, e.g., \cite{bouchaud_book,TristanThese} for a full derivation),
\be \label{DBM_fromM}
\frac{dx_i}{dt} = - \lambda x_i +  \frac{2\lambda \sigma}{N} \sum_{j (\neq i)} 
\frac{1}{x_i-x_j} + \sqrt{\frac{4\lambda \sigma}{N}} \, \xi_i(t) \;.
\ee
Fixing $\sigma=g/\lambda$, one recovers the DBM introduced in \eqref{def_DBM_Brownian} with $T=2g$, i.e., $\beta=1$. The eigenvalues of a random matrix with independent entries are thus not independent, and they tend to repulse each other. The DBM provides a natural framework to study these eigenvalues. For instance, with this connection we immediately obtain that the joint distribution of eigenvalues in the GOE is given by \eqref{DBM_Gibbs} with ``inverse temperature'' $\beta=1$. Similarly, one can show that the GUE corresponds to $\beta=2$ and the GSE to $\beta=4$. Generalizations to arbitrary values of $\beta$ have been proposed using different methods (see, e.g., \cite{GeneralBeta,BouchaudGuionnet,allez_satya}), for instance using tridiagonal matrices with non-identically distributed entries (although in this case the connection only holds in the stationary state). Independently from RMT, let us mention that the DBM for $\beta=2$ was also shown to have the same distribution as $N$ Brownian motions which are conditioned not to intersect with each other \cite{Dyson}.

\subsection{Wigner's semi-circle law} \label{sec:DBM_Wigner}

\begin{figure}
    \centering
    \includegraphics[width=0.5\linewidth,trim={5cm 3cm 5cm 8cm},clip]{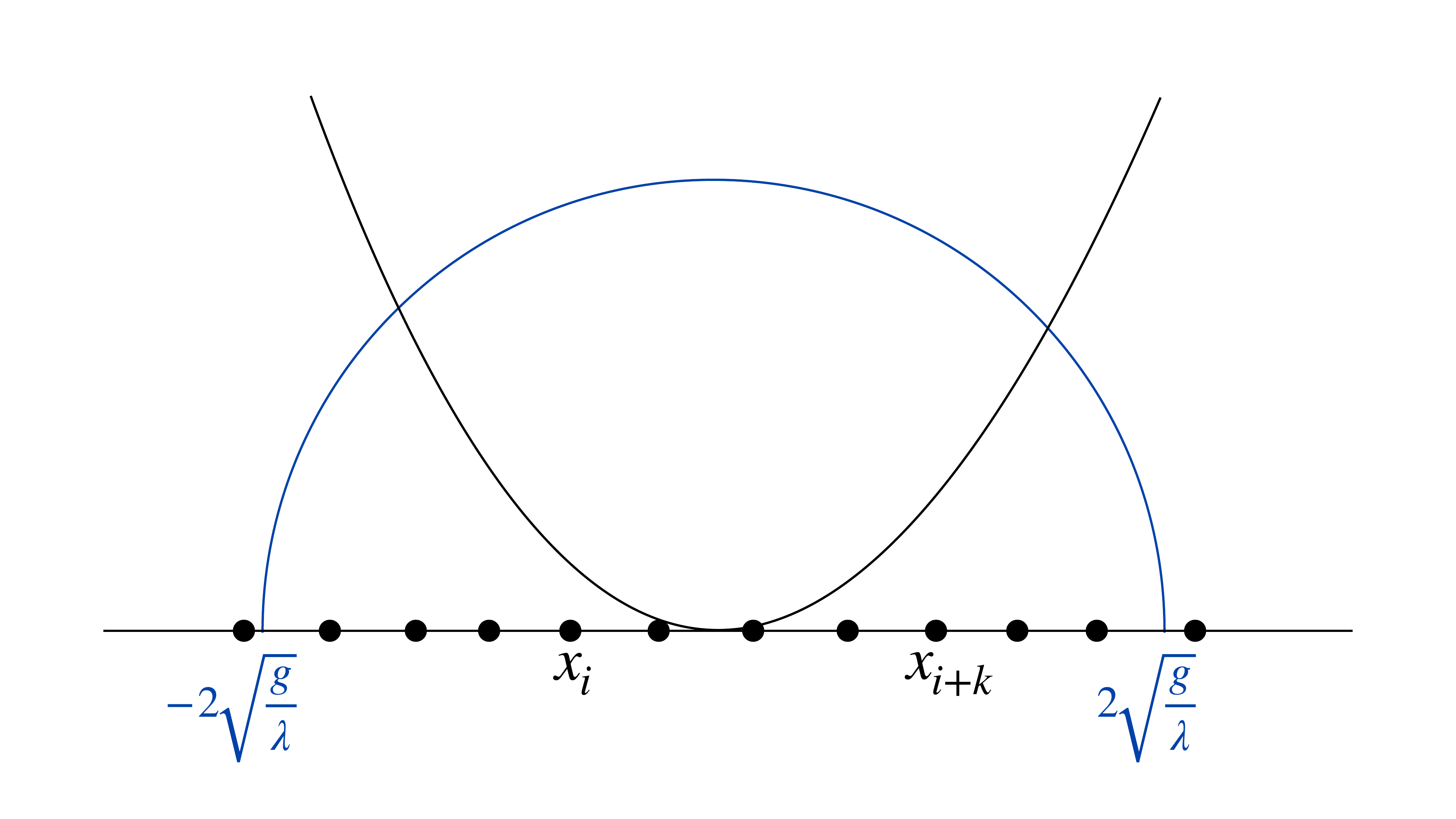}
    \caption{Schematic representation of the Dyson Brownian motion (or log-gas) in a harmonic trap, defined in \eqref{def_DBM_Brownian}, along with its equilibrium density at large $N$, the Wigner semi-circle \eqref{WignerSC}.}
    \label{fig:semicircle}
\end{figure}

As mentioned above, our first quantity of interest is the density of particles (or of eigenvalues in the RMT interpretation) defined in \eqref{def_density_Brownian}. In the stationary state, and in the limit $N\to+\infty$, it has a finite support $[-2\sqrt{g/\lambda},2\sqrt{g/\lambda}]$, and it takes a semi-circular shape (see Fig.~\ref{fig:semicircle}),
\bea \label{WignerSC}
\rho_{sc}(x) = \frac{\lambda}{2\pi g} \sqrt{\frac{4g}{\lambda}-x^2} \;.
\eea
This is the celebrated {\it Wigner semi-circle law}, which describes the distribution of eigenvalues of Gaussian matrices, and more generally Wigner matrices \cite{WignerSC}. Let us note that if we had used the definition \eqref{def_Riesz_Brownian} for $s=0$ without the $1/N$ rescaling, the support would instead scale as $\sqrt{N}$. In the absence of a confining harmonic potential, assuming all the particles start at $x=0$, the density is still a semi-circle at all times but with a support expanding as $\sqrt{t}$.

The expression \eqref{WignerSC} can be obtained through the so-called {\it resolvent method} (see, e.g., \cite{RogersShi,bouchaud_book}). This consists in deriving a differential equation for the {\it Stieltjes transform} of the density, defined as
\be \label{def_Stieltjes_DBM}
G(z,t) = \int dx \frac{\rho(x,t)}{z-x} = \frac{1}{N} \sum_i \frac{1}{z-x_i(t)} \;.
\ee
There are several ways to do this. One way is to start from the equation \eqref{Dean_eq_Brownian} for the density, with $\tilde W'(x)=-2g/x$ and $V'(x)=\lambda x$, and in the limit $N\to+\infty$, i.e., without the noise term,
\be \label{Dean_DBM_Brownian}
\partial_t \rho(x,t) = \partial_x \{ \rho(x,t) [ \lambda x - 2g
\int dy \frac{\rho(y,t)}{x-y} ]\} \;.
\ee
Note that with our choice of scaling the diffusive term also vanishes in the large $N$ limit. Multiplying by $1/(z-x)$ and integrating over $x$, and using integrations by parts as well as the identity $\partial_x \frac{1}{z-x}=-\partial_z \frac{1}{z-x}$ to rewrite the different terms, one obtains (see, e.g., the Appendix of \cite{ADBM1} for details)
\be \label{eg_resolvent_DBM}
\partial_t G(z,t) = \partial_z [\lambda zG(z,t) -g G(z,t)^2] \;.
\ee
For $\lambda=0$ this is the Burgers equation. The density can then be recovered from \eqref{eg_resolvent_DBM} using the relation
\be \label{invert_Stieltjes}
\rho(x,t) = \frac{1}{\pi} \lim_{\eps \to 0} {\rm Im} \, G(x-i\epsilon,t) \;.
\ee
This method can be used to study the time evolution of the density, or to compute the stationary state by setting the time derivative to zero and using that at large $z$, $G(z,t)\simeq1/z$, leading to
\be
G(z) = \frac{\lambda}{2g} \left(z-\sqrt{z^2-\frac{4g}{\lambda}} \right) \;.
\ee
Using \eqref{invert_Stieltjes}, one recovers the semi-circle law \eqref{WignerSC}.
\\

\noindent {\bf Small and large temperature limits.} In the limit $T\to 0$, the system converges to the minimum energy configuration. In this case, the rescaled positions of the particles are given by the zeros of the Hermite polynomial $H_N(x)$ \cite{HermiteZeros,bouchaud_book},
\be \label{eq_DBM_Hermite_zeros}
x_{{\rm eq},i} = \sqrt{\frac{2g}{\lambda N}}\, y_i \quad , \quad H_N(y_i) = 0 \quad , \quad i=1,...,N \;.
\ee 
As $N\to+\infty$, the density of Hermite zeros is known to converge to the semi-circle. This is consistent with the fact that, with the scaling used above, the role of the temperature becomes negligible at large $N$ (note that it does not appear in \eqref{WignerSC}). In other words, the density at large $N$ is given by the minimal energy configuration.

When instead the temperature in \eqref{def_DBM_Brownian} is increased and becomes of the order of $N$ (i.e., $\beta=O(1/N)$), the diffusive term in the DK equation becomes dominant and there is a crossover to a Gaussian density. This Gauss-Wigner crossover was studied in detail in several works \cite{BouchaudGuionnet,cuenca,allez_satya}. Writing $T=N \tilde T$ with $\tilde T=O(1)$, such that $\beta=\frac{2c}{N}$ with $c=\frac{g}{\tilde T}$, it takes the form
\be \label{GaussWigner}
\rho(x) = \sqrt{\frac{\tilde T}{2\pi}} \frac{1}{\Gamma(1+c)} \frac{1}{|{\cal D}_{-c}(ix)|^2} \quad , \quad
{\cal D}_{-c}(z) = \frac{e^{-\frac{z^2}{4\tilde T}}}{\Gamma(c)} \int_0^{+\infty} dy \ e^{-\frac{1}{\tilde T} (zy + \frac{y^2}{2})} \left( \frac{y}{\sqrt{\tilde T}} \right)^{c-1} \;,
\ee
where ${\cal D}_{-c}(z)$ is the parabolic cylinder function. One can check that this density indeed interpolates between the semi-circle for $c \to +\infty$ and the Gaussian for $c = 0$.

\subsection{Fluctuations of particle positions and gaps} \label{sec:DBM_micro}

\noindent {\bf Particle positions.} In the previous section we considered the DBM from a macroscopic viewpoint by looking at the total density. One may also take a more microscopic approach and consider the fluctuations at the single-particle level. Let us assume that the particles are ordered such that $x_1>x_2>...>x_N$ (which we can do at the cost of a relabeling). There is an important distinction between particles in the {\it bulk}, i.e., with label $i$ such that $i/N\to a\in(0,1)$ as $N\to+\infty$ (which we denote $i=O(N)$ for convenience), and particles located at the {\it edge}, i.e., such that $i/N\to 0$ (or 1) (which we write $i\ll N$), for which the fluctuations have a different scaling.

{\it Bulk.} For particles in the bulk, it was proved in \cite{Gustavsson} for $\beta=2$ and in \cite{ORourke2010} for $\beta=1$ and 4 that, in the large $N$ limit, the distribution of the particle position $x_i$ is Gaussian with variance
\begin{equation}
\langle (x_i-x_{{\rm eq},i})^2 \rangle = \frac{T}{\lambda} \frac{\log N}{N^2} \frac{1}{2(1-(\frac{x_{{\rm eq},i}}{2\sqrt{g/\lambda}})^2)} \;.
\label{DBMvar_largeN_ORourke}
\end{equation}
The average position $x_{{\rm eq},i}$ can be approximated at large $N$, using that the density converges to the semi-circle, as
\be \label{xeq_semiCircle}
\frac{x_{{\rm eq},i}}{2\sqrt{g/\lambda}} = \frac{y_{{\rm eq},i}}{\sqrt{2N}} \simeq \mathcal{G}^{-1}(i/N) \quad , \quad \mathcal{G}(x)= \frac{2}{\pi} \int_{-1}^x du \sqrt{1-u^2} \;,
\ee
where $\mathcal{G}(x)$ is the cumulative distribution of the semi-circle law on $[-1,1]$.

{\it Edge.} While the motion of the particles in the bulk is strongly constrained by the interactions, the edge particles are much more free to move, which leads to a different scaling of the fluctuations with $N$. The fluctuations of the rightmost particle (or the largest eigenvalue in the context of RMT) in particular has attracted a lot of attention. It was first studied by Tracy and Widom  \cite{TracyWidom1,TracyWidom2} who showed (in the case of the Gaussian ensembles, $\beta=1,2,4$, the result was later generalized to any $\beta>0$ in \cite{RamirezRiderVirag} using the stochastic Airy operator) that, at large $N$, the random variable
\be
\chi = N^{2/3}\left(\frac{x_1}{\sqrt{g/\lambda}}-2\right)
\ee
obeys the {\it Tracy-Widom distribution}, parametrized by $\beta$, with PDF $\mathcal{F}_\beta'(x)$. This is an asymmetric distribution, with tails given by
\be
\mathcal{F}_{\beta}'(x) \simeq \begin{cases} e^{-\frac{\beta}{24}|x|^3} \hspace{0.67cm} \text{as } x\to-\infty \;, \\ e^{-\frac{2\beta}{3}|x|^{3/2}} \quad \text{as } x\to+\infty \;. \end{cases}
\ee
An interesting feature of the Tracy-Widom distribution is its universality, as it was also found to appear in several apparently unrelated contexts such as growth processes or the combinatorics of longest increasing subsequences \cite{SatyaTracyWidomLecture}. 

The $N^{-2/3}$ scaling, which contrasts with the $\sqrt{\log N}/N$ scaling of the bulk, is not restricted to the rightmost particle but extends to all edge particles. For particles such that $1\ll i \ll N$, the distribution of the position $x_i$ was proved in \cite{Gustavsson,ORourke2010} to be Gaussian, as in the bulk, but with variance
\be \label{var_edge_standardDBM}
\langle (x_i-x_{{\rm eq},i})^2 \rangle = \frac{T}{\lambda N^{4/3}} \left(\frac{1}{12\pi}\right)^{2/3} \frac{\log i}{i^{2/3}} \;,
\ee
i.e., the fluctuations again scale as $N^{-2/3}$. Here, the equilibrium position can be obtained using the large $N$ asymptotics of the Hermite zeros \cite{Hermite_asymptotics},
\begin{eqnarray}
x_{{\rm eq},i} = 2\sqrt{\frac{g}{\lambda}} \left( 1  + \frac{a_i}{2} N^{-2/3} + O(N^{-1}) \right) \quad , \quad a_i = - (\frac{3 \pi}{8} (4i-1))^{2/3} + O(i^{-4/3}) \;,
\label{hermite_roots_edge_app}
\end{eqnarray}
where $a_i$ is the $i^{th}$ zero of the Airy function. Note that \eqref{var_edge_standardDBM} matches for $i\gg 1$ with \eqref{DBMvar_largeN_ORourke} using the expression of $x_{{\rm eq},i}$ near the edge \eqref{hermite_roots_edge_app}.

Let us finally mention a nice result obtained recently in \cite{GorinInfiniteBeta}, concerning the two-point two-time covariance of the position of edge particles. In the limit $\beta\to+\infty$ (i.e., $T\to0$) and $N\to+\infty$, denoting $\delta x_i(t)=x_i(t)-x_{{\rm eq},i}$, it is given by the expression
\be \label{AiryDBM}
\langle \delta x_i(t) \delta x_j(t') \rangle \simeq \frac{T}{\lambda N^{4/3}} \frac{1}{\Ai'(a_i) \Ai'(a_j)} \int_0^{+\infty} dx \ \frac{\Ai(a_i + x)\Ai(a_j + x)}{x} e^{-N^{1/3}\lambda|t-t'|x} \;,
\ee
which involves the Airy function $\Ai(x)$. This is a mathematical result, which we were able to recover through a different method presented in Chapter~\ref{chap:ADBMfluct}. In Part~\ref{part:fluctuations} we will extend some of these results to active noise and see in particular how the different scalings are affected.
\\

\noindent {\bf Interparticle distance.} Another microscopic quantity which can be studied is the relative distance between two particles, or gap. For two neighboring particles in the bulk, the gap is well approximated by {\it Wigner's surmise} distribution \cite{WignerSurmise},
\be \label{eqWignerSurmise}
p(\Delta) = a_\beta \Delta^{\beta} e^{-b_\beta \Delta^2} \quad , \quad \Delta = \frac{x_i-x_{i+1}}{\langle x_i-x_{i+1}\rangle} \;,
\ee
where $a_\beta$ and $b_\beta$ are two constants which depend on the parameter $\beta$, and the average gap is approximately given by
\be
\langle x_i-x_{i+1}\rangle = x_{{\rm eq},i}-x_{{\rm eq},i+1}\simeq 1/(N\rho_{sc}(x_{{\rm eq},i})) \;,
\ee 
where $\rho_{sc}(x)$ is the semi-circle density \eqref{WignerSC}. The distribution \eqref{eqWignerSurmise} is very different from the Poisson (i.e., exponential) distribution that one would expect for i.i.d variables. On the one hand, it vanishes for $\Delta \to 0$ with an exponent $\beta$, due to the repulsion between the particles (or the eigenvalues in RMT). On the other hand, it decays super-exponentially at large distances, meaning that two consecutive particles are never too far from each other. 

More generally, one may consider the distance between two particles $i$ and $i+k$. In the bulk, for $1\ll k \ll N$, one can show that its variance increases logarithmically with $k$ (see, e.g., the discussion in chapter 5.4 of \cite{bouchaud_book}, see also \cite{Mehta_book,Forrester_book,Bourgade2022})
\be \label{bouchaudDBM_distance} 
\langle (\delta x_i - \delta x_{i+k})^2 \rangle \simeq \frac{1}{N^2} \frac{T}{g\pi^2 \rho_{sc}(x_{{\rm eq},i})^2} \log k \;.
\ee 
This quantity is related to leading order to the variance of the number of particles contained inside a given interval, which thus also increases logarithmically with the size of the interval \cite{Dyson62,Dyson63}. This is again much slower than for a Poisson process, for which the variance of the interparticle distance is proportional to $k$. This is a sign of the {\it rigidity} of the DBM, in the sense that the long-range interaction between the particles strongly reduce the fluctuations. By contrast, active particles have been shown, both theoretically and experimentally, to exhibit {\it giant number fluctuations}, meaning that the variance increases faster than $k$ \cite{TonerTuReview,ChateGiant,GinelliGiant,DasGiant2012,Chate2010,NarayanGiant,ZhangGiant}. An interesting question would thus be to see how these two effects compete when both long-range interactions and active noise are present.

\subsection{The Calogero-Moser model} \label{sec:CM_review}

We now briefly introduce another Riesz gas model which exhibits some surprising similarities with the DBM, namely the {\it Calogero-Moser} (CM) model, which corresponds to $s=2$ in \eqref{def_Riesz_Brownian} \cite{Calogero71,Calogero75,Moser76}. As for the DBM, we will study it in the presence of a harmonic potential, scaling the parameters such that the density has a finite support as $N\to+\infty$,
\be \label{def_Calogero_Brownian}
\frac{dx_i}{dt} = - \lambda x_i + \frac{8\tilde g^2 }{N^2} \sum_{j (\neq i)} 
\frac{1}{(x_i-x_j)^3} + \sqrt{\frac{2 T}{N}} \, \xi_i(t) \;.
\ee
Although it was shown to have connections with some random matrix models \cite{BGS09}, it was mostly studied in the context of Hamiltonian dynamics (classical and quantum), due to its integrability properties \cite{Calogero71,Calogero75,Moser76,KP17,Poly06,OP81}. By contrast, little is known about its overdamped stochastic dynamics. A study of the equilibrium fluctuations in the CM model at finite temperature was recently conducted in \cite{Agarwal2019}, but it is mostly numerical.
\\

\noindent {\bf Connections with the DBM.} The first common trait between the CM model and the DBM is their minimal energy configuration. Indeed, quite surprisingly, the equilibrium positions of the particles are the rescaled zeros of the Hermite polynomial $H_N(x)$, exactly as for the DBM
\be \label{eqpos_CM}
x_{{\rm eq},i}^{CM}  = \frac{1}{\lambda^{1/4}}\sqrt{\frac{2 \tilde g}{ \, N}}\, y_i \quad , \quad H_N(y_i) = 0 \quad , \quad i=1,...,N \;.
\ee 
This implies that for $N\to+\infty$, the equilibrium density is also given by the Wigner semi-circle law
\bea \label{WignerSC_CM}
\rho_{sc}^{CM}(x) = \frac{\lambda^{1/2}}{2\pi \tilde g} \sqrt{\frac{4\tilde g}{\lambda^{1/2}}-x^2} \;,
\eea
for $x\in[-2\sqrt{\tilde g}/\lambda^{1/4},2\sqrt{\tilde g}/\lambda^{1/4}]$. The connection between the two models goes even further and extends to the fluctuations around this equilibrium configuration. Indeed, let us consider the Hessian matrices of these two models, evaluated around the equilibrium configuration $\vec{x}_{{\rm eq}} = (x_{{\rm eq},i},\dots, x_{{\rm eq},N})$,
\begin{eqnarray}
\lambda \mathcal{H}_{ij} = \frac{\partial^2 E^{DBM}}{\partial x_i \partial x_j} (\vec{x}_{\rm eq}) \quad , \quad E^{DBM}(\vec{x})=\frac{\lambda}{2} \sum_i x_i^2 -\frac{2g}{N}\sum_{i<j} \log |x_i-x_j|\; , \\
\lambda \mathcal{H}_{ij}^{CM} = \frac{\partial^2 E^{CM}}{\partial x_i \partial x_j} (\vec{x}_{\rm eq}^{CM}) \quad , \quad E^{CM}(\vec{x})=\frac{\lambda}{2} \sum_i x_i^2 + \frac{4\tilde g^2}{N^2}\sum_{i<j} \frac{1}{(x_i-x_j)^2}\; .
\label{defHessian}
\end{eqnarray}
Then the (dimensionless) Hessian matrix of the CM model is simply the square of the Hessian matrix of the DBM (see \cite{Agarwal2019} for a proof),
\be
\mathcal{H}^{CM} = \mathcal{H}^2 \;.
\ee
Fortunately, the matrix $\mathcal{H}$ can be diagonalized exactly \cite{eigenvectors}, which facilitates the study of the small temperature fluctuations in both models (see Part~\ref{part:fluctuations}).
\\

\noindent {\bf Scaling of the fluctuations.} Despite their connections, there is an important difference between the DBM and the CM model. While the DBM ($s=0$) is part of the family of long-range Riesz gases, the CM model ($s=2$) belongs to the short-range case. This has important consequences, in particular on the scaling of the fluctuations, which makes the comparison between the two models particularly interesting. In \cite{Agarwal2019}, it was shown numerically (using both Monte-Carlo simulations and numerical inversion of the Hessian matrix) that the fluctuations of the particle positions inside the bulk are Gaussian with variance scaling as $N^{-2}$ (compared to $\log N/N^2$ for the DBM). For the edge particle, the variance was found to scale as $N^{-5/3}$ (instead of $N^{-4/3}$ for the DBM), and the distribution of the rightmost particle is different from a Tracy-Widom distribution (note that our scaling implies a $1/N$ factor for the variance compared to \cite{Agarwal2019}, which we have taken into account). In Chapter~\ref{chap:ADBMfluct} we will provide an analytical confirmation of these results (in particular the scalings with $N$).

\section{Ranked diffusion} \label{sec:RD_brownian}

The last special case that we want to discuss is the 1D Coulomb interaction, corresponding to $s=-1$ in \eqref{def_Riesz_Brownian}. This is the case were the interaction force between two particles is independent of the distance. This implies that, in the absence of external potential, the total force acting on the particles only depends on their ordering, which is why this model is sometimes called {\it ranked diffusion}. In the presence of an external potential $V(x)$, the equations of motion for the positions $x_i(t)$ ($i=1,...,N$) read 
\be \label{def_1DCoulomb_Brownian}
\frac{dx_i}{dt} = \frac{\kappa}{N} \sum_{j=1}^N {\rm sgn}(x_i-x_j) - V'(x_i) + \sqrt{2 T} \, \xi_i(t) \;,
\ee 
where we have introduced the sign function
\be \label{def_sgn}
\sgn (x) = \begin{cases} 1 \quad \text{if } x>0 \;, \\ 0 \quad \text{if } x=0 \;, \\ -1 \quad \text{if } x<0 \;. \end{cases}
\ee
Here we only rescale the interaction strength as $g=\kappa/N$, as in \cite{PLDRankedDiffusion}. Contrary to the other models presented in this chapter, we will be interested in both the repulsive case $\kappa>0$ and the attractive case $\kappa=-\bar \kappa<0$.

In the repulsive case, this model was mostly studied in the presence of a harmonic confining potential, where it can be interpreted as a (classical) gas of 1D charged particles, such as electrons, inside a harmonic trap, or with a uniform background of opposite charges. In this context it is called the {\it jellium model}, or {\it one-dimensional one-component plasma} \cite{Lenard,Prager,Baxter1963,Kunz,Aizenman1980,Dean2010,Tellez,dhar2017exact,dhar2018extreme,Flack21,Flack22,Chafai_edge}. The attractive case, known as the {\it self-gravitating gas}, has also been investigated \cite{Rybicki,SireSG2002,SireSG2002bis,Sire,SireSG2004,SireSG2008}. Besides physics, this model has also been studied in the context of mathematics \cite{Pitman,OConnell} and finance \cite{Banner}. In this section, we focus on results for the particle density defined in \eqref{def_density_Brownian}.
\\

\noindent {\bf Equation for the rank field. } The density $\rho(x,t)$ was studied in \cite{PLDRankedDiffusion} for different types of external potentials, both in the repulsive and in the attractive case, using a mapping of the Dean-Kawasaki equation \eqref{Dean_eq_Brownian} at large $N$ to the Burgers equation. Starting from \eqref{Dean_eq_Brownian} with $\tilde W'(x)=-\kappa \, \sgn(x)$ and in the limit $N\to+\infty$ (i.e., without the noise term), the idea is to introduce the {\it rank field}, $r(x,t)$, which is simply the integral of the density,
\be \label{def_rank_field}
r(x,t) = \int^x_{-\infty} dy \, \rho(y,t)\, -\, \frac{1}{2} \;. 
\ee
The function $r(x,t)$ increases monotonously with $x$ from $-1/2$ at $x=-\infty$ to $+1/2$ at $x=+\infty$. Using the identity $\int dy \rho(y,t) \sgn(x-y) = 2r(x,t)$, the Dean-Kawasaki equation can be rewritten as
\be \label{eq_r_Brownian}
\partial_t r = T \partial_x^2 r - 2 \kappa r \partial_x r + V'(x) \partial_x r \;,
\ee
which is the viscous Burgers equation with an additional term due to the external potential.
\\

\begin{figure}
    \centering
    \includegraphics[width=0.35\linewidth,valign=t]{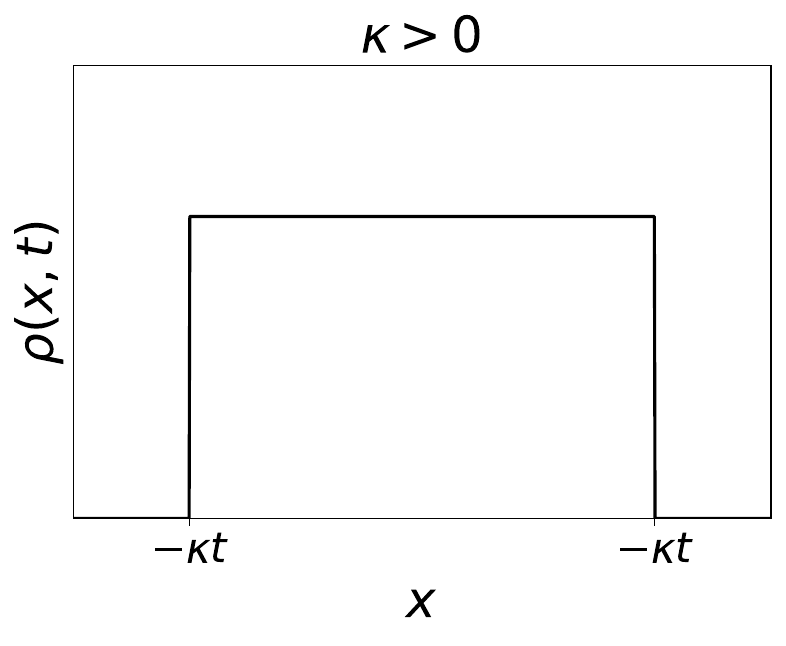}
    \hspace{1cm}
    \includegraphics[width=0.35\linewidth,valign=t]{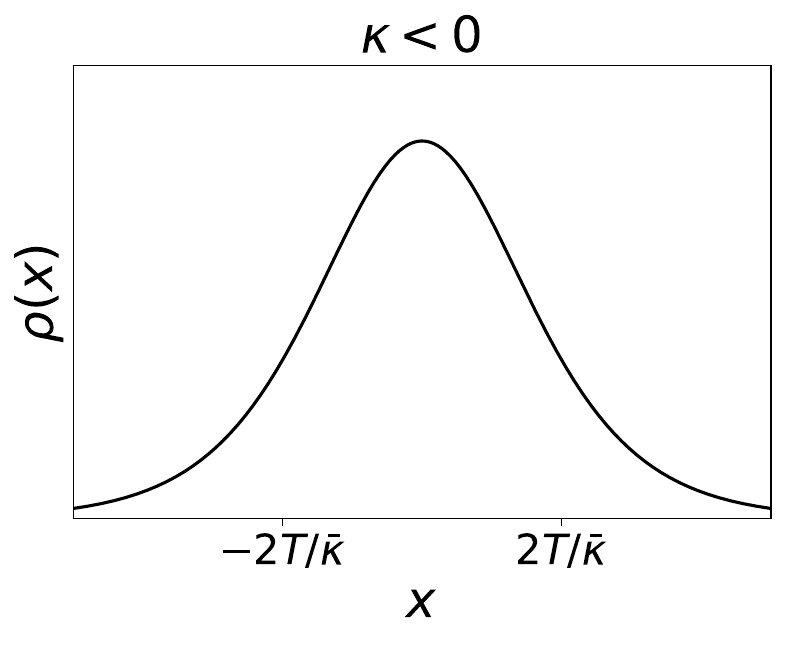}
    \caption{{\bf Left:} density at large time for the repulsive rank diffusion without external potential, given in \eqref{expanding_density_rank_repulsive_Brownian}. {\bf Right:} equilibrium density for the attractive rank diffusion without external potential, given in \eqref{soliton}.}
    \label{fig:rankBrownian}
\end{figure}

\noindent {\bf Repulsive case. } Let us start with the repulsive case $\kappa>0$. In the small temperature limit $T\to0$, and for a sufficiently confining convex potential $V(x)$, the equilibrium density has a finite support and can be easily obtained from \eqref{eq_r_Brownian} by setting the time derivative to zero,
\be
r(x) = \frac{V'(x)}{2\kappa} \quad , \quad \rho(x) = \frac{V''(x)}{2\kappa} \quad , \quad \text{for } x\in[x_e^-,x_e^+] \;,
\ee
where $x_e^\pm$ are the solutions of $V'(x_e^\pm)=\pm\kappa$. In particular, for a harmonic potential $V(x)=\frac{\mu}{2}x^2$, this gives a uniform density with support $[-\kappa/\mu,\kappa/\mu]$. The relaxation to equilibrium was also studied in \cite{PLDRankedDiffusion}. At finite temperature, $T>0$, and in the harmonic case, the equilibrium density was found to take a scaling form
\be \label{equil_density_rank_repulsive_Brownian}
\rho (x) = \frac{\mu}{2\kappa} \hat \rho_\alpha \left( x\sqrt{\frac{\mu}{T}} \right) \quad , \quad \alpha = \frac{\kappa}{\sqrt{\mu T}} \;,
\ee
where $\hat \rho_g(y)$ is a smooth function whose support is the whole real axis. As $\alpha$ varies, it interpolates between a square density in the small temperature limit $\alpha\to+\infty$ and a Gaussian in the weakly interacting limit $\alpha\to 0$.

The unconfined case $V'(x)=0$ was studied in detail in \cite{FlackRD}. In this case, one obtains at large time an expanding square density with support $[-\kappa t, \kappa t]$ (see Fig~\ref{fig:rankBrownian})
\be \label{expanding_density_rank_repulsive_Brownian}
\rho(x,t) \simeq \frac{1}{2 \kappa t} \, \Theta(\kappa t - |x|) \;,
\ee
where $\Theta(x)$ is the Heaviside theta function. For $T>0$, a boundary layer develops at the edges of the support $\pm\kappa t$, of typical size $\sim \sqrt{T\, t}$.
\\

\noindent {\bf Attractive case. } Let us now consider the attractive case $\kappa=-\bar \kappa<0$. In this case, the particles always form a bound state even in the absence of confining potential (the diffusion of the center of mass $\sim 2Tt/N$ is subleading in $N$). At $T=0$ and in the absence of potential, the density collapses in finite time and forms a {\it shock}, i.e., a delta peak. When one adds a finite temperature, this shock becomes smooth: the density is supported by the whole real axis and takes the form \cite{PLDRankedDiffusion}
\be \label{soliton} 
r(x)=\frac{1}{2} \tanh\left( \frac{\ \bar \kappa x}{2 T} \right)  \quad , \quad 
\rho(x) = \frac{\bar \kappa}{4 T \cosh( \frac{\ \bar \kappa x}{2 T} )^2} 
\ee
(see Fig~\ref{fig:rankBrownian} for a plot of the density).

\section{Microscopic fluctuations in Riesz gases} \label{sec:Riesz_brownian}

For more general Riesz gases, the equilibrium density inside a harmonic potential, at low temperature, was obtained in \cite{riesz3} for any $s>-2$, generalizing the semi-circle law for the DBM and the uniform density for the jellium. It takes the form
\be
\rho(x) = A(x_e^2-x^2)^{\alpha_s} \quad , \quad \alpha_s = \begin{cases} (s+1)/2 \quad \text{for } -2<s<1 \;, \\ 1/s \quad \text{for } s>1 \;, \end{cases}
\ee
for $x\in[-x_e,x_e]$ where $A$ is a normalization constant and $x_e$ is the edge of the support, which depends on the different parameters. The density in the expanding case (i.e., with no confining potential) was also obtained recently in the long-range case in  \cite{Riesz_expansion}.

More relevant to the present thesis is the study of the microscopic dynamics, as well as the statistics of the interparticle distances, which are less well-known. To study these quantities in the bulk, one may consider a Riesz gas as defined in \eqref{def_Riesz_Brownian}, in the absence of confinement, $V(x)=0$, with periodic conditions, as done in some mathematical works (e.g., \cite{BoursierCLT,BoursierCorrelations}). In this case, the interaction potential has to be properly periodized, which raises some technical issues. This is the approach that we will take in Part~\ref{part:fluctuations}, and we will discuss the details there. Another approach is to confine the particles inside a potential (e.g., harmonic), and to take the strength of the potential to zero while increasing the number $N$ of particles to keep the density $\rho$ constant, as done in \cite{SpohnTracer,DFRiesz23} (here and in the remainder of this section, contrary to the definition \eqref{def_density_Brownian} we do not normalize the density by $N$). In this case, one can consider the density to be uniform near the center of the trap. In the large $N$ limit both approaches are equivalent when considering microscopic quantities. In this section we thus consider a Riesz gas defined as in \eqref{def_Riesz_Brownian} with $V(x)=0$ and we consider the density $\rho$ to be a constant.

\subsection{Mean-square displacement of a tagged particle} \label{sec:Riesz_brownian_tagged}

For $s>0$, the divergence of the interaction potential $W(x)$ near $x=0$ is sufficiently fast to prevent the crossing of particle trajectories. This is also true for the DBM, $s=0$, in the case where $\beta\geq1$ (i.e., at sufficiently small $T$, including the Gaussian matrix ensembles) \cite{Lepingle07, Allez13}. When this single-file constraint holds, the displacement during time $t$ of a tagged particle is related to leading order to the current of particles $Q(t)$ flowing through the origin during time $t$ by $x_i(t)-x_i(0)\simeq Q(t)/\rho$. This allows to obtain the MSD, $C_0(t)=\langle (x_i(t)-x_i(0))^2 \rangle$, directly from the variance of $Q(t)$. 

For the DBM, this computation was performed rigorously in \cite{SpohnTracer}, leading to a logarithmic behavior at large time,
\be \label{C0_DBM_Spohn}
C_0(t) \simeq \frac{T}{\pi^2 g\rho^2} \log (t) \;.
\ee
This was extended much more recently to the long-range case with $0<s<1$ in \cite{DFRiesz23}, using a macroscopic fluctuation theory (MFT) approach \cite{MFTreview}, which in this case amounts to using the Dean-Kawasaki equation for the particle density to compute the fluctuations of the current $Q(t)$. In this case, the large time behavior was found to be
\be \label{C0_LR_MFT}
C_0(t) \simeq \frac{U_s T}{(2 g \rho^{s+2})^{\frac{1}{s+1}}} \, t^{\frac{s}{s+1}} \quad \text{with} \quad
U_s = \frac{ 4\Gamma\left(\frac{1}{s+1}\right)}{\pi s} \left[ \frac{\Gamma\left(1+\frac{s}{2}\right)}{2\sqrt{\pi} \, \Gamma\left( \frac{1-s}{2} \right)} \right]^{\frac{1}{s+1}} \;.
\ee
Finally, for $s>1$ the interaction is short-range and thus the MSD was obtained using pre-existing results on single-file diffusion \cite{SingleFileKrapivsky2014,TaggedSFD2015}
\be \label{C0_SR_MFT}
C_0(t) \simeq \frac{1}{\sqrt{1+ (1+s) \zeta(s) g\rho^s/T}} \sqrt{\frac{2Tt}{\pi \rho^2}} \;.
\ee
The marginal case $s=1$ was discussed using qualitative arguments, leading to a $\sqrt{t/\log t}$ behavior. A more controlled derivation of the result \eqref{C0_SR_MFT} was obtained more recently in \cite{Grabsch2025}. These results should of course be compared to the free diffusion case $C_0(t)\sim2Tt$. The presence of short-range interactions, which prevent particle crossings, already reduces the MSD to a $\sqrt{t}$ behavior, a well-known result of single-file diffusing systems. When the interaction becomes long-range, the movements of the particles become more and more restricted as $s$ decreases (one can say that the rigidity increases), down to a logarithmic behavior for the log-gas. The case $s<0$ cannot be studied using similar methods since in this case particles are allowed to cross, and to our knowledge it remains to be investigated.

It is important to note that the prefactors in the expressions above depend on the type of initial condition considered. The results above are given for a {\it quenched}, i.e., deterministic initial condition. One can instead consider an {\it annealed} initial condition where $x_i(0)$ is drawn randomly from the equilibrium distribution. This leads to a factor $2^{\frac{1}{s+1}}$ in the long-range case and a factor $\sqrt{2}$ in the short-range case. 

Finally, let us mention that the two-time correlations of the position $C_0(t,t')$ (defined in \eqref{def_C0tt_harmonciChain}) where also computed in \cite{DFRiesz23} for $0<s<1$. The result was found to be the same as for a fractional Brownian motion with Hurst exponent $H = s/(1+s)$ \cite{fBM}.

\subsection{Variance of the interparticle distance and counting statistics} \label{sec:gap_varRiesz_review}

We now consider another microscopic quantity, namely the variance of the distance between two particles $D_k(0)=\langle (x_i-x_{i+k})^2\rangle-\langle x_i-x_{i+k}\rangle^2$. We already discussed this quantity in the DBM case in Sec.~\ref{sec:DBM_micro}, where we mentioned that it is directly related to the variance of the number of particles contained inside a fixed interval. For the DBM, the $\log k$ behavior (see \eqref{bouchaudDBM_distance}) is another sign of the rigidity of the system. By contrast, in the short-range case, the statistics are approximately Poissonian at large scale, leading to a linear dependence $D_k(0)\propto k$. In the intermediate long-range case $0<s<1$, $D_k(0)$ was recently shown to behave as $k^s$ at large distances \cite{BoursierCLT}, showing once again that in the long-range case the rigidity increases continuously as $s$ decreases. Again, it would be interesting to see how this rigidity is affected by active noise.

\section{Conclusion}

Due to their connections with a variety of fields, including random matrix theory, Riesz gases, and especially some particular instances such as the Dyson Brownian motion and the ranked diffusion, have attracted a lot of interest both in physics and in mathematics. Yet, even for a Riesz gas of Brownian particles, there are still some open questions. For instance, a recent work points out to a possible phase transition for $-1 < s < 0$ between a crystal and a fluid phase \cite{Lelotte2023}. Some microscopic observables such as the MSD or the variance of the gaps have only been studied very recently for $0<s<1$, and more general dynamical correlation functions such as the covariance between the displacements of two particles, or the time correlations of the gaps, remain to be computed.

Beyond these remaining open problems, the question which will be of particular interest for us in the next two parts of this thesis is how these results are modified in the presence of {\it active noise}. For instance, as mentioned in the previous chapters, active particles are known to exhibit non-Boltzmann steady-states which may be very different from the equilibrium steady-states reached by Brownian particles. What do these non-Boltzmann steady-states look like in the presence of long-range interactions ? Can new phase transitions, similar to MIPS or the flocking transition, emerge in such systems ? Another interesting topic is the effect of activity on the microscopic fluctuations. While long-range interacting systems are characterized by their rigidity, it would be interesting to see how this is affected when one considers active particles, which can exhibit very strong fluctuations. These two general questions, namely the non-Boltzmann steady-states for the density and the microscopic fluctuations, will be the topic of Parts~\ref{part:density} and \ref{part:fluctuations} respectively.

\part{Non-Boltzmann steady-states for the density of interacting RTPs}
\label{part:density}

\vspace*{\fill}

\begin{center}
{\bf Abstract}
\end{center}

In this second part, we begin by introducing a general framework for studying the density of particles in models of 1D interacting RTPs, through an extension of the Dean-Kawasaki equation which we derive in Chapter~\ref{chap:DeanRTP}. We also discuss its limitations, in particular the fact that this hydrodynamic description fails in the presence of a strong short-range repulsion which prevents the particles from passing each other. We then apply this formalism to study the out-of-equilibrium stationary densities, in the limit where the number $N$ of particles is very large, for two type of interactions: a 1D Coulomb interaction, both attractive and repulsive, leading to the {\it active rank diffusion} in Chapter~\ref{chap:activeRD}, and a repulsive 2D Coulomb interaction, leading to the {\it active Dyson Brownian motion} in Chapter~\ref{chap:ADBM_Dean}. 
For the active rank diffusion, in the absence of confining potential, we obtain an exact analytical solution of the stationary DK equations in the attractive case, which allows us to shed light on a new non-equilibrium phase transition, between a phase where the density is smooth and a phase where clusters of particles (i.e., delta peaks in the density) form at the edges. While the large time behavior in the repulsive case is an expanding gas very similar to the Brownian case, we also extend our results to both a linear and a harmonic confining potentials, leading to very rich phase diagrams both for an attractive and a repulsive interaction. We also study a variation of the model involving non-reciprocal interactions.
For the active DBM, we introduce two variants of the model (both with a harmonic confining potential). For the variant where the interaction allows particles to cross, the DK equations at large $N$ allow us to perform a detailed study of the particle density and its different limiting behaviors. For the variant which forbids particle crossings, the DK approach fails but we still show, based on numerical simulations, as well as on some analytical results on the microscopic fluctuations which will be further detailed in Part~\ref{part:fluctuations}, that the density converges at large $N$ to the Wigner semi-circle for a wide range of parameters, as in the passive case. 

In this part, the Chapters~\ref{chap:DeanRTP} and \ref{chap:ADBM_Dean} are based on the Reference~\cite{ADBM1}, while Chapter~\ref{chap:activeRD} is based on the References~\cite{activeRD1} and \cite{activeRD2}. 
\vspace*{\fill}

\chapter{Dean-Kawasaki equation for interacting RTPs in 1D}
\label{chap:DeanRTP}

\section{General setting}

The aim of this chapter is to generalize the Dean-Kawasaki equation (DK equation) \cite{Dean,Kawa}, briefly introduced in Chapter~\ref{chap:Riesz_review} in the context of Riesz gases, to run-and-tumble particles in one dimension. For $N$ Brownian particles interacting via a pairwise potential, the DK equation provides an exact hydrodynamic equation describing the evolution of the particle density at large $N$, where the finite $N$ fluctuations are taken into account via multiplicative Gaussian white noise. Here we extend this approach to 1D RTPs obeying stochastic equations of motion of the form
\be \label{SDE_RTP_potential}
\frac{dx_i}{dt} = - V_{\sigma_i}'(x_i) - \frac{1}{N} \sum_{j(\neq i)} W_{\sigma_i,\sigma_j}'(x_i-x_j) + v_0 \sigma_i(t) + \sqrt{2T} \, \xi_i(t) \;,
\ee
for $i=1,\dots,N$, where the $\sigma_i(t)$ are $N$ independent telegraphic noises which switch between values $+1$ and $-1$ with rate $\gamma$, as defined in Sec.~\ref{sec:RTPdef}. At this point, both the external potential $V_\sigma(x)$ and the pairwise interaction potential $W_{\sigma,\sigma'}(x)$ are arbitrary. In particular, the interaction may be of the Riesz form (see \eqref{def_Riesz_Brownian}), as in the next two chapters, but it may also be more general. We even allow both potentials to depend on the state $\sigma_i(t)$ of the particles (we will consider some cases where this is the case in the next two chapters). Note the $1/N$ scaling for the interaction potential. For the sake of generality, we also include an additional Gaussian white noise $\xi_i(t)$. The temperature may be of order 1, or $1/N$ as in Sec.~\ref{sec:DBM_brownian} for the DBM. In the next chapters we will however mostly consider the purely active case with $T=0$.

The first important difference with the Brownian case is that we need to define two different particle densities, corresponding to the two states $\sigma=+1$ and $-1$ respectively,
\be \label{def_rho_pm} 
\rho_\sigma(x,t) = \frac{1}{N} \sum_i \delta_{\sigma_i(t),\sigma} \, \delta(x - x_i(t)) \; .
\ee
This distinction is necessary to generalize the DK equation to RTPs. We will thus obtain a set of two coupled partial differential equations, instead of a single equation as in the Brownian case. To study these equations, it is however often useful to rewrite them in terms of the total density $\rho_s(x,t)$, and the difference $\rho_d(x,t)$ (sometimes called the ``magnetization''),
\be \label{def_rho_sd}
\rho_s(x,t) = \rho_+(x,t) + \rho_-(x,t) \quad , \quad \rho_d(x,t) = \rho_+(x,t) - \rho_-(x,t) \;.
\ee
Note that the total density is the only one which is normalized to $1$, $\int dy \, \rho_s(y,t)=1$. However, if we consider the stationary state of the system in the limit $N\to+\infty$, the particles are equally split between the two states and one has in addition $\int dy \, \rho_\pm(y)=1/2$, and $\int dy \, \rho_d(y)=0$. Let us also note that, if the potentials $V_\sigma(x)$ and $W_{\sigma,\sigma'}(x)$ are even in $x$ and satisfy $V_\sigma(x)=V_{-\sigma(x)}$, $W_{\sigma,\sigma'}(x)=W_{-\sigma,-\sigma'}(x)$ (which will be the case in the next two chapters, except in Sec.~\ref{sec:non_reciprocal}), then the equation \eqref{SDE_RTP_potential} is invariant under the symmetry $(x_i,\sigma_i) \to (-x_i,-\sigma_i)$. In this case, the stationary densities for $N\to+\infty$ satisfy $\rho_-(x)=\rho_+(-x)$, i.e., $\rho_s(x)$ is even and $\rho_d(x)$ is odd. 

Below we will start by deriving the DK equations for $\rho_\pm(x,t)$ following the method of Dean \cite{Dean}, taking particular care in the treatment of the additional telegraphic noise. We will then discuss the hypotheses required for this equation to be valid. In particular, when the interaction diverges at short distance, preventing the particles from crossing each other, large clusters of particles form and the DK equation fails to properly describe the behavior of the system. To better understand this point, we will also provide an alternative derivation of a hydrodynamic equation for the mean particle density.

In the next two chapters, Chapters~\ref{chap:activeRD} and \ref{chap:ADBM_Dean}, we will use the DK equation derived here to study active versions of the ranked diffusion and of the DBM respectively. 

\section{Derivation} \label{sec:DKder}

Following the approach of \cite{Dean}, we consider an arbitrary test function $f(x)$, and we introduce,
\bea \label{defFint} 
F_\sigma(\vec x(t)) = \frac{1}{N} \sum_i f(x_i(t))\delta_{\sigma_i(t),\sigma} = \int dx f(x) \rho_\sigma(x,t) \;.
\eea
Then, using the It\=o chain rule, we have
\beq
\frac{d F_\sigma(\vec x(t))}{dt} = \frac{1}{N} \sum_i \delta_{\sigma_i(t),\sigma} f'(x_i(t)) \frac{dx_i(t)}{dt} + \frac{T}{N} \sum_i \delta_{\sigma_i(t),\sigma} f''(x_i(t)) +  \frac{1}{N} \sum_i f(x_i(t)) \frac{d\delta_{\sigma_i(t),\sigma}}{dt} \;.
\eeq
We can write $\delta_{\sigma_i(t),\sigma}=\frac{\sigma\sigma_i(t)+1}{2}$, so that $\frac{d\delta_{\sigma_i(t),\sigma}}{dt}=\frac{\sigma}{2} \frac{d\sigma_i(t)}{dt}$. Thus we get, using \eqref{SDE_RTP_potential},
\bea
\frac{d F_\sigma(\vec x(t))}{dt} = \hspace{-0.5cm} && \frac{1}{N}  \sum_i \delta_{\sigma_i(t),\sigma} f'(x_i(t)) \Big[ - V_{\sigma}'(x_i(t)) - \frac{1}{N}\sum_{j (\neq i)} W_{\sigma,\sigma_j(t)}'(x_i(t)-x_j(t)) + v_0\sigma + \sqrt{2 T} \, \xi_i(t) \Big] \nn \\
&& + \frac{T}{N} \sum_i \delta_{\sigma_i(t) ,\sigma} f''(x_i(t)) + \frac{1}{N} \frac{\sigma}{2}\sum_i f(x_i(t)) \frac{d\sigma_i(t)}{dt} \;.
\label{Dean_discret}
\eea
Let us assume for now, as in \cite{Dean}, that $W_{\sigma,\sigma'}'(0)=0$. In this case, the above equation can be rewritten as
\bea \label{integralDean0}
\frac{d F_\sigma(\vec x(t))}{dt} = \hspace{-0.5cm} && \int dx \rho_\sigma(x,t) \big[v_0\sigma f'(x) - f'(x) V_\sigma'(x) - f'(x) \sum_{\sigma'} \int dy \rho_{\sigma'}(y,t) W_{\sigma,\sigma'}'(x-y) + T f''(x) \big] \nn \\
&&  + \frac{\sqrt{2 T}}{N} \sum_i \delta_{\sigma_i(t),\sigma} f'(x_i(t)) \, \xi_i(t) + \frac{1}{N} \frac{\sigma}{2}\sum_i f(x_i(t)) \frac{d\sigma_i(t)}{dt} \;.
\eea
After integrations by parts we obtain
\bea
\int dx f(x) \partial_t  \rho_\sigma(x,t)
= \int dx f(x) && \hspace{-0.6cm} \Bigg\{ \partial_x  \Big[ \rho_\sigma(x,t)  \big(-v_0\sigma + V_\sigma'(x) + \sum_{\sigma'} \int dy \rho_{\sigma'}(y,t) W_{\sigma,\sigma'}'(x-y) \big) \Big]  \nn \\
&& + T \partial_x^2 \rho_\sigma(x,t) - \partial_x \Xi_\sigma(x,t)  +  \hat\zeta_\sigma(x,t) \Bigg\} \;. \label{integralDean}
\eea 
The last two terms correspond respectively to a passive noise $\Xi_\sigma$ (originating from the thermal white noise) and an active noise $\hat \zeta_\sigma$ (originating from the telegraphic noise). Let us examine these two terms more closely.

The passive noise term $\Xi_\sigma$ is Gaussian and reads
\be
\Xi_\sigma(x,t) = \frac{\sqrt{2 T}}{N}  \sum_i \delta_{\sigma_i(t),\sigma} \delta(x-x_i(t)) \, \xi_i(t) \;.
\ee
It is thus fully determined by its covariance which is (here and in the following we use the notation $\langle\dots\rangle$ indifferently for averages over the thermal 
and telegraphic noise) 
\be
\langle\Xi_\sigma(x,t) \Xi_{\sigma'}(x',t') \rangle = 
\frac{2 T}{N} \rho_\sigma(x,t) \delta_{\sigma,\sigma'} \delta(x-x') \delta(t-t') \;.
\ee
Hence we can write :
\beq \label{expr_thermal_noise}
\Xi_\sigma(x,t) = \sqrt{\frac{2 T}{N} \rho_\sigma(x,t)} \  \eta_\sigma(x,t) \quad , \quad \langle \eta_\sigma(x,t) \eta_{\sigma'}(x',t') \rangle = \delta_{\sigma,\sigma'} \delta(x-x')\delta(t-t') \;,
\eeq
where $\eta_\pm(x,t)$ are two independent unit Gaussian white noises. 

The active noise term reads
\beq
\hat\zeta_\sigma(x,t)=\frac{\sigma}{2N}\sum_i \delta(x-x_i(t)) \frac{d\sigma_i(t)}{dt} \;.
\eeq
To deal with the term $\frac{d\sigma_i(t)}{dt}$, we discretize time into small intervals $dt$. In the time interval $[t,t+dt]$, $\frac{d\sigma_i(t)}{dt}=-\frac{2\sigma_i(t)}{dt}$ with probability $\gamma dt$ and $0$ otherwise. Thus $\langle \frac{d\sigma_i(t)}{dt} \rangle=-2\gamma\sigma_i(t)$. Separating the mean from the fluctuations we get :
\bea
\hat\zeta_\sigma(x,t)&=& -\frac{\gamma}{N} \sum_i \sigma\sigma_i(t)\delta(x-x_i(t)) + \frac{\sigma}{\sqrt{N}} \zeta(x,t) \nn \\
&=& -\frac{\gamma}{N} \sum_i \delta_{\sigma_i(t),\sigma}\delta(x-x_i(t)) + \frac{\gamma}{N} \sum_i \delta_{\sigma_i(t),-\sigma}\delta(x-x_i(t)) + \frac{\sigma}{\sqrt{N}} \zeta(x,t) \nn \\
&=& -\gamma \rho_\sigma(x,t) +\gamma \rho_{-\sigma}(x,t) + \frac{\sigma}{\sqrt{N}} \zeta(x,t) \;,
\eea
using that $\sigma \sigma_i=\delta_{\sigma_i,\sigma} - \delta_{\sigma_i , - \sigma}$, and where we have defined
\beq
\zeta(x,t)=\frac{1}{2\sqrt{N}}\sum_i \delta(x-x_i(t)) \, r_i(t) \quad , \quad r_i(t)=\frac{d\sigma_i(t)}{dt} - \langle \frac{d\sigma_i(t)}{dt} \rangle \;.
\label{active_noise}
\eeq
The noise $\zeta(x,t)$ has zero average, and the factor $1/\sqrt{N}$ was chosen so that its variance is of order $O(1)$ at large $N$, which we will now show. We need to compute
\beq \label{cov_zeta1}
\langle \zeta(x,t)\zeta(x',t') \rangle =\frac{1}{4N}\sum_{i,j}\delta(x-x_i(t))\delta(x'-x_j(t'))\langle r_i(t)r_j(t') \rangle \;.
\eeq
Discretizing time as before, we have
\be
r_i(t) = \begin{cases} -\frac{2\sigma_i(t)}{dt}+2\gamma\sigma_i(t) \quad \rm{with \ probability \ \gamma dt} \;, \\ 2\gamma \sigma_i(t) \hspace*{1.65cm} \quad \rm{with \ probability \ 1-\gamma dt} \;. \end{cases}
\ee
For $i\neq j$, $r_i(t)$ and $r_j(t')$ are uncorrelated (and with zero average), i.e.,
\be
\langle r_i(t)r_j(t') \rangle = 0 \quad \text{if } i\neq j 
\;.
\ee
In the case where $i=j$ and $t=t'$, we find 
\be
\langle r_i(t)^2 \rangle = \gamma dt \langle \big(-\frac{2\sigma_i(t)}{dt}+2\gamma\sigma_i(t) \big)^2 \rangle + 4\gamma^2 (1-\gamma dt)
=\frac{4\gamma}{dt}-4\gamma^2 +O(dt) \;.
\ee
Finally, for $i=j$ and $t\neq t'$, one can check that
\bea
\langle r_i(t) r_i(t') \rangle &=& (\gamma dt)^2 \langle \big(-\frac{2\sigma_i(t)}{dt}+2\gamma\sigma_i(t) \big)\big(-\frac{2\sigma_i(t')}{dt}+2\gamma\sigma_i(t') \big) \rangle \\
&&+ 2\gamma dt (1-\gamma dt) \langle \big(-\frac{2\sigma_i(t)}{dt}+2\gamma\sigma_i(t) \big)2\gamma\sigma_i(t')\rangle + 4\gamma^2 (1-\gamma dt)^2 \langle \sigma_i(t) \sigma_i(t') \rangle \nn \\
&=& O(dt) \nn
\eea
(where in the second term we have used the symmetry between $t$ and $t'$). This was expected since the tumbling events are Poissonian. In the general case, we can thus write 
\be
\langle r_i(t)r_j(t') \rangle = (\frac{4\gamma}{dt}-4\gamma^2)\delta_{ij}\delta_{t,t'} +O(dt) \;.
\ee
Taking the limit $dt\rightarrow 0$, we replace $\frac{\delta_{t,t'}}{dt}$ by $\delta(t-t')$ and we obtain (as in standard calculations
for the Brownian motion) 
\beq
\langle r_i(t)r_j(t') \rangle = 4\gamma \, \delta_{ij}\delta(t-t') \;.
\eeq
Inserting into \eqref{cov_zeta1}, this heuristics derivation yields the following covariance function for $\zeta(x,t)$ 
\bea
\langle \zeta(x,t)\zeta(x',t') \rangle =\frac{\gamma}{N}\sum_{i}\delta(x-x_i(t))\delta(x-x')\delta(t-t') 
= \gamma \rho_s(x,t) \delta(x-x')\delta(t-t') \;,
\eea
where we recall that $\rho_s=\rho_++\rho_-$. This result is indeed of order $O(1)$. We can also examine the higher cumulants of $\zeta(x,t)$ using the same method. We find:
\bea
&& \hspace{-1cm} \langle\zeta(x_1,t_1)\zeta(x_2,t_2)\zeta(x_3,t_3)\rangle_c = 0 \;, \\
&& \hspace{-1cm} \langle\zeta(x_1,t_1)\zeta(x_2,t_2)\zeta(x_3,t_3)\zeta(x_4,t_4) \rangle_c =
\frac{\gamma}{N} \rho_s(x_1,t_1)  \delta(x_1-x_2)\delta(x_1-x_3)\delta(x_1-x_4) \nn \\&& \hspace{7.4cm} \times \delta(t_1-t_2)\delta(t_1-t_3)\delta(t_1-t_4) \;. \nn 
\eea
This suggests that at large $N$ the active noise becomes Gaussian. We will thus write it as
\beq \label{expr_active_noise}
\zeta(x,t) = \sqrt{\gamma \rho_s(x,t)} \  \eta_K(x,t) \quad , \quad \langle \eta_K(x,t) \eta_K(x',t') \rangle = \delta(x-x')\delta(t-t') \;,
\eeq
where $\eta_K(x,t)$ is centered delta-correlated noise with unit variance, which is Gaussian at leading order in $N$.

Let us now return to the equation \eqref{integralDean}. Using that it holds for any $f(x)$, and inserting the expressions \eqref{expr_thermal_noise} and \eqref{expr_active_noise} for the thermal and active noise, we finally obtain the stochastic evolution equation for the densities
\bea \label{dean1} 
\partial_t \rho_\sigma(x,t) =&& \hspace{-0.5cm} \partial_x \Big[ \rho_\sigma(x,t) \big(- v_0 \sigma + V_\sigma'(x) + \sum_{\sigma'} \int dy \rho_\sigma(y,t) W_{\sigma,\sigma'}'(x-y) \big) \Big] + \gamma \big( \rho_{-\sigma}(x,t) - \rho_{\sigma}(x,t) \big) \nn
\\
&& + T \partial^2_x \rho_\sigma(x,t)  
+ \frac{1}{\sqrt{N}} \partial_x [ \sqrt{2T \rho_\sigma(x,t)} \; \eta_\sigma(x,t) ] + \frac{\sigma}{\sqrt{N}} \sqrt{\gamma \rho_s(x,t)} \; \eta_K(x,t) \;.
\eea 
This equation is the main result of this chapter, and it will be our starting point for the next two chapters. It is similar to the Dean-Kawasaki equation for Brownian particles \eqref{Dean_eq_Brownian}, but with an additional drift term (with drift $v_0\sigma$), a term proportional to $\gamma$ accounting for the tumbling events, which couples the two equations, and an additional noise term proportional to $\eta_K$.

In the case where $V_{\sigma}(x)=V(x)$ and $W_{\sigma,\sigma'}(x)=W(x)$ are both independent of the state of the particles, one can rewrite these equations in terms of $\rho_s= \rho_+ + \rho_-$ and $\rho_d= \rho_+ - \rho_-$,
\bea \label{dean2rhos} 
\partial_t \rho_s(x,t) &=& \partial_x \Big[ - v_0 \rho_d(x,t)  + V'(x) \rho_s(x,t) + \rho_s(x,t) \int dy \rho_s(y,t) W'(x-y) \Big] \\
&& + T \partial^2_x \rho_s(x,t) + \frac{1}{\sqrt{N}} \partial_x [ \sqrt{ 2 T \rho_s(x,t) } \, \eta_s(x,t) ] \nonumber 
\\
\partial_t \rho_d(x,t) &=& \partial_x \Big[ - v_0 \rho_s(x,t)  + V'(x) \rho_d(x,t) + \rho_d(x,t) \int dy \rho_s(y,t) W'(x-y) \Big] - 2\gamma \rho_d(x,t) \nonumber \\
&&+ T \partial^2_x \rho_d(x,t) + \frac{1}{\sqrt{N}} \partial_x [ \sqrt{ 2 T \rho_s(x,t) } \, \eta_d(x,t) ] + \frac{2}{\sqrt{N}} \sqrt{\gamma \rho_s(x,t)} \, \eta_K(x,t) \label{dean2rhod}
\eea
where $\eta_s(x,t)$ and $\eta_d(x,t)$ are unit Gaussian white noises with covariance $\langle \eta_s(x,t) \eta_d(x,t) \rangle = \frac{\rho_d(x,t)}{\rho_s(x,t)} \delta(x-x')\delta(t-t')$. 
It is interesting to compare these equations with the equations \eqref{eqrho_lattice_noise}-\eqref{eqm_lattice_noise} obtained in \cite{Agranov2021} for RTPs on a lattice with exclusion interactions. Although the method is completely different, the equations are quite similar (taking into account the fact that the type of interaction considered is of course different). In particular, the active noise term takes exactly the same form at leading order in $N$. 
\\

\noindent {\bf Diffusive limit.} As mentioned in Chapter~\ref{chap:active}, the diffusive limit of the RTP corresponds to the limit $\gamma\to+\infty$, $v_0\to+\infty$ with the effective temperature $T_{\rm eff}=\frac{v_0^2}{2\gamma}$ fixed. Let us consider the equations \eqref{dean2rhos}-\eqref{dean2rhod} for $T=0$ and in the limit $N\to+\infty$, i.e., without the noise terms. Then, in the diffusive limit, the second equation is dominated by the balance between two terms, leading to the relation
\be
\rho_d(x,t) = -\frac{v_0}{2\gamma} \partial_x \rho_s(x,t) \;.
\ee
Inserting this into the first equation, we recover the large $N$ limit of the DK equation for Brownian particles, given in \eqref{Dean_eq_Brownian}, with temperature $T_{\rm eff}$.

\section{Discussion and comparison with the Fokker-Planck approach}
\label{sec:FPrho}

\subsection{Comparison with the numerics, self-averaging and single-file constraint}

In \eqref{def_rho_pm} we defined the density as a sum of delta functions (this is sometimes called an ``empirical density''). However, we usually like to see the density as a smooth function of the position. One way to make sense of this definition for finite $N$ is thus to perform some form of coarse-graining, e.g., by averaging over small intervals. In the limit $N\to+\infty$, we generally expect the density defined in \eqref{def_rho_pm} to converge to a smooth function (apart from a few singularities, e.g., at the edges of the support), in the sense that the size of the coarse-graining intervals can be decreased to zero. 

In the next two chapters, we will compare our analytical results derived from the large $N$ limit of the DK equation to the results of numerical simulations. However since we can only simulate systems of up to $N\sim 10^2-10^3$ particles depending on the model, the fluctuations are too strong to allow for a direct comparison with the $N\to+\infty$ limit. We will thus compute the average of the density over many realizations of the noise, or over a large time window if we only consider the stationary state (and if the system is ergodic, which is the case for the models that we will study). This procedure directly leads to a smooth density, and no coarse-graining is required in this case. The obtained density can still be compared with the one described by the $N \to +\infty$ limit of the DK equation, if we assume that the density is {\it self-averaging}, i.e., that it converges to its mean as $N\to+\infty$. This is generally the case for the type of models that we will be considering in this thesis, at least for Brownian particles, and it is reasonable to assume that it will remain true for RTPs. For more details on the numerical methods see Appendix~\ref{app:simu}. Note that for convenience, in the caption of the figures and in the discussions of the finite $N$ results, we will still denote $\rho_\sigma(x,t)$ the mean density at finite $N$. In the remaining chapters of this thesis, when discussing the ``density'' at finite $N$ we will always be implicitly referring to the mean density unless specified otherwise. 

For the active ranked diffusion, which we will discuss in Chapter~\ref{chap:activeRD} and which corresponds to an interaction potential $W(x)=-\kappa |x|$, we find that the mean densities which we obtain from numerical simulations indeed seem to converge to the density predicted by the noiseless DK equation as we increase $N$. However, this is not at all the case for the active version of the Dyson Brownian motion, $W(x)=-2g\log|x|$, which we will consider in Chapter~\ref{chap:ADBM_Dean}.

To understand why this is the case, let us recall that, in the derivation above, we have made the assumption that $W'(0)=0$~\footnote{Note that this is the case for the active rank diffusion where $W'(x)=-\kappa\sgn(x)$, since we use the convention $\sgn(0)=0$ (see \eqref{def_sgn})}. Since $W(x)$ is generally an even function of $x$, this means that the only other possibility is for $W'(x)$ to diverge at $x=0$. The assumption $W'(0)=0$ is formally required to avoid introducing an artificial self-interaction term when going from \eqref{Dean_discret} to \eqref{integralDean0}. In the case of the passive DBM (i.e., with only Brownian noise), this issue can however be resolved at the cost of an additional term of order $1/N$, which  can be written explicitly \cite{RogersShi}. If we now see the density as a smooth function, the diverging integral may then be regularized using the Cauchy principal value. For the active version of the DBM, a similar treatment of the self-interaction term does not seem possible. If we ignore this issue and simply regularize the diverging integral using the Cauchy principal value, the obtained equation fails to correctly describe the numerical observations, even as $N$ increases. This will be discussed in more detail in Chapter~\ref{chap:ADBM_Dean}. 

The physical explanation behind this phenomenon is that, for RTPs without Brownian noise, a diverging interaction force $W'(x)$ at $x=0$ always prevents particles from crossing each other (since the telegraphic noise is bounded). For Brownian particles with Riesz interaction, we recall that crossings are also forbidden for $s>0$, and for $s=0$ and $\beta \geq 1$. However, in the Brownian case the existence or not of particle crossings does not really make a difference in terms of the total density since the particles are interchangeable. This is not the case for RTPs, since they are not only characterized by their position, but also by their state $\sigma_i$ (i.e., there are two different types of particles). In the presence of a diverging interaction force, ``collisions''  occur between particles with opposite $\sigma_i$, i.e., they are prevented from crossing by the interaction and they remain in a ``jammed" configuration until one of them tumbles. This creates strong local correlations between the particles which break the hydrodynamic description. To better understand why this is the case, we now present another approach to derive an equation directly for the mean density. It does not allow for the description of the noise as the DK equation, but it helps to clarify which assumptions are broken due to the single-file constraint.

\subsection{Equation for the mean density via Fokker-Planck}

Another way to obtain a PDE describing the evolution of the mean particle density is to start from the Fokker-Planck equation. Let us briefly discuss this approach, which will allow us to better understand how the hydrodynamic description breaks down due to the formation of clusters.
\\

\noindent {\bf Derivation.} Let us consider the joint probability distribution of the positions $\vec{x}=(x_1,...,x_N)$ and states $\vec{\sigma}=(\sigma_1,...\sigma_N)$ of the $N$ particles,  ${\cal P}_t(\vec x, \vec \sigma)$. It satisfies the Fokker-Planck equation
\be 
\partial_t {\cal P}_t = \sum_k \partial_{x_k} \Big[ \big( - v_0 \sigma_k + V_{\sigma_k}'(x_k) + \frac{1}{N} \sum_{l\neq k} W'_{\sigma_k,\sigma_l}(x_k-x_l) \big) {\cal P}_t \Big] - N \gamma {\cal P}_t + \gamma \sum_k \tau_k^1 {\cal P}_t + T \sum_k \partial_{x_k}^2 {\cal P}_t \;,
\label{FP2ND}
\ee 
where $\tau_k^1 {\cal P}_t(\vec x, \vec \sigma)= {\cal P}_t(\vec x, \sigma_1,\dots,-\sigma_k,\dots,\sigma_N)$. In the previous section we have implicitly chosen a deterministic initial condition,
\be 
{\cal P}_{t=0}(\vec x, \vec \sigma)= \prod_i \delta(x_i-x_i(0)) \delta_{\sigma_i,\sigma_i(0)} \;,
\ee 
with a fixed set of $\vec x(0)$ and $\vec \sigma(0)$. However within the present method more general initial conditions may be considered. Let us define the mean density
\beq 
p_\sigma(x,t) = \langle \rho_\sigma(x,t) \rangle_{{\cal P}_t } = \langle \frac{1}{N} \sum_i \delta_{\sigma,\sigma_i} \delta(x-x_i) \rangle_{{\cal P}_t }
= \sum_{\vec \sigma} \int d\vec x \frac{1}{N} \sum_i \delta_{\sigma,\sigma_i} \delta(x-x_i) {\cal P}_t(\vec x, \vec \sigma) \;.
\eeq
Contrary to the empirical density $\rho_\sigma(x,t)$, which for finite $N$ is a stochastic function, 
$p_\sigma(x,t)$ is a deterministic quantity for any value of $N$ (which satisfies $\sum_\sigma \int dx \, p_\sigma(x,t) = 1$). As for the empirical density, we will denote $p_s=p_++p_-$ and $p_d=p_+-p_-$. We also need to introduce the two-point density function involving one particle of sign $\sigma$ and one particle of sign $\sigma'$,
\beq \label{2point_density}
p^{(2)}_{\sigma,\sigma'}(x,y,t) = \sum_{\vec \sigma} \int d\vec x \frac{1}{N(N-1)}  \sum_{i \neq j}  \delta_{\sigma,\sigma_i} \delta_{\sigma',\sigma_j} \delta(x-x_i) \delta(y-x_j) {\cal P}_t(\vec x, \vec \sigma) \;,
\eeq
which is normalized such that $\sum_{\sigma,\sigma'} \int dx dy \, p_{\sigma,\sigma'}^{(2)} (x,y,t) = 1$. 
Multiplying \eqref{FP2ND} by $\frac{1}{N} \delta_{\sigma,\sigma_i} \delta(x-x_i)$, summing over all particles $i$ as well as over all configurations $\vec{\sigma}$, and integrating over all components of $\vec{x}$, we can obtain an equation for $p_\sigma(x,t)$. The first two terms on the left-hand side become, after integrating by parts and using that $\partial_{x_k} \delta(x-x_i) = - \delta_{ik} \partial_x \delta(x-x_i)$,
\be 
\sum_{\vec \sigma} \int d\vec x \frac{1}{N} \sum_i \delta_{\sigma,\sigma_i} \delta(x-x_i) \sum_k \partial_{x_k} \Big[ \big( - v_0 \sigma_k + V'_{\sigma_k} (x_k) \big) {\cal P}(\vec x, \vec \sigma) \Big] = \partial_x [(- v_0 \sigma + V'_{\sigma} (x)) p_\sigma(x,t)] \;.
\ee
The diffusion term can be treated similarly without any particular issue. The term involving the permutation $\tau_k^1$ can be rewritten
\bea 
\gamma \sum_{\vec \sigma} && \hspace{-1cm} \int d\vec x \frac{1}{N} \sum_i \delta_{\sigma,\sigma_i} \delta(x-x_i) \sum_k  {\cal P}_t(\vec x, \sigma_1,\dots,-\sigma_k,\dots,\sigma_N) \nn \\ 
&=& \hspace{-0.2cm}  \gamma \sum_{\vec \sigma} \int d\vec x \frac{1}{N} \sum_i \delta_{-\sigma,\sigma_i} \delta(x-x_i) {\cal P}_t(\vec x, \vec \sigma) + \ \gamma \sum_{\vec \sigma} \int d\vec x \frac{1}{N} \sum_i \sum_{k (\neq i)} \delta_{\sigma,\sigma_i} \delta(x-x_i) 
   {\cal P}_t(\vec x, \vec \sigma) \nonumber\\
&=& \hspace{-0.2cm}  \gamma p_{-\sigma}(x,t)  + (N-1) \gamma p_{\sigma}(x,t) \;.
\eea 
The result above combines with the term $- \gamma N {\cal P}_t$ in Eq. (\ref{FP2ND}) to give $\gamma p_{-\sigma}(x,t)  - \gamma p_{\sigma}(x,t)$. 
Finally the interaction term gives
\bea
\frac{1}{N^2} \sum_{\vec \sigma} \int d\vec{x} \sum_i \delta_{\sigma,\sigma_i} \delta(x-x_i) \sum_k \partial_{x_k} \sum_{ l(\neq k)} W'_{\sigma_k,\sigma_l}(x_k-x_l) {\cal P}_t(\vec x, \vec \sigma) \nn \\
= \frac{1}{N^2} \partial_x \sum_{\vec \sigma} \int d\vec{x} \sum_i \sum_{l (\neq i)} \delta_{\sigma,\sigma_i} \delta(x-x_i) W'_{\sigma_i,\sigma_l}(x_i-x_l) 
{\cal P}_t(\vec x, \vec \sigma) \;.  
\eea 
We then insert $1=\sum_{\sigma'} \delta_{\sigma',\sigma_l}\int dy \, \delta(y-x_l)$ to rewrite this as
\be 
\big( 1-\frac{1}{N} \big) \sum_{\sigma'} \partial_x \int dy \, W'_{\sigma,\sigma'}(x-y) \tilde p^{(2)}_{\sigma,\sigma'}(x,y,t) \;.
\ee 
Putting everything together we finally obtain
\bea \label{eqfromFPfull}
\partial_t p_\sigma(x,t) = && \hspace{-0.55cm} \partial_x \Big[(- v_0 \sigma + V_\sigma'(x)) p_\sigma(x,t) + \big( 1-\frac{1}{N} \big) \int dy \, W_{\sigma,\sigma'}'(x-y) p^{(2)}_{\sigma,\sigma'}(x,y,t) \Big] \nn \\
&& \hspace{-0.55cm} + \gamma \big( p_{-\sigma}(x,t)  - p_{\sigma}(x,t) \big) + T \partial^2_x p_\sigma(x,t)  \;.
\eea
This is the same as the Dean-Kawasaki equation \eqref{dean1} without the noise terms, apart from the interaction term which now involves the two-point density $p^{(2)}_{\sigma,\sigma'}(x,y,t)$ (with a $(1-1/N)$ factor due to the normalization in \eqref{2point_density}). To obtain a closed equation for the mean density $p_\sigma(x,t)$, we thus need an additional assumption.
\\

\noindent {\bf Large $N$ limit.} If the empirical density is self-averaging, we expect equation \eqref{eqfromFPfull} to be the same as the Dean-Kawasaki equation \eqref{dean1} in the limit $N\to+\infty$. We see that this is true if the two-particle correlations become negligible in this limit, i.e.,
\begin{equation}
    p^{(2)}_{\sigma,\sigma'}(x,y,t) \underset{N\to+\infty}{\longrightarrow} p_\sigma(x,t) p_{\sigma'}(y,t) \;.
    \label{FPapprox}
\end{equation}
This assumption, which amounts to a mean-field limit, seems to hold in cases where the particles are allowed to cross. However, when the particles cannot cross, the clusters of particles which appear generate strong local correlations which do not disappear as $N\to+\infty$.

In Chapter~\ref{chap:ADBM_Dean} we will evaluate numerically the two-point density for the active DBM, and we will see that this de-correlation hypothesis is indeed broken in this case. We will also introduce a variant of the active DBM in which particles are allowed to cross, for which we will check that this hypothesis is indeed verified, and the Dean-Kawasaki equation can be used. Finally, in the same chapter we will also study the limit $g\to 0^+$ of this model, where only the single-file constraint remains, to better understand the formation of the clusters and how they affect the density.

This failure of hydrodynamic equations for active particles in 1D with a single-file constraint has been observed in other contexts, in particular in the case of lattice models. In \cite{KH2018}, the derivation of hydrodynamic equations only works thanks to the diffusion which allows the exchange of particles, while in \cite{lattice2lanes2025} it is the presence of two parallel lanes which allows particles to pass each other. However, in this second case they also note that their method fails to describe the fully 1D version of their model. 

\section{Conclusion}

In this chapter we have derived the Dean-Kawasaki equation for RTPs in one dimension. We have also discussed its limitations, in particular in the presence of an interaction which prevents particle crossings. In the next two chapters we will use this equation to study two different models of RTPs with long-range interaction: the {\it active ranked diffusion}, with a 1D Coulomb interaction, and the {\it active DBM}, with a logarithmic (2D Coulomb) interaction.

In this thesis, we will make a rather restrictive use of this equation, focusing mostly on the $N\to+\infty$ limit where the noise terms vanish. The use of this equation to study the fluctuations of the density, as done in \cite{Agranov2021,Agranov2022} for a lattice model, is of course an interesting direction for future work. In addition, we will mostly focus on the determination of stationary state, although the DK equation also allows to study the dynamics. Finally, for simplicity we will almost always consider the purely active case, with $T=0$, although the interplay between active and thermal noise is also an interesting question.


\chapter{Active ranked diffusion}
\label{chap:activeRD}

\section{Definition of the model and equations for the rank fields}

In this chapter, we will study an active version of the ranked diffusion model introduced in Sec.~\ref{sec:RD_brownian}. It consists in $N$ run-and-tumble particles interacting via a 1D Coulomb potential $W(x)=-\kappa|x|$. The positions $x_i(t)$ of the particles ($i=1,...,N$) obey the following equation of motion
\be \label{def_activeRD}
\frac{dx_i}{dt} = \frac{\kappa}{N} \sum_{j=1}^N {\rm sgn}(x_i-x_j) - V'(x_i) + v_0 \sigma_i(t) + \sqrt{2 T} \, \xi_i(t) \;,
\ee
where $\sgn(x)$ is the sign function defined in \eqref{def_sgn}, $V(x)$ is an external potential which at this point remains unspecified, the $\sigma_i(t)$ are independent telegraphic noises with tumbling rate $\gamma$, and the $\xi_i(t)$ are independent unit Gaussian white noises (in this chapter we will however focus on the case $T=0$). We will study both the repulsive case $\kappa>0$ and the attractive case $\kappa=-\bar \kappa<0$.

We are interested in studying the total particle density $\rho_s(x,t)$ in this model, as well as the densities $\rho_\pm(x,t)$ defined in \eqref{def_rho_pm}, in the limit of large $N$. For this, our starting point will be the DK equations in terms of $\rho_s$ and $\rho_d$, \eqref{dean2rhos}-\eqref{dean2rhod} derived in the previous chapter. In this case the interaction force vanishes at $x=0$, i.e., particle trajectories are allowed to cross, and thus we expect the DK equations to properly describe the model. Since we will not be interested in the fluctuations, we directly consider the limit $N\to+\infty$ of these equations, which in this case reads
\bea \label{deanARDrhos} 
\partial_t \rho_s &=& \partial_x \Big[ - v_0 \rho_d + \rho_s \big( V'(x) -\kappa \int dy \rho_s(y,t) \sgn(x-y) \big) \Big] + T \partial^2_x \rho_s \;,
\\
\partial_t \rho_d &=& \partial_x \Big[ - v_0 \rho_s  + \rho_d \big( V'(x) - \kappa \int dy \rho_s(y,t) \sgn(x-y) \big) \Big] + T \partial^2_x \rho_d - 2\gamma \rho_d \;.
\label{deanARDrhod}
\eea
The study of these equations is made difficult by the non-local interaction terms. For the 1D Coulomb interaction however, as in the Brownian case \cite{PLDRankedDiffusion}, this issue can be resolved by rewriting these equations in terms of the {\it rank fields},
\be \label{rankdef1}
r(x,t) = \int^x_{-\infty} dy \, \rho_s(y,t)\, -\, \frac{1}{2} \quad , \quad s(x,t) = \int_{-\infty}^x dy \, \rho_d(y,t) \; .
\ee
Since $\rho_s(x,t)$ is positive and normalized to 1, $r(x,t)$ is an increasing function with $r(-\infty,t)=- 1/2$ and $r(+\infty,t)=1/2$. For $s(x,t)$ we {\it a priori} only have the boundary condition $s(-\infty,t)=0$. However, in the stationary state for $N\to+\infty$, the particles are equally split between the $+$ and $-$ states, and thus in this case we will also have $s(+\infty,t\to+\infty)=0$. Replacing $\rho_s=\partial_xr$ and $\rho_d=\partial_x s$ in the DK equations above, and rewriting the integral in the interaction terms using integration by parts (and the fact that $\partial_y \sgn(x-y)=-2\delta(x-y)$) as
\be
\int dy \, \partial_y r(y,t) {\rm sgn}(x-y)
= 2 r(x,t) + [r(y,t) {\rm sgn}(x-y)]^{+\infty}_{-\infty} = 2 r(x,t) \;,
\ee
we obtain a set of two local PDEs for the rank fields,
\bea
&& \partial_t \partial_x r = \partial_x[ - v_0 \partial_x s - 2 \kappa r \partial_x r
+  V'(x) \partial_x r ] + T \partial_x^3 r(x,t)  \;, \\
&& \partial_t \partial_x s  =  \partial_x [ - v_0 \partial_x r - 2 \bar \kappa r \partial_x s + V'(x) \partial_x s ] +  T  \partial_x^3 s - 2 \gamma \partial_x s \;.
\eea 
These two equations can then be integrated over $x$, using that the densities $\rho_s=\partial_x r$ and $\rho_d=\partial_x s$ vanish at infinity (with the additional reasonable assumption that $V'(x)\rho_{s/d}$ also vanishes at infinity), and that $s(-\infty,t)=0$ to fix the integration constant to zero. This finally leads to the following set of equations:
\bea \label{eqrank1}
\!\!\!\!\!\!  \partial_t r &=& - v_0 \partial_x s
- 2 \kappa r \partial_x r + V'(x) \partial_x r + T \partial_x^2 r \; , \\
\!\!\!\!\!\!  \partial_t s &=& - v_0 \partial_x r  - 2 \kappa  r \partial_x s + V'(x) \partial_x s + T \partial_x^2 s - 2 \gamma s \label{eqrank2} \;.
\eea 
In the passive case $v_0 = 0$, the first equation recovers the Burgers equation which describes the usual rank diffusion \cite{PLDRankedDiffusion}. These equations are valid both in the attractive and in the repulsive case, and we will use them to study both cases in the remainder of this chapter, focusing on the stationary state when it exists, or the large time limit otherwise (and on the purely active case $T=0$). We will start with the unconfined case $V'(x)=0$. In the attractive case, we will see that a new transition occurs, completely inexistent in the passive case (i.e., in the case of Brownian noise), between a phase where the density is smooth and a phase where it has a finite support, with shocks (i.e., delta peaks) at the edges. In the absence of confinement, the large time behavior of the repulsive case is much more similar to the passive case. We will then extend our results to two types of confining potentials, first linear and then harmonic. In both cases, the model exhibits a rich phase diagram with a variety of regimes, both for an attractive and a repulsive interaction. Finally, we will briefly discuss a variant of the model where the interaction is {\it non-reciprocal}, i.e., in our case, where the force exerted by a $+$ particle on a $-$ particle is different from the force exerted by the $-$ on the $+$ particle.

The analytical results presented in this chapter were compared with the results of numerical simulations. We performed  direct simulations of the Langevin dynamics \eqref{def_activeRD} (for $T=0$), and we averaged the particle densities over large time windows after convergence to the stationary state (or over many realizations in the unconfined repulsive case), for different values of $N$. More details on the numerical methods are given in Appendix~\ref{app:simu}.

\section{Active ranked diffusion without confining potential}

\subsection{Attractive case}

In the absence of confining potential, $V'(x)=0$, but in the presence of an attractive interaction, the particles may still form a bound state at large time. Thus we will look for a stationary solution of the equations \eqref{eqrank1}-\eqref{eqrank2} in the case $\bar\kappa=-\kappa>0$. Since these equations are invariant by translation, we may choose the reference frame such that $r(0)=0$. If the density is symmetric, which will be the case in the stationary state, this amounts to fixing the position of the center of mass $\bar x(t)=\frac{1}{N}\sum_i x_i(t)$ to zero, which is what we did in the numerical simulations (by computing $\bar x(t)$ and subtracting it from all the particle positions $x_i(t)$ at each time step). Let us note that the equation of motion for the center of mass can be obtained by summing \eqref{def_activeRD} over all values of $i$. The interaction terms compensate two by two and we obtain that $\bar x(t)$ behaves as the sum of $N$ independent telegraphic noises with amplitude $v_0/N$ and $N$ independent Brownian noises with temperature $T/N^2$. Thus, at large times the center of mass diffuses as
\be
\bar x(t) \sim \sqrt{2D_N t} \quad , \quad D_N=\frac{1}{N}(T + \frac{v_0^2}{2 \gamma}) \;,
\ee
which is subleading in $N$. From now on, we focus on the purely active case $T=0$. We note that the ratio $v_0/\bar \kappa$ is the only dimensionless parameter of the model in this case.
\\

\begin{figure}
    \centering
    \includegraphics[width=0.45\linewidth,trim={0 0 1cm 0.5cm},clip]{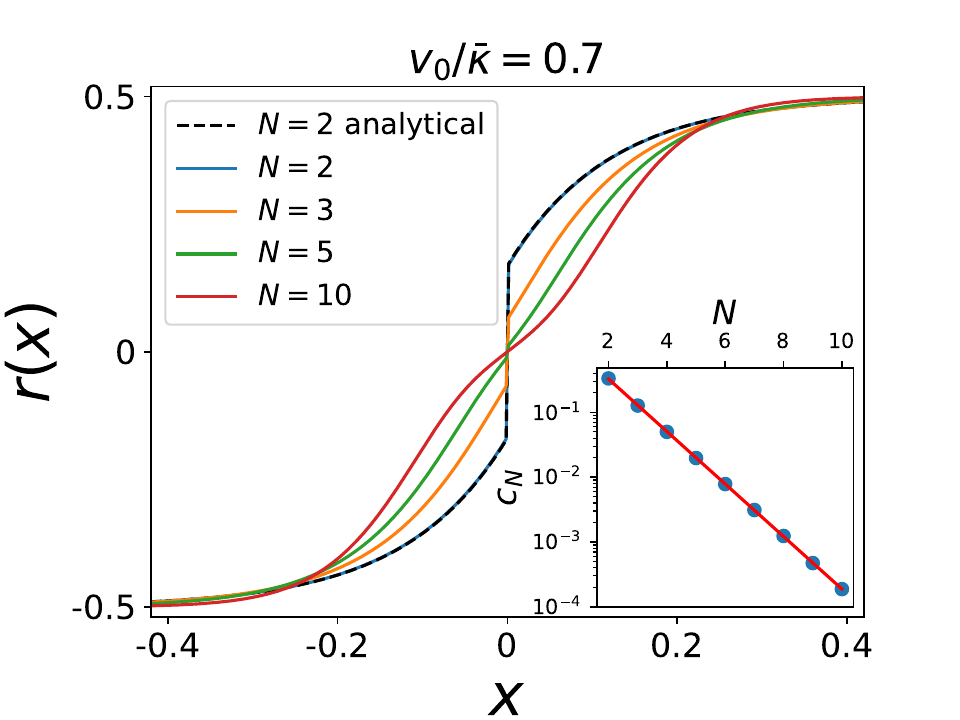}
    \includegraphics[width=0.45\linewidth,trim={0 0 1cm 0.5cm},clip]{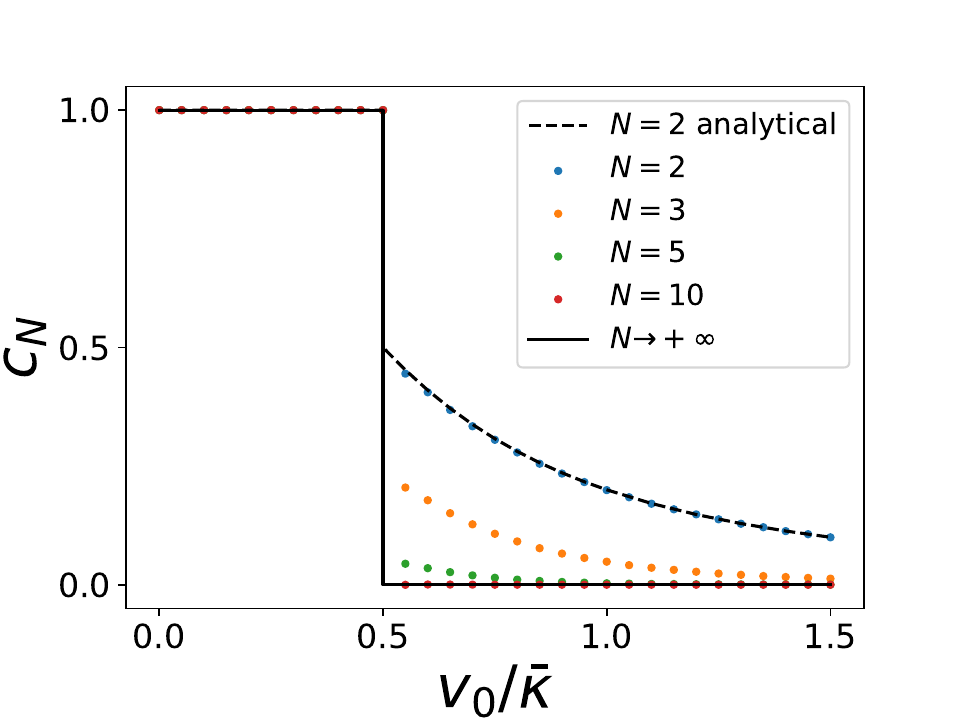}
    \caption{{\bf Left:} Rank field $r(x)=\int_{-\infty}^x dy \, \rho_s(y) -\frac{1}{2}$ in the stationary state, obtained through simulations, as a function of $x$, in the attractive case with $v_0=0.7$, $\bar \kappa=1$ and $\gamma=1$, for small values of $N$. The analytical prediction 
    from \eqref{probN2rank} for $N=2$ is also shown. The density has a delta peak at $x=0$, i.e., a jump in $r(x)$,
    with a weight $c_N$, which decreases exponentially with $N$ (see inset). {\bf Right:} Weight $c_N$ as a function of $v_0/\bar \kappa$ for different values of $N$. The black line corresponds to the limit $N \to+\infty$ (see below).}
    \label{figRD_smallN}
\end{figure}

\noindent {\bf Finite $N$.} Before studying the limit $N\to+\infty$, let us recall that the case $N=2$ was studied in \cite{LMS2021}, and was briefly discussed in Sec.~\ref{sec:interactions_exact}. The stationary PDF of the interparticle distance $y=x_1-x_2$ was given in \eqref{PDF2part_linear}. For $\bar c = \bar \kappa/2 > v_0$, the noise cannot overcome the attraction and the density is a single delta peak at $y=0$, while for $\bar c<v_0$, one finds a delta peak at zero with an exponential decay on each side. Rewritten in terms of the coordinate $z=x_1-\bar x=\frac{x_1-x_2}{2}$ and the parameter $\bar \kappa=2\bar c$ to match the present setting, it reads
\be  \label{probN2rank}
P(z) =  \frac{1-c_2}{2 \xi_2} e^{- \frac{|z|}{\xi_2}} + c_2 \delta(z) \quad , \quad  \xi_2= \frac{4 v_0^2 - \bar \kappa^2}{8 \gamma \bar \kappa} \quad  , \quad c_2=
\frac{\bar \kappa^2}{4 v_0^2 + \bar \kappa^2} \;,
\ee
for $\frac{v_0}{\bar \kappa}> \frac{1}{2}$, and $P(z)=\delta(z)$ (i.e., $c_2=1$) for $\frac{v_0}{\bar \kappa} < \frac{1}{2}$ (note that $P(z)$ can be seen as the mean of the density $\rho_s$ for $N=2$). The formation of clusters, i.e., delta peaks in the density (or ``shocks''), is a general feature of this model when the interaction is attractive. It is due to the discontinuity of the interaction force at $x=0$, and to the fact that two particles which share the same position do not interact ($\sgn(0)=0$). The first natural question that one may ask is thus how does the weight $c_N$ of the delta peak at $x=0$ vary as we increase $N$ ?

For $\frac{v_0}{\bar \kappa}<1/2$, a simple stability argument shows that the particles always form a single cluster, i.e., $c_N=1$ for any $N$. Let us denote $x_+$ (resp. $x_-$) the position of one of the rightmost (resp. leftmost) particles
and $n_+$ (resp. $n_-$) the number of particles at the same location. From the equation of motion \eqref{def_activeRD}, we have the inequalities
\be  \label{stabilityclusterRD}
\frac{d(x_+-x_-)}{dt} \leq 2 v_0 - \bar \kappa \left( 2-\frac{n_++n_-}{N} \right)
\leq 2v_0 - \bar \kappa \;,
\ee 
where the second inequality simply comes from $n_++n_- \leq N$. Thus, for $\frac{v_0}{\bar \kappa}<1/2$ the width of the support always decreases to $0$, leading to $\rho_s(x)=\delta(x)$ for any $N$.

For $\frac{v_0}{\bar \kappa}>1/2$, no such simple argument exists and we resorted to numerical simulations to see how $c_N$ evolves with $N$. The results are
shown in Fig. \ref{figRD_smallN}. We find that the delta peak in the density $\rho_s(x)$ persists for any finite $N$, but the amplitude $c_N$ decreases exponentially with $N$.
\\

\begin{figure}[t]
    \centering
    \includegraphics[width=0.95\linewidth]{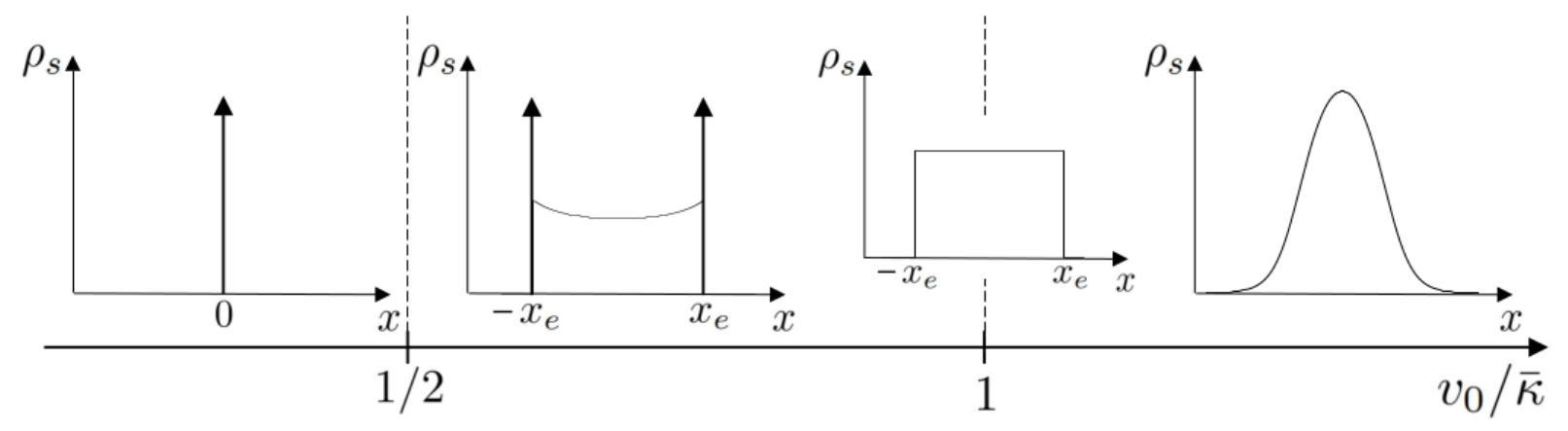}
    \caption{Phase diagram of active ranked diffusion in the attractive case (in the absence of external potential). For each phase (as well as in the marginal case $v_0=\bar \kappa$), the density $\rho_s$ in the $N \to + \infty$ limit is represented, for some values of the parameters (the arrows represent delta functions).}
    \label{phase_diagram_RD1}
\end{figure}

\noindent {\bf Large $N$ limit. } We now turn to the limit $N\to+\infty$ (which was not studied before). In this limit, the delta peak at $x=0$ is absent for $\frac{v_0}{\bar \kappa}>1/2$. However, we will see that a transition occurs at $v_0=\bar \kappa$, between a smooth density with unbounded support for $v_0>\bar \kappa$, and a density with finite support which exhibits delta peaks at the edges for $v_0 <\bar \kappa$ (see Fig.~\ref{phase_diagram_RD1}).

To study the densities in the limit $N\to+\infty$, we start from the equations \eqref{eqrank1}-\eqref{eqrank2} for the rank fields, with $V'(x)=0$ and $\bar\kappa=- \kappa>0$. We look for a stationary solution of these equations. The first equation can be integrated using the boundary conditions for $r(x)$ and $s(x)$, which yields
\be \label{integr_rank1} 
- v_0 s(x) + \bar  \kappa \big(r(x)^2 - \frac{1}{4}\big) + T r' (x) = 0 \;.
\ee
We briefly reintroduce the temperature $T>0$ to recall the solution for the Brownian case
\cite{PLDRankedDiffusion}. It can be simply obtained from \eqref{integr_rank1} by setting $v_0=0$,
\be \label{soliton2} 
r(x)=\frac{1}{2} \tanh\left( \frac{\ \bar \kappa x}{2 T} \right)  \quad , \quad 
\rho_s(x) = \frac{\bar \kappa}{4 T \cosh( \frac{\ \bar \kappa x}{2 T} )^2} \; ,
\ee
which is smooth and supported by the whole real axis. For $T \to 0$ this becomes a shock solution of Burgers equation, $r(x)= \frac{1}{2} {\rm sgn}(x)$, i.e the density
a single delta peak $\rho_s(x)=r'(x)=\delta(x)$. For $T>0$, the shock acquires a finite width of order $O(T)$. In the presence of both telegraphic and Brownian noise, we expect a similar broadening of the shocks in the solution described below, but we will not discuss this here. 
\\

\begin{figure}
    \centering
    \includegraphics[width=0.45\linewidth]{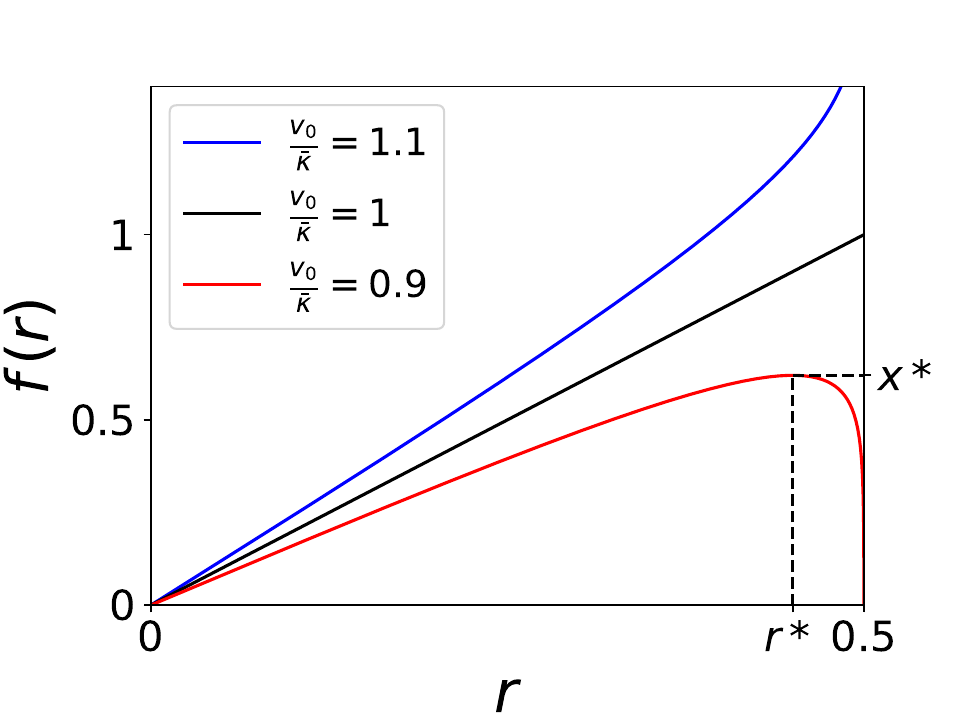}
    \includegraphics[width=0.45\linewidth]{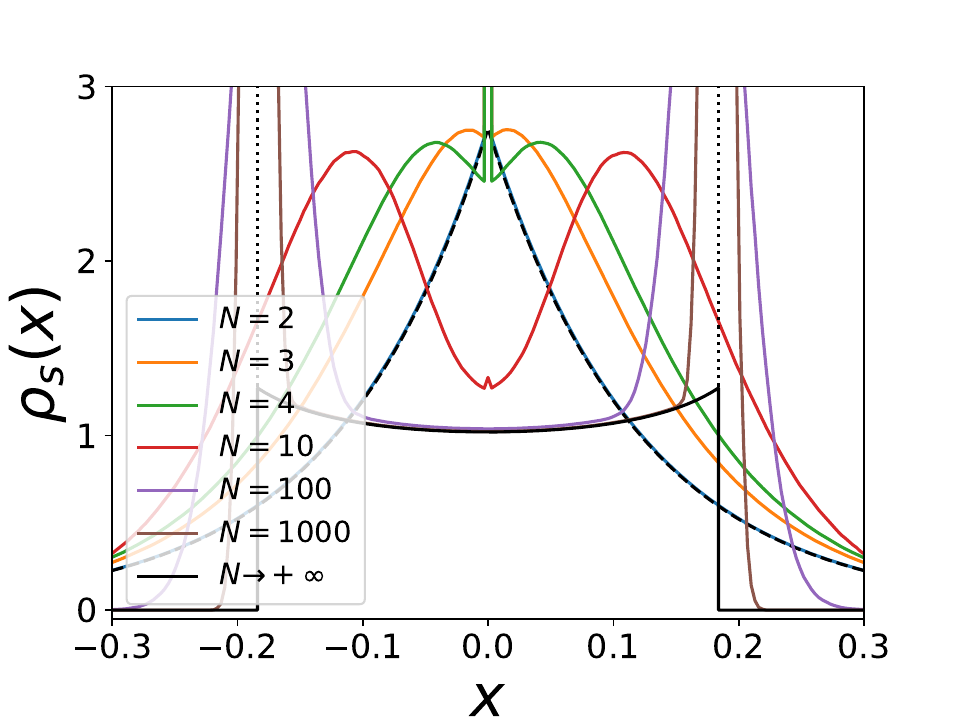}
    \caption{{\bf Left:} The (odd) function $f(r)$ in \eqref{solu_r_attractive0} for $r\geq 0$ and for 3 different values of $v_0/\bar \kappa$. For
    $v_0/\bar \kappa>1$ it diverges at $r=\frac{1}{2}$. For $v_0/\bar \kappa<1$ a maximum appears at $r=r^*<\frac{1}{2}$.
    {\bf Right:} Density $\rho_s(x)$ for $v_0=0.7$, $\bar \kappa=1$ and $\gamma=1$: as $N$ increases it takes
    a bimodal shape, with two smooth symmetric peaks which, for $N=+\infty$, become delta peaks (shocks) at the two edges (shown as dotted vertical lines). The histogram shows schematically the finite $N$ delta peak at $x=0$. The dashed
    black line shows the analytical prediction \eqref{PDF2part_linear} for $N=2$. The plot of $\rho_s$ for $N\to+\infty$ is obtained using the parametric representation \eqref{rho_parametricRD1} (on the interval $[-x_e,x_e]$).}
    \label{fig_attractive}
\end{figure}

{\it Smooth phase $v_0>\bar\kappa$.} We now focus again on the purely active case $v_0>0$ and $T=0$. In that case, we must solve the equations
\bea \label{eqrank_attractive1} 
&& v_0 s(x)  = \bar \kappa \big(r(x)^2 - \frac{1}{4} \big) \;,  \\
&& v_0 r'(x) = 2 \bar \kappa r(x) s'(x) - 2 \gamma s(x) \;,
\label{eqrank_attractive2} 
\eea 
keeping in mind that $r(x)$ may have discontinuities. Substituting the first equation into the second, we obtain
\be  \label{rp} 
\big(v_0^2 - 4 \bar \kappa^2 r(x)^2\big) r'(x) = 2 \gamma \bar \kappa \big(\frac{1}{4}-r(x)^2 \big)  \;.
\ee 
Since $|r(x)| \leq 1/2$ for all $x$ (with $r(\pm \infty)= \pm 1/2$), the r.h.s. is positive and bounded. Let us first consider the case $v_0>\bar \kappa$. Then $r'(x)$ is bounded from \eqref{rp}, hence there cannot be any shocks. In this case the stationary density is smooth, with unbounded support, and $r(x)$ is obtained by inversion of the equation \eqref{rp} (using the condition $r(0)=0$ to fix the integration constant),
\be  \label{solu_r_attractive0} 
\frac{\gamma x}{\bar \kappa} = f(r(x)) \quad , \quad f(r) = 2 \int_0^{r} du \frac{(v_0/\bar \kappa)^2-4u^2}{1-4u^2} = 2 r + \big(\frac{v_0^2}{\bar \kappa^2}- 1 \big) {\rm arctanh}(2 \, r) \;.
\ee 
One can check that for $v_0>\bar \kappa$, the function $f(r)$ is indeed
invertible (it increases monotonously from $f(-1/2)=-\infty$ to $f(1/2)=+\infty$, see Fig. \ref{fig_attractive}). The total density $\rho_s(x)=r'(x)$ is even in $x$, while $\rho_d(x)=s'(x)$ is odd in $x$, and both decay exponentially for $|x| \to +\infty$,
\bea \label{rho_asymptotics_RDattractive} 
&& \rho_s(x) \simeq A_s \, e^{- \frac{|x|-x_0}{\xi_\infty}}  \quad , \quad \rho_d(x) \simeq A_d \, {\rm sgn}(x) e^{- \frac{|x|-x_0}{\xi_\infty}} \;, \\
\text{with} && \xi_\infty =  \frac{v_0^2 - \bar \kappa^2}{ 2 \gamma \bar \kappa  }  \quad , \quad x_0=\frac{\bar \kappa}{\gamma}  \quad , \quad  A_s = \frac{1}{\xi_{\infty}}
\quad , \quad  A_d=  \frac{\bar \kappa}{v_0} A_s \nonumber \;.
\eea 
Note that in the diffusive limit $v_0,\gamma \to+\infty$ with $T_{\rm eff}=\frac{v_0^2}{2\gamma}$ fixed, we recover the decay length of the Brownian case \eqref{soliton2}, $\xi_\infty \to T_{\rm eff}/\bar \kappa$. Let us also mention that the relation \eqref{solu_r_attractive0} also provides a parametric representation for the densities as
\be \label{rho_parametricRD1}
x=\frac{\bar \kappa}{\gamma} f(r) \quad , \quad \rho_s = \frac{\gamma}{\bar \kappa f'(r)} \quad , \quad \rho_d = \frac{2\gamma r}{v_0 f'(r)} \quad , \quad r\in[-1/2,1/2],
\ee
which we use in Fig.~\ref{fig_attractive}
\\

{\it Marginal case $v_0=\bar\kappa$.} If we decrease the parameter $v_0/\bar \kappa$, we see from \eqref{rho_asymptotics_RDattractive} that the decay length $\xi_\infty$ vanishes at $v_0/\bar \kappa=1$. We thus expect a transition to occur at this point. Indeed, for $v_0/\bar \kappa=1$, one has $f(r)=2 r$, which takes finite values at $r=\pm1/2$. The solution \eqref{solu_r_attractive0} is thus only valid on a finite support $[-x_e^*,x_e^*]$, with $x_e^*=\bar \kappa/\gamma$. In this case, the function $r(x)$ is linear on this interval,
\be \label{v0=kappa_solr}
r(x) = \frac{1}{2} \frac{x}{x_e^*} \quad , \quad s(x) = - \frac{1}{4} \big(1 - \big(\frac{x}{x^*_e} \big)^2 \big) \quad , \quad |x| \leq x_e^* = \frac{\bar \kappa}{\gamma} \;,
\ee 
and since $r(x)$ is an increasing function we deduce that $r(x)=\frac{1}{2} {\rm sgn}(x)$ and $s(x)=0$ for $|x| \geq x_e$. Note that the value of $x_e^*$ coincides with $x_0$ in \eqref{rho_asymptotics_RDattractive}. We thus obtain that the densities have a finite support $[-x_e,x_e]$, and that the total density $\rho_s(x)$ is uniform on this interval,
\be \label{v0=kappa_solrho}
\rho_s(x) = \frac{\gamma}{2 \bar \kappa} \quad , \quad \rho_d(x) = \frac{ \gamma^2 x}{2 \bar \kappa^2} \quad , \quad |x| \leq x_e^*= \frac{\bar \kappa}{\gamma} \;.
\ee 
Both $\rho_s(x)$ and $\rho_d(x)$ thus vanish for $|x|>x_e^*$, with step discontinuities at the two edges. 
\\

{\it Shock phase $v_0<\bar\kappa$.} For $v_0<\bar \kappa$, $f(r)$ becomes non-invertible, as can be seen on Fig. \ref{fig_attractive}. More precisely, $f(r)$ is increasing only on an interval $[-r^*,r^*]$, with $r^*=\frac{v_0}{2\bar \kappa}<\frac{1}{2}$. Since $r(x)$ should be an increasing function of $x$, this implies that the density has a finite support $[-x_e,x_e]$, with $x_e \leq x^* = \frac{\bar \kappa}{\gamma} f(r^*)$, and that $r(x)$ exhibits shocks at $\pm x_e$. Since the r.h.s. in \eqref{rp} is bounded, $r'(x)$ can only diverge at a point where the prefactor in the l.h.s of \eqref{rp} vanishes. Naively, this would lead to $r(x_e) = r^*$, i.e., $x_e=x^*$. However, at the position of a shock, particular care should be taken in the treatment of the interaction term, as in the case of the standard Burgers equation \cite{BernardBurgers}. Indeed, let us recall that the factor $- 2\bar\kappa r(x)$ corresponds to the total force acting on a particle at $x$ due to the interactions. If the density presents a delta peak at position $x$, we should take into account the fact that ${\rm sgn}(0)=0$, meaning that the total interaction force acting on a particle only depends on the number of particles strictly at its left $n_{\rm left}$ and at its right $n_{\rm right}$. The total interaction force acting on a particle inside a cluster at position $x$ thus reads
\bea
\bar \kappa \frac{n_{\rm right}}{N} - \bar \kappa \frac{n_{\rm left}}{N} &=& \bar \kappa(\frac{1}{2} - r(x^+)) - \bar \kappa(r(x_-) + \frac{1}{2}) \nonumber \\
&=& - \bar \kappa (r(x^+)+r(x^-)) \;. \label{replaceRD} 
\eea
In the presence of a discontinuity in $r(x)$, the term $- 2\bar\kappa r(x)$ should thus be interpreted as $- \bar \kappa (r(x^+)+r(x^-))$, leading to the equations
\bea \label{eqrank_attractive1true} 
&& v_0 s'(x)  = \bar \kappa \, [r(x^+)+r(x^-)] r'(x) \;,  \\
&& v_0 r'(x) = \bar \kappa [r(x^+)+r(x^-)] s'(x) - 2 \gamma s(x) \;,
\label{eqrank_attractive2true} 
\eea 
which are a generalization of (\ref{eqrank_attractive1}-\ref{eqrank_attractive2}) which remain valid in the presence of shocks. Integrating between $x_e^-$ and $x_e^+$ then yields the two relations
\be 
\Delta r = \frac{\bar \kappa}{v_0} (r(x_e^+)+r(x_e^-)) \Delta s ~,~ 
\Delta s = \frac{\bar \kappa}{v_0} (r(x_e^+)+r(x_e^-)) \Delta r \nonumber
\ee 
where $\Delta r=r(x_e^+)-r(x_e^-)$ and similarly for $\Delta s$. A non-zero $\Delta r$ then requires $r(x_e^+)+r(x_e^-)=v_0/\bar \kappa$. Since $r(x_e^+)=1/2$, this leads to
\be \label{eq_edgeRD1}
r(x_e^-)=\frac{v_0}{\bar \kappa} -\frac{1}{2} \;,
\ee
which determines the position of the edge. For $|x|<x_e$, the prefactor in the l.h.s of \eqref{rp} is strictly positive, and thus $r(x)$ is still given by \eqref{solu_r_attractive0}, while for $|x|>x_e$ one has $r(x)=1/2$. This means that $\rho_s(x)$ has delta peaks at $\pm x_e$, each containing a fraction $1/2-r(x_e^-)=1-v_0/\bar\kappa$ of the particles. $s(x)$ is still given by \eqref{eqrank_attractive1}, leading to $s(x_e^-)=v_0/\bar \kappa-1$ and $s(x_e^+)=0$. Thus it also has jumps of amplitude $1-\frac{v_0}{\bar\kappa}$ at $\pm x_e$. This means that the cluster at $+x_e$ only contains $+$ particles, while the cluster at $-x_e$ only contains $-$ particles. Note that as $v_0/\bar \kappa \to 1/2$, $r(x_e^-)\to 0$, i.e., $x_e \to 0$, so that the density $\rho_s(x)$ converges towards a unique delta peak at $x=0$, consistent with the stability argument given above in \eqref{stabilityclusterRD}.

Let us briefly give the physical intuition behind the formation of these edge clusters. In the absence of cluster, the rightmost particle is subjected to a total force $-\bar \kappa (1-\frac{1}{N})$. Thus when $v_0<\bar \kappa$, for large enough $N$, it will always move towards the left even if its driving velocity $+ v_0$ is towards the right. It will thus aggregate with other $+$ particles, until the resulting cluster reaches a fraction $n_c/N$ of the total number of particles large enough to be at dynamical equilibrium, i.e., such that $v_0=\bar \kappa (1-n_c/N)$. By definition, $n_c/N=1/2-r(x_e^-)$, and thus we recover \eqref{eq_edgeRD1}. For finite $N$ however, the size of this cluster, and hence its position, will fluctuate, and thus we expect the delta function in the density to be replaced by a peak of finite width.

\begin{figure}
    \centering
    \includegraphics[width=0.45\linewidth,trim={0 0 1cm 0.5cm},clip]{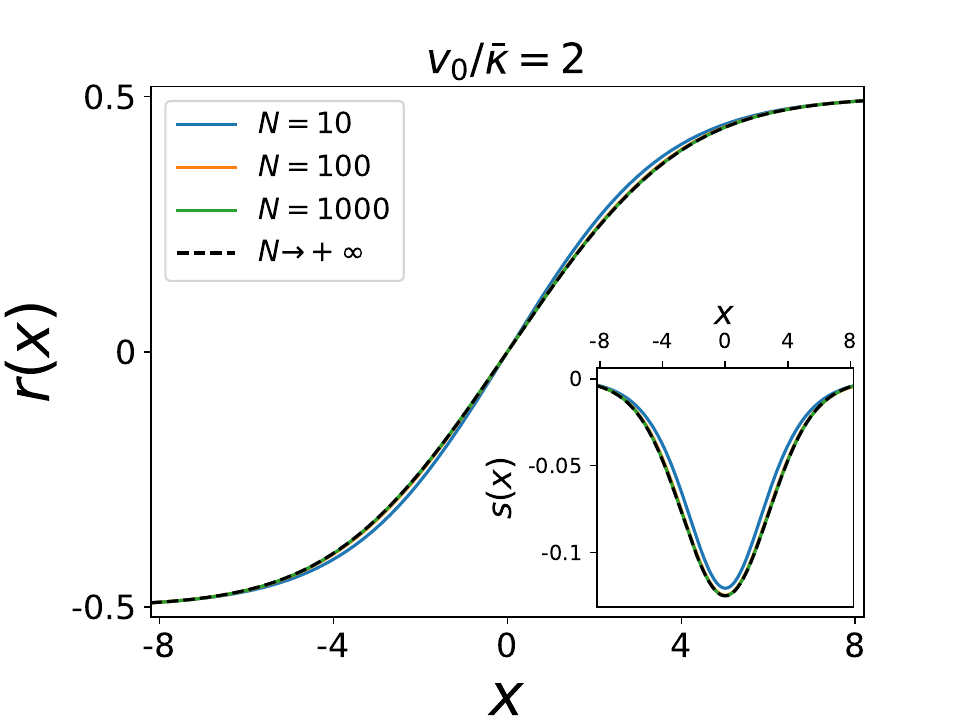}
    \hspace{0.1cm}
    \includegraphics[width=0.45\linewidth,trim={0 0 1cm 0.5cm},clip]{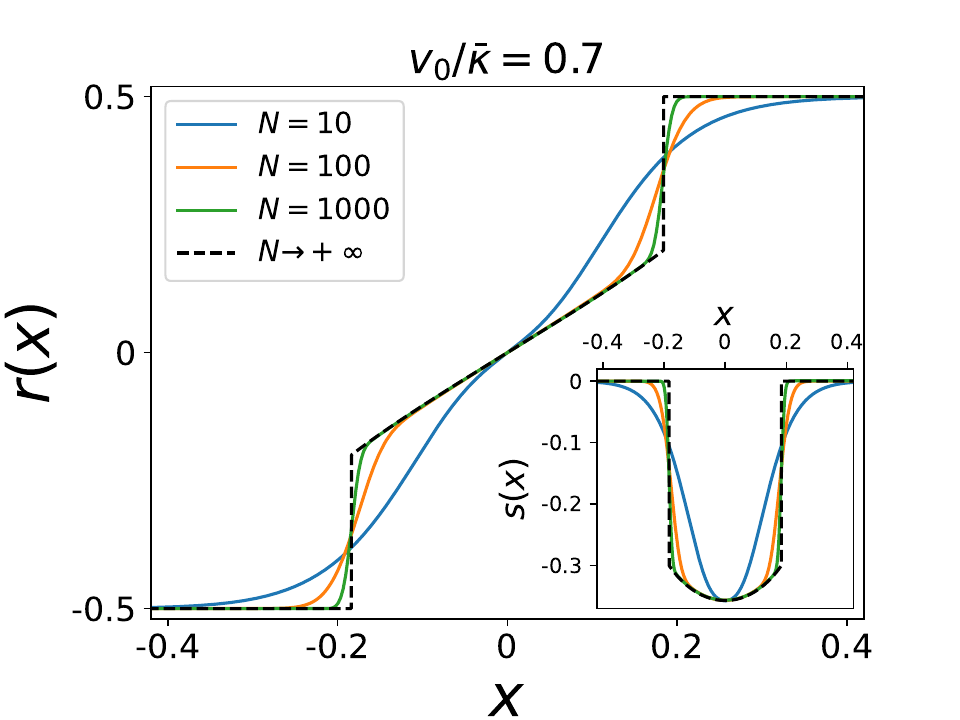}
    \caption{Rank field $r(x)$ in the stationary state, obtained through simulations, as a function of $x$, in the attractive case with $\bar \kappa=1$ and $\gamma=1$, for increasing values of $N$. The dashed black curves show the analytical prediction for $N \to +\infty$. The insets show $s(x)$ for the same set of parameters. {\bf Left:} $v_0=2$. The density is smooth and has infinite support. {\bf Right:} $v_0=0.7$. The density has finite support and exhibits $\delta$ peaks at the edges (i.e jumps in $r(x)$ and $s(x)$). }
    \label{fig_attractive_r}
\end{figure}

The predictions for $r(x)$ and $s(x)$ are compared with the results of numerical simulations in Fig.~\ref{fig_attractive_r}. For $v_0>\bar \kappa$, we find a good agreement even for very small values of $N$ ($N\sim 10$). For $\bar \kappa/2 < v_0<\bar \kappa$, although the phase transition and the delta peaks in the density are strictly speaking a special feature of the $N\to+\infty$ limit, the numerical results for finite $N$ clearly show precursor signatures of these effects, in the form of smooth peaks in the density. As one can see in Fig.~\ref{fig_attractive}, the density already takes a bimodal form for small values of $N>2$. As mentioned above, for any finite $N$ there is however a true delta peak at $x=0$, but its weight decays exponentially with $N$ such that for $N=10$ it is already almost unnoticeable in the simulations. In Fig.~\ref{fig_attractive_r}, one can see that there is a very good agreement at large $N$ between the numerics and the analytical results for $r(x)$ and $s(x)$. 

\subsection{Repulsive case}

We now turn to the repulsive case $\kappa>0$, keeping $V'(x)=0$ (in this case we keep $T>0$ as it does not make the analysis more difficult),
\bea \label{eqrank_repulsive1} 
&& \partial_t r = - v_0 \partial_x s  - 2 \kappa \, r \partial_x r + T \partial_x^2 r \;, \\
&& \partial_t s = - v_0 \partial_x r - 2 \kappa r \partial_x s - 2 \gamma s + T \partial_x^2 s \;. \label{eqrank_repulsive2} 
\eea
Since there is no confining potential to compensate the repulsion between particles, all the particles escape to infinity and there is no stationary state. However, we can still analyze the large time behavior of the system. 
As in the passive case, studied in \cite{PLDRankedDiffusion, FlackRD}, we expect the support of the density to grow linearly with time.
We thus look for a large time solution of (\ref{eqrank_repulsive1}-\ref{eqrank_repulsive2}) as a scaling function of the parameter $y=x/t$, with the $1/t$ expansion
\be \label{expansion_RDrepulsive}
r(x,t)= r_0(y)  + \frac{r_1(y)}{t} + \dots \quad , \quad 
s(x,t) = \frac{s_1(y)}{t}  + \dots \quad , \quad y=\frac{x}{t} \;.
\ee 
Inserting into the first equation yields at leading order
\be 
y r_0'(y) = 2 \kappa \, r_0(y) r_0'(y) \;.
\ee 
The solution of this equation is either $r_0'(y)=0$, or $r_0(y)=y/(2k)$, which leads to
\be 
r_0(y) = \begin{cases} -\frac{1}{2} \quad {\rm for} \ y < -\kappa \;, \\ 
\frac{y}{2 \kappa} \quad \ {\rm for} \ -\kappa < y < \kappa \;, \\
\frac{1}{2} \quad \ \ \, {\rm for} \ y > \kappa \;,
\end{cases} \quad y=\frac{x}{t} \;.
\ee
This corresponds to an expanding square density,
\be \label{rhos_repulsiveRD}
\rho_s(x,t) = \partial_x r(x,t) \simeq \frac{1}{2 \kappa t} \, \Theta(\kappa t - |x|) \;,
\ee 
where $\Theta(x)$ is the Heaviside distribution. Hence the scaled density $\rho_s(x)$ takes the form of an expanding plateau, exactly as for Brownian particles
\cite{PLDRankedDiffusion, FlackRD}. This was expected, as on large timescales and large length-scales, the active noise behaves effectively as a diffusive noise. Thus, in the repulsive case we would need to confine the particles in order to observe a behavior which is drastically different from passive particles (which we will do in the next section).

The density however exhibits non-trivial fluctuations with slow decay at large time. Indeed, the second equation gives
\be \label{repulsive_sr_rel}
s_1(y) =  - \frac{v_0}{2 \gamma}  r_0'(y) = - \frac{v_0}{4 \kappa \gamma} \, \Theta(\kappa-|y|) 
\ee 
which leads to
\bea
&& s(x,t) \simeq - \frac{v_0}{2 \gamma} \partial_x r(x,t) \simeq - \frac{v_0}{4 \kappa \gamma t} \, \Theta(\kappa t - |x|) \;, \\
&& \rho_d(x,t) \simeq - \frac{v_0}{4 \kappa \gamma t} (  \delta(x+\kappa t) - \delta(x-\kappa t) ) \;. \label{rhod_repulsiveRD1}
\eea
This result suggests that $\rho_d(x,t)=0$ inside the plateau, i.e., $\rho_+=\rho_-$ (to leading order in $1/t$), but that there is an excess of $-$ particles at the left edge, and of $+$ at the right edge, with a total weight decaying as $1/t$. However, the fact that $\rho_d$ exhibits delta peaks while $\rho_s$ does not (although in general one should always have $\rho_d\leq\rho_s$ by definition), is a sign that the present expansion does not fully capture the behavior near the edges, as we discuss below.

The higher order terms in the expansion \eqref{expansion_RDrepulsive} depend on the initial condition, and we will not study them here.  At this point we should however mention that, as in the Brownian case \cite{FlackRD}, the $1/t$ expansion is valid inside the plateau but fails at the edges of the support. In a region of width $\sim \sqrt{t}$ around the edges, a more careful analysis shows (see the SM of \cite{activeRD1}) that
the solution takes a boundary layer form (at the right edge)
\bea \label{BL_activeRD} 
\!\!\!\!\!\!\!\!\!\!\!\! &&\rho_s(x,t) = \frac{1}{\kappa t} \hat \rho_s (z)  \quad , \quad \rho_d(x,t) = -\frac{v_0}{2 \gamma \kappa \sqrt{T_{\rm eff}} \, t^{3/2}} \hat \rho_s' (z) \;, \nonumber \\
\!\!\!\!\!\!\!\!\!\!\!\! &&\hat \rho_s(z) = \frac{e^{-\frac{z^2}{2}}(2+\sqrt{\pi} e^{\frac{z^2}{4}} z \, {\rm erfc}(-\frac{z}{2}))}{2\pi \, {\rm erfc}(-\frac{z}{2})^2}  \quad , \quad z =\frac{x-\kappa t}{\sqrt{(T_{\rm eff}+T) t}} \;,
\eea
with $T_{\rm eff} = \frac{v_0^2}{2 \gamma}$. The boundary layer scaling
function $\hat \rho_s(y)$ is the same as the one obtained for the passive problem in 
\cite{FlackRD}. This confirms that, at large time, the effect of the active noise in an expanding repulsive gas is not different from that of Brownian noise at temperature $T_{\rm eff}$. This is consistent with the fact that the relation $\rho_d \simeq - \frac{v_0}{2 \gamma} \partial_x \rho_s$ holds at large time both in the plateau and in the boundary layer (we have also checked this numerically). Note also the absence of delta peak in $\rho_d(x,t)$ at the level of the boundary layer, showing that the delta functions in \eqref{rhod_repulsiveRD1} actually have a finite width, contrary to the clusters of the attractive case. We also see that the $+/-$ population imbalance of order $\sim 1/t$ is subdominant compared to the total population of particles in the boundary layer, of order $\sim 1/\sqrt{t}$. 

\begin{figure}[t]
    \centering
    \includegraphics[width=0.4\linewidth,trim={0 0 0.5cm 0.5cm},clip]{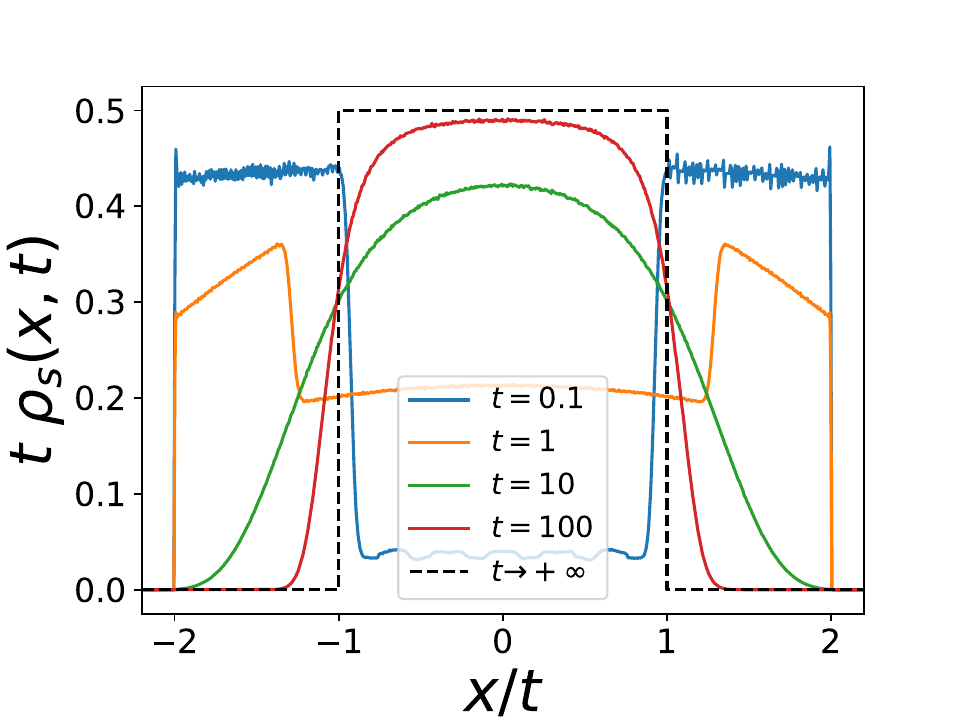}
    \includegraphics[width=0.4\linewidth,trim={0 0 0.5cm 0.5cm},clip]{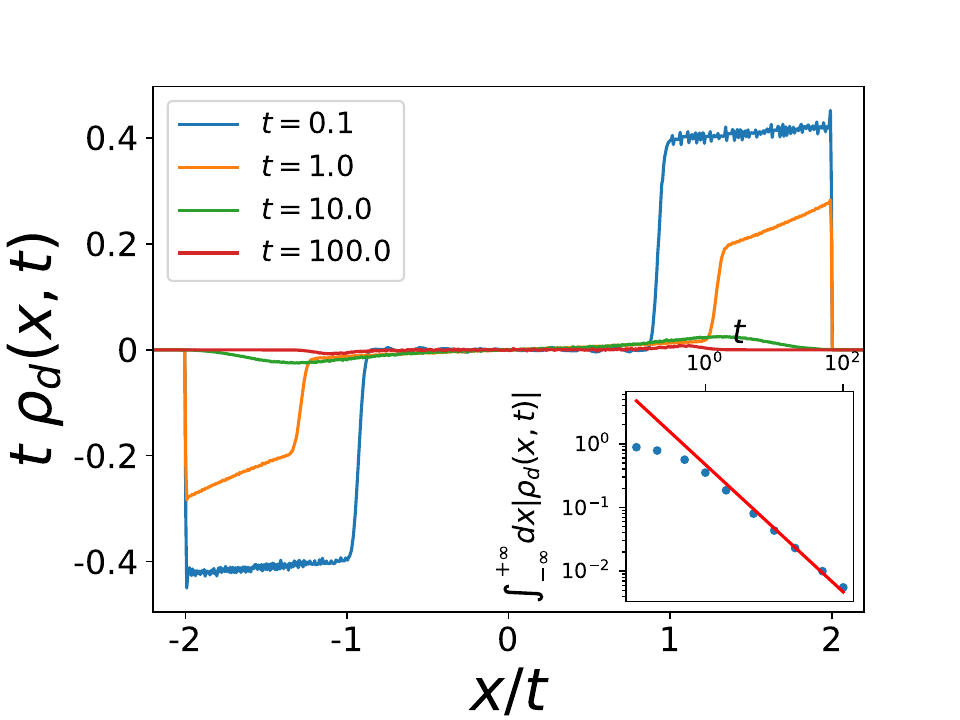}
    \includegraphics[width=0.4\linewidth,trim={0 0 0.5cm 0.5cm},clip]{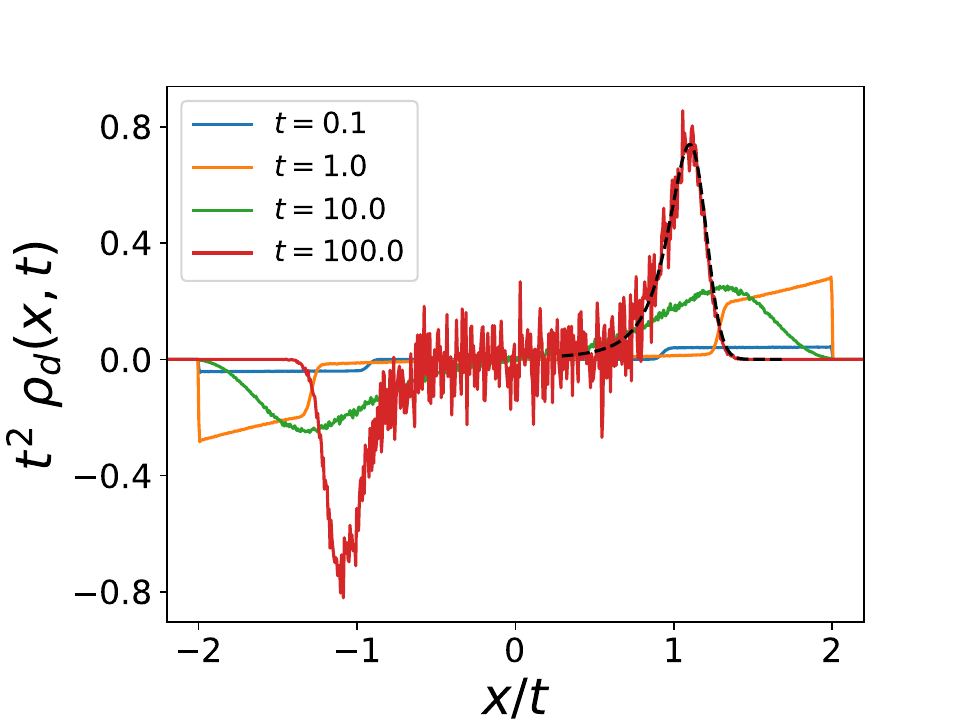}
    \includegraphics[width=0.4\linewidth,trim={0 0 0.5cm 0.5cm},clip]{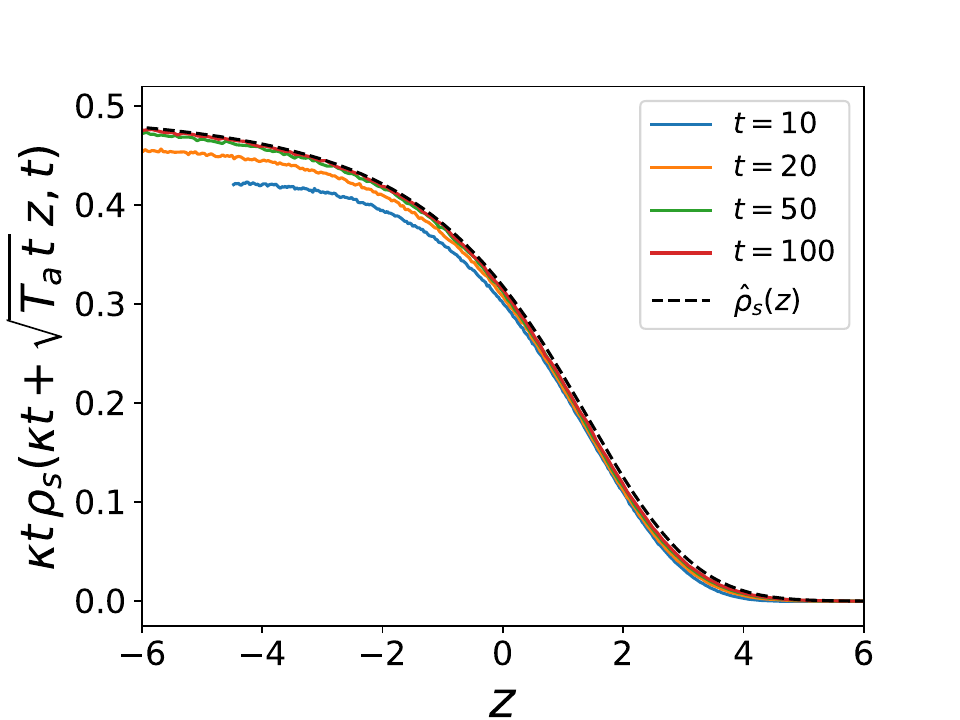}
    \caption{{\bf Top Left:} Density $\rho_s(x,t)$, obtained through simulations, plotted at different times as a function of the scaled position $x/t$, for the repulsive gas with $v_0=1$, $\kappa=1$ and $\gamma=1$, for $N=1000$ particles. At $t=0$ all particles are at $x=0$. The dashed black curve shows the analytical prediction for $t \to +\infty$. As discussed in the text, signs of the activity are clearly visible at short time where the density $\rho_s(x,t)$ features two distinct blocks composed respectively of $+$ and $-$ particles, while at large time it converges to a rectangular shape, as in the Brownian case. {\bf Top Right:} Same plot for $\rho_d(x,t)$. The inset shows the integral of the absolute value as a function of time. It is in good agreement with a $1/t$ decay (red line). {\bf Bottom left:} Plot of $\rho_d(x,t)$ multiplied by $t^2$ to compensate for the $1/t$ decay. The dashed black line shows the analytical prediction for the boundary layer \eqref{BL_activeRD} for $t=100$. {\bf Bottom right:} Density $\rho_s(x,t)$ near the edge of the plateau, plotted at different times with the boundary layer scaling \eqref{BL_activeRD}. At large times the scaled density converges to $\hat \rho_s(z)$.
    }
    \label{fig_repulsiveRD}
\end{figure}

Figure \ref{fig_repulsiveRD} shows a comparison of our analytical results with numerical simulations. In the upper left panel, we see the convergence of the density $\rho_s(x,t)$ to the expanding plateau \eqref{rhos_repulsiveRD} as time increases. At short time we can see the signature of the active noise: for $t \lesssim 1/\gamma$, the particles are split into two packets, one composed of $+$ particles with a spread of total velocities $x/t \in [v_0,v_0+ \kappa]$, and symmetrically for $-$ particles. The density thus exhibits two distinct blocks, which disappear at larger time as the active noise averages out. Figure~\ref{fig_repulsiveRD} also shows the peaks appearing in the scaled density $t^2 \rho_d(x, t)$, as well as numerical checks of the boundary layer forms \eqref{BL_activeRD}.

\section{Extensions: confining potential and non-reciprocal interaction}

In this section we briefly discuss several extensions of the model studied in the previous section. For the derivations and for more detailed analyses, see the SM of \cite{activeRD1} (in particular Section V) for the linear potential case and see \cite{activeRD2} for the harmonic potential and the non-reciprocal case. Note that for all the cases discussed below, the translational invariance is naturally broken by the external potential and we do not need to choose a reference frame as in the unconfined case above. In all of this section, we fix $T=0$.

\subsection{Linear potential}

\begin{figure}
    \centering
    \includegraphics[width=0.75\linewidth, trim={0.8cm 3.1cm 0.8cm 4.7cm}, clip]{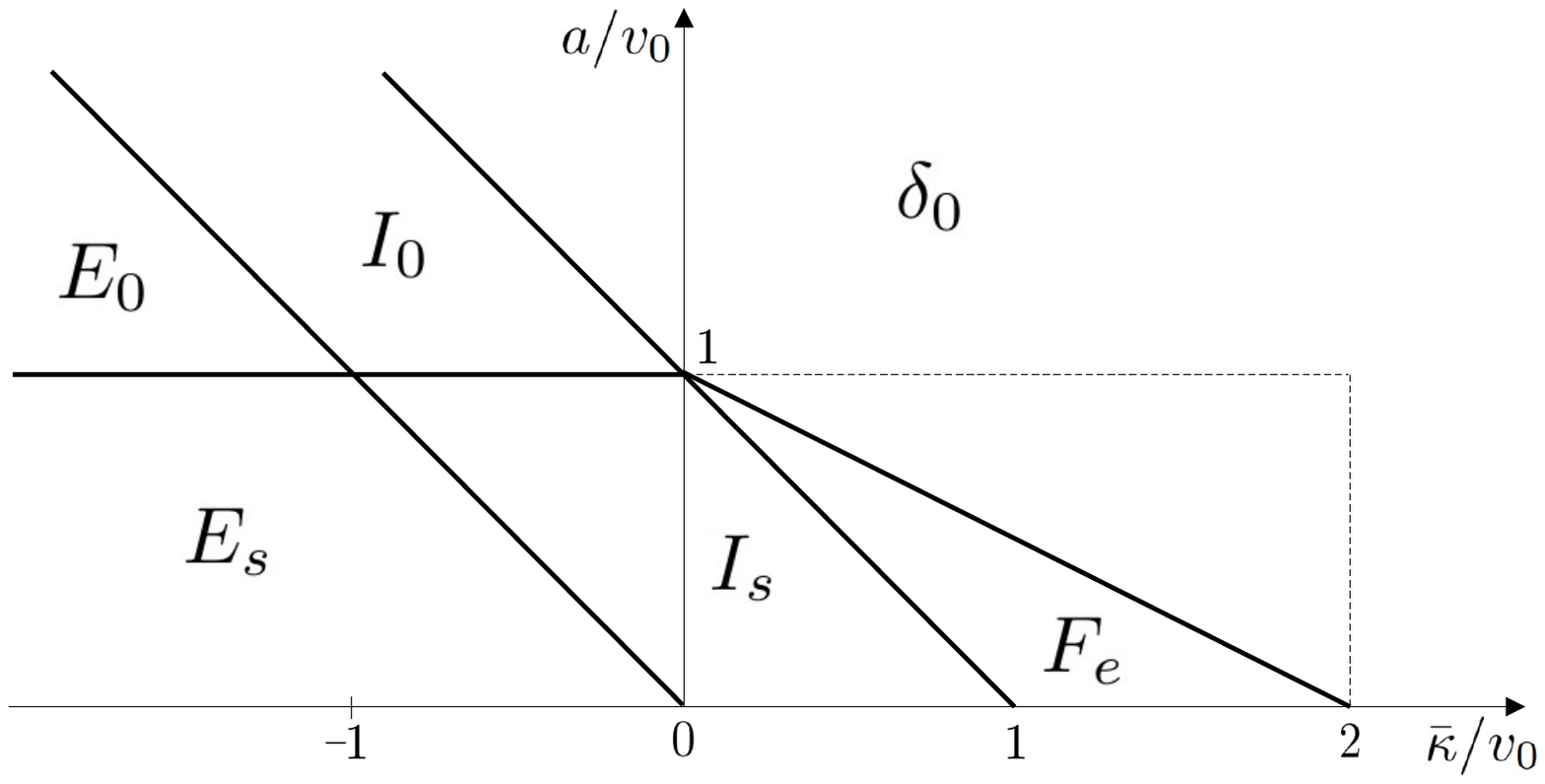}
    \caption{Phase diagram of active ranked diffusion with a linear external potential $V(x)=a|x|$. The phases are named as follows: the large letter is either $E$ for an expanding phase, $I$ for a stationary phase with infinite support, $F$ if the support is finite support, or $\delta$ for a clustered phase. The subscript indicates the position of the shocks: $0$ if there is a shock at $x=0$, $e$ if there are shocks at the edges and $s$ if the density is smooth. The dotted lines denote a change of behavior for finite $N$. Note that, in order to represent both signs of the interaction on the same diagram, we used the parameter $\bar\kappa/v_0$ on the $x$-axis instead of $v_0/\bar \kappa$ as in the unconfined case.}
    \label{phase_diagram_linpotential}
\end{figure}

We first consider what happens when we subject the particles to an attractive linear external potential $V(x)=a|x|$, with $a>0$. We recall that in the case of non-interacting particles, such a potential is enough to obtain a bound state, and that the stationary density takes the form of a double exponential with a cusp at $x=0$ for $v_0>a$, see \eqref{eqRTPlinear}, while for $v_0\leq a$ one has $\rho_s(x)=\delta(x)$ (see Sec.~\ref{sec:1particle_potential} and \cite{DKM19}). In the presence of both a linear potential and a 1D Coulomb interaction, we find that the system exhibits a rich phase diagram as we vary the dimensionless parameters $\bar \kappa/v_0$ (positive or negative) and $a/v_0$, as illustrated in Fig.~\ref{phase_diagram_linpotential}. In particular, for a repulsive interaction $\kappa=-\bar \kappa>0$, a linear potential may either allow to confine all the particles (phases $I_s$ and $I_0$ in the phase diagram of Fig.~\ref{phase_diagram_linpotential}), or only part of them (phases $E_s$ and $E_0$), as we will discuss below. Note that, as in the one particle case, the singularity of $V(x)$ may create delta peaks in the density at $x=0$. These delta peaks are present even at finite $N$ and are not an effect of the limit $N\to+\infty$, contrary to the ones at the edges.
\\

{\it Smooth phase $I_s$.} Let us consider again the equations for the rank fields for $N\to+\infty$, \eqref{eqrank1}-\eqref{eqrank2}, and look for a stationary solution,
\bea \label{eqrank1_lin}
v_0 s'(x) &=& (2 \bar \kappa r(x) + a \, {\rm sgn}(x)) \, r'(x) \; , \\
v_0 r'(x) &=& (2  \bar \kappa  r(x) + a \, {\rm sgn}(x)) \, s'(x) - 2 \gamma s(x) \;.\label{eqrank2_lin}
\eea 
Due to the symmetries of the problem, we expect $r(x)$ to be odd and $s(x)$ to be even, as in the case $a=0$ discussed above. Thus we focus on the case $x\geq0$. Since the external potential has the same form as the interaction, these equations can be solved in the same way as above, leading to
\be \label{eqres_linpot0}
\gamma x = {\sf f}_a(r)= 2 \bar \kappa r + \frac{v_0^2 - (\bar \kappa+a )^2 }{2(\bar \kappa+a)} \log\left( \frac{\bar \kappa + 2 a + 2 \bar \kappa r}{(\bar \kappa + 2 a)(1-2 r) } \right) \;.
\ee
For $a=0$ this recovers \eqref{solu_r_attractive0}, with ${\sf f}_{a=0}(r)=\bar \kappa f(r)$. From the solution for $r(x)$, we can deduce $s(x)$ using the integrated version of \eqref{eqrank1_lin} (for any $x$),
\be \label{s_function_r_lin0}
s(x) = \frac{\bar \kappa}{v_0} \big(r(x)^2 - \frac{1}{4} \big) + \frac{a}{v_0} \big(|r(x)| - \frac{1}{2} \big) \;.
\ee 
A parametric representation for the densities is given by, for $x \geq 0$ and $r \in [0,1/2]$,
\be \label{rho_parametric_linpot}
x= \frac{{\sf f}_a(r)}{\gamma} \quad , \quad \rho_s = \frac{\gamma}{{\sf f}'_a(r)} 
\quad , \quad \rho_d = \frac{(a + 2 \bar \kappa r)\gamma}{v_0 {\sf f}'_a(r)} \;.
\ee 

We now discuss the domain of validity of this solution, corresponding to the phase $I_s$ in Fig.~\ref{phase_diagram_linpotential}. As in the case $a=0$, the function ${\sf f}_a(r)$ becomes non-invertible for $v_0<\bar \kappa+a$, which leads to a finite support and shocks at the edges, as we discuss below (phase $F_e$). For $a>0$ however, this solution also extends to the repulsive case $\bar \kappa=-\kappa<0$, under the condition $\kappa < a < v_0$. In the region $I_s$, the densities have unbounded support and are smooth, except at $x=0$ where the analysis shows that $\rho_s(x)$ has a linear cusp and $\rho_d(x)$ has a jump. 
\\

{\it Phase $F_e$.} For $v_0<\bar \kappa+a$, the function ${\sf f}_a(r)$ is non-invertible, leading to a finite support $[-x_e,x_e]$ and to delta peaks at $\pm x_e$. The position of the edge $x_e$ and the weight of the delta peaks are determined in the same way as for $a=0$, by replacing $2\bar \kappa r(x) \to \bar \kappa(r(x^-)+r(x^+))$. We find
\be \label{rxe_linear} 
r(x_e^-) = \frac{v_0-a}{\bar \kappa} - \frac{1}{2}  \quad , \quad r(x_e^+) = \frac{1}{2} \quad , \quad \gamma x_e= {\sf f}_a(r(x_e^-)) \;,
\ee
which is valid for $a+\bar\kappa/2<v_0<a+\bar \kappa$. This corresponds to a cluster with a fraction $\frac{\bar \kappa + a - v_0}{\bar \kappa}$ of the particles
(all being $+$ at $x=x_e$ and $-$ at $x=-x_e$). Inside the support, $r(x)$ is again given by \eqref{eqres_linpot0},
$s(x)$ by \eqref{s_function_r_lin0} and the densities by \eqref{rho_parametric_linpot}, with the same singular behavior at $x=0$ (although $x=0$ is now a minimum for $\rho_s(x)$ instead of a maximum).
\\

{\it Phase $I_0$.} In the absence of interactions, when $a>v_0$ the particles cannot escape the point $x=0$ and the density is a single delta peak. In the presence of a repulsive interaction however, the presence of a cluster of particles at $x=0$ containing a finite fraction of the particles may generate a repulsive force strong enough for the other particles to access the whole real line. This cluster contains $+$ and $-$ particles in equal proportions and should have a total weight $\frac{a-v_0}{\kappa}$. In this case, the solution \eqref{eqres_linpot0} should be adapted as (for $x>0$)
\be
\gamma x = {\sf f}_a(r) - {\sf f}_a(r(0^+)) \quad , \quad r(0^+) = \frac{a-v_0}{2\kappa} \;,\label{rI0} 
\ee
and $s(x)$ is still given by \eqref{s_function_r_lin0} for $x\neq 0$. This solution holds in the region where $\kappa<a<v_0+ \kappa$ and $a>v_0$ and, apart from its behavior at $x=0$, it is similar to the phase $I_s$. From \eqref{rho_parametric_linpot} one can show that $\rho_s(x)$ and $\rho_d(x)$ exhibit an inverse square root divergence near $x=0$ (which involves only $+$ particles for $x>0$ and $-$ particles for $x<0$),
\be
 \rho_s(x) \simeq \frac{a-v_0}{\kappa}  \delta(x) + \frac{B}{2 \sqrt{|x|}} \quad , \quad \rho_d(x) \simeq  \frac{B}{2 \sqrt{|x|}} \, {\rm sgn}(x) \;,
\ee
with $B= \frac{1}{2\kappa} \sqrt{\frac{\gamma}{v_0} (v_0^2 - (a- \kappa)^2)}$. 
\\

{\it Phase $\delta_0$.} In \eqref{rxe_linear}, we see that $r(x_e^-)\to 0$ as $v_0 \to a +\bar\kappa/2$. Thus, for $v_0<a +\bar\kappa/2$, the density is a single delta peak at $x=0$. This phase, which we call $\delta_0$ in Fig.~\ref{phase_diagram_linpotential}, also extends to the repulsive case. Indeed, in \eqref{rI0} we see that $r(0^+) \to1/2$ as $a \to v_0+\kappa$. For $a > v_0+\kappa$, the repulsive interaction is not strong enough to allow any particle to escape $x=0$, and one has again $\rho_s(x)=\delta(x)$.

Note that, in the attractive case, contrary to the case $a=0$, a stability argument for finite $N$ as in \eqref{stabilityclusterRD} does not give the correct boundary for the $\delta_0$ phase. Indeed in the region $\min(a,\bar \kappa/2)<v_0<\bar \kappa/2+a$, a single cluster at $x=0$ is stable only in the limit $N\to+\infty$, while at any finite $N$ the densities have a finite width (see the dashed lines in Fig.~\ref{phase_diagram_linpotential}).
\\

\begin{figure}
    \centering
    \includegraphics[width=0.32\linewidth,trim={0 0 1cm 0},clip]{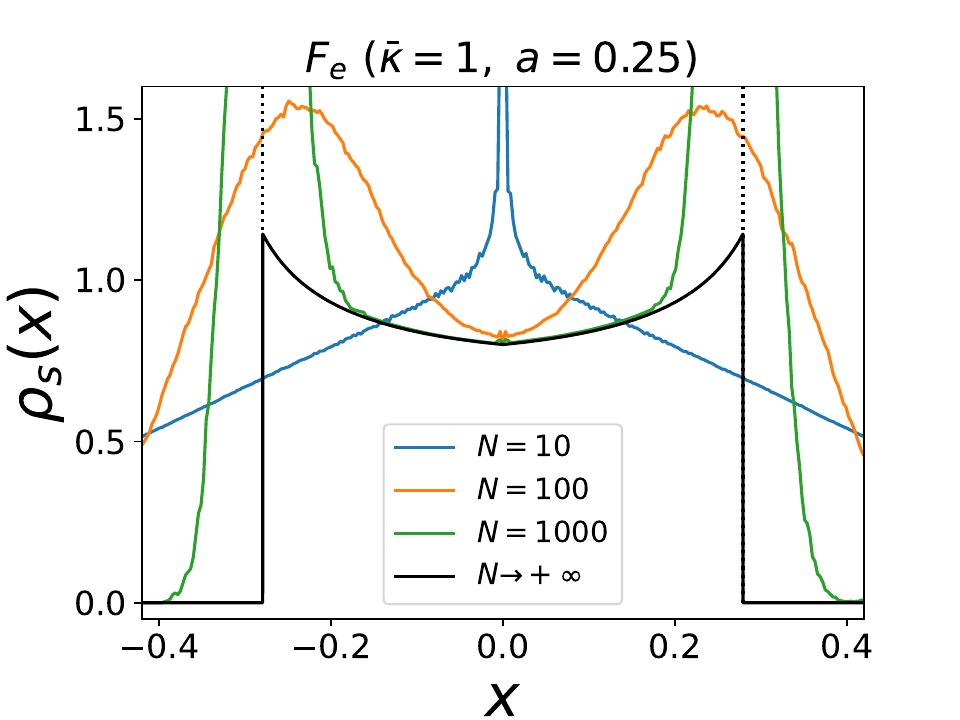}
    \includegraphics[width=0.32\linewidth,trim={0 0 1cm 0},clip]{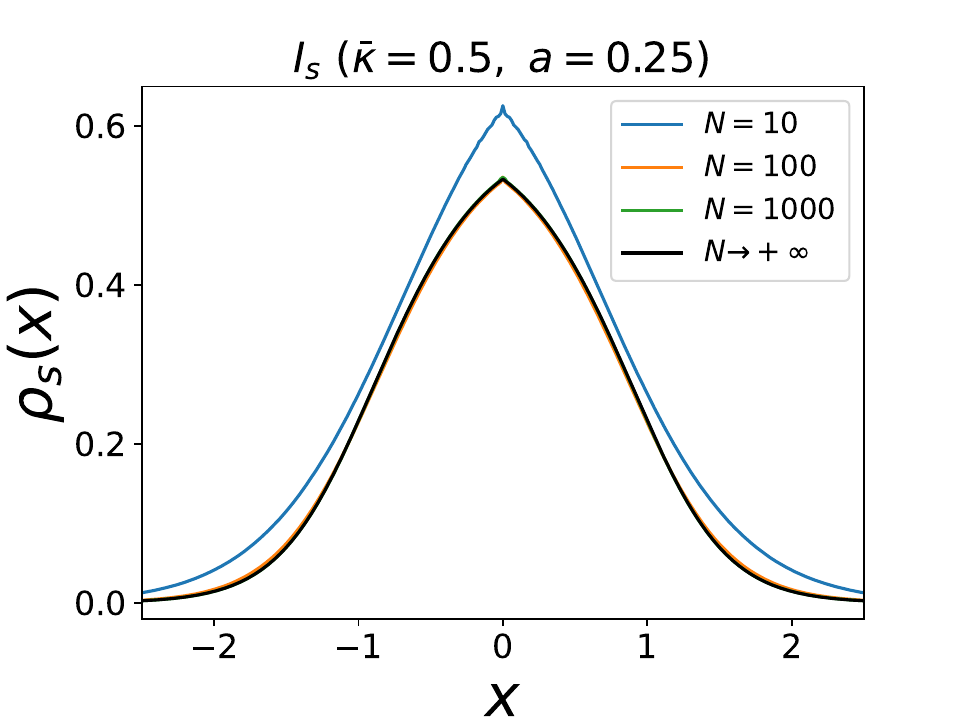}
    \includegraphics[width=0.32\linewidth,trim={0 0 1cm 0},clip]{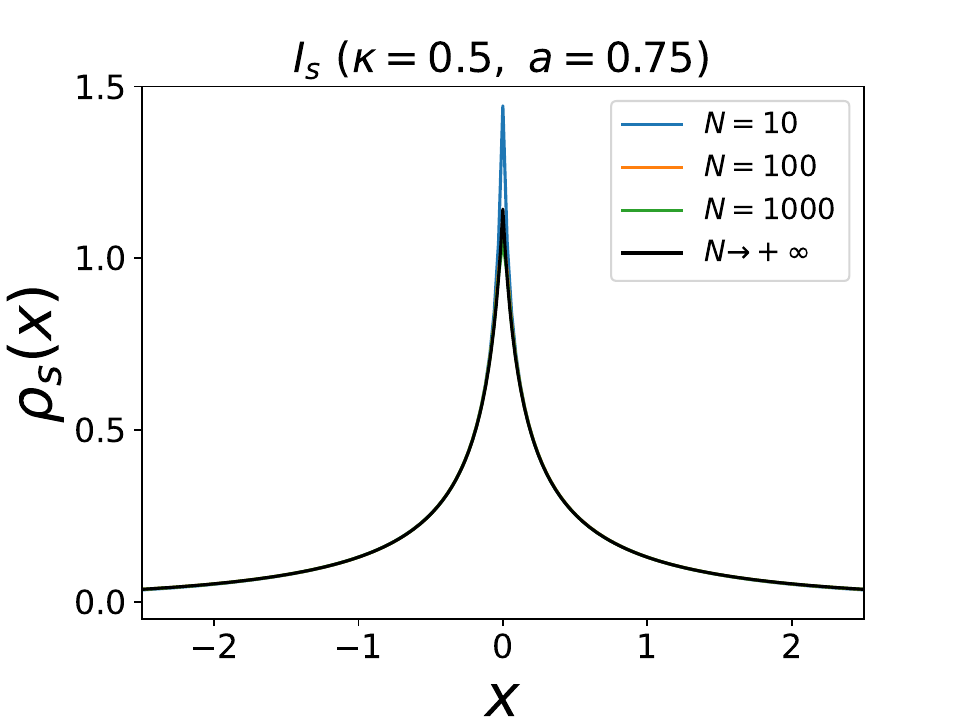}
    \includegraphics[width=0.32\linewidth,trim={0 0 1cm 0},clip]{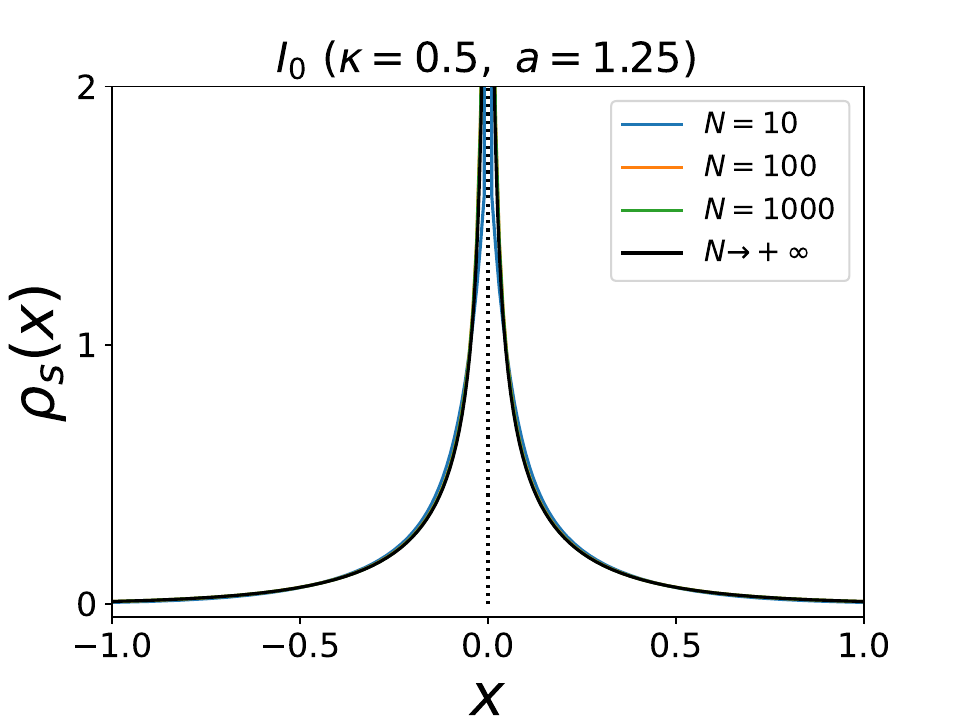}
    \includegraphics[width=0.32\linewidth,trim={0 0 1cm 0},clip]{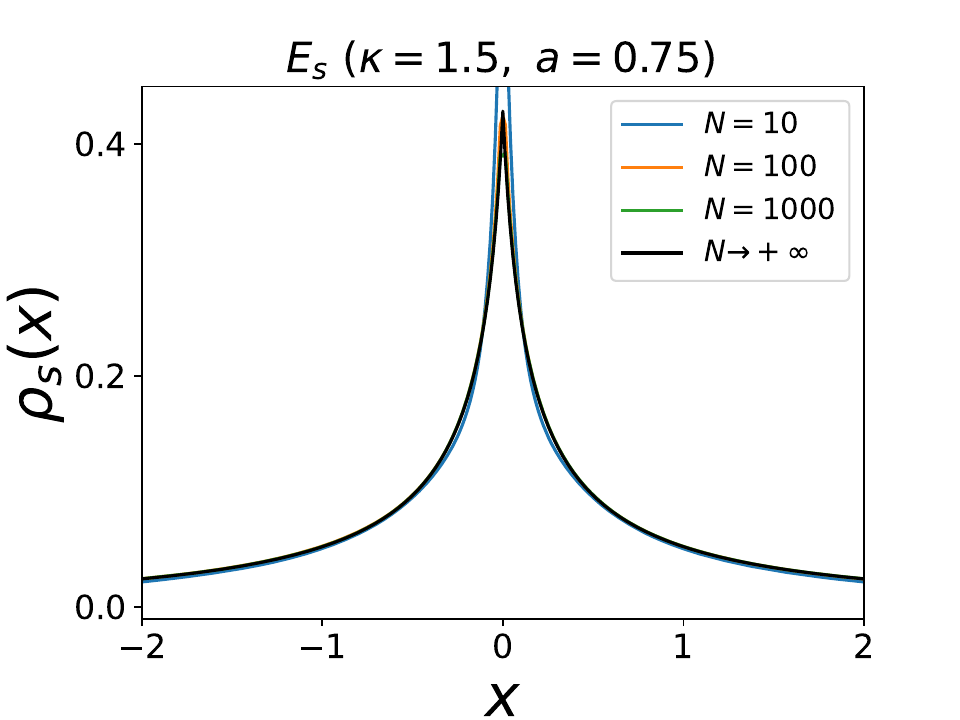}
    \includegraphics[width=0.32\linewidth,trim={0 0 1cm 0},clip]{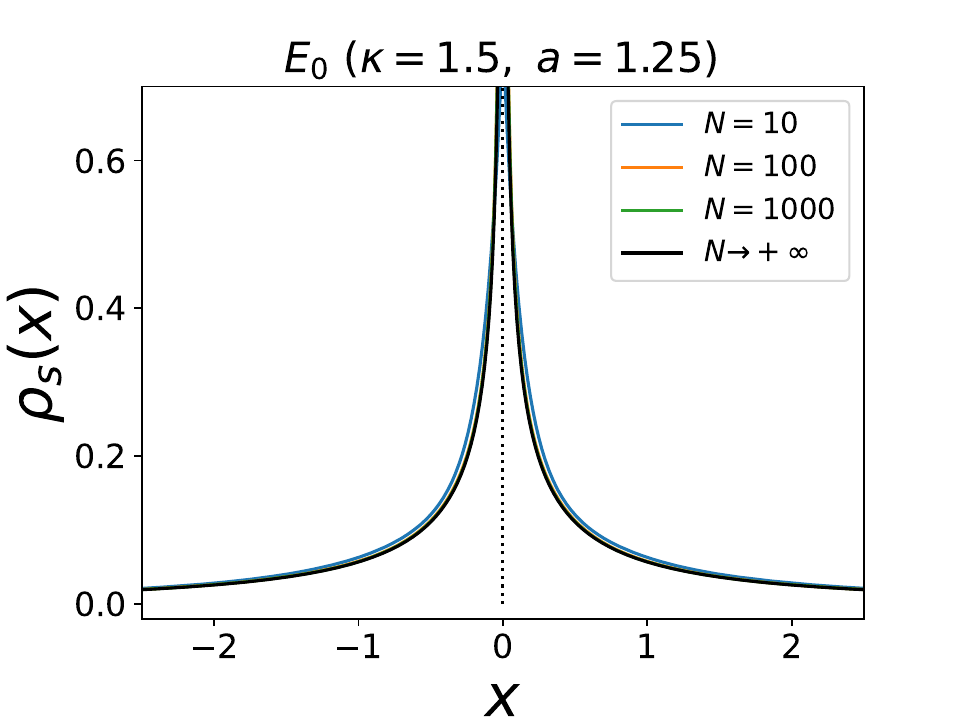}
    \caption{Density $\rho_s(x)$ in the steady state in the different phases, in the presence of a potential $V(x)=a|x|$, for different values of $N$. For all figures $v_0=1$ and $\gamma=1$. For finite $N$ it is obtained by numerical simulations. The black curves correspond to the limit $N\to +\infty$ and are computed using the parametric representation \eqref{rho_parametric_linpot} extended to all phases. The delta peaks in the density are represented by dotted vertical lines. Note that in all phases except $F_e$ the data for $N=100$ is already hardly distinguishable from the prediction at $N=+\infty$.}
    \label{fig_rhos_linpot}
\end{figure}

{\it Expanding phases $E_s$ and $E_0$.} Finally, in the repulsive case, when $\kappa>a$, the repulsion is strong enough to allow a fraction of the particles to escape to infinity, as in the unconfined case, but a fraction of the particles remains bound around $x=0$. Indeed the particles near the edges feel a total repulsive force $\sim \kappa>a$, but the particles closer to the center feel a weaker force which may not allow them to escape. The interaction force required to escape to $+\infty$ is $2\kappa r_{c}=a$, which means that a fraction $2r_c=a/\kappa$ of the particles remains bound, while a fraction $(1-a/\kappa)/2$ escapes on each side. Under the scaling $|x| \sim t$, the total density takes the form
\be \label{sol_EsE0_expanding}
\rho_s(x,t) \simeq \frac{a}{\kappa} \delta(x) + \frac{1}{2\kappa t} \Theta((\kappa-a)t - |x|) \;.
\ee
The expanding part behaves similarly as for $a=0$, i.e., it is uniform within
the support, which now spreads with velocities $\pm (\kappa-a)$. Similarly one finds
\be
\rho_d(x,t) \simeq \frac{v_0}{4\gamma\kappa t} (\delta((\kappa-a)t - x) - \delta((\kappa-a)t - x)) \;.
\ee
Both densities are smooth on scales $\sim \sqrt{t}$, with similar boundary layers 
near the edges as for $a=0$.

The delta function in \eqref{sol_EsE0_expanding} represents the bound particles, for which the densities have to be studied separately on a scale $x=O(1)$. As in the fully bound phases $I_s$ and $I_0$, we need to distinguish the case $a<v_0$, where the density has no shock (phase $E_s$), from the case $a>v_0$, where it has a shock at $x=0$ (more precisely a cluster containing a fraction $(a-v_0)/\kappa$
of the  particles,phase $E_0$). The solution for $r(x)$ is smooth away from $x=0$ and reads, for $x>0$,
\bea \label{rE}
&& \!\!\!\!\!\! \gamma x = \tilde {\sf f}_a\left(\frac{\kappa}{a} r(x)\right) - \tilde {\sf f}_a\left(\frac{\kappa}{a} r(0^+)\right) \;, \\
&& \!\!\!\!\!\! \tilde {\sf f}_a(r)= 2 a r ( \frac{v_0^2}{a^2(1-2r)} - 1 ) \quad , \quad r(0^+) = \frac{a-v_0}{2\kappa} \Theta(a-v_0) \;. \nonumber 
\eea
From \eqref{rE}, the densities $\rho_s(x)$ and $\rho_d(x)$ can be obtained explicitly. An interesting feature is that, in both phases, they decay {\it as power laws} at large distance,
\be 
\rho_s(x) \simeq \frac{v_0^2}{2 \kappa \gamma x^2} \quad , \quad \rho_d(x) \simeq \frac{v_0^3}{2 \kappa \gamma^2 x^3} \;,
\ee
while in the phases $I_s$ and $I_0$ they decay exponentially at large distance.
In that sense, the steady state of the bound particles in the phases $E_s$ and $E_0$ is always critical. Indeed, one can show that the bound particles 
are effectively described by an interaction constant $\kappa_{\rm eff}=a$. Finally, the non-analytic behavior of the densities near $x=0$ is similar to the phases $I_s$ and $I_0$ respectively. 
\\

\noindent {\bf Diffusive limit and numerical results.} To our knowledge, the Brownian version of this precise model was not studied before. It can however easily be recovered from the present results by taking the usual limit $v_0,\gamma\to+\infty$ with $T_{\rm eff}=\frac{v_0^2}{2\gamma}$ fixed. In this case the phase diagram is much simpler, and only the smooth phases $I_s$ and $E_s$ are present.

Comparisons of our analytical predictions for the density $\rho_s(x)$ in the 5 non-trivial phases with numerical results for finite $N$ are shown in Fig.~\ref{fig_rhos_linpot}, showing a very good agreement.

\subsection{Harmonic potential: the active jellium model}

\begin{figure}[t]
    \centering
    \includegraphics[width=0.49\linewidth,trim={1.8cm 3cm 0.95cm 4.4cm},clip]{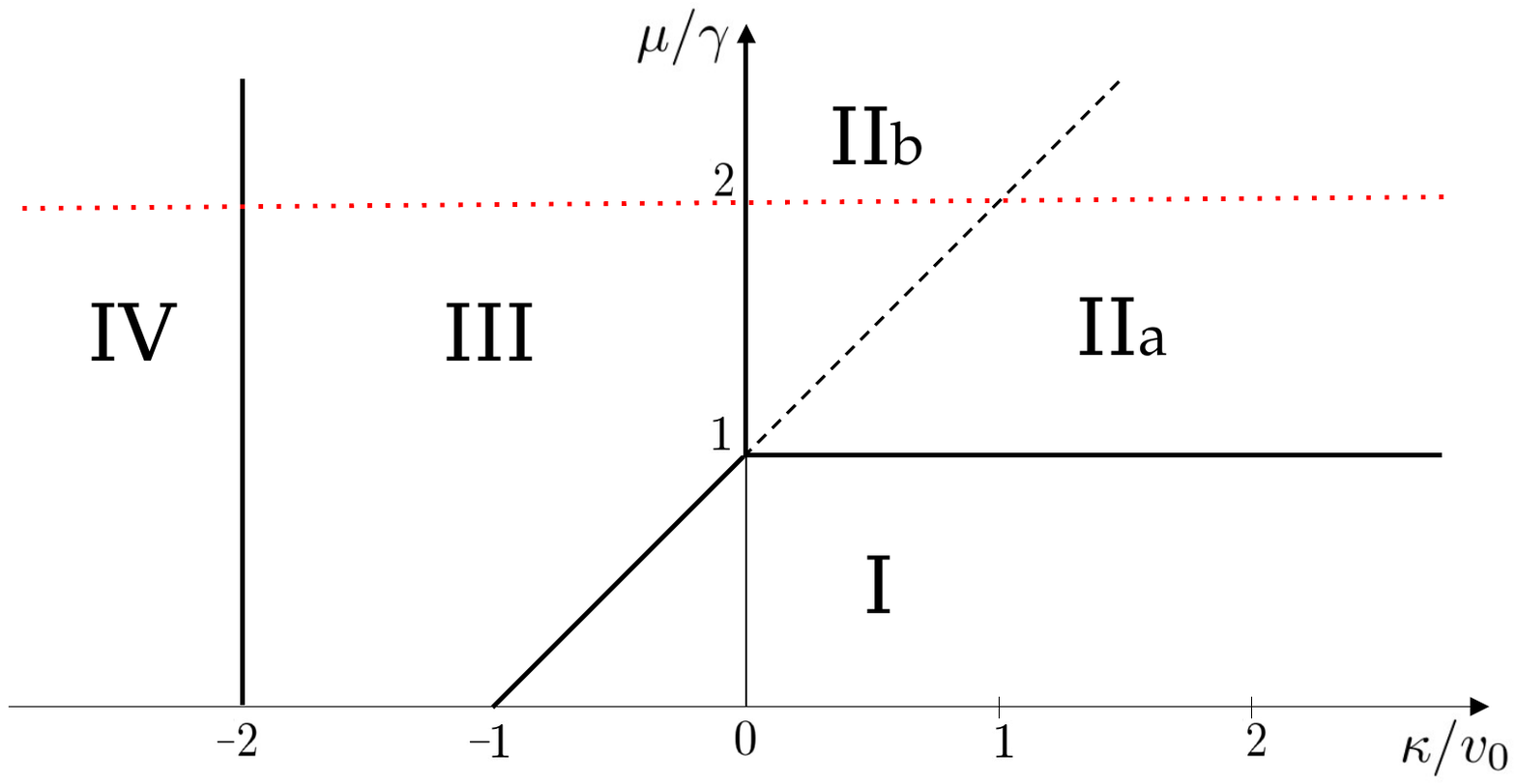}
    \includegraphics[width=0.49\linewidth,trim={0.95cm 4.55cm 1cm 3.6cm},clip]{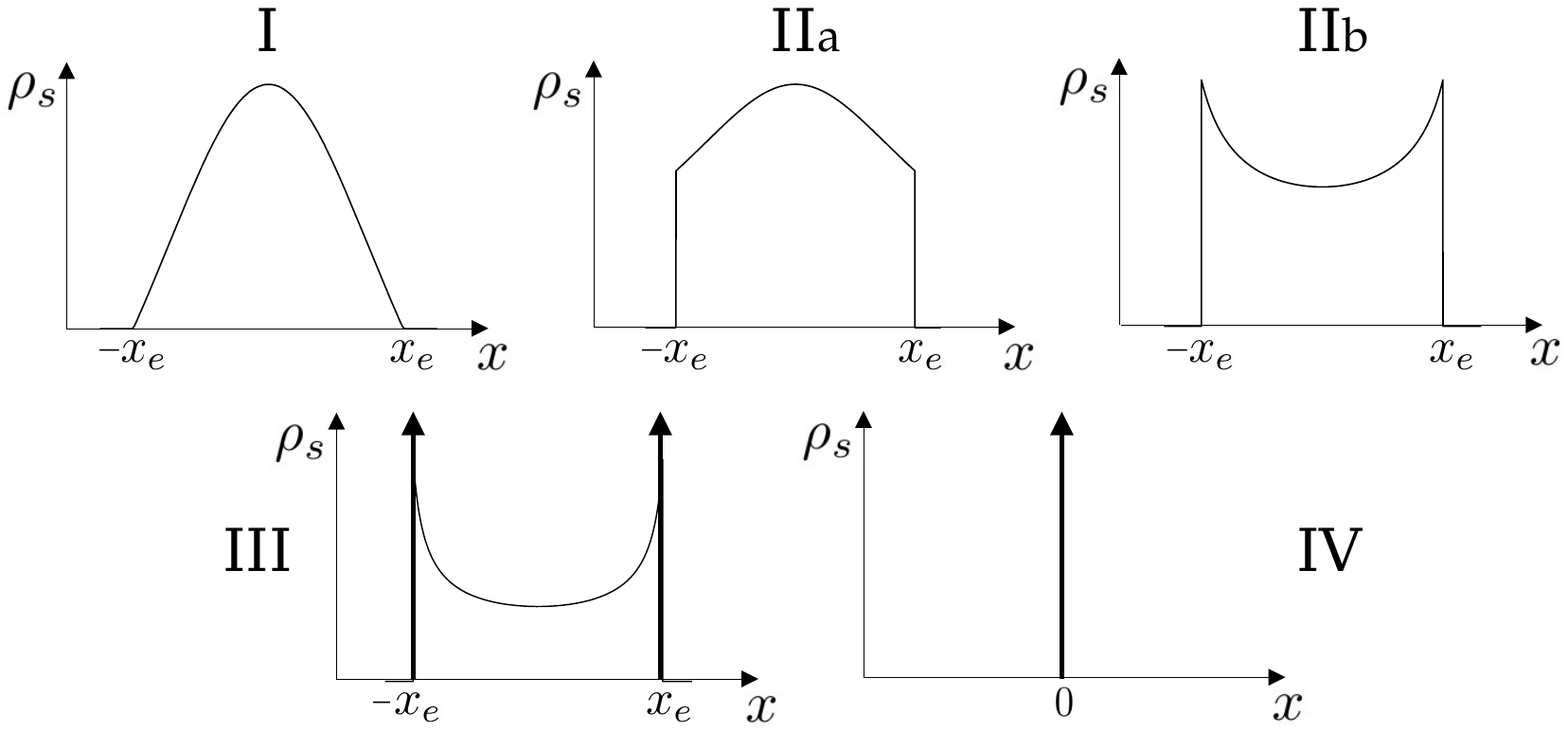}
    \caption{{\bf Left:} Phase diagram of the active jellium model, i.e., the active rank diffusion in a harmonic potential $V(x)=\mu x^2/2$ in the plane $(\kappa/v_0, \mu/\gamma)$. Note that here the parameter on the $x$-axis is $\kappa/v_0$, i.e., contrary to the previous sections the repulsive case is on the right. The dashed line between phases IIa and IIb represents only a crossover where the density changes from concave to convex (not a true phase transition). In the text, more explicit expressions are obtained on the special line $\frac{\mu}{\gamma}=2$, represented as a dotted red line on the phase diagram. {\bf Right:} Schematic representation of the total density $\rho_s(x)$ for each phase. The upward arrows represent delta functions in the density. 
    }
    \label{phase_diagram_RDharmonic}
\end{figure}

We now consider the case of a harmonic external potential $V(x)=\mu x^2/2$. In this case, since the potential is strictly convex while the amplitude of the noise and the interaction strength are finite, the density always has a bounded support $[-x_e,x_e]$ (both for an attractive and a repulsion interaction). However, we find a large variety of behaviors at the edge depending on the parameters, from continuously vanishing to jumps or even delta peaks, as illustrated in Fig.~\ref{phase_diagram_RDharmonic}.

The case of independent particles in a harmonic trap, $\kappa=0$, was discussed in Sec.~\ref{sec:1particle_potential}. In particular the total density \eqref{eqRTPharmonic} has a finite support $[-v_0/\mu,v_0/\mu]$, and may either vanish or diverge at the edges, with an exponent $\gamma/\mu-1$ \cite{DKM19}. We also recall that in the Brownian case, with repulsive interactions $\kappa>0$, one finds a smooth density which interpolates between a uniform density with support $[-\kappa/\mu,\kappa/\mu]$ at $T=0$ and a Gaussian for $T\to+\infty$ (see Sec.~\ref{sec:RD_brownian} and \cite{PLDRankedDiffusion}).
\\

\noindent {\bf Analytical method.} Let us once again study the stationary densities $\rho_s(x)$ and $\rho_d(x)$ for $N\to+\infty$, in the purely active case $v_0>0$ and $T=0$. Explicitly solving the stationary equations for the rank fields is more difficult than for the linear potential, as the external potential now has a different form from the interaction potential. However, starting from the stationary version of \eqref{eqrank_attractive1}-\eqref{eqrank_attractive2} with $T=0$ and $V'(x)=\mu x$, we obtained a representation of the solution for $r(x)$ (for $x\geq0$, using that $r(x)$ is odd) as
\be \label{param_r_RDharmonic}
\frac{1}{2} - r = \int_{0}^{U(r)} \frac{du}{\sqrt{G(u)}} \quad , \quad \mu x = 2 \kappa r - U'(r) \;,
\ee
where, introducing $a=\gamma/\mu$ and $b=4a\kappa$,
\be \label{G_expression_RDharmonic}
G(u) = \begin{dcases} C u^{1/a} + \frac{b}{a-1}u + v_0^2 \quad \text{for } a \neq 1 \;, \\
C u + b u\ln u + v_0^2 \quad \quad \ \ \text{for } a = 1 \;, \end{dcases}
\ee
and the constant $C$ is fixed by the condition $G(U(0))=0$. The second rank field is obtained by $s=-U(r)/v_0$. If we can invert the first equation in \eqref{param_r_RDharmonic} to obtain an expression of $U(r)$, the second equation then gives an explicit relation between $x$ and $r(x)$. This can be done in the particular case $a=1/2$ (i.e., $\gamma=2\mu$), and we will give the results below. In the more general case, this representation can be used to study the properties of the densities $\rho_s$ and $\rho_d$, in particular their support and edge behavior. This allowed us to obtain the full phase diagram of the model as a function of the two dimensionless parameters $\kappa/v_0$ and $\mu/\gamma$, shown in Fig.~\ref{phase_diagram_RDharmonic}, which we now describe.
\\

\noindent {\bf Discussion of the phase diagram.} In the repulsive case, there are no shocks and the density is always smooth inside its support $[-x_e,x_e]$. The position of the edge is given by
\begin{equation} \label{xe_RDharmonic}
    x_e=\frac{v_0+\kappa}{\mu} \;.
\end{equation}
In {\it phase I}, corresponding to $\mu/\gamma<1$, and which also extends to the attractive case for $\mu/\gamma<1-\bar\kappa/v_0$, the density $\rho_s(x)$ vanishes at the edges with an exponent $\frac{\gamma}{\mu}-1$, similar to the non-interacting case $\kappa=0$. In fact, the densities $\rho_\pm(x)$ also vanish with the same exponents as in the absence of interactions (see \eqref{eqRTPharmonic_sign}). In {\it phase II}, corresponding to the repulsive case with $\mu>\gamma$, 
the density instead exhibits a jump at the edge, of magnitude
\begin{equation}
\rho_s(x_e)=\frac{\mu-\gamma}{2\kappa} \;.
\end{equation}
Thus, the repulsive interactions suppress the divergence which is present for $\kappa=0$. The density $\rho_+$ still vanishes at the left edge, and $\rho_-$ at the right edge, but they vanish linearly as $|x \pm x_e|$, instead of superlinearly in the non-interacting case. In addition, phase II can be divided into two regimes IIa and IIb, depending on the convexity of the density $\rho_s(x)$ inside the support. 
On the boundary between the regimes I and IIa, i.e., for $\gamma=\mu$, the density
vanishes at the edge as 
\be
\rho_s(x) \simeq \frac{\mu}{2\kappa} \frac{1}{\big(\ln (x_e-x)\big)^2} \;.
\ee
This inverse logarithmic behavior is really an effect of the repulsive interaction, since in the non-interacting case the density is uniform on $x \in [-v_0/\mu,v_0/\mu]$ for $\gamma=\mu$.

{\it Phase III} corresponds to the attractive case with $1-\mu/\gamma<\bar\kappa/v_0<2$. In this phase, the density is still smooth inside the support but exhibits delta peaks at the edges, corresponding to a cluster of $+$ particles at $x_e$ and of $-$ particles at $-x_e$. This is similar to the phase $F_e$ for the linear potential (and for $V'(x)=0$). However, an important difference is that here the support is {\it extended} by the presence of shocks. Indeed we find
\be 
x_e = \frac{v_0-\bar \kappa}{\mu} + \frac{\bar \kappa}{\mu} \Delta r  \;,
\ee 
where $\Delta r>0$ is the fraction of particles inside each cluster. This is because the force due to the harmonic potential increases with the distance to the origin. Indeed, the position of the edge cluster is determined by a balance between the driving velocity $v_0$ on the one hand, and the attractive interaction together with the confining potential on the other hand. The larger the cluster, the more the attraction felt by the edge particles is reduced, and thus $x_e$ increases so that this is compensated by the external force.

Finally, {\it phase IV} corresponds to the case $\bar \kappa>2v_0$, for which the noise cannot overcome the attraction between the particles and one trivially has $\rho_s(x)=\delta(x)$, i.e., the particles form a single stable cluster at $x=0$. 
\\

\begin{figure}[t]
    \centering
    \includegraphics[width=0.32\linewidth,trim={0 0 1cm 0},clip]{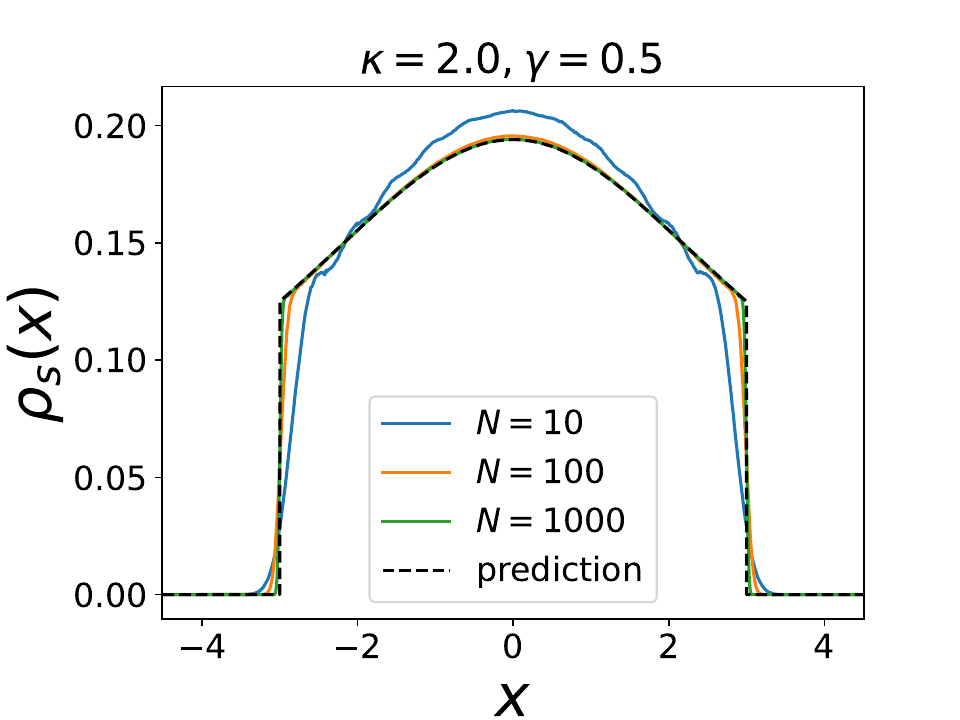}
    \includegraphics[width=0.32\linewidth,trim={0 0 1cm 0},clip,trim={0 0 1cm 0},clip]{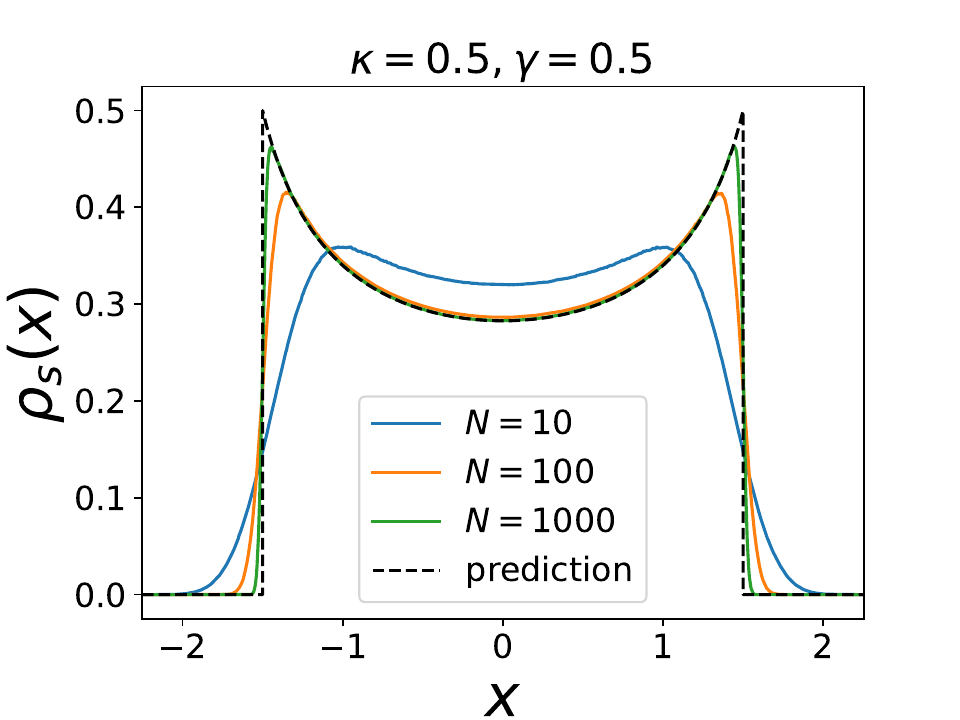}
    \includegraphics[width=0.32\linewidth,trim={0 0 1cm 0},clip]{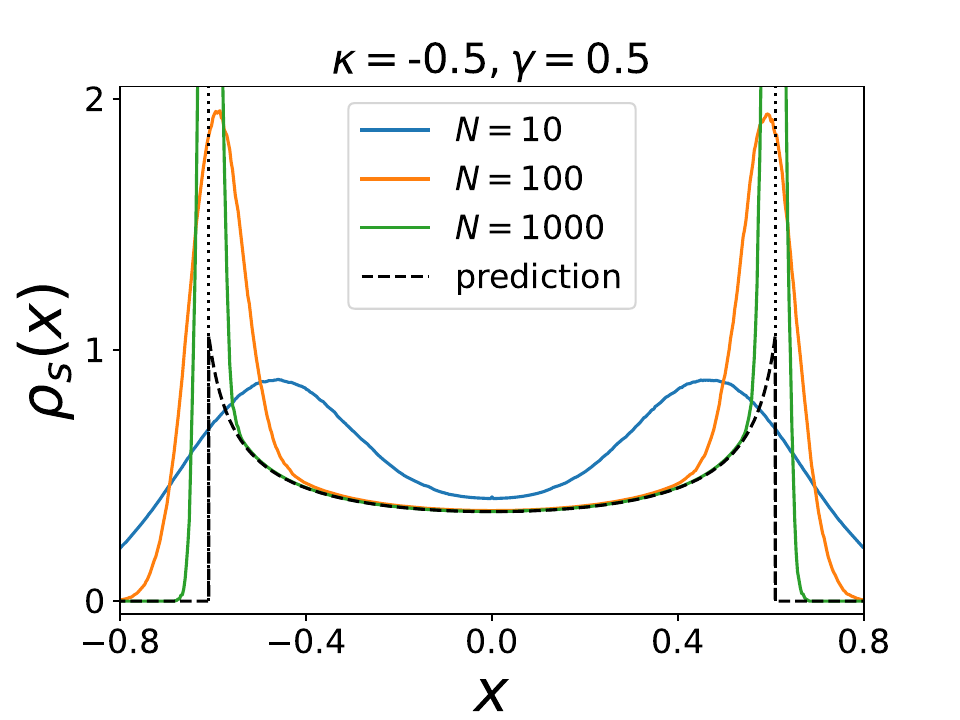}
    \caption{{\bf Top:} Comparison of the density $\rho_s(x)$ in the stationary state computed using numerical simulations for different values of $N$, with the analytical prediction for $\mu=2 \gamma$. In all cases $v_0=1$, $\mu=1$ and $\gamma=0.5$, and $\kappa$ varies to explore the 3 non-trivial regimes: $\kappa>v_0$, i.e., phase IIb (left), $0<\kappa<v_0$, i.e., phase IIa (center) and $-2v_0<\kappa<0$, i.e., phase III (right). The dashed black lines correspond to the predictions of Eqs.~\eqref{r_repulsive1_ahalf}, \eqref{r_repulsive2_ahalf} and \eqref{r_repulsive2_ahalf}-\eqref{eq_edge_attractive_RDharmonic} respectively. On the right panel the dotted vertical lines represent delta peaks in the density.}
    \label{fig_RDharmonic_ahalf}
\end{figure}

\noindent {\bf Results on the special line $\mu=2\gamma$.} In the special case $\mu/\gamma=2$ (indicated by a red line in Fig.~\ref{phase_diagram_RDharmonic}), we were able to obtain a more explicit parametric representation for $r(x)$.
For $\kappa>v_0$, i.e within phase IIa, the rank field $r(x)$ it reads
\be \label{r_repulsive1_ahalf}
\frac{\mu x}{v_0} \sinh(c/2)  = \tilde g_c(r) \quad , \quad \tilde g_c(r) :=\sinh (cr) + c \, r \cosh(\frac{c}{2}) \;,
\ee 
where $c\in[0,+\infty]$ is the solution of 
\be 
\frac{\tanh(c/2)}{c/2}= \frac{v_0}{\kappa} \;.
\ee 
Since $\tilde g_c(r)$ is an increasing function of $r$ for any value of $c$, the equation \eqref{r_repulsive1_ahalf} is invertible. Thus $r(x)$ is smooth, there are no shocks, and the edge is given by the general formula \eqref{xe_RDharmonic}. 

For $\kappa<v_0$, we find instead
\be \label{r_repulsive2_ahalf}
\frac{\mu x}{v_0} \sin(c/2)  = g_c(r) \quad , \quad g_c(r) :=\sin (cr) + c\, r \cos(\frac{c}{2}) \quad , \quad \frac{\tan(c/2)}{c/2}= \frac{v_0}{\kappa} \;.
\ee 
In the repulsive case, $\kappa>0$, i.e., in phase IIb, one has $c\in[0,\pi]$. In this case $g_c(r)$ is again an increasing function on $[0,1/2]$, so that this relation can be inverted, thus there is no shock and the edge is again given by \eqref{xe_RDharmonic}. In the attractive case, $\kappa<0$, i.e., in phase III, one has instead $c\in[\pi,2\pi]$ and $g_c(r)$ is no longer increasing on the whole interval $[0,1/2]$. This leads to a shock at the edge, i.e., a delta function in the density with weight $\frac{1}{2} - r(x_e^-)$, determined by the equation
\be \label{eq_edge_attractive_RDharmonic}
h_c(r(x_e^-)) = h_c\big(\frac{1}{2}\big) \quad , \quad h_c(r)=\sin (cr) + \frac{cr}{2} \cos \big(\frac{c}{2}\big) \;.
\ee
The position of the edge $x_e$ is then obtained by inserting the value of $r(x_e^-)$ in \eqref{r_repulsive2_ahalf}.

These analytical results are compared with numerical simulations in Fig.~\ref{fig_RDharmonic_ahalf}, showing once again a very good agreement for large $N$. Note that for finite $N$, the density is always smooth (the jumps and the delta peaks at the edges have a finite width which decreases with $N$, due in the second case to fluctuations in the size and position of the clusters).

\subsection{An example of non-reciprocal interaction} \label{sec:non_reciprocal}

We conclude this section by discussing a variation of the model \eqref{def_activeRD} where the interaction force between two particles $i$ and particle $j$ 
depends on their internal states, $\sigma_i(t)$ and $\sigma_j(t)$,
\be \label{langevin1nonreciprocal}
\frac{dx_i}{dt} = \frac{1}{N} \sum_{j=1}^N \kappa_{\sigma_i,\sigma_j}{\rm sgn}(x_i-x_j) - V'(x_i) + v_0 \sigma_i(t) 
\ee 
(where we have already set $T=0$ for simplicity). We consider the case of a non-reciprocal interaction, where the matrix $\kappa_{\sigma,\sigma'}$ is given by
\be 
\kappa_{-+} = - \kappa_{+-} = b  \quad , \quad \kappa_{++} = \kappa_{--} = 0 \;.
\ee 
Restricting to $b>0$ by symmetry, this means that the $+$ particles are attracted to the $-$ particles, while the $-$ particles are repelled by the $+$ particles (and two particles with the same sign do not interact together).

Non-reciprocal interactions, i.e., interactions which violate Newton's third law, are particularly relevant in the context of animal behavior \cite{nonreciprocal_active_book}, and have been studied experimentally in bacteria \cite{Kocabas2024}, but can also be observed at smaller scales \cite{Mandal2024}. Collective effects in systems with non-reciprocal interactions have attracted a lot of attention in recent years. Indeed such systems have been found to exhibit a variety of unusual phenomena, such as oscillating patterns (see, e.g., \cite{VitelliNature2021}). Recently the effects of non-reciprocal interactions of various forms in models of active particles (from simple forces to alignment interactions and quorum sensing) have been investigated, either numerically, or through the analysis of general field theories or hydrodynamic equations obtained by coarse-graining of a microscopic model \cite{You2020,nonreciprocalActive,KK2022,Knezevic2022,Dinelli2023,Duan2023,Duan2024,Du2024,Tucci2025,Newman2008,Peruani2016,Peruani2017,Durve2018,Negi2022,Qi2022,Stengele2022,Saavedra2024}. Non-reciprocal interactions often involve two or more distinct ``species'' of particles, meaning that the way in which a particle interacts with others is fixed once and for all by the model \cite{You2020,nonreciprocalActive,KK2022,Knezevic2022,Dinelli2023,Duan2023,Duan2024,Du2024,Tucci2025}. There are however other types of non-reciprocal interaction, which instead depend on the current ``state'' of the particle (e.g., $\sigma_i(t)$ in the present model). A common example, particularly in models of flocking, is the case of ``vision cones'', where particles only interact with other particles inside a certain angle with respect to their orientation \cite{Newman2008,Peruani2016,Peruani2017,Durve2018,Negi2022,Qi2022,Stengele2022,Saavedra2024}. The present model is closer to this second category, a choice which is motivated by the desire to keep the equations as analytically tractable as possible.
\\

\noindent {\bf Results.} For any $N \geq 2$, in the absence of confining potential, we find that the particles escape to infinity, i.e., there is no stationary bound state.
We thus studied the model in the presence of a linear external potential $V(x)=a |x| $, with $a>0$. Starting again from the Dean-Kawasaki equation \eqref{dean1}, one can re-derive the equations for the rank fields. In this case, it is simpler to first write them in terms of $h_\pm(x) = \int^x_{-\infty} dy \, \rho_\pm(y,t)$,
\bea
&& \partial_t h_+ =   ( - v_0  +  a \, {\rm sgn}(x)  +
2 b h_{-} ) h'_+  + \gamma h_{-} - \gamma h_{+} \;, \\
&& \partial_t h_- =   (  v_0  +  a \, {\rm sgn}(x)  
- 2 b h_{+} ) h'_-  + \gamma h_{+} - \gamma h_{-} \;,
\eea 
where as usual the prime denotes a spatial derivative. Taking the sum and difference, and using that $h_\pm(x) = \frac{1}{2} \left(r(x) \pm s(x) \right)$, we obtain
\bea 
&& \partial_t r = - v_0 s' + a \, {\rm sgn}(x) r' + b r s' - b s r' \;, \label{eqnr1} \\
&& \partial_t s = - v_0 r' + a \, {\rm sgn}(x) s' + b r r' - b s s' - 2 \gamma s  \;. \label{eqnr2}
\eea
From these equations, we were able to find a fully explicit solution for the densities in the stationary state.

\begin{figure}
    \centering
    \includegraphics[width=0.49\linewidth, trim={0.3cm 2.2cm 0.3cm 2.4cm}, clip]{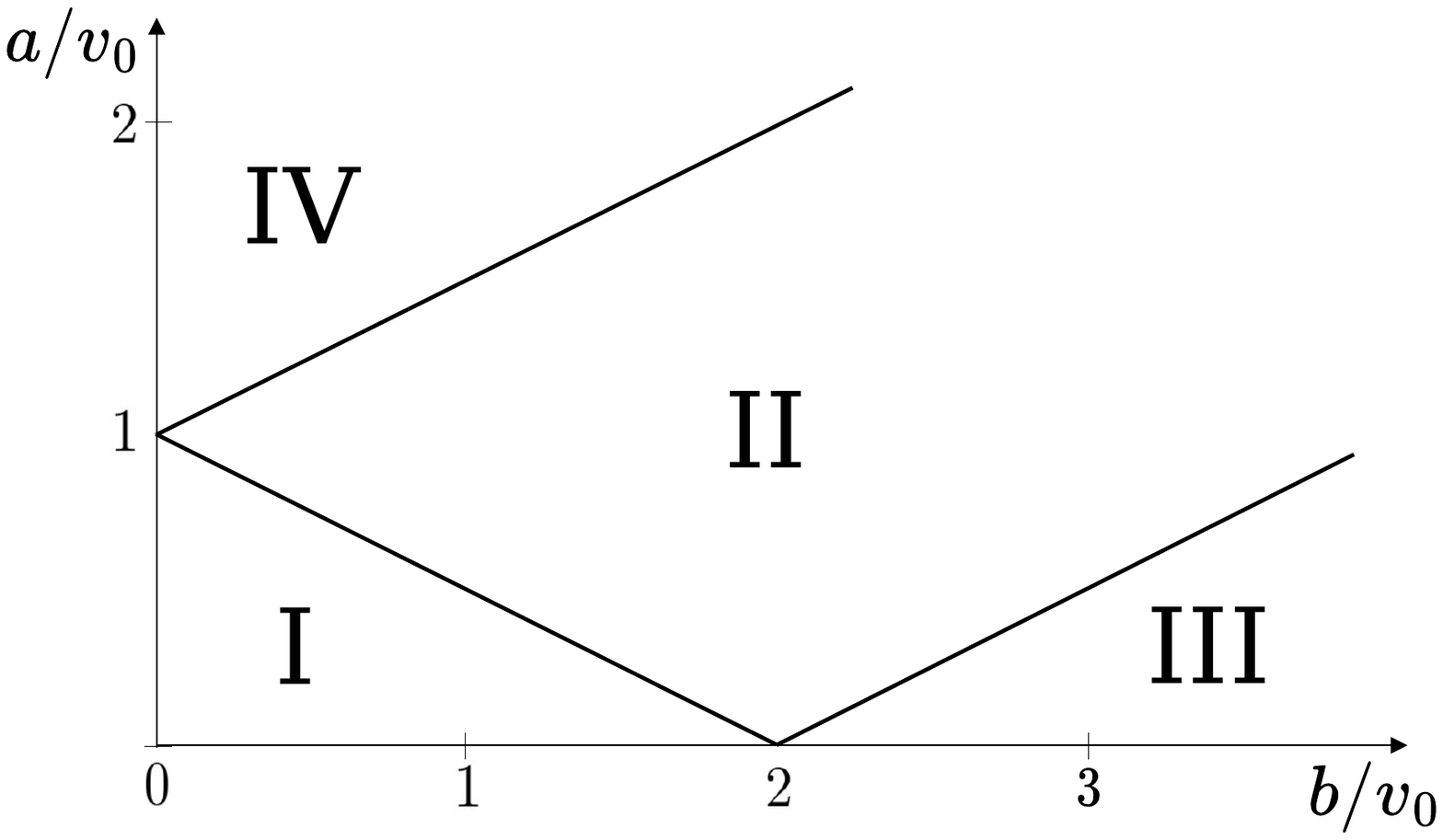}
    \includegraphics[width=0.49\linewidth, trim={0.4cm 2.2cm 0.4cm 2cm}, clip]{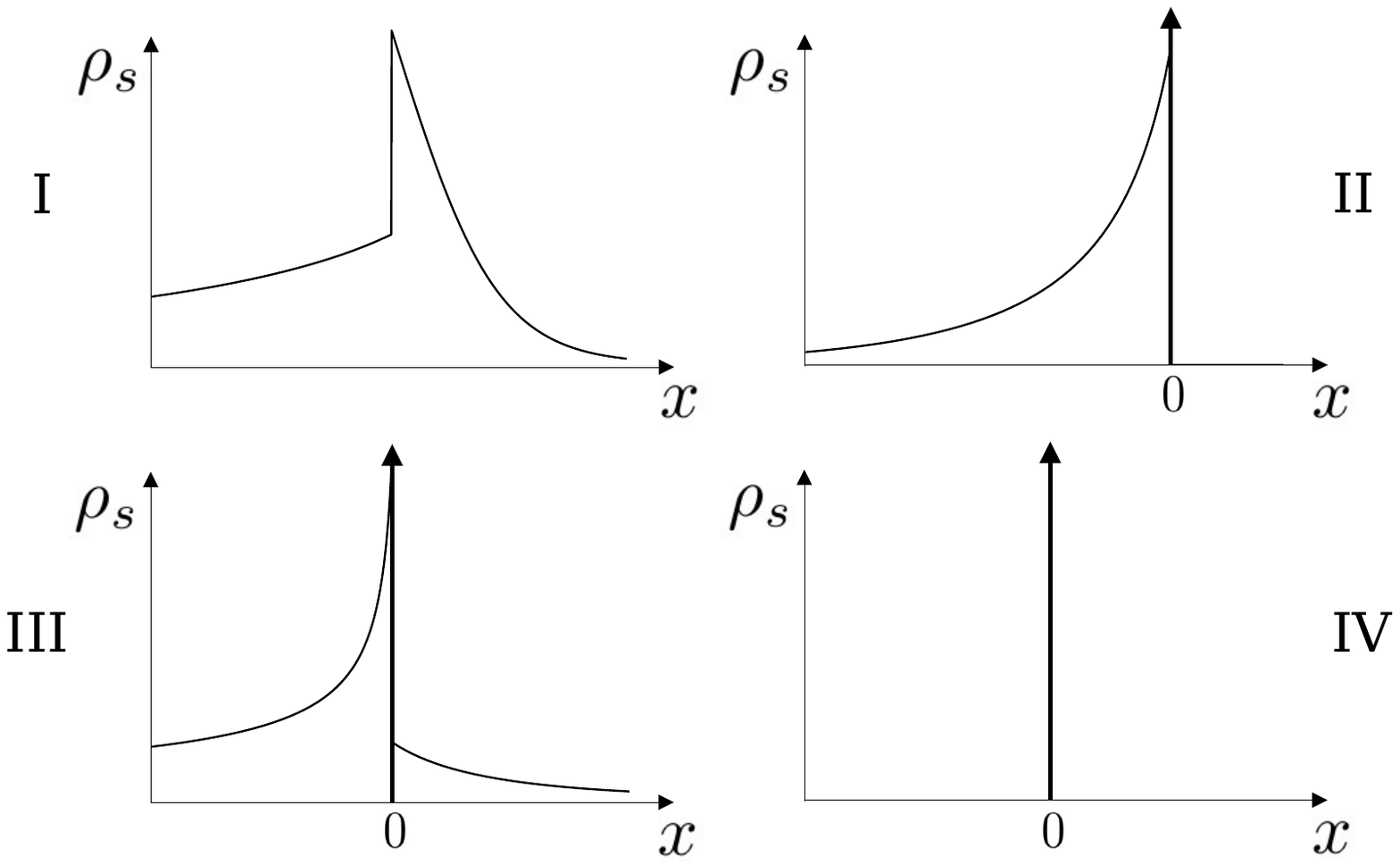}
    \caption{Left: Phase diagram of the non-reciprocal active rank diffusion. The phase diagram is
    symmetric upon $b \to - b$ and a parity transformation $x \to - x$. Right: Behavior of the total particle
    density in the four phases. The arrows denote delta function peaks in the density.}
    \label{phase_diagram_nonreciprocal}
\end{figure}

The first observation is that, as one could expect, the non-reciprocity breaks the left-right symmetry. In addition, we find that there are four different phases as we vary the dimensionless parameters $a/v_0$ and $b/v_0$, which we summarize in Fig.~\ref{phase_diagram_nonreciprocal}. Apart from {\it phase IV}, which covers the domain $a \geq v_0+b/2$ and corresponds to the trivial case where all the particles are stuck at $x=0$, leading to $\rho_s(x)=\delta(x)$, the other phases all exhibit an exponential decay of $\rho_s(x)$ as $|x|\to+\infty$ (with different rates on the right and on the left) and a discontinuity at $x=0$. However, they differ from each other through the existence or not of a delta peak at $x=0$ and through the presence or absence of particles on the half-line $x>0$.

In {\it phase I}, i.e., for $0<a<v_0-b/2$, the total density is supported on the whole real line and reads
\be \label{rhos_nr_intro}
\rho_s(x) = \begin{cases} A_+ \frac{2 v_0-b}{2 b} \left(1 - \frac{1}{1 + W\left(\frac{b}{2 v_0} e^{- A_+ x + \frac{b}{2v_0}}\right) } \right) \quad {\rm for} \ x>0  \\
A_- \frac{2 v_0+b}{2 b} \left(\frac{1}{1 + W\left(- \frac{b}{2 v_0} e^{A_- x - \frac{b}{2v_0}}\right) } -1 \right) \quad {\rm for} \ x<0  \end{cases} \hspace{-0.2cm}, \quad A_\pm=\frac{8 a \gamma}{(2 v_0\mp b)^2- 4 a^2} \;,
\ee
where $W$ is the Lambert function, i.e., the real (and first) root of $W(z) e^{W(z)} = z$. Since $W(z) \simeq z$ at small $z$, $\rho_s(x)$ decays exponentially at large $|x|$ with rates $A_\pm$. In the case $b>0$ which we consider here, one has $A_-<A_+$, i.e., the decay is slower on the negative side. The density exhibits a discontinuity at $x=0$, with $\rho_s(0^-)<\rho_s(0^+)$.

In {\it phase II}, i.e., for $v_0-b/2 < a < v_0+b/2$, we find that the density $\rho_s(x)$ vanishes for $x>0$. In addition, it has a delta peak at $x=0$ with weight
\be \label{weight2nr}
\frac{1}{2}-r(0^-)= \frac{2 a}{2 a + b + 2 v_0} \;,
\ee 
which contains only $+$ particles. For $x<0$ we find
\be \label{rhos_nr_neg2_intro}
\rho_s(x) = A_- \frac{2 v_0+b}{2 b} \left(\frac{1}{1 + W\left(- \frac{2b}{2a+b+2 v_0} e^{A_- x - \frac{2b}{2a+b+2v_0}}\right) } -1 \right) \;,
\ee
where $A_-$ is still given by \eqref{rhos_nr_intro}. 

Finally, in phase III, i.e., for $a < v_0-b/2$, the density $\rho_s(x)$ becomes once again supported by the whole real line. It also exhibits a delta peak at $x=0$, again containing only $+$ particles, with weight
\be \label{weight3nr} 
r(0^+)-r(0^-)=\frac{2 a \left(b^2-2 a b-4v_0^2\right)}{b \left(b^2-4a^2-4 v_0^2\right)} \;.
\ee
For $x \neq 0$ the density is given by
\be \label{rhos_shock_intro}
\rho_s(x) = \begin{dcases} A_+ \frac{b-2 v_0}{2 b} \left(\frac{1}{1 + W\left(-B_+ e^{- A_+ x - B_+}\right) } - 1 \right) \quad , \quad x>0 \\
A_- \frac{2 v_0+b}{2 b} \left(\frac{1}{1 + W\left(- B_- e^{A_- x -B_-}\right) } -1 \right) \quad , \quad x<0  \end{dcases} \hspace{-0.2cm}, \quad B_\pm = \frac{(b\pm 2v_0)(b-2a\mp2v_0)}{b^2-4a^2-4v_0^2} \;, 
\ee
where $A_\pm$ are the same as in \eqref{rhos_nr_intro}.

Note that $A_+$ diverges as we approach phase II both from phase I and phase III. However, when approaching from phase III the fraction of particles on the side $x>0$ vanishes continuously as we approach the line, i.e., $\frac{1}{2}-r(0^+)\to 0$, and the weight of the delta peak in the two phases, given by \eqref{weight2nr} and \eqref{weight3nr} respectively, match on the frontier between phase III and II (one finds $r(0^+)-r(0^-)=\frac{1}{2} - \frac{v_0}{b}$ in both cases). This is not the case when approaching from phase I: in this case, all the $+$ particles on the side $x>0$ suddenly form a cluster at $x=0$, leading to a non-zero weight \eqref{weight2nr} even at the border between the two phases.
\\

\begin{figure}[t]
    \centering
    \includegraphics[width=0.32\linewidth,trim={0 0 1cm 0},clip]{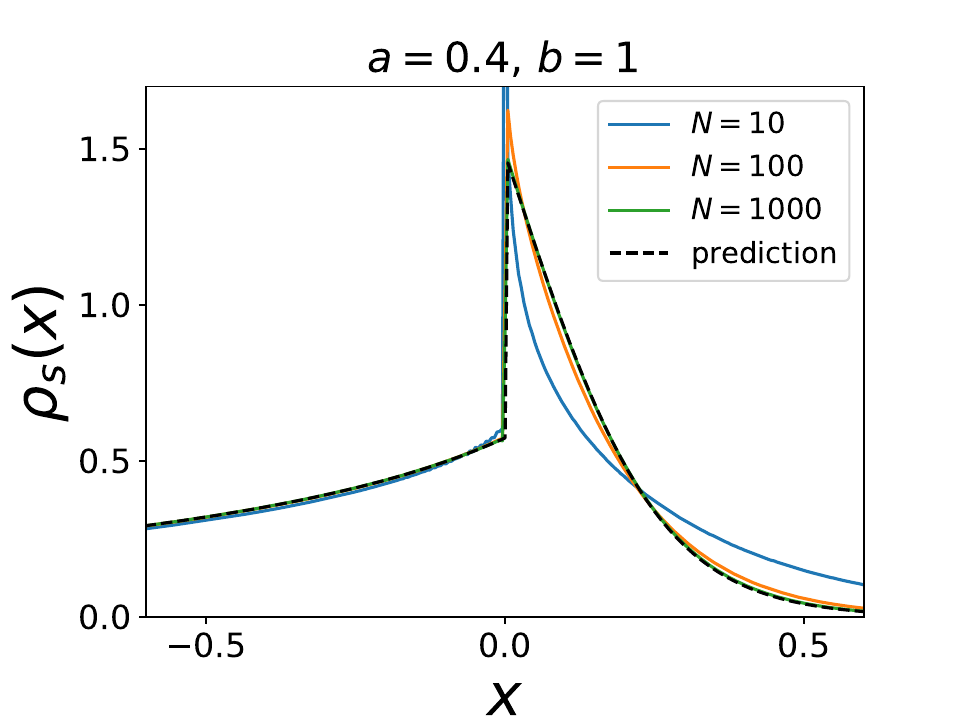}
    \includegraphics[width=0.32\linewidth,trim={0 0 1cm 0},clip]{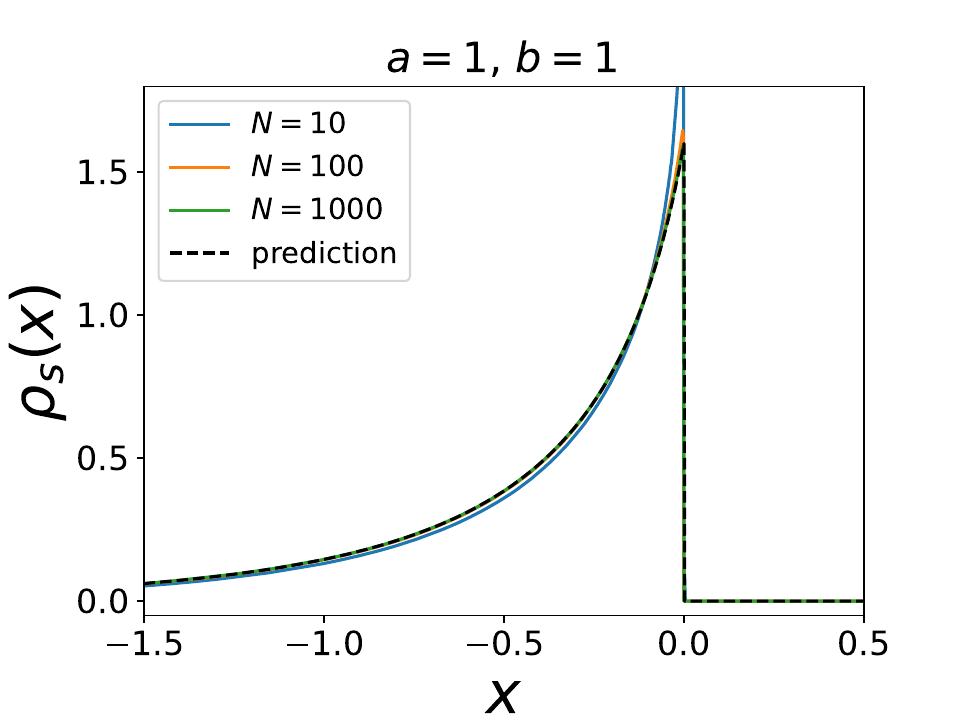}
    \includegraphics[width=0.32\linewidth,trim={0 0 1cm 0},clip]{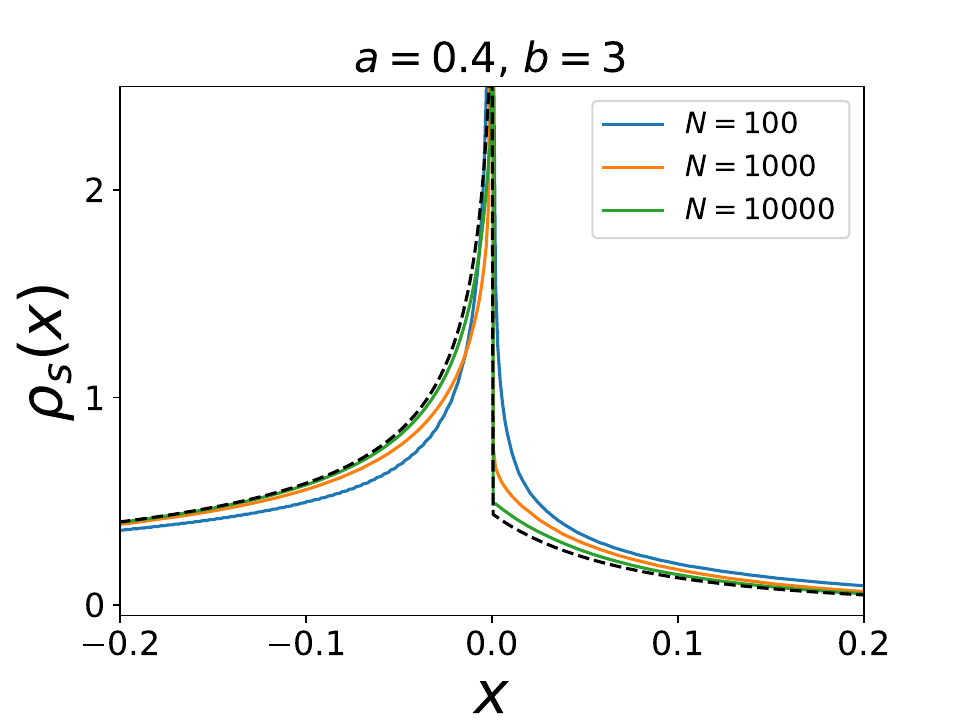}
    \caption{Total density $\rho_s(x)$ in the model with non-reciprocal interactions \eqref{langevin1nonreciprocal} obtained from simulations with $\gamma=1$ and $v_0=1$, for increasing values of $N$. The delta peaks are not shown on the densities for the phases II and III. {\bf Left:} $a=0.4$ and $b=1$ (phase I). The dashed black line corresponds to the prediction for infinite $N$ \eqref{rhos_nr_intro}. Note the discontinuity of the density at $x=0$, with
    however no shock (no delta peak). {\bf Center:} $a=1$ and $b=1$ (phase II). The dashed black line corresponds to \eqref{rhos_nr_neg2_intro}. The density is zero for $x>0$ for any $N$ (with a delta peak at $x=0$). {\bf Right:} $a=0.4$ and $b=3$ (phase III) the dashed black line corresponds to \eqref{rhos_shock_intro}. Larger values of $N$ are used here since the convergence in $N$ is slower than in the other phases.}
    \label{plots_nonreciprocal}
\end{figure}

\noindent {\bf Discussion.} The results above can be interpreted by considering the total force felt by the particles. In phase I, since $v_0>a+b/2$, all the $+$ particles move towards the right and the $-$ particles towards the left, until they tumble. Let us consider the rightmost particle. Assuming that it is a $+$ particle, it is subjected to a total force $v_0-a-b/2$. When this force becomes negative, all the $+$ particles are attracted towards $x=0$. This creates a cluster of $+$ particles at $x=0$ and marks the transition from phase I to phase II. In this phase, the $-$ particles still move towards the left, thus no particle can reach the side $x>0$. Let us now consider the leftmost particle. Assuming that it is a $-$ particle, it is subjected to a total force  $-v_0+a-b/2$. When this becomes positive, all particles are attracted towards $x=0$ indifferently of their sign and remain there, leading to phase IV. The behavior of the particles in phase III is less intuitive. In this phase, all the $+$ particles are attracted towards $x=0$, while the $-$ particles are repelled away from $x=0$ on both sides. When a $+$ particle from the cluster at $x=0$ tumbles and becomes a $-$ particle, the side towards which it is directed is determined by the fluctuations. This phase is thus more sensitive to the fluctuations than the other phases.

In Fig.~\ref{plots_nonreciprocal} we once again compare our analytical predictions for $\rho_s(x)$ to numerical results for finite $N$. For this model, the 4 phases are already present at finite $N$, with the same qualitative features (presence of a true delta peak at $x=0$, absence or presence of particles on the right...). We once again observe a good quantitative agreement at large $N$, although the convergence is slower for phase IV, confirming the peculiar role played by the fluctuations in this phase.

Let us finally add that, in the appendix of \cite{activeRD2}, we also introduced another model with non-reciprocal 1D Coulomb interactions, inspired from the concept of ``vision cones'', each particle only receives a force from the particles ``in front" of it (i.e., on the right for $+$ particles and on the left for $-$ particles). It turns out that this model can be mapped to the reciprocal active rank diffusion model studied in the rest of this chapter, which allows to easily obtain the full phase diagram.

\section{Conclusion}

In this chapter, we studied an active version of the ranked diffusion, i.e., a model of RTPs interacting via a 1D Coulomb potential. Thanks to the particular form of the interaction, we were able to rewrite the Dean-Kawasaki equations into a local form, which allowed us to compute exactly the density of particles at large times, in the limit $N\to+\infty$. For an attractive interaction, this allowed us to uncover a new type of phase transition, where the support of the stationary density becomes finite and clusters of particles with the same orientation form at the two edges. This collective effect is different from the ones discussed in Chapter~\ref{chap:interactions}, in particular MIPS as it involves attractive interactions. Nevertheless, this effect is completely absent in the Brownian version of the model, and it is thus a truly non-equilibrium phase transition. Strictly speaking, the delta peaks in the density are an effect of both the fixed amplitude of the noise and the singularity of the interaction potential at $x=0$. However, we expect the transition to survive in the presence of additional thermal noise, or if we regularize the interaction potential (e.g., ${\rm sgn}(x) \to \tanh(x/\ell)$), with sharp peaks in the density instead of delta functions. We were then able to generalize these results to different types of confining potentials, leading to very rich phase diagrams for both attractive and repulsive interactions. It would be interesting to investigate this models further, for instance by using the DK equation to study the finite $N$ fluctuations.

In the last part of this chapter, we introduced a variant of this model with non-reciprocal interactions, for which we were also able to compute exactly the stationary density in the large $N$ limit. This is however only one of the possible models that one can think of, which was chosen for its simplicity, and it leaves much room for future directions. Beside studying the effect of other confining potentials or the large time behavior in the absence of confinement, one could for instance add a reciprocal part to the interaction, to see how a small amount of non-reciprocity affects the transitions discussed in the rest of this chapter. Finally, it would be interesting to extend the present model to non-reciprocal interactions involving two different species of particles. This would probably lead to very different and interesting behaviors.


\chapter{Active Dyson Brownian motion}
\label{chap:ADBM_Dean}

\section{Definition of the model}

In this chapter we consider two active extensions of the Dyson Brownian motion introduced in Sec.~\ref{sec:DBM_brownian}. They are defined through the following equation of motion for the position of the particles $x_i(t)$ ($i=1,...,N$),
\be \label{def_ADBM} 
 \frac{dx_i}{dt} = - \lambda x_i +  \frac{2}{N} \sum_{j (\neq i)} 
\frac{g_{\sigma_i, \sigma_j}}{x_i-x_j} + v_0 \sigma_i(t) + \sqrt{\frac{2 T}{N}} \,\xi_i(t) \;,
\ee 
where the $\sigma_i(t)$ are independent telegraphic noises with tumbling rate $\gamma$ and the $\xi_i(t)$ are $N$ independent unit Gaussian white noises. As in Sec.~\ref{sec:DBM_brownian} for the standard DBM, we scale the temperature as $1/N$, although in this chapter we will focus on the purely active case where $T=0$. In both versions of the model, the particles are confined inside a harmonic potential $V(x)=\lambda x^2/2$. The two variants differ by the form of the interaction. As in the standard DBM, we consider a pairwise repulsive logarithmic potential $W_{\sigma,\sigma'}(x)=-2g_{\sigma,\sigma'} \log|x|$, but with an interaction strength $g_{\sigma,\sigma'} \geq 0$ which may depend on the states $\sigma_i$ of the two particles. More precisely, we define
\begin{eqnarray} \label{def_model12}
g_{\sigma,\sigma'}=
\begin{cases}
g \, \delta_{\sigma,\sigma'} \quad \rm{(model \; I)} \;,\\ 
g \hspace{1.2cm}  \rm{(model \; II)} \, .
\end{cases}
\end{eqnarray}
This means that in model I, a given particle only interacts with particles of the same sign, while in model II all particles interact together. Model II is the most natural extension of the DBM. In the other chapters of this thesis, when we mention the active DBM without further precision we will always be referring to model II. However, in model II the diverging interaction force prevents particles from passing each other. As we already discussed in Chapter~\ref{chap:DeanRTP}, and as we will illustrate more concretely in Sec.~\ref{sec:Dean_ADBM}, the Dean-Kawasaki equation thus does not provide a correct description of the particle densities in this model. By contrast, in model I, particles with opposite velocities do not interact with each other and are thus allowed to cross. This avoids the formation of clusters, which break the hydrodynamic description in model II, and thus we will see that the DK equation can be used in this case.
\\

\noindent {\bf Dimensionless parameters and interesting limits.} For both models, there are two dimensionless parameters (in the case $T=0$ which we consider):
\be
\frac{v_0}{\sqrt{g\lambda}} \quad \text{and} \quad \frac{\gamma}{\lambda} \;.
\ee
The aim of this chapter is to characterize the behavior of the particle densities defined in \eqref{def_rho_pm}-\eqref{def_rho_sd} in the two models, depending on these two parameters. There are 4 interesting limits that we will consider throughout this chapter:

(i) In the {\it weakly interacting limit} $g\to0^+$ (or $v_0/\sqrt{g\lambda}\gg 1$), we naively expect (although we will see that this is wrong for model II) to recover the case of independent RTPs in a harmonic trap, discussed in Sec.~\ref{sec:1particle_potential}, where the densities $\rho_s(x)$ and $\rho_d(x)$ have a finite support $[-v_0/\lambda,v_0/\lambda]$ and take the form \cite{DKM19} 
\be \label{eqRTPharmonicADBM}
\rho_s(x) = A \left[1-\left(\frac{\lambda x}{v_0}\right)^2\right]^{\frac{\gamma}{\lambda}-1} \quad , \quad \rho_d(x) = \frac{\lambda x}{v_0} \rho_s(x)
\ee
with $A$ a constant. 

(ii) In the {\it diffusive limit}, $v_0,\gamma\to+\infty$ with $T_{\rm eff}=\frac{v_0^2}{2\gamma}$ fixed (taking $\gamma\to+\infty$ with fixed $v_0$ amounts to taking $T_{\rm eff} \to 0$), we expect to recover the standard Dyson Brownian motion introduced in Sec.~\ref{sec:DBM_brownian} (at least for model II), where at small temperature the density takes the form of the Wigner semi-circle 
\be \label{WignerSC_ADBM}
\rho_{s}(x) = \frac{\lambda}{2\pi g} \sqrt{\frac{4g}{\lambda}-x^2} \;.
\ee

(iii) In the {\it weak noise limit} $v_0\to0^+$ (or $v_0/\sqrt{g\lambda}\ll 1$) we also expect to find the semi-circle density, as discussed in Sec.~\ref{sec:DBM_brownian}.

(iv) Finally, we will also consider the {\it strong persistence limit} $\gamma\to 0^+$ (i.e., $\gamma\ll\lambda$) where the particle states $\sigma_i$ vary very slowly.
\\

\begin{figure}
    \centering
    \includegraphics[width=0.5\linewidth, trim={0 5.9cm 0 2.5cm}, clip]{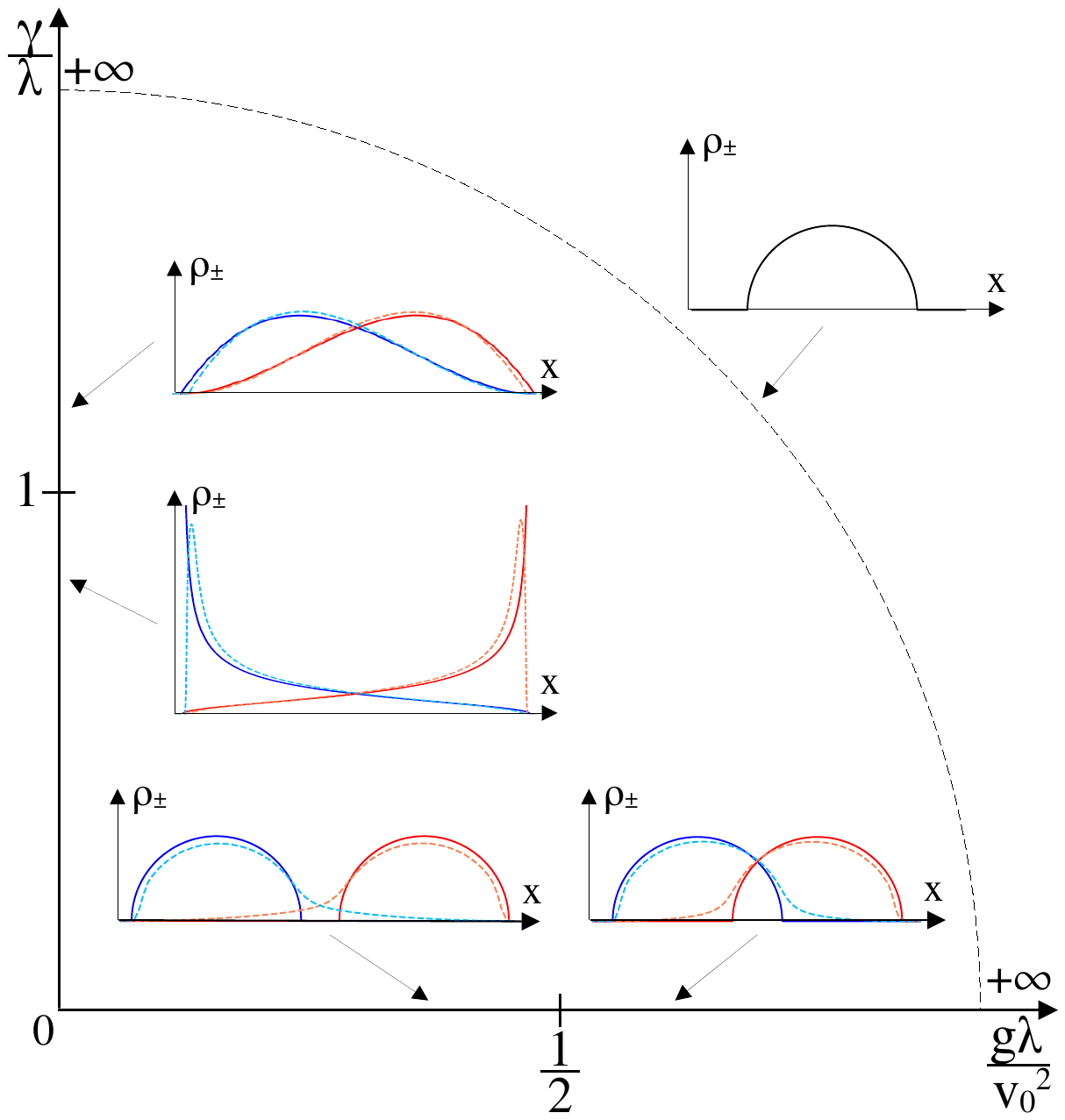}
    \caption{Shape of the particle density in model I (where the $+$ and $-$ particles do not interact together) in the plane $(g\lambda/v_0^2,\gamma/\lambda)$ in different limits. The density $\rho_+(x)$ is plotted in red and $\rho_-(x)$ in blue. When the two coincide, they are plotted in black. The light dashed curves represent the density slightly away from the limit considered. The dashed circular line in the diagram symbolizes infinity. The diffusive limit, which requires a specific scaling between $v_0$ and $\gamma$, is not shown here.}
    \label{phase_diagram_model1}
\end{figure}

\noindent {\bf Overview of the results.} 
Below, we discuss the application of the Dean-Kawasaki equation for RTPs derived in Chapter~\ref{chap:DeanRTP} to the two versions of the active DBM defined in \eqref{def_ADBM}-\eqref{def_model12}. As already announced, this approach works for model I while it fails for model II due to the non-crossing constraint, and we will propose several arguments as to why this is the case. The last two sections of this chapter focus on the large $N$ limit of the two models.

For model I, we will analyze the noiseless DK equation, using the resolvent method introduced in Sec.~\ref{sec:DBM_brownian}. Although we are not able to obtain a fully explicit solution as for the active rank diffusion, this allows us to analyze the different limits of the model and to obtain several interesting results, such as a recursion relation for the moments and a small $\gamma$ expansion of the densities. We find that in the diffusive limit we indeed recover the standard DBM (although with $g\to g/2$ since each particle only interacts with half of the other particles at any given time). However, in general the density deviates from the Wigner semi-circle, and in particular we recover the non-interacting case \eqref{eqRTPharmonicADBM} as $g\to 0^+$, while as $\gamma \to 0^+$, the $+$ and $-$ separate to form two semi-circles. The total density still vanishes with an exponent $1/2$ at the edges of the support for any parameters, but for $\rho_+(x)$ we find instead an exponent $3/2$ on the left edge (resp. $\rho_-(x)$ on the right edge).

For model II, the density is harder to study directly due to the lack of a hydrodynamic equation. The results of the last section are thus mostly based on numerical simulations. In Chapter~\ref{chap:ADBMfluct} we will however obtain exact results for the microscopic fluctuations in this model, which will provide further arguments for the observations presented here. Quite surprisingly, we find that the density is given by the Wigner semi-circle for any parameters, unless $v_0/\sqrt{g\lambda}$ scales as $\sqrt{N}$. When $v_0/\sqrt{g\lambda}\gg\sqrt{N}$, corresponding to the weakly interacting limit $g\to 0^+$, we do not recover the non-interacting case as one could have expected. This is again due to the non-crossing constraint, which still holds in this limit, and which leads to the formation of {\it true point-like clusters} of particles, since the interaction in this limit effectively becomes a contact interaction. We present an algorithm which allows to simulate this limit efficiently, which we use to obtain the stationary density and the distribution of cluster sizes.

\begin{figure}
    \centering
    \includegraphics[width=\linewidth, trim={0 2.5cm 0 8cm}, clip]{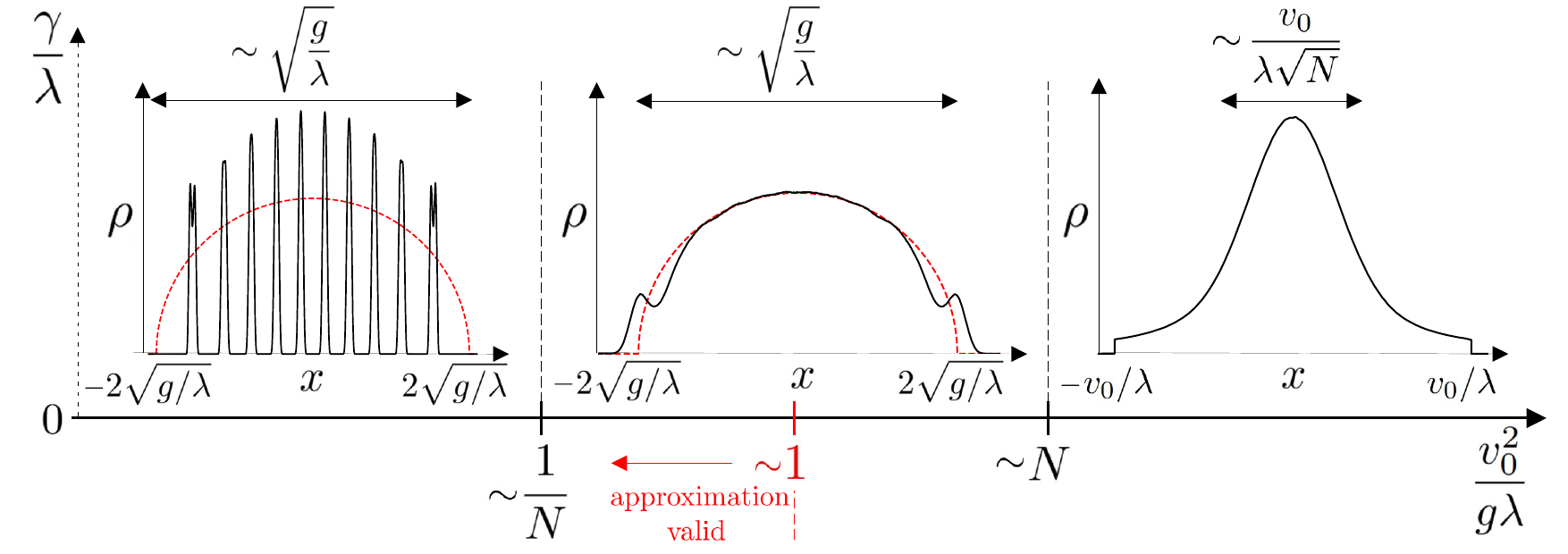}
    \caption{Different regimes of the total density $\rho_s(x)$ at large $N$ (but finite) in model II, as a function of the parameter $v_0^2/g\lambda$. The dashed red line shows the limit $N\to+\infty$ in the case $v_0^2/g\lambda\ll N$, given by the Wigner semi-circle. The spatial extension of the density as a function of the model parameters is also shown in the different regimes. The line $v_0^2/(g\lambda)\sim 1$ corresponds to the limit of validity of the approximation used in Chapter~\ref{chap:ADBMfluct}.}
    \label{phase_diagram_model2}
\end{figure}

\begin{figure}[t]
    \centering
    \includegraphics[width=0.32\linewidth,trim={0 0 1cm 1cm},clip]{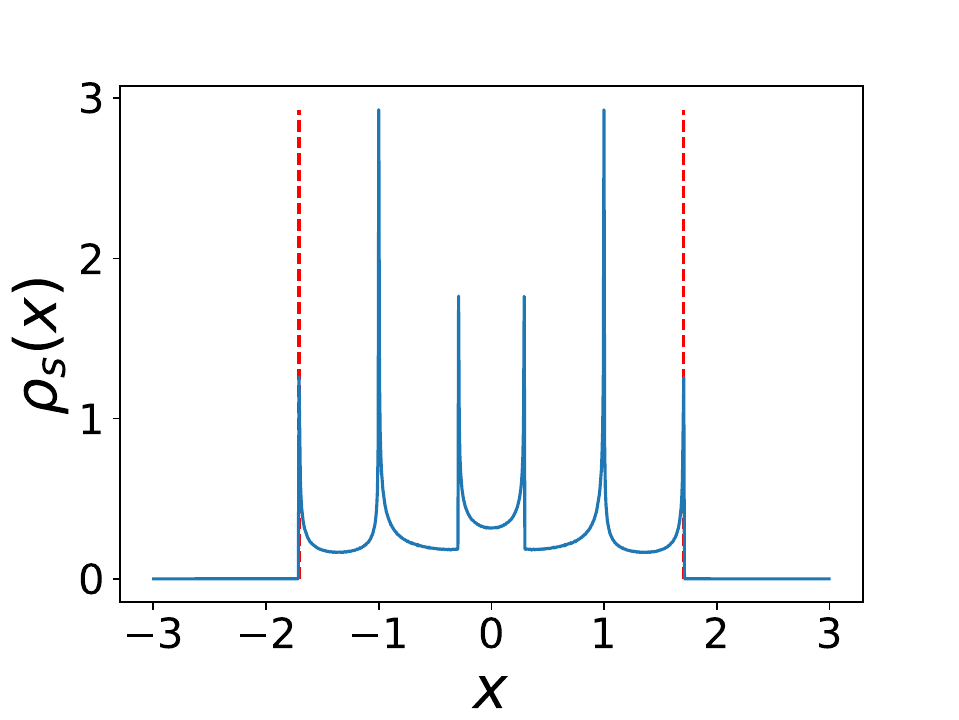}
    \includegraphics[width=0.32\linewidth,trim={0 0 1cm 1cm},clip]{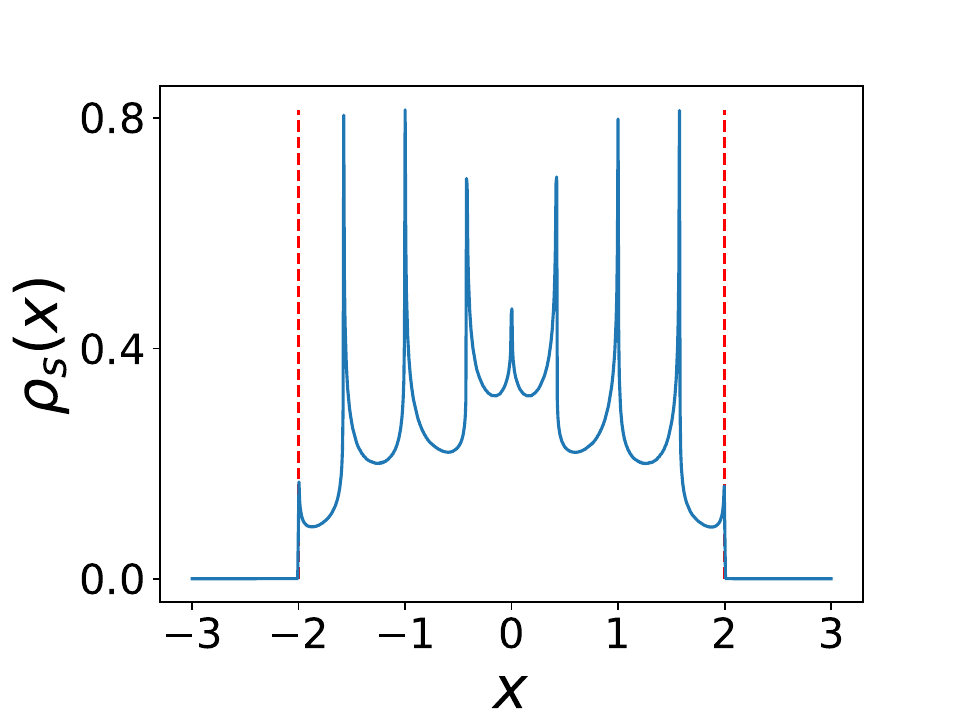}
    \includegraphics[width=0.32\linewidth,trim={0 0 1cm 1cm},clip]{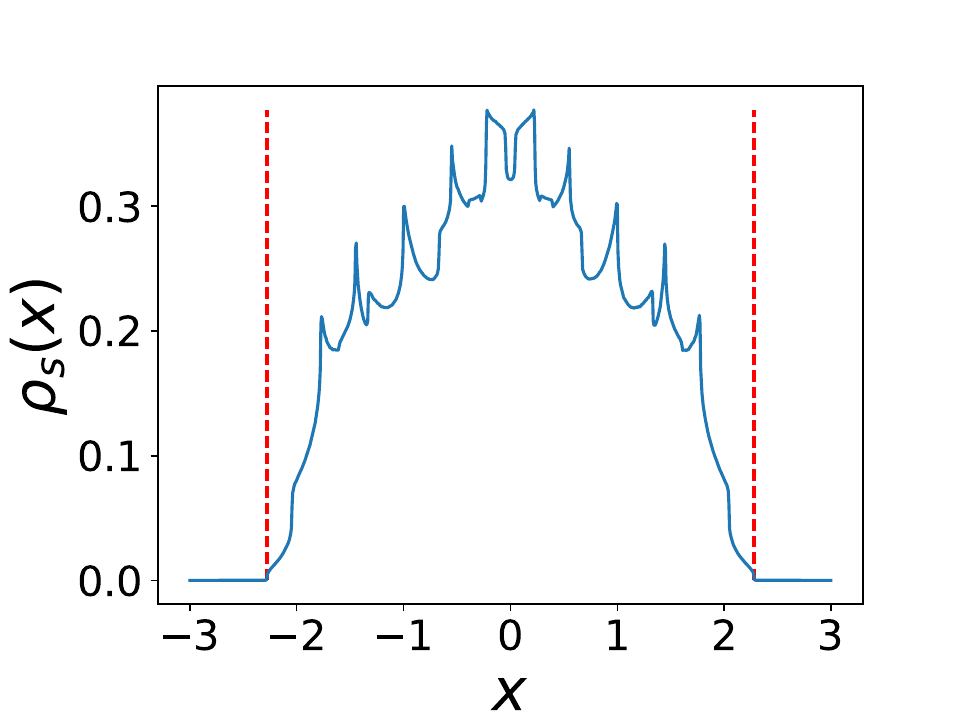}
    \includegraphics[width=0.32\linewidth,trim={0 0 1cm 1cm},clip]{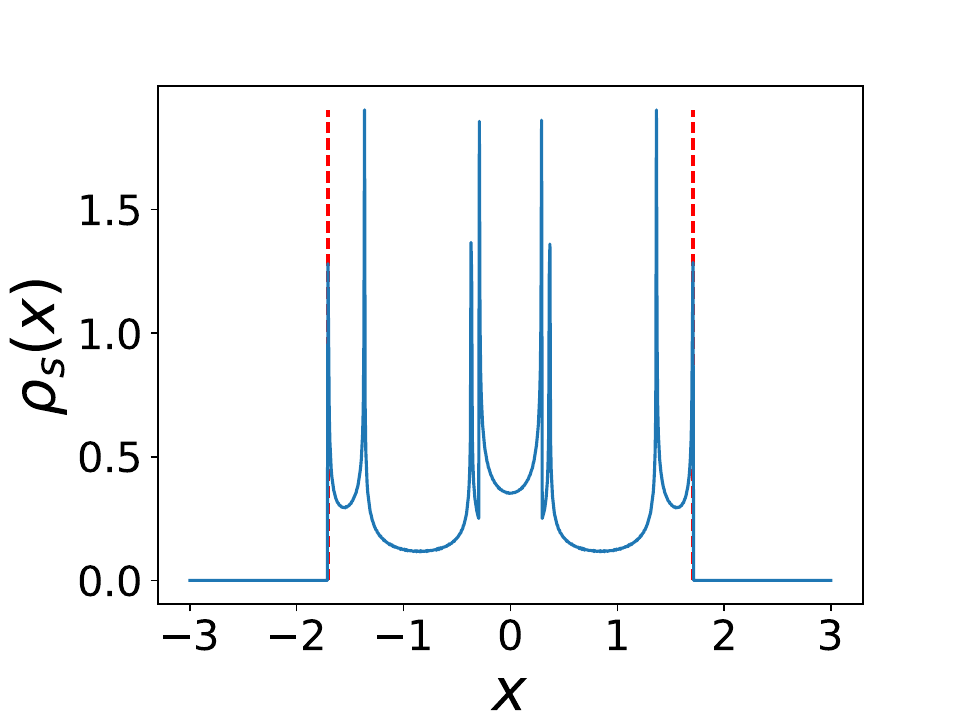}
    \includegraphics[width=0.32\linewidth,trim={0 0 1cm 1cm},clip]{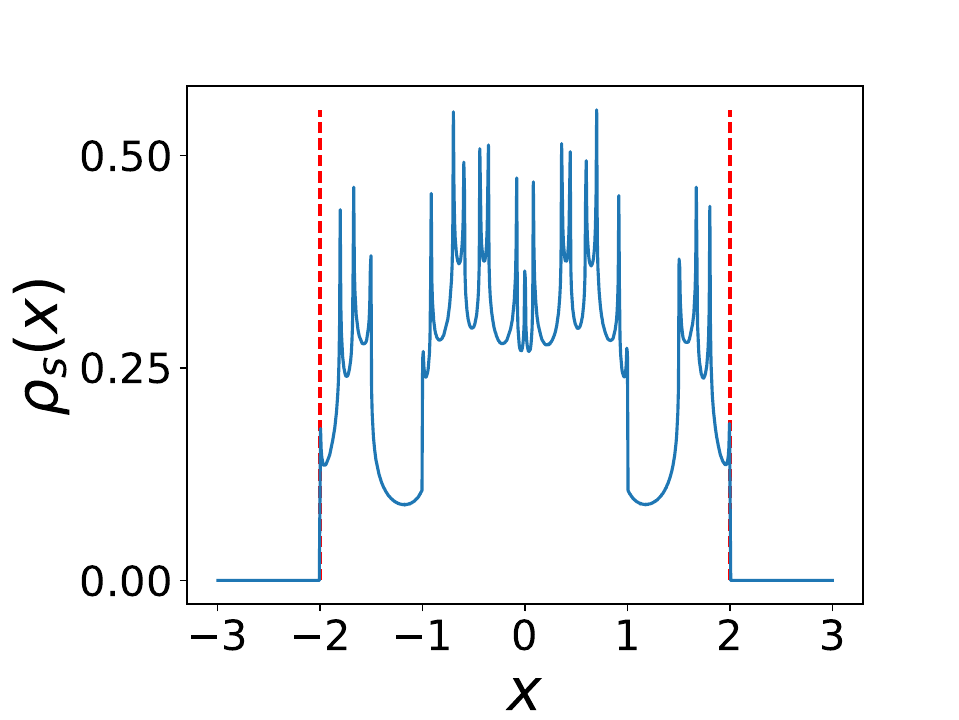}
    \includegraphics[width=0.32\linewidth,trim={0 0 1cm 1cm},clip]{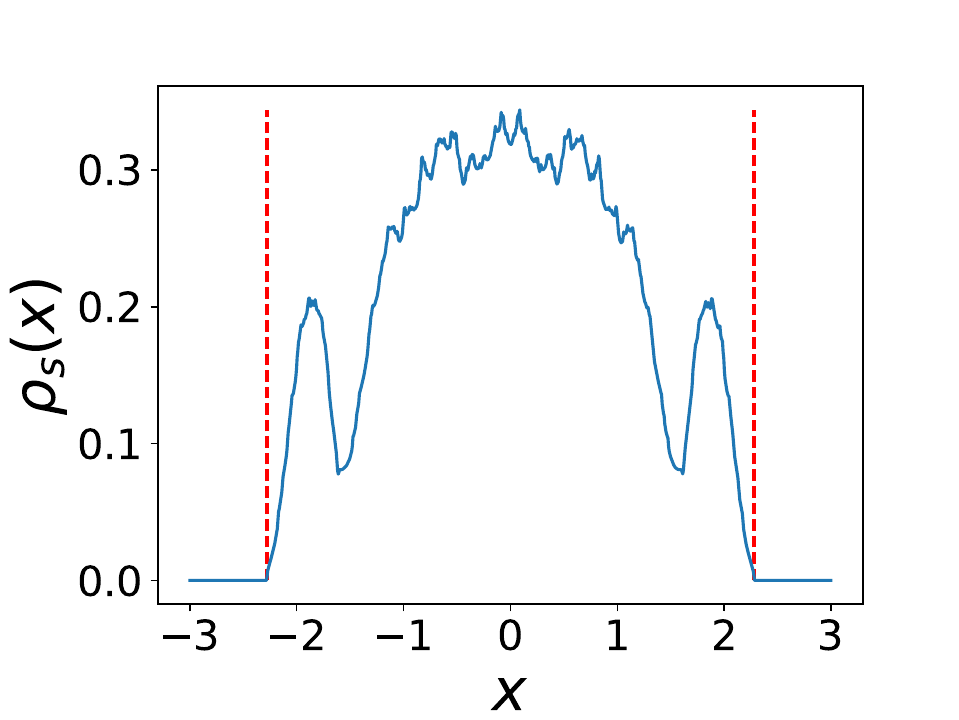}
    \caption{Total particle density $\rho_s(x)$ in model I (top) and II (bottom) for $N=2$, $3$ and $5$. The other parameters are $\lambda=1$, $g=1$, $v_0=1$ and $\gamma=0.25$. When $N\gamma/\lambda \leq 1$, we observe singularities in the density. The red lines show the predicted edges of the support.
    }
    \label{finiteN_ADBM}
\end{figure}

The different limiting behaviors of model I are summarized in Fig.~\ref{phase_diagram_model1}, while the regimes of model II are represented in Fig.~\ref{phase_diagram_model2} (the regime $v_0/\sqrt{g\lambda}\ll\sqrt{N}$ where the  particles are very localized but the limit $N\to+\infty$ is still given by the semi-circle will be discussed later). Let us add that some properties of the two models for finite $N$ were also studied in \cite{ADBM1}. In particular, we found that for small $N$ and small $\gamma$, the density exhibits singularities with an exponent $N\gamma/\lambda-1$, located at every ``fixed points'' of the dynamics, i.e., the equilibrium positions reached by the particles if we fix all the $\sigma_i$'s for an infinitely long time (see Fig.~\ref{finiteN_ADBM}). As $N$ increases, these singularities disappear leading to a smooth density for large $N$. The exponent $N\gamma/\lambda-1$ was obtained via a heuristic argument and verified numerically. For more details on these singularities, as well as on the fixed point and the support of the density for finite $N$, see \cite{ADBM1} (and in particular Sec.~II of the SM).

The results of this chapter are based on a combination of analytical computations and numerical simulations. As in the previous chapter, the numerical results are obtained through direct simulations of the Langevin dynamics \eqref{def_ADBM} of the two models. The densities are obtained by averaging over a large time window in the stationary state. For more details on the numerical simulations, see Appendix~\ref{app:simu}.

\section{Dean-Kawasaki equation for the active DBM} \label{sec:Dean_ADBM}

\subsection{Validity of the DK equation for models I and II}

\noindent {\bf Self-interaction term.} When we derived the Dean-Kawasaki equation in Sec.~\ref{sec:DKder}, we assumed that $W_{\sigma,\sigma'}'(0)=0$. This is obviously not the case for the active DBM, where $W_{\sigma,\sigma'}'(x)=-\frac{2g_{\sigma,\sigma'}}{x}$ diverges at $x=0$. When introducing the density in the step between \eqref{Dean_discret} and \eqref{integralDean0}, this leads to a diverging integral, as well as to the introduction of an artificial self-interaction term which should be removed. In the case of model I ($g_{\sigma,\sigma'}=g\delta_{\sigma,\sigma'}$), this issue can however be resolved, using the same approach as in \cite{RogersShi} for the Brownian case. We start by rewriting the interaction term in \eqref{Dean_discret} in a symmetric way,
\bea \label{ADBM_symmetryzation} 
\hspace{-0.3cm} -\frac{1}{N^2}  \sum_i \delta_{\sigma_i(t),\sigma} f'(x_i(t)) \sum_{j (\neq i)} W_{\sigma,\sigma_j(t)}'(x_i(t)-x_j(t)) \hspace{-0.2cm} &=& \hspace{-0.2cm} \frac{2g}{N^2}  \sum_i \delta_{\sigma_i(t),\sigma} f'(x_i(t)) \sum_{j (\neq i)} \frac{\delta_{\sigma,\sigma_j(t)}}{x_i(t)-x_j(t)} \\
&=& \hspace{-0.2cm} \frac{g}{N^2} \sum_i \sum_{j (\neq i)} 
\frac{f'(x_i(t)) - f'(x_j(t))}{x_i(t) - x_j(t)} \delta_{\sigma_i(t), \sigma} \delta_{\sigma_j(t), \sigma} \;. \nn
\eea  
Following \cite{RogersShi}, we rewrite this as
\bea
&& \frac{g}{N^2} \sum_{i,j} \frac{f'(x_i) - f'(x_j)}{x_i - x_j} \delta_{\sigma_i, \sigma} \delta_{\sigma_j, \sigma} - \frac{g}{N^2} \sum_i f''(x_i) \delta_{\sigma_i, \sigma} \\
&&= g \int dx dy \frac{f'(x) - f'(y)}{x - y}
\rho_\sigma(x,t)\rho_\sigma(y,t) - \frac{g}{N} \int dx f''(x) \rho_\sigma(x,t) \nonumber \\
&& = -2g \fint dx f(x) \partial_x \big[\rho_\sigma(x,t) \int \frac{dy}{x - y} \rho_\sigma(y,t) \big] - \frac{g}{N} \int dx f(x) \partial_x^2 \rho_\sigma(x,t) \;, \nn
\eea
where $\fint$ denotes the Cauchy principal value. Removing the self-interaction term thus leads to an additional diffusion term of order $1/N$. Introducing $\beta=\frac{2g}{T}$ and following the same steps as in Sec.~\ref{sec:DKder} for the other terms, we obtain the Dean-Kawasaki equation for model I:
\bea \label{DK_model1} 
\partial_t \rho_\sigma(x,t) =&& \hspace{-0.5cm} \partial_x \Big[ \rho_\sigma(x,t) \big(- v_0 \sigma + \lambda x -  2 g
\fint dy \frac{1}{x-y} \rho_\sigma(y,t) \big) \Big] + \gamma \big( \rho_{-\sigma}(x,t) - \rho_{\sigma}(x,t) \big) 
\\
&& \hspace{-0.5cm} + T (1 -  \frac{\beta}{2}) \partial^2_x \rho_\sigma(x,t)  
+ \frac{1}{N} \partial_x [ \sqrt{2T \rho_\sigma(x,t)} \; \eta_\sigma(x,t) ] + \frac{\sigma}{\sqrt{N}} \sqrt{\gamma \rho_s(x,t)} \; \eta_K(x,t) \;. \nn
\eea

We may be tempted to try to perform the same manipulations for model II. Forgetting about the self-interaction term, we would expect to arrive at the same equation \eqref{DK_model1}, replacing $\fint dy \frac{1}{x-y} \rho_\sigma(y,t)$ by $\fint dy \frac{1}{x-y} \rho_s(y,t)$. However, we cannot show this in a convincing way, since the symmetrization procedure used above fails in this case. Indeed, in \eqref{ADBM_symmetryzation} we should replace $\delta_{\sigma,\sigma_j(t)}$ by $1$. If we attempt to symmetrize this expression, we obtain the combination $\delta_{\sigma_i,\sigma} f'(x_i) - \delta_{\sigma_j,\sigma} f'(x_j)$ instead of $f'(x_i)-f'(x_j)$ in the numerator. As a result we fail to extract the self-interaction term, and the regularization through the Cauchy principal value is not really justified. Although this may seem like a technical difficulty, our numerical simulations show that the DK equation fails for model II in a more fundamental way (see below). Note that if we sum the two equations for $\rho_+(x,t)$ and $\rho_-(x,t)$, the above symmetrization procedure works, and leads to a correct equation. However this equation is not closed, as we cannot do the same for the difference. 
\\

\noindent {\bf Local correlations.} In Sec.~\ref{sec:FPrho} we derived an equation directly for the mean density, which here we denote again $p_\sigma(x,t)=\langle \rho_\sigma(x,t)\rangle$ for clarity. For the two versions of the active DBM, it reads
\be 
\partial_t p_\sigma(x,t) = \partial_x \left[(- v_0 \sigma + \lambda x) p_\sigma(x,t) - 2 g \left( 1-\frac{1}{N} \right) \fint dy \frac{1}{x-y} \tilde p^{(2)}(x,y,t;\sigma) \right] + \gamma p_{-\sigma}(x,t)  - \gamma p_{\sigma}(x,t) \;.
\label{eqfromFPfullADBM}
\ee
where
\bea  \label{defptilde}
\tilde p^{(2)}(x,y,t;\sigma) =
\begin{cases}
& p^{(2)}_{\sigma,\sigma}(x,y,t) \hspace{2.8cm} \text{for model I,} \\
& p^{(2)}_{\sigma,+}(x,y,t) + p^{(2)}_{\sigma,-}(x,y,t) \quad \text{for model II,}
\end{cases}
\eea
and the two point density $p^{(2)}_{\sigma,\sigma}(x,y,t)$ is defined in \eqref{2point_density}.

\begin{figure}
    \centering
    \includegraphics[width=0.32\linewidth,trim={0 0 1cm 1cm},clip]{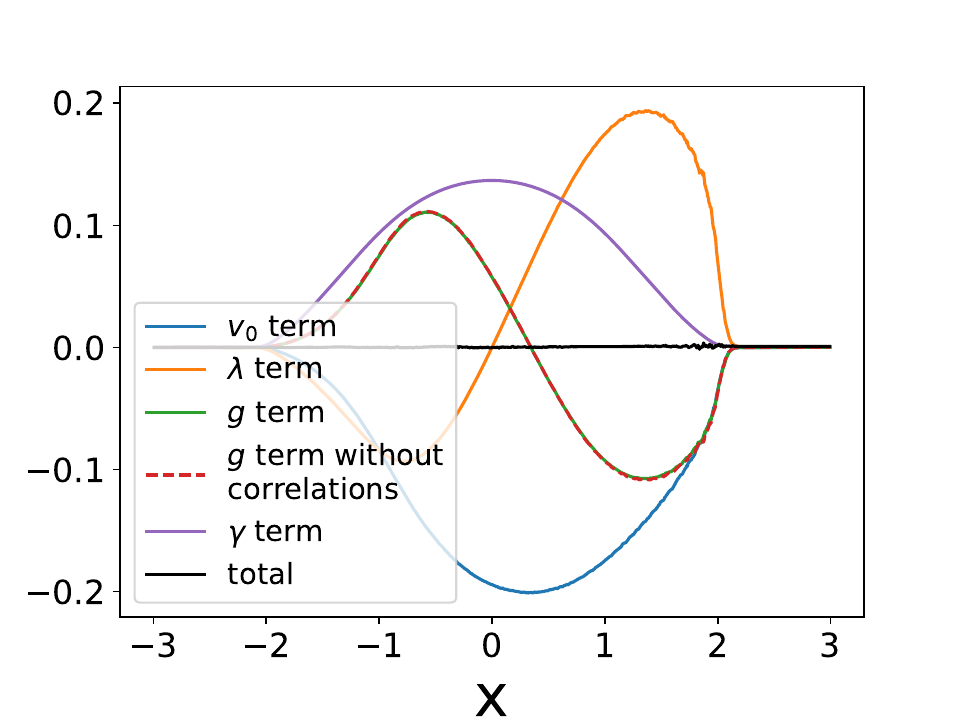}
    \includegraphics[width=0.32\linewidth,trim={0 0 1cm 1cm},clip]{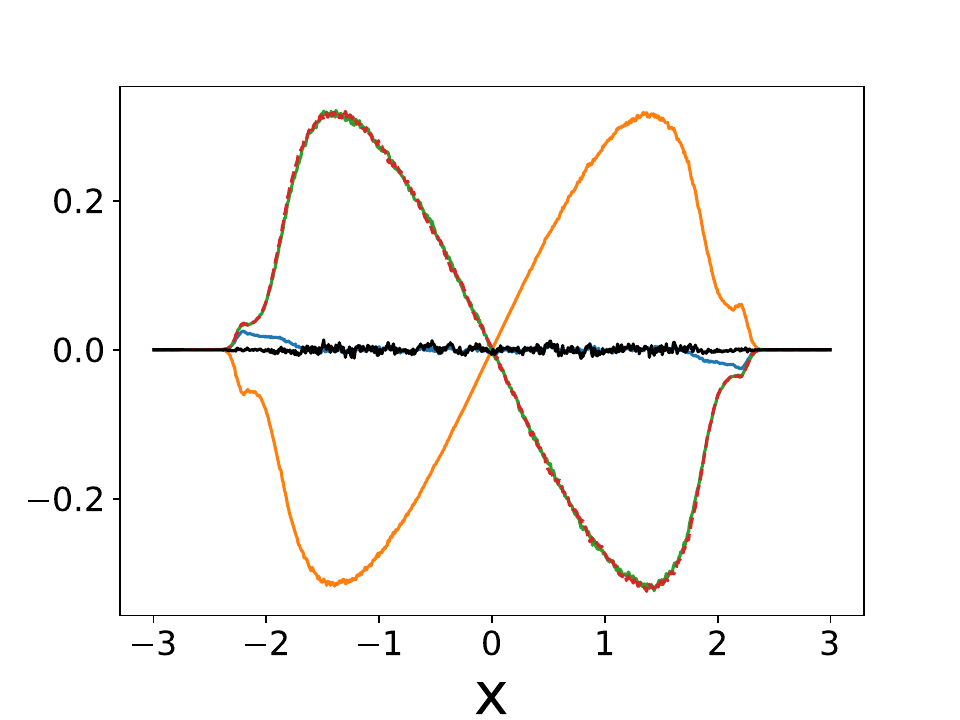}
    \includegraphics[width=0.32\linewidth,trim={0 0 1cm 1cm},clip]{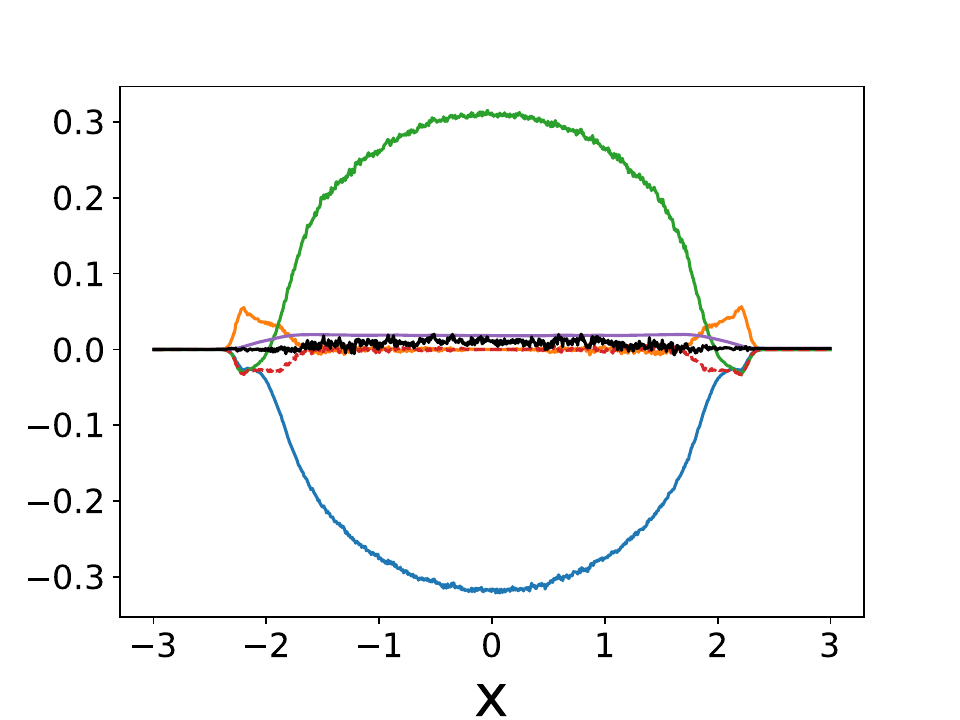}
    \caption{{\bf Left:} Different terms of the r.h.s. of Eq. \eqref{eqfromFPfullADBM} for $\sigma=+1$, for model I in the stationary state (for $N=100$ and all other parameters equal to 1). The interaction term (in red) is computed using a numerical estimation of the two-point density. The sum of all the terms is plotted in blue and is zero as expected. The interaction term obtained by neglecting correlations is also plotted in brown. It matches perfectly with the true interaction term. {\bf Center and Right:} Different terms of the r.h.s. of eq. \eqref{psfullADBM} (center) and \eqref{pdfullADBM} (right) for model II and their sum, which is zero in both cases as expected, for the same parameters. The interaction term obtained by neglecting the correlations is again plotted in brown. It matches perfectly with the true interaction term in \eqref{psfullADBM} (center) but leads to a completely wrong result in \eqref{pdfullADBM} (right).}
    \label{check_full_finiteN}
\end{figure}

As discussed in Sec.~\ref{sec:FPrho}, the DK equation is valid if in the large $N$ limit we have the decoupling condition
\begin{equation}
    p^{(2)}_{\sigma,\sigma'}(x,y,t) \underset{N\to+\infty}{\longrightarrow} p_\sigma(x,t) p_{\sigma'}(y,t) \;.
    \label{FPapprox2}
\end{equation}
For both models I and II, we estimated the two-point densities from numerical simulations, together with the usual densities. In Fig.~\ref{check_full_finiteN} (left panel), we show numerical estimations of the different terms of the equation \eqref{eqfromFPfullADBM} for model I in the stationary state, where the interaction term is evaluated both using the two-point density and using the assumption \eqref{FPapprox2}. We find that the two estimations coincide. The DK equation is thus valid for model I.

The results are however different for model II. In this case, we instead considered the equations for $p_s(x,t)=p_+(x,t)+p_-(x,t)$ and $p_d(x,t)=p_+(x,t)-p_-(x,t)$ (obtained from \eqref{eqfromFPfullADBM} by taking the sum and difference over $\sigma$),
\begin{eqnarray}
\hspace{-1.2cm} \partial_s p_s(x,t) \hspace{-0.25cm} &=& \hspace{-0.25cm}  - v_0 p_d(x,t) + \lambda x \,p_s(x,t) - 2 g \left( 1-\frac{1}{N} \right) \fint dy \frac{1}{x-y} p^{(2)}_{s,s}(x,y,t) \;, \label{psfullADBM} \\
\hspace{-1.2cm}  \partial_t p_d(x,t) \hspace{-0.25cm} &=& \hspace{-0.25cm} \partial_x [- v_0 p_s(x,t) + \lambda x \,p_d(x,t) - 2 g \left( 1-\frac{1}{N} \right) \fint dy \frac{1}{x-y} p^{(2)}_{d,s}(x,y,t)] - 2 \gamma p_d(x,t)
\label{pdfullADBM} \;,
\end{eqnarray}
where 
\be
p_{s,s}^{(2)}(x,y,t) = \sum_{\sigma,\sigma'} p_{\sigma,\sigma'}^{(2)}(x,y,t) \quad , \quad p_{d,s}^{(2)}(x,y,t) = \sum_{\sigma'} p_{+,\sigma'}^{(2)}(x,y,t) - \sum_{\sigma'} p_{-,\sigma'}^{(2)}(x,y,t) \;.
\ee
Numerical estimates of the different terms of these two equations in the stationary state are plotted in Fig.~\ref{check_full_finiteN} (center and right panel). Again we compare the interaction term computed from the two-point densities with the one obtained by neglecting the correlations. For \eqref{psfullADBM}, the two results again overlap perfectly, which means that the correlations can indeed be neglected in this equation. This is however not true for the second equation \eqref{pdfullADBM}. Indeed we find that, for model II, $p_d(x) \to 0$ as $N\to+\infty$. This could be expected: since the particles cannot cross, the $+$ and $-$ particles cannot separate. Thus, if we neglect the correlations in \eqref{pdfullADBM}, the interaction term vanishes for $N\to+\infty$. If we compute it using the numerical estimate of the two point density $p_{d,s}^{(2)}(x,y)$, we see however that it is not at all zero. In fact, it compensates exactly the term proportional to $v_0$ in the large $N$ limit, so that \eqref{pdfullADBM} indeed holds as it should when keeping the correlations. Thus, the two-point correlations cannot be neglected in this case, which explains why the DK equation fails for model II.

As already discussed in Sec.~\ref{sec:FPrho}, the strong local correlations in model II are due to the fact that particles cannot cross, which leads to the formation of clusters. More precisely, a $+$ (resp. $-$) particle at position $x$ will create an accumulation of $-$ (resp. $+$) particles immediately at its right (resp. left). For the total density, this leads to symmetric correlations $p_{s,s}^{(2)}(x,y,t)-p_s(x,t)p_s(y,t)$, which does not contribute to the integral in \eqref{psfullADBM}. The term $p_{d,s}^{(2)}(x,y,t)-p_d(x,t)p_s(y,t)$ is however antisymmetric, and thus contributes to the integral in \eqref{pdfullADBM}. This effect does not disappear as $N$ is increased, and thus the correlations cannot be neglected in this case. No such effect arises in model I since only particles of the same sign interact together, allowing particles of opposite sign to cross.

Before closing this discussion, let us however note that, taking the large $N$ limit of \eqref{psfullADBM}, for which we have shown that the correlations are indeed negligible for $N\to+\infty$, 
\begin{equation}
\partial_t p_s(x,t) = - v_0 p_d(x,t) + p_s(x,t) \Big[\lambda x - 2 g \fint dy \frac{1}{x-y} p_s(y,t) \Big] \label{ps_decoupledADBM} \;,
\end{equation}
and inserting directly $p_d(x,t)=0$, we recover the DK equation at $N\to+\infty$ of the standard DBM \eqref{Dean_DBM_Brownian}. This implies in particular that the stationary density is given by the semi-circle law \eqref{WignerSC_ADBM} for any parameters, as announced in the introduction. Note that, contrary to the standard DBM for which a temperature $T=O(1)$ is enough to deviate from the semi-circle, here the semi-circle still holds for any finite $v_0$. This result is confirmed by our numerical simulations and will be further discussed in Sec.~\ref{sec:model2}. In a sense, the Dean-Kawasaki equation thus still holds in model II for the total density $\rho_s(x,t)$, if we complement it with $\rho_d(x,t)=0$.

As a final comment, let us recall that in the DBM at finite temperature, particle crossings are allowed when $\beta=2g/T<1$ \cite{Lepingle07, Allez13}. The behavior of model II might thus be very different in the presence of a sufficiently strong diffusive noise, and the hydrodynamic description might become valid in this case. Since we focused our study on the purely active case without thermal noise, we did not investigate this further.

\subsection{Resolvent method for model I}

Let us now focus on model I, for which the DK equation \eqref{DK_model1} is valid. We would like to use this equation to study the density in the limit of large $N$. As for the active rank diffusion, this task is made difficult by the non-local interaction term, but there is a way to rewrite the equation in a local form using an appropriate transformation. Inspired by the standard DBM (see Sec.~\ref{sec:DBM_Wigner}), we introduce the Stieltjes transforms of the densities $\rho_\pm(x,t)$,
\beq \label{def_resolv}
G_\sigma(z,t) = \int \frac{dx}{z-x} \rho_\sigma(x,t) \;. 
\eeq
The function $G_\sigma(z,t)$ is defined
for $z$ in the complex plane minus the support of the density (which for finite $N$ is a collection of points on the real axis). It satisfies the boundary conditions
\bea \label{G_largez}
G_\pm(z,t) \underset{z \to \infty}{\simeq}  \frac{1}{z}  \int dx \rho_\pm(x,t) =  \frac{p_\pm(t)}{z} \;,
\eea 
where $p_\sigma(t)$ is the fraction of particles with sign $\sigma$ at time $t$. In the stationary state for $N\to+\infty$ one has $p_\pm(t\to+\infty)=1/2$. Note that the symmetry $\rho_+(x)=\rho_-(-x)$, also valid in the stationary state at large $N$, implies the relation $G_+(z)=-G_-(-z)$. 

We start from the DK equation for model I \eqref{DK_model1} in the limit $N\to+\infty$. We multiply it by $\frac{1}{z - x}$ and integrate over $x$. We then use integrations by parts (the density has finite support so there are no boundary terms) and the identity $(\partial_x + \partial_z) \frac{1}{z - x}=0$ to rewrite the different terms. The second term on the right hand side can be rewritten as
\be
\int dx \frac{1}{z - x} \partial_x ( x \rho_\sigma(x,t) )
= \partial_z \int dx \frac{x}{z - x} \rho_\sigma(x,t)
= \partial_z \big( z \int dx \frac{1}{z - x} \rho_\sigma(x,t) \big)
= \partial_z ( z G_\sigma(z,t) ) \;,
\ee
while the interaction term becomes
\bea
\int dx \frac{1}{z-x} \partial_x [\rho_\sigma(x,t) \int \frac{dy}{x-y} \rho_\sigma(y,t) ]
&=& \partial_z  \int dx dy \frac{1}{(z-x)(x - y)}
\rho_\sigma(x,t) \rho_\sigma(y,t) \\
&=& \frac{1}{2} \partial_z  \int dx dy \frac{1}{(x - y)} [\frac{1}{z-x} - \frac{1}{z-y}] \rho_\sigma(x,t) \rho_\sigma(y,t) \nn \\
&=& \frac{1}{2} \partial_z  \int dx dy \frac{1}{(z-x)(z-y)} \rho_\sigma(x,t) \rho_\sigma(y,t) \nn \\
&=& \frac{1}{2} \partial_z G(z)^2 \;. \nn
\eea 
We thus obtain the following exact equation for the Stieltjes transforms for $N\to+\infty$
\beq
\partial_t G_\sigma = \partial_z (-v_0 \sigma G_\sigma + \lambda z G_\sigma - g G_\sigma^2) + \gamma G_{-\sigma} - \gamma G_{\sigma} + \frac{T}{N} (1 -  \frac{\beta}{2})  \partial^2_z G_\sigma  \;.
\label{eqGpm_model1}
\eeq
It is useful to rewrite it in terms of $G_s(z,t)=G_+(z,t)+G_-(z,t)$ and $G_d(z,t)=G_+(z,t)-G_-(z,t)$,
\bea
&& \partial_t G_s = \partial_z (-v_0 G_d + \lambda z G_s  - \frac{g}{2} (G_s^2+G_d^2)) \;, \label{2species_Gs} \\
&& \partial_t G_d = \partial_z (-v_0 G_s + \lambda z G_d  - g G_s G_d) - 2\gamma G_d \;. \label{2species_Gd}
\eea
Finally, when looking for a stationary solution, the first equation can be integrated, using the large $z$ behaviors in (\ref{G_largez}),
\be \label{eqGs_integrated} 
-\frac{v_0}{\lambda} G_d + z G_s  - \frac{g}{2\lambda} (G_s^2+G_d^2) = 1 \;.
\ee 
In the next section, we explain how the results announced at the beginning of this chapter concerning model I can be recovered from these equations, starting with the different limiting behaviors shown in Fig.~\ref{phase_diagram_model1}.

\section{Model I in the large $N$ limit} \label{sec:model1}

\noindent {\bf Weak noise limit $v_0\to0^+$.} Fixing $v_0=0$ in \eqref{2species_Gs}-\eqref{2species_Gd}, we see that a stationary solution is given by $G_d=0$ and
$G_s(z)= \frac{\lambda z}{g} (1-\sqrt{1-\frac{2 g}{\lambda z^2}})$. We thus recover the semi-circle density as expected,
\be
 \rho_s(x) = \frac{1}{\pi} {\rm Im} \, G_s(x-i 0^+) =  \frac{\lambda}{ \pi g} \sqrt{\frac{2 g}{\lambda} - x^2} \quad , \quad x\in [-\sqrt{2 g/\lambda},\sqrt{2 g/\lambda}] \;,
\ee 
but with $g$ replaced by $g/2$, since each particle only interacts with $\sim N/2$ particles at any given time instead of $N-1$.
\\

\noindent {\bf Weakly interacting limit $g\to0^+$.} If we instead fix $g=0$, it is easier to work directly with the DK equation for the density (for $N\to+\infty$), which in this case is the same as the Fokker-Planck equation for the density of a single-particle \eqref{FP_RTP}, allowing to recover the stationary solution \eqref{eqRTPharmonicADBM}. If we instead use the stationary version of \eqref{2species_Gs}-\eqref{2species_Gd}, we find $G_s(z) = \frac{1}{z} \, _2F_1\left(\frac{1}{2},1;\frac{\gamma}{\lambda}  +\frac{1}{2};\frac{v_0^2}{\lambda^2 z^2}\right)$ and $v_0 G_d(z)= \lambda(z G_s(z)-1)$, which is consistent with this result.
\\

\noindent {\bf Diffusive limit.} We now consider the diffusive limit, $v_0,\gamma \to +\infty$ with fixed effective temperature $T_{\rm eff}=\frac{v_0^2}{2\gamma}$. In this limit \eqref{2species_Gd} gives $G_d\simeq\frac{v_0}{2\gamma} \partial_z G_s$, and inserting
in \eqref{eqGs_integrated} then leads to the following equation for $G_s$,
\beq \label{eqGs_bouchaud} 
\lambda(zG_s-1) -\frac{g}{2} G_s^2 + T_{\rm eff} \, \partial_z G_s = 0 \;.
\eeq
If we set $T_{\rm eff} \ll 1$ (e.g., by taking $\gamma$ to infinity while $v_0$ remains finite), we recover once again the semi-circle density with edge $\pm \sqrt{2g/\lambda}$. If instead $T_{\rm eff}$ is of order $O(1)$, this corresponds to the standard DBM with temperature scaling as $O(N)$ (i.e., $\beta=O(1/N)$), again with the replacement $g\to g/2$. This case, well studied in the context of RMT \cite{BouchaudGuionnet,cuenca,allez_satya}, was discussed briefly at the end of Sec.~\ref{sec:DBM_Wigner}. The density $\rho_s(x)$ is thus given by \eqref{GaussWigner} with $g\to 2g$ and $\tilde T\to T_{\rm eff}$. It interpolates between the semi-circle for $c=\frac{g}{2T_{\rm eff}} \to +\infty$ and the Gaussian for $c = 0$.
\\

\noindent {\bf Strong persistence limit $\gamma\to 0^+$.} The last interesting limit is when $\gamma\ll \lambda$, meaning that the states of the particles $\sigma_i$ are practically frozen. In this limit, it is possible to carry out a systematic expansion of the densities $\rho_\pm(x)$ in powers of $\gamma$, starting from the integrated integrated version of \eqref{eqGpm_model1} (in the stationary state)
\be
-v_0 G_+(z) + \lambda z (G_+(z) - \frac{1}{2}) - g G_{+}^2(z) = \gamma \int_{-\infty}^z [G_+(z') + G_+(-z')] dz'\;,
\label{integrated_G+0}
\ee
where we have used that $G_-(z)=-G_+(-z)$ and $G_+(z) \sim \frac{1}{2z}$ when $z \to \pm \infty$. Here we only give the main results and we refer to the SM of \cite{ADBM1} (Sec.~IV.B) for the full computation.

We begin by describing the limit $\gamma=0^+$ (i.e., the order 0 of the expansion). This is the case where the $\sigma_i$ remain completely fixed (we however assume that $\gamma t \gg1$, such that the $+$ and $-$ particles are present in equal proportions). Then, \eqref{integrated_G+0} gives $G_+(z) \simeq \frac{z-v_0/\lambda}{2} \left(1 - \sqrt{1-\frac{2 g}{\lambda(z-v_0/\lambda)^2}}\right) $, which corresponds to a semi-circle density with support $[v_0/\lambda-\sqrt{2 g/\lambda},v_0/\lambda+\sqrt{2 g/\lambda}]$,
\beq
\rho_+(x) = \frac{\lambda}{2\pi g}\sqrt{\frac{2 g}{\lambda}-\big(x-\frac{v_0}{\lambda}\big)^2} \;.
\eeq
Similarly, $\rho_-(z)$ is a semi-circle of support $[-v_0/\lambda-\sqrt{2 g/\lambda},-v_0/\lambda+\sqrt{2 g/\lambda}]$. The $+$ and $-$ particles thus separate and form two distinct semi-circles shifted by $\pm v_0/\lambda$, which do not interact with each other (and which overlap if $v_0<\sqrt{2g\lambda}$).

\begin{figure}[t!]
    \centering
    \includegraphics[width=0.45\linewidth,trim={0 0 1cm 1cm},clip]{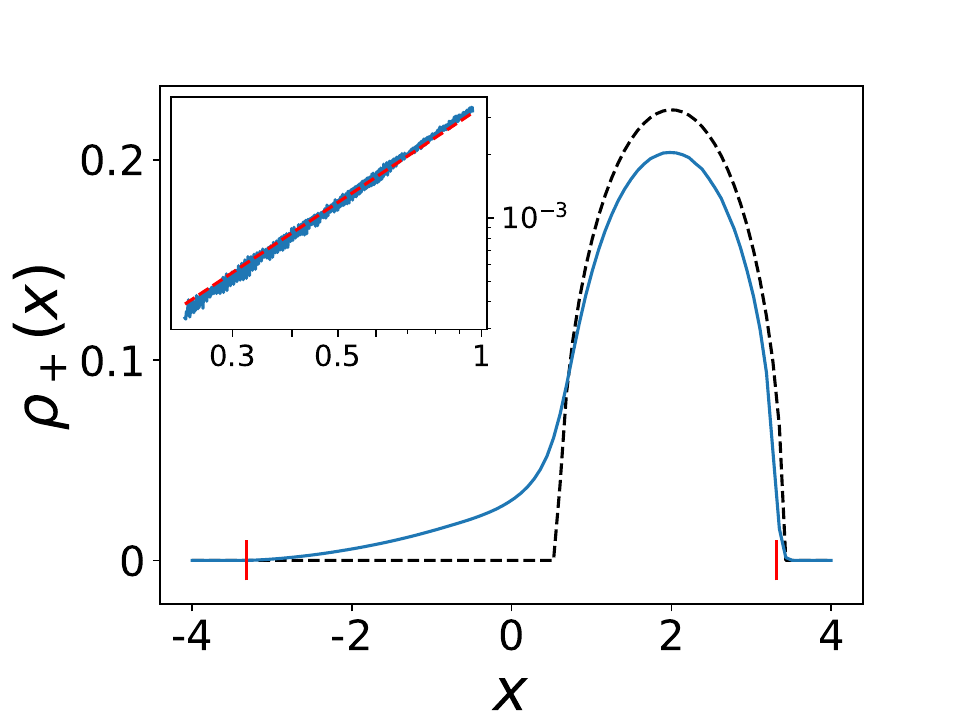}
    \includegraphics[width=0.45\linewidth,trim={0 0 1cm 1cm},clip]{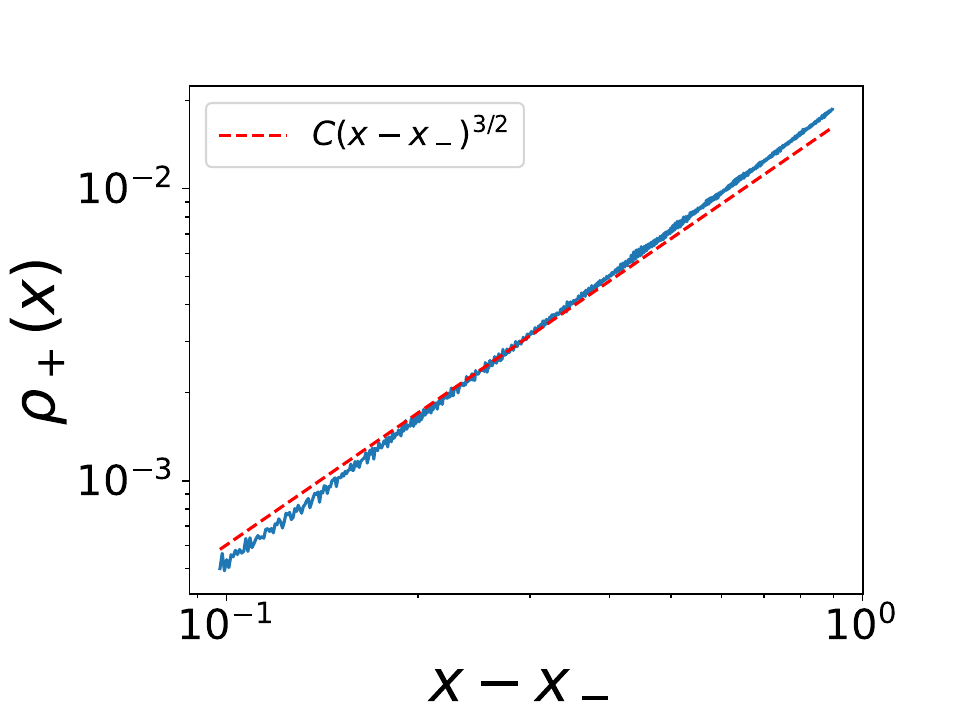}
    \caption{{\bf Left:} 
    Density $\rho_+(x)$ for  $\lambda=1$, $g=1$, $v_0=2$ and $\gamma=0.1$, for $N=100$. The dashed black line shows the limit $\gamma=0^+$ for $N\to +\infty$, i.e., a shifted semi-circle. The small red lines show the edges $\pm x_e$ of the support for $N\to +\infty$, computed numerically from \eqref{singularity_evenodd_model1}. {\it Inset:} Same density close to the left edge in log-log scale. The dashed red line has slope $3/2$. The $3/2$ exponent is observed in a small window between the the bulk regime and the regime very close to the edge where the finite $N$ effects appear, leading to an exponential decay of the density. 
    {\bf Right:} Same plot as the inset for $\gamma=1$, which shows that the $3/2$ exponent is valid beyond the limit $\gamma \to 0$.}
    \label{plot_model1_gammaexp}
\end{figure}

As soon as $\gamma>0$ however, the particles can switch sign and thus the two densities $\rho_\pm(x)$ have the same support $[-x_e,x_e]$. Going to the next order in the expansion, we find that the semi-circle exponent $1/2$ near the right edge $x_e$ survives for $\gamma>0$, i.e., $\rho_+(x) \simeq A\sqrt{x_e-x}$, with a $O(\gamma)$ shift in the position of the edge $x_e$ and in the amplitude $A$, which we computed explicitly. Interestingly, near the left edge $-x_e$ we find
\beq \label{smallgammaexp_ADBM}
\rho_+(x) \simeq \frac{\gamma \lambda^{1/4} (x-x_e)^{3/2}}{3\pi 2^{1/4} {g^{{3}/4}} \sqrt{{v_0(v_0/\lambda+\sqrt{2g/\lambda}\,)}} } \;,
\eeq
i.e., $\rho_+(x)$ vanishes with an exponent $3/2$. This exponent also survives at large values of $\gamma$, as we will argue below based on the computation of the moments, and as confirmed by the numerics, see Fig.~\ref{plot_model1_gammaexp}. This difference of 1 in the exponent between the left and right edge for $\rho_\pm(x)$ seems to be a general feature for RTPs when the density has a finite support. It also arises for non-interacting confined RTPs, e.g., in a harmonic trap \eqref{eqRTPharmonic_sign}, and is related to the fact that $+$ particles near the left edge are essentially $-$ particles which recently tumbled (and vice versa at the right edge).
\\

\noindent {\bf Moments.} For arbitrary parameters (i.e., beyond the 4 limits studied above), the equations \eqref{2species_Gs}-\eqref{2species_Gd} are more difficult to solve (even in the stationary state. They can however be used to compute the moments $m^s_k$ and $m^d_k$ of $\rho_s(x)$ and $\rho_d(x)$, as well as the moments $\langle x^k \rangle_\pm$ of the densities $2 \rho_{\pm}(x)$ (normalized to unity). They can be obtained exactly by recursion, using the large $z$ expansion 
\be
2 G_\pm(z)=\sum_{k=0}^\infty \frac{\langle x^k \rangle_{\pm}}{z^{k+1}} \;.
\ee
The symmetry $\rho_+(x)=\rho_-(-x)$ leads to $m^s_{2p}=\langle x^{2p} \rangle_+=\langle x^{2p} \rangle_-$, as well as $m^d_{2p+1}=\langle x^{2p+1} \rangle_+=-\langle x^{2p+1} \rangle_-$, and $m^s_{2p+1}=m_{2p}^d=0$. We find (see Sec.~IV.A of the SM in \cite{ADBM1})
\be
 m^d_1 = \langle x {\rangle}_+ = \frac{v_0}{\lambda+2\gamma} \quad  , \quad
 m^s_2 = \langle x^2 \rangle_+ = \frac{v_0^2}{\lambda(\lambda+2\gamma)} + \frac{g}{2\lambda} \;.
\ee 
The higher moments can be obtained from the following recursion relation  
\beq \label{moments_model1}
\langle x^k \rangle_+ = \frac{k}{k+2\frac{\gamma}{\lambda} \delta_{k,{\rm odd}}} \big(\frac{v_0}{\lambda} \langle x^{k-1} \rangle_+ + \frac{g}{2\lambda} \sum_{l=0}^{k-2} \langle x^l \rangle_+ \langle x^{k-2-l} \rangle_+ \big) \;,
\eeq
with $\delta_{k,{\rm odd}} = 1$ if $k$ is odd and $0$ otherwise. In \cite{ADBM1}, we also computed the first three moments for any finite $N$ (by applying the resolvent method to the equation \eqref{eqfromFPfullADBM} for the mean density at finite $N$). The equations \eqref{2species_Gs}-\eqref{2species_Gd} also allow to study the {\it time evolution} of the moments, beyond the stationary state (which we did, focusing on the first two moments). We find excellent agreement with our numerical simulations in all cases.

Finally, from the behavior of the high order moments, which can be computed numerically using the recursion \eqref{moments_model1}, it is possible to obtain, for $N \to +\infty$, (i) the position of the edges $\pm x_e$, and (ii) the behavior of the densities near the edges\cite{flajolet2009}. Over a wide range of parameters, we find a large $k$ behavior of the form
\beq
\langle x^k \rangle_+ \simeq \left(A k^{-\alpha_1-1} (x_e)^k + (-1)^k B k^{-\alpha_2-1} (-x_e)^k \right) \; \quad , \quad \alpha_1 \approx \frac{1}{2} \quad , \quad \alpha_2 \approx \frac{3}{2} \;.
\label{singularity_evenodd_model1}
\eeq
This indicates that the density $\rho_+(x)$ exhibits two distinct behaviors near the two edges, i.e., $\rho_{+}(x) \sim (x_e-x)^{1/2}$ at the right edge (as for the semi-circle) and $\rho_{+}(x) \sim (x+x_e)^{3/2}$ at the left edge, in agreement with the results of the small $\gamma$ expansion \eqref{smallgammaexp_ADBM} above.

\section{Model II in the large $N$ limit} \label{sec:model2}

\begin{figure}
    \centering
    \includegraphics[width=0.45\linewidth,trim={0 0 1.5cm 1cm},clip]{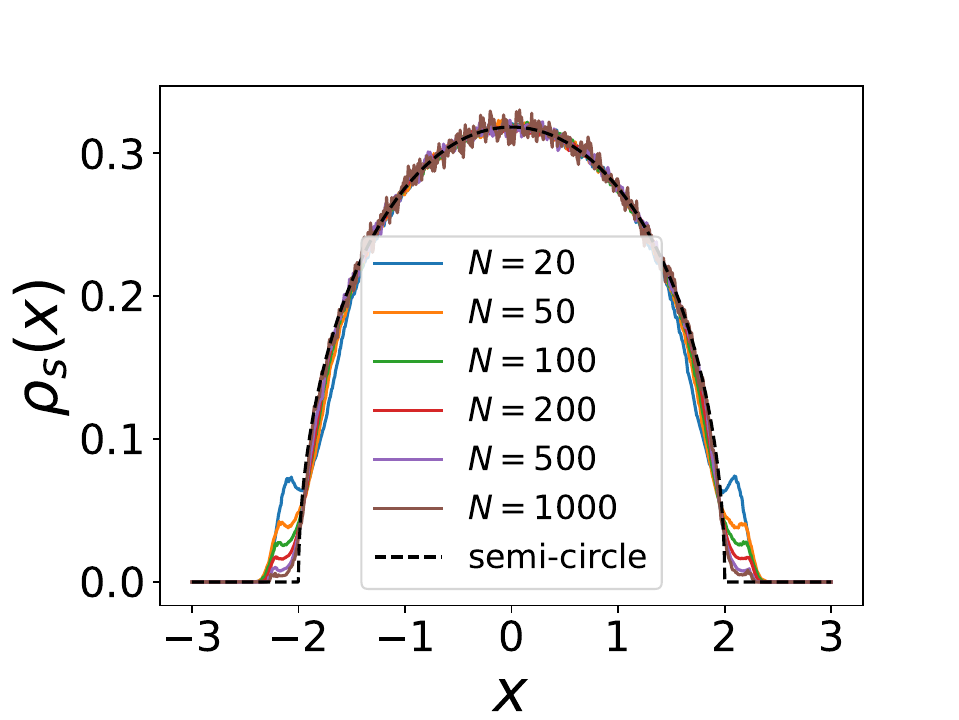}
    \hspace{0.2cm}
    \includegraphics[width=0.45\linewidth,trim={0 0 1.5cm 1cm},clip]{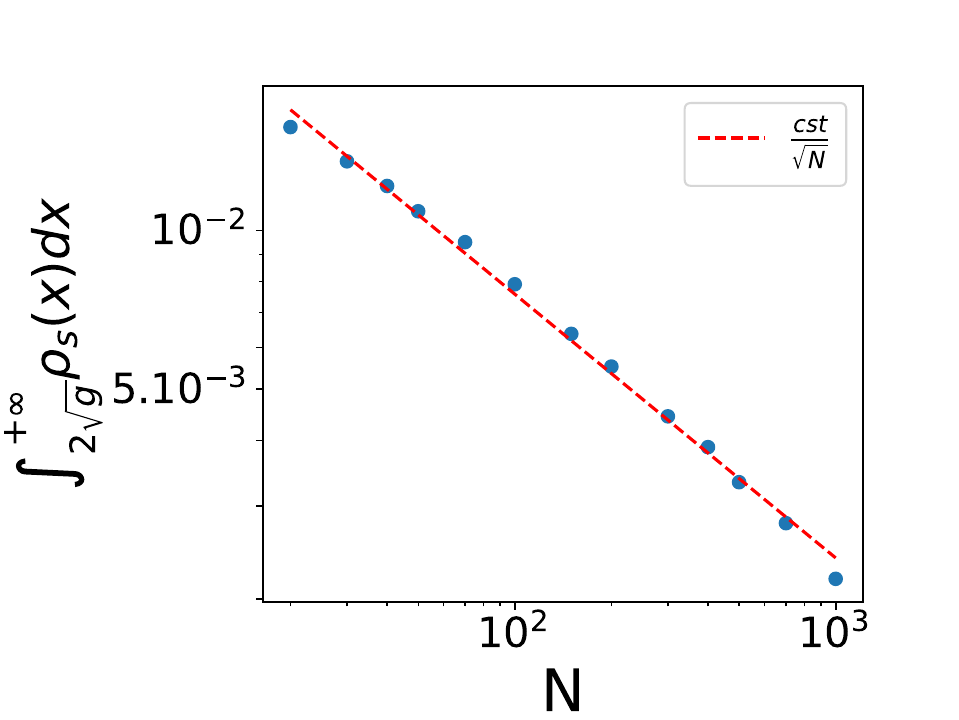}
    \includegraphics[width=0.45\linewidth,trim={0 0 1.5cm 0.5cm},clip]{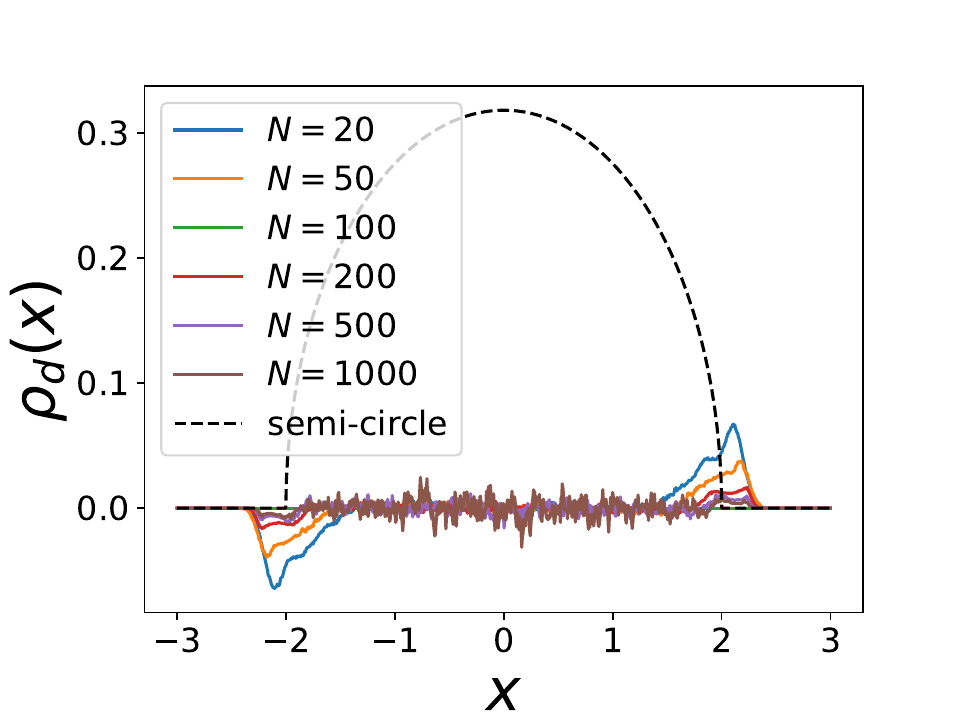}
    \hspace{0.2cm}
    \includegraphics[width=0.45\linewidth,trim={0 0 1.5cm 0.5cm},clip]{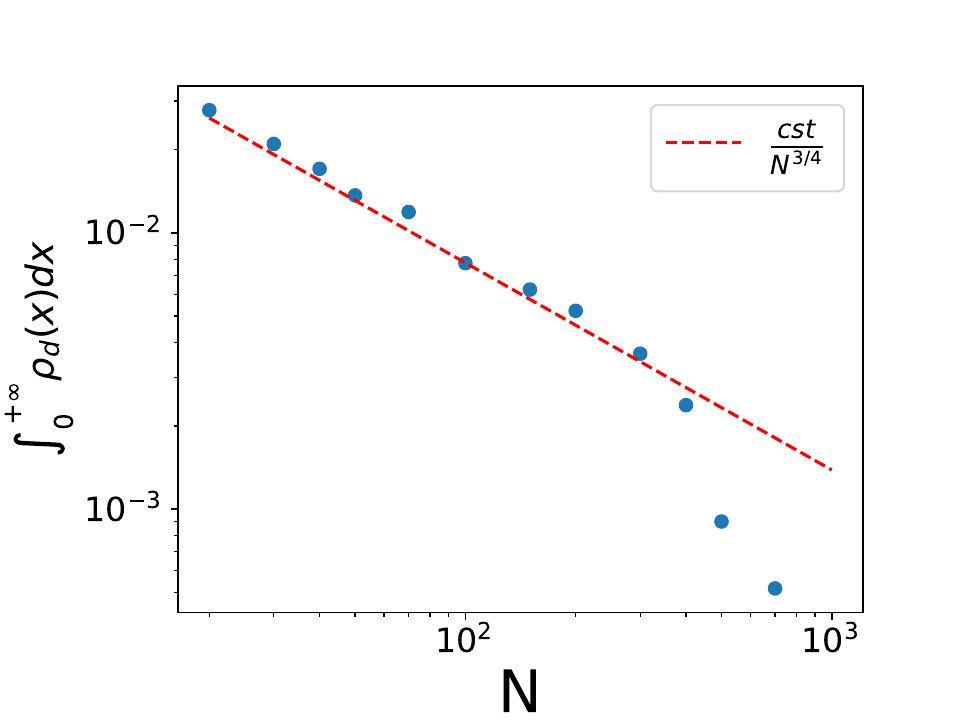}
    \includegraphics[width=0.45\linewidth,trim={0 0 1.5cm 0.5cm},clip]{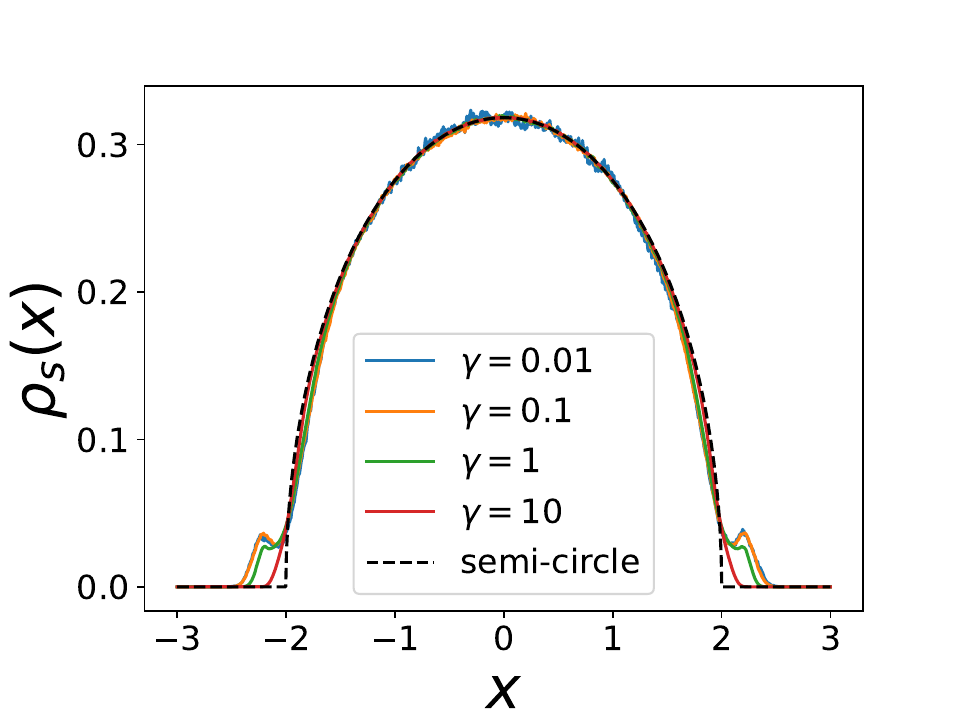}
    \caption{{\bf Top left:} Density $\rho_s(x)$ for $\lambda=1$, $g=1$, $v_0=1$ and $\gamma=1$ for  different values of $N$, showing the convergence to the semi-circle as $N$ increases. {\bf Top right:} Fraction of particles above the right edge of the semi-circle as a function of $N$, for the same set of parameters. It seems to decrease as $N^{-1/2}$. {\bf Center left:} Density $\rho_d(x)=\rho_+(x)- \rho_-(x)$ for $\lambda=1$, $g=1$, $v_0=1$ and $\gamma=1$ for different values of $N$.
    {\bf Center right:} Integral $\int_0^{+\infty} \rho_d(x)\,dx$ as a function of $N$, for the same set of parameters. It appears to decrease as $N^{-3/4}$. {\bf Bottom:} Total density of particles $\rho_s(x)$ for different values of $\gamma$, for $N=100$ and all other parameters set to 1. Increasing $\gamma$ decreases the size of the wings.}
    \label{plot_model2}
\end{figure}

\noindent {\bf Semi-circle regime $v_0/\sqrt{g\lambda}=O(1)$.} We now turn to model II, where all the particles interact together ($g_{\sigma,\sigma'}=g$). In model II, contrary to model I, the interaction prevents particles from passing each other, which leads to a very different behavior. As discussed in Sec.~\ref{sec:Dean_ADBM}, the Dean-Kawasaki equations fail in this case. The fact that the particles cannot cross imposes that $\rho_+(x)=\rho_-(x)$, i.e., $\rho_d(x)=0$ uniformly for $N\to+\infty$, as $+$ and $-$ particles cannot separate from each other. Inserting into the DK equation for $\rho_s(x)$ for $N\to+\infty$ \eqref{ps_decoupledADBM}, which as we have shown still holds in this case, we find that the total density is given by the semi-circle law with support $[-2\sqrt{g/\lambda}, 2 \sqrt{g/\lambda}]$, which we can write
\be
\rho_s(x) \sim \sqrt{\frac{\lambda}{g}} \;  f_{sc}\left( \frac{x}{\sqrt{g/\lambda}} \right) \quad , \quad f_{sc}(z) = \frac{1}{2\pi}\sqrt{4-z^2} \quad , \quad -2\leq z \leq 2 \;. \label{fsc_model2_0}
\ee
Quite surprisingly, this is completely independent of $v_0$ and $\gamma$, despite the fact that we have scaled the telegraphic noise as $O(1)$, and not $O(1/N)$ like the Brownian noise of the standard DBM. this result is confirmed by our numerical simulations. In Fig.~\ref{plot_model2} (top panels), we show the convergence of the density $\rho_s(x)$ to the semi-circle as $N$ increases. For finite $N$, the density exhibits ``wings'' on both sides of the support, with a total weight which seems to decay as $N^{-1/2}$. In the same figure we also show the convergence of $\rho_d(x)$ to zero (central panels). For finite $N$ it already almost vanishes inside the support, with mostly an accumulation of $+$ particles at the right edge and $-$ particles at the left edge, which seem to decay with an exponent $N^{-3/4}$ different from the weight of the ``wings''. For now the results concerning these finite $N$ edge effects are purely numerical.

The results above hold within the scaling of \eqref{def_ADBM}, where all the parameters are of order $O(1)$ (i.e., independent of $N$). The parameter $\gamma/\lambda$ seems to have no effect on the density for $N\to+\infty$, and only affects the finite $N$ fluctuations (the weight of the ``wings'' decreases as $\gamma$ increases, as can be seen on the bottom panel of Fig.~\ref{plot_model2}), even if we scale it with $N$. This is however not true for the other dimensionless parameter $v_0/\sqrt{g\lambda}$, as we now discuss.
\\

\begin{figure}[t!]
    \centering
    \includegraphics[width=0.32\linewidth,trim={0 0 1cm 1cm},clip]{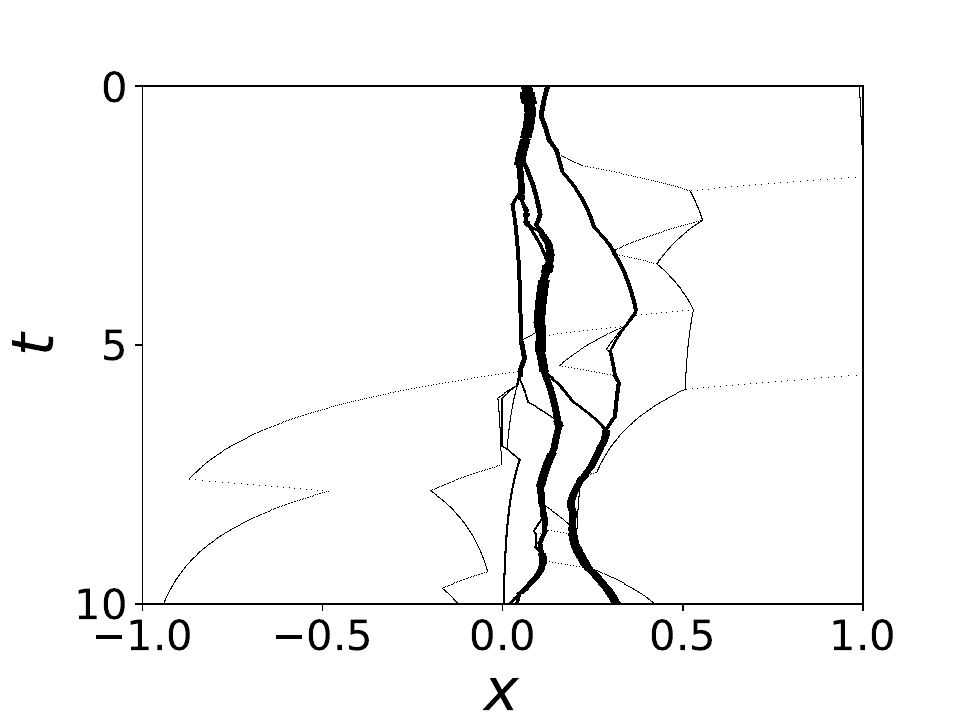}
    \includegraphics[width=0.32\linewidth,trim={0 0 1cm 1cm},clip]{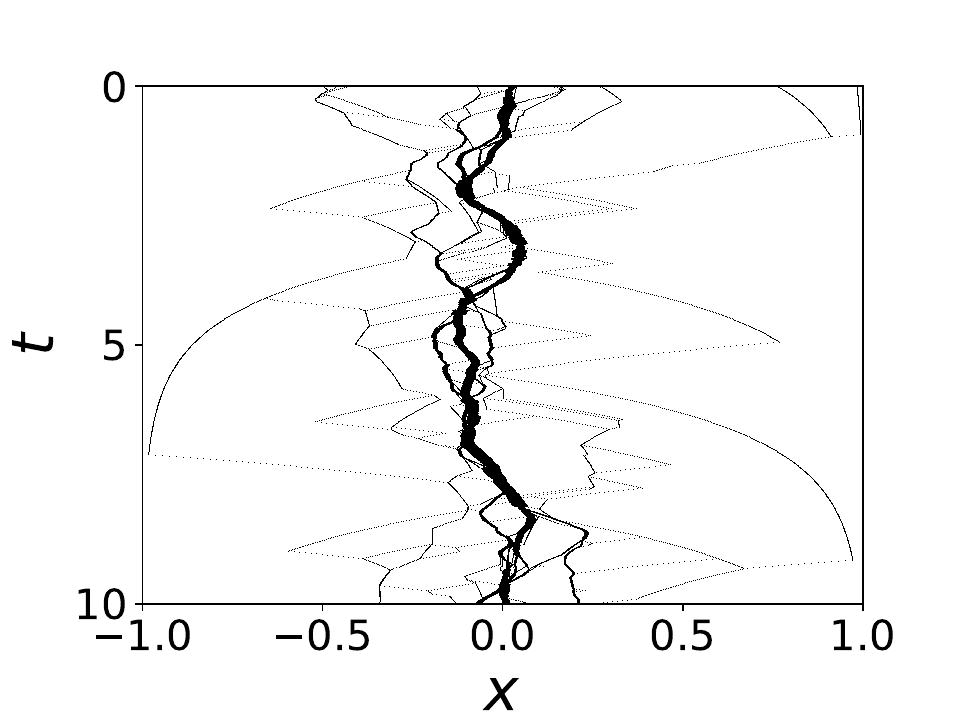}
    \includegraphics[width=0.32\linewidth,trim={0 0 1cm 1cm},clip]{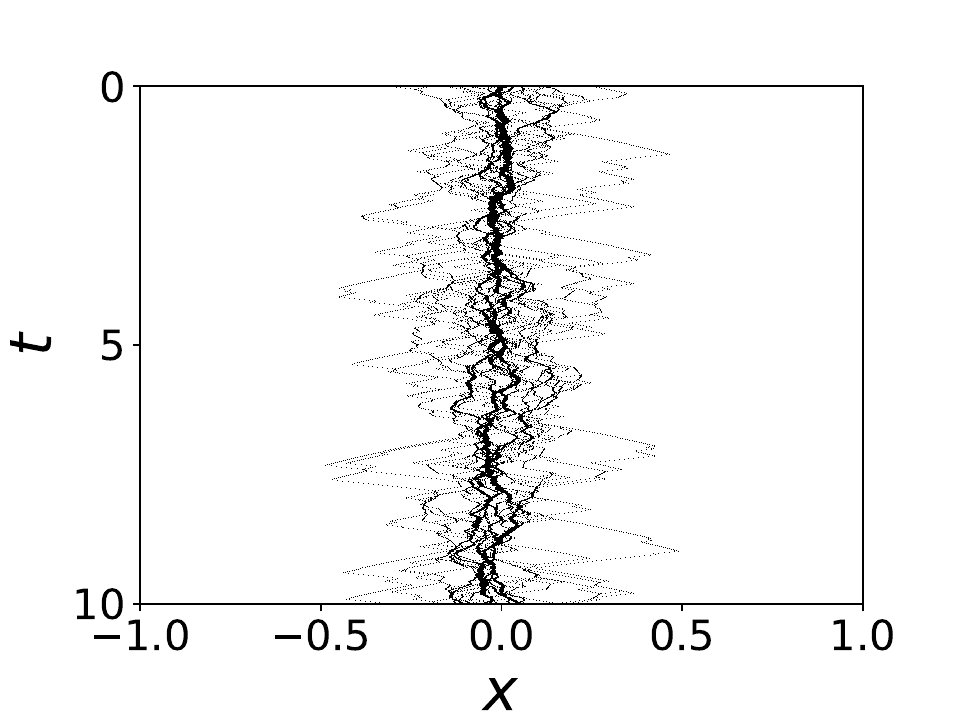}
    \caption{Time evolution of the positions of $N=100$ particles for the effective model $g\to 0^+$ of model II, for $\lambda=1$, $v_0=1$ and $\gamma=0.1$, $1$ and $10$ from left to right.}
    \label{spacetime_g0}
\end{figure}

\noindent {\bf ``Single-file'' limit $g\to 0^+$.} What happens if we make the driving velocity $v_0$ extremely large, or equivalently the interaction constant $g$ extremely small ? Said differently, can we break the semi-circle by making the noise sufficiently strong compared to the interaction ? To answer this question, let us now consider the limit where $g\to 0^+$. Contrary to model I, we do not recover the case of independent particles in this limit. Indeed, if we take $g$ continuously to zero, the long-range effect of the interaction will progressively disappear, but the non-crossing (or single-file) constraint will remain. Thus, in this limit the 2D Coulomb interaction is effectively replaced by a contact interaction, more precisely a hard-core repulsion between point-like particles. Due to the persistent motion of the particles, this hard-core repulsion will lead to the formation of point-like clusters, which may contain a large number of particles.

\begin{figure}
    \centering
    \includegraphics[width=0.32\linewidth,trim={0 0 1cm 1cm},clip]{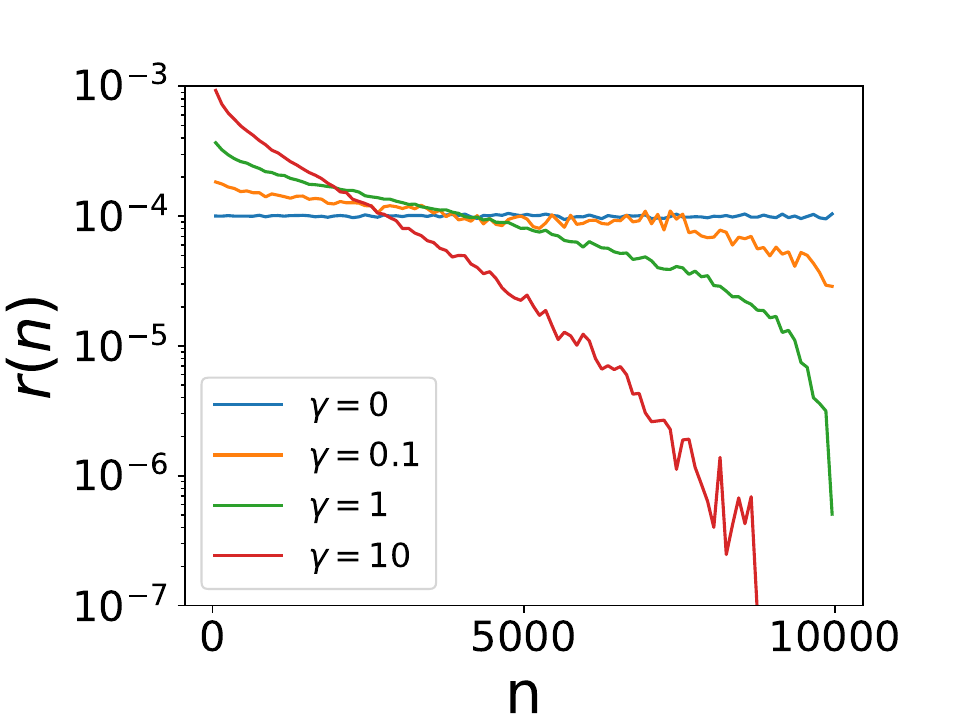}
    \includegraphics[width=0.32\linewidth,trim={0 0 1cm 1cm},clip]{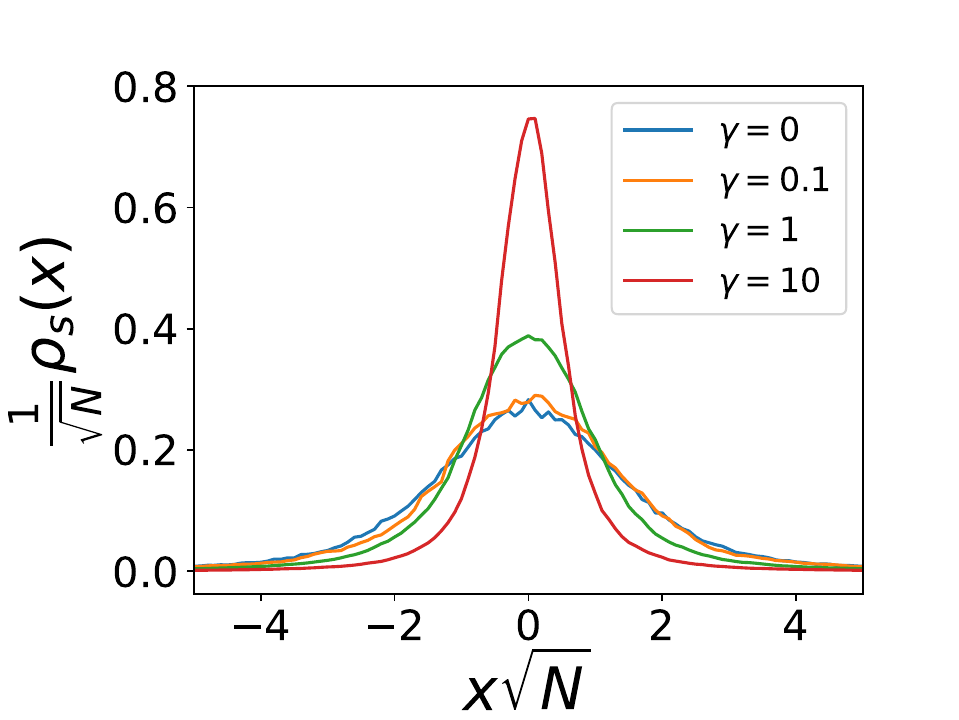}
    \includegraphics[width=0.32\linewidth,trim={0 0 1cm 1cm},clip]{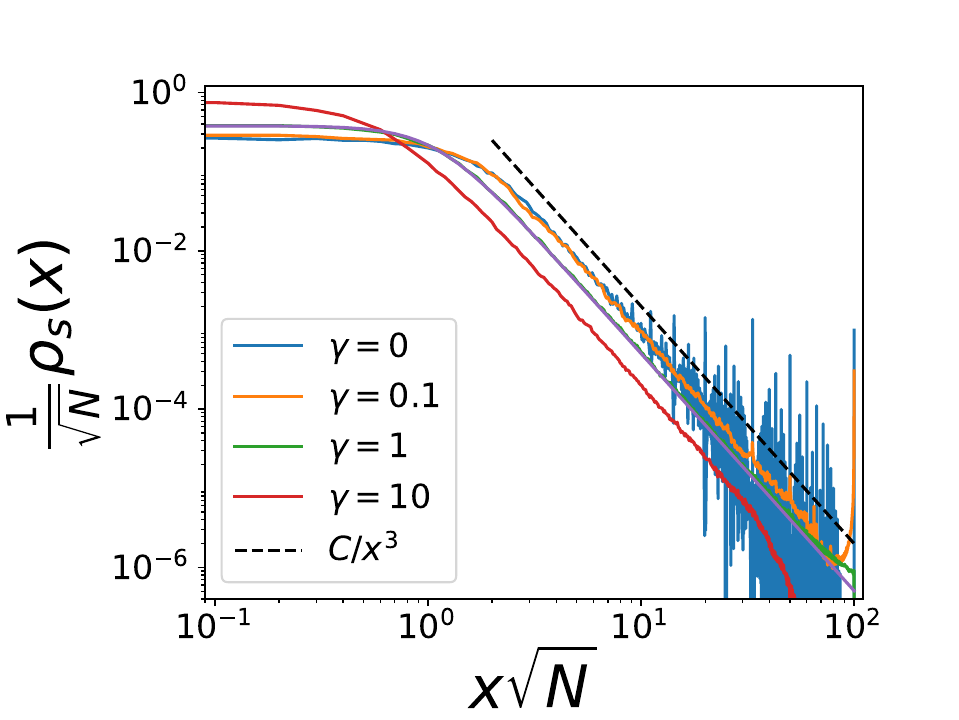}
    \caption{{\bf Left:} Fraction $r(n)$ of particles in clusters of size $n$, with $r(n)=n p(n)/\sum_{m=1}^N m p(m)$, in the limiting model $g\to 0^+$ for $N=10000$, $\lambda=1$, $v_0=1$ and  different values of $\gamma$ (averaged over $10^6$ realizations for $\gamma=0$ and over a time $10^5$ for $\gamma>0$). $r(n)$ is independent of $n$ for $\gamma=0$ and decays exponentially for $\gamma>0$. {\bf Center:} Rescaled particle density in the $g\to 0^+$ model for $N=10000$, $\lambda=1$, $v_0=1$ and different values of $\gamma$. {\bf Right:} Same plot in log-log scale. For all values of $\gamma$ the tail clearly decays as $x^{-3}$.}
    \label{g0appdx}
\end{figure}

An efficient algorithmic description of this limiting model can be obtained by viewing it in terms of clusters instead of particles. The algorithm is described in Appendix~\ref{app:simu} and can be summarized as follows. When two or more particles meet, they form a point-like cluster. The instantaneous velocity of each cluster is given by the mean velocity of all the particles forming the cluster. A cluster is characterized by the ordered list of the velocities of the particles inside it. When $\gamma>0$, each particle can change its velocity, i.e., tumble with rate $\gamma$, which may result in a breaking of the cluster into several pieces, according to precise rules 
(see Appendix~\ref{app:simu} for details). For small $\gamma$, the particles tend to form large clusters, as can be seen on Fig. \ref{spacetime_g0}. A similar phenomenon was observed in some RTP lattice models with exclusion interaction \cite{SG2014,slowman,Dandekar2020}. We have determined numerically the distribution $p(n)$ of sizes $n$ of these clusters. For $\gamma=0$ (i.e., if we simply fix all the $\sigma_i$, wait for all the clusters to form and average over many realizations instead of averaging over time), we find that $p(n)$ decays as $1/n$, for $n \leq N$, while for $\gamma>0$ the leading behavior is exponential in $n$ with a decay rate which increases with $\gamma$. In the left panel of Fig.~\ref{g0appdx}, we show the fraction $r(n)=n p(n)/\sum_{m=1}^N m p(m)$ of particles belonging to a cluster of size $n$ as a function of $n$. 

\begin{figure}
    \centering
    \includegraphics[width=0.45\linewidth,trim={0 0 1cm 1cm},clip]{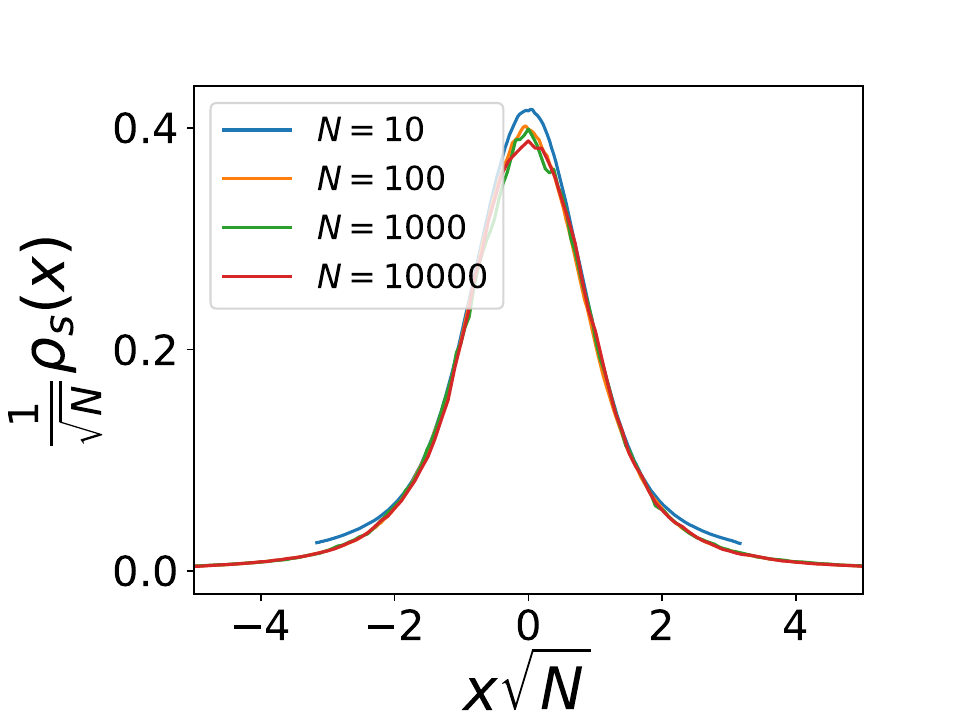}
    \hspace{0.1cm}
    \includegraphics[width=0.45\linewidth,trim={0 0 1cm 1cm},clip]{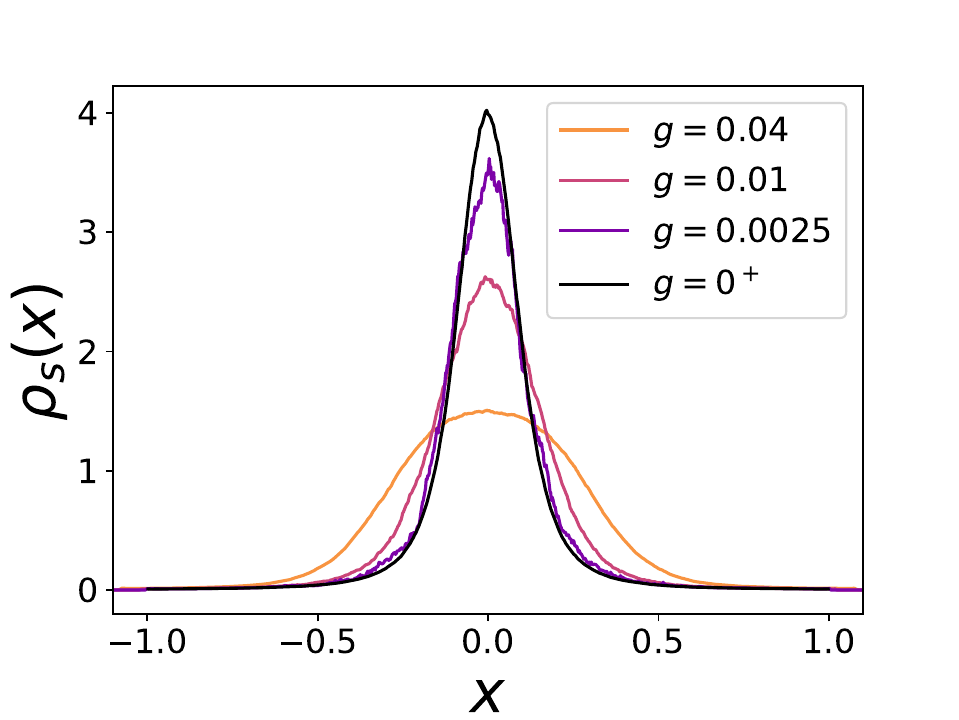}
    \caption{{\bf Left:} Rescaled particle density $\rho_s(x/\sqrt{N})/\sqrt{N}$ for different values of $N$ in the limiting model $g\to 0^+$ for $\gamma=1$, $\lambda=1$ and $v_0=1$. With this rescaling, all the plots collapse on the same curve, which is compatible with \eqref{phi_model2_0}. {\bf Right:} Density $\rho_s(x)$ in model II for small values of $g$, with $N=100$ and all other parameters set to 1. The density converges to the one of the limiting model $g\to 0^+$ (in black) as $g$ is decreased towards zero.}
    \label{g0figs}
\end{figure}

Concerning the density of particles $\rho_s(x)$, our numerical results suggest the scaling form (see the left panel of Fig.~\ref{g0figs})
\be
\rho_s(x)  \sim \frac{\lambda\sqrt{N}}{v_0} \; \phi\left( \sqrt{N} \frac{\lambda x}{v_0} \right) \;, \label{phi_model2_0}
\ee
with a scaling function $\phi(z)$ which depends on $\gamma$, see Fig.~\ref{g0appdx} (note that it seems to have a well-defined limit as $\gamma \to 0^+$). Interestingly, this scaling function seems to exhibit power law tails $\phi(z) \propto 1/|z|^3$ for large $|z|$, with a cutoff at the edges of the support $x=\pm v_0/\lambda$. 
As with the previous scaling $v_0/\sqrt{g\lambda}=O(1)$, we also find that $\rho_d(x)$ vanishes at large $N$.

Finally, we have checked that this effective model is a good description of model II in the limit $g\to0^+$ by comparing the density $\rho_s(x)$ in the model with clusters to the one of model II for decreasing values of $g$. In Fig.~\ref{g0figs} we indeed see that, for finite $N$, if we decrease $g$ to sufficiently small values the density of model II indeed converges to the one obtained from the clustering algorithm.
\\

\noindent {\bf Crossover between the two regimes.} Based on the scaling forms \eqref{fsc_model2_0} and \eqref{phi_model2_0}, we can estimate the scale of the parameter $v_0/\sqrt{g\lambda}$ for which the crossover between the semi-circle regime and the clustering regime takes place. As indicated in Fig.~\ref{phase_diagram_model2}, we see that it should occur for $v_0/\sqrt{g\lambda}\sim\sqrt{N}$. The results for the density $\rho_s(x)$ can thus be summarized as
\bea
&& \rho_s(x)  \sim \sqrt{\frac{\lambda}{g}} \;  f_{sc}\left( \frac{x}{\sqrt{g/\lambda}} \right) \quad , \quad \ \ \ \, \frac{g}{\lambda v_0^2} \gg \frac{1}{N} \;, \label{scalingforms_model2}\\
&& \rho_s(x)  \sim \frac{\lambda\sqrt{N}}{v_0} \; \phi\left( \sqrt{N} \frac{\lambda x}{v_0} \right) \quad , \quad \frac{g}{\lambda v_0^2} \ll \frac{1}{N} \;, \nn 
\eea
where $f_{sc}(z)$ is the semi-circle density given in \eqref{fsc_model2_0}. Thus, to break the semi-circle in the case of model II, we need to scale the driving velocity $v_0$ as $\sqrt{N}$, or equivalently to scale the interaction strength $g$ as $1/N$.

At this stage these results are mainly supported by numerical observations. In Chapter~\ref{chap:ADBMfluct} however, we will provide more quantitative arguments supporting this picture, by studying the fluctuations in the system at the single-particle level. We will show that the variance of the particle positions inside the bulk scale as ${\rm Var}(x_i)\sim \frac{v_0^2}{\lambda^2 N}$. By comparing the scale of these fluctuations to the size of the support of the semi-circle $2\sqrt{g/\lambda}$, we arrive at the same scaling for the crossover as in \eqref{scalingforms_model2}. In particular, as long as $v_0/\sqrt{g\lambda}\ll \sqrt{N}$, the fluctuations are too small to affect the density in the limit $N\to+\infty$, and thus $\rho_s(x)$ is the same as in the absence of noise, i.e., it is given by the Wigner semi-circle. Note that when $v_0/\sqrt{g\lambda}\ll 1/\sqrt{N}$ (very weak noise or very strong interaction), the scale of the fluctuations is even smaller than the typical distance between particles $2\sqrt{g/\lambda}/N$ and thus for high enough resolution each particle appears as a separate peak in the density (regime on the left in Fig.~\ref{phase_diagram_model2}). Note however that in this regime, the coarse-grained density is still given by the Wigner semi-circle for large $N$.

As a final remark, let us note that the diffusive limit $\gamma \to +\infty$, $v_0 \to +\infty$ and $T_{\rm eff}=\frac{v_0^2}{2 \gamma}$ fixed, we expect model II 
to converge to a variant of the DBM, where $\beta=2g/(N T_{\rm eff}) \ll 1$, with an additional hard-core repulsion between the particles. To our knowledge, this model remains to be studied. 

\section{Conclusion}

In this chapter, we introduced two versions of the active DBM, a model of RTPs interacting via a 2D Coulomb interaction, for which we studied the density of particles in the limit of large $N$. In model I where the particles are allowed to cross, we obtained several analytical results using the Dean-Kawasaki equation, including the different limits of the model and the behavior of the density at the edges of the support. For model II, where particle crossings are forbidden, we discussed the failure of the DK approach due to the formation of particle clusters, which generate strong local correlations. We found that in this case the density is given by the Wigner semi-circle for any set of parameters such that $v_0/\sqrt{g\lambda}\ll \sqrt{N}$, while in the opposite regime the model is well described by replacing the logarithmic interaction by a hard-core repulsion. In Chapter~\ref{chap:ADBMfluct}, we will come back to model II to study its fluctuations at the microscopic scale, which will provide further arguments for the results presented here. At this occasion we will also draw a parallel with the Calogero-Moser model and its active version.

Our study leaves several open questions, in particular concerning model II. The first question is of course whether there exists a way to obtain a proper hydrodynamic description for model II despite the single-file constraint, which would allow to better understand the finite $N$ fluctuations of the density (and in particular the edge effects that we observed). Second, the effect of additional thermal noise ($T>0$) 
could be important for model II since for $T>2g$ it would allow the particles to cross, and it would thus be interesting to see if a drastic change of behavior occurs as we increase the temperature above this value. Third, the model for $g\to0^+$ introduced in the last section, corresponding to $N$ point-like RTPs with hard-core repulsion in a harmonic trap, is another topic of study in itself and it would be interesting to better understand both the distribution of cluster sizes and the particle density (in particular the $1/x^3$ tail), although obtaining analytical results for this model seems quite challenging. Finally, the standard DBM was mostly studied due to its connection with RMT, and it would thus be interesting to know if there also exists a matrix model associated with the active DBM.

\part{Tagged particle fluctuations in passive and active Riesz gases}\label{part:fluctuations}

\vspace*{\fill}

\begin{center}
{\bf Abstract}
\end{center}

In this third part, we study the fluctuations at the level of the particle positions $x_i(t)$ in Riesz gases of Brownian and active particles. We begin in Chapter~\ref{chap:passiveRieszFluct} by considering a Riesz gas of $N$ Brownian particles on the circle, with a pairwise repulsive interaction $\sim|x|^{-s}$, where $s>-1$, before generalizing to active particles in Chapter~\ref{chap:activeRieszFluct}. Focusing on the limit of weak noise and linearizing the equations of motion, we obtain exact expressions for a variety of static and dynamical correlation functions, which we analyze in the limit $N\to+\infty$ with fixed density $\rho$. In the Brownian case, this allows us to recover some results obtained recently in the physics and mathematics literature via completely different methods, in particular concerning the mean squared displacement of a particle during time $t$ and the variance of the distance between two particles, but also to compute new dynamical quantities. In the active case, we find that the Brownian results are recovered at large times and large distances, but that the activity strongly affects the correlations both at short times and on small lengthscales. Finally, in Chapter~\ref{chap:ADBMfluct}, we extend these results to two special cases of the Riesz gas on the real axis inside a confining harmonic potential: the active Dyson Brownian motion ($s=0$), studied in Chapter~\ref{chap:ADBM_Dean} at the level of the particle density, and the active Calogero-Moser model, corresponding to $s=2$. For the active DBM, we show the existence of a distinct edge regime where the fluctuations have a different scaling from the bulk, as it is the case for the standard DBM. By contrast, the active CM model does not exhibit such a regime, although its Brownian version does. The results of this last part support the observations made in Chapter~\ref{chap:ADBM_Dean} concerning the stationary density in the active DBM and suggest a similar behavior for the active CM model. 

This third part is mostly based on the Reference~\cite{RieszFluct}. Chapter~\ref{chap:ADBMfluct} also uses results from \cite{ADBM2}. Concerning the CM model (discussed in the same chapter), the analytical results were derived in \cite{RieszFluct}, but it was mostly discussed in \cite{activeCM}. The simulations for the CM model were performed by Saikat Santra.

\vspace*{\fill}

\chapter{Brownian Riesz gas on the circle} \label{chap:passiveRieszFluct}

\section{Setting and main results}

\begin{figure}
    \centering
    \includegraphics[width=0.5\linewidth]{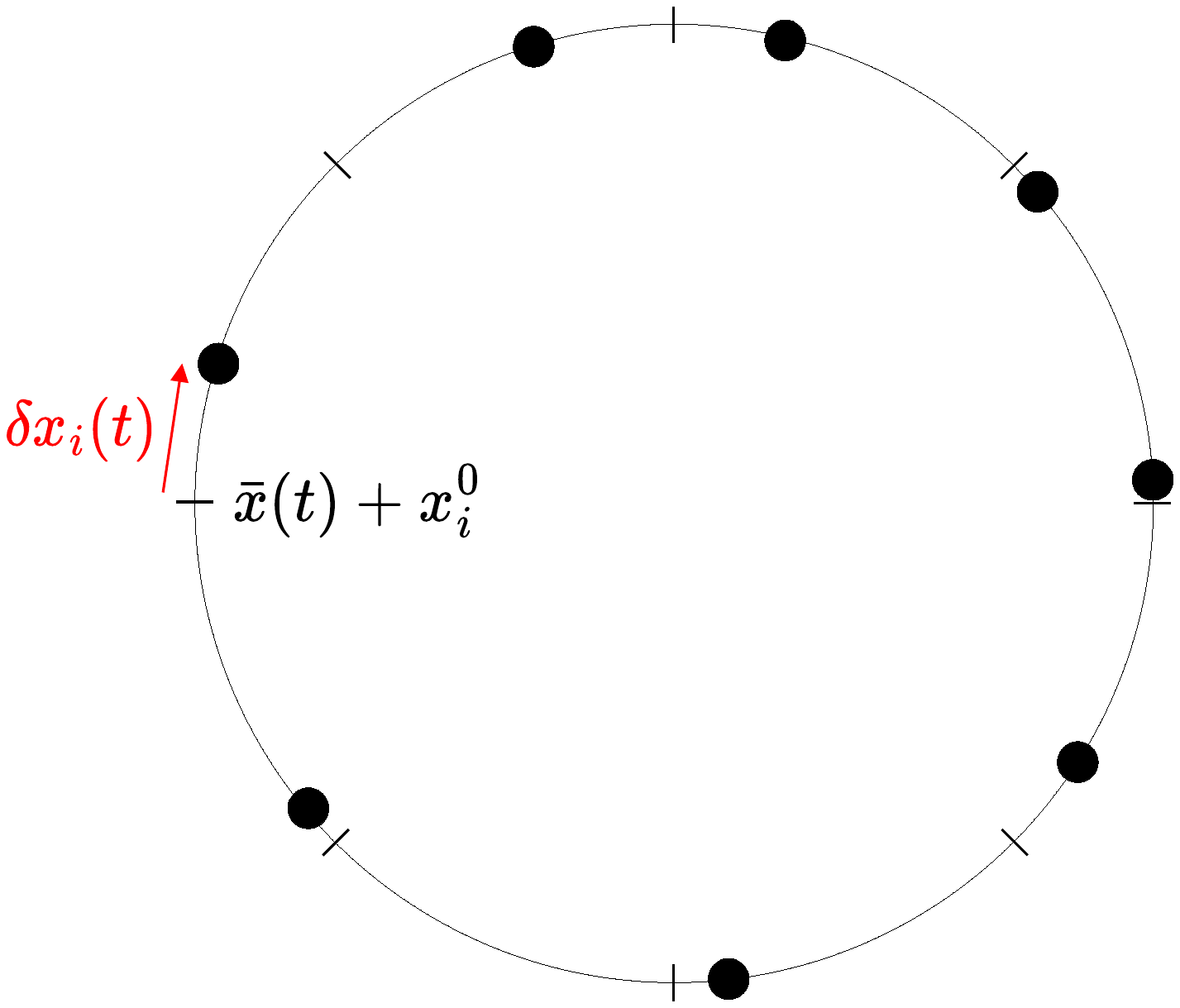}
    \caption{Schematic representation of the Riesz gas in the periodic geometry. In the ground state, the particles are equally spaced. We consider the evolution of the displacements $\delta x_i(t)$ of each particles with respect to its position in the ground state due to the Brownian noise (or active noise in the next chapter). As defined in \eqref{def_delta_x_Riesz1}, $\bar x(t)$ denotes the position of the center of mass which decouples from the $\delta x_i(t)$.}
    \label{fig:sketch_circle}
\end{figure}

In this chapter we temporarily return to Riesz gases of Brownian particles, introduced in Chapter~\ref{chap:Riesz_review}. We consider $N$ Brownian particles in 1D interacting via a pairwise repulsive power law potential $W(x)$, such that the dynamics of the particle positions $x_{i}(t)$ ($i=1,...,N$) are described by the equations of motion
\be \label{def_Riesz_chap7}
\frac{dx_i}{dt} = -\sum_{j(\neq i)} W'(x_i-x_j) + \sqrt{2T} \, \xi_i(t) \quad , \quad W'(x)=-g \, \frac{\sgn(x)}{|x|^{s+1}} \;.
\ee
Here the $\xi_i(t)$ are i.i.d. Gaussian white noises with unit variance, and we assume $s>-1$ and $g>0$. In this chapter and the next, we do not add any confining potential. Instead, we consider periodic boundary condition, i.e., we assume that the particles evolve on a ring of perimeter $L$, and we identify $x_i\equiv x_{i+N}$ (see Fig.~\ref{fig:sketch_circle}). This requires to properly periodize the interaction potential $W(x)$. This procedure may present some technical difficulties which we will discuss in the next section.

In this chapter, we study the static and dynamical correlations of the particle positions $x_i(t)$, in the limit of {\it weak noise}, i.e., for small temperature $T$. By linearizing the equations of motion \eqref{def_Riesz_chap7}, we are able to compute exactly the two-point two-time correlations in that limit. Analyzing these results in the thermodynamic limit $N,L\to+\infty$ with fixed density $\rho=N/L$, we perfectly recover some recent results obtained via completely different routes, in particular concerning the mean squared displacement (MSD) \cite{DFRiesz23} and the variance of the interparticle distance \cite{BoursierCLT,BoursierCorrelations} of long-range Riesz gases ($0<s<1$). In addition, this method also allows us to access new observables, such as the equal time covariance between the displacements of two particles or the time correlations of the interparticle distance. Besides its simplicity and the wide range of quantities that it allows us to compute, a strong advantage of the present method presented is that it can be easily extended to more complex types of noise. In the next chapter, we will extend the results of this chapter to active particles to see how the activity affects the fluctuations in the system.

We begin by introducing our method in Sec.~\ref{sec:RieszBrownian_derivation}, deriving exactly the two-point two-time covariance of the particle positions at equilibrium at the linear order for small $T$. In Sec.~\ref{sec:staticBrownian}, we use this result to study several static correlation functions in the thermodynamic limit, starting with the variance of the particle positions. We find that it diverges with $N$ for $s\geq 0$, compatible with the absence of translational order, while for $-1<s<0$ the variance remains finite for large $N$, suggesting the existence of a solid phase at small temperature, as observed recently in \cite{Lelotte2023} from numerical simulations. We also compute the variance of the distance between two particles separated by $k-1$ other particles, and show that it increases sublinearly with $k$ as $\sim k^s$ for $0<s<1$, in agreement with recent results from the mathematical literature \cite{BoursierCLT,BoursierCorrelations}. We then move on to the dynamical correlations in Sec.~\ref{sec:dynamicalBrownian}, starting with the MSD of a particle during time $t$. We obtain a subdiffusive regime at large time, where the MSD increases as $\sim \sqrt{t}$ in the short-range case $s>1$, as in single-file diffusion \cite{Harris65,Arratia83,SingleFileMajumdar1991,SingleFileKrapivsky2014,SingleFileKrapivsky2015,TaggedSFD2015,SingleFileReview}, and as $\sim t^{\frac{s}{1+s}}$ in the long-range case $0<s<1$. Remarkably, our results agree perfectly, including the prefactors, with the results obtained recently in \cite{DFRiesz23} using a completely different approach based on macroscopic fluctuation theory (MFT, see Sec.~\ref{sec:Riesz_brownian_tagged} for more details). We then analyze other dynamical quantities, including the two-time correlations of the displacement of a particle, the equal time covariance between the displacements of two particles and the time-correlations of the interparticle distance. Finally, we briefly discuss how these results are affected if we start from a deterministic (i.e., quenched) initial condition (instead of the annealed initial condition considered in the rest of this chapter). 

Throughout this chapter and the next, we assume that the particles are ordered, i.e., $x_1(t)>x_2(t)>...>x_N(t)$ at all time $t$. For $s>0$, and for $s=0$ with $\beta = g/T>1$ \cite{Lepingle07, Allez13}, the Brownian noise does not allow the particles to pass each other, and thus we can make this assumption without loss of generality (at least at small temperature for $s=0$). For $s<0$, the particle trajectories may however cross due to the noise. Since we are focusing on the low temperature regime, we assume that these crossings are negligible (within the regime of validity of our approximation discussed below), and that our method still applies in this case. This point would however require further investigation. We will not have this problem in Chapter~\ref{chap:activeRieszFluct} since RTPs cannot cross for any $s>-1$.


\section{General method} \label{sec:RieszBrownian_derivation}

\subsection{Weak noise limit}

Since the interaction is purely repulsive at all distances, the equilibrium configuration of the system is such that the particles are equally spaced. Here, we study the displacements $\delta x_i(t)$ of the particles around this equilibrium configuration. We thus decompose the positions of the particles as (see Fig.~\ref{fig:sketch_circle})
\be \label{def_delta_x_Riesz1}
x_i(t) = \bar x(t) + x_i^0  + \delta x_i(t)  \quad , \quad x_{i+1}^0-x_i^0= \frac{L}{N} \quad , \quad \bar x(t) = \frac{1}{N} \sum_i x_i(t) \;,
\ee
where the $x_i^0$ can be chosen as $x_i^0= (i - \frac{N+1}{2}) \frac{L}{N}$. Since the total energy is invariant by a global translation of all the particles, we also subtracted the displacement of the center of mass $\bar x(t)$. Summing over $i$ the equations of motion \eqref{def_Riesz_chap7}, we see that the center of mass freely diffuses with a diffusion coefficient $T/N$, i.e., $\frac{ d\bar x}{dt} = \frac{\sqrt{2T}}{N} \sum_j \xi_j(t) = \sqrt{\frac{2T}{N} } \, \xi(t)$,
where $\xi(t)$ has the same law as the $\xi_i(t)$ (since these are Gaussian variables). This behavior is subleading in $N$ and does not affect the quantities that we study here in the thermodynamic limit. By consistency, we have the relation $\sum_i \delta x_i(t)=0$.

If the temperature is sufficiently small, the particles only undergo small Gaussian fluctuations around the ground state, and thus the displacements $\delta x_i(t)$ of the particles are also small. This allows us to linearize the equations of motion \eqref{def_Riesz_chap7} as
\be \label{Eq_delta_x_RieszBrownian}
\frac{d}{dt} \delta x_i(t) = - \sum_{j=1}^N H_{ij} \delta x_j(t) + \sqrt{2T} \xi_i(t)  - \frac{\sqrt{2T}}{N} \sum_{j=1}^N \xi_j(t) \;,
\ee 
where $H$ is the Hessian matrix of the total energy $E(\{x_i\}) = \sum_{i<j} W(x_i-x_j)$,
\be \label{defHessianRiesz}
H_{ij} = \frac{\partial^2 E}{\partial x_i \partial x_j}(\{ x_i^0 \}) = \begin{cases} \sum_{k(\neq i)} W''(\frac{L}{N}(i-k)) \ {\rm for} \ i=j \;, \\ -W''(\frac{L}{N}(i-j)) \hspace{1cm} {\rm for} \ i\neq j \;. \end{cases}
\ee
At this point, the interaction potential $W(x)$ can remain arbitrary. We only assume that (i) it is periodic, with $W(x+L)=W(x)$, (ii) it is repulsive and (iii) the ground state is such that the particles are equally spaced. By taking the Taylor expansion to higher order, one can show that a reasonable criterion for the approximation \eqref{Eq_delta_x_RieszBrownian} to be valid is\footnote{This is however a necessary but not a sufficient condition in general, e.g., if $W''''(x_i^0-x_j^0)\gg W'''(x_i^0-x_j^0)$.} (see Appendix~B in \cite{RieszFluct})
\be \label{cond_approx}
\forall j\neq i, \ \delta x_i - \delta x_j \ll 2\frac{W''(x_i^0-x_j^0)}{W'''(x_i^0-x_j^0)} \;.
\ee
For the Riesz gas, we will check {\it a posteriori} the validity of this condition in Sec.~\ref{sec:gapvariance_RieszBrownian}.

The matrix $H$ is an $N \times N$ Toeplitz matrix, which can be diagonalized with eigenvectors $v^q_k=\frac{1}{\sqrt{N}} e^{2\pi i \frac{q}{N} k}$ (which form an orthonormal basis), and eigenvalues given by
\be \label{eigenvals_rieszgeneral}
\mu_q = \mu_{N-q}= 
2\sum_{\ell=1}^{N-1} W''\big(\frac{L}{N} \ell\big) \sin^2\big(\frac{\pi q\ell}{N}\big) \; , \quad q=0,1,...,N-1 \;.
\ee
Below, we first use this linear approximation to derive a general expression of the two-point two-time correlations of the particle displacements $\langle \delta x_i(t)\delta x_j(t') \rangle$, before specializing to the Riesz gas.

\subsection{Equilibrium dynamics for small deformations}

We now consider the equilibrium dynamics of \eqref{Eq_delta_x_RieszBrownian}, preparing the system at equilibrium at time $t\to-\infty$. This corresponds to an {\it annealed} initial condition. The {\it quenched} case, where we instead start from a deterministic, equally spaced configuration will be studied in Sec.~\ref{sec:quenchedBrownian} (we used the terms ``annealed" and ``quenched" by analogy with disordered systems, see, e.g., \cite{Banerjee2020} for a more detailed discussion).
Taking the Fourier transform of \eqref{Eq_delta_x_RieszBrownian} with respect to time, we obtain by inversion in the frequency domain
\be \label{eqfourier_RieszBrownian} 
\delta \hat x_j(\omega) = \sqrt{2T} \sum_{k=1}^N  [i \omega \mathbbm{1}_N + H]^{-1}_{jk} \hat \xi_k(\omega) - \frac{\sqrt{2T}}{N} \frac{1}{i\omega} \sum_{k=1}^N \hat \xi_k(\omega) \;,
\ee 
where $\mathbbm{1}_N$ is the $N \times N$ identity matrix, $\delta \hat x_i(\omega) = \int_{-\infty}^{\infty} e^{-i \omega t} \delta x_i(t)\,dt$ and $\hat \xi_i(\omega)$ is a Gaussian white noise with correlations $\langle \hat \xi_i(\omega) \hat \xi_j(\omega') \rangle= 2 \pi \delta_{ij}\,\delta(\omega+\omega')$. We have made use of the identity
\be \label{identity_hessian_fourier}
\sum_{l=1}^N [i \omega \mathbbm{1}_N + H]^{-1}_{jl} = \frac{1}{N} \sum_{q=0}^{N-1} \sum_{l=1}^N \frac{e^{2\pi i \frac{q}{N} (j-l)}}{i\omega + \mu_q} = \frac{1}{i\omega}
\ee
(all the terms in the sum over $q$ vanish except $q=0$, leading to the second equality).  From \eqref{eqfourier_RieszBrownian} we can express the two-point two-time correlation function, yielding after Fourier inversion
\be
\langle \delta x_j(t) \delta x_k(t') \rangle = 2 T \int_{-\infty}^{+\infty} \frac{d\omega}{2 \pi} e^{i \omega (t-t')} \left( [\omega^2 \mathbbm{1}_N + H^2]^{-1}_{jk} - \frac{1}{N\omega^2} \right) \;.
\ee
Using the eigensystem of $H$ given in \eqref{eigenvals_rieszgeneral}, this becomes
\bea \label{cov_brownian}
\langle  \delta x_j(t)  \delta x_k(t') \rangle &=& \frac{2 T}{N}  \sum_{q=1}^{N-1} \int_{-\infty}^{+\infty}  \frac{d\omega}{2 \pi} \frac{e^{i \omega (t-t')}}{\omega^2 + \mu_q^2 }
e^{2\pi i \frac{q}{N} (j-k)} \\
&=& \frac{2T}{N}  \sum_{q=1}^{(N-1)/2} \frac{e^{- \mu_q |t-t'|}}{\mu_q}  \cos\left({2\pi \frac{q}{N} (j-k)}\right) \;,  \nonumber
\eea
where we have used the symmetry $\mu_q=\mu_{N-q}$ in the last step. The last expression is exact only for odd values of $N$ (otherwise one simply needs to take the sum from $1$ to $N-1$ and remove the factor $2$), but since throughout the paper we will be focusing on the large $N$ limit this is irrelevant. Note that the average $\langle \delta x_i(t) \rangle$ vanishes to leading order in $T$. Hence \eqref{cov_brownian} actually gives the covariance of $x_i(t)$ and $x_j(t')$ to leading order in $T$ (after removing the center of mass).

\subsection{Specialization to the Riesz gas} \label{sec:Riesz_eigvals}

\noindent {\bf Definition of the periodized interaction.} We now specialize to the periodic Riesz gas. We define the periodized interaction potential through its derivative (regularizing the sum for $-1<s<0$, similar to, e.g., \cite{BoursierCLT,BoursierCorrelations}),
\be \label{defRiesz_periodized}
W'(x) = \begin{dcases}  
-g \lim_{n\to\infty} \left(\sum_{m=-n}^n \frac{{\rm sgn}(x+m L)}{|x+mL|^{s+1}}  \right)  \hspace{3.27cm} \text{for } -1<s < 0 \;, \\
-g \lim_{n\to\infty} \left(\sum_{m=-n}^n \frac{{\rm sgn}(x+m L)}{|x+mL|}  \right) = - g \frac{\pi}{L} \cot\big( \frac{\pi x}{L} \big)  \quad \text{for } s=0 \;, \\
- g \sum_{m=-\infty}^\infty \frac{{\rm sgn}(x+m L)}{|x+mL|^{s+1}}  \hspace{4.7cm} \text{for } s>0 \;. \\
\end{dcases}
\ee 
These sums can also be expressed using Hurwitz's zeta function, which for $r>1$ is defined by $\zeta(r,a)=\sum_{k=0}^{\infty} (k+a)^{-r}$. One has, for $0< x< 1$ and for $s>-1$,
\be \label{WHurwitz}
W'(x) = - \frac{g}{L^{s+1}} \left(\zeta(1+s,\frac{x}{L} ) - \zeta(1+s,1-\frac{x}{L}) \right) 
\ee 
(note that each term has a pole $1/s$ at $s=0$, which however cancels in the difference). Another way to define the periodized potential is to do it in Fourier space (see, e.g., \cite{Lewin,Lelotte2023}). In \cite{RieszFluct} (Appendix~E), we show that the two definitions are equivalent.

Note that the case $s=0$ corresponds to the Dyson Brownian motion on the circle with parameter $\beta=g/T$, studied, e.g., in \cite{Spohn3,ForresterCircularBM}.
Its equilibrium Gibbs measure is $\propto \prod_{i<j} |\sin( \frac{\pi}{L} (x_i-x_j)|^\beta$, and it is related with the CUE$(\beta)$ random matrix ensemble and the Calogero-Sutherland model of interacting fermions~\cite{Smith2021}.
\\

\noindent {\bf Expression of the eigenvalues $\mu_q$.} Let us now express the eigenvalues $\mu_q$ of the Hessian matrix ($q=0,...,N$), defined in \eqref{eigenvals_rieszgeneral} in the case of the Riesz gas. Taking a derivative of \eqref{defRiesz_periodized} gives (for $s>-1$)
\be \label{W2_Riesz}
W''(x) = (s+1)g \sum_{m=-\infty}^\infty \frac{1}{|x+mL|^{s+2}} \;.
\ee
Inserting into \eqref{eigenvals_rieszgeneral}, this leads to
\be 
\mu_q = \frac{2(s+1)g}{L^{s+2}} \sum_{m=-\infty}^\infty \sum_{\ell=1}^{N-1} \frac{ \sin^2(\frac{\pi q\ell}{N})}{|\frac{\ell}{N}+m|^{s+2}} = \frac{4(s+1)gN^{s+2}}{L^{s+2}} \sum_{\ell=1}^\infty \frac{ \sin^2(\frac{\pi q\ell}{N})}{\ell^{s+2}} \;,
\ee
where the second equality is obtained by combining both sums into a single sum over all integers $\ell$, and using the symmetry $\ell \to - \ell$. This can be written as
\be \label{mu_Riesz}
\mu_q = g\rho^{s+2} f_s\big( \frac{q}{N} \big) \;, \quad \text{with} \quad f_s(u)= 4(s+1) \sum_{\ell=1}^{\infty} \frac{\sin^2(\pi \ell u)}{\ell^{s+2}} \;.
\ee
This expression is exact for any $N$. The function $f_s(u)$ is defined on $[0,1]$ and satisfies $f_s(u)=f_s(1-u)$. For any $s>-1$, it is increasing on the interval $[0,1/2]$, from $f_s(0)=0$ to $f_s(1/2)=\sum_{\ell=1}^{\infty} \frac{4(s+1)}{(2\ell+1)^{s+2}}=4(1-2^{-(s+2)})(s+1)\zeta(s+2)$. When $s$ is an even integer, one can show that it takes a simple form, e.g.,
\be \label{fs_even}
f_0(u) = 2\pi^2 u(1-u) \quad , \quad f_2(u) = 2 \pi^4 u^2 (1-u)^2 \;.
\ee
We also note that as $s\to -1^+$, $f_s(u)$ has a finite limit which is independent of $u$, i.e., $f_s(u)\to2$ as $s\to -1^+$.

When $N$ is large, the sum in \eqref{cov_brownian} is dominated by small values of $q$, i.e., such that $q\ll N$. It is thus useful to analyze the behavior of $f_s(u)$ in the limit $u\ll 1$. In the long-range case $-1<s<1$, the sum over $\ell$ in \eqref{mu_Riesz}
can be replaced by an integral, leading to the asymptotic behavior
\be
f_s(u) \simeq 4(s+1) u^{s+1} \int_0^{+\infty} d\lambda \frac{\sin^2(\pi \lambda)}{\lambda^{s+2}} = 2\pi^{s+\frac{3}{2}} \frac{\Gamma(\frac{1-s}{2})}{\Gamma(1+\frac{s}{2})} u^{s+1}
\ee
(see Appendix~\ref{app:integrals} for alternative expressions using identities on the $\Gamma$-function, as well as some other useful integrals). In the short-range case $s>1$, the sum over $\ell$ in \eqref{mu_Riesz} is instead dominated by the first terms, and one can expand the sine function to obtain $f_s(u) \simeq 4\pi^2 (s+1) \zeta(s) u^2$, for $u \ll 1$. Finally, in the marginal case $s=1$, one has
\be \label{fasympt_s1}
f_1(u) = 4\pi^2 (3-2\log (2\pi u))u^2 + O(u^3) \;.
\ee
With the exception of the case $s=1$, this can be summarized as
\be \label{fasympt}
f_s(u) \underset{u\to 0}{\sim} a_s u^{z_s} \;, \quad \text{where} \quad z_s = \min(1+s,2) \quad \text{and} \quad a_s = \begin{cases} 2\pi^{s+\frac{3}{2}} \frac{\Gamma(\frac{1-s}{2})}{\Gamma(1+\frac{s}{2})} 
 \quad \hspace{0.28cm} \text{for } -1<s<1 \;, \\
4\pi^2 (s+1) \zeta(s)  \quad  \text{for } s>1 \;. \end{cases}
\ee
Below we will see that $z_s$ coincides with the dynamical exponent.
\\

\noindent {\bf Relevant timescales.} The inverse eigenvalues $1/\mu_q$ correspond to the relaxation timescales of the system at different lengthscales (respectively $L/(2q)$). Two of these timescales play a particularly important role. The smallest timescale, $1/\mu_{(N-1)/2}$, is the local relaxation time due to the interactions at the scale of the lattice spacing $L/N$. Below this timescale, the interactions do not play any role and we expect to recover free diffusion. For large $N$, it reads $1/\mu_{(N-1)/2} \sim 1/(g\rho^{s+2}) \equiv \tau $ (up to an irrelevant numerical factor). The other important timescale is the largest one, $1/\mu_1$, which corresponds to the relaxation at the scale $L$ of the full-circle (or rather $L/2$, which is the maximum possible distance between particles due to the periodicity). For large $N$ it reads $1/\mu_1 \sim  N^{z_s} \tau$, i.e., it diverges with the system size for any $s>-1$. The role of these timescales will be discussed further in Sec.~\ref{sec:dynamicalBrownian} when we will study the dynamical correlations.
\\

\noindent {\bf Comparison with the harmonic chain.} It is instructive to compare the present study with what we would obtain for a harmonic chain of particles, i.e., with only nearest neighbor harmonic interactions $W_{ij}(x_i-x_j)=\frac{1}{2}K(x_i-x_j)^2\delta_{i,j\pm1}$. For active particles this model was studied in \cite{HarmonicChainRevABP,SinghChain2020,PutBerxVanderzande2019,HarmonicChainRTPDhar} and reviewed in Sec.~\ref{sec:harmonicChain}. For Brownian particles it was found to be a good approximation for a short-range interaction with a hard-core repulsion part\cite{Lizana2010}. Indeed, in the short-range case the system is dominated by nearest-neighbor interactions and our linear approximation should be similar to the harmonic chain.

In the case of a harmonic chain, the equation \eqref{Eq_delta_x_RieszBrownian} is exact and the Hessian matrix reads $H_{ij}=K(2\delta_{i,j}-\delta_{i,j+1}-\delta_{i,j-1})$ (with periodic conditions). It is diagonalized by the same eigenvectors $v^q_k=\frac{1}{\sqrt{N}} e^{2\pi i \frac{q}{N} k}$, with eigenvalues given by
\be \label{eigvals_harmonic_chain}
\mu_q = 4K \sin^2 \left(\frac{\pi q}{N}\right) \quad , \quad q=1,...,N-1 \;.
\ee
The results of this chapter, as well as Chapter~\ref{chap:activeRieszFluct} for active particles thus also apply to the harmonic chain (even beyond the weak noise limit),
by replacing $\tau=1/(g\rho^{s+2}) \to \tau_K = 1/K$ and $f_s(u) \to f_{\rm harmo}(u)=4\sin^2(\pi u)$.
In particular, one has for $u \to 0$, $f_{\rm harmo}(u) \simeq 4\pi^2 u^2$. This means that for the asymptotic regimes which are dominated by the smallest eigenvalues, i.e., for most of the results below, the harmonic chain coincides with a short-range Riesz gas with the formal replacement $(s+1)\zeta(s) \to 1$.
\\

The results presented until now are exact at any $N$ (in the limit $T\to 0$). In the rest of this chapter, we consider several static and dynamical correlation functions and analyze them in the thermodynamic limit $N,L\to+\infty$ with fixed density $\rho=N/L$.

\section{Static correlations} \label{sec:staticBrownian}

\subsection{Variance and melting transition} \label{sec:var_Riesz_brownian}

We begin by simply computing the variance of the particle displacements $\langle \delta x_i^2\rangle$ at equilibrium\footnote{Note that by translational invariance all the quantities considered in this chapter and the next are independent of $i$.}. Evaluating \eqref{cov_brownian} for $i=j$ and $t=t'$ we obtain
\be \label{var_brownian}
\langle \delta x_i^2 \rangle = \frac{2T}{N} \sum_{q=1}^{(N-1)/2} \frac{1}{\mu_q} \simeq \frac{2T}{g\rho^{s+2}} \frac{N^{z_s-1}}{a_s} \sum_{q=1}^{\infty} \frac{1}{q^{z_s}} 
\ee
(where $a_s$ and $z_s$ are defined in \eqref{fasympt}). The second equality is valid at large $N$ for $s>0$. Indeed, in that case the sum is dominated by small values of $q$ and we can replace $\mu_q=g\rho^{s+2}f_s(u)$ by its asymptotic expression \eqref{fasympt}. Evaluating the sum and replacing $a_s$ by its expression we obtain
\be \label{var_Riesz_liquid}
\langle \delta x_i^2 \rangle = \begin{dcases} 
\frac{N^{s} T \zeta(s+1) \Gamma(1+\frac{s}{2})}{\pi^{s+\frac{3}{2}} \Gamma(\frac{1-s}{2}) g\rho^{s+2}}  \quad \text{for } 0<s<1\;, \\ 
\frac{NT}{12 (s+1) \zeta(s) g\rho^{s+2}} \quad \quad \, \text{for } s>1\;.
\end{dcases}
\ee
Thus the variance diverges with the system size, suggesting that the system is in a liquid phase. 

The limiting cases $s=1$ and $s=0$ have to be treated separately. For $s=1$, we find using the asymptotic expression \eqref{fasympt_s1}
\be \label{var_Riesz_s1}
\langle \delta x_i^2 \rangle \simeq \frac{N}{\log N} \frac{T}{24 \, g\rho^3} \;,
\ee
which again diverges with $N$. For the log-gas $s=0$, we can use the exact expression for $f_0(x)$ given in \eqref{fs_even}, which leads to
\be \label{varloggas}
\langle \delta x_i^2 \rangle = \frac{T}{2\pi^2 g\rho^2} \sum_{q=1}^{N-1} \frac{N}{q(N-q)} =\frac{T}{\pi^2 g\rho^2} \sum_{q=1}^{N-1} \frac{1}{q} = \frac{T}{\pi^2 g\rho^2} (\log N + \gamma_E + O(N^{-1})) \;,
\ee
where $\gamma_E$ is Euler's constant. Thus $\langle \delta x_i^2 \rangle$ again diverges at large $N$. For the log-gas, the absence of crystallisation transition was shown recently in \cite{leble_loggas}. In that case the system is in a liquid regime with translational quasi-order (see Sec.~\ref{sec:gapvariance_RieszBrownian}).

\begin{figure}
    \centering
    \includegraphics[width=0.45\linewidth]{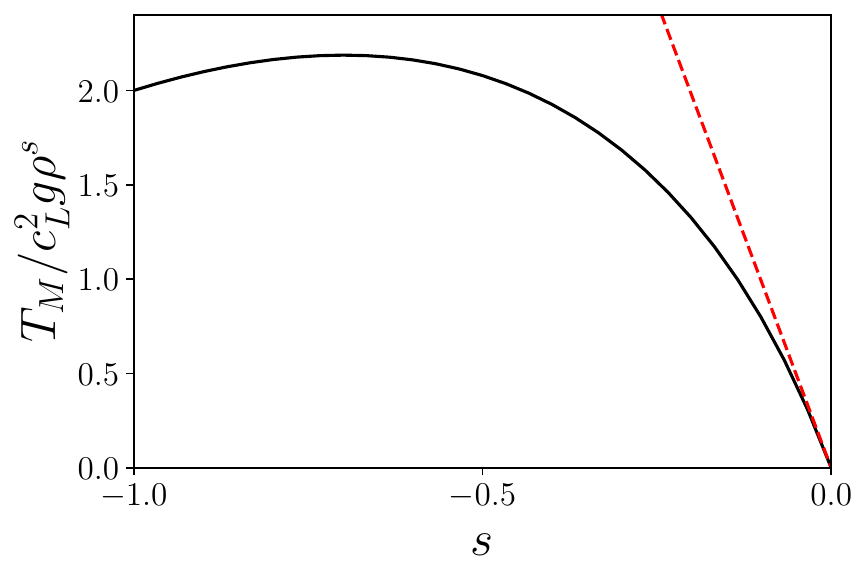}
    \caption{Plot of the ratio of the melting temperature over the Lindemann coefficient, $T_M/c_L^2$, versus $s$, for $-1<s<0$. The dashed red line shows the linear approximation in the limit $s\to 0^-$, $T_M \sim \pi^2 c_L^2 g |s|$.}
    \label{Fig_TM}
\end{figure}

For $-1<s<0$, the sum in \eqref{var_brownian} diverges. In this case, all the values of $q$ are relevant and we should instead replace the sum by an integral in the large $N$ limit,
\be \label{var_Riesz_solid}
\langle \delta x_i^2 \rangle \simeq \frac{2T}{g\rho^{s+2}} \int_0^{1/2} \frac{du}{f_s(u)} \;.
\ee 
Since $f_s(u) \sim a_s u^{s+1}$ as $u\to 0$, the integral is well-defined. Thus, in this case the variance of the displacement remains finite as $N\to+\infty$, which suggests the existence of a solid phase at low temperature, as observed recently in \cite{Lelotte2023}. The numerical results of \cite{Lelotte2023} also show evidence of a melting transition to a fluid phase at high temperature. We can obtain a rough estimate of the melting temperature $T_M$ (assuming that it exists) via a Lindemann argument by writing that at $T_M$, $\rho^2 \langle \delta x_i^2 \rangle = c_L^2$,
where $c_L$ is the phenomenological Lindemann coefficient (of order $c_L \simeq 0.05-0.2$ for 3D crystals) \cite{lindemann,navarro}. This gives
\be \label{T_melting}
T_M = \frac{1}{2} g\rho^{s} \frac{c_L^2}{\int_0^{1/2} \frac{du}{f_s(u)}} \;.
\ee 
The ratio $T_M/c_L^2$ is plotted in Fig.~\ref{Fig_TM}. It vanishes in the limit of the log-gas $s\to 0^-$, compatible with the numerical observations of \cite{Lelotte2023} and the absence of crystal phase for $s=0$. More precisely, in that limit the integral in the denominator is dominated by the edge behavior, so that we can replace $f_s(u)$ by its asymptotic expression \eqref{fasympt}, leading to 
$T_M \sim \pi^2 c_L^2 g |s|$ as $s\to 0^-$. Since our approximation holds at low temperature, we thus expect the estimate \eqref{T_melting} to be accurate for $s$ near zero. However, for $s\to -1^+$, we find that the ratio $T_M/c_L^2$ has a finite limit $2 g \rho^{-1}$. In addition, it is concave as a function of $s$ and has a local maximum on $[-1,0]$, see Fig. \ref{Fig_TM}. These observations seem incompatible with the numerical results of \cite{Lelotte2023}. In addition, for $s=-1$ (i.e., the 1D Coulomb gas) it was shown rigorously that the system is a crystal at any temperature \cite{Kunz,Aizenman1980,Lewin}. For our analysis to be compatible with these results would thus require the Lindemann coefficient $c_L$ to depend on $s$ in a singular way (at least for $s \lesssim -1/2$). 
Hence, a more predictive theory of the possible melting transition remains an open problem.

\subsection{Covariance at the macroscopic scale}

We now consider the covariance between the positions of two particles $\langle \delta x_i \delta x_{i+k}\rangle$ (here and below we take $k$ to be a positive integer). Since for $k\ll N$ this quantity coincides with the single-particle variance $\langle \delta x_i^2\rangle$ in the large $N$ limit, it is mostly relevant at the macroscopic scale, i.e., for $k = \kappa N$ with $\kappa = O(1)$. We find that, for $0<s<1$, it takes the following scaling form, for any $0<\kappa < 1$,
\be \label{cov_kappa_longrange}
\langle \delta x_i \delta x_{i+k}\rangle \simeq \frac{\Gamma(1+\frac{s}{2})}{\pi^{s+\frac{3}{2}} \Gamma(\frac{1-s}{2})} \frac{T N^s}{g\rho^{s+2}}  \sum_{q=1}^{\infty} \frac{\cos(2\pi \kappa q )}{q^{s+1}} \quad , \quad 
\kappa=\frac{k}{N} \;.
\ee
For $\kappa \to 0$ it indeed matches the on-site variance given in the first line of \eqref{var_Riesz_liquid}. In the short-range case $s>1$, we find
\be \label{cov_shortrange}
\langle \delta x_i \delta x_{i+k}\rangle \simeq \frac{NT}{12(s+1)\zeta(s) g\rho^{s+2}} (1 - 6\kappa(1-\kappa)) \quad , \quad 
\kappa=\frac{k}{N} \;,
\ee
which again matches the second line of \eqref{var_Riesz_liquid} for $\kappa \to 0$ (for $s=1$ it is similar with a scaling $N/\log N$). 
Note that the covariance decreases with $\kappa$ until it reaches a negative minimum at $\kappa=1/2$. This anti-correlation for very large separations is necessary to satisfy the condition $\sum_i \delta x_i=0$. Indeed, one can check that, considering $\kappa$ as a continuous parameter, the integral of this covariance over the whole system vanishes as it should. 
Note that, as a function of $\kappa$, the expression \eqref{cov_shortrange} coincides with the covariance of a Brownian bridge conditioned to have a zero total integral (see, e.g., \cite{BBridge}). 
The behavior near $\kappa=0$, which is non-analytic as $\sim\kappa^s$ in the long-range case, and which goes as $\sim\kappa$ in the short-range case, can
be shown to match the large $k$ behavior of the gaps, which we discuss in the next subsection.

In the case of the log-gas $s=0$ we find, for $0<\kappa<1$,
\be
\langle \delta x_i \delta x_{i+k}\rangle \simeq \frac{T}{\pi^2 g\rho^2}  \sum_{q=1}^{\infty} \frac{\cos(2\pi \kappa q )}{q} = - \frac{T}{\pi^2 g\rho^2} \log (2\sin(\pi \kappa)) \;.
\ee
This formula can be compared to the variance of the number of eigenvalues for CUE($\beta$) in a mesoscopic interval on the circle \cite{HughesCircle2001,NajnudelCircle2018}. Note that this expression diverges in the limit $\kappa \to 0$, but that by taking $k$ of order $O(1)$ we recover
the $\log N$ behavior of the variance in \eqref{varloggas}.

Finally, the case $-1<s<0$ is quite different since then the covariance remains finite as $N\to+\infty$, and we find that it decays on a scale $k=O(1)$ as $k^{-|s|}$. For $k$ of order $N$, the covariance is still given by the long-range result \eqref{cov_kappa_longrange} but is very small since it now scales as $N^{-|s|}$.

\subsection{Variance of the interparticle distance} \label{sec:gapvariance_RieszBrownian}

We now study the statistics of the gaps, i.e., of the distance between two particles $i$ and $i+k$, at equilibrium (which we have discussed very briefly in Sec.~\ref{sec:gap_varRiesz_review}). We first consider its variance, given by (using again \eqref{cov_brownian})
\be \label{gap_brownian}
D_k(0) = \langle (\delta x_{i}-\delta x_{i+k})^2 \rangle = \frac{4T}{N} \sum_{q=1}^{(N-1)/2} \frac{1-\cos\left( \frac{2\pi kq}{N} \right)}{\mu_q} = \frac{8T}{N} \sum_{q=1}^{(N-1)/2} \frac{\sin^2\left( \frac{\pi kq}{N} \right)}{\mu_q} \;.
\ee
At large $N$, for $k\ll N$, we can replace the sum by an integral for any $s>-1$, leading to
\be \label{gaps_step1_brownian}
D_k(0) \simeq \frac{8T}{g\rho^{s+2}} \int_0^{1/2} du \frac{\sin^2(\pi k u)}{f_s(u)}  \;.
\ee
This integral is always well-defined since $f_s(u)=O(u^2)$ for $u\to 0$. It is instructive to analyze this integral in the regime $1\ll k\ll N$. For $s>0$, the integral is dominated by $u\sim1/k\ll 1$, and we can use again the asymptotics \eqref{fasympt} for $f_s(u)$ to obtain
\be \label{gapss}
D_k(0) \simeq \frac{8T}{g\rho^{s+2}k} \int_0^{k/2} dv \frac{\sin^2(\pi v)}{f_s\big(\frac{v}{k}\big)} \simeq \frac{8Tk^{z_s-1}}{g\rho^{s+2}a_s} \int_0^{+\infty} dv \frac{\sin^2(\pi v)}{v^{z_s}} \simeq \frac{4T}{g\rho^{s+2}a_s} \frac{\pi^{z_s-\frac{1}{2}}}{z_s-1} \frac{\Gamma(\frac{3-z_s}{2})}{\Gamma(\frac{z_s}{2})}  k^{z_s-1} \;,
\ee
where we recall that $z_s=\min(s+1,2)$ and $a_s$ is given in \eqref{fasympt}. This can be rewritten as
\be \label{gapscases_brownian}
D_k(0) \simeq \begin{dcases} \frac{T k^s}{\pi \tan \left( \frac{\pi s}{2} \right) g \rho^{s+2}}  \quad \, \text{for } 0<s<1 \;, \\
\frac{T k}{(s+1) \zeta(s) g\rho^{s+2}} \quad \text{for } s>1 \;. \end{dcases}
\ee
For $s=1$, we find in a similar way $D_k(0) \simeq\frac{Tk}{2g\rho^3 \log (k)}$. 
The field of displacements thus exhibits a roughness exponent $\zeta=s/2$ in the long-range case, while $\zeta=1/2$ in the short-range case. This result can be compared with the recent mathematical works \cite{BoursierCLT,BoursierCorrelations}, where the same power law behavior $k^s$ was found for $0<s<1$ (and the scaled distribution of the gaps was shown to converge to a Gaussian).

In the log-gas case $s=0$, we can use the exact expression of $f_0(u)$ in \eqref{fs_even}, and we find for any $k \ll N$
\be \label{gaps_s0}
D_k(0) \simeq \frac{4 T}{\pi^2 g \rho^2} \int_0^{1/2} du \frac{\sin^2(\pi k u)}{u(1-u)} 
= \frac{2 T}{\pi^2 g \rho^2} (\log (2 \pi k)-\text{Ci}(2 \pi k)+\gamma_E) \underset{k\to+\infty}{\simeq} \frac{2 T}{\pi^2 g \rho^2} \log(k) \;,
\ee
where ${\rm Ci}(x)=-\int_x^\infty dt\frac{\cos t}{t}$ is the cosine integral. We thus recover the well-known $\log k$ behavior \cite{Mehta_book,Forrester_book,Bourgade2022}.

Finally, for $s<0$ the variance of the gaps saturates to a constant at large $k$ (obtained by replacing $\sin^2(\pi k u) \to 1/2$ in \eqref{gaps_step1_brownian}),
\be \label{gaps_sneg}
D_k(0) \simeq \frac{4T}{g\rho^{s+2}} \int_0^{1/2}  \frac{du}{f_s(u)}  = 2 \langle \delta x_i^2 \rangle \;.
\ee
Indeed in this case the displacements become independent at large distance, each with variance given in \eqref{var_Riesz_solid}.
\\


\noindent {\bf Counting statistics.} As discussed in Chapter~\ref{chap:Riesz_review}, the variance of the gaps is related to the variance of the number of particles inside a fixed interval $[a,b]$, denoted as ${\cal N}_{[a,b]}$. More precisely, if the interval $[a,b]$ is sufficiently large one has
\be 
{\rm Var} \, {\cal N}_{[a,b]} \simeq \rho^2 \langle (\delta x_i-\delta x_{i+k})^2 \rangle = \rho^2 D_k(0) \;,
\ee 
where $a=x_i^0$, $b=x_{i+k}^0$, hence $\rho(b-a)=k$. The $D_k(0)\sim \hspace{-0.05cm} k$ behavior in the short-range is thus compatible with Poissonian statistics at the scale $1\ll k\ll N$, while the $\sim \hspace{-0.05cm} k^s$ behavior for $0<s<1$ is a sign of the rigidity of the system, i.e., the fluctuations are reduced by the long-range interaction. In the log-gas case $s=0$, we recall that, for $\rho (b-a) \gg 1$,
\be \label{varinterval_DBM}
{\rm Var} \, {\cal N}_{[a,b]} \simeq \frac{2}{\pi^2 \beta} (\log (\pi \rho (b-a)) + c_\beta) \;,
\ee 
where $\beta=g/T$ and $c_\beta$ is a constant. For $\beta=1,2,4$ this is the Dyson-Mehta formula \cite{Mehta_book,Forrester_book}.
For general $\beta$, the leading term was proved in \cite{Bourgade2022} and 
the constant $c_\beta$ was computed explicitly in \cite{Smith2021}. One can check that the leading behavior is indeed compatible with our formula \eqref{gaps_s0}.
\\

\noindent {\bf Translational order correlation function.} Our results for the gap variance also allow us to study the translation order correlation function, defined as
\be 
S(k) = \langle e^{ 2 i \pi \rho (\delta x_{j+k}-\delta x_j) } \rangle \;,  
\ee 
where $1/\rho$ represents the lattice spacing. True translational order is present when $S(k)$ converges to a non-zero constant at large $k$, and absent if it
decays to zero. When it decays to zero as a power law in $k$, this is usually referred to as quasi-ordered. Since the gaps are Gaussian within our weak noise approximation, we have the relation
\be \label{gaussiandecay_brownian} 
S(k) 
\simeq e^{ - 2 \pi^2 \rho^2 \langle (\delta x_{j+k}-\delta x_j)^2 \rangle } \;,
\ee 
and we can use our results above to determine the behavior of $S(k)$ at large $k$. 
For the case of the log-gas $s=0$, we find using \eqref{gaps_s0} a power law decay, indicating quasi-order,
\be
S(k)  \underset{k \gg 1}{\simeq} \,    (e^{\gamma_E} 2 \pi \, k)^{-\frac{4 T}{g}} \;.
\ee 
The exponent $4 T/g=4/\beta$ is in agreement with the known results for the decay of the oscillating part of the density correlation \cite{Haldane,Forrester1984}. See also \cite{Lelotte2023} for a recent discussion. 


For $0<s<1$ we see, from \eqref{gaussiandecay_brownian} and \eqref{gapscases_brownian}, that $S(k)$ decays to zero at large $k$ as a stretched exponential, $\log S(k) \propto - k^s$, while the decay is exponential for $s>1$. By contrast, for $s<0$ we find that $S(k)$ saturates to a non-zero value at large $k$, indicating true translational order at low temperature, as discussed in Sec.~\ref{sec:var_Riesz_brownian}.
\\

\begin{figure}
    \centering
    \includegraphics[width=0.45\linewidth]{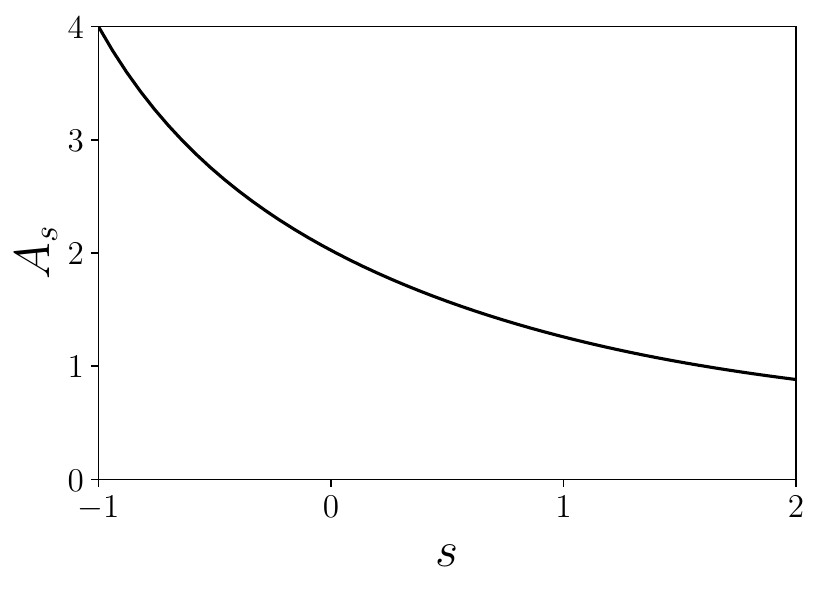}
    \caption{Plot of $A_s$ versus $s$, as defined in \eqref{validity_brownian}, which gives the temperature $T_G=A_s g \rho^s$ below which the present method should be valid.}
    \label{FigAs} 
\end{figure}

\noindent {\bf Validity of the approximation.} To linearize the equations of motion at the beginning of this chapter, we assumed that the relative displacements were sufficiently small, see \eqref{cond_approx}. We can now use our results on the gap variance to make this assumption more precise. The condition \eqref{cond_approx} can be interpreted as,
\be \label{cond_approx_Riesz1}
\forall 1\leq k\leq N-1, \ \sqrt{D_k(0)} = \sqrt{\langle (\delta x_i-\delta x_{i+k})^2\rangle} \ll 2\left|\frac{W''(x_i^0-x_{i+k}^0)}{W'''(x_i^0-x_{i+k}^0)}\right| \;.
\ee
Using the result \eqref{gaps_step1_brownian}, one can show that a sufficient condition for \eqref{cond_approx_Riesz1} is given by (see Sec.~III.E in \cite{RieszFluct})
\be \label{validity_brownian}
T \ll T_G = A_s  g\rho^s   \quad , \quad A_s^{-1} = 2(s+2)^2\int_0^{1/2} du \frac{\sin^2(\pi u)}{f_s(u)} \;.
\ee  
Our linear approximation should thus be accurate below the temperature $T_G$. The amplitude $A_s$ decreases monotonously with $s$, from $A_s\to 4$ as $s\to-1$ to $A_s\to0$ as $s\to+\infty$, with for instance $A_0=2.02441$ and $A_2=0.883237$. It is plotted in Fig. \ref{FigAs}. For $s<0$, one can show that our estimate for the melting temperature $T_M$ given in \eqref{T_melting} is always smaller than $T_G$ (assuming $c_L<1/2$). In particular, the ratio 
\be
\frac{T_M}{T_G} = c_L^2(s+2)^2 \frac{\int_0^{1/2} du \frac{\sin^2(\pi u)}{f_s(u)}}{\int_0^{1/2} \frac{du}{f_s(u)}}
\ee
vanishes as $s \to 0^-$. This confirms that our approximation should {\it a priori} be valid up to the estimated melting temperature, at least close to $s=0$.

\subsection{Spatial correlations of the gaps} 

\begin{figure}
    \centering
    \includegraphics[width=0.6\linewidth,trim={0 4.5cm 0 7cm},clip]{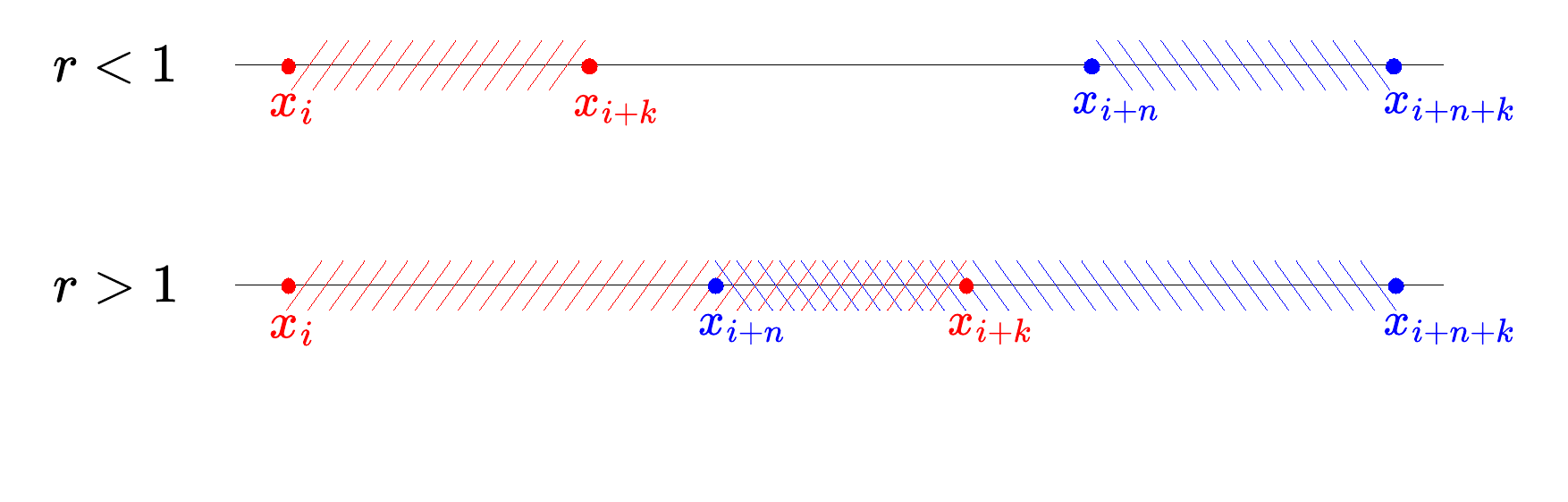}
    \includegraphics[width=0.39\linewidth,trim={0 0.3cm 0 0},clip]{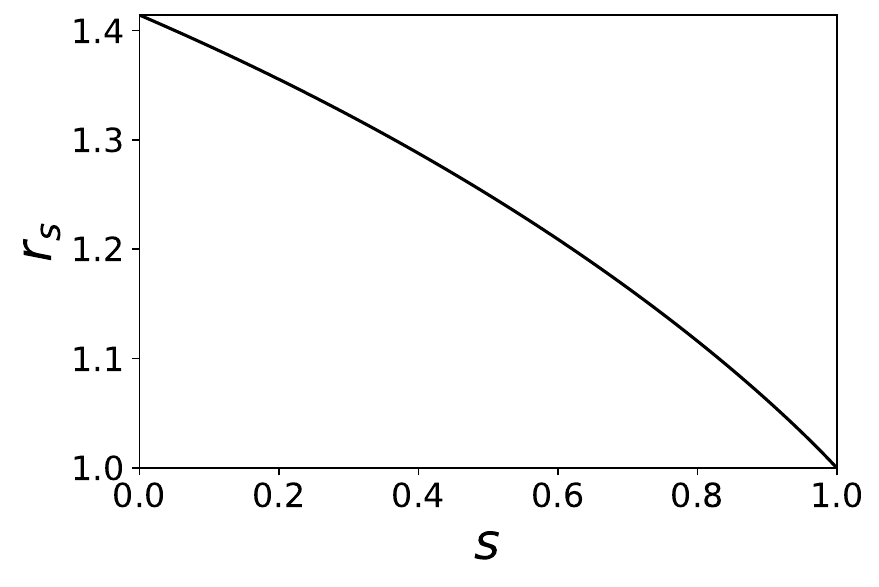}
    \caption{{\bf Left:} $D_{k,n}(0)$ (see, e.g., Eq. \eqref{LRgapcorr}) measures the covariance between the red gap and the blue gap, where the $x_i$ are the positions of the particles in the Riesz gas. For $r=k/n<1$ the two intervals are non overlapping, while for $r>1$ the two interval overlap.
    {\bf Right:} Plot of the value $r_s$ for which $D_{k,n}(0)$ changes sign, versus the Riesz interaction exponent $s$.}
    \label{gaps_sketch}
\end{figure}

To conclude our study of the static correlations, let us give some results for the spatial correlations of the gaps. More precisely, we consider the covariance $D_{k,n}$ between two gaps of sizes $k$, shifted by $n$, see Fig.~\ref{gaps_sketch} (both variables $k$ and $n$ are assumed to be positive integers),
\be \label{gapcor_brownian}
D_{k,n}(0) = \langle (\delta x_{i}-\delta x_{i+k}) (\delta x_{i+n}-\delta x_{i+n+k}) \rangle 
= \frac{8T}{N} \sum_{q=1}^{(N-1)/2} \frac{\sin^2\left( \frac{\pi kq}{N} \right)}{\mu_q}  \cos\left(2 \pi \frac{q}{N} n\right) \;.
\ee
In the long-range case $0<s<1$ we find that for $k,n,|n-k| \gg 1$, this covariance behaves as
\be \label{LRgapcorr}
D_{k,n}(0) \simeq \frac{T}{g\rho^{s+2}a_s} \frac{\pi^{s+1}}{2^{s}} \frac{|n-k|^s + (n+k)^s -2n^s}{\sin(\frac{\pi s}{2})\Gamma(1+s)} \;,
\ee
where $a_s$ is given in \eqref{fasympt}. It thus exhibits an algebraic roughness at large separations. Interestingly, this correlation is negative for $k \leq n$, 
i.e., two gaps which are disjoint are always anti-correlated. This could be expected since, if a given gap expands, the surrounding gaps are more likely to be reduced. On the contrary, if two gaps strongly overlap we expect the correlation to be positive. Indeed, denoting $r=k/n$, such that the ``overlap ratio" is $\frac{k-n}{k} = (r-1)/r$, we find that the correlations become positive at a critical value $k/N=r_s>1$, which is given by the root of the equation $(r-1)^s+(r+1)^s=2$ and is plotted in Fig.~\ref{gaps_sketch}. 

If we instead consider the regime where $n$ is large while $k$ remains fixed, we find
\be \label{Dkn_large_n}
D_{k,n}(0) \simeq -\frac{T}{g\rho^{s+2}a_s} \frac{\pi^{s+1}}{2^{s}} \frac{s(1-s) \, k^2}{n^{2-s}\sin(\frac{\pi s}{2})\Gamma(1+s)} \;.
\ee 
The dependence $\sim 1/n^{2-s}$ is consistent with the rigorous bounds obtained in \cite{BoursierCorrelations} (however, the prefactor and its $k$ dependence were not obtained there). 

For the 1D log-gas $s=0$, we find in the regime $k,n,|n-k| \gg 1$ (all three of the same order),
\be \label{gapslogDkn}
D_{k,n}(0) \simeq \frac{T}{\pi^2 g\rho^2} \log |1-\frac{k^2}{n^2}| \;.
 \ee
For $n/k \gg 1$ this gives $D_{k,n}(0) \simeq -\frac{T k^2}{\pi^2 g\rho^2 n^2}$, recovering the $1/n^2$ decay which for the log-gas was proved rigorously in \cite{ErdosCorrelations}. Note that this coincides with the limit $s \to 0$ of \eqref{Dkn_large_n}. For the log-gas, the covariance $D_{k,n}(0)$ is negative for $r=\frac{k}{n} <\sqrt{2}$ and positive for $r= \frac{k}{n} >\sqrt{2}$, in agreement with the limit $r_{s=0^+}=\sqrt{2}$. Let us add that the expression \eqref{gapslogDkn} is invalid in the regime $|n-k| = O(1)$. In this case there is another formula, given in Sec.~III.E of \cite{RieszFluct}, i.e., the divergence for $k/n=1$ in \eqref{gapslogDkn} is only apparent.

Finally, in the short-range case $s>1$, we find that the correlations are important only when the two intervals overlap. For $k,n$ large with $r=k/n$ fixed we obtain
\be
\frac{1}{n} D_{k,n}(0) \simeq  \begin{cases} 0  \hspace*{2.1cm} \quad \text{if } r\leq 1 \;, \\ \frac{4 \pi^2 T}{g\rho^{s+2}a_s} (r-1)  \quad \text{if } r>1 \;. \end{cases}
\ee
A more precise estimate of the residual correlations for $r<1$ is obtained in Sec.~III.E of \cite{RieszFluct}.

\section{Dynamical correlations} \label{sec:dynamicalBrownian}

\subsection{Mean squared displacement and two-time correlations} \label{sec:MSD_brownian}

We now move on to the study of the dynamical correlations, starting with the mean squared displacement (MSD) of a particle during time $t$. Using \eqref{cov_brownian}, it reads, within our weak noise approximation,
\be \label{disp_brownian}
C_0(t) = \langle (\delta x_i(t) - \delta x_i(0))^2 \rangle 
= \frac{4T}{N} \sum_{q=1}^{(N-1)/2} \frac{1 - e^{- \mu_q t}}{\mu_q} \;.
\ee
We recall that here we have chosen an {\it annealed} initial condition, meaning that at $t=0$ the system is at Gibbs equilibrium. The case of  a quenched initial condition will be discussed in Sec.~\ref{sec:quenchedBrownian}.

\begin{figure}
    \centering
    \includegraphics[width=0.8\linewidth]{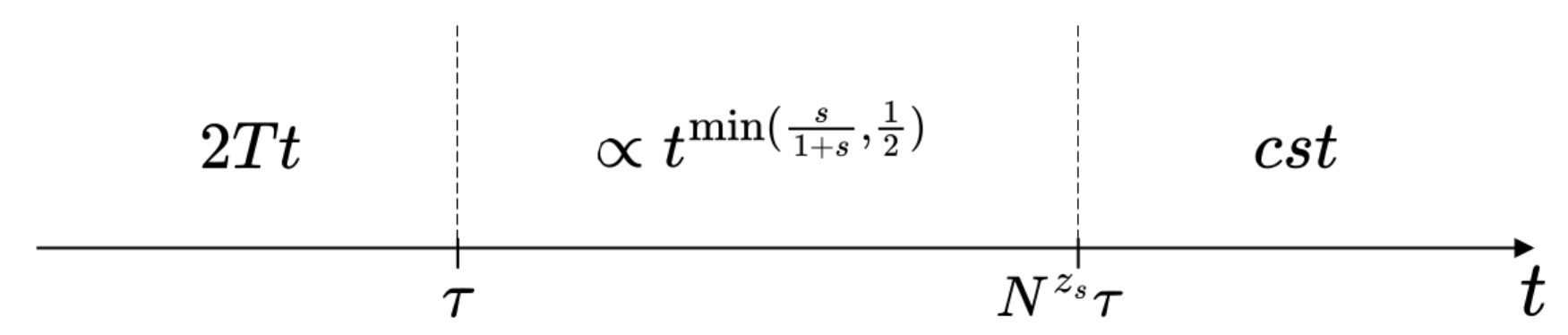}
    \caption{Representation of the three different regimes in the time evolution of the variance $C_0(t)$ of the displacement of a particle during time $t$, in a Riesz gas of Brownian particles with $s>0$. It behaves diffusively for $t\ll \tau$, followed by a crossover to a subdiffusive regime, until saturation to the value $2 \langle \delta x_i^2 \rangle$ (simply denoted as ``{\it cst}") for $t\gg N^{z_s} \tau$. }
    \label{fig:time_regimes_brownian}
\end{figure}

As we discussed briefly in Sec.~\ref{sec:Riesz_eigvals}, there are 3 different time regimes, represented in Fig.~\ref{fig:time_regimes_brownian}. For $t\ll\tau=1/(g\rho^{s+2})$, i.e., when $t$ is smaller than the smallest inverse eigenvalue $1/\mu_{(N-1)/2}$, we can expand the exponential in each term of the sum, which yields
\be
C_0(t) \simeq 2T \big( 1 - \frac{1}{N} \big) t \;,
\ee
where the missing term $2Tt/N$ comes from our subtraction of the center of mass. Thus, the interaction does not play any role on this timescale and we exactly recover free diffusion for any $s>-1$.

For $s>0$, it takes a time $t\gg N^{z_s}\tau$ for the system to equilibrate at the global scale. On this timescale, all the exponentials in \eqref{disp_brownian} converge to zero and we recover the variance computed in \eqref{var_Riesz_liquid} with an additional factor 2, i.e., $C_0(t)\to 2\langle\delta x_i^2\rangle$ (with $\langle\delta x_i^2\rangle\propto N^{z_s-1}$), meaning that $\delta x_i(t)$ has become independent of $\delta x_i(0)$. There is thus a broad intermediate time regime $\tau\ll t\ll N^{z_s}\tau$, where the time evolution of the particles strongly depends on the interaction. In the large $N$ limit, for $t\ll N^{z_s}\tau$, we can replace the sum in \eqref{disp_brownian} by an integral, yielding
\be \label{disp_step1_brownian}
C_0(t) = \langle (\delta x_i(t) - \delta x_i(0))^2 \rangle 
\simeq \frac{4T}{g\rho^{s+2}} \int_0^{1/2} du \frac{1 - e^{- g\rho^{s+2} f_s(u) t}}{f_s(u)} \;.
\ee
For $t \ll \tau=1/(g\rho^{s+2})$, this expression recovers free diffusion, $C_0(t)\simeq 2Tt$. Instead, at large time $t \gg \tau$, this integral is dominated by small values of $u$, and we can use \eqref{fasympt} to write (for $s\neq 1$)
\be
C_0(t) \simeq 4Tt \int_0^{1/2} du \frac{1 - e^{- g\rho^{s+2} a_s u^{z_s} t}}{g\rho^{s+2} a_s u^{z_s} t} \simeq \frac{4Tt}{(g\rho^{s+2} a_s t)^{1/z_s}} \int_0^{+\infty} dv \frac{1-e^{-v^{z_s}}}{v^{z_s}} = \frac{4Tt^{\frac{z_s-1}{z_s}}}{(g\rho^{s+2} a_s)^{1/z_s}} \frac{\Gamma(1/z_s)}{z_s-1}  \;.
\ee
Replacing $z_s=\min(1+s,2)$ and $a_s$ as in \eqref{fasympt}, we obtain
\be \label{displacement_Riesz_brownian}
C_0(t) \simeq \begin{dcases} U_s \frac{T \, t^{\frac{s}{s+1}}}{g^{\frac{1}{s+1}}\rho^{\frac{s+2}{s+1}}}
\hspace{2.22cm} \text{for } 0<s<1 \;, \\ 
\frac{2T}{\sqrt{\pi (s+1) \zeta(s)}} \sqrt{\frac{t}{g\rho^{s+2}}}
\quad \text{for } s>1 \;, \end{dcases}
\ee
where
\be \label{defUs}
U_s = \frac{ 4\Gamma\left(\frac{1}{s+1}\right)}{\pi s} \left[ \frac{\Gamma\left(1+\frac{s}{2}\right)}{2\sqrt{\pi} \, \Gamma\left( \frac{1-s}{2} \right)} \right]^{\frac{1}{s+1}} \;.
\ee
We thus obtain a subdiffusive regime, where in the short-range case $s>1$, we recover the $\sqrt{t}$ behavior of single-file diffusion, while in the long range case $0<s<1$ the exponent varies continuously between $0$ and $1/2$. Remarkably, these results coincide exactly, including the prefactors (at any $T$ for $0<s<1$ and at $T\ll g\rho^s$ for $s>1$) with the expressions obtained in the recent work \cite{DFRiesz23}. In this work the authors used a completely different method based on a study of the density field using macroscopic fluctuation theory (MFT). It is quite surprising that our approach, based on the linearization of the equations of motion and which is {\it a priori} restricted to small temperatures, allows to obtain the same asymptotics (in particular in the long-range case). In addition, as we have seen the present method allows to go beyond the asymptotics and also describes the crossover at small and large times, for sufficiently low temperatures.

Once again, the marginal cases $s=1$ and $s=0$ (which were not treated in detail in \cite{DFRiesz23}) need to be considered separately. For $s=1$, using the asymptotics \eqref{fasympt_s1} in \eqref{disp_step1_brownian}, we find, again for $t \gg \tau = 1/(g\rho^3)$, 
\bea \label{disp_s1}
C_0(t) \simeq 2T \sqrt{\frac{t}{\pi g\rho^3 \log(g\rho^3 t)}} \;.
\eea

In the special case of the log-gas, it is possible to compute explicitly the full crossover function between short time and large time. Indeed, for $s=0$, inserting the exact expression of $f_0(u)$ from \eqref{fs_even} into \eqref{disp_step1_brownian} yields
\be
C_0(t) \simeq 4T \int_0^{1/2} du \frac{1 - e^{- 2\pi^2 g\rho^2 u(1-u) t}}{2\pi^2 g\rho^2 u(1-u)} \;,
\ee
which can be evaluated as (e.g., by first computed its time derivative and then integrating with $C_0(t=0)=0$)
\be \label{crossoverhypergeo}
C_0(t) \simeq 2T t  \times \, _2F_2\left(1,1;\frac{3}{2},2;-\frac{\pi^2}{2} g\rho^2 t\right) \;,
\ee
which holds at large $N$ for any $t\ll N\tau$ (with $_2F_2$ denoting the hypergeometric function). This expression thus describes the full crossover from the short time diffusing regime $t\ll \tau=1/(g\rho^2)$, where it gives the correction
\be
C_0(t) \simeq 2T t - \frac{\pi^2}{3} g\rho^2 t^2 +O(t^3) \;,
\ee
to the large time subdiffusive regime $t\gg\tau$, where it yields
\be \label{disp_s0_large_t}
C_0(t) \simeq \frac{2T}{\pi^2 g\rho^2} \left( \log \big(2\pi^2 g\rho^2 t\big) + \gamma_E - \frac{1}{\pi^2g\rho^2 t} \right) + O(\frac{1}{t^2}) \;.
\ee
The leading logarithmic behavior agrees, including the prefactor, with the exact result from \cite{SpohnTracer}. Note that the corrections are power law in time.
\\

The case $s<0$ is quite different from the behavior described above for $s\geq 0$. Indeed in this case, we have seen in Sec.~\ref{sec:var_Riesz_brownian} that the variance is independent of $N$. Thus, $C_0(t)$ converges over a finite time $\sim\tau$ to its stationary value $2\langle \delta x_i^2 \rangle \sim T \tau$ given in \eqref{var_Riesz_solid}). The crossover towards this limit is given by
\be \label{decroissance}
2\langle \delta x_i^2\rangle - C_0(t) \simeq \frac{4T}{g\rho^{s+2}} \int_0^{1/2} du \frac{e^{- g\rho^{s+2} f_s(u) t}}{f_s(u)} \simeq \frac{4T}{(a_s g\rho^{s+2})^{\frac{1}{1+s}}} \frac{\Gamma(\frac{|s|}{1+s})}{1+s} t^{-\frac{|s|}{1+s}} \;,
\ee
which exhibits a power law decay in time.
\\

\noindent {\bf Two-time correlations.} Our results for the MSD $C_0(t)$ allow to directly obtain similar results for the two-time correlations of the particle displacements, given by
\be \label{multitime_brownian}
C_0(t_1,t_2) = \langle (\delta x_i(t_1) - \delta x_i(0)) (\delta x_{i}(t_2) - \delta x_{i}(0))   \rangle 
= \frac{2T}{N} \sum_{q=1}^{(N-1)/2}  \frac{1 - e^{- \mu_q t_1} -  e^{- \mu_q t_2} + e^{- \mu_q |t_1-t_2|} }{\mu_q} \;.
\ee
Indeed, we simply need to notice that (in the annealed case considered here)
\be
C_0(t_1,t_2) = \frac{1}{2} [ C_0(t_1) + C_0(t_2) - C_0(|t_1-t_2|) ] \;.
\ee
Thus, at small times $t_1,t_2,|t_1-t_2|\ll \tau = 1/(g \rho^{s+2})$ we recover the free diffusion result $C_0(t_1,t_2)=2 T \min(t_1,t_2)$, while at large times $t_1,t_2,|t_1-t_2|\gg \tau$, we find
\be \label{multitime_corr_Riesz}
C_0(t_1,t_2) \simeq \begin{dcases} \frac{U_s T}{2g^{\frac{1}{s+1}}\rho^{\frac{s+2}{s+1}}} (t_1^{\frac{s}{s+1}} + t_2^{\frac{s}{s+1}} - |t_1-t_2|^{\frac{s}{s+1}})\;,
\quad \hspace{1.2cm} \text{for } 0<s<1 \;, \\ 
\frac{T}{\sqrt{\pi (s+1) \zeta(s)g\rho^{s+2}}} (\sqrt{t_1} + \sqrt{t_2} - \sqrt{|t_1-t_2|})\;,
\quad \text{for } s>1 \;, \end{dcases}
\ee
where the prefactor $U_s$ was defined in \eqref{defUs}. As for $C_0(t)$, these results coincide exactly with the ones from \cite{DFRiesz23} obtained using MFT. As noted there, this correlation function coincides with the one of a fractional Brownian motion (fBm) of Hurst index $H=s/(2 (s+1))$ in the long-range case and $H=1/4$ in the short-range case. For $s=1$ there are additional logarithmic corrections (as for $C_0(t)$). For the log-gas $s=0$, the result \eqref{crossoverhypergeo} allows to obtain explicitly the full crossover function using hypergeometric functions. At large times $t_1, t_2, |t_1-t_2| \gg\tau$, we find
\be  \label{res_C0}
C_0(t_1,t_2) \simeq \frac{T}{\pi^2 g \rho^2} \left( \log \big( \frac{2 \pi^2 g \rho^2 t_1 t_2}{|t_1-t_2|} \big) +\gamma_E \right) \;.
\ee 
Note that, in all these cases, $C_0(t,t')$ is not simply a function of $t-t'$, similar to what is observed in aging for some simple systems \cite{CugliandoloAging}.

\subsection{Other dynamical quantities}

The method presented in this chapter also allows to compute some more complex dynamical quantities, for which we present a few results below.
\\

\noindent {\bf Equal time covariance.} A first quantity of interest is the covariance of the displacement during time $t$ of two particles $i$ and $i+k$ (where $k$ is again assumed to be a positive integer),
\be \label{Ck_def_brownian}
C_k(t) = \langle (\delta x_i(t) - \delta x_i(0)) (\delta x_{i+k}(t) - \delta x_{i+k}(0)) \rangle = \frac{4T}{N} \sum_{q=1}^{(N-1)/2}  \frac{1 - e^{- \mu_q t}}{\mu_q} \cos\big( 2 \pi \frac{q k}{N} \big) \;.
\ee 
We found that for $s>0$, this quantity takes the following scaling form for $\tau\ll t\ll N^{z_s}\tau$ and $1\ll k\ll N$, with $k \sim (t/\tau)^{1/z_s}$,
\be \label{Ck_scaling_brownian}
C_k(t) \simeq \begin{dcases} T \left(\frac{t^s}{g\rho^{s+2}}\right)^\frac{1}{s+1} F_s\left( \frac{k}{(g\rho^{s+2}t)^{\frac{1}{s+1}}} \right) \quad \text{for } 0<s<1 \;, \\
T \sqrt{\frac{t}{g\rho^{s+2}}} \, F_s\left( \frac{k}{\sqrt{g\rho^{s+2}t}} \right) \quad \hspace{1.4cm} \text{for } s>1 \;. \end{dcases}
\ee 
We see that $z_s = \min(1+s,2)$ plays the role of a dynamical exponent, as announced previously. For $0<s<1$, the scaling function $F_s(x)$ is given by
\be \label{Fs_def1}
F_s(x) = \frac{4}{a_s^{\frac{1}{s+1}}} \int_0^{+\infty} dv \frac{1-e^{-v^{s+1}}}{v^{s+1}} \cos \left( 2\pi a_s^{-\frac{1}{s+1}} x v \right)  \;.
\ee 
For small argument $x \sim k/(t/\tau)^{1/z_s} \ll 1 $ it behaves as $F_s(x) \simeq  U_s - \frac{1}{\pi \tan( \frac{\pi s}{2}) } |x|^s + \dots$, matching the result for $C_0(t)$ given in \eqref{displacement_Riesz_brownian}. In addition, this implies 
$C_0(t) - C_k(t) \simeq D_k(0) \sim k^s$ where $D_k(0)$ is the variance of the gaps given in \eqref{gapscases_brownian} (see below for the relation between these observables). For large argument $x \gg 1$, it decays as a power law $F_s(x) \sim 1/|x|^{2+s}$, leading to a decay of the correlation function as $C_k(t) \sim t^2/k^{2+s}$ for large separations $k/(t/\tau)^{1/z_s} \gg 1$. Interestingly, the behavior in time is thus ballistic at large distances. This might be due to the fact that only fast propagating excitations (i.e., ballistic) survive on these lengthscales. 

In the short range case $s>1$, we find that the scaling function $F_s(x)$ is independent of $s$ up to a rescaling,  
\be \label{Fs_SR} 
F_s(x) = \frac{1}{\sqrt{(s+1)\zeta(s)}} F_1\left( \frac{x}{\sqrt{(s+1) \zeta(s)} } \right) \quad \text{with} \quad
F_1(x) = \frac{2}{\sqrt{\pi}} e^{-\frac{x^2}{4}} + |x| \left({\rm erf}\big(\frac{|x|}{2}\big)-1\right) \;,
\ee
where $F_1(0)=2/\sqrt{\pi}$, and $F_1(x)$ decays exponentially at large $x=k/\sqrt{t/\tau}$ as $F_1(x) \simeq \frac{4}{\sqrt{\pi}} \frac{e^{-\frac{x^2}{4}}}{x^2}$, leading to a superexponential decay at large $k$, while the opposite limit $x=k/\sqrt{t/\tau}\ll 1$ again recovers $C_0(t)$ to leading order, with a correction $C_0(t) - C_k(t) \simeq D_k(0) \sim k$. For $s \to 1$, the scaling factor diverges and the correct scaling variable is $k/ \sqrt{t \log t}$. 

Finally, in the log-gas case $s=0$ we find, for large $k$ and $t$ with $k/(g\rho^2 t)$ fixed,
\be \label{Ck_s0_intro}
C_k(t) \simeq \frac{T}{g\rho^2} F_0\left(\frac{k}{g\rho^2 t}\right) \; \quad , \quad F_0(x) = \frac{1}{\pi^2} \log\left(1 + \frac{\pi^2}{x^2}\right) \;,
\ee
where $F_0(x)$ coincides with the limit $s \to 0$ of the scaling function $F_s(x)$ given in \eqref{Fs_def1}. The matching with $C_0(t)$ is more subtle in this case and is discussed in \cite{RieszFluct} (Sec.~III.H), along with the case $s<0$. 
\\

\noindent {\bf Time correlations of the gaps.} Another quantity that one may consider is the time correlation of the interparticle distance, i.e.,
\be
D_k(t) = \langle (\delta x_i(t) - \delta x_{i+k}(t)) (\delta x_{i}(0) - \delta x_{i+k}(0))  \rangle = \frac{8T}{N} \sum_{q=1}^{(N-1)/2}  \frac{e^{- \mu_q t}}{\mu_q} \sin^2\big(\pi \frac{q k}{N} \big) \;.
\ee
We recall that the gap variance $D_k(0)$, studied in Sec.~\ref{sec:gapvariance_RieszBrownian}, grows at large $k$ as $\sim\hspace{-0.05cm}k$ in the short-range case and as $\sim\hspace{-0.05cm}k^s$ in the long-range case. One can show that for an annealed initial condition, it is actually related to the correlation function $C_k(t)$ through the identity
\be \label{relationCkDk} 
C_0(t) - C_k(t) = D_k(0) - D_k(t) \;.
\ee 
This implies in particular that, for large $t$ and fixed $k$, one has $C_0(t) - C_k(t) \simeq D_k(0)$, as already noted above, while at large $k$ and fixed $t$, $D_k(0) - D_k(t) \simeq C_0(t)$, which we will use below.

We find that $D_k(t)$ takes the following scaling form for $\tau\ll t\ll N^{z_s}\tau$ and $1\ll k\ll N$, with $k \sim (t/\tau)^{1/z_s}$,
\be \label{Dk_scaling_brownian}
D_k(t) \simeq \begin{dcases} T \left(\frac{t^s}{g\rho^{s+2}}\right)^\frac{1}{s+1} G_s\left( \frac{k}{(g\rho^{s+2}t)^{\frac{1}{s+1}}} \right) \quad \text{for } 0<s<1 \;, \\
T \sqrt{\frac{t}{g\rho^{s+2}}} \, G_s\left( \frac{k}{\sqrt{g\rho^{s+2}t}} \right) \quad \hspace{1.4cm} \text{for } s>1 \;, \end{dcases}
\ee 
with the scaling function
\be \label{Gs_def1}
G_s(x) = \frac{8}{a_s^{1/z_s}} \int_0^{+\infty} dv \frac{e^{-v^{z_s}}}{v^{z_s}} \sin^2 \left( \pi a_s^{-1/z_s} x v \right) \;.
\ee
For large $x$, $G_s(x)$ behaves as 
\be 
G_s(x) \simeq \begin{dcases} \frac{4}{sa_s} (2 \pi)^s \cos(\frac{\pi s}{2}) \Gamma(1-s) x^s  \quad \text{for } 0<s<1 \;, \\ \frac{x}{(s+1)\zeta(s)} \quad \hspace{2.75cm} \text{for } s>1 \;, \end{dcases}
\ee 
where $a_s$ is given in \eqref{fasympt}. At small time we thus recover the results for $D_k(0)$ given in \eqref{gapscases_brownian}. This could be expected from the relation \eqref{relationCkDk}. In fact, this relation also gives us the next order, since for $k \gg (t/\tau)^{1/z_s}$, $C_k(t)$ decays to zero and we thus have $D_k(0) - D_k(t) \simeq C_0(t)$, where we recall that $C_0(t)\propto t^{\frac{s}{1+s}}$ for $0<s<1$ and $C_0(t)\propto\sqrt{t}$ for $s>1$. In the opposite limit $x\to 0$, one has $G_s(x) \sim x^2$ for any $s>0$. Hence at large times $t/\tau \gg k^{z_s}$, $D_k(t)$ decays as
\be \label{Dk_decay_intro}
D_k(t) \sim \begin{cases} k^2 t^{-\frac{2-s}{1+s}} \quad \text{for} \ s<1 \;, \\
k^2 t^{-1/2} \quad \; \text{for} \ s>1 \;. \end{cases}
\ee 

Finally, for the log-gas $s=0$ we find 
\be
D_k(t) \simeq \frac{T}{g\rho^2} G_0 \left( \frac{k}{g\rho^2 t} \right) \quad , \quad G_0(x) = \frac{1}{\pi^2} \log(1+\frac{x^2}{\pi^2}) \;.
\ee
Interestingly, this scaling function is related to the one for $C_k(t)$ given in \eqref{Ck_s0_intro} by $G_0(x)=F_0(1/x)$. This may be a consequence of the ``relativistic" invariance of the log-gas with dynamical exponent $z_0=1$. 
\\

\noindent {\bf Linear statistics.} A different way to characterize the space-time correlations, but now on a more macroscopic scale, is through the linear statistics. Consider a function $f(x)= \sum_{n \in \mathbb{Z}} \hat f_n e^{-2 i \pi \frac{n}{L} x}$ on the circle, of periodicity $L$. We assume that its Fourier coefficients $\hat f_n$ decay sufficiently fast, meaning that $f(x)$ varies at the scale of the circle. The linear statistics are defined as ${\cal L}_N(t) = \sum_{i=1}^N f(x_i(t))$. We find that in the large $N$ limit, for annealed initial condition, its covariance takes the form
\be
\langle {\cal L}_N(t) {\cal L}_N(t') \rangle_c \simeq \begin{dcases}  4\pi^{\frac{1}{2}-s} \frac{\Gamma(1+\frac{s}{2})}{\Gamma(\frac{1-s}{2})} \frac{ T N^s}{g\rho^s} \sum_{q=1}^{\infty} e^{-a_s q^{1+s} |\tilde t - \tilde t'|}  q^{1-s} | \hat f_q |^2  \quad \text{for } -1<s<1 \;, \\
\frac{2 T N}{(s+1) \zeta(s) g\rho^s} \sum_{q=1}^{\infty} e^{- a_s q^2 |\tilde t - \tilde t'|} | \hat f_q |^2 \quad \hspace{2.02cm} \text{for } s>1 \;, \end{dcases}
\ee
where we define $t = N^{z_s} \tau \tilde t$ with $\tau= 1/g \rho^{s+2}$, i.e., the time is expressed in units of the global relaxation time at the scale of the circle. For the log-gas $s=0$, this is exactly the formula proved in \cite{Spohn3} with $\beta=g/T$. Here we generalize to any $s>-1$. For $0<s<1$ and at equal time $t=t'$, it agrees 
with a recent result obtained in \cite{BoursierCLT}.

\subsection{Quenched initial condition} \label{sec:quenchedBrownian}

All the results of this section, concerning the dynamical correlations, were obtained for an {\it annealed} initial condition, i.e., by initializing the system at equilibrium. We now discuss how these results are modified if we consider instead a {\it quenched}, i.e., deterministic, initial condition. We assume that at $t=0$ the particles are equally spaced, i.e., $\delta x_i(0)=0$ for all $i$. This corresponds to preparing the system in the ground state at $T=0$. Going back to the linearized equation \eqref{Eq_delta_x_RieszBrownian} and integrating directly in real time with this initial condition leads to
\be
\delta x_i(t) = \sqrt{2T} \sum_{j=1}^{N} \int_0^t dt_1 [e^{(t_1-t)H}]_{ij} \left( \xi_j(t_1) - \frac{1}{N} \sum_{k=1}^N \xi_k(t_1) \right) \;,
\ee
where $H$ is the Hessian matrix defined in \eqref{defHessian}. Using that $\langle \xi_i(t) \xi_j(t')\rangle = \delta_{ij} \delta(t-t')$, we obtain
\be
\langle \delta x_j(t) \delta x_k(t') \rangle_{\rm qu} = 2T \int_0^{\min(t,t')} dt_1 \left([e^{(2t_1-t-t')H}]_{jk} -\frac{1}{N} \right) \;,
\ee
where $\langle \cdot \rangle_{\rm qu}$ denotes an average over the noise starting from a quenched initial condition, and where for the last term we have used the relation
\be \label{identity_expH}
\sum_{m,n=1}^N [e^{(t_1-t)H}]_{jm} [e^{(t_1-t')H}]_{nk} = \frac{1}{N^2} \sum_{q=0}^{N-1} \sum_{m,n=1}^N e^{\mu_q(2t_1-t-t')} e^{2\pi i \frac{q}{N}(j-m+k-n)} = 1 \;,
\ee
where only the term $q=0$ gives a non-zero contribution. Decomposing the matrix $H$ in its eigenbasis \eqref{eigenvals_rieszgeneral} and performing the integral, we obtain the equivalent of \eqref{cov_brownian} for a quenched initial condition,
\be \label{cov_brownian_quenched}
\langle \delta x_j(t) \delta x_k(t') \rangle_{\rm qu} 
= \frac{2T}{N} \sum_{q=1}^{(N-1)/2} \frac{e^{-\mu_q|t-t'|}-e^{-\mu_q(t+t')}}{\mu_q} \cos \big(2\pi \frac{q}{N}(j-k) \big) 
\ee
(note that it coincides with the annealed result \eqref{cov_brownian} for $t,t'\to+\infty$ as it should, since the static quantities do not depend on the initial condition).

Let us now compute again some of the dynamical quantities studied in this section starting from \eqref{cov_brownian_quenched} and see how they are affected by the change in the initial condition. We start with the mean squared displacement studied in Sec.~\ref{sec:MSD_brownian}. Since $\delta x_i(0)=0$, we simply have in this case
\be \label{C0_quenched}
C_0^{\rm qu}(t) = \langle \delta x_i(t)^2 \rangle_{\rm qu} = \frac{2T}{N} \sum_{q=1}^{(N-1)/2} \frac{1-e^{-2\mu_q t}}{\mu_q} = \frac{1}{2} C_0^{\rm ann}(2 t) \;,
\ee
where $C_0^{\rm ann}(t)$ denotes the MSD for an annealed initial condition \eqref{disp_brownian} which we have computed above. Thus, in the regime $\tau\ll t\ll N^{z_s}\tau$, in the leading order this simply leads to a factor $2^{-\frac{1}{2}}$ for $s\geq 1$, a factor $2^{-\frac{1}{s+1}}$ for $0<s<1$ and a factor $1/2$ for $s=0$ compared to the annealed case \eqref{displacement_Riesz_brownian}. 
This holds until $C_0^{\rm qu}(t)$ reaches its large time limit for $t\gg N^{z_s} \tau$, equal to the equilibrium variance, $ C_0^{\rm qu}(\infty) = \langle \delta x_i^2 \rangle$, i.e., half of the annealed limit. Note that the short time diffusive regime for $t\ll\tau$ is unchanged. Once again, the prefactors in the intermediate time regime match the ones obtained through MFT in \cite{DFRiesz23}.

For the two-time covariance, we now have a different relation with the MSD, given by
\be
C_0^{\rm qu}(t_1,t_2) = \langle \delta x_i(t_1) \delta x_i(t_2) \rangle_{\rm qu} = \frac{2T}{N} \sum_{q=1}^{(N-1)/2} \frac{e^{-\mu_q|t-t'|}-e^{-\mu_q(t+t')}}{\mu_q} = C_0^{\rm qu} \big(\frac{t_1+t_2}{2}\big) - C_0^{\rm qu} \big(\frac{|t_1-t_2|}{2}\big) \;.
\ee
As an example, for $0<s<1$, and for $\tau \ll t \ll N^{z_s}\tau$, this leads to
\be
C_0^{\rm qu}(t_1,t_2) \simeq \frac{U_s T}{2} \left((t_1+t_2)^{\frac{s}{s+1}} - |t_1-t_2|^{\frac{s}{s+1}}\right) \;,
\ee
with $U_s$ given in \eqref{defUs}, which again coincides with the result obtained in \cite{DFRiesz23}.

Finally, for the equal time covariance we have
\be \label{Ck_quenched_brownian}
C_k^{\rm qu}(t) = \langle \delta x_i(t) \delta x_{i+k}(t) \rangle_{\rm qu} = \frac{2T}{N} \sum_{q=1}^{(N-1)/2} \frac{1-e^{-2\mu_q t}}{\mu_q} \cos\big(2\pi \frac{qk}{N}\big) = \frac{1}{2} C_k^{\rm ann}(2 t) \;.
\ee
where $C_k^{\rm ann}(t)$ again denotes the annealed result \eqref{Ck_def_brownian}. Note that the gap correlations identically vanish in the quenched case, $D_k^{\rm qu}(t)=0$.

\section{Conclusion}

In this chapter, we studied the microscopic fluctuations in a Riesz gas of Brownian particles with periodic boundary conditions. We developed a method based on a small-temperature approximation, which allowed us to compute a wide variety of static and dynamical correlation functions of the particle positions and of the gaps. Using this approach, we recovered with a surprising precision several results previously obtained in the physics and mathematics literature via completely different methods, in particular for the mean squared displacement and the variance of the gaps. In addition, it also allowed us to derive some original results, concerning for instance the equal time covariance of particle displacements or the time correlations of the gaps.

Perhaps one of the most intriguing question raised by the present work is its perfect agreement with the MFT results from \cite{DFRiesz23}, which even seems to hold at arbitrary temperature for $0<s<1$. It would be interesting to better understand the connection between the two approaches, and more generally to have a more precise understanding of the real extent of the validity of the present method. One could also try to see if the weak noise expansion can be taken further in order to compute higher order correlation functions and their temperature dependence. On a different note, while our results seem to be compatible with the existence of a melting transition for $0<s<1$, as evidenced in \cite{Lelotte2023}, our approach does not allow us to estimate the melting temperature in a satisfactory way. Developing a more predictive theory for this transition thus remains an open problem. Finally, another question which is of course particularly relevant in the context of this thesis is the extension of these results to active particles, in order to better understand how activity affects the fluctuations in long-range interacting systems. This is the topic of the next chapter.

\chapter{Active Riesz gas on the circle} \label{chap:activeRieszFluct}

\section{Setting and main results}

Let us now return to active particles and see how we can extend the results of the previous chapters in the presence of active noise. In this chapter we consider $N$ run-and-tumble particles in 1D interacting via a pairwise Riesz potential, such that the positions of the particles $x_i(t)$ ($i=1,...,N$) obey the equations of motion
\be \label{def_Riesz_chap8}
\frac{dx_i}{dt} = -\sum_{j(\neq i)} W'(x_i-x_j) + v_0\sigma_i(t) + \sqrt{2T} \, \xi_i(t) \quad , \quad W'(x)=-g \, \frac{\sgn(x)}{|x|^{s+1}} \;,
\ee
where the $\sigma_i(t)$ are i.i.d. telegraphic noises with tumbling rate $\gamma$ (see Sec.~\ref{sec:RTPdef}) and the $\xi_i(t)$ are i.i.d. Gaussian white noises with unit variance, and we assume as before $s>-1$ and $g>0$. As in the previous chapter, we consider periodic boundary conditions with periodicity $L$, meaning that we actually use the periodized form of the interaction force $W'(x)=W'(x+L)$, defined in \eqref{defRiesz_periodized}) (and that we again identify $x_i\equiv x_{i+N}$, see Fig.~\ref{fig:sketch_circle}). As in the previous chapter we assume that the particles are ordered such that $x_1(t)>x_2(t)>...>x_N(t)$, which in this case does not pose any problem since RTPs can never cross for $s>-1$.

Contrary to the Brownian case, at large times the dynamics \eqref{def_Riesz_chap8} reaches a stationary state which is not described by the Gibbs equilibrium, and where the particle displacements are {\it a priori} non-Gaussian. However, as we have already mentioned, a strong advantage of the method presented in Chapter~\ref{chap:passiveRieszFluct} for Brownian particles is that it can easily be adapted to account for more complex types of noise. In this chapter, to fix ideas we consider RTP noise, but as we will see our results also apply to the other particle models defined in Sec.~\ref{sec:active_models} (namely AOUPs and ABPs). Throughout this chapter we will also set $T=0$ 
for simplicity. However, since we consider only the linear order in the noise, all the results that we obtain are additive when more than one type of noise are present (i.e., one has for instance $C_0^{total}(t)=C_0^{RTP}(t)+C_0^{Brownian}(t)$ at the order we consider).

For RTPs at $T=0$, there are two important parameters,
\be \label{adimparam_activeRiesz}
T_{\rm eff}= \frac{v_0^2}{2 \gamma} \quad \text{and} \quad \hat{g}= \frac{g\rho^{s+2}}{2\gamma} = \frac{1}{2 \gamma \tau} \;.
\ee 
The effective temperature $T_{\rm eff}$ measures the amplitude of the noise and can be seen as the equivalent of $T$ in the previous chapter. The second parameter $\hat g$ is the ratio of the persistence time $1/\gamma$ to the local interaction time $\tau=1/(g \rho^{s+2})$. It thus quantifies the ``activity'' of the system (i.e., the persistence of the noise). The regime $\hat g \ll 1$ corresponds to the diffusive limit, where we expect to recover the Brownian results of Chapter~\ref{chap:passiveRieszFluct} (with a temperature $T_{\rm eff}$), while in the opposite regime $\hat g \gg 1$ we expect the activity to play an important role. Indeed, we will see that, although at large times and large distances we essentially recover the results of the previous chapter, a strong activity significantly affects the behavior of the system both on short timescales and at small lengthscales. 

We will begin by showing in Sec.~\ref{sec:Riesz_derivation_rtp} how the computation of the stationary two-point two-time covariance of the particle positions is modified for RTP noise. As in the previous chapter, it will be our starting point to obtain all the other results of this chapter. We will then once again study various static and dynamical correlation functions, focusing on the thermodynamic limit $N,L\to+\infty$ with fixed density $\rho=N/L$. We begin with the static correlations in Sec.~\ref{sec:static_rtp}. We show in particular that, while for large separations the variance of the gap between two particles $i$ and $i+k$ is not different from the Brownian case, on smaller lengthscales and for sufficiently strong activity it exhibits a very distinct behavior, increasing supralinearly with $k$ for $s>0$. We then study the dynamical correlations in Sec.~\ref{sec:dynamical_rtp}, starting with the MSD during time $t$, for which we find that a new ballistic regime appears for $t\ll 1/\gamma$ compared to the Brownian case. As in the previous chapter, we also study several other dynamical quantities, namely the two-time correlations and equal time covariance of the particle displacements, as well as the time-correlations of the interparticle distance. For these quantities, we uncover a variety of new dynamical regimes, unique to the active case, both at small times and at short distances. Finally, we consider what happens if we replace the annealed initial condition by a quenched one. While in the Brownian case the initial condition mostly affects the numerical prefactors, we will see that for active particles, the short time regime for $t\ll 1/\gamma$ strongly depends on the initial condition (especially for strong activity $\hat g \gg 1$). In particular, for the MSD a new regime appears for $\tau \ll t \ll 1/\gamma$, with a superdiffusive behavior as $\sim t^{3/2}$ in the short-range case $s>1$ and $\sim t^{\frac{1+2s}{1+s}}$ in the long-range case $0<s<1$. In the short-range case, our results recover the ones on harmonic chains of active particles, studied, e.g., in \cite{HarmonicChainRevABP,SinghChain2020,PutBerxVanderzande2019,HarmonicChainRTPDhar} (and which we reviewed in Sec.~\ref{sec:harmonicChain}), with the precise mapping discussed at the end of Sec.~\ref{sec:Riesz_eigvals} (for the quantities for which prior results on the harmonic chain are available).

\section{Two-point two-time covariance in the stationary state} \label{sec:Riesz_derivation_rtp}

As in the previous chapter, we consider the limit of weak noise (i.e., small $v_0$, or more precisely small effective temperature $T_{\rm eff}$ in \eqref{adimparam_activeRiesz}), and we look at the small displacements $\delta x_i(t)$ of the particles with respect to the ground state configuration where the particles are equally spaced, see \eqref{def_delta_x_Riesz1} and Fig.~\ref{fig:sketch_circle}. As before, we also subtract the displacement of the center of mass, which in this case behaves as $\frac{ d\bar x}{dt} = \frac{v_0}{N} \sum_j \sigma_j(t)$, i.e., at large times it again diffuses with a diffusion coefficient $T_{\rm eff}/N$. We again assume that the active noise is sufficiently weak such that the condition \eqref{cond_approx} is satisfied, which allows us to linearize the equation of motion \eqref{def_Riesz_chap8} (with $T=0$) as
\be \label{Eq_delta_x_RTP}
\frac{d}{dt} \delta x_i(t) = - \sum_{j=1}^N H_{ij} \, \delta x_j(t) + v_0 \sigma_i(t) - \frac{v_0}{N} \sum_{j=1}^N \sigma_j(t)\;.
\ee 
For a generic interaction $W(x)$ (which should be periodic, repulsive and such that the particles are equally spaced in the ground state), we recall that the Hessian matrix $H$ is defined in \eqref{defHessianRiesz}, with spectrum given in \eqref{eigenvals_rieszgeneral}. As in the previous chapter, with start by considering an {\it annealed} initial condition, preparing the system (i.e., here both the positions $x_i(t)$ and the driving noise $\sigma_i(t)$) in the stationary state at $t\to-\infty$ (the quenched case will be discussed in Sec.~\ref{sec:quenched_RTP}). Taking the Fourier transform of \eqref{Eq_delta_x_RTP} with respect to time and inverting in the frequency domain we obtain
\be 
\delta \hat x_j(\omega) = v_0 \sum_{k=1}^N  [i \omega \mathbbm{1}_N + H]^{-1}_{jk} \hat \sigma_k(\omega) - \frac{v_0}{N} \frac{1}{i\omega} \sum_{k=1}^N \hat \sigma_k(\omega) \;,
\ee 
where we have used the relation \eqref{identity_hessian_fourier}. We recall that $\mathbbm{1}_N$ is the $N \times N$ identity matrix, $\delta \hat x_i(\omega) = \int_{-\infty}^{\infty} dt \, e^{-i \omega t} \delta x_i(t)$, and now $\hat \sigma_j(\omega)=\int_{-\infty}^{\infty} dt \, e^{-i \omega t} \sigma_i(t)$. As explained in Sec.~\ref{sec:RTPdef}, the correlations of $\sigma_i(t)$ are given by $\langle \sigma_i(t) \sigma_j(t') \rangle = e^{-2\gamma|t-t'|}\delta_{ij}$, which in Fourier space reads
\be \label{sigmacor_Fourier}
\langle \hat \sigma_i(\omega) \hat \sigma_j(\omega') \rangle = \frac{4\gamma}{\omega^2+4\gamma^2} 2\pi \delta(\omega+\omega') \;.
\ee
Thus, the two-point two-time covariance of the $\delta x_i(t)$ in the stationary state read (after Fourier inversion)
\be
\langle \delta x_j(t) \delta x_k(t') \rangle = v_0^2 \int_{-\infty}^{+\infty} \frac{d\omega}{2 \pi} \frac{4\gamma \, e^{i \omega (t-t')} }{\omega^2+4\gamma^2} \left( [\omega^2 \mathbbm{1}_N + H^2]^{-1}_{jk} - \frac{1}{N\omega^2} \right) \;.
\label{rtp_corr_cm}
\ee
Using the eigensystem of $H$ given in \eqref{eigenvals_rieszgeneral}, this becomes
\bea  \label{cov_rtp}
\langle  \delta x_j(t)  \delta x_k(t') \rangle &=& \frac{v_0^2}{N} \sum_{q=1}^{N-1} e^{2\pi i \frac{q}{N} (j-k)} \int_{-\infty}^{+\infty} \frac{d\omega}{2 \pi} \frac{e^{i \omega (t-t')}}{\omega^2 + \mu_q^2 } \frac{4\gamma}{\omega^2+4\gamma^2} \nn \\
&=& \frac{2v_0^2}{N} \sum_{q=1}^{(N-1)/2} \frac{\mu_q e^{-2\gamma|t-t'|} -2\gamma e^{- \mu_q |t-t'|}}{\mu_q(\mu_q^2-4\gamma^2)}  \cos\left(2\pi \frac{q}{N} (j-k)\right) \;,
\eea 
where we have used the symmetry $\mu_q=\mu_{N-q}$ in the last step (as before, the last expression is exact only for odd values of $N$ but one can make it fully general by taking the sum from $1$ to $N-1$ and removing the factor $2$). As in the Brownian case, the average $\langle \delta x_i(t) \rangle$ vanishes at leading order in $v_0$. Note however that in the RTP case (and by contrast with the Brownian case), the distributions of the displacements, even for weak noise, are {\it a priori} not Gaussian \cite{HarmonicChainRTPDhar}, so that the present computation does not give us any information on the higher order moments. 

We recall that for the periodized Riesz interaction defined in \eqref{defRiesz_periodized}, which we consider in the rest of this chapter, the eigenvalues are given by (see Sec.~\ref{sec:Riesz_eigvals})
\be \label{mu_Riesz_RTP}
\mu_q = g\rho^{s+2} f_s\big( \frac{q}{N} \big) \;, \quad \text{with} \quad f_s(u)= 4(s+1) \sum_{\ell=1}^{\infty} \frac{\sin^2(\pi \ell u)}{\ell^{s+2}} \;,
\ee
where $f_s(u)$ is an increasing function on $[0,1/2]$, with for $u\ll 1$,
\be \label{fasympt_RTP}
f_s(u) \underset{u\to 0}{\sim} a_s u^{z_s} \;, \quad z_s = \min(1+s,2) \; , \quad a_s = \begin{cases} 2\pi^{s+\frac{3}{2}} \frac{\Gamma(\frac{1-s}{2})}{\Gamma(1+\frac{s}{2})}
 \quad \hspace{0.28cm} \text{for } -1<s<1 \;, \\
4\pi^2 (s+1) \zeta(s)  \quad  \text{for } s>1 \;. \end{cases}
\ee
We also recall that the harmonic chain one has $\mu_q = 4K \sin^2 \left(\frac{\pi q}{N}\right)$, and that in the regimes dominated by the smallest eigenvalues (i.e., most of the large $N$ results below), our results for the short-range case also apply to the harmonic chain upon replacing $\tau=1/(g\rho^{s+2}) \to \tau_K = 1/K$ and $(s+1)\zeta(s) \to 1$.
\\

\noindent {\bf Extension to other active particle models.} Throughout this chapter we have decided to consider run-and-tumble particles. In the derivation above however, we see that the only ingredient which enters into the computation of the two-point two-time correlations at this order are the stationary two-time correlations of the noise $\langle \sigma(t) \sigma(t') \rangle = e^{-2\gamma|t-t'|}$. When we introduced the main models of active particles in Sec.~\ref{sec:active_models}, we saw that there are other models for which the driving noise has similar exponential correlations. In particular, we showed that for AOUPs, i.e., replacing $v_0\sigma(t)\to v(t)$ where $v(t)$ is an Ornstein-Uhlenbeck process, $\tau_{ou} \frac{dv}{dt} = -v(t) + \sqrt{2D} \, \eta(t)$ with $\eta(t)$ a unit Gaussian white noise, one has in the stationary state
\be \label{AOUPcorr_chap8}
\langle v(t) v(t') \rangle =  \frac{D}{\tau_{ou}} e^{-|t-t'|/\tau_{ou}} \;.
\ee
Thus, all the results of this chapter also apply to AOUPs upon replacing $v_0^2 \to D/\tau_{ou}$ and $2\gamma \to 1/\tau_{ou}$. Similarly, for a 2D ABP projected in 1D, corresponding to $v_0\sigma(t)\to v_0\cos\phi(t)$ where $\phi(t)$ is a Brownian motion with diffusion coefficient $D_R$, the stationary correlations read
\be \label{ABPcorr_chap8}
\langle v_0^2\cos\phi(t) \cos\phi(t') \rangle = \frac{v_0^2}{2} e^{-D_R |t-t'|} \;,
\ee
and thus our results again apply after replacing $v_0^2 \to \frac{v_0^2}{2}$ and $2\gamma \to D_R$.
\\

As in the previous chapter, we now focus on the thermodynamic limit $N,L\to+\infty$ with fixed density $\rho=N/L$, and we use the general result \eqref{cov_rtp} to study various static and dynamical correlation functions in this regime.

\section{Static correlations} \label{sec:static_rtp}

\subsection{Variance of the particle positions}

As in the Brownian case, we start with the variance of the displacements, which for any $N$ reads (using \eqref{cov_rtp})
\be \label{var_RTP}
\langle \delta x_i^2 \rangle = \frac{2 v_0^2}{N} \sum_{q=1}^{(N-1)/2} \frac{1}{\mu_q (\mu_q +2\gamma)} = \frac{2T_{\rm eff}}{N g\rho^{s+2}} \sum_{q=1}^{(N-1)/2} \frac{1}{f_s(\frac{q}{N}) (1+ \hat g f_s(\frac{q}{N}))} \;.
\ee
For $s>0$, the sum is dominated at large $N$ by the small values of $q$. Throughout this chapter we assume that $\hat g$ does not scale with $N$, i.e., $\hat g \ll N^{z_s}$, so that the last term in the denominator can be neglected. Thus, at leading order in $N$ we exactly recover the Brownian result \eqref{var_brownian} with $T\to T_{\rm eff}$. One can actually show that the leading correction compared to the Brownian case is of order $O(1)$ in $N$ (see Sec.~IV.A in \cite{RieszFluct}). This is also true in the case $s=0$ (which corresponds to the active DBM on the circle), where the leading $\log N$ behavior is unchanged compared to the Brownian case \eqref{varloggas}, while the $O(1)$ correction can be computed explicitly.

For $-1<s<0$, as in the Brownian case, the sum is dominated by $q$ of order $N$, and thus the variance has a finite limit when $N \to +\infty$,
\be \label{var_RTP_solid}
\langle \delta x_i^2 \rangle \simeq \frac{2 T_{\rm eff}}{g\rho^{s+2}} \int_0^{1/2} \frac{du}{f_s(u)\left( 1 + \hat g f_s(u) \right)} \;.
\ee
In this case, the Brownian result \eqref{var_Riesz_solid} is recovered only in the small persistence limit $\hat g \ll 1$. As in the Brownian case, the system thus exhibits translational order at low effective temperature, and we can estimate the melting temperature through the Lindemann criterion in exactly the same way as in Sec.~\ref{sec:var_Riesz_brownian},
\be  \label{TmeltingRTP}
T_{\rm eff,M} = \frac{1}{2} g\rho^{s} \frac{c_L^2}{\int_0^{1/2} \frac{du}{f_s(u)\left( 1 + \hat g f_s(u) \right)}} \;.
\ee 
where $c_L$ is a phenomenological Lindemann constant. In the limit $s \to 0^-$ one finds that $T_{\rm eff,M}$ vanishes linearly in $s$, as in the Brownian case, i.e., $ T_{\rm eff,M} \sim \pi^2 c_L^2 g |s|$.

The results for the covariance $\langle \delta x_i \delta x_{i+k}\rangle$ are similar, i.e., for $s\geq0$ the leading behavior is the same as in the Brownian case with $T\to T_{\rm eff}$, while for $s<0$ we observe an effect of the active noise (which however becomes negligible at large distances $k \gg \hat g^{1/(1+s)}$, as for the gap variance discussed below).

Let us mention that in \cite{RieszFluct}, we also considered the case where $\gamma$ scales as a negative power of $N$, such that $\hat g\sim N^{z_s}$. This limit $\gamma\to0$ where the motion of the particles is almost ballistic is related to the Jepsen gas \cite{jepsen,Bena2007}. See Sec.~IV.H in \cite{RieszFluct} for a discussion of how this scaling affects the different correlation functions.

\subsection{Variance of the interparticle distance} \label{sec:gapvarRieszactive}

Let us now consider the variance of the gaps, given by (using \eqref{cov_rtp})
\be \label{Dk0_rtp}
D_k(0) = \langle (\delta x_{i+k}-\delta x_i)^2 \rangle = \frac{4v_0^2}{N} \sum_{q=1}^{(N-1)/2} \frac{1-\cos\left( \frac{2\pi kq}{N} \right)}{\mu_q(\mu_q+2\gamma)} = \frac{8T_{\rm eff}}{N g\rho^{s+2}} \sum_{q=1}^{(N-1)/2} \frac{\sin^2\left( \frac{\pi kq}{N} \right)}{f_s(\frac{q}{N})(1+\hat g f_s(\frac{q}{N}))} \;.
\ee
At large $N$, with $k \ll N$, this becomes for any $s>-1$,
\be 
D_k(0)  \simeq \frac{8 T_{\rm eff}}{g \rho^{s+2}} \int_0^{1/2} du  \frac{\sin^2\left( \pi k u  \right)}{f_s(u)\left(1+\hat g f_s(u) \right)} \;. \label{gaprtp}
\ee
We now study the behavior of this quantity for large $k$, i.e., for $1 \ll k \ll N$. 

Let us start with the case $s>0$. For $\hat g \ll 1$, equation \eqref{gaprtp} obviously recovers the Brownian case from Sec.~\ref{sec:gapvariance_RieszBrownian}. However, in the regime of strong persistence $\hat g \gg 1$, we find that a change of behavior occurs at a characteristic lengthscale given by $k \sim \hat g^{1/z_s}$. 
In general, for $k$ and $\hat g\gg1$ with $k\sim \hat g^{1/z_s}$, the variance of the gaps takes the following scaling form
\be \label{scalinggaprtp}
D_k(0)  \simeq \frac{ T_{\rm eff} k^{z_s-1}}{g \rho^{s+2}}   \, {\sf G}_s( k/\hat g^{\frac{1}{z_s}}) \;,
\ee 
where the scaling function (obtained by noting that in this limit the integral \eqref{gaprtp} is dominated by small $u$) reads 
\be \label{eqFgap} 
{\sf G}_s({\sf x}) = \frac{8}{a_s} \int_0^{+\infty} dv  \frac{\sin^2(\pi v)}{v^{z_s}\left(1+\frac{a_s}{{\sf x}^{z_s}} v^{z_s} \right)}
\ee
(note that since $z_s=\min(2,1+s)$, the integral converges for any $s>-1/2$).
At large distances ${\sf x} = k/\hat g^{\frac{1}{z_s}} \gg 1$, one has ${\sf G}_s({\sf x}) \to {\sf G}_s(+\infty) = \frac{4}{a_s} \frac{\pi^{z_s-\frac{1}{2}}}{z_s-1} \frac{\Gamma(\frac{3-z_s}{2})}{\Gamma(\frac{z_s}{2})}$, and thus \eqref{scalinggaprtp} recovers the Brownian result \eqref{gapscases_brownian}, where $D_k(0)\propto k$ for $s>1$ and $D_k(0) \propto k^s$ for $0<s<1$, with an effective temperature $T_{\rm eff}$.
However, on smaller scales, i.e., for ${\sf x} = k / \hat g^{1/z_s} \ll 1$, the activity plays an important role. We find that there are 3 different cases depending on the value of $s$, namely
\bea
{\sf G}_s({\sf x}) \underset{{\sf x}\ll 1}{\simeq} \begin{cases} \frac{4}{a_s^2} \frac{\pi^{2s+\frac{3}{2}}}{2s+1} \frac{\Gamma(\frac{1}{2}-s)}{\Gamma(s+1)} {\sf x}^{1+2s} \hspace{2.65cm} \text{for } 0 < s < 1/2 \;, \\
\frac{8 \pi^3}{a_s} (a_s^{\frac{2-s}{1+s}}(1+s) \sin( \frac{2-s}{1+s} \pi))^{-1} {\sf x}^{2-s} \quad \text{for } 1/2 < s < 1 \;, \\
\frac{4 \pi^3}{a_s^{3/2}} {\sf x} \hspace{5cm} \; \text{for } s>1 \;. \end{cases}
\eea
Inserting into \eqref{scalinggaprtp}, this leads to
\bea \label{Dk0_rtp_cases}
D_k(0) \underset{k\ll \hat g^{1/z_s}}{\simeq} \begin{cases} \frac{4 v_0^2}{(a_s g \rho^{s +2})^2} \frac{\pi^{2s+\frac{3}{2}}}{2s+1} \frac{\Gamma(\frac{1}{2}-s)}{\Gamma(s+1)} k^{1+2s} \hspace{1.6cm} \text{for } 0 < s < 1/2 \;, \\
\frac{8\pi^3 v_0^2}{(s+1) \sin( \frac{2-s}{1+s} \pi) (a_s g\rho^{s+2})^{\frac{3}{1+s}} (2\gamma)^{\frac{2s-1}{1+s}}} k^2 \quad \text{for } 1/2 < s < 1 \;, \\
\frac{v_0^2}{(2 (s+1)\zeta(s) g \rho^{s+2})^{3/2} \gamma^{1/2}} k^2 \hspace{2.1cm} \; \text{for } s>1 \;. \end{cases}
\eea
Thus, at short distances and for sufficiently strong persistence, the gap variance increases supralinearly with $k$ for any $s>0$, and up to $\sim k^2$ for $s>1/2$. It is interesting to interpret this result in terms of counting statistics, as discussed in Sec.~\ref{sec:gapvariance_RieszBrownian}. This supralinear increase means that the variance of the number of particle inside a given interval increases faster than linearly with the size of this interval, i.e., the number fluctuations are larger than for Poissonian statistics. These {\it giant number fluctuations} are an effect of the activity, which has been predicted theoretically and observed experimentally in various types of active systems \cite{TonerTuReview,ChateGiant,GinelliGiant,DasGiant2012,Chate2010,NarayanGiant,ZhangGiant}. For $s<1$ this effect competes with the long-range interaction which tends to reduce the fluctuations. We see that for sufficiently large persistence, the effect of the activity dominates for short distances $k\ll \hat g^{1/z_s}$, while the interactions dominate on larger lengthscales $k\gg \hat g^{1/z_s}$.

Another interesting property to note is that for $0<s<1/2$, the expression \eqref{Dk0_rtp_cases} is independent of $\gamma$, and thus it remains finite in the limit $\gamma \to 0$, while for $s>1/2$ we find that $D_k(0)$ diverges as $\gamma \to 0$. As we will discuss below, for $s>1/2$ our linear approximation will thus become invalid in this limit. This suggests that in the short-range case, or for ``not too long-range'' interactions (i.e., for $s>1/2$), an very persistent noise can ``break'' the structure of the system, leading to extremely large fluctuations of the gap sizes, while this does not happen for $s<1/2$. It is particularly interesting to note that this change of behavior occurs at $s=1/2$, while in the passive case all values of $s\in(0,1)$ always lead to qualitatively similar results. 

We now briefly discuss the remaining cases. For the log-gas $s=0$, the scaling form \eqref{scalinggaprtp}
is still valid when $k, \hat g \gg 1$ with ${\sf x}=k/\hat g$ fixed. In this case, the scaling function can be written explicitly as
\be 
{\sf G}_0({\sf x}) = \frac{2}{\pi^2} 
   \left(\log(\frac{{\sf x}}{\pi })+\gamma - 
   \text{Ci}\left(\frac{{\sf x}}{\pi}\right) \cos
   \left(\frac{{\sf x}}{\pi}\right) -\text{Si}\left(\frac{{\sf x}}{\pi }\right) \sin
   \left(\frac{{\sf x}}{\pi }\right) \right) +\frac{1}{\pi} \sin(\frac{{\sf x}}{\pi}) \;,
\ee 
where ${\rm Ci}(x)=-\int_x^\infty dt\frac{\cos t}{t}$ and ${\rm Si}(x)=\int_0^x dt\frac{\sin t}{t}$. For ${\sf x}=k/\hat g \gg 1$, we find ${\sf G}_0({\sf x})\simeq  \log (\frac{{\sf x}}{\pi}) + \gamma_E $, which gives
\be \label{gap_rtp_log_log2}
D_k(0) \simeq \frac{2 T_{\rm eff}}{\pi^2 g \rho^2} \big(\log \big(\frac{k}{\pi \hat g}\big) + \gamma_E \big) \;.
\ee
In this large distance regime, the leading behavior is thus again the same as in the Brownian case with temperature $T_{\rm eff}$ (see \eqref{gaps_s0}), while the activity leads to a correction of order $O(1)$ in $k$. In the opposite limit, ${\sf x} \ll 1$, one has ${\sf G}_0({\sf x}) \simeq {\sf x}/\pi^2$, thus for $k \ll \hat g$,
\be \label{gap_rtp_log_lin}
D_k(0) \simeq \frac{v_0^2 k}{(\pi g \rho^2)^2} \;,
\ee
i.e., the formula \eqref{Dk0_rtp_cases} for $s<1/2$ still holds, leading in this case to a linear behavior.

Finally, in the case $-1<s<0$, the variance of the gaps converges to a constant at large $k$,
\be \label{gap_rtp_sneg}
D_k(0) \underset{k\gg 1}{\longrightarrow} \frac{4 T_{\rm eff}}{g \rho^{s+2}} \int_0^{1/2} \frac{du}{f_s(u)\left(1+\hat g f_s(u) \right)} = 2 \langle \delta x_i^2 \rangle \;.
\ee
However, for $s>-1/2$, the result for $s<1/2$ in \eqref{Dk0_rtp_cases} is still valid for $k \ll \hat g^\frac{1}{1+s}$ when $\hat g \gg 1$ (but in this case leads to a sublinear behavior $D_k(0)\propto k^{1+2s}$. In this case, one can show that the integral in \eqref{gap_rtp_sneg} scales as $\hat g^{\frac{s}{s+1}}$, such that the limit \eqref{gap_rtp_sneg} is indeed reached for $k \sim \hat g^\frac{1}{1+s}$ (taking into account the $1/\gamma$ coming from $T_{\rm eff}$). On the other hand, for $-1/2>s>-1$, the limit is reached much faster and the regime in $k^{1+2s}$ does not exist, even when $\hat g \gg 1$. 

\begin{table}
\begin{center}
\begin{tabular}{|c|c|c|c|c|c|c|}
\hline
 & $-1<s<-1/2$ & $-1/2<s<0$ & $s=0$ & $0<s<1/2$ & $1/2<s<1$ & $s>1$ \\
\hline
$k \ll \hat g^{\frac{1}{z_s}}$ & $\sim cst$ & $\propto k^{1+2s}$ & $\propto k$ & $\propto k^{1+2s}$ & $\propto k^2$ & $\propto k^2$ \\  
\hline
$k \gg \hat g^{\frac{1}{z_s}}$ & $\sim cst$ & $\sim cst$ & $\propto \log k$ & $\propto k^s$ & $\propto k^s$ & $\propto k$ \\
\hline
\end{tabular}
\end{center}
\caption{Different regimes for the variance of the gaps $D_k(0)$ for the RTP Riesz gas, as a function of the interparticle distance $k$ and the parameter $s$ of the interaction. For $k \gg \hat g^{\frac{1}{z_s}}$, the results coincide with the Brownian case.}
\label{table:gap_var_rtp}
\end{table}

The behavior of $D_k(0)$ in the two limiting regimes $k \ll \hat g$ and $k\gg \hat g$ for the different values of $s$ is summarized in Table~\ref{table:gap_var_rtp}. Let us finally note that our short range result $s>1$ in \eqref{Dk0_rtp_cases} also applies to the harmonic chain with the mapping $\tau\to\tau_K=1/K$ and $(s+1)\zeta(s)\to 1$ (even for stronger noise). We found only few existing results concerning the variance of the interparticle distance in harmonic chains of active particle, apart from the separation between nearest neighbors $k=1$ studied in \cite{HarmonicChainRevABP,HarmonicChainRTPDhar}. Thus, our result also allows to better understand the behavior of the gaps in the harmonic chain.
\\

\begin{figure}
\centering
    \includegraphics[width=0.45\linewidth]{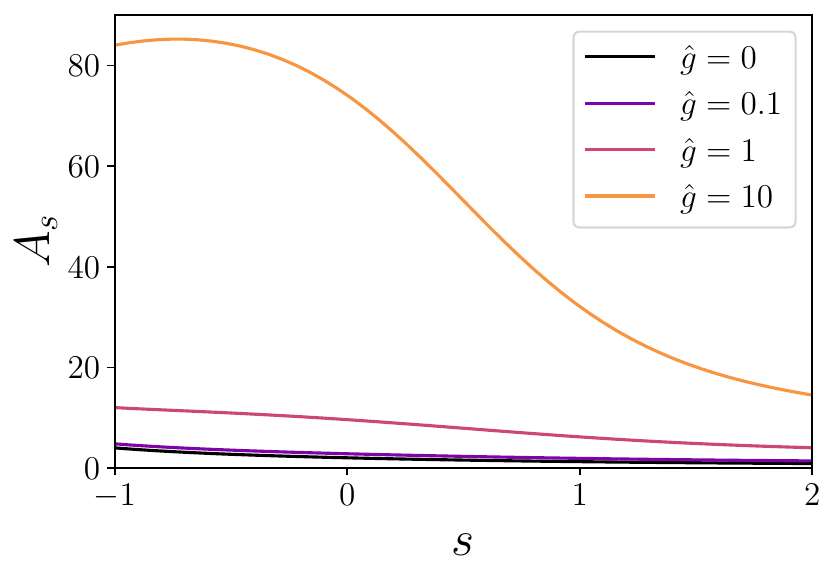}
    \caption{Plot of $A_s(\hat g)$ versus $s$, as defined \eqref{validity_rtp}, which gives the temperature $T_G=A_s(\hat g) \,g \rho^s$ below which the linear approximation should be valid for the active Riesz gas.}
    \label{FigAs_rtp} 
\end{figure}

\noindent {\bf Validity of the approximation.} These results can be used to estimate the domain of validity of our linear approximation using the same method as in Sec.~\ref{sec:gapvariance_RieszBrownian} for the Brownian case. Using the same criterion \eqref{cond_approx_Riesz1}, we find that our approximation should be valid for
\be \label{validity_rtp}
T_{\rm eff} \ll T_G = A_s(\hat g) \, g\rho^s   \quad , \quad A_s(\hat g)^{-1} = 2(s+2)^2\int_0^{1/2} du \frac{\sin^2(\pi u)}{f_s(u) (1+\hat g f_s(u))} \;.
\ee  
The constant $A_s(\hat g)$ is plotted in Fig.~\ref{FigAs_rtp} as a function of $s$, for different values of $\hat g$. For $\hat g \ll 1$, this is the same criterion as in the Brownian case. For $\hat g \gg 1$, one can show that it behaves as
\be \label{asympt_Bs}
A_s(\hat g) \simeq B_s \, \hat g^{\nu_s} \quad , \quad \text{with } \nu_s = \begin{cases} 1 \hspace{0.4cm} \text{ for } -1<s<\frac{1}{2} \;, \\
\frac{2-s}{1+s} \text{ for } \frac{1}{2}<s<1 \;, \\
\frac{1}{2} \hspace{0.35cm} \text{ for } s>1 \;, \end{cases} 
\ee
where $B_s$ is a constant which only depends on $s$. Replacing $\hat g = g\rho^{s+2}/(2\gamma)$, this leads to the following validity criterion for $\hat g \gg 1$
\bea \label{validity_rtp2}
v_0^2 \ll B_s g^2 \rho^{2s+2} \ &\text{for }& -1<s<\frac{1}{2} \;, \nn \\
\frac{v_0^2}{(2\gamma)^{\frac{2s-1}{s+1}}} \ll B_s g^{\frac{3}{1+s}} \rho^{\frac{s+4}{s+1}} \ &\text{for }& \frac{1}{2}<s<1 \;, \\
\frac{v_0^2}{\sqrt{2\gamma}} \ll B_s g^{3/2} \rho^{\frac{3}{2}s+1} \ &\text{for }& s>1 \;. \nn
\eea
As mentioned above, for $s>1/2$ this criterion becomes increasingly difficult to satisfy as $\gamma$ decreases (keeping all the other parameters fixed), while for $s<1/2$ it becomes completely independent of $\gamma$ for small $\gamma$. Thus our approximation should still hold in the limit $\gamma \to 0$, but only for $s<1/2$.

\section{Dynamical correlations} \label{sec:dynamical_rtp}

\subsection{Mean squared displacement and two-time correlations} \label{sec:MSD_rtp}

\begin{figure}
    \centering
    \includegraphics[width=0.8\linewidth,trim={0 7.5cm 0 2cm},clip]{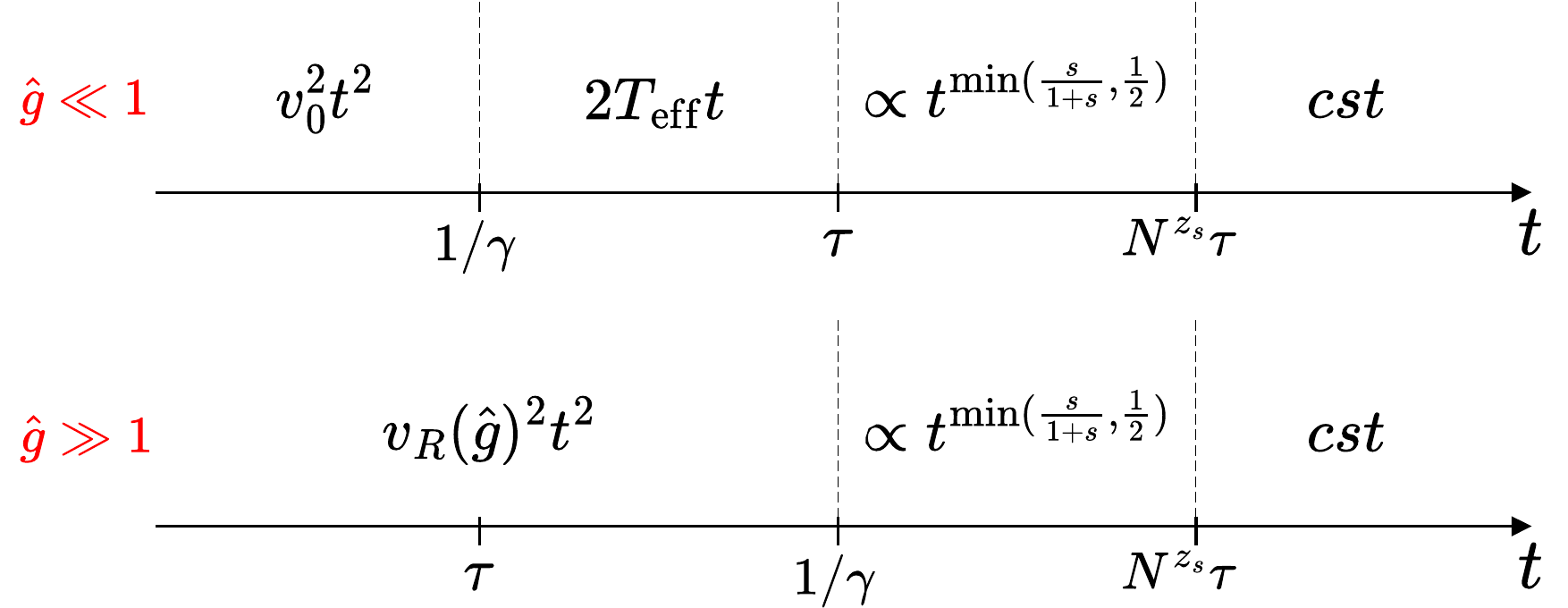}
    \caption{Sketch of the two different scenarios for the time evolution of the MSD $C_0(t)$ for the RTP Riesz gas with $s>0$, depending on the ordering of the time scales $1/\gamma$ and $\tau=1/(g\rho^{s+2})$. In the first case $\hat g \ll 1$, there is an intermediate free diffusion regime between the ballistic and the anomalous diffusion regime as time increases. In the second case $1 \ll \hat g \ll N^{z_s}$ there is a direct crossover from the ballistic to the anomalous diffusion regime. In both cases, the notation ``{\it cst}" denotes the large time saturation value $2 \langle \delta x_i^2 \rangle$. }
    \label{fig:time_regimes_rtp}
\end{figure}

Let us now move on to the dynamical correlations. As in the previous chapter, we start with the annealed case and we will discuss the quenched case in Sec.~\ref{sec:quenched_RTP}. We first consider the mean squared displacement of a tagged particle during time $t$. Using \eqref{cov_rtp}, it reads
\be  \label{disp_RTP} 
C_0(t) = \langle (\delta x_i(t) - \delta x_i(0))^2 \rangle 
= \frac{4v_0^2}{N} \sum_{q=1}^{(N-1)/2} \frac{\mu_q (1-e^{-2\gamma t}) -2\gamma (1-e^{- \mu_q t})}{\mu_q(\mu_q^2-4\gamma^2)} \;.
\ee 
As in the Brownian case, it converges to $2\langle \delta x_i^2 \rangle$ for $t \gg N^{z_s} \tau$ (and $t\gg 1/\gamma$). When $t\ll N^{z_s} \tau$, we can take the large $N$ limit by replacing the sum with an integral,
\be  \label{disp_RTP_largeN} 
C_0(t) \simeq
 4 T_{\rm eff} \tau \int_0^{1/2} du \frac{(1-e^{- f_s(u) t/\tau})-\hat g f_s(u) (1-e^{-2\gamma t})}{f_s(u)(1-\hat g^2 f_s(u)^2)} \;.
\ee
Let us now discuss the different time regimes for $t\ll N^{z_s} \tau$. Compared to the Brownian case where the only finite timescale is the interaction timescale $\tau=1/(g\rho^{s+2})$, there is an additional timescale, namely the persistence time of the active noise $1/\gamma$. The ratio of these two timescales is described by $\hat g=1/(2\gamma \tau)$. We assume that these two timescales are well separated, so that there are 3 well-defined asymptotic regimes. Depending on the value of $\hat g$, there are two possible scenarios, represented in Fig.~\ref{fig:time_regimes_rtp} for $s>0$, which we now describe: either $1/\gamma \ll \tau$ (i.e., $\hat g \ll 1$), or $1/\gamma \gg \tau$ (i.e., $\hat g \gg 1$). We note that the special case where $1/\gamma \gg N^{z_s}\tau$ was also discussed in \cite{RieszFluct}, see Sec.~IV.H therein.

We start with the large time regime where both $t\gg \tau$ and $t\gg 1/\gamma$. In this regime, one can show that the integral in \eqref{disp_RTP_largeN}
is dominated by $u\sim (t/\tau)^{-1/z_s} \ll 1$ (see Appendix~G in \cite{RieszFluct}). We thus recover the large time limit of the Brownian case, discussed in Sec.~\ref{sec:MSD_brownian}, with $T\to T_{\rm eff}$, As a reminder, this means that the leading behavior is subdiffusive for $s>0$, given in \eqref{displacement_Riesz_brownian} and logarithmic in time for $s=0$, see \eqref{disp_s0_large_t}, while for $-1<s<0$ it converges algebraically to a constant value $2\langle \delta x_i^2 \rangle$, see \eqref{decroissance}.

We now consider the regime where $1/\gamma \ll t \ll \tau$, which exists for $\hat g \ll 1$. In this case, the second term in the numerator and in the denominator of \eqref{disp_RTP_largeN} are both negligible, and we can expand the remaining exponential to first order, which leads to
\be
C_0(t) \simeq \frac{v_0^2}{\gamma} t = 2 T_{\rm eff} t \;.
\ee
Hence we recover free diffusion in this regime, with the effective diffusion coefficient $T_{\rm eff} = \frac{v_0^2}{2\gamma}$.

We thus see that on timescales which are large compared to the persistence time $1/\gamma$, we essentially recover the Brownian results. By contrast, when $t \ll 1/\gamma$ the activity plays an important role. When in addition $t\ll \tau$, it is clear that we can expand both exponentials in \eqref{disp_RTP_largeN}. One can actually show that this is also true in the other regime $\tau \ll t \ll 1/\gamma$ (see again Appendix~G in \cite{RieszFluct}). Taking this expansion to second order, we obtain a ballistic behavior
\be \label{ballistic}
C_0(t) \simeq v_R(\hat g)^2 t^2 \quad , \quad v_R(\hat g)^2 = v_0^2 \int_0^{1/2}
 \frac{2 \, du}{1 + \hat g f_s(u)}  \;,
\ee 
with a velocity $v_R=v_R(\hat g) < v_0$ which is renormalized by the interactions. This renormalized velocity is a function of the ratio $\hat g$ and tends to $v_0$ in the weak persistence limit $\hat g \ll 1$. In the opposite limit $\hat g \gg 1$, one can show that it behaves as
\bea
\frac{v_R(\hat g)^2}{v_0^2} \underset{\hat g \gg 1}{\propto} \begin{cases} \hat g^{-1} \quad \quad \text{for } -1 < s < 0 \;, \\  \hat g^{-\frac{1}{s+1}} \quad \text{for } 0 < s < 1 \;, \\  \hat g^{-1/2} \quad \; \text{for } s>1 \;. \end{cases}
\eea

The two scenarios are summarized in Fig.~\ref{fig:time_regimes_rtp}. For $\hat g\ll 1$ (top), after a ballistic behavior for $t\ll 1/\gamma$, the system becomes effectively diffusive for $t\gg \tau$, and we recover the Brownian results, with an intermediate free diffusion regime followed by a large time subdiffusive regime. On the other hand, for $\hat g\gg 1$ (bottom), there is a direct crossover from the ballistic regime to the large time subdiffusive regime on a timescale $t \sim 1/\gamma$. For $s>0$ this crossover is described by the scaling form
\be \label{C0scaling}
C_0(t) = T_{\rm eff} \tau \hat g^{1-\frac{1}{z_s}}  {\sf C}_s( \gamma t) \quad , \quad
{\sf C}_s(y) = \frac{4}{a_s^{1/z_s}} \int_0^{+\infty} dv \frac{1-e^{-2 y v^{z_s}} - v^{z_s}(1-e^{-2 y})}{v^{z_s}(1 - v^{2 z_s})} \;.
\ee
For $y \ll 1$, one has ${\sf C}_s(y) \simeq \frac{8\pi}{a_s^{1/z_s} z_s \sin(\frac{\pi}{z_s})} y^2$, which recovers the ballistic result \eqref{ballistic} in the limit $\hat g\gg 1$, while for $y \gg 1$ one has ${\sf C}_s(y) \simeq 4 a_s^{-1/z_s} \frac{\Gamma(1/z_s)}{z_s-1} (2y)^{1-\frac{1}{z_s}}$, recovering the subdiffusive regime \eqref{displacement_Riesz_brownian} (with $T\to T_{\rm eff}$). 

\begin{figure}[t]
    \centering
    \includegraphics[width=0.45\linewidth]{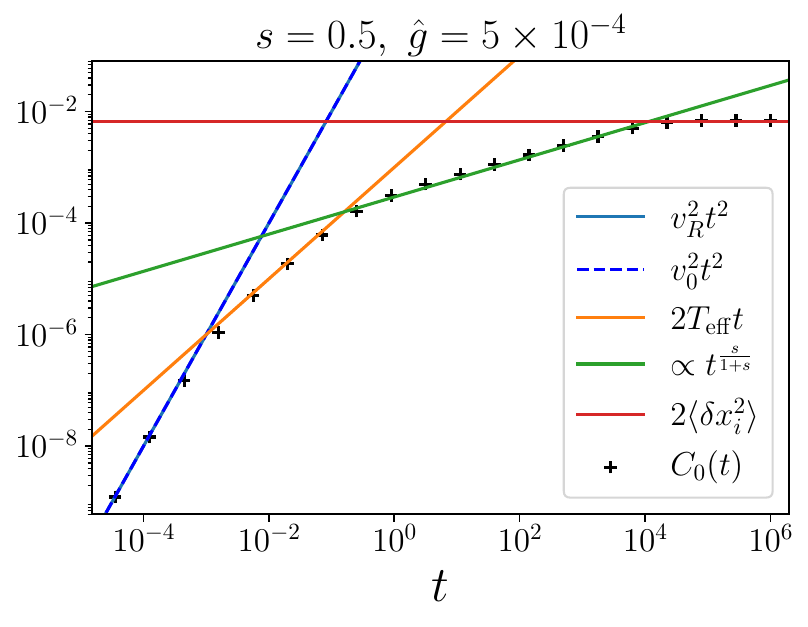}
    \hspace{0.5cm}
    \includegraphics[width=0.45\linewidth]{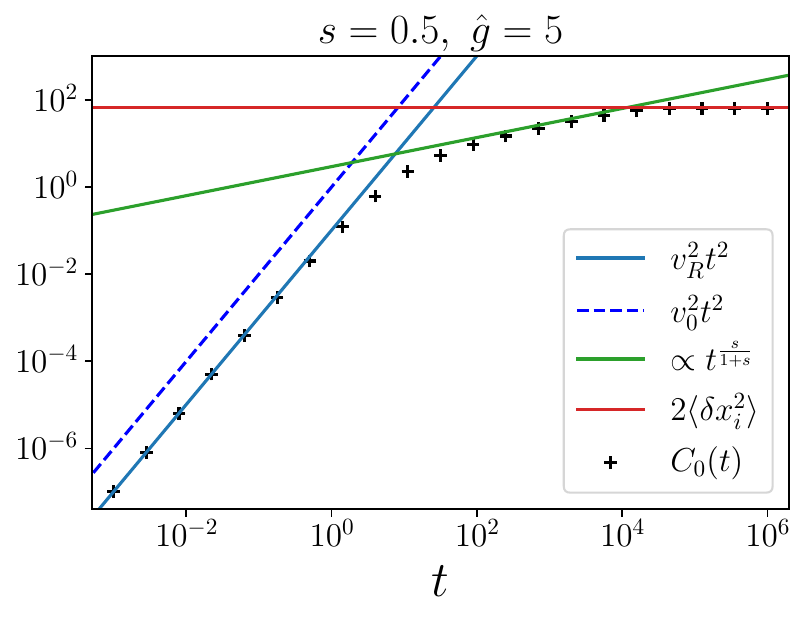}
    \caption{{\bf Left:} Evolution of the MSD $C_0(t)$ as a function of time, obtained by numerical computation of the sum \eqref{disp_RTP} for a RTP Riesz gas with $s=0.5$, $N=10001$, $g=1$, $\rho=1$, $v_0=1$ and $\gamma=1000$. The 4 time regimes of Fig.~\ref{fig:disp_rtp} (top line) are clearly visible. In the ballistic regime, the renormalized velocity $v_R(\hat g)$ is indistinguishable from the non-renormalized one $v_0$. {\bf Right:} Same plot for $\gamma=0.1$. There is no free diffusion regime in this case. In the ballistic regime, there is a clear difference between $v_R(\hat g)$ and $v_0$.}
    \label{fig:disp_rtp}
\end{figure}

The two scenarios are also illustrated in Fig.~\ref{fig:disp_rtp} for $s=0.5$, where we compare the exact expression \eqref{disp_RTP} (valid in the weak noise limit but for arbitrary $N$ and for any time $t$, here evaluated numerically for $N=10^4$) with the different asymptotic regimes in both cases $\hat g \ll 1$ (left) and $\hat g \gg 1$ (right). The time dependence of $C_0(t)$ in the different regimes for $\hat g \ll 1$ is also summarized in Table~\ref{table:disp_rtp1}, including the special cases $s=1$, $s=0$ and $-1<s<0$. As a final remark, we note that our results in the short-range coincide with what was obtained in \cite{HarmonicChainRTPDhar} for a harmonic chain of active particles, with an annealed initial condition, with the mapping discussed at the end of Sec.~\ref{sec:Riesz_eigvals}.
\\

\noindent {\bf Two-time correlations.} Let us now briefly discuss the two-time correlations $C_0(t_1,t_2)=\langle (\delta x_i(t_1) - \delta x_i(0)) (\delta x_i(t_2) - \delta x_i(0)) \rangle$. One can easily show check that the relation with $C_0(t)$ from the Brownian case still holds for RTPs (in the annealed case),
\be \label{multitime_rtp}
C_0(t_1,t_2) = \frac{1}{2} [ C_0(t_1) + C_0(t_2) - C_0(|t_1-t_2|) ]  \;.
\ee
From the above results for $C_0(t)$, when $t_1,t_2,|t_1-t_2|$ are of the same order, we thus obtain a free diffusion regime for $1/\gamma \ll t_1,t_2,|t_1-t_2| \ll \tau$ (which exists for $\hat g \ll 1$) with $C_0(t_1,t_2) \simeq 2 T_{\rm eff} \min(t_1,t_2)$, while at large time $t_1,t_2,|t_1-t_2| \gg \max(1/\gamma,\tau)$, $C_0(t_1,t_2)$ has the same form as for a fractional Brownian motion, see \eqref{multitime_corr_Riesz} (recalling however that in the RTP case the displacements are not Gaussian). Finally, in the ballistic regime $t_1,t_2,|t_1-t_2| \ll 1/\gamma$, we obtain
\be
C_0(t_1,t_2) \simeq v_R(\hat g)^2  t_1 t_2 \;,
\ee
with the same renormalized velocity $v_R(\hat g)$, as given in Eq. (\ref{ballistic}).

\begin{table}
\begin{center}
\begin{tabular}{|c|c|c|c|c|}
\hline
 & $t\ll 1/\gamma$ & $1/\gamma \ll t \ll \tau$ & $\tau \ll t \ll N^{z_s} \tau$ & $N^{z_s} \tau \ll t$ \\
\hline
$-1<s<0$ & \multirow{5}{*}{$\propto t^2$} & \multirow{5}{*}{$\propto t$} & $\sim cst$ & $\sim cst$ \\ 
\cline{1-1} \cline{4-5}
$s=0$ & & & $\propto \ln t$ & $\sim cst \times \ln N$ \\  
\cline{1-1} \cline{4-5}
$0<s<1$ & & & $\propto t^{\frac{s}{s+1}}$ & $\sim cst \times N^s$ \\
\cline{1-1} \cline{4-5}
$s=1$ & & & $\propto \sqrt{t/\ln t}$ & $\sim cst \times N/\ln N$ \\  
\cline{1-1} \cline{4-5}
$s>1$ & & & $\propto \sqrt{t}$ & $\sim cst \times N$ \\
\hline
\end{tabular}
\end{center}
\caption{Different time regimes for the MSD $C_0(t)$ as a function of the parameter $s$ of the interaction, when $1/\gamma \ll \tau$
(i.e., $\hat g \ll 1$, first line in Fig.~\ref{fig:time_regimes_rtp}). If $1/\gamma \gg \tau$ (i.e., $\hat g \gg 1$), the displacement remains ballistic 
until $t\sim 1/\gamma$ (with however a renormalized velocity $v_R(\hat g)$), and the free diffusion regime $C_0(t)\propto t$ is absent. The last regime corresponds to a saturation to a stationary limit, which depends on $N$ for $s \geq 0$.
Apart from the ballistic regime $t\ll 1/\gamma$, and the crossover away from the ballistic regime discussed in the text, see Eq.~\eqref{C0scaling}, $C_0(t)$ behaves as for the Brownian particles, discussed in Sec.~\ref{sec:MSD_brownian}, with the replacement $T \to T_{\rm eff}=\frac{v_0^2}{2\gamma}$.}
\label{table:disp_rtp1}
\end{table}

\subsection{Equal time covariance and time correlations of the gaps}

\begin{figure}[t]
    \centering
    \raisebox{-0.5\height}{\includegraphics[width=0.34\linewidth,trim={1.8cm 0cm 2cm 0.5cm},clip]{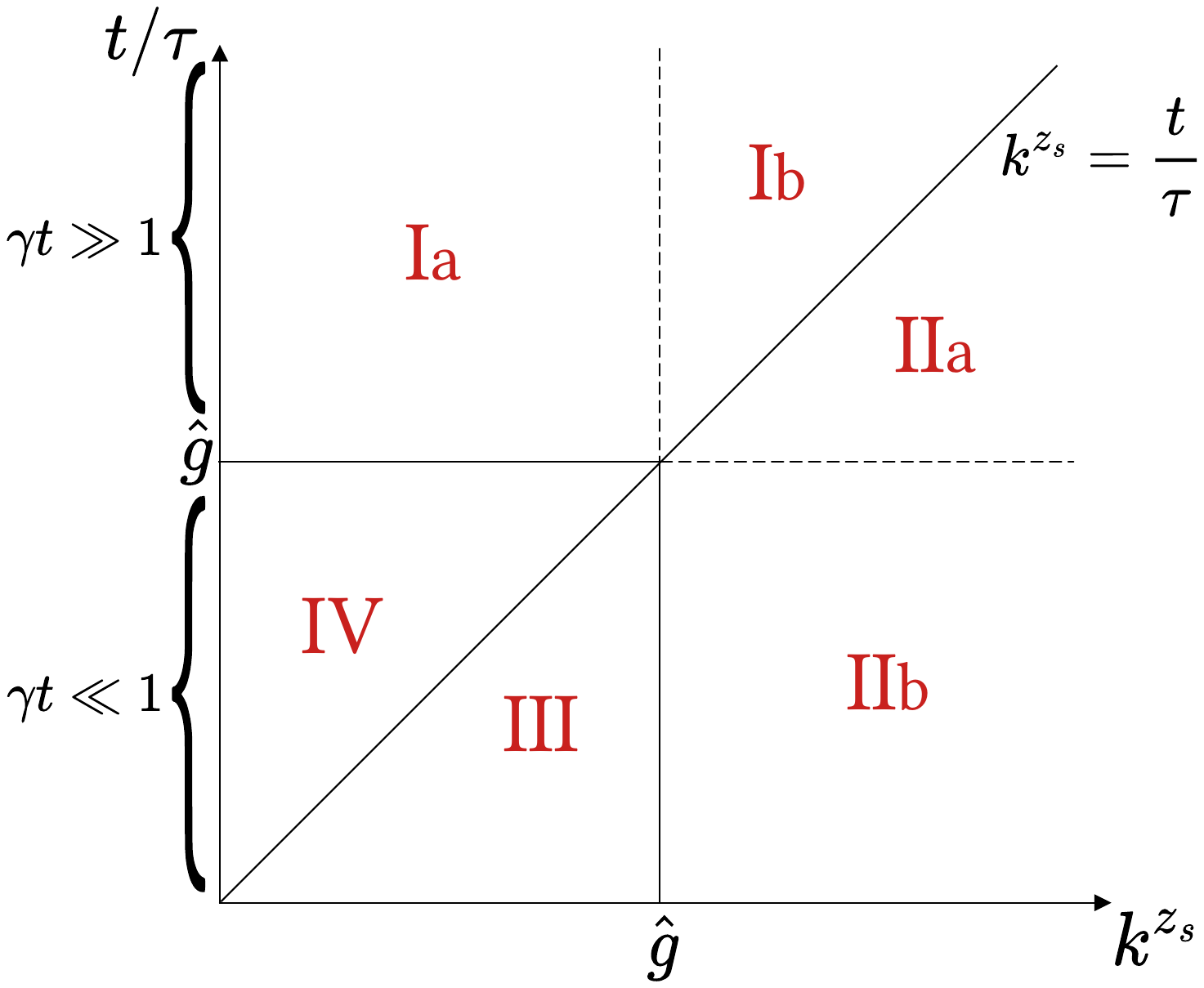}}
    \raisebox{-0.5\height}{\includegraphics[width=0.65\linewidth]{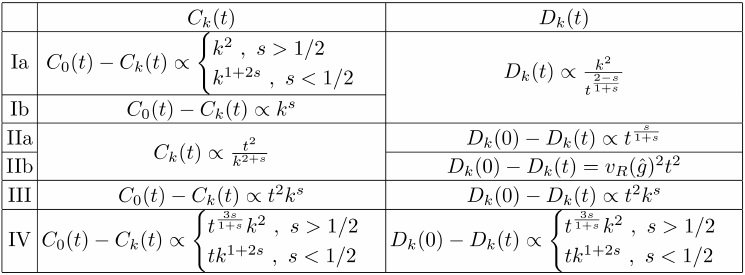}}
    \caption{{\bf Left panel:} The six asymptotic regimes of the equal time covariance $C_k(t)$ and of the two-time correlations of the gaps $D_k(t)$, defined in \eqref{defCkrtp} and \eqref{defDkrtp}, for the RTP Riesz gas, represented in the plane $(k^{z_s}, t/\tau)$, where $\hat g=1/(2 \gamma \tau)$, in the limit where $k,t/\tau,\hat g \gg 1$. {\bf Right panel:} Table of the leading behavior of $C_k(t)$ and $D_k(t)$ for $0<s<1$ in the six regions of the left panel, as a function of time $t$ and separation $k$.}
    \label{fig:diagrameCkDk_intro}
\end{figure}

As for the Brownian case, we now briefly discuss two more general dynamical quantities, namely the equal time covariance of the displacements $C_k(t)$ and the two-time correlations of the gaps $D_k(t)$, for which we recall the definitions,
\bea \label{defCkrtp}
C_k(t) &=& \langle (\delta x_i(t) - \delta x_i(0)) (\delta x_{i+k}(t) - \delta x_{i+k}(0)) \rangle \;, \\
D_k(t) &=& \langle (\delta x_i(t) - \delta x_{i+k}(t)) (\delta x_{i}(0) - \delta x_{i+k}(0))  \rangle \;. \label{defDkrtp}
\eea
As in the Brownian case, one can show that they are related through the identity $C_0(t)-C_k(t)=D_k(0)-D_k(t)$. We find that, when either $t/\tau \gg 1$, or $\hat g \gg 1$, or $k \gg 1$, $C_k(t)$ and $D_k(t)$ can each be written in two equivalent scaling forms
\be \label{Ck_scalingintro}
C_k(t) \simeq T_{\rm eff} \tau \hat g^{1-\frac{1}{z_s}}  \mathcal{\tilde F}_s( k/\hat g^{1/z_s} , \gamma t) 
= T_{\rm eff} \tau (t/\tau)^{1-\frac{1}{z_s}} \mathcal{F}_s \left(\frac{k}{(t/\tau)^{\frac{1}{z_s}}},\gamma t \right) \;,
\ee
and
\be \label{Dk_scalingintro}
D_k(t) \simeq T_{\rm eff} \tau \hat g^{1-\frac{1}{z_s}}  \mathcal{\tilde G}_s \left( \frac{k}{\hat g^{1/z_s}} , \gamma t \right) = T_{\rm eff} \tau (t/\tau)^{1-\frac{1}{z_s}} \mathcal{G}_s \left(\frac{k}{(t/\tau)^{\frac{1}{z_s}}},\gamma t \right) \;,
\ee
where we recall that the dynamical exponent is $z_s=1+s$ for $s<1$ and $z_s=2$ for $s>1$.
For each observable the two forms are equivalent, each being useful in a different regime. There are thus in total three scaling variables, which we denote $x=k/(t/\tau)^{1/z_s}$, ${\sf x} = k/\hat g^{1/z_s}$ and $y=\gamma t$. In the triple limit of large time, large separation and strong persistence, i.e., $t/\tau \gg 1$, $k \gg 1$ and $\hat g \gg 1$, this leads to six different limiting regimes, depending on whether each of the three scaling variables ${\sf x}, \, x, \, y$ is large or small. These six regimes are represented in Fig.~\ref{fig:diagrameCkDk_intro}, with the right panel giving the leading behavior of the two quantities $C_k(t)$ and $D_k(t)$ in terms of $t$ and $k$, while the left panel shows the region of the plane $(k^{z_s},t/\tau)$ in which they apply. Each line of this diagram corresponds to a crossover between two regimes, which can be described by a distinct scaling function of one of the arguments ${\sf x}, \, x, \, y$, both for $C_k(t)$ and for $D_k(t)$. Below we give a brief summary of these different regimes. Since the short range case is similar to the harmonic chain, we focus on the long-range case $0<s<1$ to simplify the discussion. For a more detailed study, including the expressions of the scaling functions in \eqref{Ck_scalingintro} and \eqref{Dk_scalingintro} (as well as the one-argument scaling functions for the different crossover regions), see Sec.~IV.E and IV.F of \cite{RieszFluct} respectively.

For both $C_k(t)$ and $D_k(t)$, the Brownian behavior of Sec.~\ref{sec:dynamicalBrownian} (with an effective temperature $T_{\rm eff}$) is recovered in the regions Ib and IIa, which correspond respectively to the regimes of large time $k^{z_s} \ll t/\tau$ and large separation $k^{z_s} \gg t/\tau$ of the Brownian case. However, the effect of the activity is visible both at short times $\gamma t \ll 1$ and at short distances $k^{z_s}\ll \hat g$. In the regions I, III and IV, i.e., when $k^{z_s}\ll t/\tau$ or $k^{z_s}\ll \hat g$, $C_k(t)$ is approximately equal to $C_0(t)$, but the difference $C_0(t)-C_k(t)$ exhibits distinct behaviors between these regions. In the region Ia, although $\gamma t \gg 1$, which implies that $C_0(t) \sim t^{\frac{s}{1+s}}$ behaves as for Brownian particles, the difference $C_0(t)-C_k(t) \simeq D_k(0)$ is given by the variance of the gaps, which in this regime where $k^{z_s}\ll \hat g$ is given by \eqref{Dk0_rtp_cases}, i.e., it is distinct from the Brownian behavior in the region Ib. In the regions II, III and IV, $C_k(t)$ is ballistic to leading order. The regions IIa and IIb are equivalent to leading order for $C_k(t)$, and correspond to large separations $k^{z_s}\gg \hat g,t/\tau$, where $C_k(t)$ is ballistic and decays to zero algebraically in $k$. Note that this ballistic behavior was already noted for Brownian particles in the previous chapter. Finally, in the regions III and IV, the leading order of $C_k(t)$ behaves as $C_0(t) \simeq v_R(\hat g)^2 t^2$, but while the difference $C_0(t)-C_k(t)$ is also ballistic in region III,
it has a non trivial power law dependence in time in region IV. For $D_k(t)$, the region I (both Ia and Ib) is identical to the large time behavior in the Brownian case, with a power law decay in time as in \eqref{Dk_decay_intro}. In the regions II, III and IV, where either $t/\tau \ll k^{z_s}$ or $\gamma t\ll 1$, the leading behavior of $D_k(t)$ is given by $D_k(0)$ (see Table \eqref{table:gap_var_rtp}), but the behavior of $D_k(0)-D_k(t)$ differs between these regions. The region IIa corresponds to the large separation limit of the Brownian case, where $D_k(0)-D_k(t) \simeq C_0(t) \sim t^{\frac{s}{1+s}}$. In the region IIb the same relation holds, but $C_0(t)$ is ballistic. The behavior of the difference $D_k(0)-D_k(t)$ in the regions III and IV can be obtained from the one of $C_0(t)-C_k(t)$ and is specific to the RTP case.

\subsection{Quenched initial condition} \label{sec:quenched_RTP}

As in the previous chapter, we have considered until now an {\it annealed} initial condition, where the system was initialized in the stationary state. Let us now see how the results of this section are modified for a {\it quenched} initial condition. As for Brownian particles, we assume that the initial density is uniform, so that $\delta x_i(0)=0$ for all $i$. Contrary to what was done in some previous works on the harmonic chain (e.g., in \cite{SinghChain2020,HarmonicChainRevABP}), we however keep a random initial condition for the telegraphic noise, i.e., the $\sigma_i(t)$ are still initialized in the stationary state ($\sigma_i(0)=\pm1$ with equal probability). As in Sec.~\ref{sec:quenchedBrownian} for the Brownian, we directly integrate the linearized equations of motion \eqref{Eq_delta_x_RTP} with this initial condition, leading to
\be
\delta x_i(t) = v_0 \sum_{j=1}^{N} \int_0^t dt_1 [e^{(t_1-t)H}]_{ij} \left( \sigma_j(t_1) - \frac{1}{N} \sum_{k=1}^N \sigma_k(t_1) \right) \;,
\ee
where $H$ is the Hessian matrix defined in \eqref{defHessian}. Using that $\langle \sigma_i(t) \sigma_j(t') \rangle = e^{-2\gamma|t-t'|}\delta_{ij}$, we obtain
\be
\langle \delta x_j(t) \delta x_k(t') \rangle_{\rm qu} = v_0^2 \int_0^{t} dt_1 \int_0^{t'} dt_2 \, e^{-2\gamma |t_1-t_2|} \left([e^{(t_1+t_2-t-t')H}]_{jk} -\frac{1}{N} \right) \;,
\ee
where we have again used the identity \eqref{identity_expH} to rewrite the last term, and where $\langle \cdot \rangle_{\rm qu}$ again denotes the average over the noise with a quenched initial condition. Finally, decomposing the matrix $H$ in its eigenbasis \eqref{eigenvals_rieszgeneral} and performing the integral we obtain,
\bea \label{covtwotime_quenched_RTP}
\langle \delta x_j(t) \delta x_k(t') \rangle_{\rm qu} 
= \frac{2v_0^2}{N} \sum_{q=1}^{(N-1)/2} &&\hspace{-0.8cm} \cos\big(2\pi \frac{q}{N}(j-k)\big) \\
&& \hspace{-1.5cm} \times \left[ \frac{e^{-2\gamma|t-t'|} - \frac{2\gamma}{\mu_q} e^{-\mu_q |t-t'|} - e^{-\mu_q t -2\gamma t'} - e^{-\mu_q t' -2\gamma t} }{\mu_q^2-4\gamma^2} + \frac{e^{-\mu_q(t+t')}}{\mu_q (\mu_q-2\gamma)} \right] \,. \nn
\eea
Again one can check that for $t,t'\to+\infty$ this coincides with the annealed result \eqref{cov_rtp}, recovering the static covariance $\langle \delta x_i \delta x_j \rangle$. Let us now see how some of the dynamical quantities computed in this section are affected. Below we only briefly describe the results, for more details see Sec.~IV.G in \cite{RieszFluct}. We note that, as in the Brownian case, $D_k^{\rm qu}(t)=0$ for the quenched initial condition.
\\

\noindent {\bf Mean squared displacement.} We first focus on the mean squared displacement during time $t$, which now reads
\be \label{C0_quenched_RTP}
C_0^{\rm qu} (t) = \langle \delta x_i(t)^2 \rangle_{\rm qu} 
= \frac{2v_0^2}{N} \sum_{q=1}^{(N-1)/2} \frac{\mu_q \left( 1 + e^{-2\mu_q t} - 2e^{-(\mu_q + 2\gamma) t} \right) - 2\gamma \left( 1 - e^{-2\mu_q t} \right)}{\mu_q(\mu_q^2-4\gamma^2)} \;.
\ee
Contrary to the Brownian case, this expression is not related to the annealed one (given in \eqref{disp_RTP}) by a simple rescaling. Thus, we will see that in this case the quenched initial condition does not only affect the numerical prefactors but actually leads to an entirely new time regime compared to the annealed case.
In the large $N$ limit, this can be written (for $t\ll N^{z_s\tau}$)
\be \label{C0_quenched_RTP_largeN}
C_0^{\rm qu} (t) \simeq 2v_0^2 \int_0^{1/2} du \frac{\frac{f_s(u)}{\tau} \left( 1 + e^{-2f_s(u)t/\tau} -2e^{-(f_s(u)/\tau+2\gamma)t} \right) -2\gamma \left(1-e^{-2f_s(u)t/\tau} \right)}{\frac{f_s(u)}{\tau} \left( \frac{f_s(u)^2}{\tau^2} -4\gamma^2 \right)}
\;.
\ee
The different time regimes that we need to consider are the same as in Sec.~\ref{sec:MSD_rtp} for the annealed case. For $\gamma t \gg 1$, we find that the results are essentially the same as in the annealed case, i.e., a free diffusion regime with diffusion coefficient $T_{\rm eff}$ in the regime $1/\gamma \ll t \ll \tau$ if it exists (i.e., if $\hat g \ll 1$), followed by a subdiffusive regime for $\tau \ll t \ll N^{z_s} \tau$, this time identical to the Brownian case with a quenched initial condition discussed in Sec.~\ref{sec:quenchedBrownian}, until convergence to the limiting value $\langle \delta x_i^2 \rangle$ given in \eqref{var_Riesz_liquid} (half of the annealed limit) for $t\gg N^{z_s}\tau$. A first interesting difference arises in the short time regime $t\ll \tau, 1/\gamma$. Indeed, in this case expanding all the exponentials to second order in \eqref{C0_quenched_RTP_largeN} leads to
\be
C_0^{\rm qu}(t) \simeq v_0^2 t^2 \;.
\ee
Thus, in this case the behavior is still ballistic, but the velocity is not renormalized by the interactions. This is in contrast with the annealed case \eqref{ballistic}, where for large $\hat g$ the effective velocity can be strongly reduced due to the interactions. The reason is that, since here we start from $\delta x_i=0$ for all $i$ at $t=0$, all the interaction forces compensate perfectly, and thus they have little effect on the short time dynamics.

\begin{figure}
    \begin{minipage}[c]{.62\linewidth}
    \centering
    \raisebox{-\height}{\includegraphics[width=\linewidth,trim={0 7.6cm 0 8.2cm}, clip]{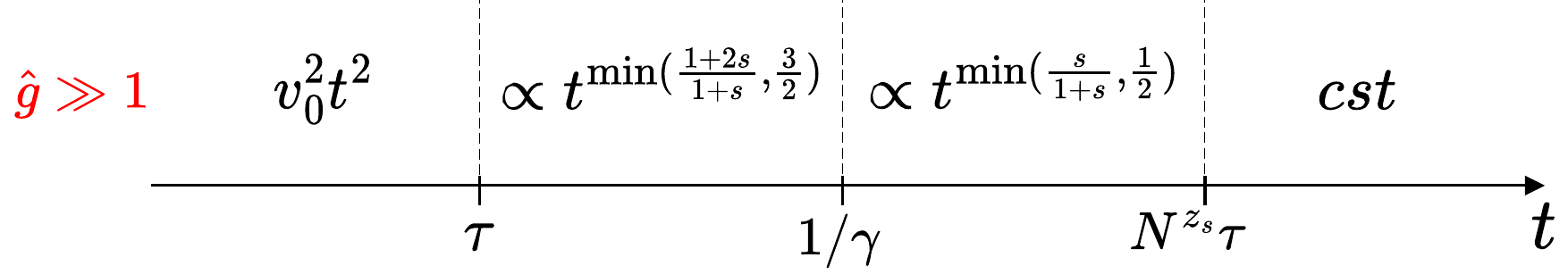}}
    \end{minipage}
    \hspace{0.2cm}
    \begin{minipage}[c]{.37\linewidth}
    \centering    
    \raisebox{-\height}{\includegraphics[width=\linewidth]{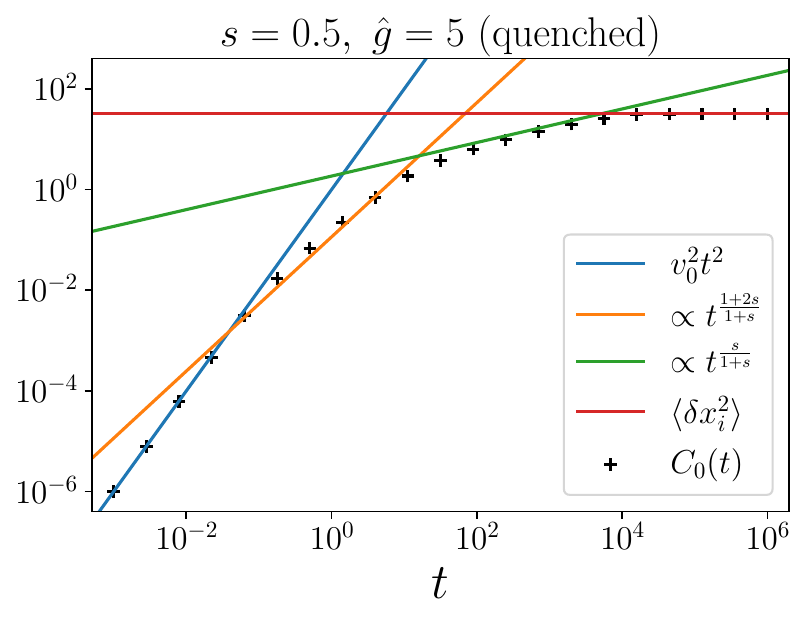}}
    \end{minipage}
    \caption{{\bf Left:} Sketch of the different time regimes for the MSD during time $t$ of a particle $C_0^{\rm qu}(t)$ for quenched initial conditions, for $s>0$. Only the case $1 \ll \hat g \ll N^{z_s}$ (corresponding to the bottom line in Fig.~\ref{fig:time_regimes_rtp}) is represented. The time evolution is first ballistic, then anomalous superdiffusive, then anomalous subdiffusive, and finally saturates to a constant $\langle \delta x_i^2 \rangle$. The anomalous superdiffusive regime has no analog in the annealed case. The case $\hat g \ll 1$ is identical to the annealed case up to numerical factors. {\bf Right:} Plot of $C_0^{\rm qu}(t)$ as a function of time, obtained by numerical computation of the sum \eqref{C0_quenched_RTP} for a RTP Riesz gas with $s=0.5$, $N=10001$, $g=1$, $\rho=1$, $v_0=1$ and $\gamma=0.1$. The 4 time regimes are clearly visible.}
    \label{fig:diagram_quenched_C0}
\end{figure}

In the strong persistence case $\hat g \gg 1$, there is one additional time regime which remains to be discussed, namely $\tau \ll t \ll 1/\gamma$. In the annealed case, we mentioned that the integral is dominated by $u \sim (t/\tau)^{-1/z_s}$, which leads to the same ballistic result $C_0(t)\simeq v_R(\hat g) t^2$ as for $t\ll \tau,1/\gamma$ (see \eqref{ballistic}). Here, one can show that the integral \eqref{C0_quenched_RTP_largeN} is again dominated by $u \sim (t/\tau)^{-1/z_s}$, but this leads to a different time dependence. Comparing the different terms in the integral, we find that, for any $s>-1/2$,
\be
C_0^{\rm qu} (t) \simeq 2v_0^2 \tau^2 \int_0^{1/2} du \frac{ \left(1 - e^{-f_s(u)t/\tau} \right)^2}{f_s(u)^2} \simeq 2v_0^2 t^{2-\frac{1}{z_s}} (\frac{\tau}{a_s})^{1/z_s} \int_0^{+\infty} dv \left(\frac{1 - e^{-v^{z_s}}}{v^{z_s}}\right)^2 \;,
\ee
which leads to
\be \label{C0_quenched_rtp_superdiff}
C_0^{\rm qu} (t) \simeq  \begin{dcases} \frac{4(2^{\frac{s}{1+s}}-1) \Gamma\left( -\frac{1+2s}{1+s} \right)}{(1+s) (a_s g\rho^{s+2})^{\frac{1}{1+s}}} v_0^2 \, t^{\frac{1+2s}{1+s}} \quad \text{for } -1/2<s<1 \;, \\ \frac{4(\sqrt{2}-1)}{3\sqrt{\pi (s+1) \zeta(s) g\rho^{s+2}}} v_0^2 \, t^{3/2}  \hspace{0.8cm} \text{for } s>1 \;. \end{dcases}
\ee
This is a new regime which was absent in the annealed setting. For $s>0$ it is superdiffusive. The crossover at $t\sim1/\gamma$ from the ballistic regime with $\hat g\gg 1$ to the large time subdiffusive regime described by the scaling form \eqref{C0scaling} in the annealed case is thus replaced by a crossover from a superdiffusive to a subdiffusive regime, with the scaling form
\bea \label{scalingC0RTP_quenched}
C_0^{\rm qu} (t) &\simeq& 2v_0^2 t^{2-\frac{1}{z_s}} \left(\frac{2\tau}{a_s}\right)^{1/z_s} {\sf C}_s^{\rm qu} (\gamma t) \;, \\
{\sf C}_s^{\rm qu} (y) &=& 2^{-2/z_s} \int_0^{+\infty} dw \frac{\frac{w^{z_s}}{2} \left( 1+e^{-w^{z_s}} -2e^{-(2y+\frac{w^{z_s}}{2})} \right) - 2y(1-e^{-w^{z_s}})}{\frac{w^{z_s}}{2} \left( \frac{w^{2z_s}}{4} -4y^2 \right)} \;. \nn
\eea

This new time regime is represented in Fig.~\ref{fig:diagram_quenched_C0} (left panel) for $s>0$, along with the other time regimes for $\hat g\gg 1$. In the right panel, we also show a comparison of the expression \eqref{C0_quenched_RTP} (exact at any $N$ and any $t$ in the weak noise limit) with the different asymptotic regimes for $s=0.5$. In the small persistence case $\hat g\ll 1$, the results are the same as in the annealed setting up to numerical prefactors, see the top line of Fig.~\ref{fig:time_regimes_rtp}. We note that in the short-range case $s>1$, our results coincides with the ones for the harmonic chain \cite{PutBerxVanderzande2019,SinghChain2020,HarmonicChainRevABP}. In particular, the expression \eqref{C0_quenched_rtp_superdiff} (bottom line) with the $t^{3/2}$ exponent coincides with the result \eqref{scaling_MSD_harmonicChain}-\eqref{scaling_MSD_harmonicChain_asympt} from \cite{SinghChain2020}, with the replacement $\tau=1/(g\rho^{s+2})\to\tau_K=1/K$ and $(s+1)\zeta(s) \to 1$ discussed in Sec.~\ref{sec:Riesz_eigvals} (the full scaling function \eqref{scalingC0RTP_quenched} also coincides with the expression (36) of \cite{SinghChain2020} for $z_s=2$ and $a_s=4\pi^2$).
\\

\noindent {\bf Equal time covariance.} In Sec.~IV.G of \cite{RieszFluct} we also discussed in detail the equal time covariance $C_k^{\rm qu}(t)$ with a quenched initial condition.
For $0<s<1$, compared with the results shown in Fig.~\ref{fig:diagrameCkDk_intro} for the annealed case, we find that the leading behavior is different in the regimes where $\gamma t \ll 1$. In particular, in the regions IIb and III, we find an algebraic decay with $k$ as $C_k^{\rm qu}(t)\propto t^3/k^{2+s}$ (versus $C^{\rm ann}_k(t)\propto t^2/k^{2+s}$ in region IIb and $C^{\rm ann}_k(t)\simeq C_0^{\rm ann}(t)$ in region III), while in region IV we now find $C_0^{\rm qu}(t)-C_k^{\rm qu}(t)\propto t^{\frac{2s-1}{1+s}}k^2$ for $s>1/2$ and $\propto k^{1+2s}$ for $s<1/2$, with the leading term $C_0^{\rm qu}(t) \propto t^{\frac{1+2s}{1+s}}$ now given by \eqref{C0_quenched_rtp_superdiff}. Again, these results generalize to the active Riesz gas previous results for the harmonic chain (which corresponds to the short-range case), in particular from \cite{SinghChain2020}.

\section{Conclusion}

In this chapter, we considered a Riesz gas of active particles with periodic boundary conditions. The method introduced in Chapter~\ref{chap:passiveRieszFluct} can be easily extended to take into account the time correlations of the noise, which allowed us to study in detail the microscopic fluctuations in this system. Although we presented our results for the case of RTPs, they are also valid for other standard models of active particles, in particular for AOUPs and ABPs. By studying quantities such as the mean squared displacement of a particle during time $t$ or the variance of the interparticle distance, we found that, while we essentially recover the Brownian results for large times and large distances, the activity leads to some unique behaviors at small times, but also on small lengthscales if the persistence time is sufficiently large. We have also compared our results with existing results for harmonic chains of active particles, which in the large $N$ limit can be directly mapped to the short-ranged case of the active Riesz gas.

As in the Brownian case, developing a more rigorous understanding of the domain of validity of our weak noise approximation, as well as taking the expansion further to compute higher order correlations and the higher order effects of the active noise are clear future directions. Similarly to the Brownian case, where we found excellent agreement between our results and some recent results obtained using MFT, another approach would be to extend MFT to the study of the active Riesz gas. This would allow us to recover some of the results of this chapter (such as the mean squared displacement and two-time correlations), and to access higher order correlation functions, as well as other interesting observables such as the current fluctuations and their large deviations. In the case of RTPs, the Dean-Kawasaki equation presented in Chapter~\ref{chap:DeanRTP} could be a good starting point for this. Another interesting direction would be to extend the present approach to higher dimensional systems. This would in particular allow us to address situations involving topological defects, such as dislocations, in two-dimensional systems 
\cite{James2021,Leticia2022,Chate2023}.

Here we have considered the case of periodic boundary conditions. One could ask if it is possible to extend the present method to interacting active particles inside a confining potential. This is indeed possible, at least for two special cases: the active DBM ($s=0$) in a harmonic potential, which we already studied in Chap.~\ref{chap:ADBM_Dean} from the point of view of the macroscopic density, as well as the active Calogero-Moser model ($s=2$) in a harmonic potential. This is the topic of the next chapter. We note that, while we have not performed numerical simulations to test the domain of validity of our weak noise approximation for the circular Riesz gas for arbitrary $s$, we have done so for the two models discussed in the next chapter. The excellent agreement between our analytical predictions and the numerical results for sufficiently small noise for these two models further confirms the validity of our method, even in the case of active particles.

\chapter{Active DBM and active Calogero-\allowbreak Moser model in a harmonic trap} \label{chap:ADBMfluct}

\section{Setting}

In Chapters~\ref{chap:passiveRieszFluct} and \ref{chap:activeRieszFluct}, the derivation of explicit expressions for the correlation functions was made possible by the periodic boundary conditions, which allowed for the exact diagonalization of the Hessian matrix for a generic interaction potential using plane waves. We may therefore wonder if it is possible to extend these results to a gas of trapped particles on the real axis, e.g., inside a harmonic potential $V(x)=\lambda x^2/2$. This is in general more difficult than in the periodic case. However, it turns out that it is possible at least in two special cases, thanks to the special structure of the Hessian matrix. The first of these two models is the active DBM in a harmonic trap, which we introduced in Chapter~\ref{chap:ADBM_Dean}, and which corresponds to a logarithmic interaction potential ($s=0$). The second model is an active version of the Calogero-Moser (CM) model discussed in Sec.~\ref{sec:CM_review}, corresponding to the case $s=2$ of the Riesz gas. 

Let us start with the active DBM. Here we focus on the variant which we called model II in Chapter~\ref{chap:ADBM_Dean}, where each particle interacts with all the other particles at all times and the particles cannot cross. We recall that the equations of motion for the particles $x_i(t)$ ($i=1,...,N$) are given by
\be \label{def_ADBM_chap9} 
\frac{dx_i}{dt} = - \lambda x_i +  \frac{2g}{N} \sum_{j (\neq i)} 
\frac{1}{x_i-x_j} + v_0 \sigma_i(t) + \sqrt{\frac{2 T}{N}} \,\xi_i(t) \;,
\ee 
where the $\sigma_i(t)$ are again independent telegraphic noises with rate $\gamma$ and the $\xi_i(t)$ are independent unit Gaussian white noises. It corresponds to the case $s=0$ of the active Riesz gas. Although the setting is different, we thus expect some connections with the results for $s=0$ from the previous chapter. We recall that, in the absence of noise, the positions of the particle converge to a ground state configuration given by \cite{HermiteZeros,bouchaud_book},
\be \label{eq_DBM_Hermite_zeros_chap9}
x_{{\rm eq},i} = \sqrt{\frac{2g}{\lambda N}}\, y_i \quad , \quad H_N(y_i) = 0 \quad , \quad i=1,...,N \;,
\ee 
where $H_N(x)$ is the $N^{th}$ Hermite polynomial. As in the previous two chapters, we consider the weak noise limit where the particles undergo small displacements around this ground state,
\be \label{dxi_adbm}
x_i(t) = x_{{\rm eq},i} + \delta x_i(t) 
\ee 
(note that contrary to the previous chapters, we do not need to remove the center of mass thanks to the confining potential). Since the particles cannot cross (at least for $T<2g$, i.e., $\beta=2g/T>1$, which we assume here), they keep the same ordering at all times and we can assume that $x_1(t) > x_2(t) >...>x_N(t)$ at all time $t$ (which also implies $x_{\rm eq,1} > x_{\rm eq,2} ...>x_{{\rm eq},N}$).

In Chapter~\ref{chap:ADBM_Dean}, we argued that for the purely active DBM ($T=0$ and $v_0>0$), the stationary density of particles (defined in \eqref{def_rho_sd}) in the large $N$ limit takes the same form as in the Brownian case (with $T=O(1)$), given by the Wigner semi-circle,
\be \label{WignerSC_chap9}
\rho_{s}(x) = \rho_{sc}(x) = \frac{\lambda}{2\pi g} \sqrt{\frac{4g}{\lambda}-x^2} \quad , \quad x\in[-\sqrt{g/\lambda},\sqrt{g/\lambda}] \;,
\ee
for a wide range of parameters (more precisely as long as $v_0/\sqrt{g\lambda}\ll\sqrt{N}$, see Fig.~\ref{phase_diagram_model2}). The results of this chapter will provide us with an additional, more quantitative argument to support this affirmation. The idea is that, since the density of the Hermite zeros converges to the semi-circle in the limit $N\to+\infty$, the stationary density of the active DBM also converges to the semi-circle (due to \eqref{eq_DBM_Hermite_zeros_chap9}) as long as the amplitude of the microscopic fluctuations vanishes for $N\to+\infty$ (see Sec.~\ref{sec:bulkFluctADBM} below).

We now introduce the active Calogero-Moser (CM) model. It corresponds to the active Riesz gas for $s=2$ on the real axis, again with a harmonic potential $V(x)=\lambda x^2/2$. Thus, contrary to the active DBM, it belongs to the short-range class. The equations of motion read
\begin{equation} \label{def_activCM}
\frac{dx_i}{dt} = - \lambda x_i + \frac{8 \tilde g^2}{N^2} \sum_{j (\neq i)} 
\frac{1}{(x_i-x_j)^3} + v_0 \sigma_i(t)  + \sqrt{\frac{2 T}{N}} \,\xi_i(t) \;.
\end{equation}
As mentioned in Sec.~\ref{sec:CM_review}, the positions of the particles in the ground state for the CM model are also described by the rescaled roots of the Hermite polynomial $H_N(x)$, as \cite{Calogero75, Moser76, Agarwal2019} 
\be \label{eq_CM}
x_{{\rm eq},i}  = \frac{1}{\lambda^{1/4}}\sqrt{\frac{2 \tilde g}{ \, N}}\, y_i \quad , \quad H_N(y_i)=0 \;.
\ee 
Once again, we consider the weak noise limit and study the small deviations $\delta x_i(t)$ around these equilibrium positions, see \eqref{dxi_adbm}. The particles can never cross and we assume again $x_1(t) > x_2(t) >...>x_N(t)$. Since the ground state is the same as for the DBM up to a rescaling, the stationary density in the limit $N\to+\infty$ is once again given by the semi-circle law in the Brownian case for $T=O(1)$, and we expect this to still be true in the active case for sufficiently weak noise,
\be \label{WignerSC_CM_chap9}
\rho_s(x)=\rho_{sc}^{CM}(x) = \frac{\lambda^{1/2}}{2\pi \tilde g} \sqrt{\frac{4\tilde g}{\lambda^{1/2}}-x^2} \quad , \quad x\in[-2\sqrt{\tilde g}/\lambda^{1/4},2\sqrt{\tilde g}/\lambda^{1/4}] \;.
\ee
Indeed, we will see that the different regimes of the active CM in the large $N$ limit are very similar to the ones of the active DBM shown in Fig.~\ref{phase_diagram_model2}. We will see however that differences between the two models appear when looking at the fluctuations, due to the fact that the active DBM is long-range while the active CM is short-range. As mentioned in Sec.~\ref{sec:CM_review}, the Hessian matrices $H^{DBM}$ and $H^{CM}$ of these two models are related through \cite{Agarwal2019}
\be \label{relHessians}
H^{CM} = \lambda^{-1} (H^{DBM})^2 \;.
\ee
They thus have the same eigenvectors, which will allow us to study both models in exactly the same way, but different eigenvalues, leading to different results for the microscopic fluctuations.

Although our method allows us to compute the dynamical correlations as we did in the previous two chapters, for this chapter we will mostly focus on the static correlations. We will begin by deriving an expression for the stationary two-point two-time covariance of the positions for the active DBM, which is exact in the weak noise limit. We will then analyze it in the limit of large $N$ to obtain scaling forms for the static variance and covariance of the particle positions as well as for the variance of the particle distance. An important new result compared to the periodic case is the distinction between a bulk regime and an edge regime, which exhibit different scalings with $N$ of the fluctuations. We will then show how these results are modified when we consider instead the active CM model. Interestingly, contrary to the active DBM, but also to the passive CM model, we will see that there is no edge regime in this case. As in the previous chapter, we will focus on the purely active case $T=0$ in most of this chapter, but we will also briefly discuss the passive (i.e., Brownian) case $v_0=0$, $T>0$, in Sec.~\ref{sec:passiveCM}, in particular for the CM model. Contrary to the passive DBM, which was extensively studied due to its connection with the Gaussian matrix ensembles (see Sec.~\ref{sec:DBM_brownian}), there are only few existing results concerning the overdamped Langevin dynamics of the passive CM model, besides the numerical study in \cite{Agarwal2019}. Our method also allows to study analytically the correlations for this case, and in particular to verify the large $N$ scalings observed numerically in \cite{Agarwal2019}. Let us recall that if both types of noise are present simultaneously, the results are additive at the linear order that we consider. We also recall that, as in the previous chapter, the RTP noise can be replaced by any other active noise with exponential time correlations, e.g., AOUP or ABP noise with the mappings given in Sec.~\ref{sec:Riesz_derivation_rtp}.

Throughout this chapter, we compare our analytical results with numerical results obtained by simulating the Langevin dynamics \eqref{def_ADBM_chap9} and \eqref{def_activCM} at finite $N$ and averaging over a large time window. Some details on the numerical simulations are given in Appendix~\ref{app:simu}. For the active Calogero-Moser model, the simulations were performed by Saikat Santra.
\\

\section{Active DBM in a harmonic trap} \label{sec:ADBM_harmonic_fluct}

\subsection{Derivation of the stationary two-point two-time covariance} \label{sec:ADBMtwotimederiv}

As in the previous two chapters, we start by linearizing the equations of motion \eqref{def_ADBM_chap9}, setting $T=0$, 
\be \label{small_dx_adbm}
\frac{d}{dt} \delta x_i(t) = - \sum_{j=1}^N H^{DBM}_{ij} \, \delta x_j(t) + v_0 \sigma_i(t) \;,
\ee
where the Hessian matrix reads
\be
H^{DBM}_{ij} = \lambda \mathcal{H}_{ij} = \lambda\left[\delta_{ij}\left(1+ \sum_{k\neq i} \frac{1}{(y_i-y_k)^2}\right) - (1-\delta_{ij}) \frac{1}{(y_i-y_j)^2}\right]\; .
\label{eqHessian_adbm}
\ee
As in the periodic case, we will estimate {\it a posteriori} the domain of validity of this approximation below. Note that the matrix $\mathcal{H}$ is only a function of the Hermite roots $y_i$, independent of the model parameters. Contrary to the circular case, we cannot use plane waves to diagonalize this matrix. However, in this particular case, the Hessian matrix can be diagonalized exactly, as proved in \cite{eigenvectors} (see also Appendix~A in \cite{ADBM2}). The eigenvalues of ${\cal H}$ are the first $N$ strictly positive integers $k=1,2,\cdots, N$, and the corresponding normalized eigenvectors read
\begin{equation}
    (\psi_k)_i = \frac{u_k(y_i)}{\sqrt{\sum_{j=1}^N u_k(y_j)^2}} \quad , \quad u_k(y) = \frac{H_N^{(k)}(y)}{H_N'(y)} 
    = 2^{k-1} \frac{(N-1)!}{(N-k)!} \frac{H_{N-k}(y)}{H_{N-1}(y)} \;.
    \label{Hermite_eigenvectors}
\end{equation}
This result is specific to the DBM $s=0$ in a harmonic trap, and the active CM $s=2$ discussed below thanks to the relation \eqref{relHessians}, which is why these are the only two cases that we can consider in this chapter. The linearized dynamics \eqref{small_dx_adbm} have exactly the same form as in the periodic case \eqref{Eq_delta_x_RTP}, except that here we do not need to subtract the motion of the center of mass since the particles are confined. Assuming that the system is in the stationary state, we can thus perform the same derivation as in Sec.~\ref{sec:Riesz_derivation_rtp}, inverting the equation \eqref{small_dx_adbm} in the frequency domain and using the expression \eqref{sigmacor_Fourier} for the correlations of the telegraphic noise to obtain the two-point two time covariance (as in the periodic case, we also have $\langle \delta x_i \rangle=0$ at this order),
\be
\langle \delta x_i(t) \delta x_j(t') \rangle = v_0^2 \int \frac{d\omega}{2 \pi} \frac{4\gamma \, e^{i \omega (t-t')}}{\omega^2+4\gamma^2} [\omega^2 \mathbbm{1}_N + \lambda^2 \mathcal{H}^2]^{-1}_{ij} \;,
\label{adbm_corr}
\ee
where $\mathbbm{1}_N$ is the $N\times N$ identity matrix. We can then use the eigendecomposition of $\mathcal{H}$ given in \eqref{Hermite_eigenvectors} to obtain
\begin{eqnarray}
\langle \delta x_i(t) \delta x_j(t') \rangle &=& v_0^2 \sum_{k=1}^N (\psi_k)_i (\psi_k)_j \int \frac{d\omega}{2 \pi} \frac{4\gamma}{\omega^2+4\gamma^2} \frac{e^{i \omega (t-t')}}{\omega^2+(\lambda k)^2} \nonumber \\
&=& \frac{v_0^2}{\lambda^2} \sum_{k=1}^N \frac{u_k(y_i)u_k(y_j)}{\sum_{l=1}^N u_k(y_l)^2} \frac{k e^{-2\gamma|t-t'|}-2\frac{\gamma}{\lambda} e^{-\lambda k|t-t'|}}{k(k^2-4\left(\frac{\gamma}{\lambda}\right)^2)} \;.
\label{corr_ADBM}
\end{eqnarray}
For the equal time covariance this gives
\be
\langle \delta x_i \delta x_j \rangle = \frac{v_0^2}{\lambda^2} \sum_{k=1}^N \frac{u_k(y_i)u_k(y_j)}{\sum_{l=1}^N u_k(y_l)^2} \frac{1}{k(k+2\frac{\gamma}{\lambda})} \;.
\label{corr_ADBM_stat}
\ee
These expressions are exact at any $N$ in the weak noise limit. They can be further simplified in the limit of large $N$, allowing in particular to understand how the fluctuations scale with $N$. However, this requires to distinguish between two regimes, which have different scalings: the {\it bulk} regime where $i\gg 1$ and $N-i \gg 1$, and the {\it edge} regime where $i=O(1)$ (or $N-i = O(1)$). We discuss both of them below.
\\

\noindent {\bf Comment on the timescales.} As in the periodic case, the inverse eigenvalues of the Hessian matrix provide the relaxation timescales of the system in the absence of active noise. The smallest timescale, corresponding to the local relaxation time, is thus $1/(\lambda N)$, while the largest one, corresponding to the global relaxation at the level of the full system, is $1/\lambda$. This is different from the previous chapters, where the local relaxation time $\tau$ was of order $O(1)$ while the global timescale $N^{z^s}\tau$ diverged with $N$. This is due to our choice of scaling: here we have scaled the interaction with $N$ such that the density retains a finite support in the limit $N\to+\infty$. This implies that the ``real'' density $N\rho_s(x)$, i.e., not rescaled by $N$ (which is the equivalent of $\rho=N/L$ from the previous two chapters), diverges as $N\to+\infty$. One implication of this change of scaling is that $\hat g=1/(2\gamma \tau) \to N \lambda/(2\gamma)\gg 1$, which suggests that the activity will have a strong effect in the present case. If we wanted to recover results closer to the previous chapter we would need to scale $\gamma\sim N$, however here we will instead keep $\gamma=O(1)$. Similarly, for the active CM model below the smallest and largest relaxation times will be given by $1/(\lambda N^2)$ and $1/\lambda$ respectively (using the relation \eqref{relHessians}), and thus we will have $\hat g= N^2 \lambda/(2\gamma)\gg 1$. Note also that, both for the active DBM and the active CM model in a harmonic trap, the relaxation times do not depend on the interaction strength or the density and are only determined by the strength $\lambda$ of the harmonic potential.

\subsection{Bulk regime} \label{sec:bulkFluctADBM}

For the particles in the bulk, the sum in \eqref{corr_ADBM} and\eqref{corr_ADBM_stat} is dominated by small values of $k$ (as long as $\gamma/\lambda=O(1)$). This allows to approximate the coefficients $u_k(y_i)$ as follows. We start from the recursion relation for the Hermite polynomials
\begin{equation}
    H_N''(x) = 2x H_N'(x) - 2N H_N(x)\; .
    \label{Hermite_equation}
\end{equation}
Differentiating this equation $k$, evaluating it at $x=y_i$ and dividing both sides by $H_N'(y_i)$, we obtain the following exact recursion relation for the $u_k(y_i)$
\begin{equation}
    u_{k+2}(y_i) = 2y_i u_{k+1}(y_i) - 2(N-k) u_{k}(y_i) \;,
    \label{recursion_exact}
\end{equation}
with initial conditions $u_0(y_i)=0$ and $u_1(y_i)=1$. Rescaling this equation by writing $y_i=\sqrt{2N}r_i$ and $u_k(y_i)=(2N)^{\frac{k-1}{2}}v_k(r_i)$, we obtain
\begin{equation}
    v_{k+2}(r_i) = 2r_i v_{k+1}(r_i) - \left(1 - \frac{k}{N} \right) v_{k}(r_i) \; .
    \label{recursion_rescaled}
\end{equation}
Since we only need to determine precisely the terms such $k\ll N$, let us now neglect the last term which is proportional to $k/N$. The simplified equation that we thus obtained is none other than the recursion relation satisfied by the Chebyshev polynomials of the second kind $U_k(r_i)$ \cite{ChebyWiki}. Since $v_1(r_i)=1=U_0(r_i)$ and $v_2(r_i) = 2r_i = U_1(r_i)$, we obtain that $v_k(r_i)=U_{k-1}(r_i)$ for all $k\geq 1$, i.e.,
\begin{equation}
    u_k(y_i) \simeq (2N)^{\frac{k-1}{2}} U_{k-1} \left( \frac{y_i}{\sqrt{2N}} \right)  \quad , \quad  U_{k-1}(r) = \frac{\sin(k \arccos(r))}{\sqrt{1-r^2}} 
    \label{uk_chebyshev}
\end{equation}
(for the second identity we have used the fact that $|y_i|<\sqrt{2N}$ for all $i$). This result can also be derived using the Plancherel-Rotach formula (see Appendix~E of \cite{ADBM2}). The denominator can then be simplified using the orthonormality of the Chebyshev polynomials $U_k(r)$ with respect to the Wigner semi-circle measure, 
\begin{equation}
    \sum_{l=1}^N U_{k-1} \left( \frac{y_l}{\sqrt{2N}} \right)^2 \simeq N \int_{-1}^1 dr \frac{2\sqrt{1-r^2}}{\pi} U_{k-1}(r)^2 = N \;.
    \label{normalization}
\end{equation}
Overall, one can check that this approximation leads to a relative error of order $O(N^{-1})$ in the bulk. We thus obtain the following expression for the two-point two-time covariance of the displacements of the bulk particles in the stationary state, in the limit $N \gg 1$ (with $\tilde \gamma=\gamma/\lambda=O(1)$ fixed),
\bea
\langle \delta x_i(t) \delta x_j(t') \rangle &\simeq& \frac{v_0^2}{\lambda^2 N} \,\mathcal{C}_b^{\gamma/\lambda}\left( \frac{x_{{\rm eq},i}}{2\sqrt{g/\lambda}}, \frac{x_{{\rm eq},j}}{2\sqrt{g/\lambda}}, \lambda |t-t'| \right) \; ,  \\
\mathcal{C}_b^{\tilde \gamma}(x,y,\tau) &=& \sum_{k=1}^\infty \frac{k e^{-2\tilde \gamma \tau}-2\tilde \gamma e^{-k\tau}}{k(k^2-4\tilde \gamma^2)} U_{k-1}(x) U_{k-1}(y) \;, \nn
\label{covADBM_largeN_time}
\eea
where the equilibrium positions $x_{{\rm eq},i}$ are given in \eqref{eq_DBM_Hermite_zeros_chap9}. For the static covariance, this reads
\begin{equation}
\langle \delta x_i \delta x_j \rangle \simeq \frac{v_0^2}{\lambda^2 N} \, \mathcal{C}_b^{\gamma/\lambda}\left( \frac{x_{{\rm eq},i}}{2\sqrt{g/\lambda}}, \frac{x_{{\rm eq},j}}{2\sqrt{g/\lambda}} \right) \quad , \quad 
\mathcal{C}_b^{\tilde \gamma}(x,y) = \sum_{k=1}^\infty \frac{U_{k-1}(x) U_{k-1}(y)}{k(k+2\tilde \gamma)}  \;.
\label{covADBM_largeN}
\end{equation}
Let us also give the expression for the variance of the displacement of a single particle,
\begin{equation}
\langle \delta x_i^2 \rangle \simeq \frac{v_0^2}{\lambda^2 N} \, \mathcal{V}_b^{\gamma/\lambda}\left( \frac{x_{{\rm eq},i}}{2\sqrt{g/\lambda}}\right) \quad , \quad \mathcal{V}_b^{\tilde \gamma}(x) = \mathcal{C}_b^{\tilde\gamma}(x,x) = \sum_{k=1}^\infty \frac{U_{k-1}(x)^2}{k(k+2\tilde \gamma)} \;.
\label{varADBM_largeN}
\end{equation}
We recall that $x_{{\rm eq},i} = \sqrt{\frac{2g}{\lambda \, N}}\, y_i \in (-2\sqrt{g/\lambda}, 2\sqrt{g/\lambda})$ and therefore $\mathcal{C}_b^{\tilde \gamma}(x,y)$ and $\mathcal{V}_b^{\tilde \gamma}(x)$ are defined on $(-1,1)^2$ and $(-1,1)$ respectively. We also recall that, inside the bulk, the equilibrium positions $x_{{\rm eq},i}$ can be approximated at large $N$ as $\frac{x_{{\rm eq},i}}{2\sqrt{g/\lambda}} = \frac{y_{{\rm eq},i}}{\sqrt{2N}} \simeq \mathcal{G}^{-1}(i/N)$, where $\mathcal{G}(x)= \frac{2}{\pi} \int_{-1}^x du \sqrt{1-u^2}$ is the cumulative distribution of the semi-circle law on $[-1,1]$. The scaling function for the variance $\mathcal{V}_b^{\tilde\gamma}(x)$ is minimal near the center of the trap $x=0$ and maximal near the edges (actually it diverges as $x\to\pm 1$, see below), see Fig.~\ref{figADBM1}. Note that these two functions are independent of $N$, and thus the covariance and variance both scale as $1/N$. This is to be compared with the case of the standard DBM, given in equation \eqref{DBMvar_largeN_ORourke} of Chapter~\ref{chap:Riesz_review} (and which we will briefly discuss again in sec.~\ref{sec:passiveCM} below), where the variance scales as $(\log N)/N^2$. While the additional $1/N$ comes from our choice of scaling, it is interesting to note the absence of the factor $\log N$ in the active case.
\\

\begin{figure}
    \centering
    \includegraphics[width=0.45\linewidth,trim={0cm 0 1cm 1cm},clip]{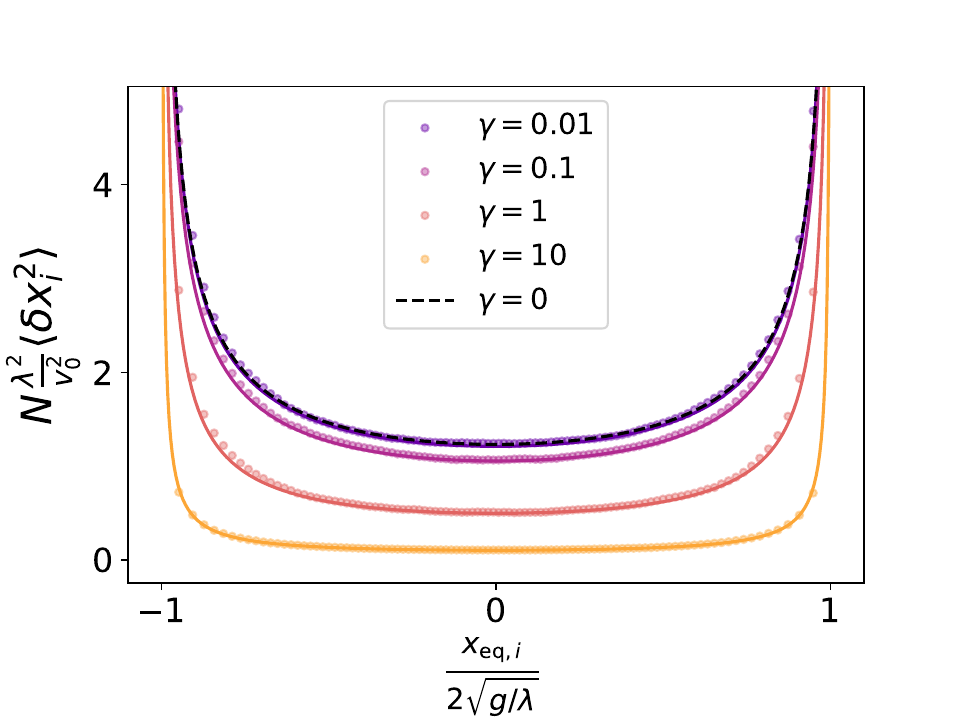}
    \includegraphics[width=0.45\linewidth,trim={0cm 0 1cm 1cm},clip]{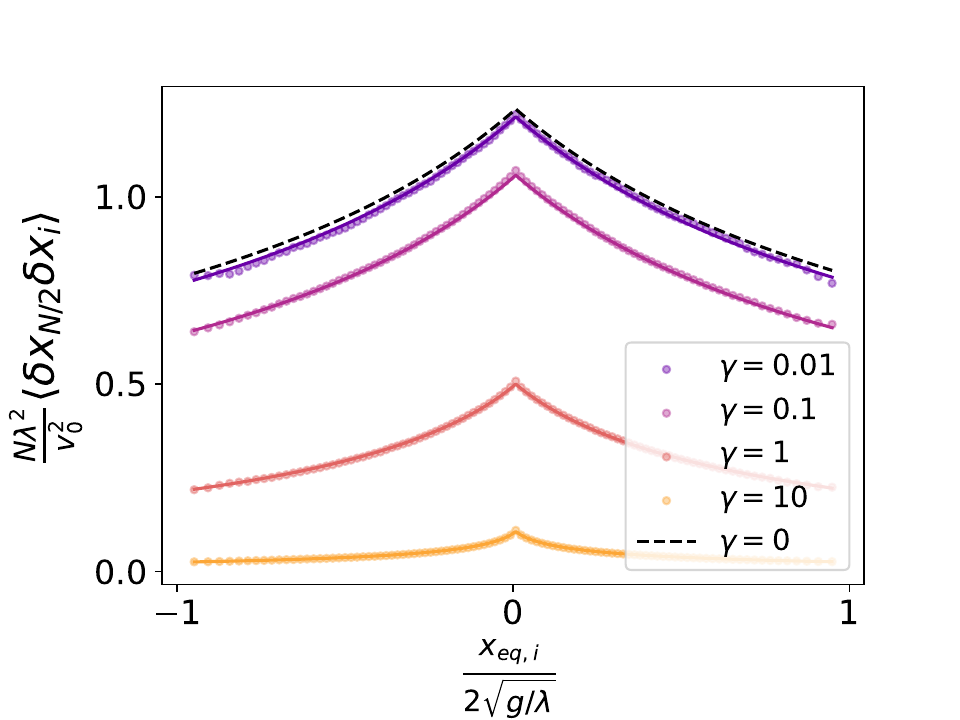}
    \caption{{\bf Left:} Variance of the displacement of a single particle $\langle\delta x_i^2 \rangle$ in the stationary state in the active DBM, plotted as a function of the equilibrium position $x_{{\rm eq},i}$, for $N=100$, $\lambda=1$, $g=1$, $v_0=0.1$ and different values of $\gamma$. The results of numerical simulations of the dynamics (dots) are compared with the the large $N$ prediction \eqref{varADBM_largeN} (lines) for each value of $\gamma$, showing a very good agreement. The dashed black line corresponds to the result for $\gamma\to 0^+$ given in \eqref{newVb}. {\bf Right:} Same plot for the covariance between particle $i$ and the central particle $i=N/2$, plotted as a function of $x_{{\rm eq},i}$. The full lines correspond to the large $N$ prediction \eqref{covADBM_largeN}, while the dashed black line correspond to the limit $\gamma\to 0^+$ \eqref{newCb}.}
    \label{figADBM1}
\end{figure}

\noindent {\bf Limit $\gamma\to 0^+$.} As can be seen from the formulas \eqref{covADBM_largeN} and \eqref{varADBM_largeN} above, both the variance and covariance are decreasing functions of $\gamma$. In addition, they have a finite limit as $\gamma \to 0^+$. In fact, for $\gamma=0$, the sums in \eqref{covADBM_largeN} and \eqref{varADBM_largeN} can be computed explicitly, leading to (see Appendix~B of \cite{ADBM2} for a derivation)
\be
\mathcal{C}_b^0(x,y) = \frac{\pi \arccos(\max(x,y)) - \arccos (x) \arccos (y)}{2\sqrt{1-x^2} \sqrt{1-y^2}} \;,
\label{newCb}
\ee
and
\be \label{newVb}
\mathcal{V}_b^0(x) = \frac{\arccos (x) (\pi-\arccos (x))}{2(1-x^2)} \;.
\ee
We note that the static covariance for $\gamma\to 0^+$ can also be obtained by considering the fixed points of the equation \eqref{small_dx_adbm} with all the $\sigma_i$ fixed, given by $\delta x_i = \frac{v_0}{\lambda} \sum_j (\mathcal{H}^{-1})_{ij} \sigma_j$ for all $i$. The idea is that, in the limit $\gamma\to 0^+$, the particles will spend a lot of time near these fixed points, thus the covariance can be obtained by averaging over all the fixed points, with the $\sigma_i$ being independent from each other and equal to $\pm1$ with equal probability. This leads to
\be \label{covADBMga0_Hessian}
\langle \delta x_i \delta x_j \rangle \simeq \frac{v_0^2}{\lambda^2} \mathcal{H}^{-2} \;,
\ee
which indeed coincides with \eqref{corr_ADBM_stat} for $\gamma=0$. This is the method which was used in \cite{ADBM2}.
\\

\noindent {\bf Comparison with numerical simulations. } We have tested numerically the validity of these analytical predictions by comparing them with numerical simulations, see Fig.~\ref{figADBM1}. For small enough ratios $v_0/\sqrt{g\lambda}$ (equal to $0.1$ in the figure), we find an excellent agreement for any value of $\gamma$.

\subsection{Edge regime} \label{sec:edgeFluctADBM}

The expressions of Sec.~\ref{sec:bulkFluctADBM} are valid in the bulk, i.e., for $i$ and $N-i$ of order $O(N)$. Looking for instance at the scaling function for the variance $\mathcal{V}_b^{\tilde \gamma}(x)$, we can see that it diverges for $x\to \pm 1$ as $\sim 1/\sqrt{1-|x|}$. Writing $\epsilon=1-x$, we have
\be \label{Vb_edge}
\mathcal{V}_b^{\tilde \gamma}(1-\epsilon) \simeq \sum_{k=1}^{\infty} \frac{\sin^2(k\sqrt{2\epsilon})}{2\epsilon k (k+2\tilde \gamma)} \simeq \int_0^\infty \frac{du}{\sqrt{2\epsilon}} \frac{\sin^2(u)}{u(u+2\tilde \gamma \sqrt{2\epsilon})} = \frac{\pi}{2\sqrt{2\epsilon}} +O(1)
\ee
(note that the leading order is independent of $\tilde \gamma=\gamma/\lambda$). This divergence suggests a different scaling at the edge. One way to see this is to consider the intermediate region where $1\ll i \ll N$ and to use the asymptotic expansion for the largest roots of the Hermite polynomials \cite{Hermite_asymptotics}, which gives
\begin{eqnarray}
x_{{\rm eq},i} = 2\sqrt{\frac{g}{\lambda}} \left( 1  + \frac{a_i}{2} N^{-2/3} + O(N^{-1}) \right) \;,
\label{hermite_roots_edge}
\end{eqnarray}
where $a_i$ is the $i^{th}$ zero of the Airy function, which for large $i$ is given by $a_i = - (\frac{3 \pi}{8} (4i-1))^{2/3} + O(i^{-4/3})$. Inserting this expansion into \eqref{varADBM_largeN} and using the asymptotic behavior in \eqref{Vb_edge}, we obtain
\begin{equation}
\langle \delta x_i^2 \rangle \simeq \frac{\pi}{2\sqrt{-a_i}} \frac{v_0^2}{\lambda^2 N^{2/3}} \underset{i \gg 1}{\simeq} \left( \frac{\pi^2}{3(4i-1)} \right)^{1/3} \frac{v_0^2}{\lambda^2 N^{2/3}} \;.
\label{rightmost_var}
\end{equation}
This suggests that the fluctuations scale as $N^{-2/3}$ for edge particles. Indeed, for $i=O(1)$, we can obtain a more precise approximation than what was done in the previous section by using the asymptotic expression of Hermite polynomials near the edge in terms of the Airy function $\Ai(x)$ (see \cite{ForresterHermiteEdge}), as well as the expansion \eqref{hermite_roots_edge} for the Hermite roots. For the static covariance, this leads to (for the derivation see Sec.~V in \cite{ADBM2})
\begin{eqnarray}
\langle \delta x_i \delta x_j \rangle &\simeq& \frac{v_0^2}{\lambda^2 N^{1/3}} \frac{1}{\Ai'(a_i) \Ai'(a_j)} \sum_{k=1}^{\infty} \frac{\Ai(a_i + kN^{-1/3})\Ai(a_j + kN^{-1/3})}{k(k+2\frac{\gamma}{\lambda})} \\
&\simeq& \frac{v_0^2}{\lambda^2 N^{2/3}} \frac{1}{\Ai'(a_i) \Ai'(a_j)} \int_0^{+\infty} dx \ \frac{\Ai(a_i + x)\Ai(a_j + x)}{x(x+2 \hat \gamma)} \quad , \quad \hat \gamma=\frac{\gamma}{\lambda}N^{-1/3} \;. \nn
\label{covADBM_edge}
\end{eqnarray}
This confirms the $N^{-2/3}$ scaling, instead of $N^{-1}$ as in the bulk, i.e., the fluctuations are larger at the edge. Note the scaling of the term $\hat \gamma=N^{-1/3}\gamma/\lambda$, which shows that if we keep $\gamma/\lambda=O(1)$ as in the rest of this chapter, the covariance and variance at the edge are actually independent of $\gamma$ at large $N$. This can be understood if we extend this result to the two-point two-time covariance, which yields the scaling form
\bea
\langle \delta x_i(t) \delta x_j(t') \rangle &\simeq& \frac{v_0^2}{\lambda^2 N^{2/3}} \, \mathcal{C}^{N^{-1/3}\gamma/\lambda}_e(a_i,a_j, N^{1/3} \lambda |t-t'|) \; , \\
{\cal C}_e^{\hat \gamma}(a_i,a_j, \hat \tau) &=& \frac{1}{\Ai'(a_i) \Ai'(a_j)} \int_0^{+\infty} dx \, \frac{x e^{-2\hat \gamma \hat \tau}-2\hat \gamma e^{-\hat \tau x}}{x(x^2-4\hat \gamma^2)} \Ai(a_i + x)\Ai(a_j + x)   \;. \nn
\label{cov_edge_integral_CM_time_result}
\eea
We see that the relaxation time scales as $N^{-1/3}$ for edge particles, i.e., it is much smaller than in the bulk where it is of order $O(1)$. Thus, the edge particles effectively behave as in the limit $\gamma\to 0^+$ if we do not scale $\gamma$ with $N$. The expressions \eqref{covADBM_edge} and \eqref{cov_edge_integral_CM_time_result} bear a strong resemblance with the recent result \eqref{AiryDBM} for the standard DBM (also obtained in the low temperature limit) \cite{GorinInfiniteBeta}. Note that the overall scaling in that case is $N^{-4/3}$ (or $N^{-1/3}$ if we do not scale the temperature as $1/N$).

One can check that the expression \eqref{rightmost_var} for $1\ll i \ll N$ can be recovered from the edge result \eqref{covADBM_edge} with $i=j$ by taking the limit $i \gg 1$, which shows that our results for the bulk and the edge indeed match in the intermediate region as one would expect (see Sec.~V of \cite{ADBM2}). As an additional comment, we note that the covariance between a bulk particle and an edge particle (i.e., $i=O(1)$ and $j\gg 1$) is still correctly given by the bulk expression \eqref{covADBM_largeN}.

\begin{figure}
    \centering
    \includegraphics[width=0.455\linewidth,trim={0cm 0 1cm 1cm},clip]{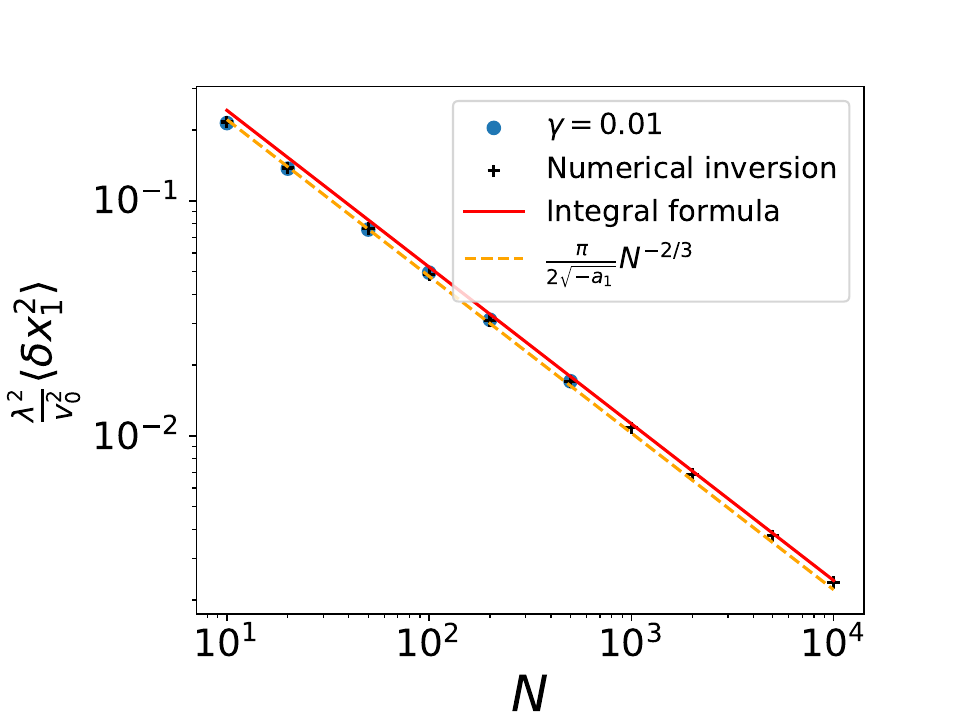}
    \includegraphics[width=0.45\linewidth,trim={0cm 0 1cm 1cm},clip]{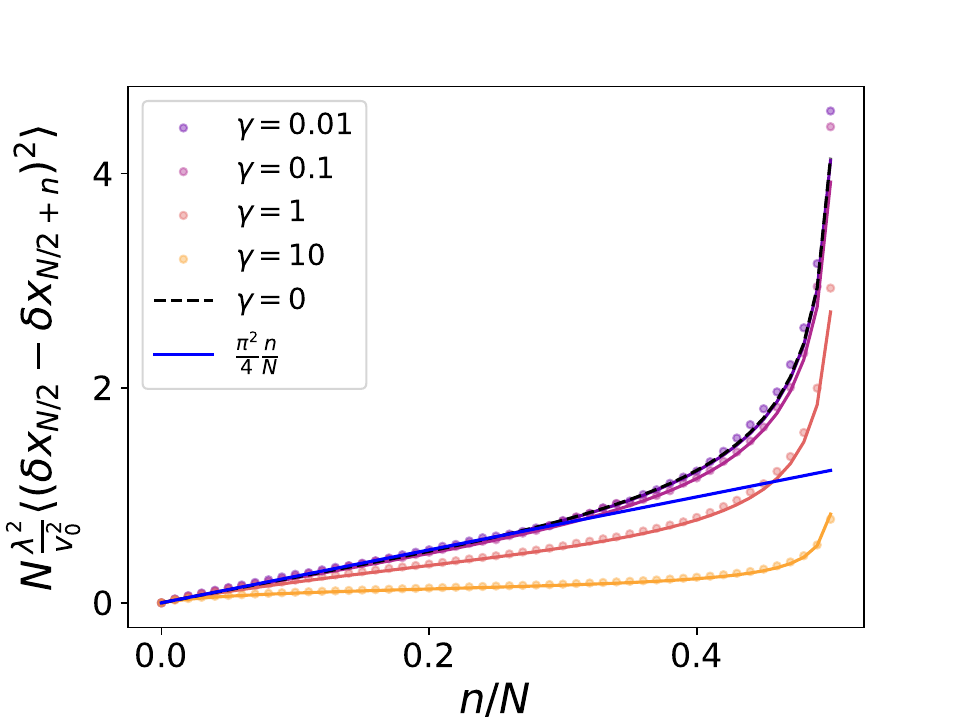}
    \caption{{\bf Left}: Variance of the position of the rightmost particle $x_1$ in the active DBM as a function of $N$, for $\lambda=1$, $g=1$, $v_0=0.1$ and $\gamma=0.01$, in log-log scale. The simulation results (blue dots) are in good agreement with the expression for large $N$ given in \eqref{covADBM_edge} (red line), which scales as $N^{-2/3}$. The black crosses show the results obtained by digonalizing numerically the Hessian matrix \eqref{eqHessian_adbm} to evaluate the finite $N$ expression \eqref{corr_ADBM_stat}. {\bf Right}: Variance of the distance between the central particle $i=N/2$ and the particle $i=N/2+n$ as a function of $n$, for $N=100$, $\lambda=1$, $g=1$, $v_0=0.1$ and different values of $\gamma$. The results of the numerical simulations (dots) are compared with the analytical prediction for large $N$ \eqref{gapvariance_largeN_ADBMharmonic} (full lines). The dashed black line shows the limit $\gamma \to 0^+$ and the blue line shows the linear approximation \eqref{gapvariance_largeN_ADBMharmonic_linear} in this limit. Note that the linear approximation breaks down as we get closer to the edge.}
    \label{variancefigedge}
\end{figure}

The edge regime is more difficult to study in simulations compared to the bulk since it requires to access larger values of $N$. We are thus not able to verify our predictions to a high precision in this case, but we still find a reasonable agreement. In particular, the $N^{-2/3}$ scaling seems to be well verified numerically, see Fig.~\ref{variancefigedge} (left panel).

\subsection{Variance of the gaps and validity of the approximation} \label{sec:gapvar_ADBMharmonic}

Our results also allow us to estimate the variance of the distance between two particles $i$ and $i+n$,
\begin{equation} \label{gapvardefADBM} 
    \langle (\delta x_i - \delta x_{i+n})^2 \rangle = \langle \delta x_i^2 \rangle + \langle \delta x_{i+n}^2 \rangle - 2 \langle \delta x_i \delta x_{i+n} \rangle \;.
\end{equation}
Let us start with the bulk regime ($i,N-i \gg 1$). In this case, we can use the results \eqref{covADBM_largeN} and \eqref{varADBM_largeN}, leading to
\bea
&&\langle (\delta x_i - \delta x_{i+n})^2 \rangle \simeq \frac{v_0^2}{\lambda^2 N} \,\mathcal{D}_b^{\gamma/\lambda} \left(\frac{x_{{\rm eq},i}}{2\sqrt{g/\lambda}}, \frac{x_{{\rm eq},i+n}}{2\sqrt{g/\lambda}}\right) \ , \\
&&\mathcal{D}_b^{\tilde \gamma} (x,y) = \mathcal{V}_b^{\tilde \gamma}(x) + \mathcal{V}_b^{\tilde \gamma}(y) - 2 \mathcal{C}_b^{\tilde \gamma}(x,y) = \sum_{k=1}^\infty \frac{\left(U_{k-1}(x)-U_{k-1}(y)\right)^2}{k(k+2\frac{\gamma}{\lambda})} \nn
\label{gapvariance_largeN_ADBMharmonic} \;.
\eea
Due to the approximation used in Sec.~\ref{sec:bulkFluctADBM}, this expression is however only valid for $n=\alpha N$ with $\alpha=O(1)$, i.e., on macroscopic scales. For $n=O(1)$, the leading order cancels due to the difference and the subleading terms can no longer be neglected. In the limit $\gamma\to 0^+$, we can analyze the behavior of \eqref{gapvariance_largeN_ADBMharmonic} for $\alpha \ll 1$ using that, for $x-y \ll 1$,
\be
\mathcal{D}_b^0(x,y) = \frac{\pi}{2} \frac{|x-y|}{(1-x^2)^{3/2}} + O\left( (x-y)^2 \right) \;.
\label{D_asympt}
\ee
Using the expression for the semi-circle density $\rho_{sc}(x)$ recalled in \eqref{WignerSC_chap9}, together with the fact that $x_{{\rm eq},i}-x_{{\rm eq},i+n}\simeq n/(N\rho_{sc}(x_{{\rm eq},i}))$, this leads to, for $1 \ll n \ll N$, and $\gamma\to 0^+$,
\begin{equation}
    \langle (\delta x_i - \delta x_{i+n})^2 \rangle \simeq \frac{v_0^2}{4 \pi^2 g^2 \rho_{sc}(x_{{\rm eq},i})^4} \frac{n}{N^2} \;.
    \label{gapvariance_largeN_ADBMharmonic_linear}
\end{equation}
The gap variance thus increases linearly with $n$ at intermediate distances inside the bulk, i.e., much faster than the $\log n$ dependence observed for the standard DBM, see \eqref{bouchaudDBM_distance} and \cite{Mehta_book,Forrester_book,bouchaud_book,Bourgade2022}. This linear dependence is in agreement with what we obtained in \eqref{gap_rtp_log_lin} of the previous chapter in the periodic case in the regime $n\ll \hat g$. Since in the present case we have $\hat g\sim N$, as noted in Sec.~\ref{sec:ADBMtwotimederiv}, we do not observe the equivalent of the large distance regime \eqref{gap_rtp_log_log2}, where the logarithmic dependence of the Brownian case is recovered. The result \eqref{gapvariance_largeN_ADBMharmonic_linear} also suggests that for $n=O(1)$, the gap variance scales as $N^{-2}$, much slower than the $N^{-1}$ scaling of the one-particle variance \eqref{varADBM_largeN}, showing that the particles move collectively. In the range $1 \ll n \ll N$ where we expect it to be valid, the prediction \eqref{gapvariance_largeN_ADBMharmonic} agrees very well with numerical simulations for small values of $v_0/\sqrt{g\lambda}$, as can be seen in the right panel of Fig.~ \ref{variancefigedge}. 

For edge particles ($i=O(1)$), we can use the expression \eqref{covADBM_edge} to evaluate the gap variance \eqref{gapvardefADBM} for large $N$,
\begin{equation}
\langle (\delta x_i - \delta x_{i+n})^2 \rangle \simeq
\frac{v_0^2}{\lambda^2 N^{2/3}} \int_0^{+\infty} \frac{dx}{x(x+2\hat \gamma)} \ \left[ \frac{\Ai(a_i + x)}{\Ai'(a_i)} - \frac{\Ai(a_{i+n} + x)}{\Ai'(a_{i+n})} \right]^2 \; , \quad \hat \gamma=\frac{\gamma}{\lambda}N^{-1/3} \;.
\label{gap_var_edge}
\end{equation}
Contrary to the bulk regime, this expression is valid for $n=O(1)$, and it has the same $N^{-2/3}$ scaling as the single particle variance. This shows that the correlations are much weaker at the edge than in the bulk, as one would expect.
\\

\noindent {\bf Validity of the weak noise approximation.} As for the Riesz gas on the circle, we can use our results on the variance of the gaps to estimate {\it a posteriori} the validity of the linear approximation used in Sec.~\ref{sec:ADBMtwotimederiv}. In this case, the general condition \eqref{cond_approx_Riesz1} implies that we should compare the standard deviation of the gaps $\sqrt{\langle (\delta x_i - \delta x_{i+n})^2 \rangle}$ with the average distance between the particle $\langle x_i - \delta x_{i+n} \rangle = x_{{\rm eq},i}- x_{{\rm eq},i+n}$. Assuming that the approximation \eqref{gapvariance_largeN_ADBMharmonic_linear} gives the correct order of magnitude for the gap variance inside the bulk for any $n$ and for arbitrary $\gamma$ (we know that for $\gamma>0$ it at least provides an upper bound), we obtain that the ratio of these two quantities decreases as $1/\sqrt{n}$, and thus the most restrictive condition is for neighboring particles, $n=1$. Using that $\rho_s\sim\sqrt{\lambda/g}$, we obtain the following validity condition inside the bulk
\begin{equation}  \label{ratioADBMvalidity} 
\frac{\sqrt{\langle (\delta x_i - \delta x_{i+1})^2 \rangle}}{x_{{\rm eq},i}- x_{{\rm eq},i+1}} \sim \frac{v_0/(\lambda N)}{\sqrt{(g/\lambda)}/N} = \frac{v_0}{\sqrt{g\lambda}} \ll 1 \; .
\end{equation}
This gives the condition under which we expect the expressions for the bulk given in Sec.~\ref{sec:bulkFluctADBM} to provide a good quantitative descriptions of the fluctuations in the model (it is indicated by a red line in Fig.~\ref{phase_diagram_model2}. Below we will however assume that the bulk scaling that we have obtained, $\langle \delta x_i^2 \rangle \sim v_0^2/(\lambda^2 N)$ remains valid beyond this regime, which seems to be the case in our numerical simulations (see Chapter~\ref{chap:ADBM_Dean}).

If we now estimate the ratio \eqref{ratioADBMvalidity} in the edge regime, we find using \eqref{gap_var_edge} that $\langle (\delta x_i - \delta x_{i+1})^2 \rangle \sim (N^{-1/3} v_0/\lambda)^2$, while $x_{{\rm eq},i}- x_{{\rm eq},i+1}\sim N^{-2/3}\sqrt{g/\lambda}$ from \eqref{hermite_roots_edge}, which leads to the condition $v_0/\sqrt{g\lambda}\ll N^{-1/3}$. We thus expect our approximation to break down more easily at the edge, which we indeed observe numerically.

\subsection{Implications for the stationary density} \label{sec:ADBMdiscussphasediagram}

We now consider the implications of our results on the microscopic fluctuations for the stationary density in the active DBM, which we studied in Chapter~\ref{chap:ADBM_Dean}. There, we saw that the stationary density at large $N$ takes the form of the Wigner semi-circle \eqref{WignerSC_chap9} for a wide range of parameters. Let us assume for now that $\gamma/\lambda=O(1)$, such that inside the bulk $\langle \delta x_i^2 \rangle \sim v_0^2/(\lambda^2 N)$ according to \eqref{varADBM_largeN}. The regime of parameters such that the stationary density converges to the semi-circle can be found by comparing the typical amplitude of the fluctuations $\sqrt{\langle \delta x_i^2 \rangle}$ to the size of the support of the semi-circle density $[-x_e,x_e]$ with $x_e=2\sqrt{g/\lambda}$,
\begin{equation} \label{ratioADBM_semicircle} 
    \frac{\sqrt{\langle \delta x_i^2 \rangle}}{x_e} \sim \frac{v_0}{\sqrt{g \lambda N}} \;.
\end{equation}
If this ratio is much smaller than unity, the total density will not deviate much from the ground state, i.e., it will remain close to the Wigner semi-circle. This implies that the semi-circle density holds when $v_0/\sqrt{g\lambda}\ll \sqrt{N}$, corresponding to the line on the right of Fig.~\ref{phase_diagram_model2}. This confirms our prediction from the scaling forms \eqref{scalingforms_model2}. For $v_0/\sqrt{g\lambda}\gg \sqrt{N}$, the density is dominated by the fluctuations and strongly deviates from the semi-circle. This regime was studied in Sec.~\ref{sec:model2}.

On the left of Fig.~\ref{phase_diagram_model2}, there is an additional regime where the coarse-grained density is still given by the Wigner semi-circle, but where each particle appears as a separate peak in the mean density. To find the limit between this regime and the one where the density is smooth, we should compare the amplitude of the fluctuations to the typical distance between neighboring particles,
\begin{equation} \label{ratio1} 
    \frac{\sqrt{\langle \delta x_i^2 \rangle}}{x_{{\rm eq},i} - x_{{\rm eq},i+1}} \sim \frac{v_0}{\sqrt{g\lambda}} \sqrt{N} \;.
\end{equation}
We thus see that the peaks will be visible when $v_0/\sqrt{g \lambda} \ll N^{-1/2}$, as indicated in Fig.~\ref{phase_diagram_model2}.

Let us now consider what happens if we strongly vary the value of the parameter $\gamma/\lambda$. As can be seen in \eqref{varADBM_largeN}, the variance is a decreasing function of $\gamma$, with a finite limit as $\gamma\to 0$. Thus, increasing the value of $\gamma$ up to $\gamma\sim N$ will delay the two transitions discussed above, i.e., from a peaked to a smooth density and from the semi-circle to the strong fluctuations regime. However, one cannot break the semi-circle simply by varying $\gamma$ while keeping $v_0/\sqrt{g\lambda}=O(1)$. To conclude this section, let us note that the larger amplitude of the fluctuations at the edges of the support found in Sec.~\ref{sec:edgeFluctADBM} may be related to the edge effects (in particular the ``wings'') observed numerically in the density in Chapter.~\ref{chap:ADBM_Dean}.

\section{Active Calogero-Moser model in a harmonic trap} \label{sec:activeCMfluct}

\subsection{Variance of the particle positions}

Let us now turn to the active Calogero-Moser model defined in \eqref{def_activCM}. Thanks to the relation \eqref{relHessians} between the Hessian matrices of the two models, we can directly adapt the result for the two-point two-time covariance of the active DBM derived in Sec.~\ref{sec:ADBMtwotimederiv} to the active CM. Indeed, since $H^{CM}=\lambda \mathcal{H}^2$, we simply need to replace $\lambda k \to \lambda k^2$ in the expression \eqref{corr_ADBM}, leading to
\be
\langle \delta x_i(t) \delta x_j(t') \rangle = \frac{v_0^2}{\lambda^2} \sum_{k=1}^N \frac{u_k(y_i)u_k(y_j)}{\sum_{l=1}^N u_k(y_l)^2} \frac{k^2 e^{-2\gamma|t-t'|}-2\frac{\gamma}{\lambda} e^{-\lambda k^2|t-t'|}}{k^2(k^4-4\left(\frac{\gamma}{\lambda}\right)^2)} \;,
\label{corr_CM}
\ee
and for the static covariance,
\be
\langle \delta x_i \delta x_j \rangle = \frac{v_0^2}{\lambda^2} \sum_{k=1}^N \frac{u_k(y_i)u_k(y_j)}{\sum_{l=1}^N u_k(y_l)^2} \frac{1}{k^2(k^2+2\frac{\gamma}{\lambda})} \;.
\label{corr_CM_stat}
\ee
These results are valid at any $N$ in the weak noise limit. An important difference with the active DBM is that, in the limit of large $N$, there is no distinct scaling for the edge particles. Thus, the approximations used in Sec.~\ref{sec:bulkFluctADBM} for the bulk regime of the active DBM are valid for all particles in the case of the active CM. For large $N$, the two-point two-time covariance thus takes the following scaling form, for any particles $i$ and $j$,
\bea
\langle \delta x_i(t) \delta x_j(t') \rangle &\simeq& \frac{v_0^2}{\lambda^2 N} \,\mathcal{\widetilde C}_b^{\gamma/\lambda}\left( \frac{x_{{\rm eq},i}}{2\sqrt{\tilde g}/\lambda^{1/4}}, \frac{x_{{\rm eq},j}}{2\sqrt{\tilde g}/\lambda^{1/4}}, \lambda |t-t'| \right) \; , \\ 
\mathcal{\widetilde C}_b^{\tilde \gamma}(x,y,\tau) &=& \sum_{k=1}^\infty \frac{k^2 e^{-2\tilde \gamma \tau}-2\tilde \gamma e^{-k^2\tau}}{k^2(k^4-4\tilde \gamma^2)} U_{k-1}(x) U_{k-1}(y) \;, \nn
\label{covCM_largeN_time}
\eea
up to a relative error of order $O(N^{-1})$ (as long as $\gamma/\lambda \ll N$), where the equilibrium positions $x_{{\rm eq},i}$ are now given in \eqref{eq_CM}. For the static covariance this becomes
\begin{equation}
\langle \delta x_i \delta x_j \rangle \simeq \frac{v_0^2}{\lambda^2 N} \,\mathcal{\widetilde C}_b^{\gamma/\lambda}\left( \frac{x_{{\rm eq},i}}{2\sqrt{\tilde g}/\lambda^{1/4}}, \frac{x_{{\rm eq},j}}{2\sqrt{\tilde g}/\lambda^{1/4}} \right) \quad , \quad 
\mathcal{\widetilde C}_b^{\tilde \gamma}(x,y) = \sum_{k=1}^\infty \frac{U_{k-1}(x) U_{k-1}(y)}{k^2(k^2+2\tilde \gamma)}  \;,
\label{covCM_largeN}
\end{equation}
and for the one-particle variance,
\begin{equation}
\langle \delta x_i^2 \rangle \simeq \frac{v_0^2}{\lambda^2 N} \, \mathcal{\widetilde V}_b^{\gamma/\lambda}\left( \frac{x_{{\rm eq},i}}{2\sqrt{\tilde g}/\lambda^{1/4}}\right) \quad , \quad \mathcal{\widetilde V}_b^{\tilde \gamma}(x) = \mathcal{\widetilde C}_b^{\tilde\gamma}(x,x) = \sum_{k=1}^\infty \frac{U_{k-1}(x)^2}{k^2(k^2+2\tilde \gamma)} \;.
\label{varCM_largeN}
\end{equation}
We recall that the expression of the Chebyshev polynomials of the second kind $U_k(r)$ is given in \eqref{uk_chebyshev}. We also recall that $x_{{\rm eq},i} \in (-2\sqrt{\tilde g}/\lambda^{1/4}, 2\sqrt{\tilde g}/\lambda^{1/4})$ so that once again $\mathcal{\widetilde C}_b^{\tilde \gamma}(x,y)$ and $\mathcal{\widetilde V}_b^{\tilde \gamma}(x)$ are defined on $(-1,1)^2$ and $(-1,1)$ respectively. As in the active DBM, the static variance and covariance of the particle positions thus scale as $1/N$. Contrary to the active DBM however, the scaling functions $\mathcal{\widetilde C}_b^{\tilde \gamma}(x,y)$ and $\mathcal{\widetilde V}_b^{\tilde \gamma}(x)$ have finite limits as $x,y\to\pm 1$ (which depend on $\tilde \gamma$). For instance, for $x\to 1^-$, one has $U_{k-1}(x) \to k$, and thus
\be
\mathcal{\widetilde V}_b^{\tilde \gamma}(x) \xrightarrow[x\to 1^-]{} \sum_{k=1}^{\infty} \frac{1}{k^2+2\tilde \gamma} = \frac{\pi}{2} \frac{\coth(\pi \sqrt{2\tilde \gamma})}{\sqrt{2\tilde \gamma}} - \frac{1}{4\tilde \gamma} \simeq \begin{cases} \frac{\pi^2}{6}  \hspace{0.38cm} \text{ for } \tilde \gamma \ll 1 \;, \\ \frac{\pi}{2\sqrt{2 \tilde \gamma}}  \text{ for } \tilde \gamma \gg 1 \;. \end{cases}
\ee
This means that the results \eqref{covCM_largeN_time}, \eqref{covCM_largeN} and \eqref{varCM_largeN}, and in particular the $1/N$ scaling, remain valid even for the edge particles. This is in contrast with the passive CM model (i.e., with Brownian particles), for which an edge regime does indeed exist, as observed numerically in \cite{Agarwal2019} and as we will confirm below.
\\

\noindent {\bf Limit $\gamma\to 0^+$.} As for the active DBM, the scaling functions $\mathcal{\widetilde C}_b^{\tilde \gamma}(x,y)$ and $\mathcal{\widetilde V}_b^{\tilde \gamma}(x)$ are decreasing functions of $\gamma$ and have a finite limit for $\gamma\to 0^+$, for which the sum can be computed explicitly, leading to
\be \label{CM_Cblimit}
\mathcal{C}_b^0(x,y) = \frac{c  (\arccos x,\arccos y ) }{\sqrt{1-x^2} \sqrt{1-y^2}} \quad , \quad  c(u,v)= \frac{v}{12}(\pi-u)(2 \pi u -u^2-v^2) \text{ for } u\geq v \;,
\ee
and for the one-particle variance,
\be \label{CM_Vblimit}
\mathcal{V}_b^0(x)=\frac{\arccos^2 x(\pi -\arccos x)^2}{6(1-x^2)} \;.
\ee
As for the active DBM, the limit $\gamma\to 0^+$ can also be studied by averaging over the fixed points of the dynamics with fixed $\sigma_i$ (see Sec.~\ref{sec:bulkFluctADBM}), as done in \cite{activeCM}.

\begin{figure}
\includegraphics[width=0.45\linewidth]{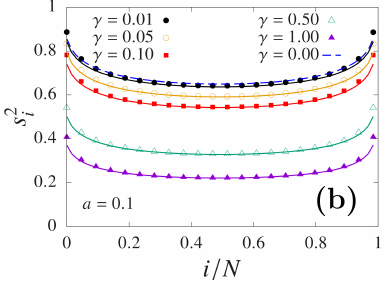}
\includegraphics[width=0.032\linewidth,trim={0 0 19.05cm 0.5cm},clip]{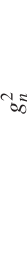}
\includegraphics[width=0.45\linewidth,trim={6.5cm 0 6.5cm 0},clip]{images/var_all_gap_128.pdf}
\caption{{\bf Left:} Variance of the displacement of a single particle $s_i^2=\langle\delta x_i^2 \rangle$ in the stationary state in the active CM model, plotted as a function of $i/N$, for $N=64$, $\lambda=1$, $\tilde g=1$, $v_0=6.4$ and different values of $\gamma$. The results of numerical simulations of the dynamics (dots) are compared with the the large $N$ prediction \eqref{varCM_largeN} (lines) for each value of $\gamma$, showing a very good agreement. The dashed blue line corresponds to the result for $\gamma\to 0^+$ given in \eqref{CM_Vblimit}. {\bf Right:} Variance of the distance between the central particle $i=N/2$ and the particle $i=N/2+n$, $g^2_n=\langle (\delta x_{N/2}-\delta x_{N/2+n})^2\rangle$, as a function of $n$, for $N=128$, $\lambda=1$, $\tilde g=1$, $v_0=6.4$ and different values of $\gamma$. The results of the numerical simulations (dots) are compared with the analytical prediction for large $N$ \eqref{gapvar_activeCM} (full lines). The dashed red line shows the limit $\gamma \to 0^+$. Simulations and figures by Saikat Santra.}
\label{figure_var_posnCM}
\end{figure}

\subsection{Variance of the gaps}

Let us now consider the variance of the distance between two particles $i$ and $i+n$, as we did for the active DBM. Using the results \eqref{covCM_largeN} and \eqref{varCM_largeN}, we obtain
\be \label{gapvar_activeCM}
\langle (\delta x_i - \delta x_{i+n})^2 \rangle = \frac{v_0^2}{\lambda^2 N} \,\mathcal{\widetilde D}_b^{\gamma/\lambda} \left(\frac{x_{{\rm eq},i}}{2\sqrt{\tilde g}/\lambda^{1/4}}, \frac{x_{{\rm eq},i+n}}{2\sqrt{\tilde g}/\lambda^{1/4}}\right) \quad , \quad \mathcal{\widetilde D}_b^{\tilde \gamma} (x,y) 
= \sum_{k=1}^{\infty} \frac{\left(U_{k-1}(x)-U_{k-1}(y) \right)^2}{k^2(k^2+2\tilde \gamma)}
\;.
\ee
Contrary to the active DBM case, this result is valid for every particle $i$, and also holds at the microscopic scale $n=O(1)$. This is due to the faster convergence of the series in \eqref{covCM_largeN}. To determine the behavior of the variance of the gaps for $n=O(1)$, we can thus expand the scaling function $\mathcal{D}_b^{\tilde \gamma}(x,y)$
for $y$ close to $x$, which leads to
\bea
&&\mathcal{D}_b^{\tilde \gamma}(x,y) \simeq A^{\tilde \gamma}(x) (x-y)^2 \; , \\
\text{where} \hspace{-0.5cm} &&A^{\tilde \gamma}(x) = \frac{1}{(1-x^2)^2} \sum_{k=1}^{\infty} \frac{(k \cos(k \arccos x) - \cot (\arccos x) \sin(k \arccos x))^2}{k^2( k^2+2\tilde \gamma) } \;. \nn
\eea
Using that $x_{{\rm eq},i}-x_{{\rm eq},i+n}\simeq n/(N\rho_{sc}^{CM}(x_{{\rm eq},i}))$, where $\rho_{sc}^{CM}(x)$ is the semi-circle density for the parameters of the CM model given in \eqref{WignerSC_CM_chap9}, we obtain, for $n \ll N$, 
\be 
\langle (\delta x_i - \delta x_{i+n})^2 \rangle \simeq \frac{v_0^2}{\lambda^2 N^3} \, B^{\gamma/\lambda}\left( \frac{x_{{\rm eq},i}}{2\sqrt{\tilde g}/\lambda^{1/4}} \right) \,   n^2 \quad , \quad 
B^{\tilde \gamma}(x) = \frac{\pi^2}{4} 
\frac{A^{\tilde \gamma}(x)}{1-x^2} \;. \label{gapvariance_largeN_gamma}
\ee 
At the center of the harmonic trap $x=0$, the sum can be computed explicitly, leading to
\be \label{gapvarCM_n2}
B^{\tilde \gamma}(0) = \frac{\pi^2}{16} \left( \frac{\pi}{\sqrt{2\tilde \gamma}} 
\coth \big( \pi \sqrt{\frac{\tilde \gamma}{2}} \big) -\frac{1}{\tilde \gamma} \right) \;.
\ee 
The variance of the gaps thus increases faster than linearly, as $n^2$. This is to be compared with the linear dependence for the passive CM model (see below). This result coincides with the one for the periodic case in the regime $n\ll \hat g^{1/2}$, given in \eqref{Dk0_rtp_cases} (bottom line). As discussed for the active DBM, here we have $\hat g \sim N^2$ (see the discussion at the end of Sec.~\ref{sec:ADBMtwotimederiv}), and thus we do not observe the linear regime at larger distances. Note the $N^{-3}$ scaling, much smaller than the $N^{-1}$ scaling for the variance of the individual particle positions, showing as for the active DBM that the particles move collectively. Finally, as we already discussed for the circular Riesz gas in Sec.~\ref{sec:gapvarRieszactive}, for $n\gg 1$ the variance of the interparticle distance is related to the variance of the number of particles inside a fixed interval (counting statistics), and thus the faster than linear increase in \eqref{gapvarCM_n2} indicates giant number fluctuations \cite{TonerTuReview,ChateGiant,GinelliGiant,DasGiant2012,Chate2010,NarayanGiant,ZhangGiant}.

\subsection{Validity of the approximation and stationary density} \label{sec:activeCMregimes}

Let us now estimate the domain of validity of the linear approximation, as done in Sec.~\ref{sec:gapvar_ADBMharmonic} for the active DBM. In this case, from \eqref{gapvariance_largeN_gamma}, we obtain the criterion
\be
\frac{\sqrt{\langle (\delta x_i - \delta x_{i+1})^2 \rangle}}{x_{{\rm eq},i}-x_{{\rm eq},i+1}} \sim \frac{v_0/(\lambda N^{3/2})}{\sqrt{\tilde g}/(\lambda^{1/4} N)} = \frac{v_0}{\lambda^{3/4} \sqrt{\tilde g N}} \ll 1 \;.
\ee
For the active CM model, the linear approximation should thus hold in a much wider range of parameters than for the active DBM (even more so since there is no edge regime, which is where the approximation first breaks down for the active DBM). Indeed, we have tested our analytical predictions by comparing them with the results of numerical simulations, both for the one particle variance \eqref{varCM_largeN} and the gap variance \eqref{gapvar_activeCM}, see Fig.~\ref{figure_var_posnCM}, and we find an excellent agreement up to higher values of the dimensionless parameter $v_0/(\lambda^{1/4}\sqrt{\tilde g})$ (equal to $6.4$ in the figure) than for the active DBM.

\begin{figure}[t]
\includegraphics[width=1.0\linewidth]{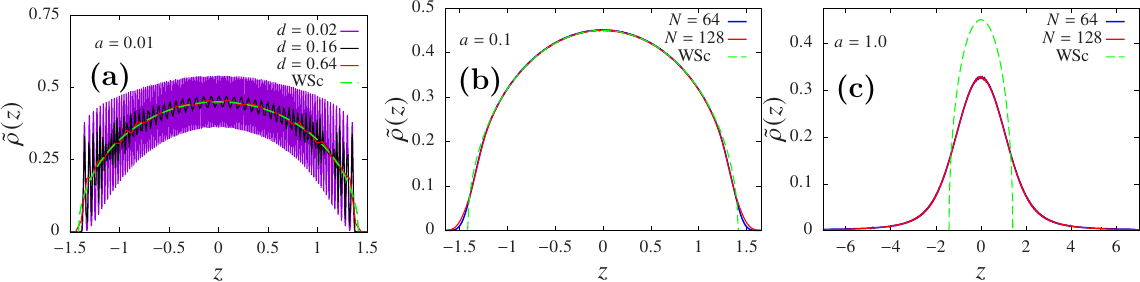}
\caption{Particle density $\rho_s(x)$ in the active CM model (rescaled) obtained from numerical simulations, showing the 3 different regimes discussed in Sec.~\ref{sec:activeCMregimes}. Here $\tilde g=1$, $\lambda=1$ and $\gamma=1$, and $v_0=a N$ with, from left to right, $a=0.01$ (strongly localized regime), $a=0.1$ (smooth semi-circle regime) and $a=1$ (strong fluctuations regime). On the left $N=128$ and the density is coarse grained with a varying bin size $d/\sqrt{N}$, showing convergence to the Wigner semi-circle as $d$ is increases. On the center and right plot, the density is plotted for $N=64$ and $N=128$. On all 3 figures the Wigner semi-circle is plotted in dashed green lines. Simulations and figures by Saikat Santra.}
\label{figure_density_large_CM}
\end{figure}

As we have already mentioned several times, the density in the ground state for the CM model is also given by the Wigner semi-circle law. We can thus use the same arguments as in Sec.~\ref{sec:ADBMdiscussphasediagram} to estimate the values of the parameters for which the transitions between the different regimes occur. Since the fluctuations have the same scaling $\langle \delta x_i^2 \rangle \sim v_0^2/(\lambda^2 N)$, we obtain similar results: (i) for very weak noise $v_0/(\lambda^{1/4}\sqrt{g})\ll N^{-1/2}$, the particles are strongly localized and appear as peaks in the density, which however takes a semi-circular shape after coarse-graining, (ii) in a broad intermediate regime $N^{-1/2} \ll v_0/(\lambda^{1/4}\sqrt{g})\ll N^{1/2}$, the mean particle density is smooth and takes a semi-circular shape, and (iii) for very strong noise $v_0/(\lambda^{1/4}\sqrt{g})\gg N^{1/2}$, the semi-circle breaks down and the density takes a bell shape. This last regime should be the same as the one studied in Sec.~\ref{sec:model2} for the active DBM, since in this regime the interaction effectively behaves as a contact interaction. The numerical simulations performed in \cite{activeCM} are compatible with these predictions, see for instance Fig.~\ref{figure_density_large_CM} which illustrates the 3 regimes for the stationary density. Differences with the active DBM however appear when looking at the finite $N$ fluctuations. In the active CM, the edge fluctuations are weaker, and accordingly the strong edge effects observed in the density of the active DBM do not appear in this case.

\section{Passive Calogero-Moser model and passive DBM in a harmonic trap} \label{sec:passiveCM}

\subsection{Passive calogero-Moser model in a harmonic trap}

We now briefly discuss the passive version of the Calogero-Moser model, described by \eqref{def_activCM} with $v_0=0$ and $T>0$. As mentioned in Sec.~\ref{sec:CM_review}, the Calogero-Moser model has mostly been studied in the context of Hamiltonian dynamics due to its integrability properties \cite{Calogero71,Calogero75,Moser76,KP17,Poly06,OP81}. By contrast, there have been until now very few studies on its overdamped dynamics in the presence of thermal noise apart from a numerical study in \cite{Agarwal2019}. The method used in this chapter allows us to study analytically the fluctuations in this method, confirming some of the numerical observations of \cite{Agarwal2019}. One way to do so is to start again from the derivation of the two-point two-time covariance in Sec.~\ref{sec:ADBMtwotimederiv} and to replace the telegraphic noise with delta-correlated noise. This is done in Sec.~III.B of \cite{ADBM2} (and the dynamical correlations are studied in Sec.~VI). Another approach would be to simply take the diffusive limit in the expressions derived for the active CM model, i.e., $v_0,\gamma\to+\infty$ with $T_{\rm eff}=\frac{v_0^2}{2\gamma}$ fixed (one can check that this indeed correctly recovers the results below). Here we instead choose to focus on the static correlations. At small temperature, the static covariance of the particle positions is given by
\be
\langle \delta x_i \delta x_j \rangle = \frac{T}{N} (H^{CM})^{-1}_{ij} = \frac{T}{\lambda N} (\mathcal{H}^{-2})_{ij}
\ee
Interestingly, this coincides exactly with the static covariance of the active DBM in the limit $\gamma\to 0^+$, see \eqref{covADBMga0_Hessian}, up to a mapping $\frac{v_0^2}{\lambda^2}\to\frac{T}{\lambda N}$ for the prefactor. Note the additional factor $1/N$ due to our choice of scaling. For the static correlations, we can thus directly reuse all the results obtained for the active DBM in the limit $\gamma\to 0^+$. In particular, forr the bulk particles this implies, using \eqref{covADBM_largeN},
\begin{equation}
\langle \delta x_i \delta x_j \rangle \simeq\frac{T}{\lambda N^2} \, \mathcal{C}_b^0\left( \frac{ \lambda^{1/4} x_{{\rm eq},i}}{2\sqrt{\tilde g}}, \frac{\lambda^{1/4} x_{{\rm eq},j}}{2\sqrt{\tilde g}} \right) \;,
\label{cov_largeN_CM_passive}
\end{equation}
with $\mathcal{C}_b^0(x,y)$ given in \eqref{newCb} and the $x_{{\rm eq},i}$ given in \eqref{eq_CM}. For the edge particles, one has from \eqref{covADBM_edge},
\be
\langle \delta x_i \delta x_j \rangle \simeq \frac{T}{\lambda N^{5/3}} \frac{1}{\Ai'(a_i) \Ai'(a_j)} \int_0^{+\infty} dx \ \frac{\Ai(a_i + x)\Ai(a_j + x)}{x^2} \;.
\label{covADBM_edge_passiveCM}
\ee
We thus recover the $N^{-2}$ and $N^{-5/3}$ scalings for the bulk and the edge respectively, which were predicted in \cite{Agarwal2019} from numerical computations (we note again that we have an additional factor $1/N$ compared to their choice of scaling). Our expressions also allow to reproduce some of there results more quantitatively, see the discussion in Sec.~II.B of \cite{ADBM2}. It is interesting to note that, while the active CM model does not exhibit an edge regime with a distinct scaling from the bulk, this is indeed the case for the passive version.

The results for the variance of the gaps in the active DBM in Sec.~\ref{sec:gapvar_ADBMharmonic} can also be directly transposed to the passive DBM. We thus obtained a linear dependence $\langle (\delta x_i - \delta x_{i+n})^2 \rangle\propto n$ at intermediate distances $1\ll n \ll N$ (see \eqref{gapvariance_largeN_ADBMharmonic_linear}), indicating Poissonian counting statistics on large scales (as for all short-range Riesz gases, see \eqref{gapscases_brownian}).

As for the active DBM and the active CM model, we can use these results to show the existence of different regimes in the equilibrium density for the passive CM model (as well as for the passive DBM below), the main difference being that in the strong noise regime the density is now Gaussian, as discussed in Sec.~\ref{sec:DBM_Wigner} for the DBM, see Fig.~2 in \cite{ADBM2}.

\subsection{Passive DBM in a harmonic trap}

For the passive DBM, corresponding to \eqref{def_ADBM_chap9} with $v_0=0$ and $T>0$ (i.e., the standard DBM discussed in Sec.~\ref{sec:DBM_brownian}), the results can be obtained as for the passive CM model either by redoing the computation of Sec.~\ref{sec:ADBMtwotimederiv} or by taking the diffusive limit of the active DBM. This simply amounts to replacing the eigenvalues $\lambda k^2\to \lambda k$ in the results for the passive CM model. The difference is that now the sum in \eqref{covADBM_largeN} does not converge, so that we need to keep the cutoff at $k_{max}=N$ for the bulk expression of the static covariance at large $N$,
\begin{equation}
\langle \delta x_i \delta x_j \rangle \simeq \frac{T}{\lambda N^2} \, \mathcal{\hat C}_{b,N}\left( \frac{x_{{\rm eq},i}}{2\sqrt{g/\lambda}}, \frac{x_{{\rm eq},j}}{2\sqrt{g/\lambda}} \right) \quad {\rm with} \quad \mathcal{\hat C}_{b,N}(x,y) = \sum_{k=1}^N \frac{1}{k} U_{k-1}(x) U_{k-1}(y) \;,
\label{DBMcov_largeN}
\end{equation}
with the $x_{{\rm eq},i}$ given in \eqref{eq_DBM_Hermite_zeros_chap9}. One can show that this still gives correct results, but the relative error is now $O(\log N)$. This allows to recover some known results from Sec.~\ref{sec:DBM_micro}, such as the expression \eqref{DBMvar_largeN_ORourke} for the one-particle variance $\langle \delta x_i^2\rangle$ in the bulk. For the edge particles, we exactly recover \eqref{AiryDBM} from \cite{GorinInfiniteBeta} for the two-point two-time covariance, as well as \eqref{var_edge_standardDBM} for the one-particle variance in the regime $1\ll i \ll N$. See Sec.~VII in \cite{ADBM2} for more details.

\section{Conclusion}

In this chapter, we have extended the method of the previous chapter, based on a linear approximation of the equations of motion in the limit of weak noise, to study the microscopic fluctuations in two special cases of the active Riesz gas in the presence of a harmonic confining potential, namely the active DBM $s=0$ and the active Caloger-Moser model $s=2$. For the active DBM, we showed that the two-particle covariance of the positions takes a different scaling form depending on whether the particles are located in the bulk or at the edges of the semi-circle, while the edge regime is absent for the active CM model. These results, in particular concerning the scaling of the fluctuations inside the bulk, provide us with an additional argument to support the observations of Chapter~\ref{chap:ADBM_Dean}, where we showed that the stationary density exhibits 3 different regimes in the active DBM. This also supports the idea that the density of the active CM exhibits a similar behavior at large $N$, as confirmed by numerical simulations in \cite{activeCM}. In addition, we also computed the variance of the interparticle distance in both models, showing the existence of giant number fluctuations in the active CM model, while in the active DBM this effect is compensated by the long-range interaction which reduces the fluctuations. Finally, this approach can also be applied to the passive versions of these two models, recovering some known results for the DBM, as well as providing an analytical confirmation of some numerical observations made recently for the overdamped Calogero-Moser model \cite{Agarwal2019}. 

The computations of this chapter where made possible thanks to the special structure of the Hessian matrix in the DBM and the CM model. An obvious question is whether one could find a way to extend this approach to other confined Riesz gases. In this case, the density at large $N$ near the ground state is not given by the Wigner semi-circle but it was determined in \cite{riesz3}. While the bulk properties can be deduced from the results of the previous chapter, such an extension would allow to study the edge regime in this more general case, and in particular to understand the criterion for its existence. Finally, as discussed at the end of the previous chapter for the active Riesz gas on the circle, it would also be interesting to see if this method can be extended further to compute higher order correlation functions, and to compare it with other approaches such as macroscopic fluctuation theory for the particle density.
\part{Siegmund duality for active particles}\label{part:siegmund}

\vspace*{\fill}

\begin{center}
{\bf Abstract}
\end{center}

This last part focuses on a different type of exact computations for models of active particles in one dimension (without interactions), namely the study of their first-passage properties. In Chapter~\ref{chap:Exitproba}, we compute explicitly the exit probability for a RTP subjected to an arbitrary external potential. By doing so, we find a surprising connection with the stationary distribution of positions of a RTP between hard walls, which is reminiscent of some results for the Brownian motion. In Chapter~\ref{chap:Siegmund}, we explain how this relation between absorbing boundary conditions and hard walls is connected to the concept of Siegmund duality, introduced in mathematics but relatively unknown in physics, which applies much beyond Brownian motion and RTPs. We then provide a new, explicit formulation of this duality for a large class of continuous stochastic processes, driven by time-correlated noise, which includes the most well-known models of active particles in 1D, as well as other stochastic processes which are relevant in physics (in particular diffusing diffusivity models and stochastic resetting). We also give a similar result in the case of random walks. We illustrate these results with numerical simulations and we discuss their relevance in the context of physics, both for analytical and numerical computations. 

Chapter~\ref{chap:Exitproba} is based on the reference \cite{SiegmundShort} while Chapter~\ref{chap:Siegmund} is based on \cite{SiegmundLong}. These works are the result of a collaboration with Mathis Gu\'eneau. The derivation of the results, the numerical simulations and the writing of the articles were all performed by both of us in equal proportions.

\vspace*{\fill}

\chapter{Exit probability of an RTP in an arbitrary potential} \label{chap:Exitproba}

\section{Context and aim of the chapter}

In these last two chapters, we leave aside the interactions and focus instead on another type of problem for a single active particle, namely the study of its first-passage properties. As we discussed in Sec.~\ref{sec:firstpassage}, the random search of a target by an active particle plays an important role in biology, e.g., for foraging or reproduction \cite{benichou1,benichou2,Targetsearch}. As such, the study of the first-passage properties of active particles has attracted considerable attention in recent years. In particular, in the case of a free RTP in one dimension, the survival probability and mean first-passage time (MFPT) have been extensively studied \cite{MalakarRTP,Masoliver92,Orsingher95,Targetsearch,Singh2020,Singh2022,MFPT1DABP}. The survival probability in the presence of a constant drift has also been computed \cite{SurvivalRTPDriftDeBruyne}. For more general external potentials however, exact results are more difficult to obtain and most studies have until now focused on the determination of the MFPT \cite{AngelaniMFPT,MathisMFPT,MathisMFPT2,Grange2025}. We refer again to Sec.~\ref{sec:firstpassage} for a more detailed review. 

\begin{figure}[t]
\centering
    \begin{minipage}[c]{1\linewidth}
        \centering
        \includegraphics[width=1.\linewidth]{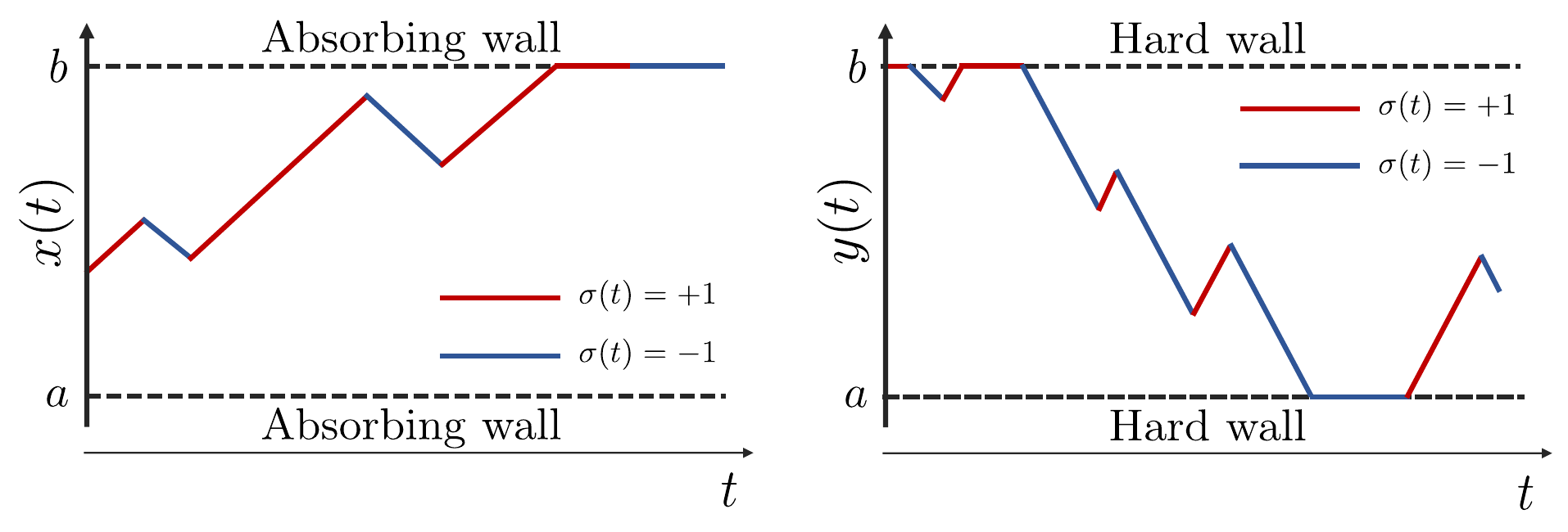}
    \end{minipage}
    \caption{We consider a RTP that evolves through the equation of motion $\dot{x}(t)=F(x) + v_0\, \sigma(t)$, where $\sigma(t)$ is a telegraphic noise which switches between the values $+1$ and $-1$ with rate $\gamma$. On the left panel we show a schematic trajectory $x(t)$ of a RTP with absorbing walls at $a$ and $b$. When the RTP reaches one of these walls, it stays there forever independently of its state $\sigma=\pm 1$. On the right panel, we show a trajectory of another RTP $y(t)$, with hard walls, i.e., if the particle reaches one of the walls and tries to cross it, it stays in the same place, but it moves away from the wall after the next tumbling event.}
\label{figureRTPwalls}
\end{figure}

Here, we focus instead on the {\it exit probability} (also called splitting or hitting probability) of a 1D RTP from an interval $[a,b]$, defined in \eqref{rel_exit_limit}. Consider a RTP evolving inside some interval $[a,b]$, with {\it absorbing walls} at $a$ and $b$, meaning that if the particle reaches one of these two points it will remain there indefinitely. The exit probability at $b$, which we denote $E_b(x)$, is the probability that, starting from some position $x\in[a,b]$ at $t=0$, the particle eventually gets absorbed at $b$ (and not at $a$). More precisely, for a RTP we need to distinguish two exit probabilities $E_b(x,+)$ and $E_b(x,-)$ depending on whether at $t=0$ it is in the state $\sigma=+1$ or $\sigma=-1$. If we assume that the particle starts with equal probability in one of these two states, we now have $E_b(x)=\frac{1}{2}(E_b(x,+)+E_b(x,-))$. For a free RTP, the exit probability is well-know \cite{MalakarRTP,Singh2020,Singh2022}. It it linear, as in the Brownian case, but with discontinuities at the boundaries such that $E_b(a^+,+)>0$ and $E_b(b^-,-)<1$,
\begin{equation} \label{exitprobafreeRTP}
    E_b(x,+) = \frac{1+\frac{\gamma}{v_0}(x-a)}{1+\frac{\gamma}{v_0}(b-a)} \quad , \quad  E_b(x,-) = \frac{x-a}{\frac{v_0}{\gamma}+(b-a)}\, .
\end{equation}
The first goal of this chapter is to extend this result to an arbitrary external force $F(x)=-V'(x)$, which we do in Sec.~\ref{sec:exitprobaRTP}. We recall that for a Brownian particle with diffusion coefficient $T$, the result is given by
\be \label{exit_brownian_potential}
E_b(x)=\frac{\int_a^x dz\, e^{\frac{V(z)}{T}}}{\int_a^b dz\, e^{\frac{V(z)}{T}}} \;.
\ee
As noted for instance in \cite{hittingProbaAnomalous}, this expression is exactly the same as the cumulative of the stationary distribution of positions of a Brownian particle with {\it hard walls} at $a$ and $b$, up to a change $V(x)\to -V(x)$. Interestingly, this surprising property also holds in the case of RTPs. We will show this in Sec.~\ref{sec:cumulativeRTPstationary} by computing explicitly the stationary distribution of an RTP between hard walls, with an arbitrary external force. This generalizes the results of \cite{AngelaniHardWalls} -- density of a RTP between hard walls but without external force -- and \cite{DKM19} -- stationary density of a RTP in an arbitrary external potential, but without walls -- see Sec.~\ref{sec:1particle_potential}. Here, the hard wall boundary condition can be understood as an infinite step of potential, meaning that a RTP which encounters the wall will remain there until its driving velocity changes sign. The two types of boundary conditions (absorbing and hard wall) are illustrated in Fig.~\ref{figureRTPwalls}. In Sec.~\ref{sec:RTPDUAL}, we summarize the relation between the two quantities and we argue, for now based on numerical simulations, that it can be generalized to the exit probability at finite time $E_b(x,t)$, defined in \eqref{defExit}. This relation is actually connected to a more general concept, known as {\it Siegmund duality}, which we will discuss in detail in the next chapter.

\section{Exit probability of a run-and-tumble particle}\label{sec:exitprobaRTP}

In this section we derive an expression for the exit probability $E_b(x)$ of a RTP under an arbitrary external force $F(x)$. Its position $x(t)$ follows the equation of motion
\begin{equation}
    \frac{dx}{dt}=F(x) + v_0 \, \sigma(t)\, ,
\label{langeRTPexit}
\end{equation}
where $\sigma(t)$ is a telegraphic noise with tumbling rate $\gamma$, as defined in Sec.~\ref{sec:RTPdef}. As explained above, the particle starts at some position $x\in [a,b]$, with absorbing walls at $a$ and $b$, and we want to determine the probability $E_b(x,\pm)$ that it gets absorbed at $b$ (after a time which is irrelevant), given its initial orientation $\sigma=\pm1$. We assume here that the external force satisfies $|F(x)|<v_0$, meaning that the whole interval $[a,b]$ is accessible to the particle. The results of this chapter, and in particular the connection with the stationary distribution between hard walls discussed below, can be extended to the case where this condition is not satisfied. We refer to Sec.~6 of \cite{SiegmundShort} for a discussion of this question.

As explained in Sec.~\ref{sec:firstpassage}, the exit probabilities $E_b(x,\pm)$ obey the stationary backward Fokker-Planck equation, i.e., \eqref{timeBFP} with the time derivatives set to zero,
\begin{eqnarray}
0 &=& \left[F(x)+v_0\right] \partial_x E_b(x,+) + \gamma\, E_b(x,-) -\gamma\,  E_b(x,+) \; , \label{RTP2gamma1}\\
0 &=& \left[F(x)-v_0\right] \partial_x E_b(x,-) + \gamma\, E_b(x,+) - \gamma\, E_b(x,-)  \label{RTP2gamma2}\; .
\end{eqnarray}
Solving these two coupled first order differential equations requires two boundary conditions. As we already explained in Sec.~\ref{sec:firstpassage}, a particle which starts near the wall $a$ with $\sigma=-1$ will always be absorbed at $a$, such that $E_b(a^+,-)=0$. If it starts in the state $\sigma=+1$ however, its velocity will drive it away from the wall and it may eventually be absorbed at $b$, such that $E_b(a^+,+)>0$ {\it a priori}. Similarly, one has $E_b(b^-,+)=1$ (but $E_b(b^-,-)>0$).

Let us now rewrite the equations (\ref{RTP2gamma1})-(\ref{RTP2gamma2}) in terms of $E_b(x)=\frac{1}{2}\, (E_b(x,+)+E_b(x,-))$ and $e_b(x)=\frac{1}{2}\, (E_b(x,+)-E_b(x,-))$, which leads to
\begin{eqnarray}
0 &=& F(x)\,  \partial_x E_b(x) + v_0\,  \partial_x e_b(x) \; ,\label{sumdifeq1} \\
0 &=& F(x)\,  \partial_x e_b(x) + v_0\,  \partial_x E_b(x)  - 2\gamma \, e_b(x) \; .\label{sumdifeq2}
\end{eqnarray}
We can then use equation (\ref{sumdifeq1}) to replace $E_b(x)$ in equation (\ref{sumdifeq2}), which gives us a simple first order differential equation for $e_b(x)$.
Solving this equation, we obtain
\be
e_b(x) = A \, \exp \left[ -2\gamma \int_{a}^x du\, \frac{F(u)}{v_0^2 - F(u)^2} \right] \;,
\ee
where $A$ is an integration constant. Deducing $E_b(x)$ from \eqref{sumdifeq1} and using the boundary conditions $E_b(a^+,-)=0$ and $E_b(b^-,+) = 1$ to fix the integration constants, we obtain
\begin{eqnarray} \label{exitprobaexplicitRTP} 
    \!\!\!\!\!\!\!\!\!\!\!\!\!\!\!\!\!\!\!  &&E_b(x) = \frac{1}{Z} \left(2\gamma v_0 \int_{a}^x dz\, \frac{\exp \left[ -2\gamma \int_{a}^z du\, \frac{F(u)}{v_0^2 - F(u)^2} \right]}{v_0^2-F(z)^2} + 1 \right)\, , \\
    \!\!\!\!\!\!\!\!\!\!\!\!\!\!\!\!\!\!\!  &&Z = 2\gamma v_0 \int_{a}^b dz\, \frac{\exp \left[ -2\gamma \int_{a}^z du\,  \frac{F(u)}{v_0^2 -F(u)^2} \right]}{v_0^2-F(z)^2} + \exp \left[ -2\gamma \int_{a}^b du\, \frac{F(u)}{v_0^2 - F(u)^2} \right] + 1\, ,
\end{eqnarray}
as well as, using that $E_b(x,\pm) = E_b(x) \pm e_b(x)$,
\begin{equation}
E_b(x,\pm) = \frac{1}{Z} \left( 2\gamma v_0 \int_{a}^x dz\, \frac{\exp \left[ -2\gamma \int_{a}^z du\, \frac{F(u)}{v_0^2 - F(u)^2} \right]}{v_0^2-F(z)^2} \pm \exp \left[ -2\gamma \int_{a}^x du \frac{F(u)}{v_0^2 - F(u)^2} \right] + 1 \right).
\label{exprEpm_final}
\end{equation}
These expressions are valid for any $x\in(a,b)$, for an arbitrary external force $F(x)$ (as long as $|F(x)|<v_0$ on the interval). One can check that they indeed satisfy the boundary conditions $E_b(a^+,-)=0$ and $E_b(b^-,+) = 1$, but that in general $E_b(a^+,+)>0$ and $E_b(b^-,-) < 1$. For $F(x)=0$, they recover the result for a free RTP given in \eqref{exitprobafreeRTP}, with a linear dependence in $x$. One can also check that in the diffusive limit $v_0,\gamma\to+\infty$ with $T_{\rm eff}=\frac{v_0^2}{2\gamma}$ fixed, the expression \eqref{exitprobaexplicitRTP} recovers the Brownian result \eqref{exit_brownian_potential} with $T\to T_{\rm eff}$ and $F(x)=-V'(x)$.

For some particular choices of the force $F(x)$, the integrals in \eqref{exitprobaexplicitRTP} and \eqref{exprEpm_final} can be computed explicitly. In the case of a constant drift $F(x)=\alpha$ with $|\alpha|<v_0$, we obtain
\begin{eqnarray}
&&E_b(x,\pm) = \frac{1}{Z} \left[ \left(1+\frac{v_0}{\alpha}\right) + \left(\pm 1 - \frac{v_0}{\alpha}\right) e^{-\frac{2\gamma \alpha}{v_0^2-\alpha^2}(x-a)} \right] \, ,\\
&&Z = 1 + \frac{v_0}{\alpha} + \left(1 - \frac{v_0}{\alpha}\right) e^{-\frac{2\gamma \alpha}{v_0^2-\alpha^2}(b-a)}\, .
\label{constantdrifteq}
\end{eqnarray}
Another particularly interesting case is that of a harmonic potential $V(x)= \mu\, x^2 / 2$, i.e., $F(x) = -\mu\, x$, with $a> -v_0/\mu$ and $b<v_0/\mu$ such that $|F(x)|<v_0$ for$x\in [a,b]$. In this case, the result \eqref{exprEpm_final} can be expressed in terms of the hypergeometric function $_2F_1(x)$,
\begin{equation}\label{explicithramonicEb}
\begin{split}
    \!\!\!\! E_b(x,\pm)=\frac{1}{Z}\left\{1 \pm \left(\frac{{v_0^2 - a^2 \mu^2}}{{v_0^2 - \mu^2 x^2}}\right)^{\frac{\gamma}{\mu}} + \frac{2\, \gamma}{v_0} \left(1 - \frac{a^2 \mu^2}{v_0^2}\right)^{\frac{\gamma}{\mu}} \left[x \, _2F_1\left(\frac{1}{2}, 1+\frac{\gamma}{\mu}, \frac{3}{2}, \frac{\mu^2 x^2}{v_0^2}\right) \right.\right.\\ 
    \left.\left.- a \, _2F_1\left(\frac{1}{2},1+ \frac{\gamma}{\mu}, \frac{3}{2}, \frac{a^2 \mu^2}{v_0^2}\right)\right]\right\}\, ,
\end{split}
\end{equation}  
with
\begin{equation}
\begin{split}
      \!\!\!\!\!\!\!\!\!\!\!\! Z = 1 + \left(\frac{{v_0^2 - a^2 \mu^2}}{{v_0^2 - b^2 \mu^2}}\right)^{\frac{\gamma}{\mu}} +
\frac{2\, \gamma}{v_0} \left(1 - \frac{a^2 \mu^2}{v_0^2}\right)^{\frac{\gamma}{\mu}}\left[b \, _2F_1\left(\frac{1}{2}, 1 + \frac{\gamma}{\mu}, \frac{3}{2}, \frac{b^2 \mu^2}{v_0^2}\right) \right. \\
\left.- a \, _2F_1\left(\frac{1}{2}, 1+ \frac{\gamma}{\mu}, \frac{3}{2}, \frac{a^2 \mu^2}{v_0^2}\right)\right]  \, .
\end{split}
\end{equation}
One can check from Eq.~(\ref{explicithramonicEb}) that $E_b(x,-)$ vanishes linearly as $x\to a^+$, and that $E_b(b^-,-)<1$ (and similarly $E_b(a^+,+)>0$, and $1-E_b(x,+)$ vanishes linearly as $x\to b^-$).

\begin{figure}
\centering
    \includegraphics[width=0.45\linewidth]{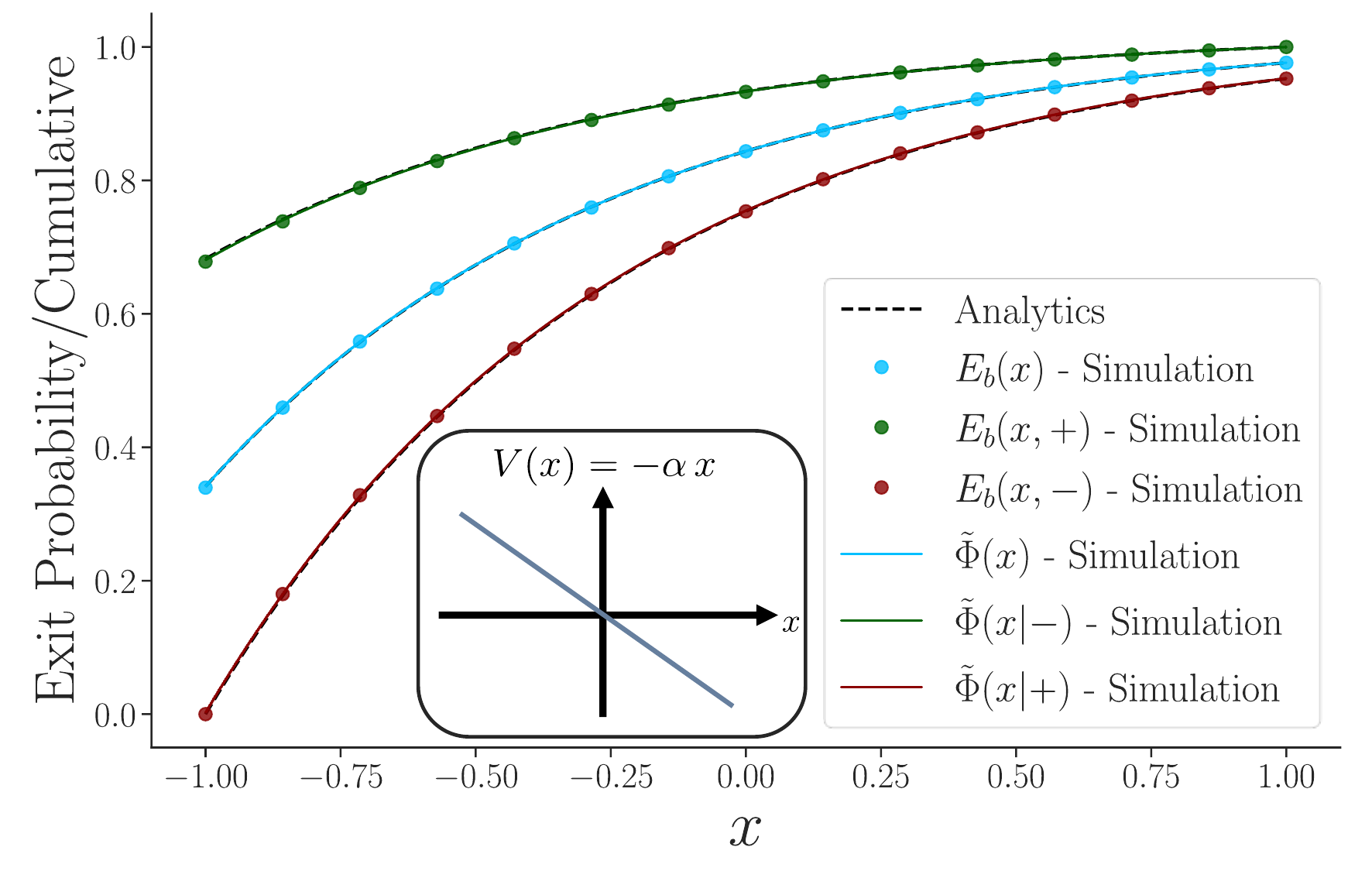}
    \hspace{0.2cm}
    \includegraphics[width=0.45\linewidth]{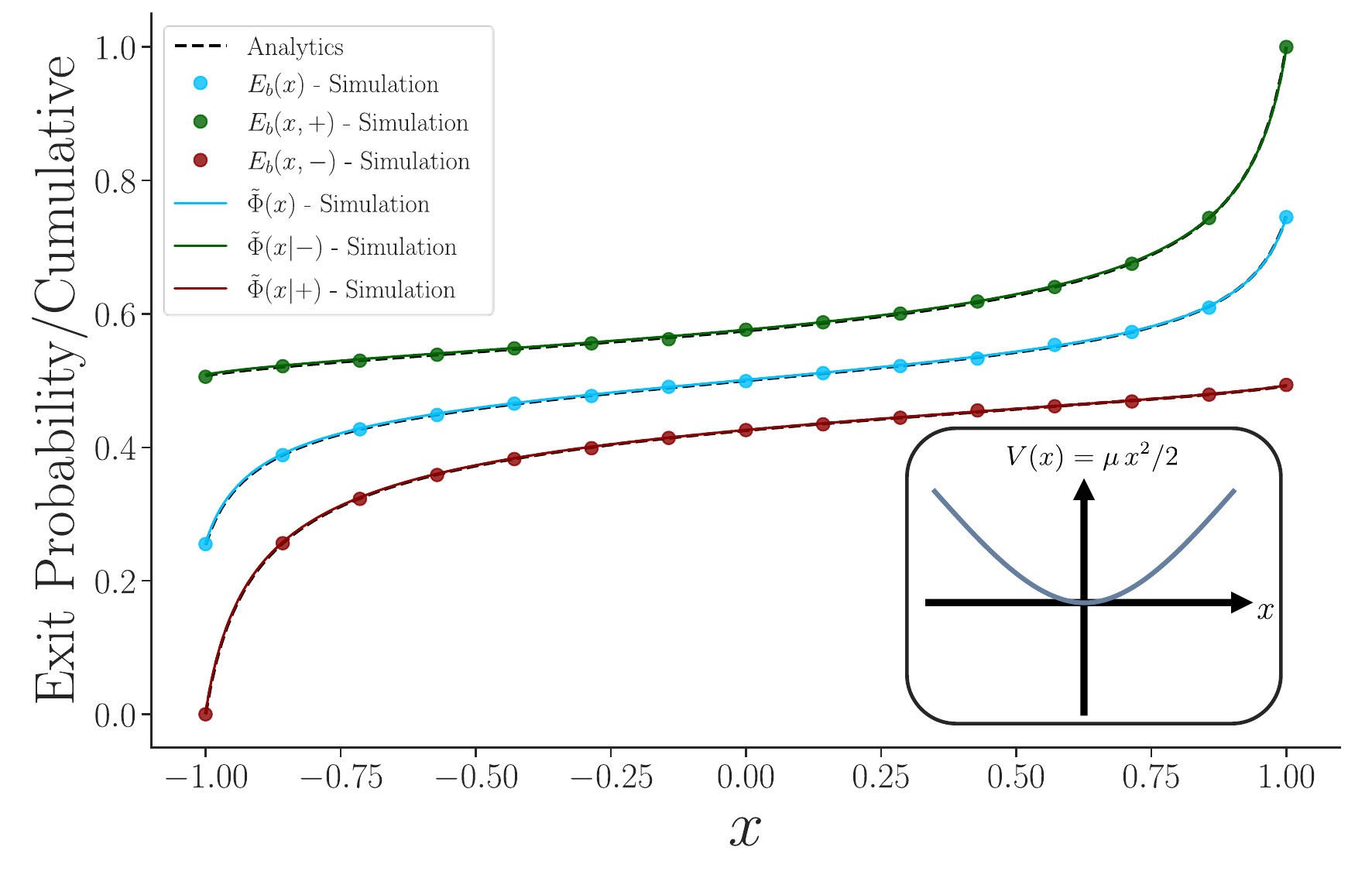}
    \caption{This figure illustrates the duality relations $E_b(x) = \tilde{\Phi}(x)$ (\ref{duality1RTPstatio}) and $E_b(x,\pm) = \tilde{\Phi}(x|\mp)$ (\ref{duality2RTPstatio}) for a run-and-tumble particle (RTP). We compute the exit probability $E_b$ (resp. the stationary cumulative distribution between hard walls $\tilde \Phi$) of a RTP for two different choices of the external force $F(x)$ (resp. $-F(x)$), with absorbing walls (resp. hard walls) located at positions $a=-1$ and $b=1$. The dots and the solid lines show the results obtained by numerical simulations of the Langevin dynamics \eqref{langeRTPexit} for the exit probability $E_b$ and for the stationary cumulative distribution $\tilde{\Phi}$ respectively. The data from  $E_b$ and $\tilde{\Phi}$ overlap exactly. We also compare with our analytical predictions from \eqref{exprEpm_final} and \eqref{cumuPMRTP}, in dashed black lines, which also overlap perfectly with the numerical data. In both cases, we made sure that the condition $|F(x)|<v_0$ is satisfied inside $[a,b]$. On the left, $F(x)=\alpha$ with are $\alpha = 0.5$, $v_0=1$ and $\gamma=1$. On the right, $F(x)=-\mu x$ with $\mu=1.9$, $v_0=2$ and $\gamma=1$.}
\label{figureRTPdualitystatio}
\end{figure}

In the Appendix of \cite{SiegmundShort}, we also provide expressions for the case of a linear potential $V(x)=\alpha |x|$ and of a double-well potential $V(x) = \frac{\mu}{2} \left(|x| - x_0\right)^2$. In Figure \ref{figureRTPdualitystatio}, we compare the expressions \eqref{constantdrifteq} and \eqref{explicithramonicEb} to the results of numerical simulations, showing a perfect agreement.

\section{Distribution of the position of a run-and-tumble particle with hard walls}\label{sec:cumulativeRTPstationary}

Let us now consider a different problem, which in appearance seems unrelated, namely the computation of the stationary density of a RTP subjected to an external force $F(x)$, and confined between two hard walls (i.e., infinite potential steps) at positions $a$ and $b$ (see right panel of Fig.~\ref{figureRTPwalls}). The solution without walls was derived in \cite{DKM19} for an arbitrary $F(x)$, and is presented in Sec.~\ref{sec:confiningPot} (see \eqref{eqRTPpotential}), while the solution in the presence of walls but without the external force was obtained in \cite{AngelaniHardWalls} and is discussed in Sec.~\ref{sec:HardWalls} (see \eqref{HardwallsNoforce}). Here we take inspiration from these two computations. In the following we again assume that $|F(x)|<v_0$ for any $x$ in the interval $[a,b]$ (see Sec.~6 of \cite{SiegmundShort} for the more general case).

Let us denote $P(x,+)$ and $P(x,-)$ the stationary densities of a RTP respectively for $\sigma=+1$ and $\sigma=-1$, which are normalized such that $\int_a^b dx \left[P(x,+) + P(x,-)\right] = 1$.
The steady-state Fokker-Planck equations for these densities are given by (see \eqref{FP_RTP}),
\begin{eqnarray}
0&=&\partial_x \left[\left(F(x)+v_0\right)P(x,+)\right] + \gamma\,  P(x,+) - \gamma\,  P(x,-) \; , \\
0&=&\partial_x \left[\left(F(x)-v_0\right)P(x,-)\right] - \gamma\,  P(x,+) + \gamma\,  P(x,-) \; .
\end{eqnarray}
Introducing $P(x)=P(x,+)+P(x,-)$ and $Q(x)=P(x,+)-P(x,-)$, we obtain
\begin{eqnarray}
0&=&\partial_x \left[F(x)\, P(x) + v_0\, Q(x)\right]  \label{eq1Density}\, ,\\
0&=&\partial_x \left[F(x)\, Q(x) + v_0\, P(x)\right] +2\gamma\,  Q(x) \, , \label{eq2Density}
\end{eqnarray}
as in \cite{DKM19}. The difference arises when considering the boundary conditions. Due to its persistent motion, the RTP may remain stuck at either wall for a finite time. This means that the density $P(x,-)$ has a finite mass $\kappa_a$ at $x=a$, while $P(x,+)$ has a finite mass $\kappa_b$ at $x=b$. Since $\kappa_a$ and $\kappa_b$ are stationary, the total current $J(x)=J(x,+)+J(x,-)$, where $J(x,\pm) = (F(x) \pm v_0)P(x,\pm)$, vanishes at the boundaries $a$ and $b$. In addition, the probability current of a $+$ (resp. $-$) particle at $x=a$ (resp. $x=b$) arises entirely from a $-$ (resp. $+$) particle stuck at the wall which tumbles. Therefore, we can write $J(a,+) = - J(a,-) = \gamma\,  \kappa_a$ and $J(b,+) = - J(b,-) = \gamma\,  \kappa_b $, which translates to 
\begin{eqnarray}
&&\left[v_0+F(a)\right]P(a,+) = \left[v_0-F(a)\right]P(a,-) = \gamma \, \kappa_a \, ,\label{leftedge}\\
&&\left[v_0+F(b)\right]P(b,+) = \left[v_0-F(b)\right]P(b,-) = \gamma \, \kappa_b \, ,\label{rightedge}
\end{eqnarray}
and implies
\begin{equation}
F(a)\, P(a)+v_0\,  Q(a) = 0 \quad , \quad 
F(b)\, P(b)+v_0\,  Q(b) = 0\, .
\end{equation}
Integrating \eqref{eq1Density} and taking these boundary conditions into account yileds, for all $x\in(a,b)$,
\begin{equation}
F(x)\, P(x) + v_0\,  Q(x) = 0\, ,
\end{equation}
which we can use to replace $Q(x)$ in \eqref{eq2Density}. We thus obtain the equation
\begin{equation}
\partial_x [(v_0^2-F(x)^2)\, P(x)] - 2\gamma\, F(x) P(x) = 0\, ,
\end{equation}
which is solved by
\begin{equation}
P(x) = \frac{1}{Z}\, \frac{2\gamma\, v_0}{v_0^2-F(x)^2} \exp \left[ 2\gamma \int_a^x dz\, \frac{F(z)}{v_0^2-F(z)^2} \right]\, ,
\label{exprPbulk}
\end{equation}
for any $x\in(a,b)$, with $Z$ a normalization constant. This is the same expression as the one obtained in the absence of walls in \cite{DKM19}. The difference comes from the normalization and from the presence of delta functions at the walls. The expressions for $P(x,\pm)$ are then simply obtained from
\begin{equation}
P(x, \pm) = \frac{1}{2}(P(x) \pm Q(x)) = \frac{1}{2} \Big(1 \mp \frac{F(x)}{v_0} \Big) P(x) 
\;.
\label{exprPpmbulk}
\end{equation}
Using the conditions (\ref{leftedge})-(\ref{rightedge}), we deduce the weights of the delta peaks at $a$ and $b$,
\begin{equation}
\kappa_a = \frac{1}{Z} \quad , \quad \kappa_b = \frac{1}{Z} \exp \left[ 2\gamma \int_a^b dz\, \frac{F(z)}{v_0^2-F(z)^2} \right] \, .\label{leftmass}
\end{equation}
Finally, the constant $Z$ is fixed by the normalization condition 
$\int_{a^+}^{b^-} dx\, P(x) + \kappa_a + \kappa_b = 1$. We can thus write the full expression for the total density $P(x)$ for any $x\in[a,b]$ as
\begin{eqnarray} 
P(x) &=& \frac{1}{Z} \left( 2\gamma v_0 \frac{\exp \left[ 2\gamma \int_{a}^x du\, \frac{F(u)}{v_0^2 - F(u)^2} \right]}{v_0^2-F(x)^2} + \delta(x-a) + \exp \left[ 2\gamma \int_a^b du\, \frac{F(u)}{v_0^2-F(u)^2} \right] \delta(x-b) \right) \; , \nonumber \\
Z &=& 2\gamma v_0 \int_{a}^b dz\, \frac{\exp \left[ 2\gamma \int_{a}^z du\, \frac{F(u)}{v_0^2 - F(u)^2} \right]}{v_0^2-F(z)^2} + \exp \left[ 2\gamma \int_{a}^b du\, \frac{F(u)}{v_0^2 - F(u)^2} \right] + 1\; .
\end{eqnarray}
The expressions for $P(x,\sigma)$ can be deduced from Eq.~(\ref{exprPpmbulk}), adding Kronecker symbols $\delta_{\sigma, \pm}$ to account for the fact that only $-$ particles accumulate at the left wall $a$, and only $+$ particles at the right wall $b$,
\begin{equation}
P(x,\sigma) = \frac{1}{Z} \left( \gamma \frac{\exp \left[ 2\gamma \int_{a}^x dy \frac{F(y)}{v_0^2 - F(y)^2} \right]}{v_0 + \sigma F(x)} + \delta_{\sigma,-}\delta(x-a) + \delta_{\sigma,+}\exp \left[ 2\gamma \int_a^b dy \frac{F(y)}{v_0^2-F(y)^2} \right] \delta(x-b) \right)\, .
\label{exprPpm_final}
\end{equation}
Let us now write the associated cumulative distributions. For the total density we obtain, for any $x\in[a,b]$,
\begin{equation}
\Phi(x) = \int_{a^-}^x dz\, P(z) = \frac{1}{Z} \left(2\gamma v_0 \int_{a}^x dz\, \frac{\exp \left[ 2\gamma \int_{a}^z du\, \frac{F(u)}{v_0^2 - F(u)^2} \right]}{v_0^2-F(z)^2} + 1 \right)\; .
\label{cumulativeRTP}
\end{equation}
As announced at the beginning of this chapter, this is exactly the same expression as \eqref{exitprobaexplicitRTP} for the exit probability $E_b(x)$, but with an opposite force $-F(x)$. This connection still holds when we condition on the state of the particle. Indeed, integrating \eqref{exprPpm_final} over $x$ and dividing by $P(\sigma)=\frac{1}{2}$ (the probability for the particle to be in the state $\sigma$ in the stationary state), we obtain the cumulative distribution of the positions conditioned on the internal state $\sigma$, for $x\in [a,b]$,
\begin{equation}
\Phi(x|\sigma) = \int_{a^-}^x dz\, \, \frac{P(z, \sigma)}{P(\sigma)} = \frac{2}{Z} \left(\gamma \int_{a}^x dz\, \frac{\exp \left[ 2\gamma \int_{a}^z du\, \frac{F(u)}{v_0^2 - F(u)^2} \right]}{v_0 + \sigma F(z)} + \delta_{\sigma,-} \right)\label{cumulative_step}
\end{equation}
This can be rewritten by writing $\frac{1}{v_0 + \sigma F(z)}=\frac{1}{v_0^2 - F(z)^2}-\frac{\sigma F(z)}{v_0^2 - F(z)^2}$, which leads to
\begin{equation}
\Phi(x|\pm) = \frac{1}{Z} \left(2\gamma v_0 \int_{a}^x dz\, \frac{\exp \left[ 2\gamma \int_{a}^z du\, \frac{F(u)}{v_0^2 - F(u)^2} \right]}{v_0^2 - F(z)^2} \mp \exp \left[ 2\gamma \int_{a}^x dz\, \frac{F(z)}{v_0^2 - F(z)^2} \right] + 1 \right)\, . 
\label{cumuPMRTP}
\end{equation}
Once again, this coincide exactly with the expression \eqref{exprEpm_final} for $E_b(x,\pm)$ with $F(x)\to -F(x)$, and with an additional exchange of the sign of the particles $\sigma \to -\sigma$. Note in particular that the weight of the delta peak at $x=a$ in the distribution of $-$ particles, given by $\Phi(a^+|-)=2\kappa_a = 2/Z$, coincides with the probability $E_b(a^+,+)$ that a particle starting at position $a^+$ with $\sigma=+1$ eventually exits at $b$, and similarly at the right wall $\Phi(b^-|+)=2\kappa_b=E_b(b^-,-)$. These two effects, namely the accumulation of particles at a hard wall and the ability to escape an absorbing boundary given the right initial orientation, are consequences of the persistent motion of the RTP which are completely inexistent for Brownian particles. It is thus interesting to see that the connection between the two quantities $E_b(x)$ and $\Phi(x)$ even extends to these peculiar behaviors at the boundaries. The agreement of the analytical expressions \eqref{cumulativeRTP} and \eqref{cumuPMRTP} with numerical simulations is tested in Fig.~\ref{figureRTPdualitystatio}, which also illustrates the relation with the exit probability.

\section{A duality relation between the exit probability and the distribution of positions with hard walls} \label{sec:RTPDUAL}

Let us formulate more precisely the connection between the two results \eqref{exprEpm_final} and \eqref{cumuPMRTP}. Consider the process $x(t)$ defined in \eqref{langeRTPexit}, i.e., a RTP on the interval $[a,b]$, subjected to an external force $F(x)$ and with {\it absorbing walls} at $a$ and $b$. We can define a ``dual'' process $y(t)$ as
\begin{equation} \label{defRTP_dual}
    \dot{y}(t)=-F(y) + v_0\, \tilde{\sigma}(t)\, ,
\end{equation}
where $\tilde{\sigma}(t)$ is a different realization of the same telegraphic noise with tumbling rate $\gamma$ as in \eqref{cumuPMRTP}. The process $y(t)$ describes the motion of a RTP subjected to the reversed force $-F(y)$. In addition, the process $y(t)$ has {\it hard walls} at $a$ and $b$.

We denote by $\tilde{\Phi}(y)$ the cumulative distribution of the dual process $y(t)$ in the stationary state. The equation \eqref{cumulativeRTP}), shows that the exit probability of the process $x(t)$ identifies with the stationary cumulative distribution of its dual, i.e., for all $x\in[a,b]$,
\begin{equation}
    E_b(x)=\tilde{\Phi}(x)\, .
\label{duality1RTPstatio}
\end{equation}
One can also write a more general relation by conditioning on the state of the RTP, as can be seen from \eqref{exprEpm_final} and \eqref{cumuPMRTP}). This relation reads, again for any $x\in[a,b]$,
\begin{equation}
E_b(x,\pm) = \tilde \Phi(x|\mp)\, .
\label{duality2RTPstatio}
\end{equation}
Here we have shown this identity in the case where $|F(x)|<v_0$ for all $x\in[a,b]$. In \cite{SiegmundShort} we showed that it also holds when this condition is not satisfied. See again Fig.~\ref{figureRTPdualitystatio} for an illustration of this duality using numerical simulations. 
\\

\begin{figure}[t]
\centering
    \includegraphics[width=.45\linewidth]{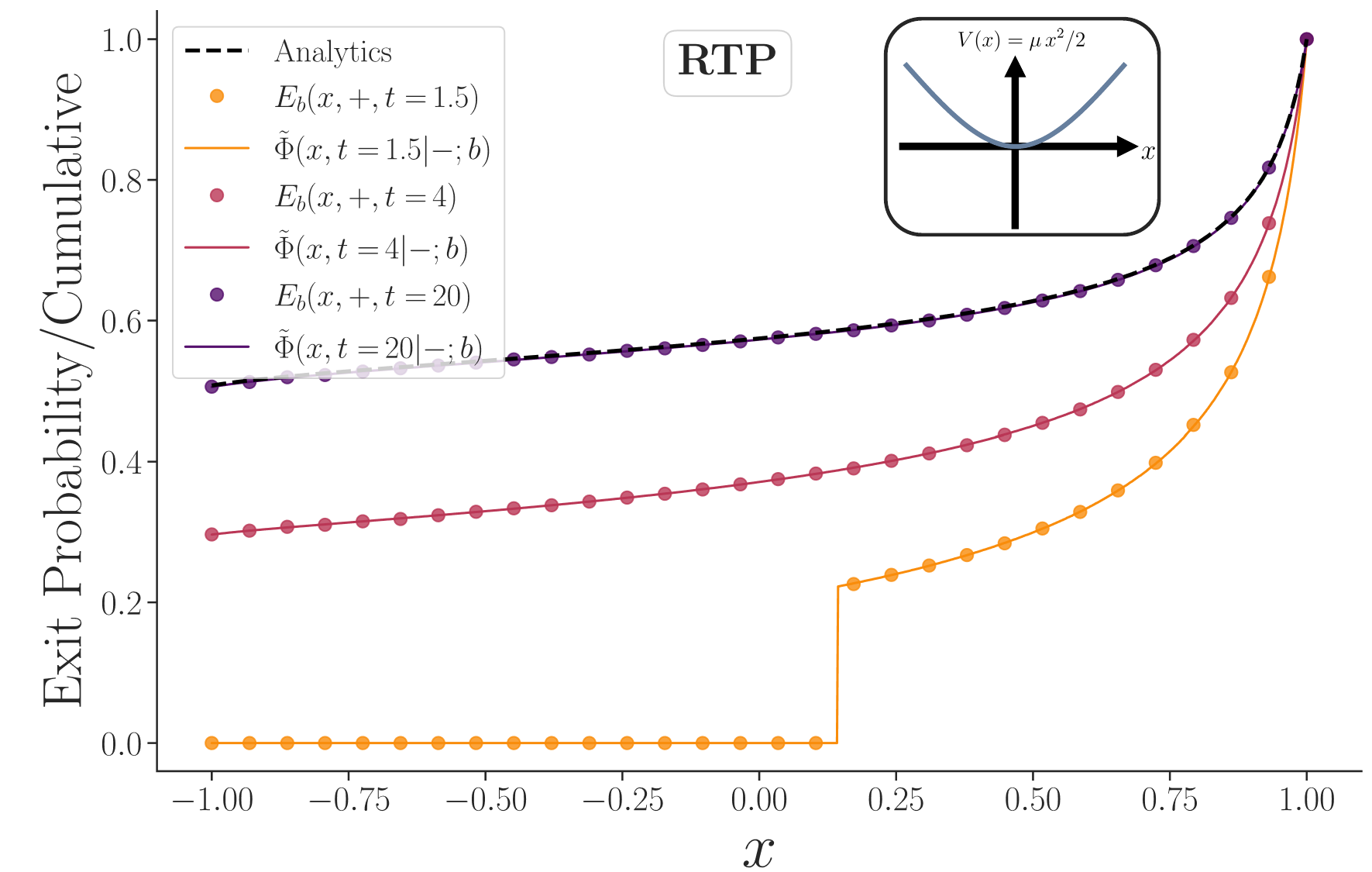}
    \hspace{0.2cm}
    \includegraphics[width=.45\linewidth]{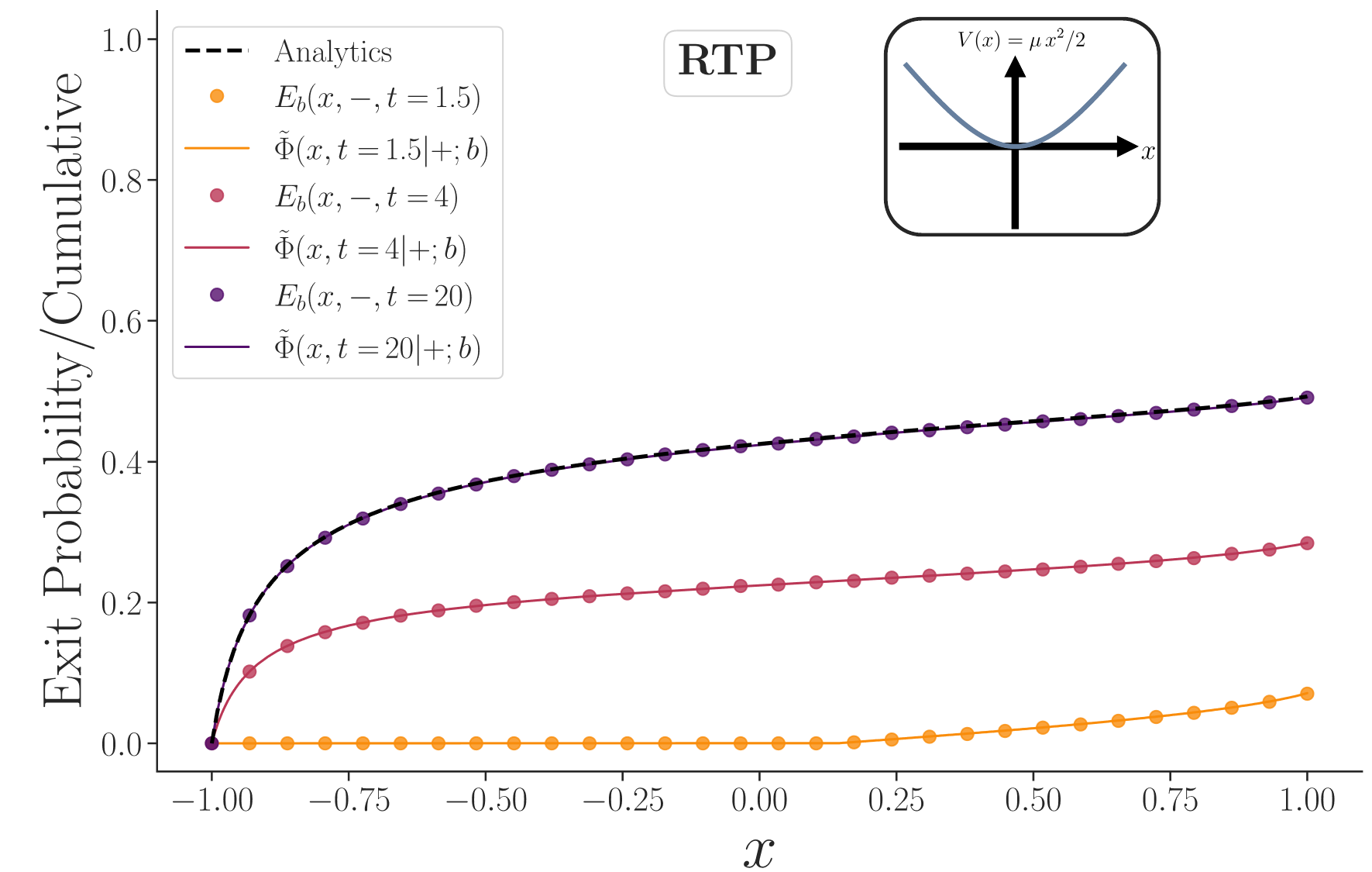}
    \caption{Illustration of the duality relation \eqref{duality_finite_time_pm} at finite time $t$ for a RTP in a harmonic potential. {\bf Right:} The dots represent the exit probability $E_b(x,+,t)$ in the presence of a potential $V(x)=\mu\,  x^2/2$, with absorbing walls at $a=-1$ and $b=1$, while the lines show the cumulative distribution of the dual $\tilde \Phi(x,t|-;b)$, with a potential $-V(x)$ and hard walls at $a$ and $b$. {\bf Left:} Same plot with the $+$ and $-$ particles exchanged. All the results were obtained by averaging over $10^6$ simulated trajectories, with parameters $\mu=1.9$, $v_0=2$ and $\gamma=1$. The dashed black lines show the analytical predictions for the stationary state \eqref{exprEpm_final}-\eqref{cumuPMRTP}. The discontinuities observed in the yellow curve on the left can be attributed to the fact that particles starting on the left side of these discontinuities do not have enough time to exit the interval, even if they remain in the positive state throughout the entire simulation.}
\label{figureRTPtime}
\end{figure}

One may wonder about the generality of this duality relation. First of all, here we have related the probability that the particle is absorbed at $b$ after an arbitrary long time, $E_b(x)$, to the stationary cumulative distribution of the dual $\tilde \Phi(x)$. One may ask if it is possible to relate in the same way the probability that the particle is absorbed at $b$ {\it after a finite time} $t$, $E_b(x,t)$, as defined in \eqref{defExit}, with the cumulative distribution of the dual at finite time $\tilde \Phi(x,t)$, for some choice of initial condition. As discussed in sec.~\ref{sec:firstpassage}, the quantity $E_b(x,t)$ is particularly interesting since it contains both the information on the exit probability $E_b(x)$ and on the survival probability (see \eqref{rel_survival_exit}). One can show that this more general identity indeed holds if we initialize the dual process $y(t)$ at the position $y(0)=b$, and with $\sigma(0)=\pm1$ with equal probability, i.e., one has at any time $t$ and for any $x\in [a,b]$,
\begin{equation} \label{duality_finite_time}
E_{b}(x,t)= \tilde \Phi(x,t|y(0)=b) \;.
\end{equation}
If we want to generalize in the same way the relation \eqref{duality2RTPstatio} for the exit probability conditioned on the initial state of the particle, the conditioning for the dual is on the final value $\tilde \sigma(t)$, while the initial condition remains the same, i.e., $y(0)=b$ and $\sigma(0)=\pm1$,
\begin{equation} \label{duality_finite_time_pm}
E_{b}(x,\pm,t)= \tilde \Phi(x,t|\tilde \sigma(t) = \pm1;y(0)=b) 
\end{equation}
(here and in the next chapter we use a semicolon to separate the conditioning on the events at time $t$ and at time $0$). Computing these two quantities explicitly at finite time is much more difficult than for the finite time quantities, especially in the presence of an external force. However, we have confirmed through numerical simulations that the relations \eqref{duality_finite_time} and \eqref{duality_finite_time_pm} indeed hold at any time $t$ for a wide variety of external forces $F(x)$. An example is shown in Fig.~\ref{figureRTPtime} for a harmonic external potential. In Sec.~5 of \cite{SiegmundShort}, we proved this relation by showing that the two quantities on the right and on the left of \eqref{duality_finite_time_pm} obey the same partial differential equation, with the same boundary and initial conditions. In the next chapter we will provide a derivation in a more general setting.

Beyond this extension to finite time, the fact that the relation \eqref{duality1RTPstatio} holds both for Brownian particles (see \eqref{exit_brownian_potential}) and for RTPs suggests that it could apply to a wider class of stochastic processes. Indeed, in the next chapter we will see that this identity is related to a more general concept, known in the mathematics literature as Siegmund duality, and that it is known to hold for a large family of stochastic processes, including Brownian motion and random walks with i.i.d steps. We will then extend it to a large class of stochastic processes, driven by time-correlated noise (e.g., $\sigma(t)$ for RTPs), which includes other models of active particles, but also other types of stochastic processes which are particularly relevant in physics, such as diffusing diffusivity models. We will see that this duality can be intuitively understood as a form of time-reversal symmetry and we will discuss its potential applications.

\chapter{Siegmund duality for stochastic processes driven by stationary noise} \label{chap:Siegmund}

\section{Siegmund duality: history and known results}

\subsection{General definition}

In the previous chapter we have shed light on a surprising connection between the exit probability and the distribution of positions with hard walls for a run-and-tumble particle. To better understand this result, we need to replace it in a more general context. Although it is not widely known in physics, and in particular in the field of active matter, the relation between absorbing and hard wall (or reflective) boundary conditions has been studied for a long time in mathematics. It was first investigated by L\'evy in the case of Brownian motion \cite{Levy}, and by Lindley for discrete random walks \cite{Lindley}. Later on this relation was put into a more general framework by Siegmund, who gave his name to what is now called {\it Siegmund duality}. Generally speaking, two processes $x(t)$ and $y(t)$ (where the time $t$ is either discrete or continuous), such that $x(0)=x$ and $y(0)=y$, are said to be Siegmund duals if, at any time $t$~\footnote{Throughout this chapter, we use the notation $\mathbb{P}(x(t)\in...)$ to refer to the probability mass function that $x(t)$ belongs to a given set, while the notation $P$ or $p$ refers to a probability density.},
\be\label{siegmundintro}
\mathbb{P}(x(t) \geq y | x(0)=x) = \mathbb{\tilde P}(y(t) \leq x | y(0)=y) \;.
\ee 
For the sake of clarity we will denote with a tilde all quantities associated to the dual throughout this chapter. Siegmund showed the existence of a Siegmund dual for any one-dimensional stochastically monotone Markov process (meaning that $\mathbb{P}(x(t) \geq y | x(0)=x)$ is a non-decreasing function of $x$) \cite{Siegmund}. Extensions to more general Markov processes have later been considered \cite{SiegmundDualityClifford, KolokoltsovDuality, partiallyOrdered, DualityZhao}. In the simple cases where $x(t)$ is a Brownian motion or a random walk with i.i.d steps, with absorbing boundary conditions, then the dual process $y(t)$ has the exact same dynamics as $x(t)$, but with hard walls. For more general processes however, it is not always clear how to explicitly construct the dual. 

Throughout this chapter, $x(t)$ denotes a stochastic process on an interval $[a,b]$ (potentially with $a\to-\infty$) with absorbing walls at $a$ and $b$, while $y(t)$ denotes its dual, with hard walls at $a$ and $b$. We recall our definitions of these two types of processes, which are illustrated in Fig.~\ref{figureRTPwalls} for the RTP. By {\it absorbing wall} we mean that if the particle reaches either $a$ or $b$, it ``sticks'' to the wall and remains there forever (i.e., it does not disappear, contrary to what is sometimes considered in the physics literature \cite{redner, Metzler_book, Bray2013}). On the other hand, a {\it hard wall} can be seen as an infinite step of potential. In the case of Brownian motion, this is equivalent to a reflecting boundary condition, where the direction of motion of the particle is reversed when it touches the wall. However, for processes driven by time-correlated noise it can lead to the particle spending a finite time at the wall, until its velocity changes sign.

In this setting, the relation between the exit probability and the cumulative distribution with hard walls introduced in the previous chapter can be seen as a special case of the Siegmund duality relation \eqref{siegmundintro}. Indeed, let us recall our definition of the exit probability at finite time, i.e., the probability that the process $x(t)$ is absorbed at $b$ before or at time $t$, starting from $x(0)=x$,
\be  \label{defEb_general}
E_b(x,t) = \mathbb{P}(x(t) = b | x(0)=x) \;.
\ee
On the other hand, we consider the cumulative distribution of the dual process $y(t)$ at time $t$, initialized at $y(0)=b$,
\be \label{defPhi_general}
\tilde \Phi(x,t|b)=\mathbb{\tilde P}(y(t) \leq x | y(0)=b) \;.
\ee
Then, specializing \eqref{siegmundintro} to $y=b$, we recover the relation between these two quantities introduced in the previous chapter for the RTP \eqref{duality_finite_time_pm}, i.e., at any time $t$ and for any $x\in[a,b]$,
\be\label{siegmundExitIntro}
E_b(x,t) = \tilde \Phi(x,t|b) 
\ee 
(since, due to the absorbing wall, $x(t)\geq b$ is simply equivalent to $x(t)=b$). As a side comment, note that both $E_b(x,t)$ and $\tilde \Phi(x,t|b)$ are always increasing functions of $x$ and $t$ for the type of processes that we consider here.

\subsection{Examples in the infinite time limit}

In the limit $t\to+\infty$, the identity \eqref{siegmundExitIntro} relates the exit probability $E_b(x)=E_b(x,t\to+\infty)$ to the stationary cumulative distribution with hard walls $\tilde \Phi(x|b) = \tilde \Phi(x,t\to+\infty|b)$, as we discussed in detail for the RTP in the previous chapter. We also mentioned how this duality relation at infinite time applies for a Brownian particle, where the exit probability is given by \eqref{exit_brownian_potential}. In both cases, we saw that the sign of the external force $F(x)=-V'(x)$ is reversed in the dual $y(t)$ compared to the initial process $x(t)$. This change in the sign of the external force between the initial process and its dual is a result which holds more generally. Another interesting class of processes for which we can verify explicitly that \eqref{siegmundExitIntro} holds in the infinite time limit (here without an external force) is L\'evy flights. In the continuum limit, if $x(t)$ is a L\'evy flight whose increments obey a L\'evy stable symmetric law of index $0<\mu\leq 2$, the probability that it exits the interval $[a,b]$ at $b$ after an infinite time is given by \cite{WidomExitprobaLevyflight, Levy3, hittingProbaAnomalous}
\begin{equation} \label{exitprobalevy}
    E_b(x,{\normalcolor t\to +\infty}) = \frac{\Gamma(2\phi)}{\Gamma(\phi)^2} (b-a)^{1-2\phi} \int_a^x du\, \left[(u-a)(b-u)\right]^{\phi - 1}\, ,
\end{equation}
where $\phi = \mu/2$. In this case, the dual $y(t)$ is also a Lévy flight with the same increments. Indeed, the stationary distribution of a L\'evy flight between two hard walls, $\tilde P_{st}(x)=\partial_x \tilde \Phi(x,t\to+\infty|b)$, was obtained more recently via a completely independent computation in \cite{DenisovLevy}, and reads
\begin{equation}
    \tilde P_{st}(x) =\frac{\Gamma(2\phi)}{\Gamma(\phi)^2} (b-a)^{1-2\phi} \left[(x-a)(b-x)\right]^{\phi - 1} = \frac{dE_b(x,{\normalcolor t\to \infty})}{dx} \, .
\end{equation}
We can clearly see that $\tilde P_{st}(x)$ is simply the derivative of $E_b(x)$, in agreement with the duality relation \eqref{siegmundExitIntro} at infinite time\footnote{Our derivation for discrete time random walks discussed in Sec.~\ref{sec:discreteSiegmund}, as well as the original results by Lindley, apply in particular to Lévy flights in discrete time. Taking the continuous time limit would require a rigorous analysis on its own, but it is reasonable to assume that the duality still holds in that case.}. The possibility to derive one quantity directly from the other is a strong motivation to try to better understand the duality relation \eqref{siegmundExitIntro} and its range of application.

\subsection{An example at finite time: the Brownian case} \label{sec:SiegmundFinitetime}

Concerning the relation \eqref{siegmundExitIntro} at finite time, an example is given in Appendix~\ref{survivalBM} for a Brownian particle without external force and with only a single wall (i.e., for $a\to-\infty$), where we again compute explicitly the two sides of the identity and verify that they indeed coincide. More generally, let us consider a Brownian particle subjected to an external force $\tilde F(x)$, and with a diffusion coefficient $T(x)$ which may be space-dependent. In this case, the duality relation at finite time \eqref{siegmundExitIntro} can be derived as follows. We start from the Fokker-Planck equation for the particle density $\tilde P(x,t)$ (with the It\=o convention), and we assume the presence of hard walls at $a$ and $b$,
\begin{equation}
\partial_t \tilde P = - \partial_x [\tilde F(x) \tilde P] + \partial^2_{xx}[T(x)\tilde P] \;.
\end{equation}
Introducing the cumulative distribution $\tilde \Phi(x,t|b)=\int_a^x dy \tilde P(y,t|b)$ (where the conditioning on $b$ denotes the initial condition, as in\eqref{defPhi_general}) and replacing $\tilde P(x,t|b)=\partial_x \tilde \Phi(x,t|b)$, we obtain
\begin{equation}
\partial_x \{-\partial_t \tilde \Phi - \tilde F(x)\partial_x \tilde \Phi + \partial_x [T(x) \partial_x \tilde \Phi] \} = 0 \;.
\label{cumudiffut}
\end{equation}
The zero-flux boundary condition at $x=a$ reads (and similarly at $x=b$)
\begin{equation}
0 = \tilde F(a)\tilde P(a,t|b) -\partial_{x}[T(a)\tilde P(a,t|b)] = \tilde F(a)\partial_x \tilde \Phi(a,t|b) -\partial_x [T(x) \partial_x\tilde \Phi(a,t|b)] \;,
\end{equation}
and in addition $\partial_t \tilde \Phi(a,t|b)=\partial_t \tilde \Phi(b,t|b)=0$ (since $\tilde \Phi(a,t|b)=0$ and $\tilde \Phi(b,t|b)=1$ at any time $t$). Thus we can integrate Eq. \eqref{cumudiffut} to obtain
\begin{equation}
\partial_t \tilde \Phi = [- \tilde F(x)+\partial_x T(x)]\partial_x \tilde \Phi + T(x)\, \partial^2_{xx}\tilde \Phi \;.
\end{equation}
We may now notice that this coincides exactly with the backward-Fokker Planck equation satisfied by $E_b(x,t)$, recalled in \eqref{BFP}, but with an external force $F(x)=- \tilde F(x)+\partial_x T(x)$. In addition, $\tilde \Phi(x,t|b)$ and $E_b(x,t)$ satisfy the same boundary conditions $E_b(a,t)=\tilde \Phi(a,t|b)=0$ and $E_b(b,t)=\tilde \Phi(b,t|b)=1$, as well as the same initial condition $E_b(x,0)=\tilde \Phi(x,0|b)=\mathbbm{1}_{x\geq b}$. Since this is enough to specify these functions completely, this proves the identity \eqref{siegmundExitIntro}, where now $x(t)$ is a Brownian motion with external force $F(x)$ and diffusion coefficient $T(x)$ (and absorbing walls at $a$ and $b$), while its dual $y(t)$ is a Brownian motion with the same diffusion coefficient but with an external force $\tilde F(x)=- F(x)+\partial_x T(x)$ (and with hard walls at $a$ and $b$). This derivation can actually be extended to show the full Siegmund duality relation \eqref{siegmundintro} (here we simply focused on the case $y=b$ to simplify the notations). In Sec.~5 of \cite{SiegmundShort}, we extended this derivation to the case of a RTP. The main goal of this chapter will be to generalize this derivation to processes with time-correlated noise, including active particle models.

\subsection{Aim of this chapter and overview}

The original results by Siegmund \cite{Siegmund} were derived for one-dimensional Markov processes. Active particle models are by definition non-Markovian if we only consider the position of the particle, but they are Markovian if we consider both the position and the driving velocity, which is itself a Markov process. The existence of a Siegmund dual for stochastic processes driven by a stationary process was proved in \cite{AsmussenDiscrete} for discrete time and in \cite{SigmanContinuous} for continuous time (this has even been extended to higher dimensions in \cite{DualityMultidimensions}). In this setting, inspired by applications to finance, the exit probability (called ruin probability in this context) is related to the cumulative distribution of some dual process, defined in very general terms.

Siegmund duality can be seen as a particular case of Markov duality (see \cite{JansenDualityReview, DualityGenerators, CoxEntranceExitLaws, PathwiseDualSturm} for general reviews and other examples). Such duality relations are frequently studied in the mathematics literature and have been applied in various contexts, including queuing theory, finance and population genetics, as well as interacting particle systems and systems with a reservoir of particles \cite{Kolotsovkthorder, DualityLevyProcessesGoffard, MohleDualityGenetics, DualityBranchingFoucart, LiggettInteractingParticles, DualityBoundaryDriven}. However, they generally attract less attention among the physics community, although similar relations have sometimes been pointed out \cite{hittingProbaAnomalous, Comtet2011, Comtet2020, ThibautDual}. Some connections between different types of boundary conditions for various Markov processes have also been studied in the context of physics. For instance, in \cite{Szabo, Spouge}, the authors showed that the propagator of a diffusion process with partially absorbing boundary conditions can be obtained from the propagator of the same process with reflective boundary conditions. This approach was later extended to other situations \cite{Scher1, Scher2, Guerin}. Another example is the defect technique \cite{defect1, defect2}, which allows for instance to relate the first-passage time distribution of a diffusing particle in the presence of an absorbing wall to the cumulative distribution without walls (in Laplace space) \cite{defect3}.

In the previous chapter, we studied the connection between absorbing boundaries and hard walls for a specific model of active particle, namely the run-and-tumble particle, by computing explicitly the two sides of \eqref{siegmundExitIntro} in the infinite time limit, for an arbitrary external force. The aim of this chapter is to generalize the relation \eqref{siegmundExitIntro} (at any time $t$) to other models of active particles, and more generally to other stochastic processes driven by time-correlated noise which are commonly studied in physics. We consider two different settings. In Sec.~\ref{sec:continuousSiegmund} we introduce a general model of a continuous stochastic process in 1D driven by a stationary noise, which includes the most well-know models of active particles (RTP, AOUP and ABP), but also diffusing diffusivity models (see below). We give an explicit formulation of the dual process and show numerical evidence of the duality relation \eqref{siegmundExitIntro} for several models of interest. A derivation of the duality relation \eqref{siegmundExitIntro} based on the Fokker-Planck equation is provided in Appendix~\ref{app:ProofSiegmund}, generalizing the one given in \eqref{sec:SiegmundFinitetime} for the Brownian motion. We then consider in Sec.~\ref{sec:discreteSiegmund} the case of a 1D random walk with stationary increments, illustrating it with the example of a RTP model on lattice. We again define the dual explicitly, and we briefly give the idea of the proof which is detailed in \cite{SiegmundLong}. We then discuss some additional extensions in Sec.~\ref{sec:SiegmundExtensions}, namely continuous time random walks and stochastic resetting.


Although some general mathematical results already existed (see, e.g., \cite{AsmussenDiscrete, SigmanContinuous}), our achievement was to provide an explicit construction of the dual process for a wide range of physically relevant models (including active particles). We provided original and intuitive derivations of the duality relation \eqref{siegmundExitIntro} for both the continuous time and the discrete time setting, and illustrated its application to some well-known models through numerical simulations. As we will discuss more in detail in Sec.~\ref{sec:SiegmundDiscussion}, we see two main applications for the duality studied in this chapter. The first one is analytical: by studying either the first-passage properties of a stochastic process, we can immediately obtain new results for its distribution of positions in the presence of hard walls and vice-versa using the relation \eqref{siegmundExitIntro}. The second motivation is numerical. Indeed, in many cases the distribution of positions is often much simpler to compute numerically than the first-passage properties. In particular, if the system is ergodic, the stationary distribution can be obtained from a single run of the simulation by averaging over time, while to compute the exit probability one has to restart the simulation a large number of times.

\section{Siegmund duality for continuous stochastic processes driven by stationary noise} \label{sec:continuousSiegmund}

\subsection{Definition of the model}

In this section, we consider a one-dimensional stochastic process $x(t)$, with absorbing walls at $a$  and $b$, which evolves according to the following stochastic differential equation (SDE),
\begin{equation}
    \dot{x}(t) = f\left(x(t),\bm{\theta}(t)\right) + \sqrt{2\mathcal{T}\left(x(t),\bm{\theta}(t)\right)}\, \xi(t)\, ,
\label{LangevinIntroduction}
\end{equation}
where $\xi(t)$ is a Gaussian white noise with zero mean and unit variance, and $f\left(x,\bm{\theta}\right)$ and $\mathcal{T}\left(x,\bm{\theta}\right)$ are two arbitrary functions which correspond respectively to a force and to a temperature. Throughout this chapter, we use the It\=o prescription for the multiplicative noise \cite{handbookSM,riskenBook}. Here $\bm{\theta}(t)$ is a vector of arbitrary dimension whose components obey the following Markovian dynamics (which is independent of $x(t)$),
\begin{equation}
    \bm{\dot{\theta}}(t) = \bm{g}\left(\bm{\theta}(t)\right) + \left[2\underline{\mathcal{D}}(\bm{\theta}(t))\right]^{1/2} \cdot \bm{\eta}(t)\, ,
\label{SDEtheta_Introduction}
\end{equation}
where $\underline{\mathcal{D}}$ is a positive matrix.
The components $\eta_i(t)$ of $\bm{\eta}(t)$ are again independent Gaussian white noises with zero mean and unit variance. In addition, we allow $\bm{\theta}(t)$ to jump from a value $\bm{\theta}$ to $\bm{\theta}'$ with a transition kernel $\mathcal{W}(\bm{\theta}'|\bm{\theta})$. This means that, during a time interval $dt$, ${\bm \theta}(t)$ either evolves according to \eqref{SDEtheta_Introduction} with probability $1-dt \int d{\bm \theta}' \mathcal{W}(\bm{\theta}'|\bm{\theta})$, or jumps to some value $\bm{\theta}'$ with probability $\mathcal{W}(\bm{\theta}'|\bm{\theta}) d\bm{\theta}' \, dt$. We assume that $\bm \theta(t)$ admits an equilibrium distribution $p_{eq}(\bm{\theta})$ which satisfies the local detailed balance conditions\footnote{When initialized in its equilibrium distribution $p_{eq}({\bm \theta})$, ${\bm \theta}(t)$ satisfies $P({\bm \theta}(t_1), ..., {\bm \theta}(t_n)) = P({\bm \theta}(t_1+\tau), ..., {\bm \theta}(t_n+\tau))$ for any times $t_1,...,t_n$ and any time-shift $\tau$, and is called a stationary process.}
\begin{eqnarray}\label{detailed_balance_Introduction1}
&&- g_i(\bm{\theta}) p_{eq}(\bm{\theta}) + \sum_{j} \partial_{\theta_j}[\mathcal{D}_{ij}(\bm{\theta}) p_{eq}(\bm{\theta})]=0 \,, \, \forall \ i, \\
&& \mathcal{W}(\bm{\theta}|\bm{\theta}')p_{eq}(\bm{\theta}') = \mathcal{W}(\bm{\theta}'|\bm{\theta})p_{eq}(\bm{\theta}) \,. 
\label{detailed_balance_Introduction2}
\end{eqnarray}
The first equation \eqref{detailed_balance_Introduction1} corresponds to the vanishing of the probability current in \eqref{SDEtheta_Introduction}. Since it does not account for the discrete jumps, we need an additional detailed balance condition which is given by \eqref{detailed_balance_Introduction2}.

This definition is quite formal, but it encompasses a wide variety of stochastic processes which are relevant to physics. Below we will show how it can be specialized to recover the most well-known models of active particles. The general idea is that in this case ${\bm \theta}(t)=v(t)$ represents the driving velocity of the particle. For the AOUP and the ABP, $v(t)$ evolves according to a Langevin equation of the form \eqref{SDEtheta_Introduction} (e.g., an Ornstein-Uhlenbeck process for the AOUP), while for the RTP $v(t)$ takes discrete values, hence the transition kernel $\mathcal{W}(\bm{\theta}'|\bm{\theta})$. A combination of these two types of evolution is also possible, such as for the direction reversing active Brownian particle (DRABP) \cite{DRABP1,DRABP2}. This definition is however much more general than active particles. In particular, the fact that the temperature $\mathcal{T}\left(x,\bm{\theta}\right)$ may also depend on the process $\bm{\theta}$ allows us to include in this definition another important class of models, known as {\it diffusing diffusivity models} \cite{DiffDiffChubynsky, DiffDiffChechkin, DiffDiffJain, DiffDiffFPTSposini}. These models were recently introduced in order to describe the ``non-Gaussian normal diffusion" observed in several soft matter systems \cite{DDsoftmatter1,DDsoftmatter2,DDsoftmatter3}, i.e., the fact that some quantities exhibit a diffusive scaling but with a distribution which is not Gaussian. In this case, $\bm \theta$ is a $d$-dimensional Ornstein-Uhlenbeck process (i.e., $\underline{\mathcal{D}}$ is a constant and $\bm g$ is a harmonic force in \eqref{SDEtheta_Introduction}, and $\mathcal{W}=0$), the force $f\left(x,\bm{\theta}\right)=F(x)$ is independent of $\bm{\theta}$, but the temperature is $\bm{\theta}$-dependent, e.g., $\mathcal{T}\left(x,\bm{\theta}\right)=\bm \theta^2$.

\subsection{Statement of the duality} \label{sec:statementSiegmundContinuous}

For the process $x(t)$ defined by \eqref{LangevinIntroduction}, we showed in \cite{SiegmundLong} (see Appendix~\ref{app:ProofSiegmund} for a reproduction of the derivation) that the Siegmund dual $y(t)$, with hard walls at $a$ and $b$, is defined by the SDE
\begin{equation}
    \dot{y}(t) = \tilde f\left(y(t),\bm{\theta}(t)\right) + \sqrt{2\mathcal{T}\left(y(t),\bm{\theta}(t)\right)}\, \xi(t) \quad , \quad \tilde f(x,{\bm \theta})=-f\left(x,\bm{\theta}\right) + \partial_x \mathcal{T}\left(x,\bm{\theta}\right) \;,
\label{LangevinIntroductionDUAL}
\end{equation}
where $\bm{\theta}(t)$ and $\xi(t)$ denote different realizations of the processes appearing in \eqref{LangevinIntroduction}. If the temperature is independent of the position, $y(t)$ has exactly the same dynamics as $x(t)$, but with the sign of $f$ reversed. In the case where the temperature is space-dependent, the force also has an additional term $\partial_x \mathcal{T}$. Note that the transformation $f \to -f+\partial_x \mathcal{T}$ is its own inverse.

In Appendix~\ref{app:ProofSiegmund} we actually prove a more precise statement, which relates the exit probability at finite time for the process $x(t)$ {\it conditioned on the initial value of} $\bm{\theta}(t)$,
\begin{equation}\label{defExitProba}
    E_b(x,\bm{\theta},t) = \mathbb{P}(x(t)=b|x(0)=x,\bm{\theta}(0)=\bm{\theta}) \;,
\end{equation}
to the cumulative distribution of positions of the dual $y(t)$, {\it conditioned on the final value of} $\bm{\theta}(t)$, and where at $t=0$, $\bm{\theta}(t)$ is initialized in its equilibrium distribution $p_{eq}(\bm{\theta})$, with initial position $y(0)=b$,
\begin{equation}
    \tilde \Phi(x,t|\bm{\theta};b) = \int d\bm{\theta}_0 \, p_{eq}(\bm{\theta}_0) \, \tilde{ \mathbb{P}}(y(t)\leq x | \bm{\theta}(t)=\bm{\theta};y(0)=b, \bm{\theta}(0)=\bm{\theta}_0) \;.
    \label{Phi_def_MainResults1}
\end{equation}
For the two processes defined in \eqref{LangevinIntroduction} and \eqref{LangevinIntroductionDUAL} above, these two quantities are equal
\begin{equation} \label{mainRelation}
    E_b(x,\bm{\theta},t) = \tilde \Phi(x,t|\bm{\theta};b) \;.
\end{equation}
This relation of course has an equivalent for the exit probability at $x=a$, $E_a(x,\bm{\theta},t)$, which can be immediately deduced by symmetry (i.e., we only need to revert the inequality in the definition of $\tilde \Phi$ and to choose $y(0)=a$). The relation (\ref{mainRelation}) remains valid in the limit where there is only one wall, i.e., for $a\to - \infty$. As illustrated in the previous chapter for the RTP, a particular case where this relation can be useful is in the limit $t\to+\infty$, where it relates the exit probability at infinite time $E_b(x,\bm{\theta})$ and the cumulative distribution of its dual in the stationary state $\tilde \Phi(x|\bm{\theta};b)$ (if it exists), which is independent of the initial condition if the system is ergodic. But the finite time identity \eqref{mainRelation} contains additional information, in particular on the survival probability which can be recovered through the relation \eqref{rel_survival_exit}. The equivalent of \eqref{siegmundExitIntro} can be obtained by averaging over the equilibrium distribution of $\bm{\theta}$, i.e.,
\begin{equation} \label{mainRelation_averaged}
    E_b(x,t) = \tilde \Phi(x,t|b) \;.
\end{equation}
where
\begin{equation}
    E_b(x,t) = \int d\bm{\theta} \, p_{eq}(\bm{\theta}) E_b(x,\bm{\theta},t) \;,
\end{equation}
and
\begin{equation}
    \tilde \Phi(x,t|b) = \int d\bm{\theta} \, p_{eq}(\bm{\theta}) \tilde \Phi(x,t|\bm{\theta};b) \, .
    \label{Phi_def_MainResults2}
\end{equation}
Note that in this case it is important that $\bm{\theta}$ is initialized in its equilibrium distribution for both $x(t)$ and $y(t)$. This relation can be useful in particular when the initial value of $\bm{\theta}$ is unknown.

In \cite{SiegmundLong} (see Appendix~A therein), we also proved a more general statement, namely the equivalent of the ``full'' Siegmund duality \eqref{siegmundintro}, which reads
\begin{equation} \label{Siegmund}
    \mathbb{P}(x(t)\geq y|x(0)=x,\bm{\theta}(0)^{eq}) = \tilde{\mathbb{P}}(y(t)\leq x|y(0)=y,\bm{\theta}(0)^{eq})\, ,
\end{equation}
where the conditioning on $\bm{\theta}(0)^{eq}$ indicates that $\bm{\theta}(t)$ is initialized in its equilibrium distribution. This relation connects the full probability density of $x(t)$ at any time $t$, in the presence of absorbing walls, to the probability density of $y(t)$, with hard walls, for an arbitrary initial position. It recovers the relation \eqref{mainRelation_averaged} for $y=b$. It is worth mentioning that this more general relation still holds in the absence of walls.

\begin{figure}
\centering
    \includegraphics[width=0.9\linewidth]{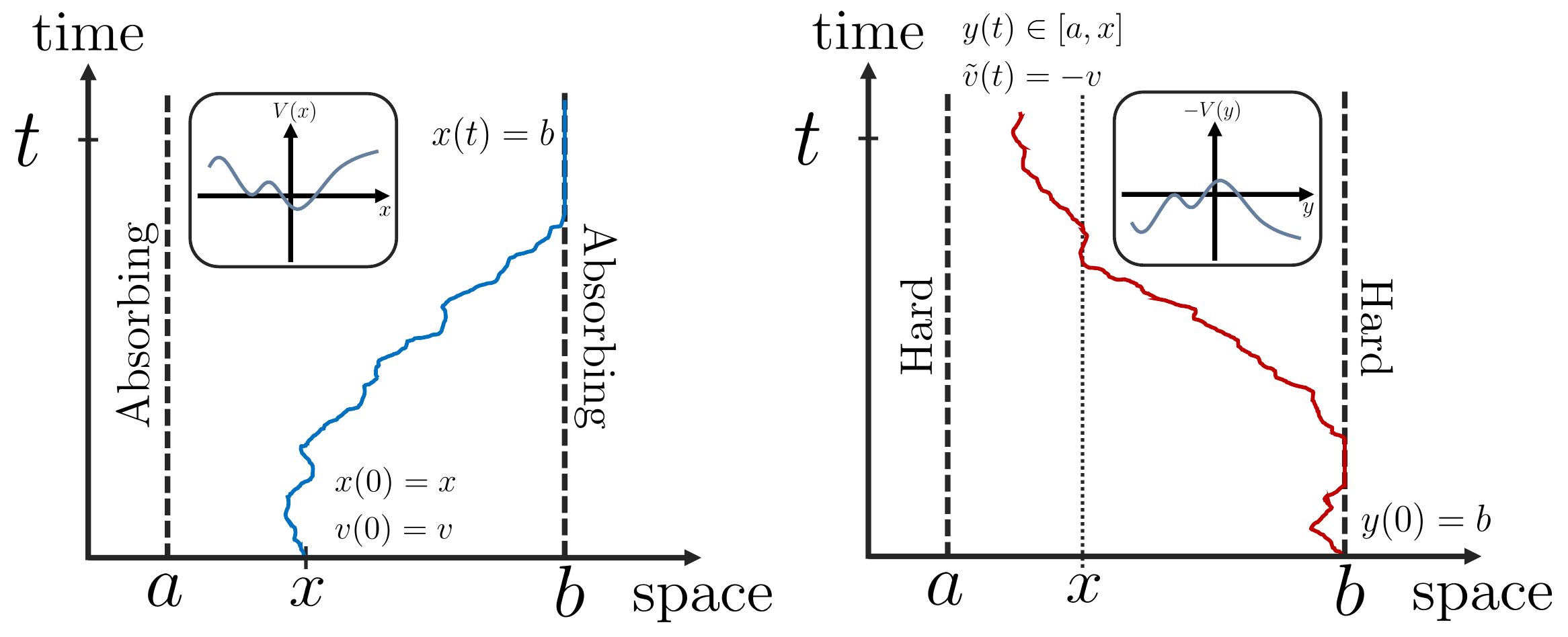} 
    \caption{Schematic representation of typical trajectories which contribute to the probabilities $E_b(x,v,t)$ and $\tilde \Phi(x,t|-v;b)$ for an active particle. \textbf{Left}:~The process $x(t)$ following the Langevin dynamics \eqref{Langevin_Active_Introduction} initiates its motion at position $x(t=0)=x$ with velocity $v(t=0)=v$. It is subjected to an external potential $V(x)$. Two absorbing walls are located at $a$ and $b$. The trajectory shown is absorbed at $b$ before time $t$ and hence contributes to the exit probability $E_b(x,v,t)$. \textbf{Right}:~The dual process of $x(t)$, namely $y(t)$, with Langevin dynamics given in \eqref{Langevin_Active_Introduction_Dual}, with hard walls at $x=a$ and $x=b$. The dual particle initiates its motion at $y(t=0)=b$ with an initial velocity drawn from the equilibrium distribution of the process $v(t)$, and in the presence of the reversed potential $-V(x)$. The trajectory shown contributes to $\Phi(x,t|\tilde v(t)=-v;b)$, i.e., the probability that the dual particle i located within the interval $[a,x]$ with a velocity $-v$ at time $t$. In this chapter we discuss the identity $E_b(x,v,t) = \tilde \Phi(x,t|-v;b)$, which we proved in \cite{SiegmundLong}.}
\label{figureIntro_duality}
\end{figure}

The duality relation \eqref{mainRelation} is illustrated schematically in Fig.~\ref{figureIntro_duality} (for the special case of active particles described below), where we show a typical trajectory contributing to each of the probabilities $E_b(x,\bm{\theta},t)$ and $\tilde \Phi(x,t|\bm{\theta};b)$. Intuitively, it can be seen as a form of time reversal symmetry (hence the minus sign in front of the force in \eqref{LangevinIntroductionDUAL}, and the additional term $\partial_x{\mathcal T}$ compensating for the flux of probability generated by the gradient of diffusion coefficient). The detailed balance conditions \eqref{detailed_balance_Introduction1}-\eqref{detailed_balance_Introduction2} for the driving process $\bm{\theta}(t)$ play a crucial role in this symmetry. This interpretation becomes more clear when we consider the discrete time variant of this duality relation, as we will briefly discuss in Sec.~\ref{sec:discreteSiegmund}.
\\

\subsection{Specialization to active particle models} \label{sec:SiegmundActive}

We now explain how the case of active particles can be recovered from the general model \eqref{Langevin_Active_Introduction}. In general, active particle models can be described by taking $\bm{\theta}(t)=v(t)$ to be a scalar, which corresponds to the driving velocity of the particle, and choosing a force $f(x,v) = F(x) + \alpha(x)\, v(t)$ which depends linearly on $v$. Here $F(x)$ is an additional external force, and we also allow for the presence of a space-dependent prefactor $\alpha(x)$ (generally positive) which modulates the velocity of the particle. In this case, the temperature $\mathcal{T}(x,v) =T(x)$ is independent of $v$. The equation then becomes
\begin{equation}
    \dot{x}(t) = F(x(t)) + \alpha(x(t))\, v(t) + \sqrt{2 T(x(t))}\, \xi(t)\, .
\label{Langevin_Active_Introduction}
\end{equation}
In this case, we can redefine the dual process \eqref{LangevinIntroductionDUAL} as
\begin{equation}
    \dot{y}(t) = \tilde F(y(t)) + \alpha(y(t))\, \tilde{v}(t) + \sqrt{2T(y(t))}\, \xi(t) \quad , \quad \tilde F(x) = -F(x) + \partial_x T(x) \;,
\label{Langevin_Active_Introduction_Dual}
\end{equation}
where $\tilde{v}(t)$ has the same law as $-v(t)$. This redefinition allows us to stay in line with the interpretation of $v$ as a velocity. Indeed, changing the sign of $\alpha(x)$ would mean that a positive value of $v(t)$, pushes the particle towards the negative direction, which would be counterintuitive. For all the examples that we consider here (RTP, AOUP and ABP), the equation describing the evolution of $v(t)$ is invariant under the change $v \to -v$, so that $\tilde v(t)$ has in fact exactly the same law as $v(t)$. In this setting, the relation \eqref{mainRelation} reads
\begin{equation}  \label{mainRelationActive}
    E_b(x,v,t) = \tilde \Phi(x,t|\tilde v(t)=-v;b) \;,
\end{equation}
i.e., the exit probability of a particle with initial velocity $v$ is now equal to the cumulative distribution of the dual particle, conditioned on it having a velocity $-v$ at time $t$. For a RTP, this simply amounts to exchanging $+$ and $-$ particles in the conditioning, and we thus recover the relation \eqref{duality_finite_time_pm} from the previous chapter.
\\

Let us briefly discuss how the different models introduced in Sec.~\ref{sec:active_models} can be recovered:
\begin{itemize}
    \item The RTP, defined in \eqref{defRTP}, can be recovered by setting $g=0$, $\mathcal{D}=0$, and $\mathcal{W}(v'|v)=\gamma\, \delta(v+v')$ in Eq. \eqref{SDEtheta_Introduction}, with the initial condition $v(0)=\pm v_0$, meaning that the driving velocity $v(t)$ simply jumps between the values $\pm v_0$ with a rate $\gamma$ (or equivalently $v(t)=v_0\sigma(t)$ where $\sigma(t)$ jumps between the values $\pm1$). In this case, the equilibrium distribution of $v$ is simply given by $p_{eq}(v)=\frac{1}{2}\delta(|v|-v_0)$, i.e., $v$ takes the values $\pm v_0$ with equal probability.  
    \item The AOUP, defined in \eqref{defAOUP}, corresponds to $g(v)=-v/\tau$ and $\mathcal{D}=D/\tau^2$ (constant), with $\mathcal{W}=0$, i.e., $v(t)$ is an Ornstein-Uhlenbeck process. In this case $p_{eq}(v)$ is Gaussian, as given by \eqref{AOUP_statv}.
    \item Finally, for a 2D ABP projected into one dimension, as defined in \eqref{defABP}, it is simpler to choose $\bm{\theta}$ in \eqref{LangevinIntroduction} as the angle $\varphi$ between the particle's orientation and the $x$-axis, and to set $f(x,\varphi) = F(x)+\alpha(x)\cos \varphi$ (we assume that the external force along the $x$-direction does not vary along the other space direction). Then we can simply set $g=0$ and $\mathcal{D}$ a constant in \eqref{SDEtheta_Introduction}, with $\mathcal{W}=0$, and the distribution $p_{eq}(\varphi)$ is uniform on $[0,2\pi)$. This can however easily be rewritten under the form \eqref{Langevin_Active_Introduction} by writing $v(t)=\cos\varphi$.
\end{itemize}

We now illustrate how the duality relation \eqref{mainRelationActive} applies to these models through numerical simulations.

\subsection{Illustration through numerical simulations}

\begin{figure}
\centering
        \includegraphics[width=0.45\linewidth]{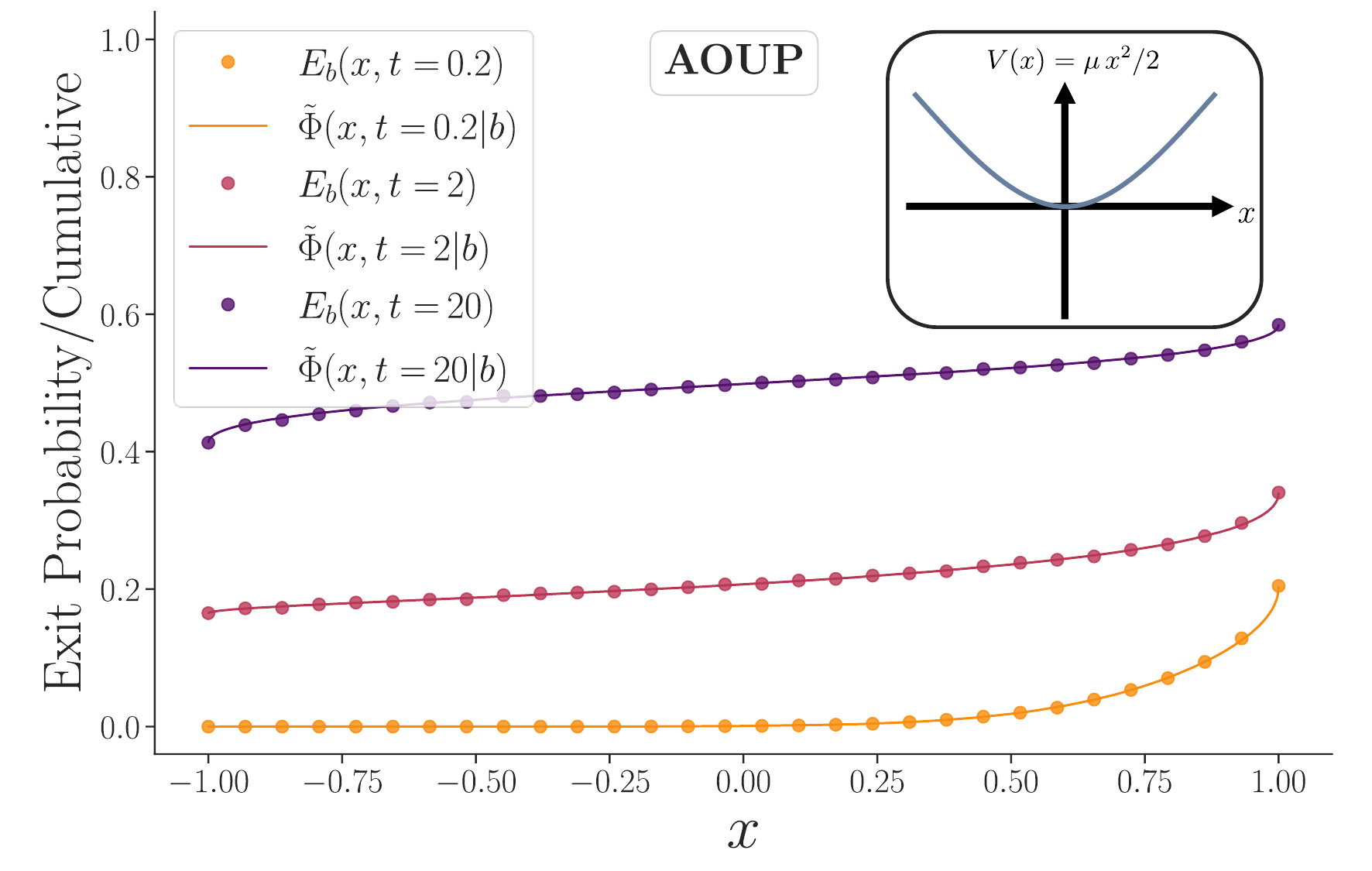}
        \hspace{0.2cm}
        \includegraphics[width=0.45\linewidth]{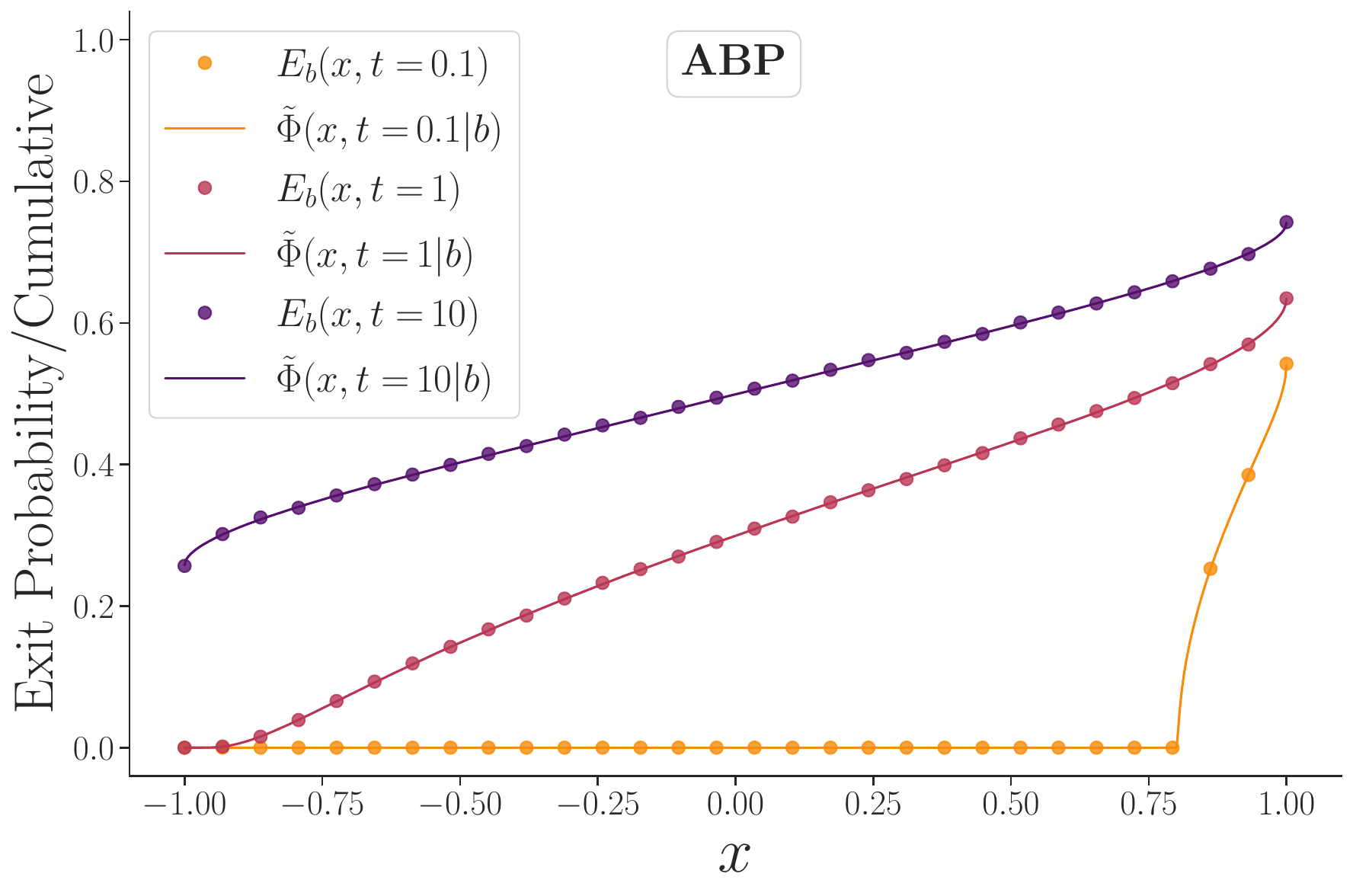}
    \caption{Illustration of the duality relation \eqref{mainRelation_averaged} at finite time for the AOUP and the ABP through numerical simulations. The dots represent the exit probability $E_b(x,t)$, while the solid lines show the cumulative distribution of the dual process $\tilde \Phi(x,t|b)$, both computed numerically through averages over many trajectories. \textbf{Left:} AOUP with $D=4$, $\tau=1$, in the presence of a harmonic potential $V(x)=\frac{\mu}{2}x^2$ with $\mu=1$ for $E_b$, and a potential $-V(x)$ for $\tilde \Phi$. \textbf{Right:} ABP with $v_0=2$ and $D=1$, in the absence of external potential. For both plots, $a=-1$ and $b=1$. In both cases, the exit probability overlaps the cumulative of the dual process perfectly.}
\label{FiniteTimeFig}
\end{figure}

We have confirmed the validity of \eqref{mainRelation_averaged} for several models of interest by performing direct numerical simulations of the Langevin dynamics \eqref{LangevinIntroduction} and \eqref{LangevinIntroductionDUAL} and computing the two quantities $E_b(x,t)$ and $\tilde \Phi(x,t|b)$ for each process. In the previous chapter, we have tested the relation \eqref{mainRelation}, with the conditioning on $\bm{\theta}=v_0\sigma$, in the case of the RTP, see Fig.~\ref{figureRTPtime}. Here we consider cases where the parameter ${\bm \theta}$ takes continuous values, and thus it is simpler to only test the duality relation integrated over $\bm{\theta}$ \eqref{mainRelation_averaged}. For $E_b(x,{\bm \theta},t)$, each point corresponds to an average over $N$ independent trajectories with initial position $x$ and $\bm{\theta}(0)$ drawn from $p_{eq}(\bm{\theta})$ (where we simply count the fraction of these trajectories which get absorbed at $b$ before time $t$). For the cumulative $\tilde \Phi(x,t|b)$, each curve is a histogram of the positions at time $t$ obtained from $N$ independent trajectories initialized at $y(0)=b$, again with $\bm{\theta}(0)$ drawn from $p_{eq}(\bm{\theta})$. In practice we used $N\sim 10^5 \hspace{-0.05cm}-\hspace{-0.05cm}10^7$ depending on the model.

\begin{figure}
\centering
        \includegraphics[width=0.45\linewidth]{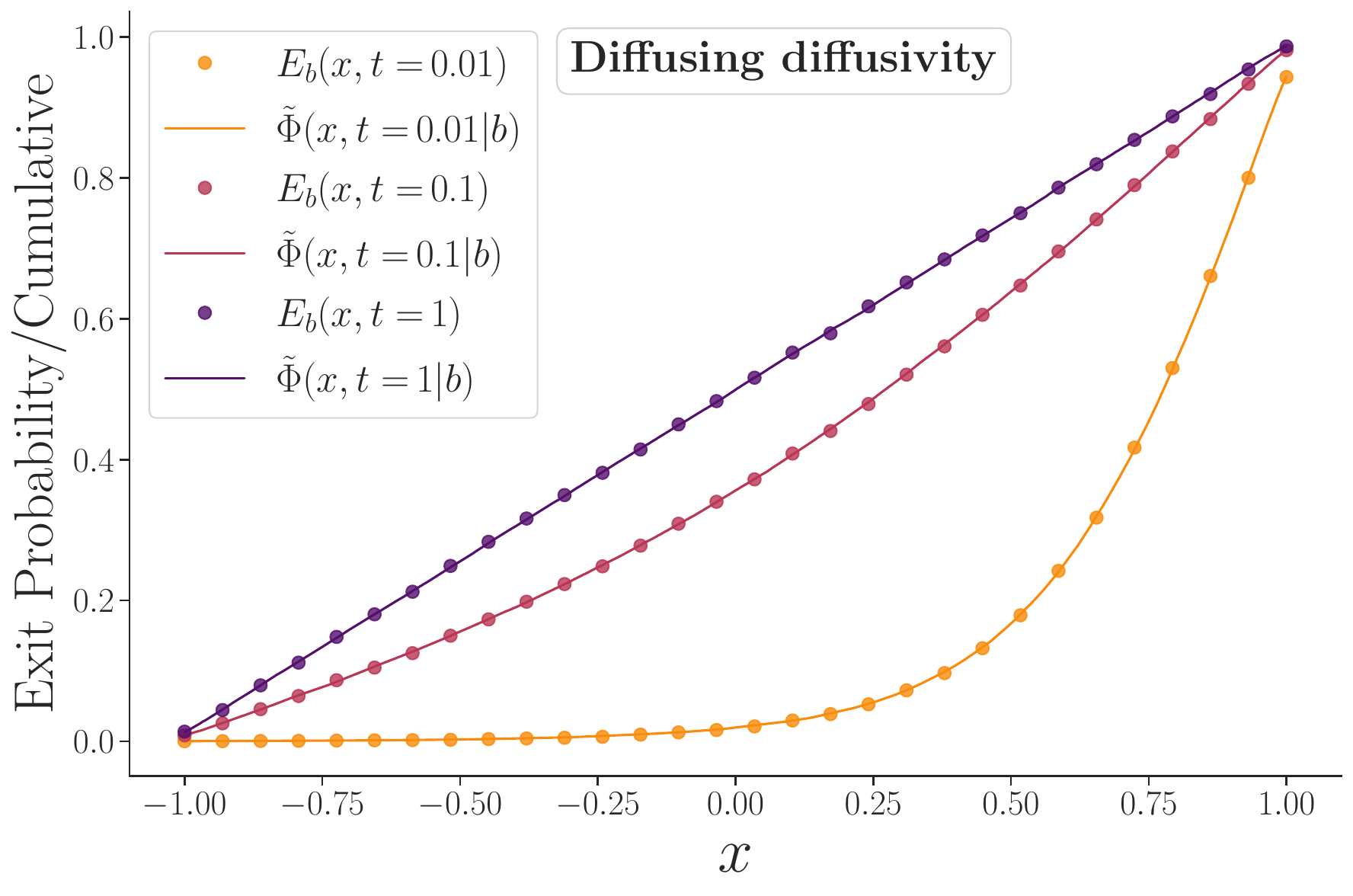}
        \hspace{0.2cm}
        \includegraphics[width=0.45\linewidth]{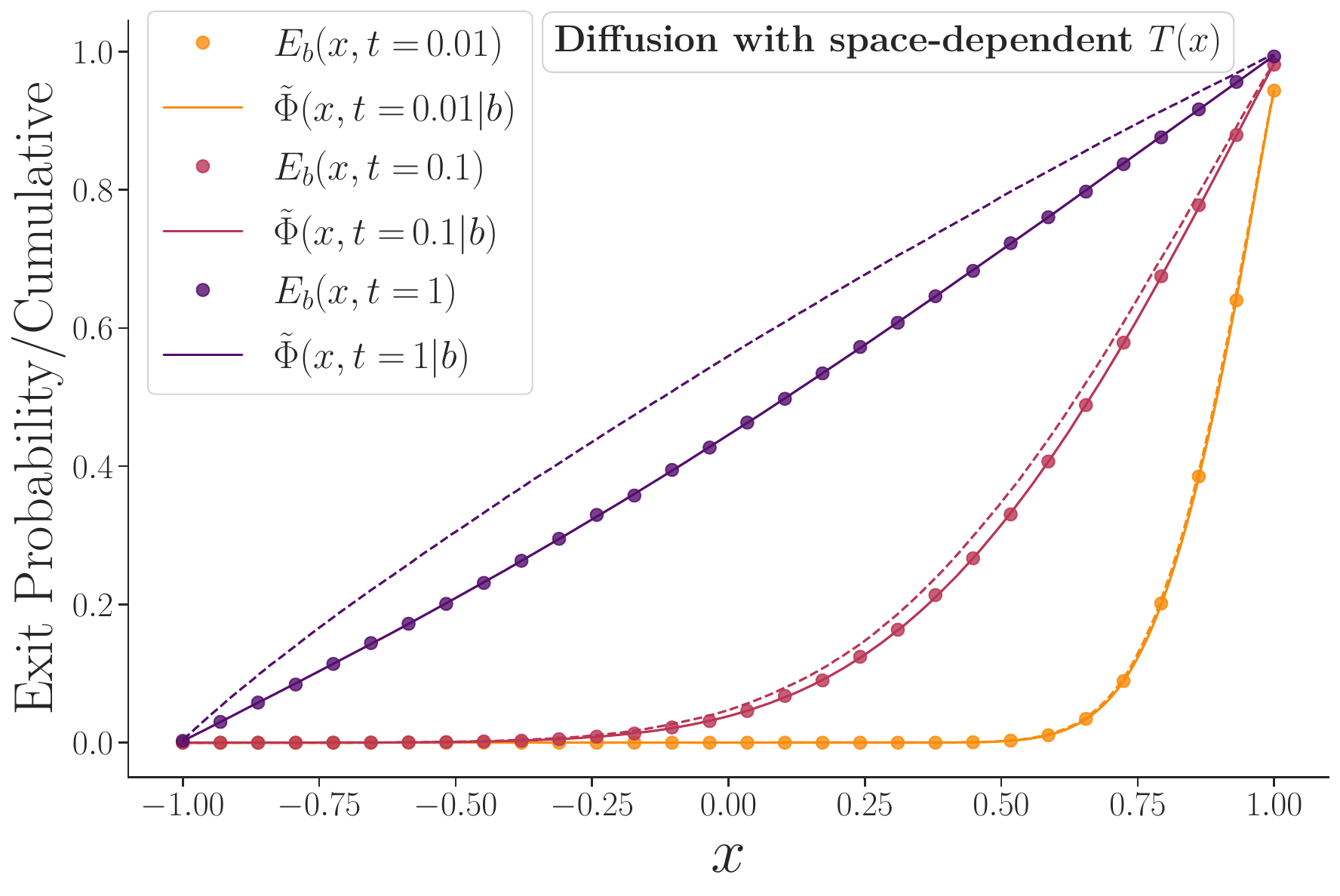}
    \caption{Illustration of the duality relation \eqref{mainRelation_averaged} as in Fig.~\ref{FiniteTimeFig}, for two different models, with $a=-1$ and $b=1$ in both cases. \textbf{Left:} Diffusing diffusivity model defined in \eqref{defDiffDiff} (for a one-dimensional $\theta$) with diffusion coefficient $D=4$ and relaxation time $\tau=1$. \textbf{Right:} Brownian particle with space-dependent diffusion coefficient $T(x)=1+\frac{x}{2}$ (and no external potential). To compensate the variations in temperature, the dual process is subjected to a constant force $\tilde f=\partial_x T=\frac{1}{2}$. The dashed line shows the cumulative distribution in the absence of this force. For both models, the dots which correspond to $E_b(x,t)$ overlap perfectly with the cumulative of the dual $\tilde \Phi(x,t|b)$.}
\label{FiniteTimeFig2}
\end{figure}

The first two models which we consider in Fig.~\ref{FiniteTimeFig} are the AOUP and the ABP, which we have just discussed in Sec.~\ref{sec:SiegmundActive}. In Fig.~\ref{FiniteTimeFig2} (left panel), we also consider a diffusing diffusivity model, described by the equations \cite{DiffDiffChubynsky, DiffDiffChechkin, DiffDiffJain, DiffDiffFPTSposini}
\begin{equation}\label{defDiffDiff}
\frac{dx}{dt} = F(x) +  \sqrt{2T(t)} \ \xi(t) \quad , \quad T(t)=\bm{\theta}^2(t) \quad , \quad \tau \frac{d\bm{\theta}}{dt} = -\bm{\theta}(t) + \sqrt{2 D} \ \bm{\eta}(t) \;,
\end{equation}
where $F(x)$ is an external force, $D$ is a diffusion coefficient, and $\xi(t)$ and $\bm{\eta}(t)$ are Gaussian white noises. Here $\bm{\theta}(t)$ is therefore an Ornstein-Uhlenbeck process in d dimensions. For the simulations we only considered the case $d=1$, in which case the distribution $p_{eq}(\theta)$ is a simple Gaussian in 1D (as for the AOUP),
\be
p_{eq}(\theta)= \sqrt{\frac{\tau}{2\pi D}} \, e^{-\frac{\tau \theta^2}{2D}}\, .
\ee
Finally, in the same figure \ref{FiniteTimeFig2}, we also show a simple case where the diffusion coefficient $T(x)$ is space-dependent. In all cases, we find a perfect overlap between the exit probability $E_b(x,t)$ of the process $x(t)$, and the cumulative distribution $\tilde \Phi(x,t|b)$ of its dual $y(t)$.

\section{Siegmund duality for random walks with correlated steps} \label{sec:discreteSiegmund}

\subsection{Definition of the model} \label{sec:discreteDef}

In Sec.~\ref{sec:continuousSiegmund}, we gave an explicit construction of the Siegmund dual for a family of continuous stochastic processes driven by a stationary Markov process. In this subsection, we present a similar result in the case of discrete-time random walks. We consider a random walk whose position at time $n$ is denoted $X_n$. At the $n^{th}$ time step, it performs a jump $W_n$ according to some distribution, which may be either discrete or continuous. The values of $W_n$ at different steps $n$ may be correlated (see below). The walker starts at some position $X_0$ inside the interval $[a,b]$, and there are absorbing boundary conditions at $x=a^-$ and $x=b$. Here, the notation $a^-$ should be understood as $a-\epsilon$ with $\epsilon \to 0$, meaning that if the walker is located exactly at $X_n=a$ it is not in an absorbed state and can still move towards the right, but any step towards the left, however small, will lead to absorption (the reason for this choice will be clarified below when we will discuss the proof of the duality). The evolution of $X_n$ can thus be summarized as follows,
\begin{eqnarray}
    && X_{n} = \begin{cases} (X_{n-1} + W_{n})_{[a^-,b]} \quad {\rm if} \ X_n \in ]a^-,b[ \\
    X_{n-1} \quad {\rm if} \ X_{n-1} = a^- \ {\rm or} \ b \end{cases}\, ,
\label{defXt}
\end{eqnarray}
where we have introduced the notation
\begin{eqnarray}
    && (x)_{[a,b]} = \begin{cases} a \quad {\rm if}\  x \leq a \\
    x \quad {\rm if} \ a < x < b \\
    b \quad {\rm if} \ x \geq b \end{cases}\, .
\end{eqnarray}
Here $W_n$ is a stationary stochastic process with stationary distribution $p_{st}(w)$, which satisfies the following time reversal property:
\begin{equation}
    p_{st}(w_1) P(W_2=w_2,...,W_T=w_T|W_1=w_1) = p_{st}(w_T) P(W_2=w_{T-1},...,W_T=w_1|W_1=w_T) \label{DBstatement1} \, .
\end{equation}
As mentioned above, the process $W_n$ can take either discrete or continuous values, but its evolution should not depend on the position $X_n$. Note that a realization of the process $\{X_n,W_n\}=(X_0,...,X_T,W_1,...,W_T)$ is completely determined by the initial position $X_0$ and the set of $W_n$'s.

This definition is very general and encompasses a wide variety of one-dimensional models with correlated steps. Below we will illustrate our results with the example of a RTP on a lattice with discrete time, defined in Sec.~\ref{sec:PRWdef}. Note that the discrete RTP is an example where $W_n$ is a Markov process, but this does not have to be the case in general, i.e., the distribution of $W_n$ may depend on all the previous values $W_1,...,W_{n-1}$. In that regard, this model is more general than the continuous one defined in \eqref{LangevinIntroduction}\footnote{Note however that it is also less general in the sense that here the $W_n$ should be completely independent of the position $X_n$, while in \eqref{LangevinIntroduction} the force $f(x,\bm{\theta})$ and the diffusion coefficient $\mathcal{T}(x,\bm{\theta})$ could depend on the position.}. A particular case which is closer to the hypotheses of the continuous setting is if $W_n=W(\bm{\Theta}_n)$, with $W(\bm{\Theta})$ an arbitrary function from $\mathbb{R}^d$ to $\mathbb{R}$, and $\bm{\Theta}_n$ a Markov process on $\mathbb{R}^d$ with transition probability $\pi(\bm{\Theta}_n|\bm{\Theta}_{n-1})$. In this case, the time reversal property \eqref{DBstatement1} reduces to a simple detailed balance condition for $\bm{\Theta}_n$, i.e., $\bm{\Theta}_n$ should admit an equilibrium distribution $p_{eq}(\bm{\Theta})$, satisfying
\begin{equation}
\pi(\bm{\Theta}_n|\bm{\Theta}_{n-1}) p_{eq}(\bm{\Theta}_{n-1}) = \pi(\bm{\Theta}_{n-1}|\bm{\Theta}_n) p_{eq}(\bm{\Theta}_n)\; .
\label{detailed_balance}
\end{equation}
Indeed, applying \eqref{detailed_balance} recursively straightforwardly leads to \eqref{DBstatement1}. In this specific setting, the conditioning on $W_1$ below could be replaced by a conditioning on ${\bm\Theta}_1$, as in the continuous case. The connection between the discrete and continuous settings is discussed in more details in Sec.~IV.C of \cite{SiegmundLong}.

Before defining the dual, let us note that, in the case where the distribution of the jumps $W_n$ is continuous, the absorbing wall can be placed at $x=a$ instead of $a^-$, since the probability that the process reaches exactly $x=a$ at some time $n$ is zero. However, if we consider instead a lattice model, such as the lattice RTP discussed below, then the shift of the absorbing wall $a\to a^-$ amounts to adding an additional site to the left of the lattice (e.g., the sites could be labeled $a-1,a, a+1, ..., b$, with $a-1$ and $b$ being the absorbing site. This is important since for the dual process, the hard walls are located exactly at $a$ and $b$, as we now discuss (which leads to one less site compared to the process $X_n$ for a lattice model).

\subsection{Statement of the duality} \label{sec:discreteStatement}

We consider the process $X_n$ defined in \eqref{defXt} on a finite time interval $\llbracket 0,T \rrbracket$. On this interval, we define its {\it dual process} $Y_n$, which starts at some value $Y_0 \in[a,b]$, with hard walls at $a$ and $b$, and evolves according to
\begin{equation}
Y_n =  (Y_{n-1} - \tilde W_{n})_{[a,b]} \;,
\label{defdual}
\end{equation}
where the $\tilde W_n$'s follow the same stochastic dynamics as the $W_n$'s. This can be seen as a time-reversed version of the process $X_n$, but with an arbitrary initial position $Y_0$, and with hard walls at $a$ and $b$ instead of absorbing walls. Note that for the dual $Y_n$, the left wall is located at $x=a$ and not at $a^-$.

With this definition of the dual process $Y_n$, we showed the identity
\begin{equation}
E_b(x,w,T) = \tilde{\Phi}(x,T|w;b) \; ,
\label{ExitRW_precise}
\end{equation}
between the exit probability at $b$ at time $T$ for the process $X_n$, conditioned on the initial step $W_1$,
\begin{equation}
    E_b(x,w,T) = \mathbb{P}(X_T=b|X_0=x, W_1=w)\, , 
    \label{exit_def_discrete}
\end{equation}
and the cumulative of its dual $Y_n$ conditioned on the last step $\tilde W_T$, where $\tilde W_1$ is drawn from $p_{st}(w)$,
\begin{equation}
    \tilde{\Phi}(x,T|w;y) = \int dw \ p_{st}(w_1) \, \mathbb{\tilde P}(Y_T\leq x|\tilde W_T=w, Y_0=y,\tilde W_1=w_1)\; .
    \label{cumul_def}
\end{equation}
Note that $E_b(x,w,T)$ is conditioned on the first step {\em after} time $n=0$ (i.e., the jump that occurs between $X_0$ and $X_1$), while the cumulative is conditioned on the last step {\em before} time $T$ (i.e., between $Y_{T-1}$ and $Y_T$).

All the remarks we have made in Sec.~\ref{sec:statementSiegmundContinuous} for the continuous case also apply here. First, the equivalent of \eqref{exit_def_discrete} for the exit probability at $a$ can be deduced by symmetry. Note however that in this case, the absorbing walls should be placed at positions $a$ and $b^+$ (instead of $a^-$ and $b$) for the process $X_n$. Second, the relation \eqref{exit_def_discrete} is still valid for $a\to-\infty$. Third, it also holds in the limit $T\to+\infty$ if it is well-defined, providing a relation between the exit probability $E_b(x,w)=E_b(x,w,T\to+\infty)$ and the stationary cumulative distribution of the dual $\tilde{\Phi}(x|w;y)=\tilde{\Phi}(x,T\to+\infty|w;y)$. Fourth, the relation \eqref{exit_def_discrete} can also be averaged over the stationary distribution $p_{st}(w)$, which yields the relation
\begin{eqnarray}
    E_b(x,T) &=& \tilde{\Phi}(x,T|b)\, , \\
    {\rm with}\, \,  \quad E_b(x,t) &=& \int dw \ p_{st}(w) \mathbb{P}(X_t=b|X_0=x, W_1=w) \, , \nn \\
   {\rm and} \quad \tilde{\Phi}(x,T|b) &=& \int d\tilde w_1 \ p_{st}(\tilde w_1) \tilde{P}(Y_T \leq x|Y_0=y, \tilde W_1=\tilde w_1) \, , \nn
\label{Siegmund_averaged}
\end{eqnarray}
which relates the exit probability at time $T$, $E_b(x,T)$ and the cumulative distribution of the dual $\tilde{\Phi}(x,T|b)$, where for both processes the jumps $W_n$ and $\tilde W_n$ are initialized from the stationary distribution $p_{st}(w)$. Finally, we also have the ``full'' Siegmund duality relation
\begin{equation}
\mathbb{P}(X_T \geq y|X_0=x,W_1^{st}) = \tilde{\mathbb{P}}(Y_T \leq x|Y_0=y, \tilde W_1^{st}) \; ,
\label{Siegmund_symmetric}
\end{equation}
where the conditioning on $W_1^{st}$ again means that $W_1$ is initialized from $p_{st}(w)$ (and similarly for $\tilde W_1$).

\subsection{Sketch of the proof}

\begin{figure}
  \centering
    \includegraphics[width=0.45\textwidth]{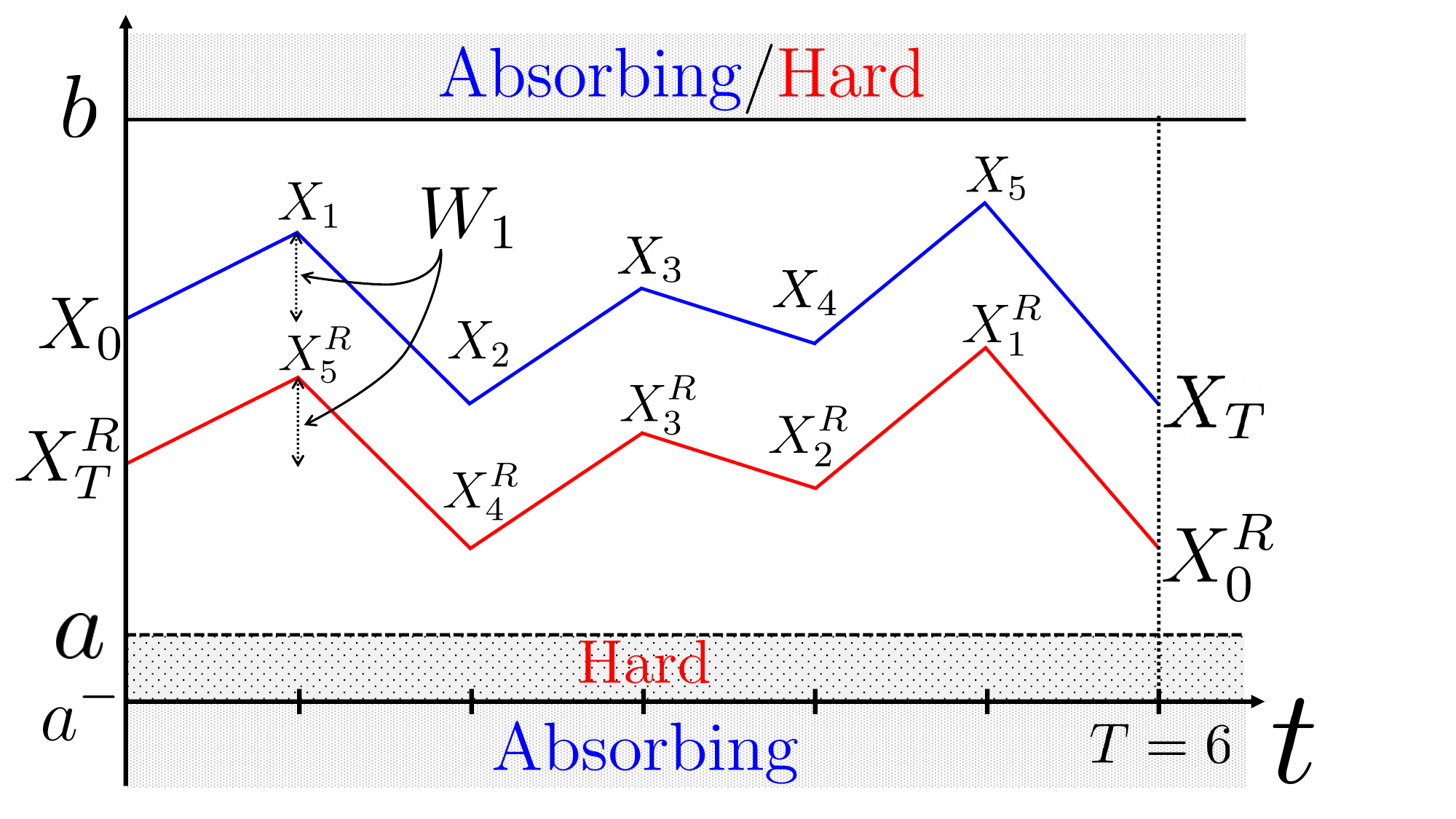}
    \label{fig:subfig1_process}
    \includegraphics[width=0.45\textwidth]{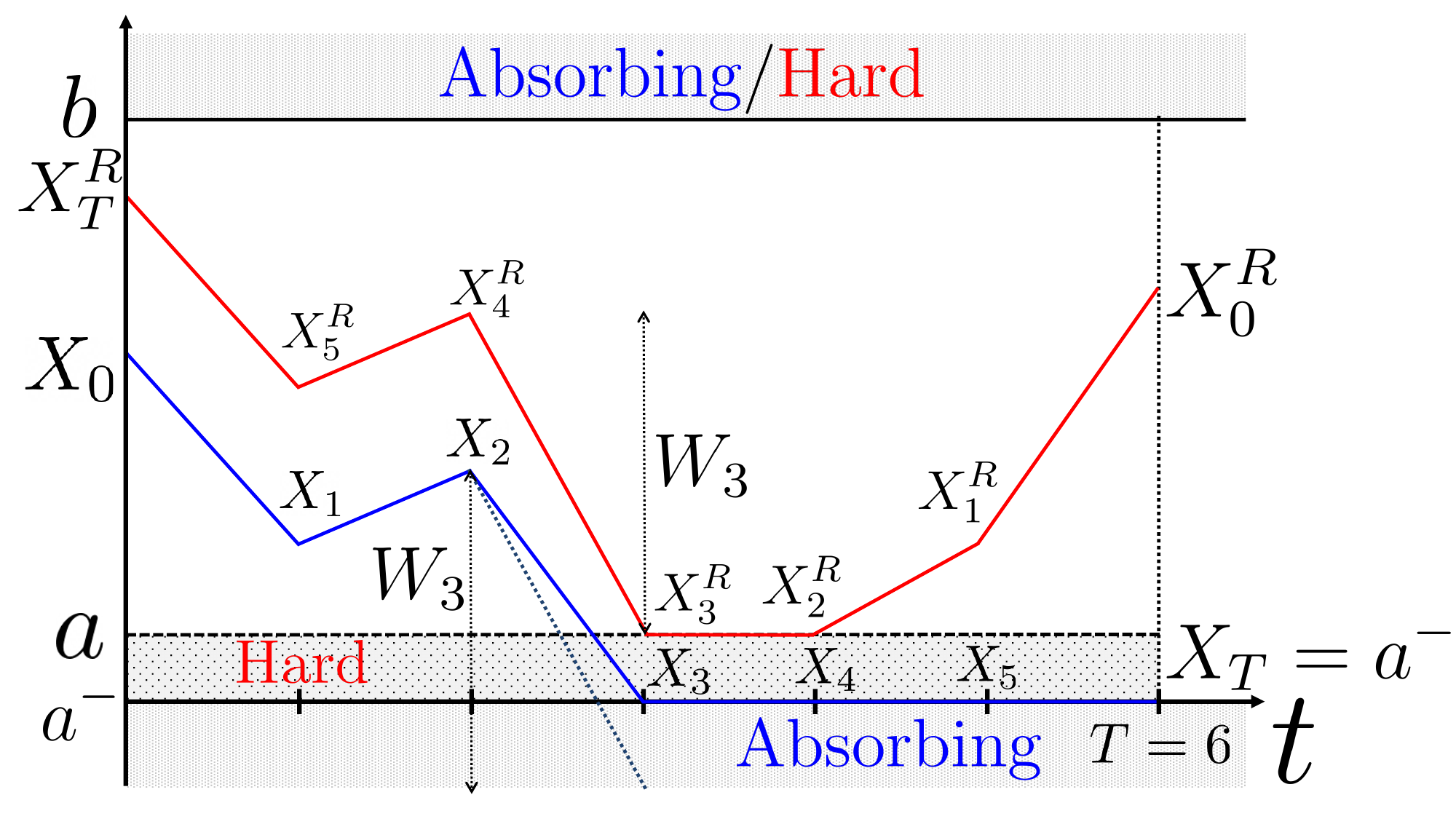}
    \label{fig:subfig2_process}
  \caption{Here we show in blue two examples of trajectories of a process $X_n$ (defined in \eqref{defXt}), and in red $\hat{X}^R_n= X^R_{T-n}$, where $X_n^R$ is a dual trajectory of $X_n$. The {\bf left} panel illustrates the fact that when $X_n$ never reaches a wall, $\hat{X}_n^R$ is parallel to $X_n$ and never crosses it. On the {\bf right}, $X_n$ reaches a wall at time $t_0=3$ and is absorbed at $x=a^-$. On the other hand, for $\hat{X}^R_n$, $a$ and $b$ are hard walls and it can still move inside $[a,b]$. For both examples, the equivalence (\ref{equivalence}) holds.}
  \label{fig:two_subfigs_process}
\end{figure}

We now give the main ideas of the proof of \eqref{ExitRW_precise}, which helps to clarify the interpretation of the Siegmund duality as a time-reversal symmetry. A complete version of the proof is given in Sec.~IV.A of \cite{SiegmundLong}. It relies on a mapping between the trajectories of the process $X_n$ and its dual $Y_n$, through the notion of dual trajectory $X_n^R$. For a fixed value of $X_0^R\in [a^-,b]$, we define the {\it dual trajectory} (or time-reversed trajectory) $X^R_n$ of a given realization of $X_n$ with jump sequence $(W_1,W_2,\ldots,W_T)$, as the realization of the dual process $Y_n$, defined in \eqref{defdual}, which starts at the initial position $Y_0=X_0^R$, and whose jump sequence is $(W_T,W_{T-1},\ldots,W_1)$, i.e., $\tilde W_n = W_{T+1-n}$ for all $n\in\llbracket 0,T \rrbracket$. Examples of trajectories along with their dual trajectories are shown in Fig.~\ref{fig:two_subfigs_process}. For the visualization it is convenient to also define $\hat{X}^R_n=X^R_{T-n}$.

The first part of the derivation consists in proving the following equivalence
\begin{equation}
    X_0 \geq X^R_T \Leftrightarrow X_T \geq X^R_0 \, ,
    \label{equivalence}
\end{equation}
which simply means that a trajectories $X_n$ and $\hat{X}^R_n$ can never cross. As can be seen on the left panel of Fig.~\ref{fig:two_subfigs_process}, when the two trajectories do not interact with the walls, these two trajectories are parallel and the equivalence \eqref{equivalence} is straightforward. The difficulty is to show that it still holds when we take into account the different boundary conditions. This step is the reason why the shift $a\to a^-$ of the absorbing wall is required. We refer again to Sec.~IV.A of \cite{SiegmundLong} for more details. The second part of the proof is then to use the time reversal property \eqref{DBstatement1}, and to integrate it over all the realizations where the events in \eqref{equivalence} are realized (seeing $X^R_T$ as a realization of $Y_n$), which leads to an intermediate identity from which both \eqref{ExitRW_precise} and \eqref{Siegmund_symmetric} can be obtained.

This one-to-one mapping between the trajectories of $X_n$ and $Y_n$, shows that the two processed can be seen as some form of time-reversal of each other with different boundary conditions, which provides an intuitive interpretation of the Siegmund duality.

\subsection{An example: the discrete RTP}

Let us now illustrate the results above with a discrete version of the RTP model, (sometimes called a persistent random walk) which we already briefly introduced in Sec.~\ref{sec:PRWdef} (see, e.g., \cite{Larralde2020,PRWSurvivalLacroixMori} for studies of this model). It consists in a particle, with position at the discrete time $n$ given by $x_n$, evolving on a 1D lattice with $L+2$ sites labeled $i=0,...,L+1$, with absorbing sites at $0$ and $L+1$ (i.e., if the particle reaches one of these two sites it stays there forever), according to the equation
\begin{equation} \label{PRWx}
    x_{n} = x_{n-1} + \sigma_n\, ,
\end{equation}
where the steps $\sigma_n = \pm 1$ follow a Markov dynamics defined by
\begin{equation}
\begin{aligned} \label{PRWsigma}
\sigma_n = \begin{cases}
\sigma_{n-1} &\text{with probability } q \\
-\sigma_{n-1} &\text{with probability } 1-q
\end{cases} \, .
\end{aligned}
\end{equation}
The parameter $q \in [0,1]$ controls the ``persistence" of the random walk. For $q=1/2$, we recover a simple symmetric random walk with uncorrelated steps. For $q>\frac{1}{2}$, the steps are correlated positively, leading to a persistent motion similar ti the continuous RTP, while for $q<\frac{1}{2}$ the steps are negatively correlated. In Sec.~\ref{sec:PRWdef} we explained how the continuous RTP can be recovered from this model in the continuous limit.

Since the evolution of the jumps $\sigma_n$ is invariant by parity (i.e., we have $p(\sigma_{n+1}=\sigma_{n}|\sigma_{n}) = p(\sigma_{n+1}=-\sigma_{n}|-\sigma_{n}) = q$), the dual process of $x_n$, which we denote $y_n$, is simply defined by the same recursive relation (see \eqref{defdual}),
\begin{equation} \label{PRWy}
    y_{n} = y_{n-1} + \sigma_n\, ,
\end{equation}
but now with hard walls at sites $1$ and $L+1$, i.e., if the particle is on site 1 and jumps to the left it stays at 1, and similarly at the other end of the lattice (as explained in Sec.~\ref{sec:discreteDef}, the left wall is shifted by one site compared to the process $x_n$). Here we fix $y_0=L+1$.

For this model, the duality relation \eqref{exit_def_discrete} provides a relation between the exit probability of the process $x_n$, defined as
\begin{equation} \label{PRWEdef}
    E_i^\pm(n) = \mathbb{P}(x_n = L+1| x_0=i ,\sigma_1 \pm 1) \;,
\end{equation}
and the cumulative of the distribution of positions of $y_n$ conditioned on $\sigma_n$,
\be
\tilde \Phi_i^{\pm}(n) = \mathbb{P}(y_n \leq i| \sigma_n \pm 1; y_0=L+1) \;,
\ee
i.e., we have for any $n\geq0$ and any $i\in \llbracket 1,L+1 \rrbracket$,
\begin{equation} \label{dualityPRW}
    E_i^\pm(n) = \tilde \Phi_i^{\mp}(n)  \, .
\end{equation}

\begin{figure}
\centering
        \includegraphics[width=0.45\linewidth]{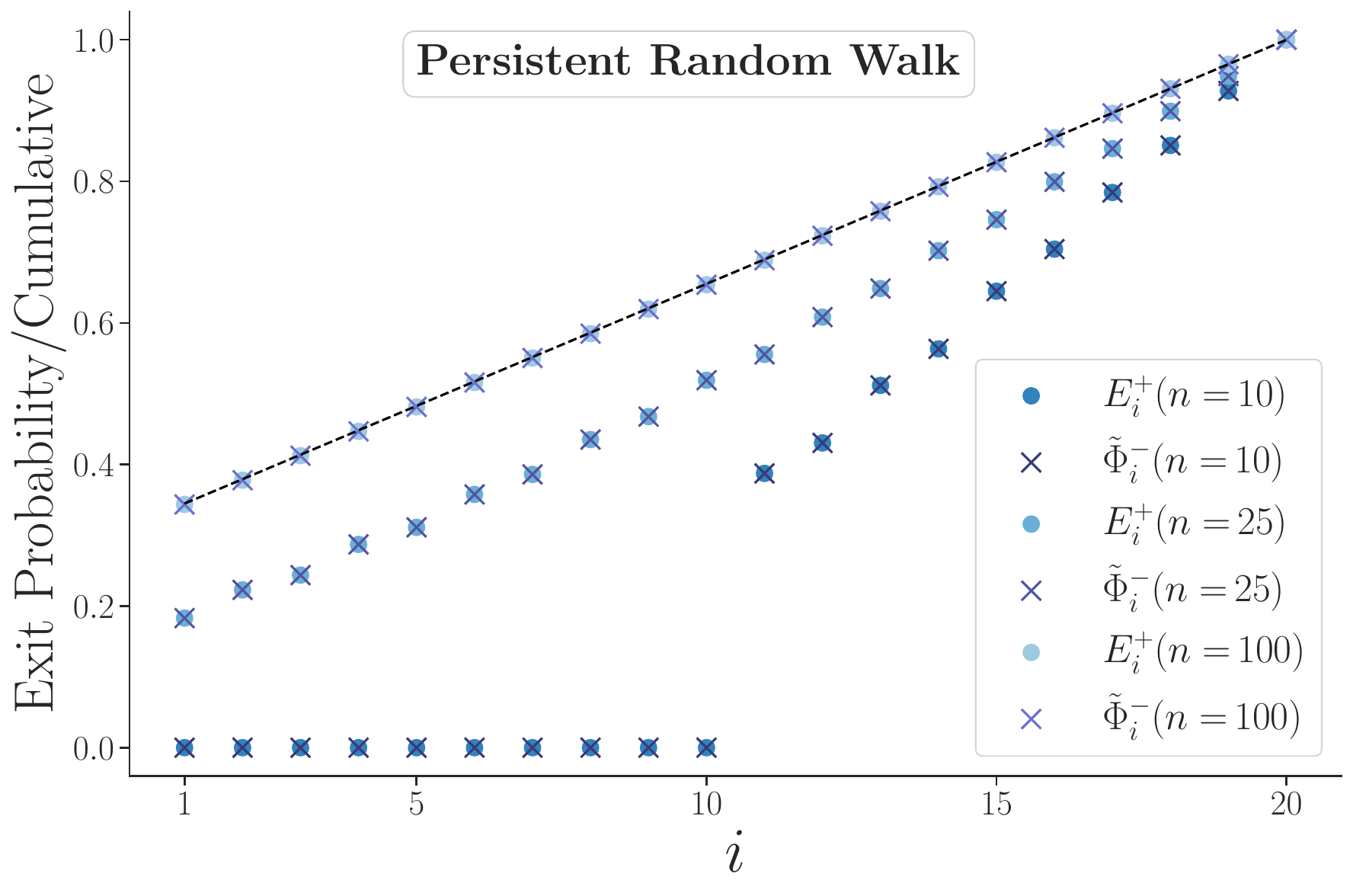}
        \hspace{0.2cm}
        \includegraphics[width=0.45\linewidth]{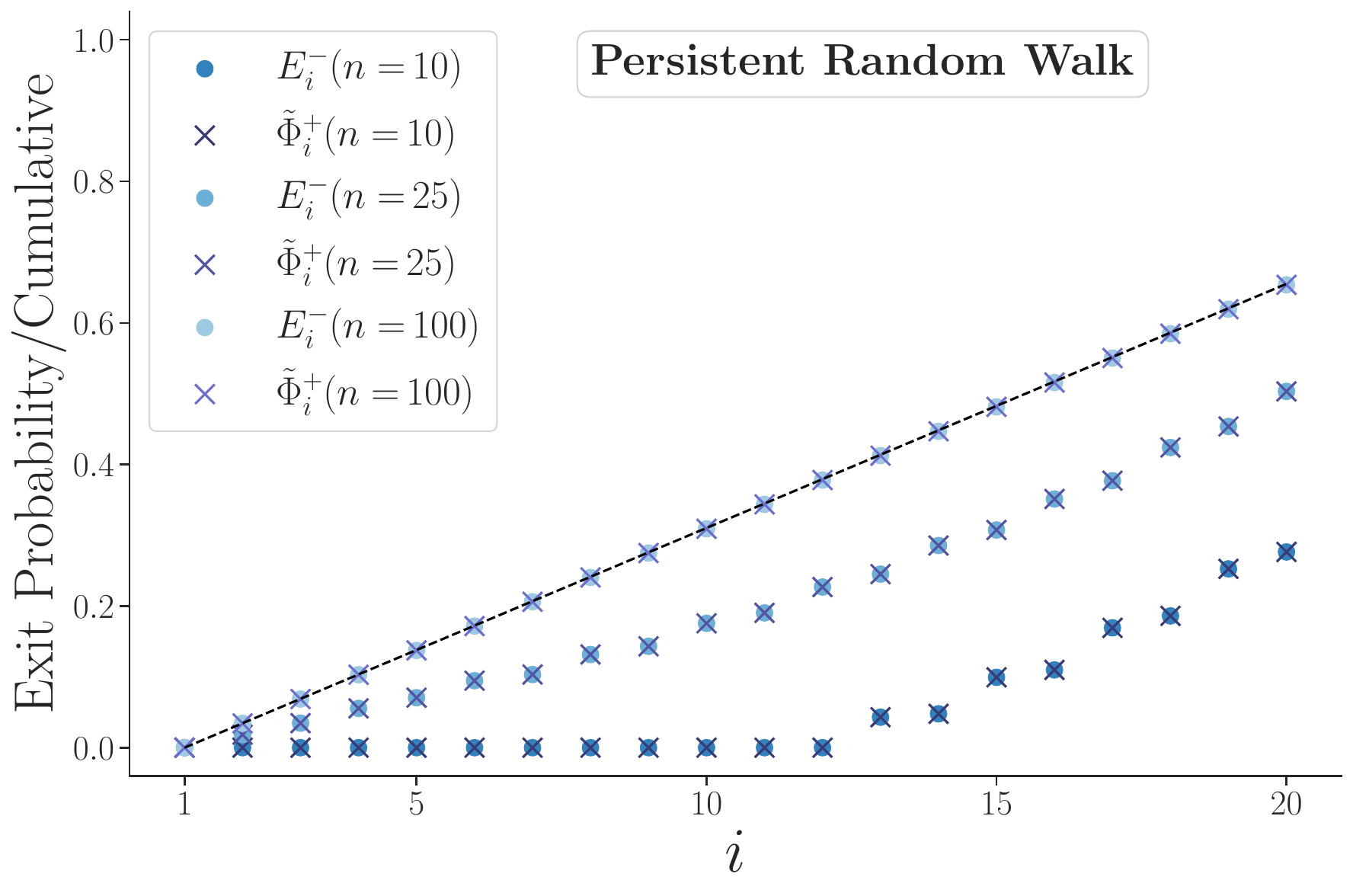}
    \caption{\textbf{Left}: For a discrete RTP, also called persistent random walk, we computed numerically the exit probability $E_i^+(n)$ and the cumulative distribution with hard walls $\tilde \Phi_i^-(n)$ using transfer matrices, for $q=0.9$ and $L=20$. The dashed black line corresponds to the analytical result for the stationary state. \textbf{Right}: Same plot for $E_i^-(n)$ and $\tilde \Phi_i^+(n)$. On both plots, the to quantities overlap perfectly. The details of the transfer matrix method and the computation of the stationary state are detailed in Appendix~D of \cite{SiegmundLong}.}
    \label{PRWfig}
\end{figure}

We have computed these two quantities at different times using transfer matrices, see Appendix~D of \cite{SiegmundLong} for the details of the method. As can be seen on Fig.~\ref{PRWfig}, there is indeed a perfect overlap. In the same appendix, we also computed explicitly the stationary state for both quantities and verify that they indeed match as expected.

\section{Additional extensions} \label{sec:SiegmundExtensions}

\subsection{Continuous time random walks}

The results of Sec.~\ref{sec:discreteSiegmund} can be generalized to continuous-time random walks (CTRW) \cite{CTRW1,CTRW2, CTRWA, CTRWB, CTRWC}. The only difference is that time is now a continuous variable, and the random walk $X_t$ performs jumps at random times. The time intervals $\tau_1, \tau_2 ...$ between successive jumps are drawn from an arbitrary probability distribution $\omega(\tau)$, until $\tau_1 + ... + \tau_N<T$ and $\tau_1 + ... + \tau_{N+1} \geq T$. Denoting $t_i=\sum_{j=1}^{i} \tau_j$ (with $t_0=0$) we then define the process $X_t$ (with absorbing walls at $a^-$ and $b$) as
\begin{eqnarray}
    && X_{t_{i}} = \begin{cases} (X_{t_{i-1}} + W_{i})_{[a^-,b]} \quad {\rm if} \ X_{t_{i-1}} \in ]a^-,b[ \\
    X_{t_{i-1}} \quad {\rm if} \ X_{t_{i-1}} = a^- \ {\rm or} \ b \end{cases}\, ,
\label{defXtCTRW}
\end{eqnarray}
where the $W_i$ satisfy the same hypotheses as in Sec.~\ref{sec:discreteSiegmund}, and $X_t$ remains fixed between $t_{i-1}$ and $t_i$. In this case, the dual process $Y_t$, (with hard walls at $a$ and $b$) is defined as
\begin{equation}
Y_{\tilde t_{i}} =  (Y_{\tilde t_{i-1}} - \tilde W_{i})_{[a,b]} \;,
\label{defdualCTRW}
\end{equation}
where the $\tilde W_i$'s and the $\tilde t_i$'s follow the same laws as the $W_i$'s and the $t_i$'s. All the results can then be directly transposed from the ones of Sec.~\ref{sec:discreteStatement} (in particular the duality relation \eqref{ExitRW_precise} still holds).

\subsection{Stochastic resetting} \label{sec:SiegmundResetting} 

The last extension of our results that we would like to mention in this chapter is stochastic resetting \cite{resettingPRL,resettingReview, resettingBriefReview}. The idea is to consider a stochastic process $x(t)$ which evolves according to some Langevin equation, but with a certain rate $r$ it is reset to a given position $X_r$. The most-known case is the resetting Brownian motion \cite{resettingPRL}, but many other examples have been studied, such as the resetting RTP \cite{resettingRTP, resettingRTP2,resettingRTP2D}. There are two main interests to these models. First, due to the resetting events which violate detailed balance, they constitute an example of out-of-equilibrium dynamics, which is convenient to study analytically thanks to techniques such as renewal theory and can also be realized in experiments \cite{Besga20, Faisant21, IntermittentReset}. Second, from the point of view of first-passage problems, they provide a way to optimize search processes by tuning the resetting rate $r$ \cite{resettingPRL, resettingRTP, Besga20, Faisant21, resettingInInterval,randomwalkresetPRL}.

One can imagine adding stochastic resetting to all the processes discussed in this chapter, both in the continuous setting \eqref{LangevinIntroduction} and for discrete and continuous time random walks \eqref{defXt} and \eqref{defXtCTRW}. For all these cases, it is possible to extend Siegmund duality to the case where stochastic resetting is present (as long as the time between resets is distributed exponentially). The dual process should however be adapted in a non-trivial way. To get an intuition of how the resetting procedure affects the dual process, it is useful to think about non-instantaneous resetting, which has been studied in the context of experiments \cite{Besga20, Faisant21,IntermittentReset}. The idea is that instead of resetting instantaneously to the position $X_r$, the particle is pushed towards this position by the application of a strong external force during a short period of time. We now need to remember that, in the dual process, the sign of any external force is reversed, which means that the particle will now be pushed away from the position $X_r$ and towards the walls. Going back to instantaneous resetting, this means that, with rate $r$, the dual process $y(t)$ will be reset to $a$ if at the time $t$ of the resetting, $y(t)<X_r$, and at $b$ if $y(t)>X_r$. Note that this duality transformation of the resetting events works both ways, i.e., if a process $y(t)$ with hard walls is subjected to resetting, its dual $x(t)$ with hard walls will be immediately absorbed with rate $r$ at the wall $a$ if $x(t)<X_r$ and at the wall $b$ if $x(t)>X_r$. In Sec.~V of \cite{SiegmundLong} we discuss in detail the extension of Siegmund duality to stochastic resetting. There we extend the proofs for both the continuous and the discrete case and we also show some results of numerical simulations for the resetting Brownian motion.

\section{Discussion} \label{sec:SiegmundDiscussion}

In this chapter, we explored the connection between the first-passage properties of a stochastic process in 1D and the spatial distribution of a related process in the presence of hard walls. This relation has a long history in mathematics, where it is known as Siegmund duality. We provided an explicit and practical formulation of this duality for a variety of models which are relevant for physics, and for which this duality had not been investigated before, including active particles, diffusing diffusivity models and stochastic resetting, both in a continuous and in a discrete setting.

As we see it, the practical applications of this duality relation are twofold. First, from an analytical standpoint, this means that deriving a new result for the first-passage properties of a stochastic process gives us direct information on the behavior of the dual process (which in many cases is very close to the original process) in the presence of hard walls, and vice versa. While both are generally hard to compute analytically, being able to derive one from the other is a significant advantage. Sometimes, one of the two settings might be easier to study than the other, e.g., because a given method or approximation might appear more naturally in one case than in the other (such as the approximation of the hard wall by a steep harmonic potential as done in \cite{hardWallsJoanny, hardWallsCaprini}). Second, from the point of view of numerical simulation, there are situations where a given quantity is clearly simpler to evaluate than its equivalent in the dual process. In particular, in order to compute the exit probability at infinite time directly, we need to run a large number of simulations starting from each position $x\in[a,b]$ and wait for the particle to reach one of the absorbing walls in every one of them. Using the duality, we can instead run a single long simulation of the dual process with hard walls and average over time to obtain the stationary cumulative distribution (provided that the system is ergodic). This could also be useful in experiments, where obtaining a single long time series of data is often simpler \cite{singletraj}. As another example, it can be quite difficult in a simulation to condition the probability density on a given value of ${\bm\theta}$, while computing the exit probability for a given initial value of ${\bm \theta}$ is quite straightforward. The gains provided by this duality are particularly important in the context of active particles, for which both the first-passage properties and the behavior near a hard boundary are of particular relevance.

To conclude, let us discuss a few possible extensions. Let us first mention that, in \cite{SiegmundLong}, we also performed some simulations for the fractional Brownian motion \cite{fBM}. Although this model does not satisfy the assumptions of our Sec.~\ref{sec:continuousSiegmund}, since it is strongly non-Markovian, we find once again that the Siegmund duality seems to apply perfectly (the dual being a fBm with the same Hurst index and simply different boundary conditions). We also stress that our framework requires the driving noise ${\bm \theta}$ to admit a stationary distribution, and it therefore does not apply to processes such as the random acceleration process (where the acceleration $\ddot x(t)$ is a Gaussian white noise). Going beyond our assumptions to include more models such as these two examples could be an interesting future direction. Another important question concerns the extension to higher dimensions, which is particularly relevant for applications. Although abstract formulations of such a duality in more than one dimension have been given in other contexts \cite{DualityMultidimensions}, it remains unclear whether a practical formulation as the one we provided here is possible beyond the 1D case. Finally, another interesting extension would be the case of $N$-particle systems (which is particularly relevant for active particles as we have seen in the rest of this thesis), for which a formulation of Siegmund duality was recently proposed in \cite{Nparticleduality}.

\newpage
\chapter*{\vspace*{-0.4cm}Conclusion and perspectives}
\addcontentsline{toc}{chapter}{Conclusion and perspectives}%

\vspace*{0.85cm}

In this thesis, we have introduced a number of exact methods to study active particle models, in particular in the presence of long-range interactions. We began in Part~\ref{part:context} by giving a broad overview of the existing literature concerning both active particles, with or without interactions, and Riesz gases (i.e., Brownian particles with power law interactions), with a strong emphasis on exact results. In Part~\ref{part:density}, we introduced an exact hydrodynamic description, including noise terms, of the particle density for run-and-tumble particles (RTPs) in one dimension, interacting via a pairwise potential, which generalizes the Dean-Kawasaki equation. We then used this equation to characterize the non-equilibrium stationary states in several models of RTPs with long-range interactions, in particular for 1D Coulomb (active rank diffusion) and 2D Coulomb (active Dyson Brownian motion or active DBM) interaction potentials. Focusing on the limit where the number of particles $N$ is infinitely large, so that the fluctuations vanish, we obtained an exact explicit expression for the stationary density in several examples (in particular for an attractive or repulsive 1D Coulomb interaction, with or without confining potential), revealing unique behaviors, such as new out-of-equilibrium phase transitions. In Part~\ref{part:fluctuations}, we studied the fluctuations at the level of the particle positions for both Brownian and active particles on a circle with a generic power law interaction, obtaining exact expressions for the space-time correlations in the limit of weak noise. This allowed us to shed light on the prevalent role of the activity on both short timescales and small lengthscales in these systems, as well as to derive some new results for Riesz gases of Brownian particles in the process. For some particular cases these results can be extended to particles in a confining potential, revealing the existence of a distinct edge regime with a different scaling of the fluctuations. Finally, in Part~\ref{part:siegmund}, we discussed a surprising relation between absorbing and hard wall boundary conditions for one-dimensional stochastic processes, known in mathematics under the name of Siegmund duality. We proposed an explicit formulation of this relation for a variety of physically relevant stochastic processes in 1D, including active particle models.

This work leaves open a number of interesting questions, some of which we have already mentioned throughout this thesis. Concerning the Dean-Kawasaki equation for RTPs introduced in Chapter~\ref{chap:DeanRTP}, its main limitation in the present state is certainly the fact that it fails in the presence of a single-file constraint, i.e., when the interaction prevents particle crossings. Throughout Part~\ref{part:density}, we have discussed in detail this effect, which originates in the strong local correlations generated by the coexistence of the persistent motion of the active particles with the single-file constraint. Although a fully coarse-grained description involving only the one-point density is probably not possible in such cases, finding a way to derive exact closed hydrodynamic equations for such systems is an exciting direction for future investigation. Another point which might stand out to the reader is the fact that, although the Dean-Kawasaki equation allows to describe the noise in the particle density, this direction has not really been explored in the present thesis, where we have mostly considered the large $N$ limit. This is of course one of the main directions for future works. In particular, the similarity of our equations with the ones derived in \cite{Agranov2021} encourages us to pursue a similar approach to study the fluctuations of the density and of the current of particles, both at the typical level and at the level of large deviations using macroscopic fluctuation theory (MFT) \cite{Agranov2022}. Concerning the method presented in Part~\ref{part:fluctuations} for the computation of the space-time fluctuations, there are also several directions in which our results could be extended, such as taking the expansion further to uncover higher order effects of the active noise and computing the higher order correlations, as well as investigating the edge behavior for a general power law interaction in the absence of boundary conditions. Another interesting question concerns the convergence between the results obtained by this method and those of MFT (the study of the particle current allowing for instance to compute the mean square displacement of a tagged particle \cite{DFRiesz23}). It would be interesting to see if the remarkable agreement that we observed in the Brownian case, even beyond the weak noise limit, still holds for active particles. 

Generally speaking, the extension of all these results beyond the one-dimensional case constitutes another interesting challenge, as it would allow for the study of additional effects which cannot be observed in 1D, such as dislocations in active crystals \cite{James2021,Leticia2022,Chate2023}. Finally, another quantity which is of particular relevance in active particle systems but which we have not at all discussed here is the entropy production rate, which can be seen as a way to quantify the distance to equilibrium. As shown recently in \cite{EntropyAgranov}, MFT could also constitute a good starting point to study this quantity for interacting active particles. Concerning the duality between absorbing and hard wall boundary conditions which we discussed in the last part of this thesis, the extension of our results to higher dimensions, to many particle systems or to other processes which do not satisfy the assumptions of chapter~\ref{chap:Siegmund}, such as the fractional Brownian motion, are all interesting possible directions.

We hope that the work presented in this thesis will spark more interest for the study of long-range interacting active particle systems, as well as for the role of Siegmund duality in statistical physics models, and that it will contribute to the development of new ideas in the ever-growing field of active matter.
\appendix
\part*{Appendix}
\addcontentsline{toc}{chapter}{Appendix}
\renewcommand{\thesection}{\Alph{section}}

\numberwithin{proposition}{section}
\numberwithin{lemma}{section}
\numberwithin{theorem}{section}
\numberwithin{corollary}{section}

\section{Survival probability of a Brownian motion} \label{survivalBM}

In this Appendix, we illustrate the Siegmund duality through a simple example, for which the probability density is known both in the presence of absorbing walls and hard walls. Let us consider a 1D Brownian motion $x(t)$, with diffusion coefficient $T$, on the interval $]-\infty,b]$ with an absorbing boundary condition at $b$, and with initial condition $x(0) =x_0 \in ]-\infty,b]$. We want to compute the survival probability $Q_b(x_0,t)$, i.e., the probability that the particle $x(t)$ stays inside the interval $]-\infty,b[$ up to time $t$. It reads
\begin{equation}
    Q_b(x_0,t) = \int_{-\infty}^b dx\, p_b(x,t|x_0)\, ,
\end{equation}
with $p_b(x,t|x_0)$ the Brownian propagator with an absorbing wall at $x=b$. This propagator can be calculated using the {\it method of images} (see, e.g., \cite{redner}). The idea is to start from a Brownian propagator with initial position $x_0$, and to subtract a second Brownian propagator - the image with respect to the wall - at initial position $2b -x_0$ such that $p_b(b,t|x_0) = 0$. Over time, this image propagator subtracts the mass of the particles absorbed by the wall. This leads to
\begin{equation}
    p_b(x,t|x_0) = \frac{1}{\sqrt{4\pi\, T\, t}}\left(e^{-\frac{(x-x_0)^2}{4T\, t}}-e^{-\frac{(x-(2b-x_0))^2}{4T\, t}}\right)\, .
\end{equation}
Integrating from $-\infty$ to $b$, we obtain the survival probability \cite{redner}
\begin{equation}
    Q_b(x_0,t) = \text{erf}\left(\frac{b - x_0}{\sqrt{4T\, t}}\right)\, .
\end{equation}

Let us now consider instead a hard wall (or reflective) boundary condition at $x=b$. We denote the corresponding process $y(t)$. In this case, the propagator $\tilde p_b(x,t|y_0)$ can again be computed using the method of images, but now one needs to add the image propagator, such that the flux at the wall is zero, $\partial_y \tilde{p}_b(y,t|y_0)|_{y=b} = 0$. This gives
\begin{equation}
    \tilde{p}_b(y,t|b) = \frac{1}{\sqrt{4\pi\, T\, t}}\left(e^{-\frac{(y-y_0)^2}{4T\, t}}+e^{-\frac{(y-(2b-y_0))^2}{4T\, t}}\right) \, .
\end{equation}
We may now notice that, specializing to $y_0=b$ and integrating up to $x_0$, we find
\begin{equation}
  1-  \tilde{\mathbb{P}}(y(t) \leq x_0 |y_0 =b) = 1-\int_{-\infty}^{x_0} dy \, \tilde{p}_b(y,t|b) = \text{erf}\left(\frac{b-x_0}{\sqrt{4T\, t}}\right) = Q_b(x_0,t)\, ,
\end{equation}
which leads to the relation
\begin{equation}
E_b(x_0,t) = 1- Q_b(x_0,t) =\tilde{\mathbb{P}}(y(t) \leq x_0 |y_0 =b)\, .
\end{equation}
This is an instance of the Siegmund duality which is the subject of part \ref{part:siegmund} of this thesis.

\section{Numerical methods} \label{app:simu}

\subsection{Details on the numerical simulations}

The numerical simulations for the active ranked diffusion and the active DBM were performed by integrating the stochastic differential equation using a simple Euler scheme. To compute the densities in the stationary state, we run a simulation long enough so that the system reaches stationarity, and then build the histogram of positions (after subtracting the position of the center of mass if there is no confining potential) over a large time window. For both models we use a time step $dt=0.001$ and run the dynamics for $10^7$ to $10^9$ steps, depending on the value of $N$. To look at the escaping particles in the expanding phases of the active rank diffusion, we instead run the dynamics $10^5$ times and build a distinct histogram for each time step. In this case we use $dt=0.01$.

For the active rank diffusion, we took advantage of the fact that the total interaction force acting on a given particle only depends on its rank to optimize the simulations, which allows us to access larger values of $N$. We keep track of the rank of each particle if they were ordered by position (taking into account the fact that some can have the same position if a cluster forms), and then update simultaneously the positions of all particles at each time step according to
\be
x_i(t+dt) = x_i(t) + dt \left(\frac{\bar \kappa}{N}(N_i^{\rm right}(t)-N_i^{\rm left}(t)) -V'(x_i(t))+v_0 \sigma_i(t) \right) 
\ee
where $N_i^{\rm right}(t)$ (resp. $N_i^{\rm left}(t)$) is the number of particles strictly at the right (resp. left) of particle $i$ at time $t$, and the $\sigma_i$'s switch sign independently with probability $\gamma dt$ at each time step. In the presence of a linear external potential, some additional care should be taken to simulate the dynamics near $x=0$. For details, see Sec. II of the SM in \cite{activeRD1}.

\subsection{Algorithm for the limit $g\to0^+$ of model II}

In this appendix, we reproduce the discrete-time algorithm given in Sec.~V.C of the SM of \cite{ADBM1}, which allows to simulate the limit $g\to 0^+$ of model II of the active DBM (discussed in Chapter~\ref{chap:ADBM_Dean}), where the logarithmic interaction is replaced by a hard-core repulsion between particles. In this limiting model, we consider that particles which collide form a point-like cluster for which we will describe the dynamics below. It is more convenient to define the whole system as a set of clusters (possibly containing a single particle), each of them characterized by :
\begin{itemize}[noitemsep,nolistsep]
    \item its position $x$,
    \item the vector $\vec{\sigma}$ containing the spins of all the particles in the cluster, ordered from left to right.
\end{itemize}
Between two collisions, such a cluster follows the equation of motion (obtained by summing the equations of motion for all the particles in the cluster, which have the same position, and dividing by the size $n$ of the cluster) :
\beq
\dot{x} = - \lambda x + \frac{v_0}{n} \sum_i \sigma_i \;.
\eeq
The parameters of the algorithm are the number of particles $N$, the tumbling rate $\gamma$, the driving velocity $v_0$ and the strength of the harmonic potential $\lambda$ (which can both be set to $1$ without loss of generality) and the time-step $dt$. For infinitely small time-steps $dt$, the algorithm below should coincide exactly with the limit $g\to 0^+$ of model II (see Fig. \ref{g0figs} right panel for a numerical check of this statement).
\\

{\bf Main algorithm:}

\noindent We start with only clusters of size $1$ distributed uniformly on the interval $[-v_0/\lambda,v_0/\lambda]$ (for $g\to0^+$ the particles are confined to this interval), which we sort according to their position ($x_k<x_{k+1} \ \forall k$). Then at each time-step we perform the following steps in order :
\begin{itemize}[noitemsep,nolistsep]
    \item Flip each spin independently with probability $\gamma dt$.
    \item For each cluster containing more than one particle for which at least one spin has flipped, determine if it breaks into several clusters (see below).
    \item For each cluster compute its speed $v=- \lambda x + \frac{v_0}{n} \sum_i \sigma_i$ and its new position $x^{new}=x+v dt$.
    \item For $k=1...N_{clusters}$: if $x_k^{new}>x_{k+1}^{new}$ (it means there has been a collision), then:
    \begin{itemize}
        \item compute the collision time $dt_{col}=\frac{x_{k+1}-x_{k}}{v_{k}-v_{k+1}}$,
        \item compute the position of the collision $x'=x_k+v_k dt_{col}$,
        \item create a new cluster with $\vec{\sigma}$ being the concatenation of the $\vec{\sigma}$'s of the two clusters and :
        \begin{itemize}
            \item $v=-\lambda x'+\frac{v_0}{n} \sum_i \sigma_i$,
            \item $x^{new}=x'+v(dt-dt_{col})$,
            \item $x=x'-v dt_{col}$ (position of the cluster before the update if it had already existed - useful in case of a new collision at the same step).
        \end{itemize}
    \end{itemize}
    \item Repeat the previous step until there are no collisions.
    \item Update the position of all clusters $x\leftarrow x^{new}$.
\end{itemize}

\

To determine if a cluster breaks, the idea is that a cluster breaks if it can be divided into smaller clusters which have individual speeds driving them apart from each other. The most natural way to do this is the following:
\\

{\bf Cluster decomposition 1:}
\begin{itemize}[noitemsep,nolistsep]
    \item Decompose the cluster into a list of mini-clusters for which all $+$ particles are on the left and all $-$ particles on the right.
    \item Compute for each mini-cluster $k$ its number of particles $n_k$ and its average spin $\bar{\sigma}_k=\frac{\sum_i \sigma_i}{n_k}$
    \item For $k=1...N_{miniclusters}$: if $\bar{\sigma}_k>\bar{\sigma}_{k+1}$, merge mini-clusters $k$ and $k+1$.
    \item Repeat until the $\bar{\sigma_k}$'s are ordered.
\end{itemize}

\

Another way to obtain the same decomposition is to use the following characterization: a list of spins $\vec \sigma$ of size $n$ forms a cluster iff the running average 
$S_k=\frac{1}{k} \sum_{i=1}^k \sigma_i$ reaches its minimum at the end of the list, i.e., $k_{\min}=n$. Indeed in this case, if we try to divide the list in two at any position, the part on the left will have a larger speed than the part on the right, so they will form a cluster. If the global minimum is reached before, i.e., if $k_{\min}<n$, and if we cut the list just after this minimum, the part on the left will have a smaller speed than the part on the right, hence will be a genuine cluster, and the two parts will separate. One then repeats the operation for the part on the right. We can therefore apply the following algorithm instead of the one above:
\\

{\bf Cluster decomposition 2:}
\begin{itemize}[noitemsep,nolistsep]
    \item Compute the running average at each position.
    \item Find the global minimum.
    \item The part on the left of this minimum (including the minimum) forms an independent cluster.
    \item Repeat the process removing the independent cluster from the computation of the running average, until the global minimum is at the end of the list.
\end{itemize}

\

The second method is faster than the first one if the clusters do not decompose too often (i.e., when $\gamma$ is not too large), as we have to go through the list only once if there is a single cluster (which is not the case with the first algorithm).

As for model I and II, the quantities of interest (here the particle density and the distribution of cluster sizes) are computed by running the dynamics described above and averaging over a large time window. With this algorithm, the simulations are much faster than the simulations of model II for small non-zero values of $g$ using the Langevin dynamics, since we can use larger time-steps (and if there are a lot of large clusters, e.g., for small $\gamma$, there are few positions to update), allowing to access larger values of $N$ more easily. As shown in Fig.~\ref{g0figs} (right panel), the total particle density in model II seems to correctly converge to the density in this effective model as $g\to 0^+$.
\\

{\bf Limit $\gamma\to0^+$:} \nopagebreak

\noindent Finally, we also studied the double limit $g\to0^+$ and $\gamma\to0^+$. In this case the results were simply obtained by drawing a random list of $N$ independent spins (each equal to $\pm1$ with equal probability), each one being $1$ or $-1$ with equal probability, and decomposing this list into clusters as described above. Since in this case there are no spin flips, each cluster simply converges to a stationary position given by $x_{eq} = \frac{v_0}{n\lambda} \sum_i \sigma_i$, where the sum is over the spins in the cluster of size $n$. By definition of the clusters, these positions form a strictly increasing sequence.
The quantities of interest are then obtained by averaging over a large number of such realizations. The idea is that for very small $\gamma$, the clusters will spend a lot of time near their equilibrium positions before a tumbling occurs. The results obtained by this method are very close to what we obtain for small values of $\gamma$ using the dynamics described above, suggesting that the $\gamma \to 0$ limit is well defined for the $g\to0^+$ model.

\section{Useful integrals and identities for the gamma function} \label{app:integrals}

We recall here some known integrals and identities for the Gamma function which are useful in Chapters \ref{chap:passiveRieszFluct} and \ref{chap:activeRieszFluct}. For any $1<\alpha\leq 2$,
\be
\int_0^{+\infty} du \frac{\sin^2(\pi u)}{u^{\alpha}} = \frac{\pi^{\alpha-\frac{1}{2}}}{2(\alpha-1)} \frac{\Gamma(\frac{3-\alpha}{2})}{\Gamma(\frac{\alpha}{2})} = -2^{\alpha-2} \pi ^{\alpha-1} \sin \left(\frac{\pi  \alpha}{2}\right) \Gamma (1-\alpha) = -\frac{2^{\alpha-3} \pi^\alpha}{\cos(\frac{\pi\alpha}{2})\Gamma(\alpha)} \;.
\ee
This leads to, for $0<s<1$,
\be
a_s = 2\pi^{s+\frac{3}{2}} \frac{\Gamma(\frac{1-s}{2})}{\Gamma(1+\frac{s}{2})} = -2^{2+s}\pi^{s+1} \sin(\frac{\pi s}{2}) \Gamma(-s) = \frac{2^{1+s} \pi^{2+s}}{\cos(\frac{\pi s}{2})\Gamma(1+s)} \;.
\ee
We also have, for $\alpha>1$,
\be
\int_0^{+\infty} dv \frac{1-e^{-v^{\alpha}}}{v^{\alpha}} = \frac{\Gamma(1/\alpha)}{\alpha-1}
\quad \text{and} \quad \int_0^{+\infty} dv \left(\frac{1 - e^{-v^{\alpha}}}{v^{\alpha}}\right)^2 = \frac{2(2^{\frac{\alpha-1}{\alpha}}-1)}{\alpha} \Gamma\left(-\frac{2\alpha-1}{\alpha}\right) \;,
\ee
as well as
\be
\int_0^{+\infty} \frac{dv}{1+v^{\alpha}} =\Gamma(\frac{\alpha-1}{\alpha}) \Gamma(\frac{\alpha+1}{\alpha}) = \frac{\pi}{\alpha \sin(\frac{\pi}{\alpha})} \;.
\ee

\section{Derivation based on the Fokker-Planck equation} \label{app:ProofSiegmund}

In this appendix, we reproduce the derivation from \cite{SiegmundLong} of the duality relation \eqref{mainRelation} between $E_b(x,\bm{\theta},t)$ and $\tilde \Phi(x,t|\bm{\theta};b)$, for the continuous stochastic process defined in Sec.~\ref{sec:continuousSiegmund}. We recall that it is defined through the SDE
\begin{equation}
    \dot{x}(t) = f\left(x(t),\bm{\theta}(t)\right) + \sqrt{2\mathcal{T}\left(x(t),\bm{\theta}(t)\right)}\, \xi(t)\, ,
\label{SDEgeneral}
\end{equation}
where $\xi(t)$ is Gaussian white noise with zero mean and unit variance. $\bm{\theta}(t)$ is a vector of parameters which follows a stochastic evolution independent of $x$, of the form 
\begin{equation}
    \bm{\dot{\theta}}(t) = \bm{g}\left(\bm{\theta}(t)\right) + \left[2\underline{\mathcal{D}}(\bm{\theta}(t))\right]^{1/2} \cdot \bm{\eta}(t)\, ,
\label{SDEtheta}
\end{equation}
where $\underline{\mathcal{D}}$ is a positive matrix and the $\eta_i(t)$'s are independent Gaussian white noises with zero mean and unit variance. In addition, it can jump from the value $\bm{\theta}$ to $\bm{\theta}'$ with a transition kernel $\mathcal{W}(\bm{\theta}'|\bm{\theta})$.

The Fokker-Planck equation for the joint probability density $P(x,\bm{\theta},t)$ reads (with the It\=o convention)
\bea \label{FPgeneral}
\partial_t P = && \hspace{-0.6cm} - \partial_x [f(x,\bm{\theta}) P] + \partial_x^2 [\mathcal{T}(x,\bm{\theta}) P] - \sum_i \partial_{\theta_i}[g_i(\bm{\theta}) P] + \sum_{i,j} \partial_{\theta_i\theta_j}^2 [\mathcal{D}_{ij}(\bm{\theta}) P] \\
&& \hspace{-0.5cm} + \int d\bm{\theta}' \left[\mathcal{W}(\bm{\theta}|\bm{\theta}')P(x,\bm{\theta}',t) - \mathcal{W}(\bm{\theta}'|\bm{\theta})P(x,\bm{\theta},t)\right] \,. \nn
\eea
We will denote $P_{\bm{\theta}}(x,t)=P(x,t|\bm{\theta})$ the probability density of the positions conditioned on $\bm{\theta}$ and $p(\bm{\theta},t)$ the density probability of $\bm{\theta}$, such that $P(x,\bm{\theta},t) = p(\bm{\theta},t) P_{\bm{\theta}}(x,t)$. Integrating \eqref{FPgeneral} over $x$ we obtain the Fokker-Planck equation for $p(\bm{\theta},t)$ (we will consider densities with a finite support $[a,b]$, thus the boundary terms vanish)
\begin{equation}
\partial_t p = - \sum_i \partial_{\theta_i}[g_i(\bm{\theta}) p] + \sum_{i,j} \partial_{\theta_i\theta_j}^2 [\mathcal{D}_{ij}(\bm{\theta}) p] + \int d\bm{\theta}' \, [\mathcal{W}(\bm{\theta}|\bm{\theta}')p(\bm{\theta}',t) - \mathcal{W}(\bm{\theta}'|\bm{\theta})p(\bm{\theta},t)] \, .
\label{FPtheta}
\end{equation}
Note that we can also derive Eq.~(\ref{FPtheta}) directly from Eq.~(\ref{SDEtheta}).
We now assume that equation \eqref{FPtheta} has an equilibrium solution $p_{eq}(\bm{\theta})$, i.e., a stationary solution which satisfies the local detailed balance conditions
\begin{eqnarray}
&&0 = - g_i(\bm{\theta}) p_{eq}(\bm{\theta}) + \sum_{j} \partial_{\theta_j}[\mathcal{D}_{ij}(\bm{\theta}) p_{eq}(\bm{\theta})] \,, \, \forall \ i,\label{detailed_balance1_proof_continuous} \\
&& \mathcal{W}(\bm{\theta}|\bm{\theta}')p_{eq}(\bm{\theta}') = \mathcal{W}(\bm{\theta}'|\bm{\theta})p_{eq}(\bm{\theta}) \,. 
\label{detailed_balance2_proof_continuous}
\end{eqnarray}

Let us first consider the dynamics \eqref{SDEgeneral}-\eqref{SDEtheta} on an interval $[a,b]$ with absorbing boundary conditions at $x=a$ and $x=b$. In this section we assume that the whole interval is accessible to the particle\footnote{This may not be the case, e.g., when the noise terms have a finite amplitude (for instance in the case of RTPs) and the external force is too strong. See \cite{SiegmundShort} to see how this can be dealt with in the particular case of RTPs at infinite times.}. We are interested in the probability that a particle starting at position $x$ at time $t=0$, with a certain initialization value of $\bm{\theta}$, is absorbed at $x=b$ before time $t$, denoted $E_b(x,\bm{\theta},t)$. Since the joint process $(x,\bm{\theta})$ is Markovian, one has
\begin{eqnarray}
E_b(x,\bm{\theta},t+dt) = && \hspace{-0.6cm} \mathbb{E}_{\xi,\bm \eta}\Big[E_b\big(x+dt[f(x,\bm{\theta})+\sqrt{2\mathcal{T}(x,\bm{\theta})} \, \xi(t)],\bm{\theta}+dt[\bm{g}(\bm{\theta}) + (2\underline{\mathcal{D}}(\bm{\theta}))^{1/2} \cdot \bm{\eta}(t)],t\big)\Big] \nonumber \\
&& \hspace{-0.5cm} +\, dt \int d\bm{\theta}' \, \mathcal{W}(\bm{\theta}'|\bm{\theta})[E_b(x,\bm{\theta}',t) - E_b(x,\bm{\theta},t)] \;,
\end{eqnarray}
which leads to the backward Fokker-Planck equation for $E_b(x,\bm{\theta},t)$,
\bea \label{backwardFPgeneral}
\partial_t E_b = && \hspace{-0.6cm} f(x,\bm{\theta}) \partial_x E_b + \mathcal{T}(x,\bm{\theta}) \partial_x^2 E_b + \sum_i g_i(\bm{\theta}) \partial_{\theta_i} E_b + \sum_{i,j} \mathcal{D}_{ij}(\bm{\theta}) \partial_{\theta_i\theta_j}^2 E_b \\
&& \hspace{-0.5cm}+ \int d\bm{\theta}'\, \mathcal{W}(\bm{\theta}'|\bm{\theta})[E_b(x,\bm{\theta}',t) - E_b(x,\bm{\theta},t)] \;, \nn
\eea
which is complemented by the boundary conditions\footnote{In Eq.~(\ref{exit_bc}), by ``$\mathcal{T}(x,\bm{\theta})>0 \text{ in the vicinity of } a$'' we mean that there does not exist $\epsilon>0$ such that $\mathcal{T}(x,\bm{\theta})=0$ for every $x\in[a,a+\epsilon]$.} 
\begin{eqnarray} \label{exit_bc}
&&E_b(a^+, \bm{\theta},t) = 0 \quad \text{for all } f(a^+,\bm{\theta})<0, \text{ or if } \mathcal{T}(x,\bm{\theta})>0 \text{ in the vicinity of } a, \\
&&E_b(b^-, \bm{\theta},t) = 1 \quad \text{for all } f(b^-,\bm{\theta})>0, \text{ or if } \mathcal{T}(x,\bm{\theta})>0 \text{ in the vicinity of } b, \nonumber
\end{eqnarray}
and the initial conditions
\begin{eqnarray}\label{exit_ic}
&&E_b(x,\bm{\theta},0) = 0 \text{ for } x<b, {\color{red}}\\
&&E_b(b,\bm{\theta},0) = 1. \nonumber
\end{eqnarray}
The boundary conditions \eqref{exit_bc} require some explanation. Although the values of $E_b(x,\bm{\theta},t)$ exactly at $x=a$ and $x=b$ are fixed by the absorbing conditions, there can be discontinuities in some cases. Indeed if $\mathcal{T}(x,\bm{\theta})=0$ and the force $f(x,\bm{\theta})$ is driving the particle away from the wall at time $t=0$, then a particle starting at an infinitesimal distance from the wall will not be absorbed immediately (in fact it may even escape and reach the opposite wall). However if the force is driving the particle towards the wall, then it will be absorbed with probability 1 and there will be no discontinuity. Additionally, if a Brownian term is present, i.e., $\mathcal{T}(x,\bm{\theta})>0$, then there can be no discontinuity independently of the sign of $f$. Indeed at infinitely small times, only the Brownian term is relevant in this case. Since a Brownian motion always goes back to its starting point infinitely many times before moving away, the particle will be absorbed.

Let us now consider the probability density in the presence of hard walls at $x=a$ and $x=b$. In addition, we replace the force $f(x,{\bm \theta})$ by some $\tilde f(x,{\bm \theta})$. All the probabilities and probability densities related to this new process will be denoted with a tilde (however the distribution $p_{eq}({\bm \theta})$ remains the same). In this case, if $\mathcal{T}=0$, the density may have some delta peaks at $x=a$ and $x=b$ (see examples in section \ref{sec:continuousSiegmund}). Here we assume that at $t=0$ the parameter $\bm{\theta}$ is initialized in its equilibrium distribution $p_{eq}(\bm{\theta})$ (and thus it keeps the same distribution at all times). Starting from \eqref{FPgeneral}, we derive an equation for the conditional density $\tilde P_{\bm{\theta}}(x,t)$ (using $\tilde P(x,\bm{\theta},t) = p_{eq}(\bm{\theta}) \tilde P_{\bm{\theta}}(x,t)$),
\begin{eqnarray}
p_{eq}(\bm{\theta}) \partial_t \tilde P_{\bm{\theta}} &=& - p_{eq}(\bm{\theta}) \partial_x \left[\tilde f(x,\bm{\theta}) \tilde P_{\bm{\theta}}\right] + p_{eq}(\bm{\theta}) \partial_x^2 [\mathcal{T}(x,\bm{\theta}) \tilde P_{\bm{\theta}}] \\
&& + \tilde P_{\bm{\theta}} \sum_i \partial_{\theta_i}\Big\{-g_i(\bm{\theta}) p_{eq}(\bm{\theta}) + \sum_j \partial_{\theta_j}[\mathcal{D}_{ij}(\bm{\theta}) p_{eq}(\bm{\theta})]\Big\} \nonumber \\
&& + \sum_{i} \Big\{-g_i(\bm{\theta}) p_{eq}(\bm{\theta}) + 2 \sum_j \partial_{\theta_j} [\mathcal{D}_{ij}(\bm{\theta}) p_{eq}(\bm{\theta})]\Big\} \partial_{\theta_i} \tilde P_{\bm{\theta}} + \sum_{i,j} \mathcal{D}_{ij}(\bm{\theta}) p_{eq}(\bm{\theta}) \partial_{\theta_i\theta_j}^2 \tilde P_{\bm{\theta}} \nonumber \\
&&+ \int d\bm{\theta}' \left[\mathcal{W}(\bm{\theta}|\bm{\theta}') p_{eq}(\bm{\theta}') \tilde P_{\bm{\theta}'} - \mathcal{W}(\bm{\theta}'|\bm{\theta}) p_{eq}(\bm{\theta}) \tilde P_{\bm{\theta}}\right] \nonumber \\
&=&  p_{eq}(\bm{\theta}) \Big\{  -\partial_x [\tilde f(x,\bm{\theta}) \tilde P_{\bm{\theta}}] + \partial_x^2 [\mathcal{T}(x,\bm{\theta}) \tilde P_{\bm{\theta}}] + \sum_{i} g_i(\bm{\theta}) \partial_{\theta_i} \tilde P_{\bm{\theta}} + \sum_{i,j} \mathcal{D}_{ij}(\bm{\theta}) \partial_{\theta_i\theta_j}^2 \tilde P_{\bm{\theta}} \nonumber \\
&&+ \int d\bm{\theta}' \mathcal{W}(\bm{\theta}'|\bm{\theta}) [\tilde P_{\bm{\theta}'} - \tilde P_{\bm{\theta}}] \Big\} \nonumber
\end{eqnarray}
where we have made an extensive use of the detailed balance conditions \eqref{detailed_balance1_proof_continuous} and \eqref{detailed_balance2_proof_continuous} to obtain the second identity. We can then eliminate $p_{eq}(\bm{\theta})$ to obtain
\begin{equation}
\partial_t \tilde P_{\bm{\theta}} = -\partial_x [\tilde f(x,\bm{\theta}) \tilde P_{\bm{\theta}}] + \partial_x^2 [\mathcal{T}(x,\bm{\theta}) \tilde P_{\bm{\theta}}] + \sum_{i} g_i(\bm{\theta}) \partial_{\theta_i} \tilde P_{\bm{\theta}} + \sum_{i,j} \mathcal{D}_{ij}(\bm{\theta}) \partial_{\theta_i\theta_j}^2 \tilde P_{\bm{\theta}} + \int d\bm{\theta}' \mathcal{W}(\bm{\theta}'|\bm{\theta}) [\tilde P_{\bm{\theta}'} - \tilde P_{\bm{\theta}}] \;.
\label{FPconditional}
\end{equation}

We now introduce the cumulative distribution of the conditional density $\tilde P_{\bm{\theta}}$,
\begin{equation}
    \tilde \Phi(x, t|\bm{\theta}) = \int_{a^-}^x dy \, \tilde P_{\bm{\theta}}(y,t) = \int_{a^-}^x dy \, \tilde P(y,t|\bm{\theta}) \quad , \quad \tilde P_{\bm{\theta}}(x,t) = \partial_x \tilde \Phi(x, \bm{\theta}, t)
\end{equation}
where the integral starts at $a^-$ to include a potential delta peak at $x=a$. Writing \eqref{FPconditional} in terms of $\tilde \Phi$ yields
\bea
\partial_x \partial_t \tilde \Phi = && \hspace{-0.6cm} \partial_x \Big\{-\tilde f(x,\bm{\theta}) \partial_x \tilde \Phi + \partial_x [\mathcal{T}(x,\bm{\theta}) \partial_x \tilde \Phi] + \sum_{i} g_i(\bm{\theta}) \partial_{\theta_i} \tilde \Phi + \sum_{i,j} \mathcal{D}_{ij}(\bm{\theta}) \partial_{\theta_i\theta_j}^2 \tilde \Phi \nn \\
&& \hspace{-0.5cm} + \int d\bm{\theta}' \mathcal{W}(\bm{\theta}'|\bm{\theta}) [\tilde \Phi(x, t|\bm{\theta}') - \tilde \Phi(x, t|\bm{\theta})] \Big\} \;.  \label{preFPphi}
\eea
We can then integrate this equation between $x=-\infty$ and $x$, using that $\tilde \Phi(-\infty,t|\bm{\theta})=0$ as well as its derivatives. Using that $\partial_x [\mathcal{T}(x,\bm{\theta}) \partial_x \tilde \Phi] = \partial_x \mathcal{T}(x,\bm{\theta}) \partial_x \tilde \Phi + \mathcal{T}(x,\bm{\theta}) \partial_x^2 \tilde \Phi$, this finally yields the differential equation satisfied by $\tilde \Phi(x, t|\bm{\theta})$,
\bea
\partial_t \tilde \Phi = && \hspace{-0.6cm} [-\tilde f(x,\bm{\theta}) + \partial_x \mathcal{T}(x,\bm{\theta})] \partial_x \tilde \Phi + \mathcal{T}(x,\bm{\theta}) \partial_x^2 \tilde \Phi + \sum_{i} g_i(\bm{\theta}) \partial_{\theta_i} \tilde \Phi + \sum_{i,j} \mathcal{D}_{ij}(\bm{\theta}) \partial_{\theta_i\theta_j}^2 \tilde \Phi \nn \\
&& \hspace{-0.5cm}+ \int d\bm{\theta}' \mathcal{W}(\bm{\theta}'|\bm{\theta}) [\tilde \Phi(x, t|\bm{\theta}') - \tilde \Phi(x, t|\bm{\theta})] \;.
\label{FPphi}
\eea

Let us now fix $\tilde f(x,{\bm \theta}) = -f(x,{\bm \theta}) + \partial_x \mathcal{T}(x,\bm{\theta})$. Then this is exactly the same as the equation \eqref{backwardFPgeneral} for the exit probability at $E_b(x,\bm{\theta},t)$. Note that to go from Eq.~(\ref{FPconditional}) to Eq.~(\ref{FPphi}), it is essential that $\bm g$ and $\underline{\mathcal{D}}$ do not depend on $x$. The boundary conditions are
\begin{eqnarray}\label{cumul_bc}
&&\tilde \Phi(a^+,t|\bm{\theta}) = 0 \quad \text{for all } \tilde f(a^+,\bm{\theta})>0, \text{ or if } \mathcal{T}(x,\bm{\theta})>0 \text{ in the vicinity of } a, \\
&&\tilde \Phi(b^-,t|\bm{\theta}) = 1 \quad \text{for all } \tilde f(b^-,\bm{\theta})<0, \text{ or if } \mathcal{T}(x,\bm{\theta})>0 \text{ in the vicinity of } b, \nonumber
\end{eqnarray}
which are also the same as \eqref{exit_bc} when writing $\tilde f(x,{\bm \theta}) = -f(x,{\bm \theta}) + \partial_x \mathcal{T}(x,\bm{\theta})=0$ (indeed if $\mathcal{T}(x,\bm{\theta})=0$ near the wall then one simply has $\tilde f(x,{\bm \theta}) = -f(x,{\bm \theta})$ in this region). These conditions translate the fact that there can be an accumulation of particles at the boundaries (i.e., a delta in the density at $a$ or $b$), but only if the velocity is oriented towards the wall, and if there is no Brownian noise. Finally, one can choose an initial condition which matches \eqref{exit_ic} 
by assuming that at $t=0$ all particles are at $x=b$,
\begin{eqnarray}\label{cumul_ic}
&& \tilde \Phi(x,0|\bm{\theta};b) = 0 \text{ for } x<b, \\
&& \tilde \Phi(b,0|\bm{\theta};b) = 1 \nonumber
\end{eqnarray}
(remember that $\bm{\theta}$ is initialised at equilibrium). With this choice of initial condition, the two quantities $E_b(x,\bm{\theta},t)$ and $\tilde \Phi(x, t|\bm{\theta};b)$ 
follow the same differential equation with the same boundary and initial condition. One can thus reasonably assume that\footnote{In theory, one would need to show the unicity of the solution of the PDE with these initial and boundary conditions. Here we choose to leave aside these considerations.}
\begin{equation}
    E_b(x,\bm{\theta},t) = \tilde \Phi(x, t|\bm{\theta};b) \;.
\label{identity_general}
\end{equation}

In Appendix~A of \cite{SiegmundLong}, using the same derivation procedure, we show the more general result
\begin{equation}
     \mathbb{P}(x(t)\geq y|x,{\bm \theta}) = \tilde{\mathbb{P}}(y(t)\leq x|{\bm \theta}(t)=-{\bm \theta};y,{\bm \theta}(0)^{eq})  \, ,
\end{equation}
from which \eqref{Siegmund} can be deduced by averaging over $p_{eq}({\bm \theta})$. This relates the cumulative distribution of the process with absorbing wall with the cumulative of its dual initialized at position $y$. It yields back \eqref{identity_general} when taking $y=b$.

\renewcommand\bibname{Bibliography}

\newpage{\pagestyle{empty}\cleardoublepage}


\begin{thebibliography}{99}
\setcounter{enumiv}{\value{firstbib}}

\bibitem{ADBM1}
L. Touzo, P. Le Doussal, G. Schehr, {\it Interacting, running and tumbling: the active Dyson Brownian motion}, \doidoi{10.1209/0295-5075/acdabb}{EPL {\bf 142}, 61004 (2023)} and \href{https://arxiv.org/abs/2302.02937}{arXiv:2302.02937}.

\bibitem{activeRD1} L. Touzo, P. Le Doussal, {\it Non-equilibrium phase transitions in active rank diffusions}, \doidoi{10.1209/0295-5075/ad222b}{EPL {\bf 145}, 41001 (2024)} and \href{https://arxiv.org/abs/2308.06118}{arXiv:2308.06118}.

\bibitem{activeRD2} L. Touzo, P. Le Doussal, {\it Run-and-tumble particles with 1D Coulomb interaction: the active jellium model and the non-reciprocal self-gravitating gas}, \doidoi{10.1088/1742-5468/ade86d}{J. Stat. Mech. (2025) 073204} and \href{https://arxiv.org/abs/2502.09466}{arXiv:2502.09466}.

\bibitem{ADBM2}
L. Touzo, P. Le Doussal, G. Schehr, {\it Fluctuations in the active Dyson Brownian motion and the overdamped Calogero-Moser model},
\doidoi{10.1103/PhysRevE.109.014136}{Phys. Rev. E {\bf 109}, 014136 (2024)} and \href{https://arxiv.org/abs/2307.14306}{arXiv:2307.14306}.

\bibitem{RieszFluct} L. Touzo, P. Le Doussal, G. Schehr, {\it Spatio-temporal fluctuations in the passive and active Riesz gas on the circle}, \doidoi{10.1007/s10955-025-03452-7}{J. Stat. Phys. {\bf 192}, 79 (2025)} and \href{https://arxiv.org/abs/2411.01355}{arXiv:2411.01355}.

\bibitem{activeCM}
S. Santra, L. Touzo, C. Dasgupta, A. Dhar, S. Dutta, A. Kundu, P. Le Doussal, G. Schehr, P. Singh, {\it Crystal to liquid cross-over for active particles with inverse-square power-law interaction}, \doidoi{10.1088/1742-5468/adbb5d}{J. Stat. Mech. (2025) 033203} and \href{https://arxiv.org/abs/2411.13478}{arXiv:2411.13478}.


\bibitem{SiegmundShort} M. Guéneau, L. Touzo, {\em Relating absorbing and hard wall boundary conditions for active particles}, \doidoi{10.1088/1751-8121/ad4753}{J. Phys. A: Math. Theor. {\bf 57} 225005 (2024)} and \href{https://arxiv.org/abs/2312.13200}{arXiv:2312.13200}.

\bibitem{SiegmundLong} M. Guéneau, L. Touzo, {\em Siegmund duality for physicists: a bridge between spatial and first-passage properties of continuous- and discrete-time stochastic processes}, \doidoi{10.1088/1742-5468/ad6134}{J. Stat. Mech. (2024) 083208} and \href{https://arxiv.org/abs/2404.10537}{arXiv:2404.10537}.


\setcounter{firstbib}{\value{enumiv}}
\end{thebibliography}

\begin{thebibliography}{100}

\setcounter{enumiv}{8}




\bibitem{Ramaswamy2010} S. Ramaswamy, {\it The mechanics and statics of active matter}, Annu. Rev. Condens. Matter Phys. {\bf 1}, 323–345 (2010).

\bibitem{Bechinger}
C. Bechinger, R. Di Leonardo, H. L\"owen, C. Reichhardt, G. Volpe, G. Volpe, {\it Active particles in complex and crowded environments}, Rev. Mod. Phys. {\bf 88}, 045006 (2016).


\bibitem{Marchetti2018} \'{E}. Fodor, M. Cristina Marchetti, {\em The statistical physics of active matter: From self-catalytic colloids to living cells}, Physica A: Statistical Mechanics and its Applications {\bf 504}, 106-120 (2018). 

\bibitem{Vicsek} T. Vicsek, A. Czirók, E. Ben-Jacob, I. Cohen, O. Shochet, {\it Novel Type of Phase Transition in a System of Self-Driven Particles}, Phys. Rev. Lett. {\bf 75}, 1226 (1995).


\bibitem{CavagnaGiardina} A. Cavagna, I. Giardina, {\it Bird Flocks as Condensed Matter}, Annu. Rev. Condens. Matter Phys. {\bf 5}, 183 (2014).

\bibitem{Calovi2018} D. S. Calovi, A. Litchinko, V. Lecheval, U. Lopez, A. P\'erez Escudero, H. Chat\'e, C. Sire, G. Theraulaz, {\it Disentangling and modeling interactions in fish with burst-and-coast swimming reveal distinct alignment and attraction behaviors}, PLoS Comput. Biol. {\bf 14}(1): e1005933 (2018).


\bibitem{Berg2004} H. C. Berg, {\it E. Coli in Motion}, (Springer Verlag, Heidelberg, Germany) (2004).

\bibitem{ErbeExperiments} A. Erbe, M. Zientara, L. Baraban, C. Kreidler, P Leiderer, {\it Various driving mechanisms for generating motion of colloidal particles}, J. Phys.: Condens. Matter {\bf 20}, 404215 (2008).


\bibitem{JanusReview} A. Walther, A. H. E. Müller, {\it Janus particles: Synthesis,
self-assembly, physical properties, and applications}, Chem. Rev. {\bf 113}, 5194–5261 (2013).

\bibitem{Quincke} G. E. Pradillo, H. Karanib, P. M. Vlahovska, {\it Quincke rotor dynamics in confinement: rolling and hovering}, Soft Matter {\bf 15}, 6564 (2019). 

\bibitem{Chate2010} J. Deseigne, O. Dauchot, H. Chaté, {\it Collective Motion of Vibrated Polar Disks}, Phys. Rev. Lett. {\bf 105}, 098001 (2010).


\bibitem{Schranz}
M. Schranz, M. Umlauft, M. Sende, W. Elmenreich, {\it Swarm robotic behaviors and current applications}, Frontiers in Robotics and AI {\bf 7}, 36 (2020).



\bibitem{Yang2014} X. Yang, M. L. Manning, M. C. Marchetti, {\em Aggregation and segregation of confined active particles}, Soft Matter, {\bf 10}, 6477 (2014).

\bibitem{Uspal2015} W. E. Uspal, M. N. Popescu, S. Dietrich, M. Tasinkevych, {\em Self-propulsion of a catalytically active particle near a planar wall: from reflection to sliding and hovering}, Soft Matter {\bf 11}, 434 (2015).


\bibitem{TailleurCates2009} J. Tailleur, M. E. Cates, {\it Sedimentation, trapping, and
rectification of dilute bacteria}, EPL {\bf 86}, 60002 (2009).

\bibitem{DKM19} 
A. Dhar, A. Kundu, S. N. Majumdar, S. Sabhapandit, G. Schehr, {\it Run-and-tumble particle in one-dimensional confining potentials: Steady-state, relaxation, and first-passage properties}, Phys. Rev. E {\bf 99}, 032132 (2019).


\bibitem{LMS2020} 
P. Le Doussal, S. N. Majumdar, G. Schehr, {\it Velocity and diffusion constant of an active particle in a one-dimensional force field}, EPL {\bf 130}, 40002 (2020).


\bibitem{SzamelAOUP} G. Szamel, {\it Self-propelled particle in an external potential: Existence of an effective temperature}, Phys. Rev. E {\bf 90}, 012111 (2014).


\bibitem{Bonilla2019} 
L. L. Bonilla, {\em Active Ornstein-Uhlenbeck particles}, Phys. Rev. E \textbf{100}, 022601 (2019).

\bibitem{Wijland21} 
D. Martin, J. O'Byrne, M. E. Cates, E. Fodor, C. Nardini, J. Tailleur, F. van Wijland, {\it Statistical mechanics of active Ornstein-Uhlenbeck particles}, Phys. Rev. E {\bf 103}, 032607 (2021).

\bibitem{ABM2} U. Basu, S. N. Majumdar, A. Rosso, G. Schehr, {\it Long-time position distribution of an active Brownian particle in two dimensions}, Phys. Rev. E {\bf 100}, 062116 (2019). 


\bibitem{AngelaniHardWalls} L. Angelani, {\em Confined run-and-tumble swimmers in one
dimension}, J. Phys. A: Math. Theor. {\bf 50}, 325601 (2017).


\bibitem{Targetsearch} U. Basu, S. Sabhapandit, I. Santra, {\em Target search by active particles}, In: D. Grebenkov, R. Metzler, G. Oshanin (eds), {\em Target Search Problems}, Springer, Cham (2024). 


\bibitem{MalakarRTP} K. Malakar, V. Jemseena, A. Kundu, K. Vijay Kumar, S. Sabhapandit, S. N. Majumdar, S. Redner, A. Dhar, {\em Steady state, relaxation and first-passage properties of a run-and-tumble particle in one-dimension}, J. Stat. Mech. (2018) 043215.

\bibitem{Singh2020} P. Singh, S. Sabhapandit, A. Kundu, {\em Run-and-tumble particle in inhomogeneous media in one dimension}, J. Stat. Mech. (2020) 083207.

\bibitem{Singh2022} P. Singh, S. Santra, A. Kundu, {\em Extremal statistics of a one-dimensional run and tumble particle with an absorbing wall}, J. Phys. A: Math. Theor. {\bf 55} 465004 (2022).

\bibitem{MFPT1DABP} S. A. Iyaniwura, Z. Peng, {\em Asymptotic analysis and simulation of mean first passage time for active Brownian particles in 1-D}, SIAM Journal on Applied Mathematics {\bf 84}(3), 1079-1095 (2024). 


\bibitem{RTPsurvivalMori} F. Mori, P. Le Doussal, S. N. Majumdar, G. Schehr, {\em Universal Survival Probability for a d-Dimensional Run-and-Tumble Particle}, Phys. Rev. Lett. {\bf 124}, 090603 (2020).

\bibitem{TVB12}
V. Tejedor, R. Voituriez, O. B{\' e}nichou, {\em Optimizing persistent random searches}, Phys. Rev. Lett. {\bf 108}, 088103 (2012).

\bibitem{RBV16}
J.-F. Rupprecht, O. B{\'e}nichou, R. Voituriez, {\em Optimal search strategies of run-and-tumble walks}, Phys. Rev. E {\bf 94}, 012117 (2016).


\bibitem{SurvivalRTPDriftDeBruyne} B. De Bruyne, S. N. Majumdar, G. Schehr, {\em Survival probability of a run-and-tumble particle in the presence of a drift}, J. Stat. Mech. (2021) 043211.


\bibitem{AngelaniMFPT} L. Angelani, R. Di Leonardo, M. Paoluzzi, {\it First-passage time of run-and-tumble particles}, Eur. Phys. J. E {\bf 37}, 59 (2014).

\bibitem{MathisMFPT} M. Gu\'eneau, S. N. Majumdar, G. Schehr, {\em Optimal mean first-passage time of a run-and-tumble particle in a class of one-dimensional confining potential}, EPL {\bf 145}, 61002 (2024).

\bibitem{MathisMFPT2} M. Gu\'eneau, S. N. Majumdar, G. Schehr, {\em Run-and-tumble particle in one-dimensional potentials: Mean first-passage time and applications}, Phys. Rev. E {\bf 111}, 014144 (2025).

\bibitem{Grange2025} P. Grange, L. Yuan, {\it Mean first-passage time at the origin of a run-and-tumble particle with periodic forces}, arXiv:2411.11601 (2025).


\bibitem{TonerTu95} J. Toner, Y. Tu, {\it Long-range order in a two-dimensional dynamical XY model: how birds fly together}, Phys. Rev. Lett. {\bf 75}, 4326 (1995).

\bibitem{TonerTu98} J. Toner, Y. Tu, {\it Flocks, herds, and schools: A quantitative theory of flocking}, Phys. Rev. E {\bf 58}, 4828 (1998).

\bibitem{TonerTuReview} J. Toner, Y. Tu, S. Ramaswamy, {\em Hydrodynamics and phases of flocks}, Ann. of Phys. {\bf 318}, 170 (2005).

\bibitem{CT2015}
M. Cates, J. Tailleur, {\it Motility-induced phase separation}, Annu. Rev. Condens. Matter Phys. {\bf 6}, 219 (2015).

\bibitem{OByrne2021}
J. O'Byrne, A. Solon, J. Tailleur, Y. Zhao, {\it An introduction to motility-induced phase separation}, in {\it Out-of-equilibrium Soft Matter}, The Royal Society of Chemistry (2023). 


\bibitem{slowman}
A. B. Slowman, M. R. Evans, R. A. Blythe, {\it Jamming and attraction of interacting run-and-tumble random walkers}, Phys. Rev. Lett. {\bf 116}, 218101 (2016).

\bibitem{slowman2}
A. B. Slowman, M. R. Evans, R. A. Blythe, {\it Exact solution of two interacting run-and-tumble random walkers with finite tumble duration}, J. Phys. A: Math. Theor. {\bf 50}, 375601 (2017).

\bibitem{Mallmin2019} 
E. Mallmin, R. A. Blythe, M. R. Evans, {\it Exact spectral solution of two interacting run-and-tumble particles on a ring lattice}, J. Stat. Mech. (2019) 013204.

\bibitem{KunduGap2020}
A. Das, A. Dhar, A. Kundu, {\it Gap statistics of two interacting run and tumble particles in one dimension}, J. Phys. A: Math. Theor. {\bf 53}, 345003 (2020).

\bibitem{Hahn2023} L. Hahn, A. Guillin, M. Michel, {\it Jamming pair of general run-and-tumble particles: Exact results and universality classes}, arXiv:2306.00831.


\bibitem{Metson2022}
M. J. Metson, M. R. Evans, R. A. Blythe, {\it Tuning attraction and repulsion between active particles through persistence}, EPL {\bf 141}, 41001 (2023).

\bibitem{MetsonLong}
M. J. Metson, M. R. Evans, R. A. Blythe, {\it From a microscopic solution to a continuum description of interacting active particles}, Phys. Rev. E {\bf 107}, 044134 (2022).


\bibitem{LMS2021} 
P. Le Doussal, S. N. Majumdar, G. Schehr, {\it Stationary nonequilibrium bound state of a pair of run and tumble particles}, Phys. Rev. E {\bf 104}(4), 044103 (2021).

\bibitem{Hahn2025} L. Hahn, {\it Steady state and mixing of two run-and-tumble particles interacting through jamming and attractive forces}, arXiv:2501.11379.


\bibitem{KH2018} 
M. Kourbane-Houssene, C. Erignoux, T. Bodineau, J. Tailleur, {\it Exact hydrodynamic description of active lattice gases}, Phys. Rev. Lett. {\bf 120}, 268003 (2018).

\bibitem{Erignoux_derivation} C. Erignoux, {\it Hydrodynamic Limit For An Active Exclusion Process}, arXiv:1608.04937 (2016).

\bibitem{Agranov2021}
T. Agranov, S. Ro, Y. Kafri, V. Lecomte, {\it Exact fluctuating hydrodynamics of active lattice gases -- typical fluctuations}, J. Stat. Mech. (2021) 083208.

\bibitem{Agranov2022}
T. Agranov, S. Ro, Y. Kafri, V. Lecomte, {\it Macroscopic Fluctuation Theory and Current Fluctuations in Active Lattice Gases}, SciPost Phys. {\bf 14}, 045 (2023). 

\bibitem{lattice2lanes2025} R. Mukherjee, S. Saha, T. Sadhu, A. Dhar, S. Sabhapandit, {\it Hydrodynamics of a hard-core active lattice gas}, Phys. Rev. E {\bf 111}, 024128 (2025).


\bibitem{PutBerxVanderzande2019}
S. Put, J. Berx, C. Vanderzande, {\it Non-Gaussian anomalous dynamics in systems of interacting run-and-tumble particles}, J. Stat. Mech. (2019) 123205.


\bibitem{SinghChain2020} 
P. Singh, A. Kundu, {\it Crossover behaviours exhibited by fluctuations and correlations in a chain of active particles}, J. Phys. A: Math. Theor. {\bf 54}, 305001 (2021).

\bibitem{HarmonicChainRevABP} S. Prakash, U. Basu, S. Sabhapandit, {\em Tagged particle behavior in a harmonic chain of direction reversing active Brownian particles}, J. Stat. Mech. (2024) 083211.

\bibitem{HarmonicChainRTPDhar} S. Paul, A. Dhar, D. Chaudhuri, {\em Dynamical crossovers and correlations in a harmonic chain of active particles}, Soft Matter (2024).


\bibitem{ActiveMeltingNature2024} 
H. Massana-Cid, C. Maggi, N. Gnan, G. Frangipane, R. Di Leonardo, {\it Multiple temperatures and melting of a colloidal active crystal}, Nat. Commun. {\bf 15}, 6574 (2024).


\bibitem{ishikawa2007} T. Ishikawa, G. Sekiya, Y. Imai, T. Yamaguchi, {\it Hydrodynamic Interactions between Two Swimming Bacteria}, Biophysical Journal {\bf 93} (6), 2217 - 2225 (2007).

\bibitem{zottl2023} A. Zöttl, H. Stark, {\it Modeling Active Colloids: From Active Brownian Particles to Hydrodynamic and Chemical Fields}, Annu. Rev. Cond. Mat. Phys.  {\bf 14}, 109-127 (2023). 

\bibitem{Filella2018} A. Filella, F. Nadal, C. Sire, E. Kanso, C. Eloy, {\it Model of collective fish Behavior with hydrodynamic interactions}, Phys. Rev. Lett. {\bf 120}, 198101 (2018).



\bibitem{Riesz}
M. Riesz, {\it Riemann Liouville integrals and potentials}, Acta Sci. Math. Univ. Szeged {\bf 9}, 1 (1938).


\bibitem{Lewin}
M. Lewin, {\it Coulomb and Riesz gases: The known and the unknown}, J. Math. Phys. {\bf 63}, 061101 (2022).

\bibitem{SerfatyBook} S. Serfaty, {\it Lectures on Coulomb and Riesz gases}, arXiv:2407.21194 (2024).

\bibitem{Mehta_book}
M. L. Mehta, {\it Random matrices}, Elsevier (2004).

\bibitem{Forrester_book}
P. J. Forrester, {\it Log-gases and random matrices}, Princeton university press (2010).

\bibitem{bouchaud_book}
M. Potters, J. P. Bouchaud, {\it A First Course in Random Matrix Theory: For Physicists, Engineers and Data Scientists}, Cambridge University Press (2020).

\bibitem{Dyson} F. J. Dyson, {\it A Brownian-motion model for the eigenvalues of a random matrix}, J. Math. Phys. {\bf 3}:1191–1198 (1962).

\bibitem{PLDRankedDiffusion} 
P. Le Doussal, {\it Ranked diffusion, delta Bose gas and Burgers equation}, Phys. Rev. E {\bf 105}, L012103 (2022).

\bibitem{Agarwal2019} 
S. Agarwal, M. Kulkarni, A. Dhar, {\it Some connections between the Classical Calogero-Moser model and the Log Gas}, J.~Stat.~Phys. {\bf 176}, 1463 (2019).



\bibitem{Dean}
D. S. Dean, {\it Langevin Equation for the density of a system of interacting Langevin processes}, J. Phys. A: Math. Gen. {\bf 29}, L613 (1996). 

\bibitem{Kawa}
K. Kawasaki, {\it Microscopic analyses of the dynamical density functional equation of dense fluids}, J. Stat. Phys. {\bf 93}, 527 (1998).


\bibitem{DFRiesz23} 
R. Dandekar, P. L. Krapivsky, K. Mallick, {\em Dynamical fluctuations in the Riesz gas}, Phys. Rev. E {\bf 107}, 044129 (2023).

\bibitem{BoursierCLT} J. Boursier, {\em Optimal local laws and CLT for the circular Riesz gas}, arXiv:2112.05881 (2021).


\bibitem{BoursierCorrelations}
J. Boursier, {\em Decay of correlations and thermodynamic limit for the circular Riesz gas}, arXiv:2209.00396 (2022).


\bibitem{ChateGiant} H. Chat\'e, F. Ginelli, G. Gr\'egoire, F. Raynaud, {\it Collective motion of self-propelled particles interacting without cohesion}, Phys. Rev. E {\bf 77}, 046113 (2008).

\bibitem{GinelliGiant} F. Ginelli, F. Peruani, M. Bär, H. Chat\'e, {\it Large-Scale Collective Properties of Self-Propelled Rods}, Phys. Rev. Lett. {\bf 104}, 184502 (2010).

\bibitem{DasGiant2012} 
S. Dey, D. Das, R. Rajesh, {\it Spatial structures and giant number fluctuations in models of active matter}, Phys. Rev. Lett. {\bf 108}, 238001 (2012).


\bibitem{NarayanGiant} V. Narayan, S. Ramaswamy, N. Menon, {\it Long-Lived Giant Number Fluctuations in a Swarming Granular Nematic}, Science {\bf 317}, 105 (2007).

\bibitem{ZhangGiant} H. P. Zhang, A. Be'er, E.-L. Florin, H. L. Swinney, {\it Collective motion and density fluctuations in bacterial colonies}, Proc. Natl. Acad. Sci. U.S.A. {\bf 107}, 13626 (2010).


\bibitem{Siegmund} D. Siegmund, {\em The Equivalence of Absorbing and Reflecting Barrier Problems for Stochastically Monotone Markov Processes}, Ann. Probab. {\bf 4}(6): 914-924 (1976).



\bibitem{DiffDiffChubynsky} M. V. Chubynsky, G. W. Slater, {\em Diffusing Diffusivity: A Model for Anomalous, yet Brownian, Diffusion}, Phys. Rev. Lett. {\bf 113}, 098302 (2014).

\bibitem{DiffDiffChechkin} A. V. Chechkin, F. Seno, R. Metzler, I. M. Sokolov, {\em Brownian yet Non-Gaussian Diffusion: From Superstatistics to Subordination of Diffusing Diffusivities}, Phys. Rev. X {\bf 7}, 021002 (2017).

\bibitem{DiffDiffJain} R. Jain, K. L. Sebastian, {\em Diffusing diffusivity: a new derivation and comparison with simulations}, Journal of Chemical Sciences {\bf 129}, 929–937 (2017).

\bibitem{DiffDiffFPTSposini} V. Sposini, A. Chechkin, R. Metzler, {\em First passage statistics for diffusing diffusivity}, J. Phys. A: Math. Theor. {\bf 52} 04 (2019).


\bibitem{resettingPRL} M. R. Evans, S. N. Majumdar, {\em Diffusion with stochastic resetting}, Phys. Rev. letters, 106(16), 160601 (2011).

\bibitem{resettingReview} M. R. Evans, S. N. Majumdar, G. Schehr, {\em Stochastic resetting and applications}, J. Phys. A: Math. Theor. {\bf 53} 193001 (2020).

\bibitem{resettingBriefReview} S. Gupta, A. M. Jayannavar, {\em Stochastic resetting: A (very) brief review}, Frontiers in Physics {\bf 10} 789097 (2022).








\bibitem{cellsMarchetti} D. Bi, X. Yang, M. C. Marchetti, M. L. Manning, {\it Motility-Driven Glass and Jamming Transitions in Biological Tissues},
Phys. Rev. X {\bf 6}, 021011 (2016).


\bibitem{HJ95} 
P. H\"anggi, P. Jung, {\it Colored Noise in Dynamical Systems}, Adv. Chem. Phys. {\bf 89}, 239 (1995).



\bibitem{Lowen2020} H. Löwen, {\it Inertial effects of self-propelled particles: From active Brownian to active Langevin motion}, J. Chem. Phys. {\bf 152}, 040901 (2020).

\bibitem{Cates2012} M. E. Cates, {\it Diffusive transport without detailed balance: Does microbiology need statistical physics ?}, Rep. Prog. Phys. {\bf 75}, 042601 (2012).


\bibitem{CMPT2010}
M. E. Cates, D. Marenduzzo, I. Pagonabarraga, J. Tailleur, {\it Arrested phase separation in reproducing bacteria creates a generic route to pattern formation}, Proc. Natl. Acad. Sci. U.S.A. {\bf 107}, 11715 (2010).

\bibitem{Chaudhuri2014} D. Chaudhuri, {\it Active brownian particles: Entropy production and fluctuation response}, Phys. Rev. E {\bf 90}, 022131 (2014).

\bibitem{Mandal2017} D. Mandal, K. Klymko, M. R. DeWeese, {\it Entropy Production and Fluctuation Theorems for Active Matter}, Phys. Rev. Lett. {\bf 119}, 258001 (2017).

\bibitem{Pietzonka2018} P. Pietzonka, U. Seifert, {\it Entropy production of active particles and for particles in active baths}, J. Phys. A: Math. Theor. {\bf 51}, 01LT01 (2018).

\bibitem{Razin2020} N. Razin, {\it Entropy production of an active particle in a box}, Phys. Rev. E {\bf 102}, 030103(R) (2020).


\bibitem{Solon15} A. P. Solon, Y. Fily, A. Baskaran, M. E. Cates, Y. Kafri, M. Kardar, J. Tailleur, {\it Pressure is not a state function for generic active fluids}, Nature Phys. {\bf 11}, 673 (2015).






\bibitem{Kac1974} 
M. Kac, {\it A stochastic model related to the telegrapher's equation}, Rocky Mountain J. Math. {\bf 4}, 497 (1974).

\bibitem{Orsingher90} 
E. Orsingher, {\it Probability law, flow function, maximum distribution of wave-governed random motions and their connections with Kirchoff's laws}, Stoch. Process. Their Appl. {\bf 34}, 49 (1990).

\bibitem{Masoliver1993} J. Masoliver, J. M. Porrà, G. H. Weiss, {\it Solution to the telegrapher’s equation in the presence of reflecting and partly reflecting boundaries}, Phys. Rev. E {\bf 48}, 939 (1993).

\bibitem{PRWWeiss} 
G. H. Weiss, {\it Some applications of persistent random walks and the telegrapher's equation}, Physica A {\bf 311}, 381-410 (2002).

\bibitem{Masoliver2017} 
J. Masoliver, K. Lindenberg, {\it Continuous time persistent random walk: a review and some generalizations}, Eur. Phys. J. B {\bf 90}, 107 (2017).

\bibitem{Schnitzer} 
M. J. Schnitzer, {\it Theory of continuum random walks and application to chemotaxis}, Phys. Rev. E {\bf 48}, 2553 (1993).

\bibitem{Tailleur_RTP} 
J. Tailleur, M. E. Cates, {\it Statistical mechanics of interacting run-and-tumble bacteria}, Phys. Rev. Lett. {\bf 100}, 218103 (2008).


\bibitem{2dRTPSantra} I. Santra, U. Basu, S. Sabhapandit, {\it Run-and-tumble particles in two dimensions: Marginal position distributions}, Phys. Rev. E {\bf 101}, 062120 (2020).





\bibitem{3statesBasu}
U. Basu, S. N. Majumdar, A. Rosso, S. Sabhapandit, G. Schehr, {\it Exact stationary state of a run-and-tumble particle with three internal states in a harmonic trap}, J. Phys. A: Math. Theor. {\bf 53}, 09LT01 (2020).

\bibitem{Sun2024} A. Sun, F. Ye, R. Podgornik, {\it Exact moments for a run and tumble particle in a harmonic trap with a finite tumble time}, arXiv:2409.00578 (2024).



\bibitem{Korobkova2006} E. A. Korobkova, T. Emonet, H. Park, P. Cluzel, {\it Hidden Stochastic Nature of a Single Bacterial Motor}, Phys. Rev. Lett. {\bf 96}, 058105 (2006).

\bibitem{Xie2011} L. Xie, T. Altindal, S. Chattopadhyay, X.-L. Wu, {\it Bacterial flagellum as a propeller and as a rudder for efficient chemotaxis}, Proc. Natl. Acad. Sci. USA {\bf 108}, 2246 (2011).

\bibitem{Theves2013} M. Theves, J. Taktikos, V. Zaburdaev, H. Stark, C. Beta, {\it A Bacterial Swimmer with Two Alternating Speeds of Propagation}, Biophys. J. {\bf 105}, 1915 (2013).

\bibitem{natureruntime} E. Korobkova, T. Emonet, J.M. Vilar, T.S Shimizu, P. Cluzel, {\em From molecular noise to behavioural variability in a single bacterium}, Nature {\bf 428}(6982):574-8 (2004).

\bibitem{runtime2} N. Figueroa-Morales, R. Soto, G. Junot, T. Darnige, C.
Douarche, V. A. Martinez, A. Lindner, E. Cl\'ement, {\it 3D spatial exploration by E. coli echoes motor temporal variability}, Phys. Rev. X {\bf 10}, 021004 (2020).



\bibitem{levywalks} V. Zaburdaev, S. Denisov, J. Klafter, {\em L\'evy walks}, Reviews of Modern Physics, {\bf 87}, 483 (2015).

\bibitem{Detcheverry} F. Detcheverry, {\it Generalized run-and-turn motions: From bacteria to L\'evy walks}, Phys. Rev. E {\bf 96}, 012415 (2017).

\bibitem{Naftali2024} O. Farago, N. R. Smith, {\it Confined run and tumble particles with non-Markovian tumbling statistics}, Phys. Rev. E {\bf 109}, 044121 (2024).


\bibitem{Larralde2020} H. Larralde, {\it First-passage probabilities and mean number of sites visited by a persistent random walker in one- and two-dimensional lattices}, Phys. Rev. E {\bf 102}, 062129 (2020).

\bibitem{PRWSurvivalLacroixMori} B. Lacroix-A-Chez-Toine and F. Mori, {\em Universal survival probability for a correlated random walk and applications to records}, J. Phys. A: Math. Theor. {\bf 53}, 495002 (2020).



\bibitem{ABP2012} P. Romanczuk, M. B\"ar, W. Ebeling, B. Lindner, L. Schimansky-Geier, {\em Active Brownian particles}, Eur. Phys. J. Spec. Top. {\bf 202}, 1–162 (2012).

\bibitem{ABM} U. Basu, S. N. Majumdar, A. Rosso, G. Schehr, {\em Active Brownian motion in two dimensions}, Phys. Rev. E {\bf 98}(6), 062121 (2018).



\bibitem{Kurzthaler2018} C. Kurzthaler, C. Devailly, J. Arlt, T. Franosch, W. C. K. Poon, V. A. Martinez, A. T. Brown, {\it Probing the Spatiotemporal Dynamics of Catalytic Janus Particles with Single-Particle Tracking and Differential Dynamic Microscopy}, Phys. Rev. Lett. {\bf 121}, 078001 (2018).


\bibitem{DRABP1} I. Santra, U. Basu, S. Sabhapandit, {\em Active Brownian motion with directional reversals}, Phys. Rev. E {\bf 104}, L012601 (2021).

\bibitem{DRABP2} I. Santra, U. Basu, S. Sabhapandit, {\em Direction reversing active Brownian particle in a harmonic potential}, Soft Matter {\bf 17}, 10108 (2021).




\bibitem{Takatori2016} S. C. Takatori, R. De Dier, J. Vermant, J. F. Brady, {\it Acoustic trapping of active matter}, Nature Comm. {\bf 7}, 10694 (2016).

\bibitem{Buttinoni2022} I. Buttinoni, L. Caprini, L. Alvarez, F. J. Schwarzendahl, H. Löwen, {\it Active colloids in harmonic optical potentials}, Europhys. Lett. {\bf 140}, 27001 (2022).

\bibitem{Dauchot2019} O. Dauchot, V. D\'emery, {\it Dynamics of a Self-Propelled Particle in a Harmonic Trap}, Phys. Rev. Lett. {\bf 122}, 068002 (2019).


\bibitem{Klyatskin1977} V. I. Klyatskin, {\it Dynamic systems with parameter fluctuations of the telegraphic-process type}, Radiophys. Quantum Electron. {\bf 20}, 382–392 (1977).

\bibitem{Lefever1980} K. Kitahara, W. Horsthemke, R. Lefever, Y. Inaba, {\it Phase Diagrams of Noise Induced Transitions: Exact Results for a Class of External Coloured Noise}, Prog. Theor. Phys. {\bf 64} (4), 1233 (1980).

\bibitem{Sevilla}
F. J. Sevilla, A. V. Arzola, E. P. Cital, {\em Stationary superstatistics distributions of trapped run-and-tumble particles}, Phys. Rev. E {\bf 99}, 012145 (2019).

\bibitem{Smith2DRTP} N. R. Smith1, P. Le Doussal, S. N. Majumdar, G. Schehr, {\it Exact position distribution of a harmonically confined run-and-tumble particle in two dimensions}, Phys. Rev. E {\bf 106}, 054133 (2022).







\bibitem{Malakar2020} K. Malakar, A. Das, A. Kundu, K. V. Kumar, A. Dhar, {\it Steady state of an active Brownian particle in a two-dimensional harmonic trap}, Phys. Rev. E {\bf 101}, 022610 (2020).

\bibitem{Chaudhuri2021} D. Chaudhuri, A. Dhar, {\it Active Brownian particle in harmonic trap: exact computation of moments, and re-entrant transition}, J. Stat. Mech. (2021) 013207.

\bibitem{Franosch2022} M. Caraglio, T. Franosch, {\it Analytic Solution of an Active Brownian Particle in a Harmonic Well}, Phys. Rev. Lett. {\bf 129}, 158001 (2022).



\bibitem{AngelaniReflecting} L. Angelani, {\it Run-and-tumble particles, telegrapher's equation and absorption problems with partially reflecting boundaries}, J. Phys. A: Math. Theor. {\bf 48}, 495003 (2015).


\bibitem{Lee2013} C. F. Lee, {\em Active particles under confinement: aggregation at the wall and gradient formation inside a channel}, New J. Phys. {\bf 15}, 055007 (2013).

\bibitem{PireySphere2023} 
T. Arnoulx de Pirey, F. van Wijland, {\it A run-and-tumble particle around a spherical obstacle: the steady-state distribution far-from-equilibrium}, J. Stat. Mech. (2023) 093202.



\bibitem{hardWallsJoanny} C. Sandford, A. Y. Grosberg, J.-F. Joanny, {\em Pressure and flow of exponentially self-correlated active particles},
Phys. Rev. E {\bf 96}, 052605 (2017).

\bibitem{hardWallsCaprini} L. Caprini, U. M. B. Marconi, {\em Active particles under confinement and effective force generation among surfaces}, Soft matter {\bf 14}, 9044-9054 (2018).


\bibitem{Duzgun2018} A. Duzgun, J. V. Selinger, {\em Active Brownian particles near straight or curved walls: Pressure and boundary layers}, Phys. Rev. E {\bf 97}, 032606 (2018).





\bibitem{redner}
S. Redner, {\it A guide to first-passage processes}, Cambridge university press (2001).

\bibitem{Bray2013} 
A. J. Bray, S. N. Majumdar, G. Schehr, {\it Persistence and first-passage properties in nonequilibrium systems}, Adv. Phys. {\bf 62}, 225 (2013).

\bibitem{Metzler_book}
R. Metzler, S. Redner, G. Oshanin, {\it First-passage phenomena and their applications}, Vol. 35, World Scientific, (2014).



\bibitem{Comtet2005} A. Comtet, S. N. Majumdar, {\it Precise asymptotics for a random walker’s maximum}, J. Stat. Mech. (2005) P06013.

\bibitem{reviewEVSPal} S. N. Majumdar, A. Pal, G. Schehr, {\it Extreme value statistics of correlated random variables: a pedagogical review}, Physics Reports {\bf 840} 1-32 (2020).

\bibitem{livreSG} S. N. Majumdar, G. Schehr, {\em Statistics of Extremes and Records in Random Sequences}, Oxford University Press, (2024).

\bibitem{SirePersistence1} S. N. Majumdar, C. Sire, {\it Survival Probability of a Gaussian Non-Markovian Process: Application to the $T=0$ Dynamics of the Ising Model}, Phys. Rev. Lett. {\bf 77}, 1420 (1996).

\bibitem{SirePersistence2} S. N. Majumdar, C. Sire, A. J. Bray, S. J. Cornell, {\it Nontrivial Exponent for Simple Diffusion}, Phys. Rev. Lett. {\bf 77}, 2867 (1996).

\bibitem{DerridaPersistence} B. Derrida, V. Hakim, R. Zeitak, {\it Persistent Spins in the Linear Diffusion Approximation of Phase Ordering and Zeros of Stationary Gaussian Processes}, Phys. Rev. Lett. {\bf 77}, 2871 (1996).







\bibitem{benichou1} O. B\'enichou, C. Loverdo, M. Moreau, R. Voituriez, {\it Intermittent search strategies}, Reviews of Modern Physics, {\bf 83}(1), 81. (2011).

\bibitem{benichou2} O. B\'enichou, M. Coppey, M. Moreau, P. H. Suet, R. Voituriez, {\it Optimal search strategies for hidden targets} Phys. Rev. letters, {\bf 94}(19), 198101 (2005).




\bibitem{Masoliver92} J. Masoliver, J. M. Porrà, G. H. Weiss, {\it Solutions of the telegrapher’s equation in the presence of traps}, Phys. Rev. A {\bf 45}, 2222 (1992).

\bibitem{Orsingher95} E. Orsingher, {\it Motions with reflecting and absorbing barriers driven by the telegraph equation}, Random Operator Stoch. Equ. {\bf 3} (1), 9-21 (1995).

\bibitem{MukherjeeOccupationTime} S. Mukherjee, P. Le Doussal, N. R. Smith, {\it Large deviations in statistics of the local time and occupation time for a run and tumble particle}, Phys. Rev. E {\bf 110}, 024107 (2024).


\bibitem{LMS2019}
P. Le Doussal, S. N. Majumdar, G. Schehr, {\it Noncrossing run-and-tumble particles on a line}, Phys. Rev. E {\bf 100}, 012113 (2019).








\bibitem{BressloffStickyBoundary} P. C. Bressloff, {\it Encounter-based model of a run-and-tumble particle}, J. Stat. Mech. (2022) 113206.

\bibitem{BressloffStickyBoundaries} P. C. Bressloff, {\em Encounter-based model of a run-and-tumble particle II: absorption at sticky boundaries}, J. Stat. Mech. (2023) 043208.

\bibitem{AngelaniGenericBC} L. Angelani, {\em One-dimensional run-and-tumble motions with generic boundary conditions}, J. Phys. A: Math. Theor. {\bf 56}, 455003 (2023).

\bibitem{AngelaniOptimalEscapes} L. Angelani, {\it Optimal escapes in active matter}, Eur. Phys. J. E 4{\bf 7}, 9 (2024).

\bibitem{RTPpartiallyAbsorbingTarget} E. Jeon, B. Go, Y. W. Kim, {\em Searching for a partially absorbing target by a run-and-tumble particle in a confined space}, Phys. Rev. E {\bf 109}, 014103 (2024).






\bibitem{Baouche2025} Y. Baouche, M. Le Goff, C. Kurzthaler, T. Franosch, {\it First-passage-time statistics of active Brownian particles: A perturbative approach}, Phys. Rev. E {\bf 111}, 054113 (2025). 


\bibitem{BressloffABP} P. C. Bressloff, {\it Trapping of an active Brownian particle at a partially absorbing wall}, Proc. R. Soc. A. {\bf 479}, 20230086 (2023).

\bibitem{Caraglio2024} M. Caraglio, {\it Two-Dimensional Active Brownian Particles Crossing a Parabolic Barrier: Transition-Path Times, Survival Probability, and First-Passage time}, arXiv:2410.07226 (2024).



\bibitem{AOUPEscapeLecomte} E. Woillez, Y. Kafri, V. Lecomte, {\em Nonlocal stationary probability distributions and escape rates for an active Ornstein–Uhlenbeck particle}, J. Stat. Mech. (2020) 063204.





\bibitem{Marchetti2013} 
M. C. Marchetti, J. F. Joanny, S. Ramaswamy, T. B. Liverpool, J. Prost, M. Rao, R. A. Simha, {\it Hydrodynamics of soft active matter}, Rev. Mod. Phys. {\bf 85}, 1143 (2013).

\bibitem{Ramaswamy2017}
S. Ramaswamy, {\it Active Matter}, J. Stat. Mech. (2017) 054002.


\bibitem{Poon2012} J. Schwarz-Linek, C. Valeriani, A. Cacciuto, M. E. Cates, D. Marenduzzo, A. N. Morozov, W. C. K. Poon, {\em Phase separation and rotor self-assembly in active particle suspensions}, Proc. Natl. Acad. Sci. USA {\bf 109}, 4052 (2012). 

\bibitem{Stenhammar2015} J. Stenhammar, R. Wittkowski, D. Marenduzzo, M. E. Cates, {\em Activity-Induced Phase Separation and Self-Assembly in Mixtures of Active and Passive Particles}, Phys. Rev. Lett. {\bf 114}, 018301 (2015).


\bibitem{Buttinoni2013}
I. Buttinoni, J. Bialké, F. Kümmel, H. Löwen, C. Bechinger, T. Speck, {\it Dynamical Clustering and Phase Separation in Suspensions of Self-Propelled Colloidal Particles}, Phys. Rev. Lett. {\bf 110}, 238301 (2013).

\bibitem{Palacci2013} J. Palacci, S. Sacanna, A. P. Steinberg, D. J. Pine, P. M. Chaikin, {\it Living Crystals of Light-Activated Colloidal Surfers}, Science {\bf 339}, 936-940 (2013).



\bibitem{Liu2019} G. Liu, A. Patch, F. Bahar, D. Yllanes, R. D. Welch, M. C. Marchetti, S. Thutupalli, J. W. Shaevitz, {\it Self-Driven Phase Transitions Drive Myxococcus xanthus Fruiting Body Formation}, Phys. Rev. Lett. {\bf 122}, 248102 (2019).



\bibitem{FM2012}
Y. Fily, M. C. Marchetti, {\it Athermal phase separation of self-propelled particles with no alignment}, Phys. Rev. Lett. {\bf 108}, 235702 (2012).


\bibitem{Redner2013} G. S. Redner, M. F. Hagan, A. Baskaran, {\em Structure and Dynamics of a Phase-Separating Active Colloidal Fluid}, Phys. Rev. Lett. {\bf 110}, 055701 (2013).

\bibitem{Levis2014} D. Levis, L. Berthier, {\it Clustering and heterogeneous dynamics in a kinetic Monte Carlo model of self-propelled hard disks}, Phys. Rev. E {\bf 89}, 062301 (2014).

\bibitem{FHM2014}
Y. Fily, S. Henkes, M. C. Marchetti, {\it Freezing and phase separation of self-propelled disks}, Soft matter {\bf 10}, 2132 (2014).

\bibitem{SG2014}
R. Soto, R. Golestanian, {\it Run-and-tumble dynamics in a crowded environment: Persistent exclusion process for swimmers}, Phys. Review E {\bf 89}, 012706 (2014).


\bibitem{Bialke2013} J. Bialké, H. Löwen, T. Speck, {\it Microscopic theory for the phase separation of self-propelled repulsive disks}, Europhys. Lett. {\bf 103}, 30008 (2013).


\bibitem{Thom2011}
A. G. Thompson, J. Tailleur, M. E. Cates, R. A. Blythe, {\it Lattice models of nonequilibrium bacterial dynamics}, J. Stat. Mech. (2011) P02029.

\bibitem{Wittkowski2014} R. Wittkowski, A. Tiribocchi, J. Stenhammar, R. J. Allen, D. Marenduzzo, M. E. Cates, {\it Scalar $\phi^4$ field theory for active-particle phase separation}, Nat. Commun. {\bf 5}, 4351 (2014).


\bibitem{Solon2018_1} A. P. Solon, J. Stenhammar, M. E. Cates, Y. Kafri, J. Tailleur, {\it Generalized thermodynamics of motility-induced phase separation: phase equilibria, Laplace pressure, and change of ensembles}, New J. Phys. {\bf 20}, 075001 (2018).

\bibitem{Solon2018_2} A. P. Solon, J. Stenhammar, M. E. Cates, Y. Kafri, J. Tailleur, {\it Generalized thermodynamics of phase equilibria in scalar active matter}, Phys. Rev. E {\bf 97}, 020602 (2018).



\bibitem{QuorumSensing} M. B. Miller, B. L. Bassler, {\it Quorum sensing in bacteria}, Annu. Rev. Microbiol. {\bf 55}, 165–199 (2001).


\bibitem{Gregoire2004} G. Gr\'egoire, H. Chat\'e, {\it Onset of Collective and Cohesive Motion}, Phys. Rev. Lett. {\bf 92}, 025702 (2004).




%

\bibitem{Bertin2006} E. Bertin, M. Droz, G. Grégoire, {\it Boltzmann and hydrodynamic description for self-propelled particles}, Phys. Rev. E {\bf 74} 022101 (2006).

\bibitem{Bertin2009} E. Bertin, M. Droz, G. Grégoire, {\it Hydrodynamic equations for self-propelled particles: microscopic derivation and stability analysis}, J. Phys. A {\bf 42}, 445001 (2009).

\bibitem{Peshkov2014} A. Peshkov, E. Bertin, F. Ginelli, H. Chaté, {\it Boltzmann–Ginzburg–Landau approach for continuous descriptions of generic Vicsek-like models}, Eur. Phys.J. E Soft Matter {\bf 223}, 1315 (2014).


\bibitem{Aranson2005} I. S. Aranson, L. S. Tsimring, {\it Pattern formation of microtubules and motors: Inelastic interaction of polar rods}, Phys. Rev. E {\bf 71}, 050901 (2005).

\bibitem{Baskaran2008} A. Baskaran, M. C. Marchetti, {\it Enhanced Diffusion and Ordering of Self-Propelled Rods}, Phys. Rev. Lett. {\bf 101}, 268101 (2008).


\bibitem{Mishra2010} S. Mishra, A. Baskaran, M. C. Marchetti, {\it Fluctuations and pattern formation in self-propelled particles}, Phys. Rev. E {\bf 81}, 061916 (2010).



\bibitem{Kudrolli2008} A. Kudrolli, G. Lumay, D. Volfson, L. S. Tsimring, {\it Swarming and swirling in self-propelled polar granular rods}, Phys. Rev. Lett. {\bf 100}, 058001 (2008).

\bibitem{Kumar2014} N. Kumar, H. Soni, S. Ramaswamy, A.~K. Sood, {\em Flocking at a distance in active granular matter}, Nature Comm. {\bf 5},  4688 (2014). 


\bibitem{Ballerini2008} M. Ballerini, N. Cabibbo, R. Candelier, A. Cavagna, E. Cisbani, I. Giardina, V. Lecomte, A. Orlandi, G. Parisi, A. Procaccini, M. Viale, V. Zdravkovic, {\it Interaction ruling animal collective behavior depends on topological rather than metric distance: Evidence from a field study}, Proc. Natl. Acad. Sci. U.S.A. {\bf 105} (4), 1232-1237 (2008).

\bibitem{Cavagna2010} A. Cavagna, A. Cimarelli, I. Giardina, G. Parisi, R. Santagati, F. Stefanini, M. Viale, {\it Scale-free correlations in starling flocks}, Proc. Natl. Acad. Sci. USA {\bf 107}, 11865 (2010).

\bibitem{Ginelli2010} F. Ginelli, H. Chaté, {\it Relevance of metric-free interactions in flocking phenomena}, Phys. Rev. Lett. {\bf 105}, 168103 (2010).




\bibitem{Hatwalne2004} Y. Hatwalne, S. Ramaswamy, M. Rao, R. A. Simha, {\it Rheology of Active-Particle Suspensions}, Phys. Rev. Lett. {\bf 92}, 118101 (2004).

\bibitem{Brotto2013} 
T. Brotto, J. B. Caussin, E. Lauga, D. Bartolo, {\it Hydrodynamics of confined active fluids}, Phys. Rev. Lett. {\bf 110}, 038101 (2013).


\bibitem{Liverpool2006} T. B. Liverpool, M. C. Marchetti, {\it Rheology of Active Filament Solutions}, Phys. Rev. Lett. {\bf 97}, 268101 (2006).


\bibitem{Saintillan2008} D. Saintillan, M. J. Shelley, {\it Instabilities and Pattern Formation in Active Particle Suspensions: Kinetic Theory and Continuum Simulations}, Phys. Rev. Lett. {\bf 100}, 178103 (2008).


\bibitem{Pahlavan2011} A. A. Pahlavan, D. Saintillan, {\it Instability regimes in flowing suspensions of swimming micro-organisms}, Physics of Fluids {\bf 23}, 011901 (2011).


\bibitem{Yoshinaga2017} N. Yoshinaga, T. B. Liverpool, {\it Hydrodynamic interactions in dense active suspensions: From polar order to dynamical clusters}, Phys. Rev. E {\bf 96}, 020603(R) (2017).


\bibitem{Baskaran2009} A. Baskaran, M. C. Marchetti, {\it Statistical mechanics and hydrodynamics of bacterial suspensions}, Proc. Natl. Acad. Sci. U.S.A. {\bf 106} (37), 15567-15572 (2009).

\bibitem{Leoni2010} M. Leoni, T. B. Liverpool, {\it Swimmers in Thin Films: From Swarming to Hydrodynamic Instabilities}, Phys. Rev. Lett. {\bf 105}, 238102 (2010).


\bibitem{Debnath2018} 
T. Debnath, Y. Li, P. K. Ghosh, F. Marchesoni, {\it Hydrodynamic interaction of trapped active Janus particles in two dimensions}, Phys. Rev. E, {\bf 97}(4), 042602 (2018).

\bibitem{Maes_bound_state}
P. Dolai, S. Krekels, C. Maes, {\it Inducing a bound state between active particles}, Phys. Rev. E {\bf 105}, 044605 (2022).


\bibitem{Saha2014} S. Saha, R. Golestanian, Sriram Ramaswamy, {\it Clusters, asters, and collective oscillations in chemotactic colloids}, Phys. Rev. E {\bf 89}, 062316 (2014).

\bibitem{Pohl2014} O. Pohl, H. Stark, {\it Dynamic Clustering and Chemotactic Collapse of Self-Phoretic Active Particles}, Phys. Rev. Lett. {\bf 112}, 238303 (2014).






\bibitem{Reichhardt2014} C. Reichhardt, C.J. Olson Reichhardt, {\it Absorbing phase transitions and dynamic freezing in running active matter systems}, Soft Matter {\bf 10}, 7502 (2014).

\bibitem{Bialke2014} J. Bialké, T. Speck, H. Löwen, {\it Crystallization in a dense suspension of self-propelled particles}, Phys. Rev. Lett. {\bf 108}, 168301 (2012).

\bibitem{Menzel2014} A.M. Menzel, T. Ohta, H. Löwen, {\it Active crystals and their stability}, Phys. Rev. E {\bf 89}, 022301 (2014).

\bibitem{James2021}
M. James, D. A. Suchla, J. Dunkel, M. Wilczek, {\it Emergence and melting of active vortex crystals}, Nat. Commun. {\bf 12}, 5630 (2021).

\bibitem{Leticia2022}
P. Digregorio, D. Levis, L. F. Cugliandolo, G. Gonnella, I. Pagonabarraga, {\it Unified analysis of topological defects in 2D systems of active and passive disks}, Soft Matter {\bf 18}, 566 (2022).


\bibitem{Chate2023} 
X. Q. Shi, F. Cheng, H. Chaté, {\it Extreme spontaneous deformations of active crystals}, Phys. Rev. Lett. {\bf 131}(10), 108301 (2023).





\bibitem{Bi2014} D. Bi, J.H. Lopez, J.M. Schwarz, M.L. Manning, {\it Energy barriers and cell migration in densely packed tissues}, Soft Matter {\bf 10}, 1885 (2014).

\bibitem{Bi2015} D. Bi, J.H. Lopez, J.M. Schwarz, M.L. Manning, {\it A density-independent rigidity transition in biological tissues}, Nat. Phys. {\bf 11}, 1074 (2015).




\bibitem{SussmanVertex} D. M. Sussman, J. M. Schwarz, M. C. Marchetti, M. L. Manning, {\it Soft yet Sharp Interfaces in a Vertex Model of Confluent Tissue}, Phys. Rev. Lett. {\bf 120}, 058001 (2018).

\bibitem{ClaussenVertex} N. H. Claussen, F. Brauns, B. I. Shraiman, {\it A geometric-tension-dynamics model of epithelial convergent extension}, Proc. Natl. Acad. Sci. U.S.A. {\bf 121} (40), e2321928121 (2024). 










\bibitem{MFTreview} L. Bertini, A. De Sole, D. Gabrielli, G. Jona-Lasinio, C. Landim, {\it Macroscopic fluctuation theory}, Rev. Mod. Phys. {\bf 87}, 593 (2015).


\bibitem{EntropyAgranov} T. Agranov, M. E. Cates, R. L. Jack, {\it Entropy production and its large deviations in an active lattice gas}, J. Stat. Mech. (2022) 123201.


\bibitem{Dandekar2020}
R. Dandekar, S. Chakraborti, R. Rajesh, {\it Hard core run and tumble particles on a one-dimensional lattice}, Phys. Rev. E {\bf 102}, 062111 (2020).






\bibitem{Samanta2016} N. Samanta, R. Chakrabarti, {\it Chain reconfiguration in active noise}, J. Phys. A: Math. Theor. {\bf 49}, 195601 (2016).

\bibitem{Chaki2019} S. Chaki, R. Chakrabarti, {\it Enhanced diffusion, swelling, and slow reconfiguration of a single chain in non-Gaussian active bath}, J. Chem. Phys. {\bf 150}, 094902 (2019).



\bibitem{Lizana2010} L. Lizana, T. Ambjörnsso, A. Taloni, E. Barkai, M. A. Lomholt, {\it Foundation of fractional Langevin equation: Harmonization of a many-body problem}, Phys. Rev. E {\bf 81}, 051118 (2010).


\bibitem{Galanti2013} M. Galanti, D. Fanelli, F. Piazza, {\it Persistent random walk with exclusion}, Eur. Phys. J. B {\bf 86}, 456 (2013).

\bibitem{Teomy1} E. Teomy, R. Metzler, {\it Transport in exclusion processes with one-step memory: density dependence and optimal acceleration}, J. Phys. A: Math. Theor. {\bf 52}, 385001 (2019).

\bibitem{Teomy2} E. Teomy, R. Metzler, {\it Correlations and transport in exclusion processes with general finite memory}, J. Stat. Mech. (2019) 103211.

\bibitem{Dolai2020} P. Dolai, A. Das, A. Kundu, C. Dasgupta, A. Dhar, K. V. Kumar, {\it Universal scaling in active single-file dynamics}, Soft Matter {\bf 16},
7077 (2020).

\bibitem{Banerjee2022} T. Banerjee, R. L. Jack, M. E. Cates, {\it Tracer dynamics in one dimensional gases of active or passive particles}, J. Stat. Mech. (2022), 013209.


\bibitem{Harris65} T. E. Harris, {\it Diffusion with “collisions” between particles}, J. Appl. Probab. {\bf 2}, 323 (1965).

\bibitem{Arratia83} R. Arratia, {\it The motion of a tagged particle in the simple symmetric exclusion system on Z}, Ann. Probab. {\bf 11}, 362–373 (1983).

\bibitem{SingleFileMajumdar1991} 
S. N. Majumdar, M. Barma, {\it Tag diffusion in driven systems, growing interfaces, and anomalous fluctuations}, Phys. Rev. B {\bf 44}, 5306 (1991).

\bibitem{SingleFileKrapivsky2014} P. L. Krapivsky, K. Mallick, T. Sadhu, {\it Large deviations in single-file diffusion}, Phys. Rev. Lett. {\bf 113}, 078101 (2014).

\bibitem{SingleFileKrapivsky2015} 
P. L. Krapivsky, K. Mallick, T. Sadhu, {\it Dynamical properties of single-file diffusion}, J. Stat. Mech. (2015) P09007.

\bibitem{TaggedSFD2015}
P. L. Krapivsky, K. Mallick, T. Sadhu, {\it Tagged Particle in Single-File Diffusion}, J. Stat. Phys. {\bf 160}, 885–925 (2015).

\bibitem{SingleFileReview} O. Bénichou, P. Illien, G. Oshanin, A. Sarracino, R. Voituriez, {\it Tracer diffusion in crowded narrow channels}, J. Phys.: Condens. Matter {\bf 30}, 443001 (2018).



\bibitem{HarmonicBasu}
I. Santra, U. Basu, {\it Activity driven transport in harmonic chains}, SciPost Physics {\bf 13}(2), 041 (2022).

\bibitem{CugliandoloAging} L. F. Cugliandolo, J. Kurchan, G. Parisi, {\it Off equilibrium dynamics and aging in unfrustrated systems}, J. Phys. I France {\bf 4}, 1641-1656 (1994).


\bibitem{Pad90} T. Padmanabhan, {\it Statistical mechanics of gravitating systems}, Physics Reports {\bf 188} (5), 285 (1990).

\bibitem{Chavanis} P.H. Chavanis, {\it Phase transitions in self-gravitating systems}, International Journal of Modern Physics B {\bf 20}, 3113 (2006).

\bibitem{Plasma} Y. Elskens, D. Escande, {\it Microscopic Dynamics of Plasmas and Chaos}, IOP Publishing, Bristol (2002).

\bibitem{Miller90} J. Miller, {\it Statistical mechanics of Euler equations in two dimensions}, Phys. Rev. Lett. {\bf 65}, 2137 (1990).





\bibitem{DauxoisPhysRep2009} 
A. Campa, T. Dauxois, S. Ruffo, {\it Statistical mechanics and dynamics of solvable models with long-range interactions}, Phys. Rep. {\bf 480}, 57 (2009).

\bibitem{Dauxois_book}
A. Campa, Th. Dauxois, D. Fanelli, S. Ruffo, {\it Physics of long-range interacting systems}, Oxford University Press, Oxford, (2014).


\bibitem{GB2012} E. Sandier, S. Serfaty, {\it From the Ginzburg-Landau model to vortex lattice problems}, Commun. Math. Phys. {\bf 313}, 635–743 (2012).

\bibitem{Cohn2017} H. Cohn, {\it A conceptual breakthrough in sphere packing}, Notices Amer. Math. Soc. {\bf 64}, 102–115 (2017).

\bibitem{Cohn2022} H. Cohn, A. Kumar, S. D. Miller, D. Radchenko, M. Viazovska, {\it Universal optimality of the $E_8$ and Leech lattices and interpolation formulas}, Ann. of Math. (2) {\bf 196}(3), 983-1082 (2022).

\bibitem{Petrache2020} M. Petrache, S. Serfaty, {\it Crystallization for Coulomb
and Riesz interactions as a consequence of the Cohn-Kumar conjecture}, Proc. Amer. Math. Soc. {\bf 148}, 3047–3057 (2020).



\bibitem{leble2017}  
T. Lebl{\'e}, S. Serfaty, {\it Large deviation principle for empirical fields of Log and Riesz gases}, Invent. Math. {\bf 210}, 645 (2017).

\bibitem{leble2018}
D. P. Hardin, T. Lebl{\'e}, E. B. Saff, S. Serfaty, {\it Large deviation principles for hypersingular Riesz gases}, Constr. Approx. {\bf 48}, 61 (2018).

\bibitem{riesz3}
S. Agarwal, A. Dhar, M. Kulkarni, A. Kundu, S. N. Majumdar, D. Mukamel, G. Schehr, {\it Harmonically confined particles with long-range repulsive interactions}, Phys. Rev. Lett. {\bf 123}, 100603 (2019).

\bibitem{jit2021}
J. Kethepalli, M. Kulkarni, A. Kundu, S. N. Majumdar, D. Mukamel, G. Schehr, {\it Harmonically confined long-ranged interacting gas in the presence of a hard wall }, J. Stat. Mech. (2021) 103209. 


\bibitem{leble_loggas}
M. Erbar, M. Huesmann, T. Lebl\'e, {\it The One‐Dimensional Log‐Gas Free Energy Has a Unique Minimizer}, Commun. Pur. Appl. Math. {\bf 74}, 615 (2021).





\bibitem{jit2022}  
J. Kethepalli, M. Kulkarni, A. Kundu, S. N. Majumdar, D. Mukamel, G. Schehr, {\it Edge fluctuations and third order phase transition in harmonically confined long-range systems}, J. Stat. Mech. (2022) 033203. 


\bibitem{santra2022}   
S. Santra, J. Kethepalli, S. Agarwal, A. Dhar, M. Kulkarni, A. Kundu, {\it Gap statistics for confined particles with power-law interactions}, Phys. Rev. Lett. {\bf 128}, 170603 (2022).

\bibitem{dereudre2023number}
D. Dereudre, T. Vasseur, {\it Number-rigidity and $\beta$-circular Riesz gas}, Ann. Probab. {\bf 51}, 1025, (2023).


\bibitem{Lelotte2023} R. Lelotte, {\em Phase transitions in one-dimensional Riesz gases with long-range interaction}, arXiv:2309.08951 (2023).

\bibitem{Beenakker_riesz}
C. W. J. Beenakker, {\it Pair correlation function of the one-dimensional Riesz gas}, Phys. Rev. Res. {\bf 5}, 013152 (2023).


\bibitem{Riesz_FCS}
J. Kethepalli, M. Kulkarni, A. Kundu, S. N. Majumdar, D. Mukamel, G. Schehr, {\it Full counting statistics of 1d short-range Riesz gases in confinement}, J. Stat. Mech. (2024) 083206.


\bibitem{UsRieszCumulants} 
P. Le Doussal, G. Schehr, {\it Cumulants and large deviations for the linear statistics of the one-dimensional trapped Riesz gas}, J. Stat. Phys. {\bf 192}, 47 (2025). 



\bibitem{Huse_riesz}
D. A. Huse, M. Kulkarni, {\it Spatiotemporal spread of perturbations in power-law models at low temperatures: Exact results for classical out-of-time-order correlators}, Phys. Rev. E {\bf 104}, 044117 (2021).


\bibitem{SerfatyDynamics2022} 
Q. H. Nguyen, M. Rosenzweig, S. Serfaty, {\it Mean-field limits of Riesz-type singular flows}, Ars Inven. Anal. No. {\bf 4}, 45 (2022).


\bibitem{SerfatyDynamics2023} 
M. Rosenzweig, S. Serfaty, {\it Global-in-time mean-field convergence for singular Riesz-type diffusive flows}, Ann. Appl.
Probab. {\bf 33}, 954 (2023).


\bibitem{Riesz_expansion} P. L. Krapivsky, K. Mallick, {\it Expansion into the vacuum of stochastic gases with long-range interactions}, Phys. Rev. E {\bf 111}, 064109 (2025). 







\bibitem{Spohn2}
H. Spohn, {\it Interacting Brownian particles: A study of Dyson’s model}, in Hydrodynamic Behavior and Interacting Particle Systems, edited by G. Papanicolaou (Springer, New York, 1987).

\bibitem{TristanThese} T. Gauti\'e, {\it Stochastic and quantum dynamics of repulsive particles : from random matrix theory to trapped fermions}, PhD thesis, Universit\'e Paris sciences et lettres (2021).


\bibitem{Wigner1} E. P. Wigner, {\it On the statistical distribution of the widths and spacings of nuclear resonance levels}, Mathematical Proceedings of the Cambridge Philosophical Society {\bf 47}, 790–798 (1951).


\bibitem{OxfordRMT} G. Akemann, J. Baik, P. Di Francesco, {\it The Oxford handbook of
random matrix theory}, Oxford University Press (2011).


\bibitem{Erdos}
L. Erd\"os, {\it Universality of Wigner random matrices: a survey of recent results}, Russian Mathematical Surveys, {\bf 66}(3):507 (2011).

\bibitem{ErdosYau} 
L. Erd\"os, HT. Yau, J. Yin, {\it Bulk universality for generalized Wigner matrices}, Probability Theory and Related Fields, {\bf 154}(1-2):341–407 (2012).

\bibitem{Pastur}
L.A. Pastur, M. Shcherbina, {\it Eigenvalue distribution of large random matrices} {\bf 171}, American Mathematical Soc. (2011).

\bibitem{Soshnikov} 
A. Soshnikov, {\it Level spacings distribution for large random matrices: Gaussian fluctuations}, Annals of mathematics, pages 573–617 (1998).

\bibitem{Sosh99}
A. Soshnikov, {\it Universality at the edge of the spectrum in Wigner random matrices}, Commun. Math. Phys. {\bf 207}, 697 (1999).

\bibitem{Tao_book}
T. Tao, {\it Topics in random matrix theory}, Vol. {\bf 132}, American Mathematical Society, (2023).



\bibitem{GeneralBeta} I. Dumitriu, A. Edelman, {\it Matrix models for beta ensembles}, J. Math. Phys. {\bf 43}, 5830 (2002).

\bibitem{BouchaudGuionnet}
R. Allez, J.P. Bouchaud, A. Guionnet, {\it Invariant $\beta$-ensembles and the Gauss-Wigner crossover}, Phys. Rev. Lett. {\bf 109}, 094102 (2012).

\bibitem{allez_satya}
R. Allez, J. P. Bouchaud, S. N. Majumdar, P. Vivo, {\it Invariant $\beta$-Wishart ensembles, crossover densities and asymptotic corrections to the Marčenko–Pastur law}, J. Phys. A: Math. Theor., {\bf 46}(1), 015001 (2012).


\bibitem{WignerSC} E. P. Wigner, {\it On the Distribution of the Roots of Certain Symmetric Matrices}, The Annals of Mathematics {\bf 67}, 325 (1958).




\bibitem{RogersShi} 
L.C.G. Rogers, Z. Shi, {\it Interacting Brownian particles and the Wigner law}, Prob. Th. Rel. Fields {\bf 95}, 555-570 (1993).

\bibitem{HermiteZeros}
P. Forrester, J. Rogers. {\it Electrostatics and the zeros of the classical polynomials}, SIAM Journal on Mathematical Analysis, {\bf 17}(2):461–468, 1986. 




\bibitem{cuenca}
F. Benaych-Georges, C. Cuenca, V. Gorin, {\it Matrix addition and the Dunkl transform at high temperature}, Commun. Math. Phys. {\bf 394}, 735–795 (2022).




\bibitem{Gustavsson} J. Gustavsson, {\it Gaussian fluctuations of eigenvalues in the GUE}, In Annales de l'IHP Probabilités et statistiques {\bf 41} (2), 151-178 (2005).

\bibitem{ORourke2010}
Sean O’Rourke, {\it Gaussian Fluctuations of Eigenvalues in Wigner Random Matrices}, J. Stat. Phys. {\bf 138}, 1045–1066 (2010).



\bibitem{TracyWidom1} C. A. Tracy, H. Widom, {\it Level-spacing distributions and the Airy kernel}, Communications in Mathematical Physics {\bf 159}(1), 151–174 (1994).

\bibitem{TracyWidom2} C. A. Tracy, H. Widom, {\it On orthogonal and symplectic ensembles}, Comm. Math. Phys. {\bf 177}, 727–754 (1996).


\bibitem{RamirezRiderVirag}
J. Ramirez, B. Rider, B. Virag, {\it Beta ensembles, stochastic Airy spectrum and a diffusion}, J. Amer. Math. Soc. {\bf 24} 919-944 (2011).


\bibitem{SatyaTracyWidomLecture} S. Majumdar, {\it Random matrices, the Ulam problem, directed polymers \& growth models, and sequence matching}, In Les Houches Lecture Notes “Complex Systems”, volume {\bf 85}, 179-216, Elsevier Science (2007).


\bibitem{Hermite_asymptotics}
\url{https://dlmf.nist.gov/18.16}


\bibitem{GorinInfiniteBeta}
V. Gorin, V. Kleptsyn, {\it Universal objects of the infinite beta random matrix theory}, . Eur. Math. Soc. {\bf 26}, 3429–3496 (2024).



\bibitem{WignerSurmise} E. P. Wigner, {\it Statistical properties of real symmetric matrices with many dimensions}, Princeton University (1957).



\bibitem{Bourgade2022} P. Bourgade, K. Mody, M. Pain, {\it Optimal local law and central limit theorem for $\beta$-ensembles}, Commun. Math. Phys. {\bf 390}, 1017 (2022).

\bibitem{Dyson62} F. J. Dyson, {\it Statistical Theory of the Energy Levels of Complex Systems}, J. Math. Phys. {\bf 3}, 140 (1962); {\bf 3}, 157 (1962); {\bf 3}, 166 (1962).

\bibitem{Dyson63} F. J. Dyson, M. L. Mehta, {\it Statistical Theory of the Energy Levels of Complex Systems. IV}, J. Math. Phys. {\bf 4}, 701 (1963).

%
%
%
%
%
%
%



\bibitem{Calogero71} F. Calogero, {\it Solution of the one-dimensional N-body problem with quadratic and/or inversely quadratic pair potentials}, J. Math. Phys. {\bf 12}:419–436 (1971).

\bibitem{Calogero75}
F. Calogero, {\it Exactly solvable one-dimensional many-body problems}, Lett. Nuovo Cimento {\bf 13}, 411 (1975).

\bibitem{Moser76}
J. Moser, {\it Three integrable Hamiltonian systems connected with isospectral deformations}, Adv. Math.
{\bf 16}, 197 (1975).

\bibitem{BGS09}
E. Bogomolny, O. Giraud, C. Schmit, {\it Random matrix ensembles associated with lax matrices}, Phys. Rev. Lett. {\bf 103}, 054103 (2009).

\bibitem{KP17}
M. Kulkarni, A. Polychronakos, {\it Emergence of the Calogero family of models in external potentials: duality, solitons and hydrodynamics}, J. Phys. A {\bf 50}, 455202 (2017).

\bibitem{Poly06}
A. P. Polychronakos, {\it The physics and mathematics of Calogero particles}, J. Phys. A {\bf 39}, 12793 (2006).

\bibitem{OP81}
M. Olshanetsky, A. M. Perelomov, {\it Classical integrable finite-dimensional systems related to Lie algebras}, Phys. Rep. {\bf 71}, 313 (1981).




\bibitem{eigenvectors}
S. Ahmed, M. Bruschi, F. Calogero, M. A. Olshanetsky, A. M. Perelomov, {\it Properties of the zeros of the classical polynomials and of the Bessel functions}, Nuovo Cimento B {\bf 49}, 173 (1979).




\bibitem{Lenard} 
A. Lenard, \emph{Exact statistical mechanics of a one-dimensional system with Coulomb forces},  J. Math. Phys. \textbf{2}, 682 (1961).

\bibitem{Prager}
S. Prager, {\it The One-Dimensional Plasma}, Adv. Chem. Phys. {\bf 4}, 201 (1962).


\bibitem{Baxter1963} 
R. J. Baxter, \emph{Statistical mechanics of a one-dimensional Coulomb system with a uniform charge background}, {Proc. Camb. Phil. Soc. \textbf{59}, 779 (1963)}.

\bibitem{Kunz}
H. Kunz, {\it The one-dimensional classical electron gas}, Ann. Phys. {\bf 85}, 303 (1974).

\bibitem{Aizenman1980}
M. Aizenman, P. A. Martin, {\it Structure of Gibbs states of one dimensional Coulomb systems}, Commun. Math. Phys. {\bf 78}, 99 (1980).

\bibitem{Dean2010}
D. S. Dean, R. R. Horgan, A. Naji, R. Podgornik, {\it Effects of dielectric disorder on van der Waals interactions in slab geometries}, Phys. Rev. E {\bf 81}, 051117 (2010).

\bibitem{Tellez}
G. Tellez, E. Trizac, {\it Screening like charges in one-dimensional Coulomb systems: Exact results}, Phys. Rev. E {\bf 92}, 042134 (2015).

\bibitem{dhar2017exact}
A. Dhar, A. Kundu, S. N. Majumdar, S. Sabhapandit, G. Schehr, {\it Exact extremal statistics in the classical 1d Coulomb gas}, Phys. Rev. Lett. {\bf 119}, 060601 (2017). 

\bibitem{dhar2018extreme}
A. Dhar, A. Kundu, S. N. Majumdar, S. Sabhapandit, G. Schehr,
{\it Extreme statistics and index distribution in the classical 1d Coulomb gas}, J. Phys. A: Math. and Theor. {\bf 51}, 295001 (2018).

\bibitem{Flack21}
A. Flack, S. N. Majumdar, G. Schehr, {\it Truncated linear statistics in the one dimensional one-component plasma},  J. Phys. A: Math. Theor. {\bf 54}, 435002 (2021).

\bibitem{Flack22}
A. Flack, S. N. Majumdar, G. Schehr, {\it Gap probability and full counting statistics in the one-dimensional one-component plasma}, J. Stat. Mech. (2022) 053211.

\bibitem{Chafai_edge}
D. Chafai, D. Garcia-Zelada, P. Jung, {\it At the edge of a one-dimensional jellium}, Bernoulli {\bf 28}, 1784 (2022).

\bibitem{Rybicki}
G. B. Rybicki, {\it Exact statistical mechanics of a one-dimensional self-gravitating system}, Astrophys. Space Sci. {\bf 14}, 56 (1971). 

\bibitem{SireSG2002} P. H Chavanis, C. Rosier, C. Sire, {\it Thermodynamics of self-gravitating systems}, Phys. Rev. E {\bf 66}, 036105 (2002).

\bibitem{SireSG2002bis} P. H. Chavanis, C. Sire, {\it Thermodynamics and collapse of self-gravitating Brownian particles in D dimensions}, Phys. Rev. E {\bf 66}, 046133 (2002).

\bibitem{Sire} P. H. Chavanis, C. Sire, {\it Anomalous diffusion and collapse of self-gravitating Langevin particles in $D$ dimensions}, Phys. Rev. E {\bf 69}, 016116 (2004).

\bibitem{SireSG2004} P. H. Chavanis, C. Sire, {\it Postcollapse dynamics of self-gravitating Brownian particles and bacterial populations}, Phys. Rev. E {\bf 69}, 066109 (2004).

\bibitem{SireSG2008} P. H. Chavanis, C. Sire, {\it Critical dynamics of self-gravitating Langevin particles and bacterial populations}, Phys. Rev. E {\bf 78}, 061111 (2008).

\bibitem{Pitman} 
S. Pal, J. Pitman, {\it One-dimensional Brownian particle systems with rank-dependent drifts}, Ann. Appl. Probab. {\bf 18}, 2179 (2008).

\bibitem{OConnell}
N. O'Connell, J. Ortmann, {\it Product-form invariant measures for Brownian motion with drift satisfying a skew-symmetry type condition}, ALEA, Lat. Am. J. Probab. Math. Stat. {\bf 11}, 307 (2014).

\bibitem{Banner} 
A. D. Banner, R. Fernholz, I. Karatzas, {\it Atlas models of equity markets}, 
Ann. Appl. Probab. {\bf 15}, 2296 (2005).




\bibitem{FlackRD} 
A. Flack, P. Le Doussal, S. N. Majumdar, G. Schehr, {\it Out of equilibrium dynamics of repulsive ranked diffusions: the expanding crystal}, Phys. Rev. E {\bf 107}, 064105 (2023).





\bibitem{SpohnTracer} 
H. Spohn, {\it Tracer dynamics in Dyson’s model of interacting Brownian particles}, J. Stat. Phys. {\bf 47}, 669-679 (1987).

\bibitem{Lepingle07}
E. Cépa, D. L{\'e}pingle, {\it No multiple collisions for mutually repelling Brownian particles}, In Séminaire de Probabilités XL (pp. 241-246). Springer, Berlin, Heidelberg (2007).

\bibitem{Allez13}
R. Allez, A. Guionnet, {\it A diffusive matrix model for invariant $\beta $-ensembles}, Electronic Journal of Probability, {\bf 18}, 1-30 (2013).

\bibitem{Grabsch2025} A. Grabsch, D. Venturelli, O. Bénichou, {\it Exact large-scale correlations in diffusive systems with general interactions: explicit characterisation without the Dean--Kawasaki equation}, arXiv:2504.08560 (2025).

\bibitem{fBM} B. Mandelbrot, J. W. Van Ness, {\it Fractional Brownian motions, fractional noises and applications}, SIAM Review {\bf 10}, 422–437 (1968).









\bibitem{BernardBurgers}
M. Bauer, D. Bernard, {\it Sailing the deep blue sea of decaying Burgers turbulence}, J. Phys. A: Math. Gen. {\bf 32}(28), 5179 (1999).



\bibitem{VitelliNature2021}
M. Fruchart, R. Hanai, P. B. Littlewood, V. Vitelli, {\it Non-reciprocal phase transitions}, Nature {\bf 592}(7854), 363-369 (2021).


\bibitem{nonreciprocal_active_book} A. Haluts, D. Gorbonos, N.S. Gov, {\it Models of Animal Behavior as Active Particle Systems with Nonreciprocal Interactions}, in J.A. Carrillo, E. Tadmor, (eds) Active Particles, Volume 4: Modeling and Simulation in Science, Engineering and Technology, Springer Nature Switzerland, Birkhäuser, Cham (2024).

\bibitem{Kocabas2024} A. Kocabas, S. Ozdemir, M. Basaran, T. Yüce, A. Kecebas,
B. Altin, Y. Yaman, E. Demir, C. Kocabas, {\it Non-reciprocal phase transition enables swarming motility in biological active matter}, preprint available at Research Square [https://doi.org/10.21203/rs.3.rs-3956047/v1] (2024).

\bibitem{Mandal2024} N. S. Mandal, A. Sen, R. D. Astumian, {\it A molecular origin of non-reciprocal interactions between interacting active catalysts}, Chem. {\bf 10} (4), 1147-1159 (2024).


\bibitem{You2020} Z. You, A. Baskaran, M. C. Marchetti, {\it Nonreciprocity as a generic route to traveling states}, Proc. Natl. Acad. Sci. U.S.A. {\bf 117} (33) 19767-19772 (2020). 

\bibitem{nonreciprocalActive} D. Martin, D. Seara, Y. Avni, M. Fruchart, V. Vitelli, {\it The transition to collective motion in nonreciprocal active matter: coarse graining agent-based models into fluctuating hydrodynamics}, arXiv:2307.08251 (2023).


\bibitem{KK2022} K. L. Kreienkamp, S. H. L. Klapp, {\it Clustering and flocking of repulsive chiral active particles with non-reciprocal couplings}, New J. Phys. {\bf 24}, 123009 (2022).

\bibitem{Knezevic2022} M. Knezevic, T. Welker, H. Stark, {\it Collective motion of active particles exhibiting non-reciprocal orientational interactions}, Sci. Rep. {\bf 12}, 19437 (2022).

\bibitem{Dinelli2023} A. Dinelli, J. O’Byrne, A. Curatolo, Y. Zhao, P. Sollich, J. Tailleur, {\it Non-reciprocity across scales in active mixtures}, Nat. Commun. {\bf 14}, 7035 (2023).

\bibitem{Duan2023} Y. Duan, J. Agudo-Canalejo, R. Golestanian, B. Mahault, {\it Dynamical Pattern Formation without Self-Attraction in Quorum-Sensing Active Matter: The Interplay between Nonreciprocity and Motility}, Phys. Rev. Lett. {\bf 131}, 148301 (2023).


\bibitem{Duan2024} Y. Duan, J. Agudo-Canalejo, R. Golestanian, B. Mahault, {\it Phase Coexistence in Nonreciprocal Quorum-Sensing Active Matter}, Phys. Rev. Research {\bf 7}, 013234 (2025). 

\bibitem{Du2024} M. Du, S. Vaikuntanathan, {\it Hidden nonreciprocity as a stabilizing effective potential in active matter}, arXiv:2401.14690 (2024).

\bibitem{Tucci2025} G. Tucci, G. Pisegna, R. Golestanian, S. Saha, {\it Hydrodynamic stresses in a multi-species suspension of active Janus colloids}, arXiv:2502.07744 (2025).


\bibitem{Newman2008} J. P. Newman, H. Sayama, {\it Effect of sensory blind zones on milling behavior in a dynamic self-propelled particle model}, Phys. Rev. E {\bf 78}, 011913 (2008).

\bibitem{Peruani2016} L. Barberis, F. Peruani, {\it Large-Scale Patterns in a Minimal Cognitive Flocking Model: Incidental Leaders, Nematic Patterns, and Aggregates}, Phys. Rev. Lett. {\bf 117}, 248001 (2016).

\bibitem{Peruani2017} F. Peruani, {\it Hydrodynamic Equations for Flocking Models without Velocity Alignment}, J. Phys. Soc. Jpn. {\bf 86}, 101010 (2017).

\bibitem{Durve2018} M. Durve, A. Saha, A. Sayeed, {\it Active particle condensation by non-reciprocal and time-delayed interactions}, Eur. Phys. J. E {\bf 41}, 49 (2018). 

\bibitem{Negi2022} R. Singh Negi, R. G. Winkler, G. Gompper, {\it Emergent collective behavior of active Brownian particles with visual perception}, Soft Matter {\bf 18}, 6167 (2022).

\bibitem{Qi2022} J. Qi, L. Bai, Y. Xiao, Y. Wei, W. Wu,
{\it The emergence of collective obstacle avoidance based on a visual perception mechanism}, Information Sciences {\bf 582}, 850-864 (2022).

\bibitem{Stengele2022} P. Stengele, A. Lüders, P. Nielaba, {\it Group formation and collective motion of colloidal rods with an activity triggered by visual perception}, Phys. Rev. E {\bf 106} (2022).

\bibitem{Saavedra2024} R. Saavedra, F. Peruani, {\it Self-trapping of active particles with non-reciprocal interactions in disordered media}, Phys. Rev. E {\bf 110}, 064602 (2024). 




\bibitem{flajolet2009}
P. Flajolet, R. Sedgewick, {\it Analytic combinatorics}, Cambridge University Press (2009).



\bibitem{Banerjee2020} T. Banerjee, S. N. Majumdar, A. Rosso, G. Schehr, {\it Current fluctuations in noninteracting run-and-tumble particles in one dimension}, Phys. Rev. E {\bf 101}(5), 052101 (2020).

\bibitem{Spohn3}
H. Spohn, {\it Dyson’s model of interacting Brownian motions at arbitrary coupling strength}, Markov Process. Relat. {\bf 4}, 649 (1998).

\bibitem{ForresterCircularBM}
P. J. Forrester, T. Nagao,{\em Correlations for the circular Dyson Brownian motion model with Poisson initial conditions}, Nucl. Phys. B {\bf 532}, 733 (1998).

\bibitem{Smith2021} 
N. Smith, P. Le Doussal, S. N. Majumdar, G. Schehr, {\it Full counting statistics for interacting trapped fermions}, SciPost Physics {\bf 11}(6), 110 (2021).



\bibitem{lindemann}
F. A. Lindemann, {\it The calculation of molecular vibration frequencies}, Phys. Z. {\bf 11}, 609-612 (1910).

\bibitem{navarro}
R. Guardiola, J. Navarro, {\it On the Lindemann Criterion for quantum clusters at very low temperature}, J. Phys. Chem. A, {\bf 115}, 6843 (2011).


\bibitem{BBridge} S. N. Majumdar, A. Comtet, {\it Airy Distribution Function: From the Area Under a Brownian Excursion to the Maximal Height of Fluctuating Interfaces}, J. Stat. Phys. {\bf 119}, 777–826 (2005).

\bibitem{HughesCircle2001}
C. P. Hughes, J. Keating, N. O'Connell, {\it On the Characteristic Polynomial of a Random Unitary Matrix}, Commun. Math. Phys. {\bf 220}, 429 (2001).

\bibitem{NajnudelCircle2018}
R. Chhaibi, T. Madaule, J. Najnudel, {\it On the maximum of the C$\beta$E field}, Duke Math. J. {\bf 167}, 2243 (2018).



\bibitem{Haldane}
F. D. M. Haldane, {\em Effective harmonic-fluid approach to low-energy properties of one-dimensional quantum fluids}, Phys. Rev. Lett. {\bf 47}, 1840 (1981).

\bibitem{Forrester1984}
P. J. Forrester, {\em Analogues between a quantum many body problem and the log-gas}, J. Phys. A: Math. Gen. {\bf 17}, 2059  (1984).








\bibitem{ErdosCorrelations} 
L. Erd\"os, H.-T. Yau, {\em Gap universality of generalized Wigner and beta-ensembles}, J. Eur. Math. Soc. {\bf 17}, 1927 (2015).





\bibitem{jepsen}
D. W. Jepsen, {\it Dynamics of a simple many‐body system of hard rods}, J. Math. Phys. {\bf 6}, 405 (1965).

\bibitem{Bena2007} 
I. Bena, S. N. Majumdar, {\it Universal Extremal Statistics in a Freely Expanding Jepsen Gas}, Phys. Rev. E {\bf 75}, 051103 (2007).




















\bibitem{ChebyWiki}
\url{https://en.wikipedia.org/wiki/Chebyshev_polynomials}



\bibitem{ForresterHermiteEdge}
P. J. Forrester, N. E. Frankel, T. M. Garoni, {\it Asymptotic form of the density profile for Gaussian and Laguerre random matrix ensembles with orthogonal and symplectic symmetry}, J. Math. Phys. {\bf 47}, 023301 (2006)

















\bibitem{hittingProbaAnomalous} S. N. Majumdar, A. Rosso, A. Zoia, {\em Hitting probability for anomalous diffusion processes}, Phys. Rev. Lett. {\bf 104}, 020602 (2010).


\bibitem{Levy} P. L\'evy, {\em Processus stochastiques et mouvement brownien}, Gauthier-Villars, Paris (1948).

\bibitem{Lindley} D. Lindley, {\em The theory of queues with a single server}, Mathematical Proceedings of the Cambridge Philosophical Society {\bf 48}(2), 277-289 (1952).


\bibitem{SiegmundDualityClifford} P. Clifford, A. Sudbury, {\em A Sample Path Proof of the Duality for Stochastically Monotone Markov Processes}, Ann. Probab. {\bf13}(2): 558-565 (1985).

\bibitem{KolokoltsovDuality} V. N. Kolokoltsov, {\em Stochastic monotonicity and duality for one-dimensional Markov processes}, Mathematical Notes {\bf 89}, 652–660 (2011).

\bibitem{partiallyOrdered} P. Lorek, {\em Siegmund duality for Markov Chains on partially ordered state spaces}, Probability in the Engineering and Informational Sciences, {\bf 32}(4), 495-521 (2018).

\bibitem{DualityZhao} P. Zhao, {\em Siegmund Duality for Continuous Time Markov Chains on $\mathbb{Z}_+^d$}, Acta Mathematica Sinica. English Series; Heidelberg {\bf 34}, 9: 1460-1472 (2018). 




\bibitem{WidomExitprobaLevyflight} H. Widom  {\em Stable processes and integral equation}, Trans. Amer. Math. Soc. {\bf 98}, 430 (1961).

\bibitem{Levy3} R. M. Blumenthal, R. K. Getoor, D. B. Ray, {\em On the Distribution of First Hits for the Symmetric Stable Processes}, Trans. Am. Math. Soc. {\bf 99}, 540 (1961).


\bibitem{DenisovLevy} S. I. Denisov, W. Horsthemke, P. Hänggi {\em Steady-state L\'evy flights in a confined domain}, Phys. Rev. E, {\bf77}, 061112 (2008).



\bibitem{AsmussenDiscrete} S. Asmussen, K. Sigman, {\em Monotone Stochastic Recursions and their Duals}, Probability in the Engineering and Informational Sciences {\bf 10}(1), 1-20 (1996).

\bibitem{SigmanContinuous} K. Sigman, R. Ryan, {\em Continuous-time monotone stochastic recursions and duality}, Advances in Applied Probability {\bf 32}(2), 426-445 (2000).


\bibitem{DualityMultidimensions} B. Blaszczyszyn, K. Sigman, {\em Risk and duality in multidimensions}, Stochastic Processes and their Applications {\bf 83} (2), 331-356 (1999).


\bibitem{JansenDualityReview} S. Jansen, N. Kurt, {\em On the notion(s) of duality for Markov processes}, Probab. Surveys {\bf 11}: 59-120 (2014).

\bibitem{DualityGenerators} V. Kolokoltsov, R. Lee, {\em Stochastic duality of Markov processes: a study via generators}, Stochastic Analysis and Applications {\bf 31}(6), 992–1023 (2013). 

\bibitem{CoxEntranceExitLaws} J. T. Cox, U. Rösler, {\em A duality relation for entrance and exit laws for Markov processes}, Stochastic Processes and their Applications {\bf 16} 2, 141-156 (1984).

\bibitem{PathwiseDualSturm} A. Sturm, J. M. Swart, {\em Pathwise Duals of Monotone and Additive Markov Processes}, Journal of Theoretical Probability {\bf 31}, 932–983 (2018).



\bibitem{Kolotsovkthorder} V. Kolokoltsov, {\em Stochastic Monotonicity and Duality of kth Order with Application to Put-Call Symmetry of Powered Options}, Journal of Applied Probability, {\bf 52}(1), 82-101 (2015).

\bibitem{DualityLevyProcessesGoffard} P.-O. Goffard, A. Sarantsev, {\em Exponential convergence rate of ruin probabilities for level-dependent Lévy-driven risk processes}, Journal of Applied Probability, {\bf 56}(4), 1244-1268 (2019).

\bibitem{MohleDualityGenetics} M. Möhle, {\em The concept of duality and applications to Markov processes arising in neutral population genetics models}, Bernoulli {\bf 5}(5): 761-777 (1999).

\bibitem{DualityBranchingFoucart} C. Foucart, {\em Local explosions and extinction in continuous-state branching processes with logistic competition}, arXiv:2111.06147 (2021).

\bibitem{LiggettInteractingParticles} T. M. Liggett, {\em Interacting Particle Systems}, Springer-Verlag, Berlin (1985).

\bibitem{DualityBoundaryDriven} G. Carinci, S. Floreani, C. Giardin{\`a}, F. Redig, {\em Boundary driven Markov gas: duality and scaling limits}, Ensaios Matem{\'a}ticos (2021).




\bibitem{Comtet2011} A. Comtet, Y. Tourigny, {\em Excursions of diffusion processes and continued fractions}, Annales de l'Institut Henri Poincar\'e -- Probabilit\'es et Statistiques, {\bf 47}, 850-874 (2011).

\bibitem{Comtet2020} A. Comtet, F. Cornu, G. Schehr, {\em Last-Passage Time for Linear Diffusions and Application to the Emptying Time of a Box}, J. Stat. Phys. {\bf 181}, 1565-1602 (2020).

\bibitem{ThibautDual} T. Arnoulx de Pirey, {\em Extreme value statistics of non-Markovian processes from a new class of integrable nonlinear differential equations}, arXiv:2402.05091 (2024).



\bibitem{Szabo} A. Szabo, G. Lamm, G. H. Weiss, {\it Localized Partial Traps in Diffusion Processes and Random Walks}, J. Stat. Phys. {\bf 34}, 225 (1984).

\bibitem{Spouge} J.L. Spouge, A. Szabo, G.H. Weiss, {\it Single-particle survival in gated trapping}, Phys. Rev. E, {\bf 54}(3), 2248 (1996).


\bibitem{Scher1} Y. Scher, S. Reuveni, {\it Unified approach to gated reactions on
networks}, Phys. Rev. Lett. {\bf 127}(1), 018301 (2021).

\bibitem{Scher2} Y. Scher, A. Kumar, M.S. Santhanam, S. Reuveni, {\it Continuous gated first-passage processes}, Rep. Prog. Phys. {\bf 87}, 108101 (2024).

\bibitem{Guerin} T. Guérin, M. Dolgushev, O. Bénichou, R. Voituriez, {\it Universal kinetics of imperfect reactions in confinement}, Communications chemistry {\bf 4}(1), 157 (2021).


\bibitem{defect1} Elliott W. Montroll, Renfrey B. Potts, {\it Effect of Defects on Lattice Vibrations}, Phys. Rev. {\bf 100}, 525 (1955).

\bibitem{defect2} V. M. Kenkre, {\it Memory Functions, Projection Operators, and the Defect Technique: Some Tools of the Trade for the Condensed Matter Physicist}, Springer Nature (2021).

\bibitem{defect3} T. Kay, T.J. McKetterick, L. Giuggioli, {\it The defect technique for partially absorbing and reflecting boundaries: application to the
Ornstein–Uhlenbeck process}, International Journal of Modern Physics B, {\bf 36}(07-08), 2240011 (2022).



\bibitem{riskenBook} H. Risken, {\em The Fokker-Planck equation: Methods of solution and applications}, Springer Series in Synergetics, Springer Berlin, Heidelberg (1996).

\bibitem{handbookSM} C. W. Gardiner, {\em Handbook of stochastic methods}, Berlin: Springer (1985).




\bibitem{DDsoftmatter1} T. J. Lampo, S. Stylianidou, M. P. Backlund, P. A. Wiggins, A. J. Spakowitz, {\em Cytoplasmic RNA-protein particles exhibit non-Gaussian subdiffusive behavior}, Biophysical journal {\bf 112} (3), 532-542 (2017). 

\bibitem{DDsoftmatter2} G. Kwon, B. J. Sung, A. Yethiraj, {\em Dynamics in crowded environments: is non-Gaussian Brownian diffusion normal?}, J. Phys. Chem. B, {\bf 118} (28), 8128-8134 (2014). 

\bibitem{DDsoftmatter3} B. Wang, J. Kuo, S. C. Bae, S. Granick, {\em  When Brownian diffusion is not Gaussian}, Nature materials {\bf 11} 6, 481-485 (2012). 




\bibitem{CTRW1} E. W. Montroll, G. H. Weiss, {\it Random walks on lattices. II.}, J. Math. Phys. {\bf 6}(2), 167-181 (1965).

\bibitem{CTRW2} H. Scher, M. Lax, {\it Stochastic transport in a disordered solid. I. Theory} Phys. Rev. B  {\bf 7}, 10 4491 (1973).

\bibitem{CTRWA} J. P. Bouchaud, A. Georges, {\em Anomalous diffusion in disordered media: Statistical mechanisms, models and physical applications}, Physics Reports, {\bf 195}, 127-293 (1990).

\bibitem{CTRWB} R. Metzler, J. Klafter, {\em The random walk's guide to anomalous diffusion: a fractional dynamics approach}, Physics Reports, {\bf339}, 1-77 (2000).

\bibitem{CTRWC} R. Kutner, J. Masoliver, {\em The continuous time random walk, still trendy: fifty-year history, state of art and outlook}, Eur. Phys. J. B {\bf90}, 50 (2017).







\bibitem{resettingRTP} M. R. Evans, S. N. Majumdar, {\em Run and tumble particle under resetting: a renewal approach}, J. Phys. A: Math. Theor. {\bf 51}, 475003 (2018). 

\bibitem{resettingRTP2} G. Tucci, A. Gambassi, S. N. Majumdar, G. Schehr, {\em First-passage time of run-and-tumble particles with noninstantaneous resetting}, Phys. Rev. E, {\bf 106}(4), 044127 (2022).

\bibitem{resettingRTP2D} I. Santra, U. Basu, S. Sabhapandit, {\it Run-and-tumble particles in two dimensions under stochastic resetting conditions}, J. Stat. Mech. (2020) 113206.


\bibitem{Besga20} B. Besga, A. Bovon, A Petrosyan, S. N. Majumdar, S. Ciliberto, {\em Optimal mean 
first-passage time for a Brownian searcher subjected to resetting: experimental and theoretical results},   
Phys. Rev. Research, {\bf 2}, 032029 (R) (2020). 

\bibitem{IntermittentReset} G. Mercado-V\`asquez, D. Boyer, S. N. Majumdar, G. Schehr  {\em Intermittent resetting potentials},  J. Stat. Mech. (2020) 113203.

\bibitem{Faisant21} F. Faisant, B. Besga, A. Petrosyan, S. Ciliberto, S. N. Majumdar, {\em Optimal mean first-passage time of a Brownian searcher with resetting in one and two dimensions: 
experiments, theory and numerical tests},
J. Stat. Mech. (2021) 113203.


\bibitem{randomwalkresetPRL} L. Kusmierz, S. N. Majumdar, S. Sabhapandit, G. Schehr, {\em First order transition for the optimal search time of Lévy flights with resetting}, Phys. Rev. Lett. {\bf 113} (22), 220602 (2014). 


\bibitem{resettingInInterval} A. Pal, V. V. Prasad, {\em First passage under stochastic resetting in an interval}, Phys. Rev. E {\bf 99}, 032123 (2019).








\bibitem{singletraj} V. Sposini, R. Metzler, G. Oshanin, {\em Single-trajectory spectral analysis of scaled Brownian motion}, New J. Phys. {\bf 21}, 073043  (2019).




\bibitem{Nparticleduality} T. Assiotis, N. O'Connel, J. Warren, {\em Interlacing Diffusions}, Séminaire de Probabilités L, 301-380 (2016).







\end{thebibliography}
\end{document}